\documentclass[a4paper,12pt,twoside,openright]{report} % anche book

\usepackage{slashed}
\usepackage{axodraw4j}
\usepackage{pstricks}
\usepackage{color}
\usepackage[T1]{fontenc}
\usepackage[latin1]{inputenc}
\usepackage[english]{babel}
\usepackage{epsfig}
\usepackage{fancyvrb}
\usepackage{indentfirst}
\usepackage[]{graphicx}
\usepackage{fancyhdr}
\usepackage{pstricks}
\usepackage{amssymb}  %   MATH
\usepackage{amsmath}  %   MATH
\usepackage{latexsym} %   MATH
\usepackage{amsthm}   %   MATH
\usepackage[retainorgcmds]{IEEEtrantools}
 \usepackage{dsfont}   
 \usepackage{braket}
 \usepackage[toc,page]{appendix}

\begin{document}        

\pagestyle{fancy}
\renewcommand{\chaptermark}[1]{\markboth{#1}{}}
\renewcommand{\sectionmark}[1]{\markright{\thesection\ #1}}
\fancyhf{}
\fancyhead[LE,RO]{\bfseries\thepage}
\fancyhead[LO]{\bfseries\rightmark}
\fancyhead[RE]{\bfseries\leftmark}
\renewcommand{\headrulewidth}{0.5pt}
\renewcommand{\footrulewidth}{0pt}
\addtolength{\headheight}{0.5pt}
\fancypagestyle{plain}{\fancyhead{} \renewcommand{\headrulewidth}{0pt}}

%Frontespizio
\thispagestyle{empty}
\begin{titlepage}
\begin{figure}[h]
\begin{center}
\includegraphics[width=5.0cm]{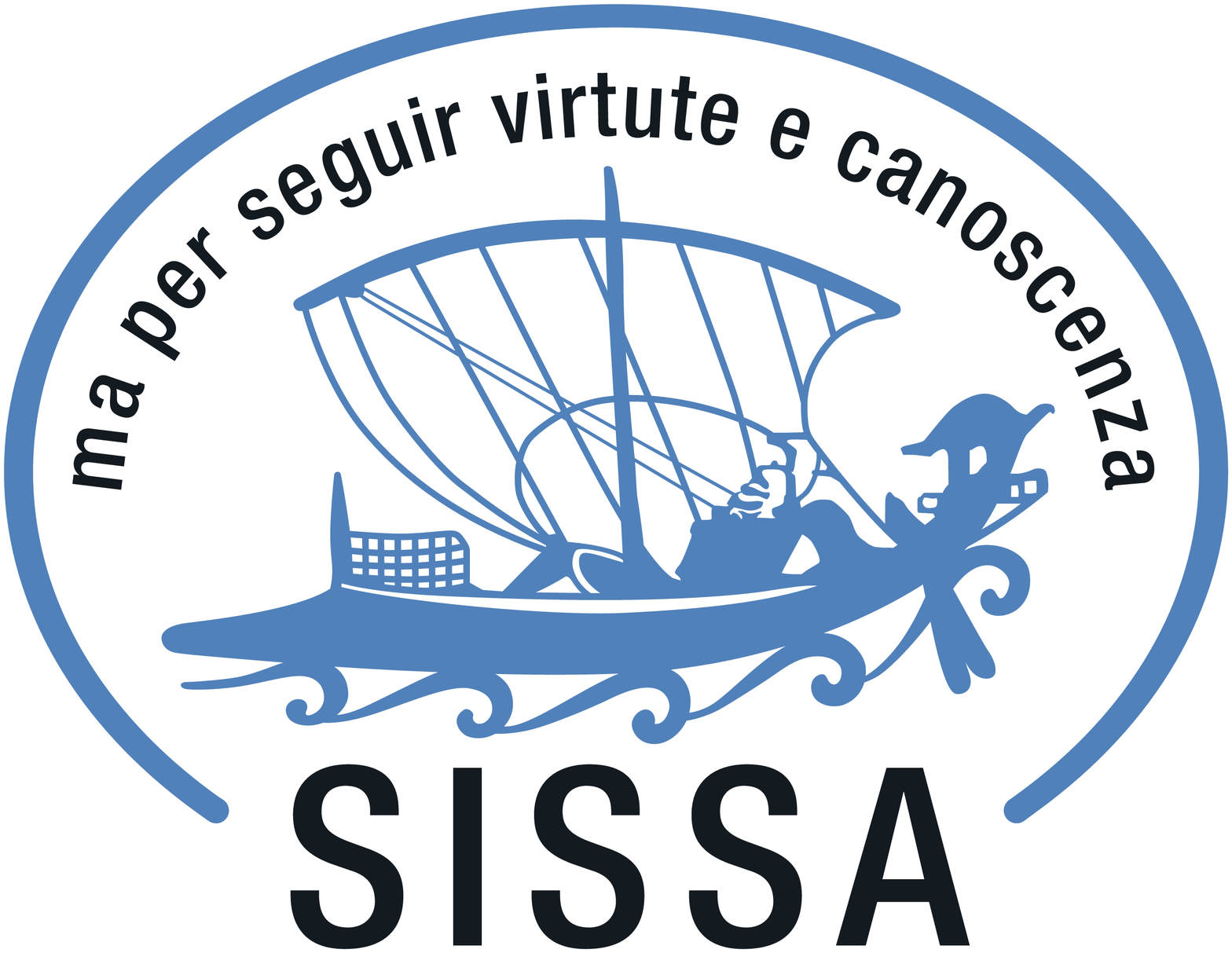}
\end{center}
\end{figure}

\begin{center}
\Large
SISSA-ISAS \\
Scuola Internazionale Superiore di Studi Avanzati\\
International School for Advanced Studies\\
\vspace{0.2cm}
\large 
- Astroparticle physics curriculum -
\end{center}

\vspace{1.2cm}

\begin{center}
\textbf{{\LARGE Ab initio calculations on nuclear matter}}

\vspace{0.4cm}

\textbf{{\LARGE properties including the effects }}

\vspace{0.2cm}

\textbf{{\LARGE of three-nucleons interaction }}

\end{center}

\vspace{0.8cm}

\begin{center}
\Large
Thesis submitted for the degree of\\
\textbf{\Large Doctor Philosophie}
\end{center}

\vspace{1.8cm}

\begin{tabular}{lcl}
\textbf{\large Supervisors} & \phantom{aaaaaaaaaaaaaaaaaaaaaaaaaaaaa} & \textbf{\large Candidate:} \\
{\large Prof. Stefano Fantoni} & \phantom{aaaaaaaaaaaaaaaaaaaaaaaaaaaaa} & {\large Alessandro Lovato}\\
{\large Prof. Omar Benhar} & \phantom{a}
\end{tabular}

\vspace{0.8cm}

\begin{center}
Trieste, 20 September 2012
\end{center}

\vspace*{1.5 cm}

\begin{titlepage}
\thispagestyle{empty}
\topmargin=1.0cm
\raggedleft
\small %\sc{
[Uliva:] <<Malgrado tutto, ho una certa stima di te. Da molti anni ti vedo impegnato in una specie di vertenza cavalleresca con la vita, o, se preferisci, col creatore: la lotta della creatura per superare i suoi limiti. Tutto questo, lo dico senza ironia, \`e nobile; ma richiede un'ingenuit\`a che a me manca>>.

<<L'uomo non esiste veramente che nella lotta contro i propri limiti>>, disse Pietro.\\
\vspace{2cm}

\begin{flushright}
\it Ignazio Silone (Vino e pane, 1938;1955)
\end{flushright}

\clearpage{\pagestyle{empty}\cleardoublepage}
\end{titlepage}
\end{titlepage}

\thispagestyle{empty}
\cleardoublepage
\pagenumbering{roman}

% Indice
\tableofcontents 
\newpage
\pagenumbering{arabic}

% Introduzione

\addcontentsline{toc}{chapter}{Introduction}
\pagenumbering{arabic}
\chapter*{Introduction}
\markboth{Introduction}{}
\markright{Introduction}{}
{\it Ab initio} nuclear many-body approaches are based on the premise that the dynamics can be modeled studying exactly solvable few-body systems. This is a most important feature since, due to the complexity of strong interactions and to the prohibitive difficulties associated with the solution of the quantum mechanical many-body problem, theoretical calculations of nuclear observables generally involve a number of approximations. Hence, models of nuclear dynamics extracted from analyses of the properties of complex nuclei are plagued by the systematic uncertainty associated with the use of a specific approximation scheme.

Highly realistic two-nucleon potentials, either purely phenomenological \cite{lacombe_80,stoks_94,wiringa_95,machleidt_01} or  based on chiral perturbation theory (ChPT) \cite{entem_03,epelbaum_05}, have been obtained from accurate fits of the properties of the bound and scattering states of the two-nucleon system \cite{stoks_93,bergervoet_90,vanderleun_82, ericson_83,rodning_90,simon_81,bishop_79}. Unfortunately, however, the extension to the case of the three-nucleon potential, the inclusion of which is needed to account 
for the properties of the three-nucleon systems, is not straightforward. 

The definition of the potential describing three-nucleon interactions is a central issue of nuclear few- and many-body theory. Three-nucleon forces (TNF) are long known to provide a sizable contribution to the energies of the ground and low-lying excited states of light-nuclei, and play a critical role in determining the equilibrium properties of isospin-symmetric nuclear matter (SNM). In addition, their effect is expected to become large, or even dominant, in high density neutron matter, the understanding of which is required for the theoretical description of compact stars. 

Phenomenological models, such as the Urbana IX (UIX) potential, that reproduce the observed binding energy of $^3$H by construction, fail to explain the measured $nd$ doublet scattering length, $^2a_{nd}$ \cite{schoen_03}, as well as the proton analyzing power in $p$-$^3$He scattering, $A_y$ \cite{shimizu_95}.

The investigation of uniform nuclear matter may shed light on both the nature and the parametrization of the TNF. The equation of State (EoS) of SNM
is constrained by the available empirical information on saturation density, $\rho_0$, binding energy per nucleon at equilibrium, $E_0$, and compressibility, $K$. Furthermore, the recent observation of a neutron star of about two solar masses \cite{demorest_10} puts a constraint on the stiffness of the EoS of beta-stable matter, closely related to that of pure neutron matter (PNM).

Nuclear matter calculations are carried out using a variety of many-body approaches. The scheme referred to as Fermi-Hyper-Netted-Chain/Single-Operator-Chain (FHNC/SOC), based on correlated basis functions and the cluster expansion technique, has been first used to perform accurate nuclear matter calculations with realistic three body potentials in Ref.~ \cite{carlson_83}. This analysis included early versions of both the Urbana (UIV, UV) and Tucson Melbourne (TM) three body interactions with the set of parameters reported in  Ref. \cite{coon_81}. The results indicate that the UV model, the only one featuring a phenomenological repulsive term, provides a reasonable nuclear matter saturation density, while the UIV and TM potentials fail to predict saturation. In addition, none of the considered models yields reasonable values of the SNM binding energy and compressibility. 

The findings of Ref.~\cite{carlson_83} are similar to those obtained in Ref.~\cite{wiringa_88}, whose authors took into account additional diagrams of the cluster expansion and used the UVII model. The state-of-the-art variational calculations discussed in Ref.~\cite{akmal_98}, carried out using the Argonne $v_{18}$ \cite{wiringa_95} and UIX \cite{pudliner_95} potentials, also sizably underbinds SNM. 

While the authors of Ref.~\cite{akmal_98} ascribed this underbinding to deficiencies of the variational wave function, 
a signal of the limitations of the UIX three-nucleon potential has been provided by the authors of Ref. \cite{gandolfi_07b}, who carried out a study of symmetric nuclear matter within 
the Auxiliary Field Diffusion Monte Carlo (AFDMC) approach. Their results, obtained using the Argonne $v_{6}^\prime$ NN interaction, show that AFDMC simulations do not lead to an increase of the binding energy predicted by Fermi-Hyper-Netted-Chain (FHNC) and Brueckner-Hartree-Fock (BHF) calculations \cite{bombaci_05}. 

Different three-nucleon potential models \cite{pudliner_95,pieper_01} when applied to the calculation of nuclear properties, predict sizably different equations of state (EoS) of pure neutron matter at zero temperature and densities exceeding the nuclear matter saturation density, $\rho_0 = 0.16$ fm$^{-3}$ \cite{sarsa_03}. In this region, the three-nucleon force contribution to the binding energy becomes very large, the ratio between the potential energies associated with two- and the three-body interactions being $\sim 20 \%$ at density $\rho \sim 2 \rho_0$ (see, e.g. Ref. \cite{akmal_98b}). 

In Refs. \cite{lovato_11,lovato_11b}, we have followed a strategy which is somewhat along the line of the Three-Nucleon-Interaction (TNI) model proposed by Lagaris and Pandharipande \cite{lagaris_81} and Friedman and Pandharipande \cite{friedman_81} in the 1980s.  

The authors of Refs. \cite{lagaris_81,friedman_81} suggested that the main effects of three- and many-nucleon forces  can be taken into account through an effective, density-dependent two-nucleon potential. However, they adopted a purely 
phenomenological procedure, lacking a clearcut interpretation based on the the analysis of many-nucleon interactions at microscopic level. 

The TNI potential consists of two density-dependent functions involving three free parameters, whose values were determined through a fit of the saturation density, binding energy per nucleon and compressibility of SNM, obtained from FHNC variational calculations. The numerical values of the three model parameters resulting from recent studies performed by using AFDMC simulations turn out to be only marginally different from those of the original TNI potential \cite{gandolfi_10}.

The TNI potential has been successfully applied to obtain a variety of nuclear matter properties, such as  the nucleon momentum distribution \cite{fantoni_84b}, the linear response \cite{fantoni_87,fabrocini_89}, and the Green's function \cite{benhar_89, benhar_92}.

The strategy based on the development of two--body density-dependent potentials has been later abandoned, because their application to the study of finite nuclei involves a great deal of complication, mainly stemming from the breakdown of translation invariance. While in uniform matter the density is constant and the expansion of the effective potential in powers of $\rho$ is straightforward, in nuclei different powers of the density correspond to different operators, whose treatment is highly non trivial.

However, the recent developments in numerical methods for light nuclei seem to indicate that the above difficulties may turn out to be much less severe then those implied in the modeling of explicit many--body forces and, even more, in their use in {\sl ab initio} nuclear calculations.

The approach described in this Thesis, described in Ref. \cite{lovato_11,lovato_11b} is based on the assumption that $n$-body potentials ($n\geq3$) can be replaced by an effective  two-nucleon potential, obtained  through  an average over the degrees of freedom of $n-2$ particles. Hence, the effective potential can be written as a sum of contributions ordered according to powers of density, the $p$-th order term being associated with $(p+2)$-nucleon forces.

Obviously, such an approach requires that the average be carried out using a formalism suitable to account for the  
full complexity of nuclear dynamics. Our results show that, in doing such reduction, of great importance is the proper inclusion of both dynamical and statistical NN correlations. Therefore, we have used  the Correlated Basis Function (CBF) approach and the Fantoni Rosati (FR) cluster expansion formalism to perform the calculation of the terms linear in density of the effective potential, arising from the irreducible three-nucleon interactions modeled by the UIX potential.

It should be noticed that our approach significantly improves on the TNI model, as the resulting potential is obtained from a realistic microscopic three-nucleon force, which provides an accurate description of the properties of light nuclei.

While being the first step on a long road, the effective potential we have derived is valuable in its own right, as it can be used to include the effects of three-nucleon interactions in the calculation of the nucleon-nucleon scattering cross section in the nuclear medium. The knowledge of this quantity is required to obtain a number of nuclear matter properties of astrophysical interest, ranging from the transport coefficients to the neutrino emission rates \cite{benhar_07,benhar_08}.

Moreover, the density dependent potential has been implemented in the AFDMC computational scheme to obtain the EoS of SNM. Similar calculations using the UIX potential are not yet possible, due to the complexities arising from the commutator term. Monte Carlo calculations does not show an increase of the binding energy with respect to the variational FHNC/SOC calculations. This suggests that UIX potential model may not be adequate to properly describe three-nucleon interactions. 

In recent years, the scheme based on  ChPT has been extensively employed to obtain three-nucleon potential models \cite{epelbaum_02,{bernard_08}}. The main advantage of this approach is the possibility of treating the nucleon-nucleon (NN) potential and the TNF in a more consistent fashion, as some of the parameters fixed by NN and $\pi N$ data, are also used in the definition of the TNF. In fact,  the next-to-next-to-leading-order (NNLO) three-nucleon interaction only involves two parameters, namely $c_D$ and $c_E$, that do not appear in the NN potential and have to be determined fitting low-energy three-nucleon (NNN) observables. Unfortunately, however, $\pi N$ and $NN$ data still leave some uncertainties that can not be completely determined by NNN observables.

A comprehensive comparison between purely phenomenological and {\em chiral inspired} TNF, which must necessarily involve the analysis of both pure neutron matter and symmetric nuclear matter, is made difficult by the fact that chiral TNF are derived in momentum space, while many theoretical formalisms are based on the coordinate space representation.    

The local, coordinate space, form of the chiral NNLO three nucleon potential, hereafter referred to as  NNLOL, can be found in Ref. \cite{navratil_07}. However, establishing a connection between momentum and coordinate space representations involves some subtleties.  

The authors of Ref. \cite{epelbaum_02} have shown that the NNLO (momentum space) three body potential obtained from the chiral Lagrangian, when operating on a antisymmetric wave function, gives rise to contributions that are not all independent of one another. To obtain a local potential in coordinate space one has to regularize using the momenta transferred among the nucleons. This regularization procedure makes all the terms of the chiral potential independent, so that, in principle, all of them have to be taken into account. The potential would otherwise be somewhat inconsistent, as it becomes apparent in nuclear matter calculations, which involve larger momenta.

Momentum space chiral three-body interaction have been also employed in nuclear matter \cite{bogner_05,hebeler_11, hebeler_10}. In these studies, the NNNLO chiral two-body potential has been evolved to low momentum interaction $V_{low\,k}$, suitable for use in standard perturbation theory in the Fermi gas basis. The results, showing that the TNF is essential to obtain saturation and realistic equilibrium properties of SNM \cite{bogner_05,hebeler_11}, exhibit a sizable cutoff dependence. At densities around the saturation point this effect is $\sim 4\,\text{MeV}$. In addition, different values of the constants $c_i$ lead to different Equations of State for SNM \cite{hebeler_11} and PNM \cite{hebeler_10}. 

A comparative study of different three-nucleon local interactions (Urbana UIX (UIX), chiral inspired revision of Tucson-Melbourne (TM$^\prime$) and chiral NNLOL three body potential), used in conjunction with the local  Argonne $v_{18}$ NN potential, has been recently performed by the authors of Ref. \cite{kievsky_10}. 
They used the hyperspherical harmonics formalism to compute the binding energies of $^3$H and $^4$He, as well as the $nd$ doublet scattering length, and found that the three body potentials do not simultaneously reproduce these quantities. Selecting different sets of parameters for each TNF, they were able to obtain results compatible with experimental data, although a unique parametrization for each potential has not been found.  This problem is a consequence of the fact that the three low-energy observables considered are not enough to completely fix the set of parameters entering the definition of the potentials.

In Ref. \cite{lovato_12} we have analyzed the different parametrizations of the TM$^\prime$ and chiral NNLOL three body potential in nuclear matter discussed in  Ref. \cite{kievsky_10},  
carrying out nuclear matter calculations within both the FHNC/SOC variational scheme and the AFDMC formalism. The unphysical cutoff dependence of the contact terms of the NNLOL three body potential has been discussed. This feature, that clearly emerges already in the illustrative example of noninteracting FG, leads to sizable effects in the more refined calculations of the Equation of State of both SNM and PNM.

We have found that none of the parametrizations simultaneously reproduces the equilibrium properties of nuclear matter. Nevertheless, one of the TM$^\prime$ three-body potentials and one of the chiral NNLOL potentials provide values of SNM saturation density close to the experimental one. This is a remarkable feature of these potentials, as, unlike the UIX model, they do not involve any parameter adjusted to reproduce $\rho_0$.

Over the past few years, the CBF approach and the cluster expansion formalism have been also used to develop well behaved effective interactions, which take into account
the main effects of NN correlations and are suitable for use in standard perturbation theory in the Fermi gas basis, thus
allowing for a consistent treatment of equilibrium and non-equilibrium properties of nuclear matter \cite{cowell_04, benhar_07}. In view of the critical role played by 
interactions involving more than two nucleons, the implementation of the results discussed in this Thesis in the CBF effective interaction should be 
regarded as one of the most interesting applications of our analysis.  

As a first step in this direction, we have employed the effective interaction approach to compute the weak response of symmetric nuclear matter including three-body cluster 
contribution. Although the density, transverse and longitudinal responses of nuclear matter at momentum transfer around $\sim$ 1 fm$^{-1}$ have been already computed 
within CBF using the chain summation scheme \cite{fantoni_87,fabrocini_89,fabrocini_97}, our approach, based on effective weak operators and effective potentials, is more 
general, as it allows for a consistent description of the nuclear response in the regions of both low and high momentum transfer, where long- and short-range correlations
are known to be dominant. 

%The consistent description of the nuclear response at low and high momentum transfer requires a unified dynamical model, suitable to account for both short- and long-range correlation effects. We report the results of a study of the charged current weak response of symmetric nuclear matter, carried out using an effective interaction obtained from a realistic model of the nucleon-nucleon force within the formalism of correlated basis functions.
%Long-range correlations, leading to the excitation of collective modes which are known to be dom, become dominant in the region of low momentum transfer where the resolution of the probe $\lambda\sim\mathbf{q}^{-1}$, becomes much larger than the average NN separation. The Tamm-Dancoff approximation (TDA) is well suited to account for the long range correlation. 

%The definition of an effective interaction based on realistic model of NN force and , suitable to be used in standard perturbation theories, allow for a consistent treatment of short- and long-range correlations \cite{cowell_04,benhar_09}. The former are encompassed by the effective interaction, that, together with the effective operators, allow for employing the TDA in the Fermi gas uncorrelated basis. 

%We have included the UIX three-body force in the effective interaction, and three body clusters term have been considered in the evaluation of the effective weak operators. 

%{\color{red} Devo anticipare dei risultati??? O li metto nelle conclusioni? }

In Chapter \ref{chapt:nm_ni}, we introduce the concept of nuclear matter, emphasizing its features and relations with physical systems. The derivation of two- and three- nucleon interactions and their capability of reproducing experimental data are discussed and the formalism of chiral perturbation theory is outlined. 

In Chapter \ref{chapt:mb}, after discussing the limitations of the independent particle model, we describe CBF theory and the FHNC/SOC summation scheme, as well as the 
Monte Carlo many-body formalisms, pointing out that these approaches are able to encompass the correlation structure of nuclear matter, originating from the nuclear interactions. 

The first part of Chapter \ref{chapt:tbp} is devoted to the derivation of the density dependent potential, obtained from an average of the UIX three-nucleon force, while in 
the rest of the Chapter we discuss a comparative analysis of the chiral inspired three-nuclear forces in nuclear matter. 

Chapter \ref{chapt:resp} is focused on the inclusion of the effects of three-nucleon interactions in the CBF effective interaction, and the application of this approach to the 
calculation of the weak response of nuclear matter.

\newpage             
\thispagestyle{empty}

% Chapter 1 %%%%%%%%%%%%%%%%%%%%%%%%%%%%%%
\chapter{Nuclear matter and nuclear interactions}
\label{chapt:nm_ni}
\section{Bulk properties of nuclear matter}
Nulcear matter is uniform system of nucleons interacting through strong interactions only. While being a theoretical construct, it provides an extremely useful model to investigate the properties of both atomic nuclei and neutron star matter. Note that in these systems, with the exception of neutron stars in the early stages of their life, the temperature can be safely set to zero, as thermal energies are negligible compared to Fermi energies.

Two quantities that characterize nuclear matter are the density $\rho$ and the proton fraction $x_p$
\begin{align}
\rho&=\rho_p+\rho_n\nonumber \\
x_p&=\frac{\rho_p}{\rho}
\end{align}
where $\rho_p$ and $\rho_n$ indicate the proton and neutron densities, respectively. PNM is the limiting case in which $x_p=0$, while for SNM $x_p=1/2$. Strong interactions do not bind PNM, which in neutron stars is packed by gravitational attraction. SNM on the other hand is bound, and its equilibrium properties can be deduced deduced from the analysis of nuclear data.

The nuclear charge distribution, $\rho_{ch}(r)$ is almost constant within the nuclear volume and its central value is basically the same for all stable nuclei. It can be parametrized by
\begin{equation}
\rho_{ch}(r)=\frac{\rho_0}{1+e^{(r-R)/D}}\, .
\label{eq:rhoch}
\end{equation}
Elastic electron-nucleus scattering experiments have shown that the nuclear charge radius, $R$, is proportional to $A^{1/3}$
\begin{equation}
R=r_0A^\frac{1}{3}\, ,
\label{eq:ch_rad}
\end{equation}
implying that the volume increases linearly with the mass number. The parameters $r_0=1.15$ fm and $D=0.54$ fm have been extracted from experimental data, see for instance \cite{vries_87}. Equation (\ref {eq:ch_rad}) and the nuclear mass formula, to be discussed below, imply that the equilibrium
\begin{equation}
\rho_0=\frac{3}{4\pi r_{0}^3}=\simeq 0.16\pm 0.02\,\text{fm}^{-3}\, .
\end{equation}

\begin{figure}[!h]
\begin{center}
\vspace{0.5cm}
\includegraphics[angle=0,width=10cm]{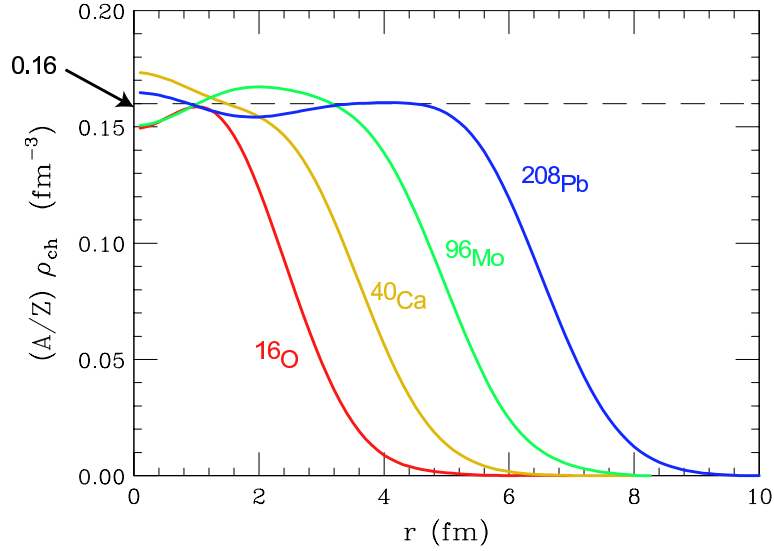}
\caption{Saturation of central nuclear densities of medium-heavy nuclei as measured by electron-nucleus scattering, from \cite{farina_09b}.\label{fig:sat_dens}}
\end{center}
\end{figure}

In addition, one can observe that the central charge density of atomic nuclei, measured by elastic electron-nucleon scattering, does not depend upon $A$ for large $A$. As shown in Fig. \ref{fig:sat_dens}, the limiting value does not differ from the one resulting from $r_0$.

The curves of Fig. \ref{fig:sat_dens}, parametrized by Eq. (\ref{eq:rhoch}), show that the charge density drops from $90\%$ to $10\%$ of its value over a distance $R_T\approx 2.5$ fm, independent on A,  called {\it surface thickness}. 

The (positive) binding energy per nucleon is defined as the difference between the mass of the bound nucleus and that of its constituents
\begin{equation}
B(Z,A)=\frac{1}{A}[Zm_p+(A-Z)m_n+Zm_e-M(Z,A)]\, .
\end{equation}
In Table \ref{tab:nuc_be} the masses and the binding energies for $^{16}O$, $^{56}$Fe, $^{62}$Ni and $^{120}$Sn are shown. The dependence of the binding energy on the atomic and mass numbers can be parametrized according to the {\it semiempirical mass formula} \cite{krane_87,fetter_03}, based on the liquid drop model and the shell model 
\begin{equation}
B(Z,A)=\frac{1}{A}\Big[a_VA-a_SA^{2/3}-a_CZ(Z-1)A^{1/3}-a_A\frac{(A-2Z)^2}{4A}+a_P\lambda A^{1/2}\Big]\, .
\label{eq:semf}
\end{equation}
The first term in square brackets, proportional to $A$ is the {\it volume term} and describes the bulk energy of nuclear matter. It is due to the strong nuclear interaction, that does not distinguish between neutrons and protons. Because strong interactions are short ranged, a given nucleon may only interact strongly with its nearest neighbors and next nearest neighbors, this explaining the scaling with $A$ instead of the $A(A-1)$ characteristic of the long-ranged interaction. The term proportional to $A^{2/3}$, denoted as {\it surface term} is also due to the strong interactions. It is actually a correction to the volume term, arising from the the fact that nucleons close to the surface have fewer neighbors than the inner ones.
The third term accounts for the Coulomb repulsion between protons. Since the electrostatic interaction is long ranged the scaling is given by $Z(Z-1)$. The fourth term, proportional to $[(A-Z)-Z]^2$, goes under the name of {\it symmetry energy}. Its origin can be justified on the basis of the Pauli exclusion principle explaining the experimental evidence that stable nuclei tend to have the same number of protons and neutrons. The last term, {\it pairing term}, which captures the effect of the spin coupling, can be exhaustively explained in the framework of the the shell model. It accounts for the fact that even-even nuclei (i. e. nuclei having even $Z$ and even $A-Z$) are likely to be more stable with respect to even-odd or odd-odd nuclei. Hence, the value of the constant $\lambda$ is $-1$, $0$ and $+1$ for even-even, even-odd and odd-odd nuclei, respectively.

\begin{table}[h!]
\begin{center}
\caption{Mass and binding energies of some stable nuclei. \label{tab:nuc_be}}
\vspace{0.3cm}
\begin{tabular}{c  c c c} 
\hline 
   &  $Z$  &  $M(Z,A)$ (amu)  &  $B(Z,A)$ (MeV) \\ 
\hline
$^{16}$O     & $8$   & $15.9949$ & $7.9765 $ \\ 
$^{56}$Fe   & $26$ & $55.9349$ & $8.7906   $ \\ 
$^{62}$Ni    & $28$ & $61.9283$ & $8.7948 $ \\
$^{120}$Sn & $50$ & $119.9022$ & $8.5048 $\\
$^{208}$Pb & $82$ & $207.9767$ &  $7.8677$\\
\hline
\end{tabular} 
\vspace{0.1cm}
\end{center}
\end{table}

SNM is described by the semiempirical mass formula by putting $Z=A/2$ and taking the limit for $A\to \infty$. Neglecting the Coulomb repulsion, the volume term is the only one surviving in this limit. Therefore, the coefficient $a_V$ can be identified with the binding energy per particle of SNM. Typical fits of Eq. (\ref{eq:semf}) gives \cite{mackie_77}
\begin{equation}
E_0\equiv E(\rho_0)=a_V=-15.6\pm 0.2\,\text{MeV}
\end{equation}

In the vicinity of the equilibrium density, the energy per particle of SNM can be expanded according to
\begin{equation}
\frac{E_{\text{SNM}}(\rho)}{A}=E_0+\frac{K_0}{2}\Big(\frac{\rho-\rho_0}{3\rho_0}\Big)^2+\mathcal{O}(\rho-\rho_0)^3\, .
\end{equation}
The coefficient 
\begin{equation}
K_0=\frac{9\rho_{0}^2}{A}\Big(\frac{\partial^2 E_{\text{SNM}}(\rho)}{\partial \rho^2}\Big)
\end{equation}
is the (in) compressibility module that can be extracted from isoscalar breathing modes \cite{blaizot_76,colo_04}, and isotopic differences in charge densities of large nuclei \cite{cavedon_87}
\begin{equation}
K_0=220\pm 30 \,\text{MeV}\, .
\end{equation}

\section{Nuclear hamiltonian}
\label{sec:nucl_ham}
Although impressive steps have been done in this directions \cite{yamazaki_12,beane_12,aoki_12,savage_11}, the description of nuclear matter properties at finite density and zero temperature within the framework of quantum crhomo-dynamics (QCD) still seems to be out of reach for the present computational techniques. 

For this reason, in this work we rely on dynamical models in which non relativistic nucleons interacts by means of instantaneous potentials, describing both the short and the long range interactions, the latter given by meson (mainly pions) exchange. 

Within this picture, nuclei can be described in terms of point like nucleons of mass $m$, whose dynamics are dictated by the hamiltonian
\begin{equation}
\hat{H} = \sum_i -\frac{{\nabla}^2_i}{2m} + \sum_{j>i} \hat{v}_{ij} + \sum_{k>j>i} \hat{V}_{ijk}  \ .
\label{eq:hamiltonian}
\end{equation}
where $\hat{v}_{ij}$ and $\hat{V}_{ijk}$ are the two- and three- body potentials, respectively. In principle four- and more- body potential could be included in the hamiltonian, but there are convincing indications that their contribution is negligibly small.

\subsection{Two-body interaction}
Highly realistic two-nucleon potentials, either purely phenomenological or based on chiral perturbation theory (ChPT)  have been obtained from accurate fits of the properties of the bound and scattering states of the two-nucleon system \cite{stoks_93,bergervoet_90,vanderleun_82,ericson_83,rodning_90,simon_81,bishop_79}. 

Besides the technical complexities involved in the fits, from the analysis of nuclear experimental data, the following main features of the NN interaction can be inferred.
\begin{itemize}
\item[$\bullet$] The saturation of nuclear densities, see Fig. \ref{fig:sat_dens}, showing that the density in the inner part of nuclei is almost constant and independent on $A$, indicates that nucleons cannot be packed together too tightly.  In a non relativistic picture, a coordinate space potential with a repulsive core is able to reproduce this experimental evidence. Denoting by $\mathbf{r}_{12}$ the inter particle distance, one has
\begin{equation}
v(\mathbf{r}_{12})>0 \quad , \quad r_{12}<r_c\, .
\end{equation}
It is worth mentioning that the authors of Ref. \cite{bogner_05}, using the renormalization group equations to smear off the repulsive core, were able to develop a class of low-momentum potentials. In this framework, the saturation properties is explained in terms of many-nucleons interactions.

\item[$\bullet$] The nuclear binding energy is almost the same for nuclei with $A\geq 12$. Together with the proton neutron scattering data, this is an indication for the NN interaction to be short-ranged 
\begin{equation}
v(\mathbf{r}_{12})=0 \quad , \quad r_{12}>r_0\, .
\end{equation}

\item[$\bullet$] Combining the neutron-proton cross section data with the properties of the deuteron ($^2$H) it is found that the singlet state does not have bound states, as opposite to the triplet state. In particular, the deuteron is the only NN bound state, consisting of a neutron and a proton coupled to total spin $S=1$ and isospin $T=0$. This is a clear manifestation of the spin-dependence of the NN.

\item[$\bullet$] From the observation that the deuteron exhibits a non vanishing quadrupole moment, it can be deduced that the angular momentum does not commute with the hamiltonian. Consequently the NN potential cannot be invariant under the rotation of the spatial coordinates alone.
\item[$\bullet$] The {\it mirror} nuclei are pairs of nuclei such that the proton number in one equals
the neutron number in the other and vice versa. This obviously implies that mirror nuclei have the same mass number but atomic number differing by one unit. Examples of mirror nuclei are $^{13}_{7}$N and $^{13}_{6}$C or $^{15}_{7}$N and $^{15}_{8}$O. The spectra of mirror nuclei show striking similarities, as the energy of the levels with the same parity and angular momentum are the same, beside small electromagnetic corrections, showing that the nuclear interactions are charge symmetric. This is a manifestation of a more general symmetry of the underlying theory, the {\it isospin invariance}.
\end{itemize}

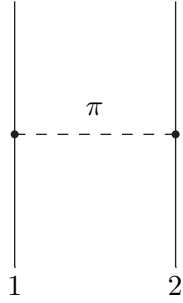
\begin{figure}[!ht]
\begin{picture}(100,100)(-180,20)
\Line(20,0)(20,100)
\Line(80,0)(80,100)
\DashLine(20,50)(80,50){4}
\Vertex(20,50){1.5}
\Vertex(80,50){1.5}
\Text(50,60)[]{\small $\pi$}
\Text(20,-7)[]{\small $1$}
\Text(80,-7)[]{\small $2$}
\end{picture}
\vspace{1.0cm}
\caption{Feynman diagram describing the one-pion exchange process between two nucleons. \label{fig:ope}}
\end{figure}

The first attempt of a theoretical description of NN scattering data is due to Yukawa \cite{yukawa_35}. He made the hypothesis of nucleons interacting through the exchange of a particle of mass $\mu$, related with the range of the interaction $r_0$ through 
\begin{equation}
r_0\sim \frac{1}{\mu}\, .
\end{equation}
For $r_0\sim 1.0$fm the above equation gives $\mu\sim 200$ MeV, which is of the same order of magnitude of the pion mass $m_\pi\simeq 140$ MeV. The most simple parity conserving vertex between the nucleons and the pseudo scalar pion has the form $i g \gamma_5 \vec{\tau}$, where $g$ is a coupling constant and $\vec{\tau}$ accounts for the isospin of the nucleons. Hence, the non relativistic limit of the scattering amplitude described by the Feynman diagram of Fig. \ref{fig:ope}, leads to the definition of a NN potential, whose expression in coordinate space reads 
\begin{equation}
v_\pi(r_{12})=\frac{1}{3}\frac{g^2}{4\pi}\frac{m_{\pi}^3}{4m^2}\tau_{12}\Big[T_\pi(r_{12})S_{12}+\Big(Y_\pi(r_{12})-\frac{4\pi}{m_{\pi}^3}\delta(\mathbf{r}_{12})\Big)\sigma_{12}\Big]\, ,
\label{eq:ope_NN}
\end{equation}
In the above equation,  $\sigma_{ij}=\vec{\sigma}_i \cdot \vec{\sigma}_j$ and $\tau_{ij}=\vec{\tau}_i \cdot \vec{\tau}_j$, where $\vec{\sigma}_i$ and $\vec{\tau}_i$ are Pauli matrices acting on the spin or isospin of the $i$-th, while
\begin{equation}
S_{ij}=(3\hat{r}_{ij}^\alpha\hat{r}_{ij}^\beta-\delta^{\alpha\beta})\sigma_{i}^\alpha\sigma_{j}^\beta \ ,
\label{eq:tens_def}
\end{equation}
with $\alpha, \ \beta= 1, \ 2, \ 3$, is the tensor operator. The radial functions associated with the spin and tensor components read
\begin{align}
Y(x)&=\frac{e^{-x}}{x}\xi_Y(x)\\
T(x)&=\Big(1+\frac{3}{x}+\frac{3}{x^2}\Big)Y(x)\xi_T(x)\, ,
\label{eq:YT}
\end{align}
The phase shift analysis of the high angular momentum neutron-neutron ($nn$) and neutron-proton ($np$) scattering states  shows that for $g^2/(4\pi)\simeq 14$ the one pion exchange potential, $v_{\pi}$, provides an accurate description of the long range part ($r_{12}\geq 1.4$ fm) of the NN interaction. High angular momentum in fact gives rise to a high centrifugal barrier, preventing the nucleons from coming very close to each other.
In order to describe NN interactions at intermediate range one should consider processes in which more two and more pions, possibly interacting among themselves, are exchanged. In the short range region, exchange of heavier mesons and more complicated processes, that can be modeled through, e.g., contact interactions, are expected to be dominant. 

\subsubsection{Chiral perturbation theory (ChPT)}
A framework in which the above processes are taken into account in a systematic way is provided by the chiral perturbation theory. 

By analyzing the spectra of hadrons having up and down valence quarks, it can be deduced that the approximate chiral symmetry of QCD is spontaneously broken. The corresponding quasi-Goldstone bosons can be identified with pions, that in turn are much lighter than all other hadrons. They would be exactly massless if the masses of $u$ and $d$ quarks were vanishing. Goldstone's theorem states that the interactions of Goldstone bosons become weak for small momenta, of the order of the pion mass.

The natural cutoff of ChPT is the rho-meson mass, providing the high-energy scale of the theory, $\Lambda_{\chi}\simeq 800$MeV. Therefore it is possible to expand the scattering amplitude in powers of the small ratios between either the external momenta or the pion mass and $\Lambda_{\chi}$. Pion loops are naturally incorporated and all corresponding ultraviolet divergences can be absorbed at each fixed order in the chiral expansion by counter terms of the most general Lagrangian, involving pions and nucleons, consistent with spontaneously broken chiral symmetry and other known symmetries. Is it worth noting that the Lagrangian also contains contact interaction terms among nucleons, needed to renormalize loop integrals, make results fairly independent of the regulators, and parametrize the unresolved short-distance dynamics of the nuclear force \cite{machleidt_11}.

In the pion-nucleon sector ChPT work well as the interaction vanishes at vanishing external momenta in the chiral limit. In the pure nucleonic sector the situation is more complicated, since the strong interaction does not become perturbative even in the chiral limit at vanishing three-momenta of the external nucleons. In his seminal papers \cite{weinberg_90,weinberg_91} Weinberg proposed to apply ChPT to the ``effective NN potential'', defined as the sum of connected diagrams for the scattering matrix, generated by old fashioned time-ordered perturbation theory. Following this idea Weinberg was able to demonstrate the validity of the well-established intuitive hierarchy of the few-nucleon forces: the two-nucleon interactions are more important than the three-nucleon ones, which are more important than the four-nucleon interactions and so on.

Within ChPT it also possible to explain why the first order diagram of Fig. \ref{fig:ope} is able to provide an accurate description of the long-range part of the NN potential, although the coupling constant $g$ is much larger than one. The one-pion exchange is in fact the leading contribution in the chiral parameter $|\mathbf{q}|/\Lambda_\chi$, where $|\mathbf{q}|\sim m_\pi$ is the spatial momentum of the nucleons.

Two-body potential at next-to-next-to-next leading order (NNNLO) in the chiral expansion has been derived independently by Entem and Machleidt \cite{entem_03} and by the Julich group \cite{epelbaum_05}. Both these potentials are able to reproduce the Nijmegen phase shifts $\chi^2\simeq 1$. Unfortunately, a coordinate space expression for such potentials is not available. Therefore, they can be employed neither in FHNC/SOC nor in AFDMC formalism yet.

On the other hand, a local version of the three-body potential at NNLO does exist and will be discussed later in this Chapter

\subsubsection{The Argonne potential}
Phenomenological NN potential \cite{lacombe_80,stoks_94,wiringa_95,machleidt_01} are generally written as
\begin{equation}
\hat{v}=\hat{v}_\pi(r_{12})+\hat{v}_{I}(r_{12})+\hat{v}_{S}(r_{12})\, ,
\end{equation}
where $\hat{v}_\pi(r_{12})$ is given by Eq. (\ref{eq:ope_NN}) stripped of the delta function contribution, $\hat{v}_{I}(r_{12})$ describes the intermediate range attraction attributed to two-pion exchange, and $\hat{v}_{S}(r_{12})$ accounts for the short range repulsion, which may be due to the exchange of heavier mesons and/or to the overlap of the quarks distributions of the nucleons. Comparison with ChPT suggests that $\hat{v}_{S}(r_{12})$ is strictly related to the contact terms in the chiral Lagrangian.
The highly realistic Argonne $v_{18}$ potential (AV18) \cite{wiringa_95} can be written in the form
\begin{equation}
v_{18}(r_{12})=\sum_{p=1}^{18}v^p(r_{12})\hat{O}^{p}_{12}\, .
\label{eq:av18}
\end{equation}
The static part of AV18, given by the first six operators 
\begin{equation}
\hat{O}^{p=1-6}_{ij}=(1,\sigma_{ij},S_{ij})\otimes(1,\tau_{ij})\,.
\end{equation}
sufficient to describe deuteron properties and the phase shifts corresponding to S and D states. In order to explain the 
P-wave phase shifts, the spin-orbit term has to be introduced  
\begin{equation}
\hat{O}^{p=7-8}_{ij}=\mathbf{L}_{ij}\cdot\mathbf{S}_{ij} \otimes(1,\tau_{ij})\,.
\end{equation}
In the above equation $\mathbf{L}_{ij}$ is the relative angular momentum 
\begin{equation}
\mathbf{L}_{ij}=\frac{1}{2i}(\mathbf{r}_i-\mathbf{r}_j)\times (\boldsymbol{\nabla}_i-\boldsymbol{\nabla}_j)
\end{equation}
and $\mathbf{S}_{ij}$ is the total spin of the pair
\begin{equation}
\mathbf{S}_{ij}=\frac{1}{2}(\boldsymbol{\sigma}_i+\boldsymbol{\sigma}_j)\, .
\end{equation}
The remaining $10$ operators are required to achieve the description of the Nijmegen scattering data with $\chi^2\simeq 1$. They are given by
\begin{align}
\hat{O}^{p=9-14}_{ij}&=(\mathbf{L}^2,\mathbf{L}^2\sigma_{ij},\mathbf{L}\cdot\mathbf{S})\otimes (1,\tau_{ij})\nonumber\\
\hat{O}^{p=15-18}_{ij}&=(1,\sigma_{ij},S_{ij})\otimes (T_{ij},\tau_{i}^z+\tau_{j}^z)\, ,
\end{align}
where
\begin{equation}
T_{ij}=(3\hat{r}_{ij}^\alpha\hat{r}_{ij}^\beta-\delta^{\alpha\beta})\tau_{i}^\alpha\tau_{j}^\beta\, .
\end{equation}
The last four operators account for the charge symmetry breaking effect, due to the different masses and coupling constants of charged and neutral pions.

Instead of the full AV18, we will be using the so called Argonne $v_{8}^\prime$ and Argonne $v_{6}^\prime$ potentials, which are not simple truncations of the original model, but rather ``reprojections'' \cite{wiringa_02}. 

\begin{figure}[!!h]
\begin{center}
\includegraphics[width=7.9cm,angle=0]{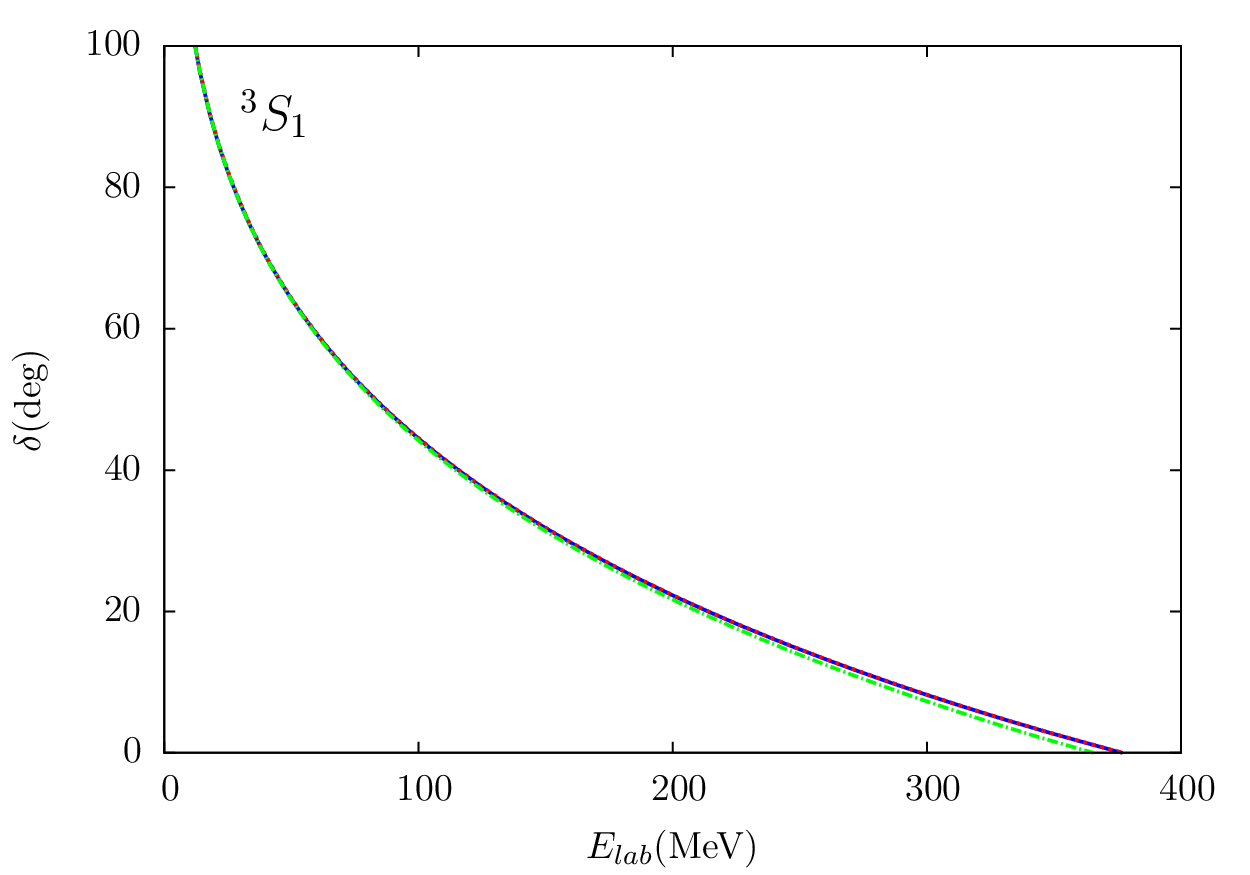}
\includegraphics[width=7.9cm,angle=0]{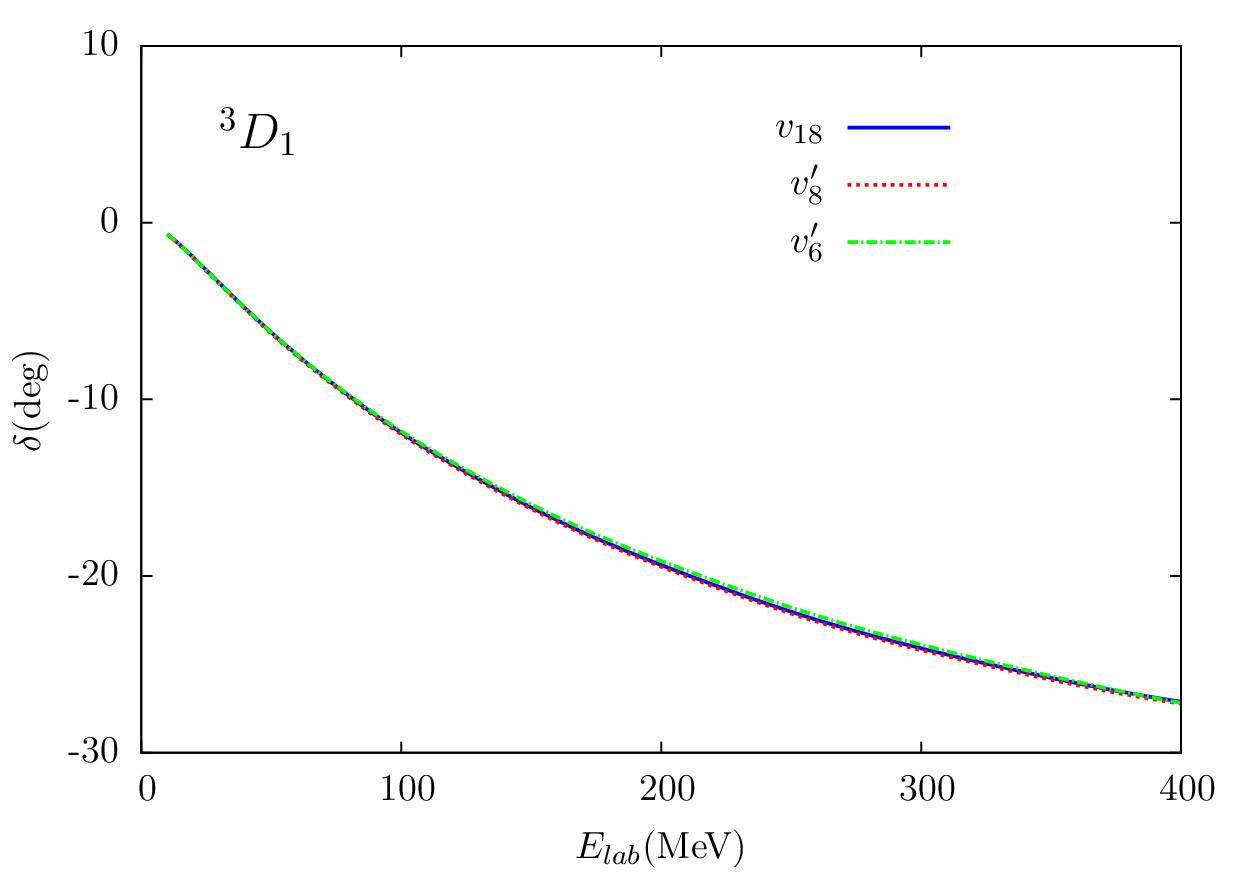}
\includegraphics[width=7.9cm,angle=0]{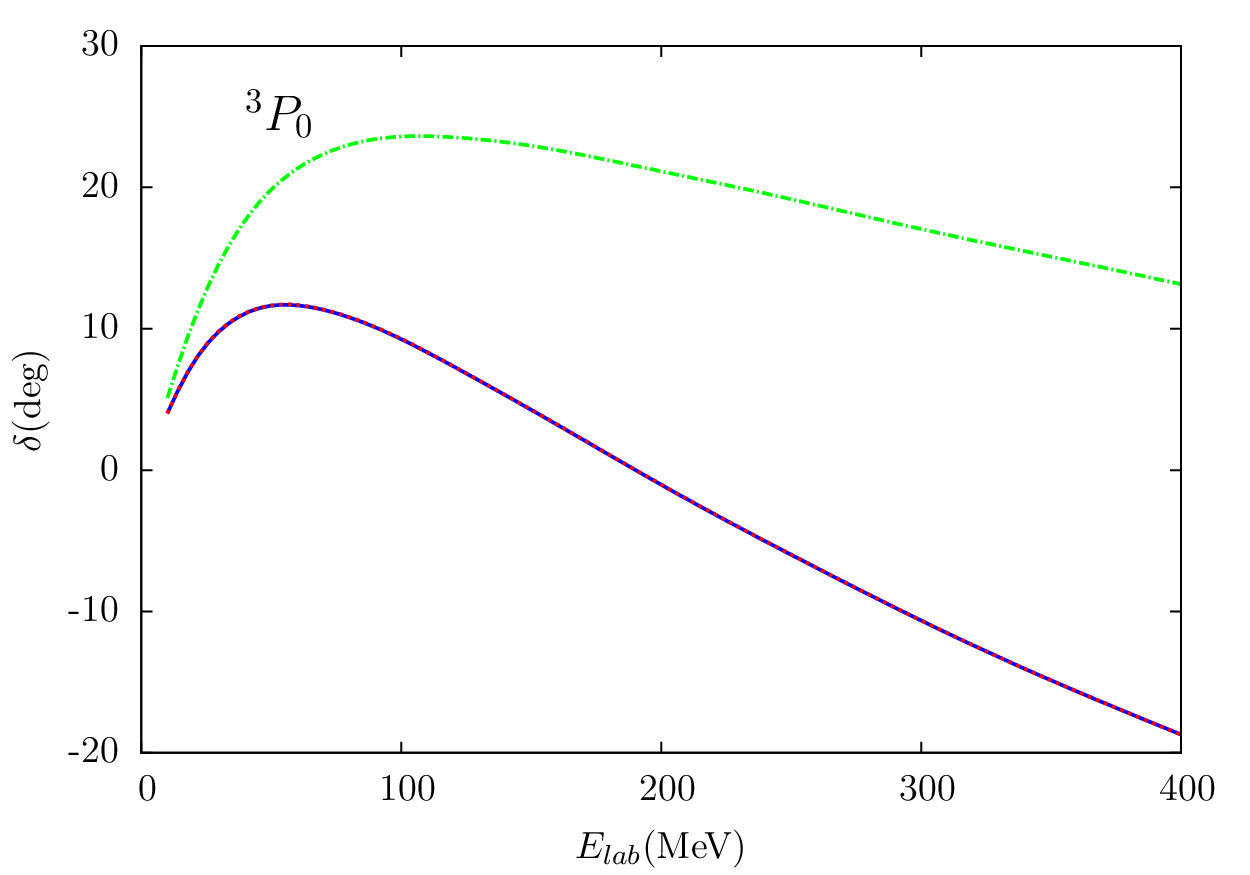}
\includegraphics[width=7.9cm,angle=0]{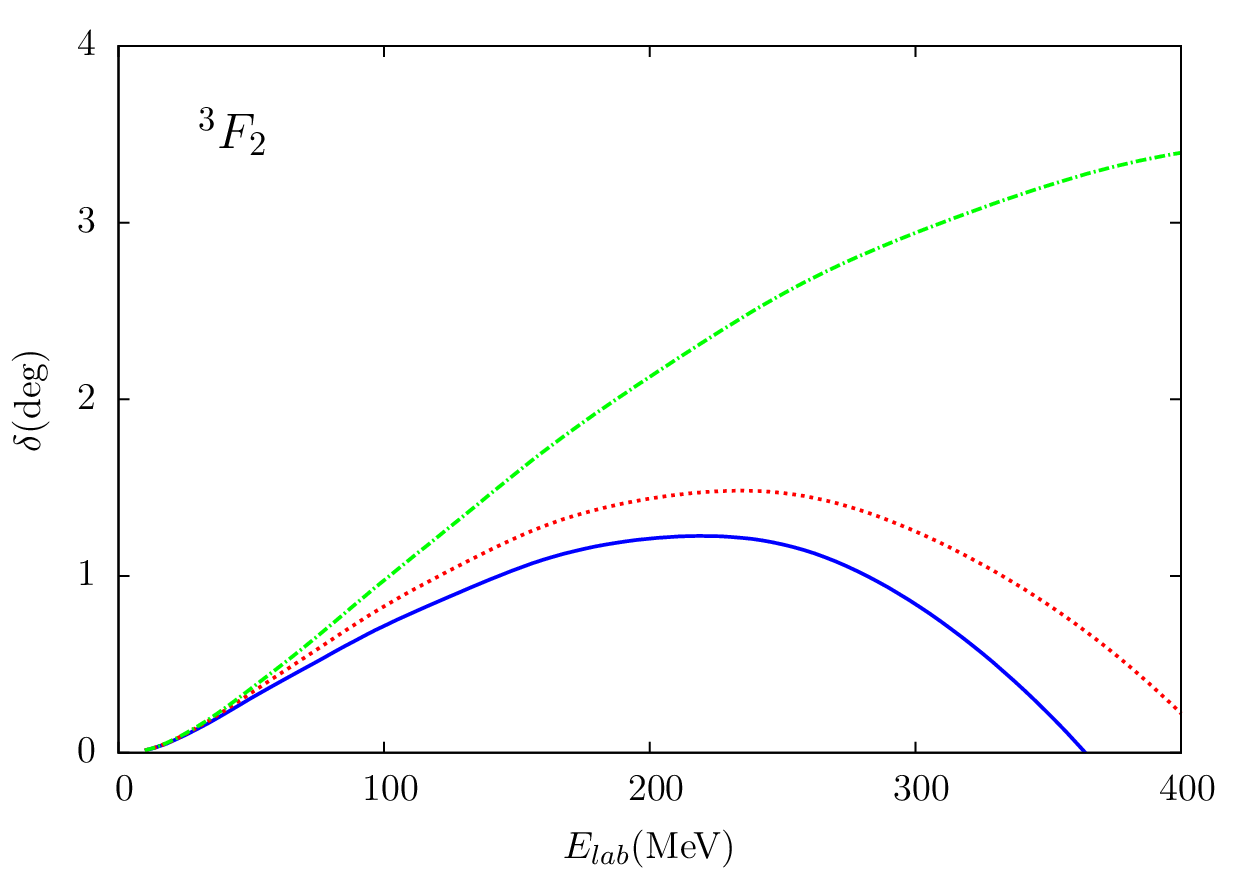}
\end{center}
\caption{$^3 S_1$, $^3 D_1$, $^3 P_0$ and $^3 F_2$ nucleon-nucleon phase shifts for Argonne $v_{18}$ (solid line), Argonne $v_{8}^\prime$ (dotted line) and  Argonne $v_{6}^\prime$ (dotted dashed line) potentials, from \cite{baldo_12}. \label{fig:avps}}
\end{figure}

For the purpose of the following discussion, the $^3 S_1$, $^3 D_1$, $^3 P_0$ and $^3 F_2$ phase shift calculated using the Argonne $v_{18}$ potential and its reprojected versions $v_{6}^\prime$ and $v_{8}^\prime$ are displayed in Fig. \ref{fig:avps}. 

The Argonne $v_{8}^\prime$ potential is obtained by refitting the scattering data in such a way that all $S$ and $P$ partial waves as well as the $^3 D_1$ wave and its coupling to $^3 S_1$ are reproduced equally well as in Argonne $v_{18}$. The differences with the full AV18 starts appearing in higher partial waves' phase shift, like the $^3F_2$ plotted of Fig. \ref{fig:avps}. In all light nuclei and nuclear matter calculations the results obtained with the $v_{8}^\prime$ are very close to those obtained with the full $v_{18}$, and the difference $v_{18}-v_{8}^\prime$ can be safely treated perturbatively. 

The Argonne $v_{6}^\prime$ is not just a truncation of $v_{8}^\prime$, as the radial functions associated with the first six operators are adjusted to preserve the deuteron binding energy. Our interest in this potential is mostly due to the fact that AFDMC simulations of nuclei and nuclear matter can be performed most accurately with $v_6$--type of two--body interactions. Work to include the spin--orbit terms in AFDMC calculations is in progress. On the other hand we need to check the accuracy of our proposed density-dependent reduction with both FHNC and AFDMC many--body methods before proceeding to the construction of a realistic two--body density-dependent model potential and comparing with experimental data.

\subsection{Three-body interaction}
\label{sec:TBF}
It is well known that using a nuclear Hamiltonian including only two-nucleon interactions leads to the underbinding of light nuclei and overestimating the equilibrium density of nuclear matter. Hence, the contribution of three-nucleon interactions must necessarily be taken into account.

In order for the three-body potential to be symmetric under the exchange of particles $1$, $2$ and $3$ (remember that the sum of Eq. (\ref{eq:hamiltonian}) has the constraint $k>j>i$), it has to written as a cyclic sum. For all the potentials we are considering in this Thesis, it turns out that there are only three independent cyclic permutations
\begin{equation}
\hat{V}_{123}=\hat{V}(1:23)+\hat{V}(2:13)+\hat{V}(3:12)\, , 
\label{eq:cycl_sum}
\end{equation}
with $\hat{V}(i:jk)=\hat{V}(i:kj)$.

\subsubsection{UIX three-body potential}
One of the most widely used three-body potential is the Urbana IX (UIX) \cite{pudliner_95}, that consists of two terms. The attractive two-pion ($2\pi$) exchange  interaction $V^{2\pi}$ turns out to be helpful in fixing the problem of the underbinding in light nuclei, but makes the nuclear matter energy worse. The purely phenomenological repulsive term $V^R$ prevents nuclear matter from being overbound at large density.

\begin{figure}[!h]
\begin{center}
\fcolorbox{white}{white}{
  \begin{picture}(150,100)(-25,20)
	\SetWidth{0.5}
	\SetColor{Black}
	\Line(0,0)(0,100)
	\Line(50,0)(50,30)
	\SetWidth{2.0}
	\Line(50,30)(50,70)
	\SetWidth{0.5}	
	\Line(50,70)(50,100)
	\Line(100,0)(100,100)	
	\DashLine(0,30)(50,30){5}
	\DashLine(50,70)(100,70){5}
	\Text(25,20)[]{$\pi$}
	\Text(75,80)[]{$\pi$}
	\Text(40,50)[]{$\Delta$}
    \end{picture}
}
\vspace{1.5cm}
\caption{Feynman Diagram associated with the Fujita Miyazawa three-nucleon potential term $\hat{V}^{2\pi}(3:12)$.}
\label{fig:Fujita_Miyazawa}
\end{center}
\end{figure}
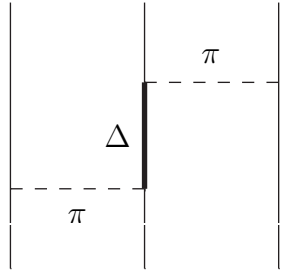

The $V^{2\pi}$ term was first introduced by Fujita and Miyazawa \cite{fujita_57} to describe the process 
whereby two pions are exchanged among nucleons and a $\Delta$ resonance is excited in the intermediate state, as 
shown in the Feynman diagram of Fig.~\ref{fig:Fujita_Miyazawa}. It can be conveniently written as a sum of an anticommutator and a commutator term
\begin{equation}
\hat{V}^{2\pi}(3:12)=A_{2\pi}\{\hat{X}_{13},\hat{X}_{23}\}\{\tau_{13},\tau_{23}\}+C_{2\pi}[\hat{X}_{13},\hat{X}_{23}][\tau_{13},\tau_{23}]\,,
\label{eq:fm_uix}
\end{equation}
where
\begin{equation}
\hat{X}_{ij}=Y(m_\pi r)\sigma_{ij}+T(m_\pi r)S_{ij}\, .
\end{equation}
The $\xi(x)$ are short-range cutoff functions defined by
\begin{equation}
\xi_{Y}(x)=\xi_{T}(x)=1-e^{-cx^2}\, .
\end{equation}
In the UIX model, the cutoff parameter 
is kept fixed at $c=2.1$ fm$^{-2}$, the same value as in the
cutoff functions appearing in the one-pion exchange term of the Argonne $v_{18}$ two-body potential. On the other hand, $A_{2\pi}$ is varied to fit the observed binding energies of $^3$H. The three-nucleon interaction depends on the choice of the NN potential; for example, using the Argonne $v_{18}$ model one gets $A_{2\pi}=-0.0293\,\,\text{MeV}$.

The repulsive term $V^R$ is spin-isospin independent and can be written in the simple form
\begin{equation}
V^R(3:12)=U_0T^2(m_\pi r_{13})T^2(m_\pi r_{23})\, ,
\label{eq:VR_uix}
\end{equation}
with $T(x)$ defined in Eq. (\ref{eq:YT}). The strength $U_0$,  adjusted to reproduce the empirical nuclear matter 
saturation density, is $U_0=0.0048\,\,\text{MeV}$ with $v_{18}$.

\begin{figure}[!h]
\begin{center}
\vspace{0.5cm}
\includegraphics[angle=0,width=10cm]{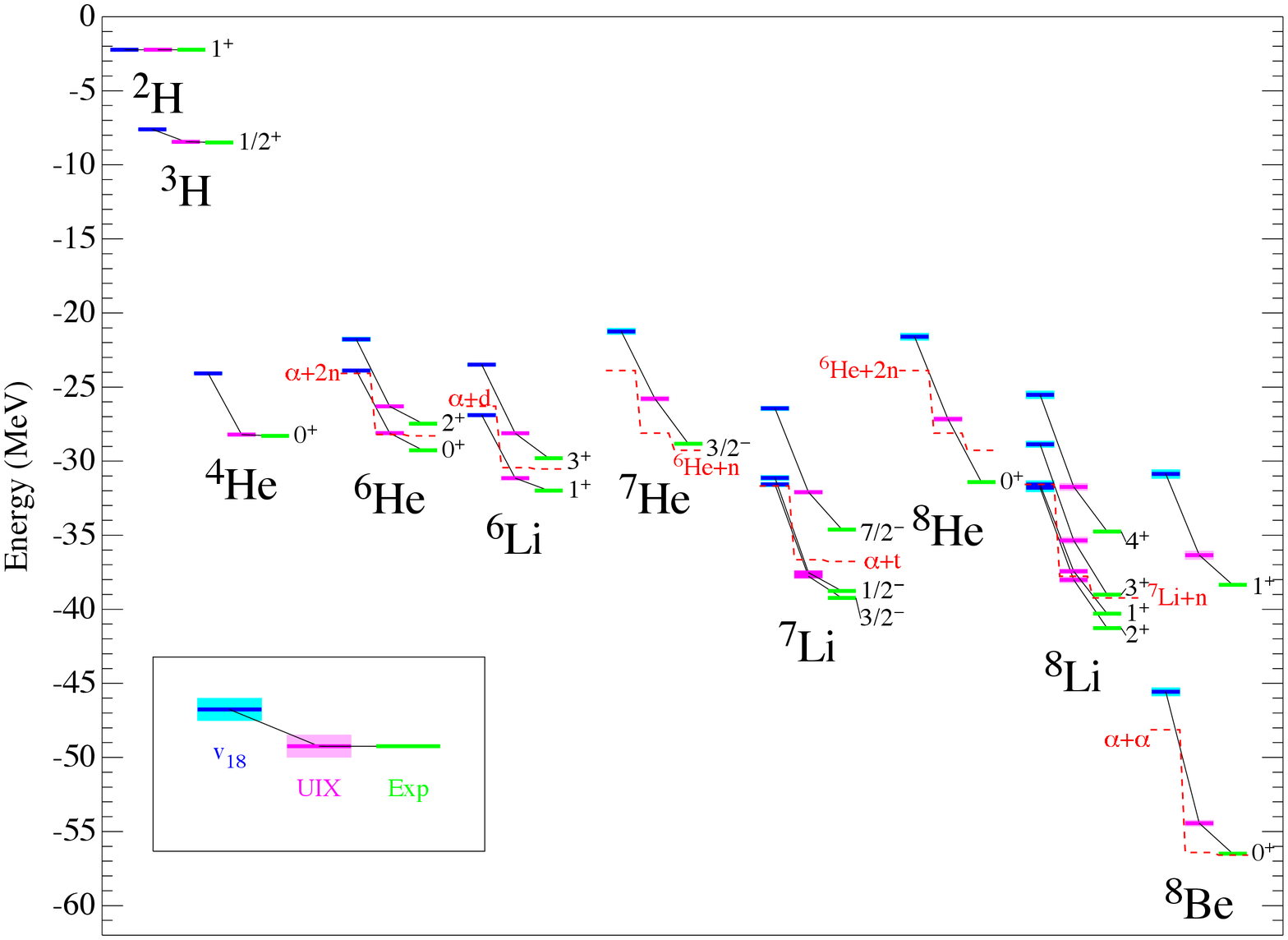}
\caption{Energies per particle of ground and low-lying excited states of light nuclei, resulting from GFMC calculations, computed with the AV18 and AV18+UIX interactions, compared to experiment \cite{pieper_01}. The Monte Carlo statistical error are represented by the light shaded region. The red dashed lines indicate the breakup thresholds  for each model or experiment. \label{fig:pieper_nucl_UIX}}
\end{center}
\end{figure}

The two parameters $A_{2\pi}$ and $U_0$ have different values  for $v_{8}^\prime$ and $v_{6}^\prime$. We disregard such small differences in this analysis, mostly aimed at testing the quality of the density-dependent reduction of the UIX three--body potential, rather than reproducing empirical data. 

As displayed in Fig. \ref{fig:pieper_nucl_UIX}, showing the results of the GFMC calculations of Ref. \cite{pieper_01}, when the UIX potential is used the binding energy of $^3$H is exactly reproduced by construction, and that of $^4$He turns out to be very close to the experimental value. A significant improvement is also observed for the binding of the p-shell nuclei. However, more and more underbinding is provided by the AV18+UIX for increasing $A$ and $A-Z$. In particular a problem with the isospin dependence of the interaction model is revealed by the fact that  $^8$He is more underbound than $^8$Be. With respect to the pure AV18 case, the relative stability of the lithium nuclei is improved, but the Borromean helium nuclei are still unbound. Additional GFMC calculations of higher-lying excited states, not shown in Fig. \ref{fig:pieper_nucl_UIX}, indicate that the AV18+UIX model underestimate the spin-orbit splittings among spin-orbit partners such as the $3/2^-$ and $1/2^-$ states in $^5$He.

Moreover, the phenomenological UIX model fails to explain the measured $nd$ doublet scattering length, $^2a_{nd}$ \cite{schoen_03}, as well as the proton analyzing power in $p$-$^3$He scattering, $A_y$ \cite{shimizu_95}.

\subsubsection{Chiral inspired models of three nucleon forces}
\label{sec:nitbp}
In recent years, the scheme based on  ChPT has been extensively employed to obtain three-nucleon potential models \cite{epelbaum_02,bernard_08,bernard_11,krebs_12}. The main advantage of this approach is the possibility of treating the nucleon-nucleon (NN) potential and the TNF in a more consistent fashion, as the parameters $c_1$, $c_3$ and $c_4$, fixed by NN and $\pi N$ data, are also used in the definition of the TNF. In fact,  the next-to-next-to-leading-order (NNLO) three-nucleon interaction only involves two parameters, namely $c_D$ and $c_E$, that do not appear in the NN potential and have to be determined fitting low-energy three-nucleon (NNN) observables. Unfortunately, however, $\pi N$ and $NN$ data still leave some uncertainties on the $c_i$'s, that can not be completely determined by NNN observables.

A comprehensive comparison between purely phenomenological and {\em chiral inspired} TNF, which must necessarily involve the analysis of both pure neutron matter and symmetric nuclear matter, is made difficult by the fact that chiral TNF are derived in momentum space, while many theoretical formalisms are based on the coordinate 
space representation.    

The local, coordinate space, form of the chiral NNLO three nucleon potential, hereafter referred to as  NNLOL, can be found in Ref. \cite{navratil_07}. However, establishing a connection between momentum and coordinate space representations involves some subtleties.  

The authors of Ref. \cite{epelbaum_02} have shown that the NNLO (momentum space) three body potential obtained from the chiral Lagrangian, when operating on a antisymmetric wave function, gives rise to contributions that are not all independent of one another. To obtain a local potential in coordinate space one has to regularize using the momenta transferred among the nucleons. This regularization procedure makes all the terms of the chiral potential independent, so that, in principle, all of them have to be taken into account. The potential would otherwise be somewhat inconsistent, as it becomes apparent in nuclear matter calculations, which involve larger momenta.

A comparative study of different three-nucleon local interactions (Urbana UIX (UIX), chiral inspired revision of Tucson-Melbourne (TM$^\prime$) and chiral NNLOL three body potential), used in conjunction with the local  Argonne $v_{18}$ NN potential, has been recently performed \cite{kievsky_10}. 
The authors of Ref.~\cite{kievsky_10} used the hyperspherical harmonics formalism to compute the binding energies of $^3$H and $^4$He, as well as the $nd$ doublet scattering length, and found that the three body potentials do not
simultaneously reproduce these quantities. Selecting different sets of parameters for each TNF they were able to obtain results compatible with experimental data, although a unique parametrization for each potential has not been found.  This problem is a consequence of the fact that the three low-energy observables considered are not enough to completely fix the set of parameters entering the definition of the potentials.

\begin{figure}[!ht]
\begin{center}
\begin{picture}(300,120)(0,0)

\Line(10,0)(10,100)
\Line(40,0)(40,100)
\Line(70,0)(70,100)
\DashLine(10,50)(70,50){4}
\Vertex(10,50){1.5}
\Vertex(70,50){1.5}
\Vertex(40,50){3}
\Text(25,40)[]{\small $q_1$}
\Text(55,40)[]{\small $q_2$}
\Text(10,-7)[]{\small $1$}
\Text(40,-7)[]{\small $3$}
\Text(70,-7)[]{\small $2$}
\Text(40,-25)[]{TPE}
\Line(120,0)(150,100)
\Line(150,0)(120,100)
\Line(180,0)(180,100)
\DashLine(135,50)(180,50){4}
\Vertex(180,50){1.5}
\Vertex(135,50){3}
\Text(160,40)[]{\small $q_2$}
\Text(150,-25)[]{OPE}
\Text(120,-7)[]{\small $1$}
\Text(150,-7)[]{\small $3$}
\Text(180,-7)[]{\small $2$}
\Line(230,0)(290,100)
\Line(260,0)(260,100)
\Line(290,0)(230,100)
\Vertex(260,50){3}
\Text(260,-25)[]{NNN contact}
\Text(230,-7)[]{\small $1$}
\Text(260,-7)[]{\small $3$}
\Text(290,-7)[]{\small $2$}
\end{picture}
\end{center}
\vspace{0.7cm}
\caption{TPE, OPE and NNN contact interactions of the chiral three-body force at NNLO. For the OPE, the diagrams in which $1\leftrightarrow 2$ also needs to be considered.}
\label{fig:NNLO_diag}
\end{figure}
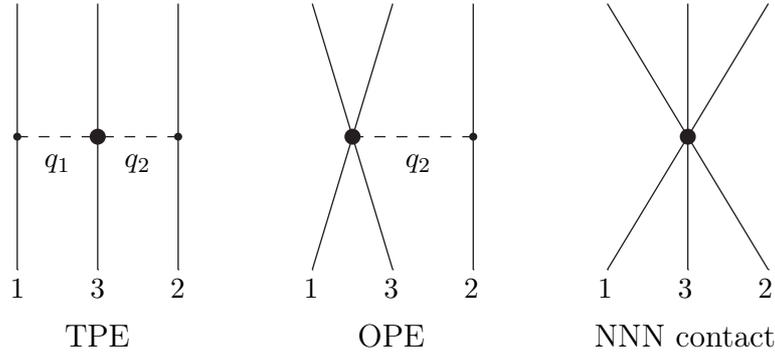

In a chiral theory without $\Delta$ degrees of freedom, the first nonvanishing three-nuclon interactions appear at NNLO in the Weinberg power counting scheme \cite{weinberg_90,weinberg_91}. The interaction  is described by three different physical mechanisms, corresponding to three different topologies of Feynman diagrams, drawn in Fig. \ref{fig:NNLO_diag} 
\cite{epelbaum_02}. The first two diagrams correspond to two-pion exchange (TPE) and one-pion exchange (OPE) with the pion emitted (or absorbed) by a contact NN interaction. The third diagram represents a contact three-nucleon interaction.

As shown in Eq. (\ref{eq:cycl_sum}), the full expression for the TNF is obtained by summing all possible permutations of the three nucleons. The Feynman diagrams of Fig. \ref{fig:NNLO_diag} refer to the permutation $(3:12)$ of the chiral potential $V^\chi$, that can be written as
\begin{align}
\hat{V}^{\chi}(3:12)&=c_1\hat{V}_1(3:12)+c_3\hat{V}_3(3:12)+c_4\hat{V}_4(3:12)\nonumber\\
             &+c_D\hat{V}_D(3:12)+c_E\hat{V}_E(3:12)\, .
\end{align}
The first three terms $\hat{V}_1$, $\hat{V}_3$ and $\hat{V}_4$ come from the TPE diagram and are related to $\pi N$ scattering. In particular, $\hat{V}_1$ describes the $S$-wave contribution, while $\hat{V}_3$ and $\hat{V}_4$ are associated with the $P$-wave. The other terms, $\hat{V}_D$ and $\hat{V}_E$, are the OPE and contact contributions, respectively. 
Their momentum space expressions are \cite{epelbaum_02} 
\begin{align}
\tilde{V}_1(3:12)&=-V_0m_{\pi}^2\,\tau_{12}\frac{(\vec{\sigma}_1\cdot\mathbf{q}_1)}{(q_{1}^2+m_{\pi}^2)}
\frac{(\vec{\sigma}_2\cdot\mathbf{q}_2)}{(q_{2}^2+m_{\pi}^2)}\nonumber\\
\tilde{V}_3(3:12)&=\frac{V_0}{2}\,\tau_{12}\frac{(\vec{\sigma}_1\cdot\mathbf{q}_1)}{(q_{1}^2+m_{\pi}^2)}
\frac{(\vec{\sigma}_2\cdot\mathbf{q}_2)}{(q_{2}^2+m_{\pi}^2)}\mathbf{q}_1\cdot\mathbf{q}_2\nonumber\\
\tilde{V}_4(3:12)&=\frac{V_0}{4}\,\vec{\tau}_3\cdot(\vec{\tau}_1\times\vec{\tau}_2)\,
\vec{\sigma}_3\cdot(\mathbf{q}_1\times\mathbf{q}_2)
\frac{(\vec{\sigma}_1\cdot\mathbf{q}_1)}{(q_{1}^2+m_{\pi}^2)}
\frac{(\vec{\sigma}_2\cdot\mathbf{q}_2)}{(q_{2}^2+m_{\pi}^2)}
\nonumber\\
\tilde{V}_D(3:12)&=-V_0^{D}\tau_{12}\Big[\frac{(\vec{\sigma}_2\cdot\mathbf{q}_2)}{(q_{2}^2+m_{\pi}^2)}(\vec{\sigma}_1\cdot\mathbf{q}_2)
+\frac{(\vec{\sigma}_1\cdot\mathbf{q}_1)}{(q_{1}^2+m_{\pi}^2)}(\vec{\sigma}_2\cdot\mathbf{q}_1)\Big]\nonumber\\
\tilde{V}_E(3:12)&=V_0^{E}\tau_{12}\, .
\end{align}
The strengths of the TPE, OPE and contact terms, $V_0$, $V_{0}^D$ and $V_{0}^E$ are given by
\begin{equation}
V_0=\Big(\frac{g_A}{F_{\pi}^2}\Big)^2\qquad V_{0}^D=\frac{g_A}{8 F_{\pi}^4 \Lambda_\chi}\qquad V_{0}^E=\frac{1}{F_{\pi}^4 \Lambda_\chi} 
\end{equation}
where $g_A=1.29$ is the axial-vector coupling constant, $F_\pi=92.4\,\text{MeV}$ is the weak pion decay constant and $\Lambda_\chi$ is
the chiral symmetry-breaking scale, of the order of the $\rho$ meson mass. 

The low energy constants (LEC) $c_1$, $c_3$ and $c_4$ also appear in the sub-leading two-pion exchange term of the chiral NN potential and are fixed by $\pi N$ \cite{fettes_98,buttiker_00} and/or $NN$ \cite{entem_03} data.
The parameters $c_D$ and $c_E$ are specific to the three-nucleon interaction and have to be fixed using NNN low energy observables, such as the $^3$H binding energy and the $nd$ doublet scattering length $^2a_{nd}$ \cite{epelbaum_02}.

The many-body methods employed in our work, namely FHNC/SOC and AFDMC, require a local expression of the three-body potential in coordinate space, that can be obtained performing  the Fourier transform \cite{navratil_07} 
\begin{align}
\hat{V}^{\chi}(3:12)&=\int \frac{d^3q_1}{(2\pi)^3}\frac{d^3q_2}{(2\pi)^3} \tilde{V}^{\chi}(3:12)
 F_{\Lambda}(q_{1}^2)F_{\Lambda}(q_{2}^2) {\rm e}^{i{\bf q}_1 \cdot {\bf r}_{13}}{\rm e}^{i{\bf q}_2 \cdot {\bf r}_{23}} \ ,
\end{align}
where the cutoff functions $F_{\Lambda}$, defined as
\begin{equation}
F_{\Lambda}(q_{i}^2)=\text{exp}\Big(-\frac{q_{i}^4}{\Lambda^4}\Big) \ ,
\label{eq:chi_cut}
\end{equation}
can depend on the momenta transferred among the nucleons, $q_i$, only. This feature has important consequences for the OPE and contact terms, that 
will be discussed at a later stage. 

The cutoff $\Lambda$ in the previous equation, while not being required to be the same as  $\Lambda_\chi$,  is of the same order of magnitude. 
Choosing the fourth power of the momentum in Eq.~(\ref{eq:chi_cut}) is therefore convenient, as the regulator generates powers of $q/\Lambda$ which 
are beyond NNLO in the chiral expansion. 

The Fourier transform can be readily computed, and provides the following coordinate-space representation of the chiral three-body potential:
\begin{align}
\hat{V}_1(3:12)&=W_0\,\tau_{12}(\vec{\sigma}_1\cdot\mathbf{r}_{13})(\vec{\sigma}_2\cdot\mathbf{r}_{23})y(r_{13})y(r_{23})\nonumber\\
\hat{V}_3(3:12)&=W_0\,\tau_{12}[\sigma_{12}y(r_{13})y(r_{23})
+(\vec{\sigma}_{1}\cdot\mathbf{r}_{23})(\vec{\sigma}_{2}\cdot\mathbf{r}_{23})t(r_{23})y(r_{13})\nonumber\\
&+(\vec{\sigma}_{1}\cdot\mathbf{r}_{13})(\vec{\sigma}_{2}\cdot\mathbf{r}_{13})t(r_{13})y(r_{23})
+(\mathbf{r}_{13}\cdot\mathbf{r}_{23})(\vec{\sigma}_{1}\cdot\mathbf{r}_{13})(\vec{\sigma}_{2}\cdot\mathbf{r}_{23})t(r_{13})t(r_{23})]\nonumber\\
\hat{V}_4(3:12)&=W_0\,(\vec{\tau}_3\cdot\vec{\tau}_1\times\vec{\tau}_2)
[(\vec{\sigma}_{3}\cdot\vec{\sigma}_{2}\times\vec{\sigma}_{1})y(r_{13})y(r_{23})\nonumber\\
&+(\vec{\sigma}_{3}\cdot\mathbf{r}_{23}\times\vec{\sigma}_1)(\vec{\sigma}_2\cdot\mathbf{r}_{23})
t(r_{23})y(r_{13})
+(\vec{\sigma}_2\cdot\mathbf{r}_{13}\times\vec{\sigma}_3)(\vec{\sigma}_1\cdot\mathbf{r}_{13})
t(r_{13})y(r_{23})\nonumber\\
&+(\vec{\sigma}_3\cdot\mathbf{r}_{23}\times\mathbf{r}_{13})(\vec{\sigma}_{1}\cdot\mathbf{r}_{13})(\vec{\sigma}_{2}\cdot\mathbf{r}_{23})
t(r_{13})t(r_{23})]\nonumber\\
\hat{V}_D(3:12)&=W_{0}^D\tau_{12}[\sigma_{12}y(r_{23})z_{0}(r_{13})
+(\vec{\sigma}_1\cdot\mathbf{r}_{23})(\vec{\sigma}_2\cdot\mathbf{r}_{23})t(r_{23})z_{0}(r_{13})\nonumber\\
&+\sigma_{12}y(r_{13})z_{0}(r_{23})
+(\vec{\sigma}_2\cdot\mathbf{r}_{13})(\vec{\sigma}_1\cdot\mathbf{r}_{13})t(r_{13})z_{0}(r_{23})]\nonumber\\
\hat{V}_E(3:12)&=W_{0}^E\tau_{12}z_{0}(r_{13})z_{0}(r_{23})\ , 
\label{eq:2pi_3body_conf}
\end{align}
where $W_0$, $W_{0}^D$ and $W_{0}^E$ are obtained multiplying the corresponding $V_0$, $V_{0}^D$ and $V_{0}^E$ by a factor $m_{\pi}^6/(4\pi)^2$.
The radial functions appearing in the above equations are defined as 
\begin{align}
y(r)&=\frac{z_{1}'(r)}{r}\nonumber\\
t(r)&=\frac{1}{r}y'(r)=\frac{1}{r^2}\Big(z_{1}''(r)-\frac{z_{1}'(r)}{r}\Big)
\label{eq:def_y_t}
\end{align}
while $z_n$, proportional to $Z_n$ introduced in Ref.~\cite{coon_81}, is given by
\begin{align}
z_n(r)&=\frac{4\pi}{m_{\pi}^3}\int\frac{d^3q}{(2\pi)^3}\frac{F_\Lambda(q^2)}{(q^2+m_{\pi}^2)^n}e^{i\mathbf{q}\cdot\mathbf{r}}
=\frac{2}{\pi m_{\pi}^3}\int dq q^2\frac{F_{\Lambda}(q^2)}{(q^2+m_{\pi}^2)^n}j_0(qr)\, ,
\label{eq:def_zn}
\end{align}
with $j_0(x)=\sin(x)/x$. 
Note that, due to the form of the cutoff function of Eq.~(\ref{eq:chi_cut}), the radial functions are not known in analytic form, and must be 
obtained from a numerical integration.

Recently, the authors of Ref.\cite{kievsky_10} have studied the low energy NNN observables using the hyperspherical harmonics formalism and a nuclear hamiltonian including the NNLOL potential and the Argonne $v_{18}$ \cite{wiringa_95} two-body interaction. 

This mixed approach requires a fit of all the LECs appearing in the chiral  three-body interaction, not $c_D$ and $c_E$ only. Hence, consistency in the 
treatment of two- and three- nucleon interactions, that would be achievable using a hamiltonian in which all potentials are derived from chiral effective theory, is lost.  
Nevertheless, it is possible to exploit chiral perturbation theory to assess the importance of the different terms contributing to the TNF. This procedure allows one 
to select the most relevant spin-isospin structures  entering the three-nucleon potential, as well as the shape of the corresponding radial functions,. 

Within the chiral approach, to obtain a potential yielding a fit to the experimental data of accuracy comparable to that achieved by the Argonne $v_{18}$ model, 
one has to include terms up to NNNLO [9, 10]. As a consequence, a fully consistent calculation in principle requires a NNNLO three-body interaction, 
the expression of which has been only recently derived in Ref. \cite{bernard_08,bernard_11}. It turns out that some of the terms appearing at NNNLO 
can be taken into account shifting the constants $c_i$ of about  20-30\% with respect to their original values \cite{bernard_08}. This procedure has been followed 
in precision studies of TNF.
By fitting all the LECs of the NNLOL interaction,  the authors of Ref.~\cite{kievsky_10} have improved upon the NNLO approximation, as they have effectively included the corrections to the $c_i$ appearing at NNNLO level.

The best fit parameters for the $^3$H and $^4$He binding energies and for the $nd$ scattering length, $^2a_{nd}$, are listed in Table \ref{tab:chi_parameters}. For all the different parametrizations, denoted by NNLOL$_i$, $c_1$ and $\Lambda_\chi$ have been fixed to their original values $0.00081\, \text{MeV}^{-1}$  and $700\,\text{MeV}$, respectively \cite{epelbaum_02}. 
The momentum cutoff of Eq.~(\ref{eq:chi_cut}) has been set to $500\,\text{MeV}$.

\begin{table}[h!]
\begin{center}
\caption{Parameters of the NNLOL interactions of Ref.~\cite{kievsky_10}. \label{tab:chi_parameters}}
\vspace{0.3cm}
\begin{tabular}{c c c c c c} 
\hline 
 Potential &  $c_3\,(\text{MeV}^{-1})$  &  $c_4\,(\text{MeV}^{-1})$  &  $c_D$  &  $c_E$ \\ 
\hline
NNLOL$_1$ & -0.00448 & -0.001963 & -0.5 & 0.100  \\ 
NNLOL$_2$ & -0.00448 & -0.002044 & -1.0 & 0.000  \\ 
NNLOL$_3$ & -0.00480 & -0.002017 & -1.0 &-0.030 \\
NNLOL$_4$ & -0.00544 & -0.004860 & -2.0 &-0.500 \\
\hline
\end{tabular} 
\vspace{0.1cm}
\end{center}
\end{table}

As noticed in Ref.~\cite{friar_99}, despite the different underlying physical mechanisms, both TM and UIX three-nucleon interactions can be written as a sum of terms of the same form as those appearing in Eq.~(\ref{eq:2pi_3body_conf}). The differences among NNLOL, TM and UIX lie in the constants and in the radial functions. 

The TM$^\prime$ potential only involves the $V_1$, $V_3$ and $V_4$ contributions \cite{coon_01}. The cutoff function for this potential is not the same as in Eq.~(\ref{eq:chi_cut}), but 
\begin{equation}
F_{\Lambda}(q^2)=\Big(\frac{\Lambda^2-m_{\pi}^2}{\Lambda^2+q^2}\Big)^2\,.
\label{eq:TM_cut}
\end{equation}
The above form allows for the analytical integration of Eq.~(\ref{eq:def_zn}), yielding the radial functions
\begin{align}
y(r)&=\frac{e^{-r\Lambda}}{2m_{\pi}^3 r^3}\Big[2-m_{\pi}^2r^2-2(1+m_\pi r)e^{r(\Lambda-m_\pi)}+r\Lambda(2+r\Lambda)\Big]\nonumber\\
t(r)&=\frac{e^{-r\Lambda}}{2m_{\pi}^3 r^5}\Big[-6+2(3+3m_\pi r+m_{\pi}^2r^2)e^{r(\Lambda-m_\pi)}\nonumber \\
&\quad+m_{\pi}^2r^2(1+r\Lambda)-r\Lambda[6+r\Lambda(3+r\Lambda)]\Big]\, .
\end{align}

The TM$^\prime$ potential corresponds to the following choice of the strength constants (compare to Eq.~(\ref{eq:2pi_3body_conf}))
\begin{align} 
W_{0}=\Big(\frac{g m_\pi}{8\pi m_{N}}\Big)^2m_{\pi}^4
\end{align}
and
\begin{equation}
c_1 = \frac{a}{m_{\pi}^2} \quad,\quad c_3=2b\quad,\quad c_4=-4d \ ,
\end{equation}
$a$, $b$ and $c$ being the parameters entering the definition of the TM$^\prime$ potential \cite{coon_01}. 
The authors of Ref.\cite{kievsky_10} have determined the parameters of the TM$^\prime$ potential fitting the same set of low energy NNN observables
employed for the NNLOL potential.
In order to get a better description of the experimental data, they introduced a repulsive three-nucleon contact term, similar to the 
chiral $V_E$, but with $\tau_{12}$ omitted
\begin{equation}
\hat{V}_{E}(3:12)=W_{0}^{E}z_0(r_{13})z_0(r_{23})\, ,
\end{equation}
where
\begin{equation}
W_{0}^{E}=\Big(\frac{g m_\pi}{8\pi m_{N}}\Big)^2\frac{9 m_{\pi}^2}{\Lambda_\chi}\, .
\label{eq:V0ETM}
\end{equation}
The corresponding radial function can be computed analytically from Eq.~(\ref{eq:def_zn})
\begin{equation}
z_{0}(r)=\frac{e^{-r\Lambda}}{8\pi\Lambda}(m_{\pi}^2-\Lambda^2)^2\, .
\end{equation}

As in the original paper \cite{coon_81}, in Ref.~\cite{kievsky_10} the value of the pion-nucleon coupling constant is set to $g^2=179.7\,\text{MeV}$,  the pion mass is  $m_\pi=139.6\,\text{MeV}$ and the nucleon mass is defined through the ratio $m_N/m_\pi=6.726$. The symmetry breaking scale $\Lambda_\chi$ of Eq.~(\ref{eq:V0ETM}) has the same value,  $700\,\text{MeV}$,  
used for the NNLOL potential.

The parameters of the TM$^\prime$ potentials, TM$^\prime_i$, that according to Ref.\cite{kievsky_10} reproduce the binding energies of $^3$H and  $^4$He and $^2a_{nd}$, are listed in Table \ref{tab:tm_parameters}. It turns out that $V_1$, gives a very small contribution to the low energy NNN observables. Therefore, the parameter $a$ has been 
kept to its original value $-0.87\,m_{\pi}^{-1}$. 

\begin{table}[h!]
\begin{center}
\caption{Parameters of the TM$^\prime$ potential reproducing low energy the NNN experimental data with $a=-0.87\,m_{\pi}^{-1}$
 \cite{kievsky_10}. \label{tab:tm_parameters}}
\vspace{0.3cm}
\begin{tabular}{c c c c c c} 
\hline 
 Potential & $b(m_{\pi}^{-3})$ & $d(m_{\pi}^{-3})$ & $c_E$ & $\Lambda(m_\pi)$\\ 
\hline
%$TM_0$  & -2.580 & -0.753 & 0.0 & 4.8 \\
TM$^\prime_1$  & -8.256 & -4.690 & 1.0 & 4.0 \\ 
TM$^\prime_2$  & -3.870 & -3.375 & 1.6 & 4.8 \\ 
TM$^\prime_3$  & -2.064 & -2.279 & 2.0 & 5.6 \\
\hline
\end{tabular} 
\vspace{0.1cm}
\end{center}
\end{table}

It can be shown that the anticommutator and commutator terms of the UIX potential, displayed in Eq. (\ref{eq:fm_uix}), correspond to $V_3$ and $V_4$ of Eq.~(\ref{eq:2pi_3body_conf}), provided the following relations between the constants
\begin{align}
c_3W_0&=4A_{2\pi}\nonumber\\
c_4W_0&=4C_{2\pi}\, 
\label{eq:const_rel}
\end{align}
and the radial functions
\begin{equation}
\left\{
\begin{array}{rl}
Y(r)&=y(r)+\frac{r^2}{3}t(r)\\
T(r)&=\frac{r^2}{3}t(r)
\end{array} \right.
\label{eq:rad_gen_UIX}
\end{equation}
are satisfied.

On the other hand, the repulsive term of the UIX potential of Eq. (\ref{eq:VR_uix}) is equivalent to the $V_E$ term appearing in the TM$^\prime$ potential and (aside from the $\tau_{12}$ factor) in the NNLOL chiral potential if the following relations 
hold
\begin{equation}
T^2(m_\pi r)=z_0(r)\quad,\quad U_0=c_E W_{0}^E\, .
\label{eq:E_equiv}
\end{equation}

In Ref. \cite{kievsky_10} it has been found that the original parametrization of the UIX potential underestimates $^2a_{nd}$ and slightly overbinds  of $^4$He.

The authors of Ref.~\cite{kievsky_10} have calculated the differential cross section and the vector and tensor analyzing powers of $p-d$ scattering at $E_{lab} = 3$ MeV for the different parametrizations of NNLOL and TM$^\prime$ potentials. They found that all of them lead to underestimating $A_y$ (the so-called $A_y$ puzzle remains unsolved) and $T_{11}$, while the central minimum in $T_{21}$  is always overestimated.  However, NNLOL model provides a slight improvement with respect to the UIX potential in the description of the polarization observables. On the other hand, no substantial modifications from the UIX results are given by the TM$^\prime$ interactions.

\newpage             
\thispagestyle{empty}     

% Chapter 2 %%%%%%%%%%%%%%%%%%%%%%%%%%%%%%
\chapter{Many body description of nuclear matter}
\label{chapt:mb}
{\em Ab initio} nuclear many-body approaches are based on the premise that nuclear dynamics can be modeled studying exactly solvable systems, having mass number $A \leq 3$. This is a most important feature since, due to the complexity of strong interactions and to the prohibitive difficulties associated with the solution of the quantum mechanical many-body problem, theoretical calculations of nuclear observables generally involve a number of approximations. Hence, models of nuclear dynamics extracted from analyses of the properties of complex nuclei are plagued by the systematic uncertainty associated with the use of a specific approximation scheme.

In ab initio approaches, the hamiltonian entering the time-independent many-body Schoredinger equation
\begin{equation}
\hat{H}\Psi_n(x_1,\dots,x_A)=E_n\Psi_n(x_1,\dots,x_A)
\label{eq:mb_sch}
\end{equation}
is the one defined in Section \ref{sec:nucl_ham}, without any additional adjustable parameters. 

In the first Section of this Chapter we discuss the independent particle model, and argue that it is not suitable to encompass correlation structure induced by the nuclear hamiltonian. The following Sections are devoted to more advanced  approaches, allowing one to take into account correlation effects. We will focus on the variational method, based on correlated basis function (CBF) theory, and the diffusion Monte Carlo technique.

\section{Mean field approach: the Hartree-Fock method}
\label{sec:hfa}
D. R. Hartree \cite{hartree_28}, V. A. Fock \cite{fock_30} and J. C. Slater \cite{slater_30}, proposed to use as a starting point toward the solution of the many-body Schroedinger equation describing atomic electrons, the {\it central field approximation}. Within this approximation, based on the {\it independent particle model}, each nucleon moves in a single-particle effective potential representing the average effect of the interactions with the other $A-1$ nucleons. Each nucleon is described by its own wave function, $\psi_{n_i}(x_i)$ eigenfunction of the hermitian operator $\hat{h}^{HF}$
\begin{align}
\hat{h}^{HF}(x_i)\psi_{n_i}(x_i)&=\epsilon_{n_i}\psi_{n_i}(x_i)\,,
\label{eq:hf_intro}
\end{align}
where the generalized coordinate $x_i=\{\mathbf{r}_i,\sigma_i,\tau_i\}$ represents both the position and the spin- isospin variables of the i-th nucleon.
The operator $\hat{h}^{HF}$, called the one-particle Fock hamiltonian, is given by
\begin{equation}
\hat{h}^{HF}(x_i)=-\frac{\nabla^{2}_i}{2m}+\hat{v}^{HF}(x_i)\, ,
\end{equation}
where $\hat{v}^{HF}(r)$ is the single particle Hartree-Fock potential, that is built from the states $\psi_{n_i}(x_i)$ using a {\it self-consistent} iterative procedure, based on the variational principle, described in Appendix \ref{app:hf}. 

Within this approximation, the many particle ground-state for a system made of $A$ nucleons is a single Slater determinant $\Psi$ of one-nucleon states
\begin{equation}
\Psi=\mathcal{A}[\psi_{n_1}(x_1)\dots\psi_{n_a}(x_a)]\, .
\label{eq:gs_slater}
\end{equation}
where $\mathcal{A}$ is the antisymmetrization operator.

As shown in Appendix \ref{app:hf}, the {\it single particle energy} $\epsilon_i$ is given by
\begin{equation}
\epsilon_{n_i}=\frac{\mathbf{k}_{i}^2}{2m}+\sum_{n_j=1}^A \int dx_j\psi_{n_i}^*(x_i)\psi_{n_j}^*(x_j)\hat{v}_{ij}[\psi_{n_i}(x_i)\psi_{n_j}(x_j)-\psi_{n_j}(x_i)\psi_{n_i}(x_j)]\,,
\label{spe_nm1}
\end{equation}
where $\int dx_j$ stands for integration over the coordinate $\mathbf{r}_j$ and trace over the spin and isospin variables of the $j$--th nucleon.

The total energy of the system, $E[\Psi]$, is {\it not} the sum of the single particle energies, but rather 
\begin{equation}
E[\Psi]=\sum_{n_i}\epsilon_{n_i}-\frac{1}{2}\sum_{n_i,n_j}\int dx_j\psi_{n_i}^*(x_i)\psi_{n_j}^*(x_j)\hat{v}_{ij}[\psi_{n_i}(x_i)\psi_{n_j}(x_j)-\psi_{n_j}(x_i)\psi_{n_i}(x_j)]\, .
\end{equation}

A physical meaning to the single particle energies can be given through {\it Koopmans' theorem}. Assuming that the spin orbitals of the $A-1$ system are the same as those of the $A$ system, from the previous equation it can be shown that $\epsilon_{n_i}$ is the {\it separation energy} of the nucleon in the state $\psi_{n_i}$
\begin{equation}
\epsilon_{n_i}=E_A-E_{A-1}(i)\, .
\end{equation}

As explained before, the self-consistent field method allows for the determination of the spin-orbitals of the $A$ occupied states, $\{\psi_1,\dots,\psi_A\}$, with single-particle energies $\epsilon_1,\dots,\epsilon_A$, $E_A$ being the Fermi energy of the system. The remaining eigenfunctions of $\hat{h}^{HF}$, which satisfy Eq. (\ref{eq:hf}), are associated with unoccupied (virtual) states having single particle energies larger than the Fermi energy. Unlike $\{\psi_1,\dots, \psi_A\}$, they are not determined in a self-consistent fashion, as they do not enter the definition of the Fock hamiltonian. 

The key point of the Hartree Fock approach is that occupied and virtual states provide a natural basis to describe the many-body system \cite{lipparini_08}. While the many-body ground state is the Slater determinant of occupied single-particle states, Eq. (\ref{eq:gs_slater}),excited many-body states are constructed by removing $n$ occupied states from the Slater determinant and replacing them with $n$ virtual states. Such excited states are called  $n-particle\ n-hole$ ($np\ nh$) states and are eigenstates of the Hartree Fock hamiltonian, also known as ``Fokian''
\begin{align}
\hat{H}^{HF}&=\sum_{i=1}^A \hat{h}^{HF}_i \nonumber \\
\hat{H}^{HF}|\Psi_{h_1,\dots,h_n;p_1,\dots,p_n}\rangle&=
\Big[\sum_{n_i=1}^A\epsilon_{n_i}+\sum_{i=1}^n (\epsilon_{p_i}-\epsilon_{h_i})\Big]|\Psi_{h_1,\dots,h_n;p_1,\dots,p_n}\rangle\, .
\label{eq:fokian_eigen}
\end{align}

The Hartree-Fock procedure is the basis, for instance, of the nuclear shell model, that has been successfully applied to explain many nuclear properties. \cite{mayer_49,mayer_50,haxel_49}. As far as nuclear matter is concerned, the single particle wave functions are known to be plane waves, as dictated by translation invariance. Therefore, a uniform system can be conveniently described within a box of volume $V$ with periodic boundary conditions \cite{fetter_03}, using the wave functions
\begin{equation}
\psi_{n_i}(x_i)=\frac{e^{i\mathbf{k}_i \cdot \mathbf{r}_i}}{\sqrt{V}}\eta_{\alpha_i}\, ,
\label{eq:nm_wf}
\end{equation}
where $\eta_{\alpha_i}\equiv\chi_{\sigma_i}\chi_{\tau_i}$ represents the product of Pauli spinors describing the spin and the isospin of particle $i$.
In order to satisfy the periodic boundary conditions, the wave vector $\mathbf{k}$ is discretized; for a cubic box of side $L$, it turns out that
\begin{equation}
k_i=\frac{2\pi}{L}n_i\qquad i=x,y,z \qquad n_i=0,\pm 1,\pm 2,\dots  \, .
\label{eq:discrete_mom}
\end{equation}
The momentum of the occupied states is smaller than the Fermi momentum $\mathbf{k}_F$, which is related to the density of the system, $\rho$, through $k_F=(6\pi^2\rho/\nu)^{1/3}$, and $\nu$ is the spin-isospin degeneracy ($\nu=2$ for PNM, $\nu=4$ for SNM).

The plane waves of Eq. (\ref{eq:nm_wf}) are already solutions to the Hartree Fock equations; in other words they are the best single-particle wave functions for uniform systems. A remarkable feature of nuclear matter is that the starting single particle wave functions are known and simple, unlike what happens, for instance, in finite nuclei. Due to the lack of translation invariance, even generating the single particle wave functions is a difficult task, as it requires the solution of the Hartree-Fock equations \cite{fetter_03}.

The single particle energy of nuclear matter can be easily derived substituting the wave function of Eq. (\ref{eq:nm_wf}) in Eq. (\ref{eq:spe_1}). 
In the case of SNM ($\nu=4$) for potentials of the form of Argonne $v_{18}$, carrying out the summation over the occupied states with $|\mathbf{k}_j|\leq k_F$ yields
\begin{equation}
\epsilon_{n_i}=\frac{\mathbf{k}_{i}^2}{2m}+\rho \int d\mathbf{r}_{ij}\Big[v^{c}_{ij}-\frac{1}{4}\ell(k_Fr_{ij})e^{-i\mathbf{k}_i\cdot \mathbf{r}_{ij}}(v^{c}_{ij}+3v^{\tau}_{ij}+3v^{\sigma}_{ij}+9v^{\sigma\tau}_{ij})\Big]\,.
\label{eq:spe_nm2}
\end{equation}
where the Slater function is given by
\begin{equation}
\ell(k_F r)=3\Big[\frac{\sin(k_Fr)-3k_Fr \cos(k_Fr)}{(k_Fr)^3}\Big]\, .
\end{equation}
Summing over spin-isospin states of Eq. (\ref{spe_nm1}) amounts to tracing over the spin-isospin variables of the nucleon $j$. Such a trace is normalized, as it incorporates the factor $1/\nu$ coming from the summation  over the momentum $\mathbf{k}_j$. The factor $A$ arising from the same sum, divided by wave function normalization factor $V$ produces the factor $\rho$, appearing in Eq. (\ref{eq:spe_nm2}).

Standard perturbation theory performed in the basis of the Hartree-Fock solutions can not cope with the repulsive core of the nuclear force, which cause individual terms of the perturbative expansion to diverge \cite{akmal_98}. 

As an example\cite{jastrow_55}, consider the scalar repulsive potential 
\begin{equation}
v(\mathbf{r})= 
\begin{cases}
      |v_0| &  |\mathbf{r}| \leq r_0\\
      0      &   |\mathbf{r}| > r_0\, .
    \end{cases}
\end{equation}
The single particle energy computed from Eq. (\ref{eq:spe_nm2}) using this potential is seen to be of order $\rho\,r_{0}^3 |v_0|$; if the potential approaches the hard sphere interaction, similar to the strong repulsive core of the nuclear interaction, the single particle energy keeps increasing. 

In other words, since the eigenfunctions of the Fock hamiltonian are the same as those of the non interacting Fermi gas, the many-body wave function largely differs from the exact ground state associated with the nuclear hamiltonian. Standard perturbation theory in such a basis can not be expected to be convergent as the matrix elements of the nuclear hamiltonian between $np\ nh$ states are not perturbative corrections to the ground state expectation value.

To circumvent this problem, one can follow two different strategies, leading to either G-matrix or correlated basis function (CBF) perturbation theory. 

Within the former approach proposed by Brueckner \cite{brueckner_54,brueckner_54b,brueckner_55,brueckner_55b}, the bare potential $v_{ij}$, is replaced by a well behaved effective interaction, the G-matrix, which is obtained by summing up the series of particle-particle ladder diagrams. The physical basis of this theory was elucidated by Bethe \cite{bethe_56}, while Goldstone introduced the linked cluster expansion \cite{goldstone_57}. For a more recent review of the G-matrix approach, also known as Brueckner-Bethe-Goldstone expansion, see Refs. \cite{muther_00,baldo_07}.

In this Thesis  we have been using the CBF approach, to which the following Section is devoted. 

\section{Correlated basis functions theory}
Theories of Fermi liquids based on correlated basis functions are a natural extension of variational approaches in which the trial ground state wave function is written in the form
\begin{equation}
|\Psi_0)=\frac{\hat{\mathcal{F}}|\Psi_0\rangle}{\langle \Psi_0|\hat{\mathcal{F}}^\dagger \hat{\mathcal{F}} | \Psi_0\rangle},
\end{equation}
where $\hat{F}$ is suitable many-body correlation operator. The simplest choice suitable for dealing with the strong short-range repulsions is the scalar correlator of the form
\begin{equation}
\hat{\mathcal{F}}=\prod_{j>i=1}^Af(r_{ij})\, ,
\label{eq:jastrow_corr}
\end{equation}
known as Jastrow correlator \cite{jastrow_55}. However, this choice for the correlation operator is only suitable for purely central potentials, such as those describing the interaction between $^3$He atoms. For state-dependent potentials, like the Argonne nuclear interaction, spin-isospin dependent correlations, to be introduced at a later stage, are needed.  

The variational approach consists in the minimization of the expectation value of the hamiltonian
\begin{equation}
E_V=(\Psi_0|\hat{H} | \Psi_0)
\end{equation}
which is an upper bound to the true ground-state energy $E_0$. For instance, in the pure central Jastrow case, minimizing $E_V$ allows for finding the radial function $f(r_{ij})$. Apart from the technical difficulties involved in finding the optimal radial function, it is clear that the resulting correlation function is small within the repulsive region of the NN potential.

As noted in the review of Clark \cite{clark_79}, historically, the development of the variational approach has been somewhat discouraged not only by the difficulties involved in the calculation of $E_V$, potentially leading to violations of the variational principle, but also by a psychological obstacle: the embarrassing conceptual simplicity of the method, in other words, its lack of ``snob appeal''.  

Nevertheless, the variational approach succeeded in treating the atomic helium in both the liquid and solid phases \cite{feenberg_69,croxton_78}. Although nowadays the numerical problem of solving the many-body Schr\"{o}dinger equation for the ground state has been resolved, to a large extent, by the Green's function Monte Carlo method \cite{kalos_81}, this approach does not provide a quantitative understanding of the ground-state wave function \cite{usmani_82}. However, the knowledge of the analytic form of the ground-state function would be particularly useful to extend the microscopic theory to treat the elementary excitations and finite-temperature properties of helium liquids. A successful approach in this direction has been provided by the variational theory \cite{moroni_95}, including also the {\it back flow correlation}, proposed by Feynman and Cohen \cite{feynman_56}: a velocity dependent correlation, arising from the flow induced by a moving atom. 

As far as the nuclear many body problem in concerned, this method is supported by a variety of experimental evidence \cite{benhar_93,pandha_97} showing that short range NN correlations are a fundamental feature of nuclear structure. The description of nuclear dynamics in terms of interactions derived in coordinate space appears to be the most appropriate, for both conceptual and technical reasons. First of all, correlations between nucleons are predominantly of spatial nature, in analogy with what happens in all known strongly correlated systems, like liquid $^4$He. In addition, one needs to clearly distinguish the effects due to the short-range repulsion from those due to relativity.

The correlated basis theories of Fermi liquids are a natural extension of the variational approach. A non orthogonal but complete set of correlated basis states can be defined as \cite{clark_66,fantoni_98}
\begin{equation}
|\Psi_n)\equiv\frac{\hat{\mathcal{F}}|\Psi_n\rangle}{\langle \Psi_n|\hat{\mathcal{F}}^\dagger \hat{\mathcal{F}} | \Psi_n\rangle},
\end{equation}
where $|\Psi_n\rangle$ is the $n-particle\ n-hole$ state of Eq. (\ref{eq:fokian_eigen}). The correlation operator,$\hat{\mathcal{F}}$, is determined by the variational calculation of the ground state energy. The variational energies $E_{n}^v$, although only  $E_{0}^v$ has been variationally estimated, are given by the diagonal matrix elements of the hamiltonian between correlated states
\begin{equation}
E_{n}^v=(\Psi_n|\hat{H}|\Psi_n)\, .
\label{eq:varen_def}
\end{equation}
The energies $E_{n}^v$ are extensive quantities, as they are of order $A$, while excitation energies $E_{n}^v-E_{0}^v$ are of order 1. 

In order to compute the perturbative corrections to the variational energies, the hamiltonian $H$ is decomposed in two terms
 \begin{align}
\hat{H}=\hat{H}^0+\hat{H}^1\, .
\label{eq:H0H1}
\end{align}
where, as will became clear in the following, neither $\hat{H}_0$ nor $\hat{H}_1$ are hermitian operators. 
The ``unperturbed'' hamiltonian $\hat{H}^0$ is defined through the correlated basis states and the variational energies, in such a way that
\begin{equation}
\hat{H}^0|\Psi_n)=E_{n}^v|\Psi_n)\, 
\label{eq:H0_def}
\end{equation}
Notice that, since the correlated states are not orthogonal, $\hat{H}^0$ is not diagonal in this basis
\begin{equation}
(\Psi_n|\hat{H}^0|\Psi_m)=E^{v}_m N_{nm}\, .
\label{eq:H0_matel}
\end{equation}
The metric matrix $N_{nm}$ is defined by
\begin{equation}
N_{nm}\equiv\delta_{nm}+S_{nm}\equiv(\Psi_n|\Psi_m)\, ,
\label{eq:N_matel}
\end{equation}
where $S_{nm}$ is the overlap matrix, with
\begin{equation}
S_{nn}=0\, .
\end{equation}

It is convenient to distinguish the diagonal part from the not diagonal part of the hamiltonian 
\begin{equation}
H_{nm}\equiv (\Psi_n|\hat{H}|\Psi_m)=E^{v}_m\delta_{nm}+H^{\prime}_{nm}\, ,
\end{equation}
where $H^\prime_{mm}=0$. 
The closer the CBF states are to the true eigenstates of the hamiltonian, the smaller $H^{\prime}_{nm}$ becomes. Using
Eq. (\ref{eq:H0_matel}) and (\ref{eq:N_matel}) it turns out that the off-diagonal part of the hamiltonian matrix element is
\begin{equation}
H^{\prime}_{nm}=E^{v}_mS_{nm}+H^{1}_{nm}\ ,.
\end{equation}
This is consistent with the fact that $H^{1}_{nn}=0$, resulting from Eqs. (\ref{eq:varen_def}),(\ref{eq:H0H1}) and (\ref{eq:N_matel}).

Assuming that the nondiagonal elements of both the metric and the hamiltonian be small, there have been two fundamental ways of treating the problem pertubatively. One way, consists in diagonalizing $H_{nm}$ as it is, without bothering to orthogonalize the basis, using the {\it nonorthogonal perturbation theory}. The other way is employing some procedure to orthogonalize the basis first, and then apply the standard perturbation theory. 

Within the former approach, the authors of Ref. \cite{kulas_73}, introducing the so called {\it diagonal metric}, were able to show that the perturbative corrections to the variational energy $E_{n}^v$ can be casted in a way that is formally identical to standard perturbation theory in an orthogonal basis 
\begin{align}
E_n-E_{v}^n=&V_{nn}+\sum_{k\neq n}\frac{V_{nk}V_{kn}}{E_{n}^v-E_{k}^v}+\nonumber \\
&\Big(
\sum_{k\neq n,m\neq n}\frac{V_{nm}V_{mk}V_{kn}}{(E_{n}^v-E_{m}^v)(E_{n}^v-E_{k}^v)}-V_{nn}\sum_{k\neq n} \frac{V_{nk}V_{kn}}{(E_{k}^v-E_{n}^v)^2}\Big)+\, \dots
\label{eq:energy_corr}
\end{align}
where $E_n$ is the exact eigenvalue of the full hamiltonian $\hat{H}$, the eigenstates of which are denoted as 
$|\Psi_n \}$
\begin{equation}
\hat{H} | \Psi_n \} = E_n |\Psi_n \}\, .
\label{eq:exact_h}
\end{equation}
The differences with respect to the orthogonal case are enclosed in the matrix $V_{nm}$, the perturbative expansion of which reads 
\begin{align}
V_{nm}=&(H^{\prime}_{nm}-S_{nm}E_{n}^v)-\sum_k S_{nk}(H^{\prime}_{km}-S_{km}E^{v}_n)+\nonumber \\
&\sum_{k,l}S_{nk}S_{kl}(H^{\prime}_{lm}-S_{lm}E^{v}_n)+\, \dots\,.
\end{align}

Replacing $V_{nm}$ in Eq. (\ref{eq:energy_corr}) with its expansion leads to
\begin{align}
E_n-E_{v}^n=\sum_{k\neq n}\frac{(H^\prime_{nk}-E_nS_{nk})(H^\prime_{kn}-E_nS_{kn})}{E_{n}^v-E_{k}^v}+\,\dots
\label{eq:energy_corr2}
\end{align}

Earlier derivations of the latter result, not involving the diagonal metric formalism, can be found in Refs. \cite{morse_53,clark_59}.
Like in ordinary many-body perturbation theory, each order of the perturbative expansion diverges with the number of particle, $A$. However in Ref. \cite{fantoni_84} it has been shown that divergent terms appearing at different orders cancel each other.

A major difference with respect to ordinary many-body perturbation theory is that there is an energy dependence in the matrix element $V_{nm}$ of Eq. (\ref{eq:energy_corr}), arising from the non orthogonality of the CBF state. Another peculiar feature of CBF perturbation theory is the fact that $V_{nm}$ is a many-body operator, as, through $\hat{F}$, it incorporates the effect of the correlations among all the particles of the system.

In the earlier calculations \cite{clark_75,jackson_82}, where the correlator was taken to be of the simple Jastrow form of Eq. (\ref{eq:jastrow_corr}), the second order term of the perturbative corrections has been found to be large. The NN potential has indeed a complicate spin-isospin structure, that can not be encompassed considering radial correlations only. In particular, since this wave function is spherically symmetric, the expectation value of the tensor component of the NN interaction averages to zero. In the pure Jastrow case, the CBF states are not sufficiently close to the exact eigenstates of the hamiltonian, and more terms in the perturbative series need to be calculated. In liquid $^3$He or in electron gas, where the potential is purely central, Jastrow CBF is much more justified \cite{friman_82}.

A generalization of the Jastrow correlation operator whose structure reflects the complexity of the NN interaction has been proposed in Ref. \cite{pandharipande_71,ristig_71,pandharipande_72}
\begin{equation}
\hat{\mathcal{F}}(X)=\Big(\mathcal{S} \prod_{j>i=1}^A \hat{F}_{ij} \Big)\ ,
\label{eq:Foperator}
\end{equation}
with
\begin{equation}
\hat{F}_{ij} = \sum_{p=1}^6 f^{p}(r_{ij})\hat{O}^{p}_{ij} \ .
\label{eq:Foperator6}
\end{equation}

Note that the symmetrization operator $\mathcal{S}$ is needed to fulfill the requirement of antisymmetrization of the state $|\Psi_n\rangle$, since, in general, $[\hat{O}^{p}_{ij},\hat{O}^{q}_{ik}] \neq 0$.

Since the first six operators present in the NN potential form a closed set, this choice for the correlation operator has a tremendous advantage in analytic manipulation necessary to compute the energy per particle. As will be shown in Section \ref{sec:fhnc_soc_op}, the product of any two of the $O^p$, $p<6$, can be reduced to a linear combination of elements from this set. 

In this Thesis we will stick to this choice for $\hat{\mathcal{F}}$ although in Ref. \cite{lagaris_80} the correlation operator has been extended including spin-orbit correlations. In fact, the variational choice of Eq. (\ref{eq:Foperator6}) ($F_6$ model) implies that spin orbit correlations are neglected. We motivate this choice mainly with the technical difficulties of consistently including spin-orbit correlations; in spite of the calculations performed of Ref. \cite{lagaris_80}, we believe that the contribution of the spin-orbit correlation is still an open problem. In several FHNC/SOC calculations of the binding energy of SNM the spin orbit terms of the potential have been included only pertubatively. Moreover in all the FHNC/SOC calculations of the linear response \cite{fantoni_87,fabrocini_89}, optical potential and Green's function \cite{benhar_89,benhar_92} of SNM, whose results have been used to explain a variety of experimental data, spin-orbit correlations have been neglected.

Before moving to the cluster expansion technique, which has been developed to compute the matrix elements of the hamiltonian, it is worth spending few words on orthogonal CBF theory. As a matter of fart, a clear analysis of the convergence properties of the non-orthogonal CBF perturbation theory has not been performed yet. For instance, the truncation of the series at some perturbative order leads to non-orthogonality spuriousities, whose effects may not be negligible. Moreover, the calculation of quantities other than the ground-state energy, like the response function, is made difficult by the fact that properly orthogonalized eigenvectors cannot be easily extracted the nonorthogonal CBF basis.

If one attempts to diagonalize the CBF states using the standard L\"owdin transformation \cite{lowdin_50}, the resulting states are worst than the original one. For instance, the expectation value of the hamiltonian in the L\"owdin orthogonal ground-state is larger than $E_{0}^v$. To avoid this inconvenient, a two-step orthogonalization procedure which preserves the variational diagonal matrix elements of the hamiltonian and allows for using ordinary orthogonal perturbation theory in zero temperature calculations, has been developed \cite{fantoni_88}.

\section{Cluster expansion formalism}
\label{sec:cluster_exp}
Both correlation operators of Eq. (\ref{eq:jastrow_corr}) and (\ref{eq:Foperator6}) are defined in such a way to possess the cluster property. This means that if the system is split in two (or more)  subset of particles that are moved far away from each other, the correlation operator factorizes into a product of two (or more) correlation operators in such a way that only particles belonging to the same subset are correlated. For instance, consider two subsets, say $i_1,\dots\,i_m$ and $i_{m+1},\dots\,i_A$; the cluster property implies
\begin{equation}
\hat{\mathcal{F}}(x_1,\dots,x_A)\to \hat{\mathcal{F}}(x_{i_1},\dots,x_{i_m})\times \hat{\mathcal{F}}(x_{i_{m+1}},\dots,x_{i_A})\, .
\end{equation}

The above property allows for expanding the matrix elements of the hamiltonian, (or of any other many-body operator), between CBF states in sum of terms involving an increasing number of particles, knowns as {\it clusters}. In the literature are present both analytic \cite{clark_79,jackson_82} and diagrammatic cluster expansion formalisms \cite{fantoni_74,fantoni_75,krotscheck_75}; moreover different classification schemes have been adopted, corresponding to different choices for the smallness parameters of the perturbative expansion.

In the calculation of the expectation value of any many-body operator it is convenient to perform separate cluster expansion for the numerator and the denominator, the latter arising from the normalization of CBF states
\begin{equation}
O_{nm}\equiv(n|\hat{O}|m)\equiv \frac{\langle n |\hat{\mathcal{F}} \hat{O} \hat{\mathcal{F}} |m\rangle }{\langle n |\hat{\mathcal{F}} \hat{\mathcal{F}} |m\rangle}\equiv \frac{N_{nm}}{D_{nm}}\, .
\end{equation}
It is a general property of the cluster expansion, to be discussed in detail below, that divergent terms coming from the expansion of the numerator and of denominator cancel. 

The two following subsections will be devoted to the description of the original Fantoni Rosati (FR) diagrammatic cluster expansion formalism \cite{fantoni_74,fantoni_75} and to its generalization \cite{pandha_76,pandha_79}, developed to deal with spin-isospin dependent correlation operators. 

\section{Fantoni Rosati cluster expansion and FHNC summation scheme}
\label{sec:FR_exp}
The FR cluster expansion has been obtained through a generalization of the concepts underlying the Mayer expansion scheme, originally developed to describe classical liquids \cite{mayer_40}, to the case of quantum Bose and Fermi systems.  

Within the FR approach, both the term $\hat{\mathcal{F}}\hat{H}\hat{\mathcal{F}}$ of the numerator $N_{nm}$ and the term $\hat{\mathcal{F}}\hat{\mathcal{F}}$ of the denominator $D_{nm}$ associated with the expectation value of the hamiltonian are expanded in terms of
\begin{equation}
h(r_{ij})\equiv f^c(r_{ij})^2-1\, .
\label{eq:hscalar}
\end{equation}

Notice that for scalar correlation operator of Eq. (\ref{eq:jastrow_corr}) to respect the cluster properties one can impose
\begin{equation}
f^c(r\geq d^c)= 1\, .
\end{equation}
The variational parameter $d^c$, to be fully explained later on, is the {\it central healing distance} encompassing the fact that when two-particles are further apart than $d_c$ they are not anymore correlated. Hence, the quantity $h(r_{ij})$ can be seen as a smallness parameter for the cluster expansion, as it is indeed in the case of the ``power-series'' (PS) expansion scheme. 

\subsubsection{Two-body distribution function}
In the calculation of the ground-state expectation value of any two-body scalar operator, it is very useful to employ the scalar two-body distribution function, $g^c(\mathbf{r}_1,\mathbf{r}_2)$, defined as
\begin{equation}
g^c(\mathbf{r}_1,\mathbf{r}_2)=\frac{A(A-1)}{\rho^2}\frac{\text{Tr}_{12}\int dx_{3,\dots,A} \Psi_0^*(X)\hat{\mathcal{F}}^\dagger \hat{\mathcal{F}} \Psi_0(X)}{\int dx_{1,\dots,A}  \ \Psi_0^*(X)\hat{\mathcal{F}}^\dagger \hat{\mathcal{F}} \Psi_0(X)}\, .
\label{eq:g2bc_def}
\end{equation}

In terms of $g^c(\mathbf{r}_1,\mathbf{r}_2)$, the expectation value of the two-body potential reads
\begin{equation}
\langle v \rangle\equiv\sum_{i<j}(0|v_{ij}|0)=\frac{A(A-1)}{2}(0|v_{12}|0)=\frac{\rho^2}{2}\int d\mathbf{r}_{1,2}\,g^c(\mathbf{r}_1,\mathbf{r}_2)v(r_{12})\, .
\label{eq:pot_exp}
\end{equation}
For the first equality we have exploited the symmetry property of the wave function, which is due to the fact that $\mathcal{F}$ is symmetric and $\Psi_0$ antisymmetric in the generalized particle coordinates. 
Since nuclear matter is uniform, $g^c(\mathbf{r}_1,\mathbf{r}_2)=g(r_{12})$, implying that $\langle v \rangle$ diverges with the number of particles. However, the potential energy per particle is finite and reads
\begin{equation}
\frac{\langle v \rangle}{A}=\frac{\rho}{2}\int dr_{12}g^c(r_{12})v(r_{12})\, .
\label{eq:potpparcent}
\end{equation}
Particles $1$ and $2$ are denoted as {\it interacting particles} and are distinguished from the other particles in the medium. 

The scalar two-body distribution function obeys the following sum rule
\begin{equation}
\rho\int d\mathbf{r}_{12}[g^c(r_{12})-1]=-1\,,
\label{eq:gc_sumrule}
\end{equation}
that can be easily derived by integrating Eq. (\ref{eq:g2bc_def}) and using translation invariance of $g^c(r_{12})$. Note that the latter results is a consequence of the fact that the scalar two-body distribution function can interpreted as the joint probability of finding two particles with coordinates $\mathbf{r}_1$ and $\mathbf{r}_2$.

Following \cite{fantoni_98}, in the following subsections we will provide a detailed description of the FR cluster expansion of the scalar two-body distribution function. 

\subsubsection{Cluster decomposition of $\hat{\mathcal{F}}^\dag\hat{\mathcal{F}}$}
In the scalar Jastrow case the product of the correlation operator reduces to
\begin{equation}
\hat{\mathcal{F}}^\dag\hat{\mathcal{F}}=\prod_{i<j}f^2(r_{ij})=\prod_{i<j}[1+h(r_{ij})]\, .
\label{eq:FF_exp}
\end{equation}
It is convenient to put aside the correlation between the interacting particles, denoted as ``active correlation'', while the others are called ``passive correlations''. Without loss of generality we can write
\begin{align}
\hat{\mathcal{F}}^\dag\hat{\mathcal{F}}&=f^2(r_{12})\prod_{i<j\neq1,2}[1+h(r_{ij})]\nonumber\\
&=f^2(r_{12})\Big(1+\sum_{i\neq1,2}X^{(3)}(\mathbf{r}_1,\mathbf{r}_2;\mathbf{r}_i)+\sum_{i<j\neq1,2}X^{(4)}(\mathbf{r}_1,\mathbf{r}_2;\mathbf{r}_i,\mathbf{r}_j)+\dots\Big)\, .
\label{eq:FF_ce}
\end{align}
The generic cluster term $f^2(r_{12})X^{(n)}(\mathbf{r}_i,\dots,\mathbf{r}_n)$ correlates the positions of the two interacting particles and of the $n-2$ medium particles and should be considered as an $n$-body operator. For the sake of clarity we give the explicit expression of the first cluster terms\begin{align}
X^{(3)}(\mathbf{r}_1,\mathbf{r}_2;\mathbf{r}_i)&=h(r_{1i})+h(r_{2i})+h(r_{1i})h(r_{2i})\nonumber\\ 	
X^{(4)}(\mathbf{r}_1,\mathbf{r}_2;\mathbf{r}_i,\mathbf{r}_j)&=h(r_{ij})+h(r_{1i})h(r_{2j})+h(r_{1i})h(r_{1j})+h(r_{2i})h(r_{2j})
+h(r_{1i})h(r_{ij})\nonumber \\
&+h(r_{2i})h(r_{ij})+h(r_{1i})h(r_{2j})h(r_{ij})+\dots
\end{align}

\subsubsection{Expansion of the numerator in cluster diagrams}
The cluster expansion of the numerator can be performed by substituting the rhs of Eq. (\ref{eq:FF_ce}) in Eq. (\ref{eq:g2bc_def}) 
\begin{align}
\text{num}=&\frac{A(A-1)}{\rho^2}{\text{Tr}_{12}\int dx_{3,\dots,A} f^2(r_{12})\Big(1+\sum_{i\neq1,2}X^{(3)}(\mathbf{r}_1,\mathbf{r}_2;\mathbf{r}_i)+\dots\Big)|\Psi_{0}(X)|^2}\,.
\label{eq:num_1}
\end{align}

The integration of the cluster term on the squared modulus of the Fermi-gas wave function, which is invariant under the exchange of any particles, gives rise to a combinatory factor:
\begin{align}
&\sum_{i<j<,\dots\neq1,2}\int dx_{3,\dots,A} X^{(N)}(\mathbf{r}_1,\mathbf{r}_2;\mathbf{r}_i,\mathbf{r}_j,\dots)|\Psi_{0}(X)|^2\nonumber \\
&\qquad =\frac{(A-2)!}{(A-N)!(N-2)!}\int dx_{3,\dots,A} X^{(N)}(\mathbf{r}_1,\mathbf{r}_2;\mathbf{r}_3,\mathbf{r}_4,\dots,\mathbf{r}_N)|\Psi^{0}(X)|^2\, .
\end{align}
Using the above result in Eq. (\ref{eq:num_1}) leads to 
\begin{align}
\text{num}=&f^2(r_{12})\Big[\sum_{N=2}^{A}\frac{\rho^{N-2}}{(N-2)!}\int d\mathbf{r}_{3,\dots,N}  X^{(N)}(\mathbf{r}_1,\mathbf{r}_2;\mathbf{r}_3,\dots,\mathbf{r}_N) g_{N}^{MF}(\mathbf{r}_1,\dots,\mathbf{r}_N)\Big]\, ,
\label{eq:num_last}
\end{align}
where $X^{(2)}$=1. Note that we have introduced the mean-field $N$-body correlation function, defined as:
\begin{equation}
g_{N}^{MF}(\mathbf{r}_1,\dots,\mathbf{r}_N)=
\frac{A!}{(A-N)!}\frac{1}{\rho^N}\,\text{Tr}_{1,\dots,N}\int dx_{N,\dots,A}|\Psi_{0}(X)|^2\,.
\label{eq:def_corrMF}
\end{equation}

We now proceed by integrating out the variables $x_{N+1},\dots, x_{A}$ from $g_{N}^{c\,MF}$ by using the orthogonality of single particle states. As shown in Appendix \ref{app:slat}, extracting $N$ particles from the Slater determinant of the ground-state $\Psi_0$ yields 
\begin{align}
\Psi_0&=\sqrt{\frac{(A-N)!}{A!}}\sum_{n_1<\dots<n_N}(-1)^{n_1+\dots+n_N+1}\,\times\nonumber\\
&\qquad \mathcal{A}[\psi_{n_1}(x_1)\dots\psi_{n_N}(x_N)]\Psi_{0}^{m \neq n_1,\dots,n_N}(x_{N+1}\,\dots\,x_A)\,.
\label{eq:np_extracted}
\end{align}
where $\Psi_{0}^{m \neq n_1,\dots,n_N}(x_{N+1}\,\dots\,x_A)$ is the minor $\psi_0$, describing a system of $A-N$ particles with holes $n_1<,\dots,<n_N$. The minors satisfy the following orthonormality condition
\begin{align}
\int dx_{N+1,\dots,A}{\Psi_{0}^\dagger}^{\,m \neq n_1,\dots,n_N}(x_{N+1}\,\dots\,x_A)\Psi_{0}^{m \neq l_1,\dots,l_N}(x_{N+1}\,\dots\,x_A)=\delta_{n_1l_1}\dots\delta_{n_Nl_N}\, .
\end{align}
With the help of the above equation, the mean-field $N$-body distribution function can be written in the form
\begin{align}
&g_{N}^{MF}(\mathbf{r}_1,\dots,\mathbf{r}_N)=\nonumber\\
&\qquad \frac{1}{\rho^{\,N}}\sum_{n_1<\dots<n_N}\text{Tr}_{1,\dots,N} \Big[\mathcal{A}[\psi_{n_1}^\dagger(x_1)\dots\psi_{n_N}^\dagger(x_N)] \mathcal{A}[\psi_{n_1}(x_1)\dots\psi_{n_N}(x_N)]\Big]=\nonumber\\
&\qquad \frac{1}{\rho^{\,N}}\sum_{n_1,\dots,n_N}\text{Tr}_{1,\dots,N}\Big[ \psi_{n_1}^\dagger(x_1)\dots\psi_{n_N}^\dagger(x_N) \mathcal{A}[\psi_{n_1}(x_1)\dots\psi_{n_N}(x_N)]\Big]\, ,
\label{eq:gmf_tr}
\end{align}
implying that if the number of particles, $N$, is larger than the number of quantum states, $A$, the mean-field $N$-body distribution function vanishes, i. e.
\begin{equation}
g_{N}^{c\,MF}(\mathbf{r}_1,\dots,\mathbf{r}_N)=0\qquad \text{if}\qquad N>A\, .
\label{eq:gram_gN}
\end{equation}
We preliminary remark that this property is crucial for the exact cancellation of the unlinked diagrams of the numerator with the denominator to take place.

The antisymmetrization operator $\mathcal{A}$ can be written in the form 
\begin{equation}
\mathcal{A}=1-\sum_{i<j} \hat{P}_{ij}+\sum_{i<j<k}(\hat{P}_{ij}\hat{P}_{jk}+\hat{P}_{ik}\hat{P}_{kj})+\dots\,  .
\label{eq:anti_exch}
\end{equation}
For a uniform system like nuclear matter the single particle states are normalized plane waves, see Eq. (\ref{eq:nm_wf}). Thus, the two-particle exchange operator, defined by the relation
\begin{equation}
\hat{P}_{ij} \psi_{n_i}(x_i) \psi_{n_j}(x_j) = \psi_{n_i}(x_j) \psi_{n_j}(x_i) \ .
\end{equation}
can be written as
\begin{equation}
\hat{P}_{ij}=\hat{P}_{ij}^{\sigma\tau}\times P_{ij}^\mathbf{r}\, ,
\label{def:Pij}
\end{equation}
where
\begin{equation}
\hat{P}_{ij}^{\sigma\tau}=\frac{1}{4}(1+\sigma_{ij})(1+\tau_{ij})\equiv\sum_{p=1}^4 \Delta^p \hat{O}^{p}_{ij}
\label{eq:exch_op}
\end{equation}
acts on the spin-isospin degrees of freedom of the nucleons' wave function, while
\begin{equation}
P_{ij}^\mathbf{r}=\exp[-i(\mathbf{k}_i-\mathbf{k}_j)\cdot \mathbf{r}_{ij}]
\end{equation}
exchanges the radial coordinates of particles $i$ and $j$.

Because Pauli matrices are traceless, in the pure Jastrow case, when the traces of Eq. (\ref{eq:gmf_tr}) are carried out, the exchange operator reduces to its central part 
\begin{equation}
\hat{P}_{ij}\to\frac{1}{\nu}\exp[-i(\mathbf{k}_i-\mathbf{k}_j)\cdot \mathbf{r}_{ij}]\,  .
\label{eq:exch_central}
\end{equation}
and one is left with
\begin{align}
&g_{N}^{c\,MF}(\mathbf{r}_1,\dots,\mathbf{r}_N)=\left(\frac{\nu}{\rho}\right)^{\,N}\sum_{k_1,\dots,k_N} \phi_{\mathbf{k}_1}^*(\mathbf{r}_1)\dots\phi_{\mathbf{k}_N}^*(\mathbf{r}_N) \mathcal{A}[\phi_{\mathbf{k}_1}(\mathbf{r}_1)\dots\phi_{\mathbf{r}_N}(\mathbf{k}_N)]\Big]\, .
\end{align}

Therefore, $g_{N}^{MF}$ can be written in terms of the Slater function, defined by
\begin{align}
\ell(r_{ij})&=\frac{\nu}{\rho}\sum_{|\mathbf{k}|<k_F}\phi_{\mathbf{k}}^*(\mathbf{r}_i)\phi_{\mathbf{k}}(\mathbf{r}_j)\, .
\end{align}
In the limit of infinite volume, the sum over the discrete momentum can be replaced by an integral. Hence it can be easily shown that
\begin{align}
\ell(r_{ij})&=3\Big[\frac{\sin(k_Fr_{ij})-k_Fr_{ij}\cos(k_Fr_{ij})}{(k_Fr_{ij})^3}\Big]\,.
\label{eq:slat_def}
\end{align}

The first terms of the mean-field $N$-body distribution function reads
\begin{align}
g_{n}^{MF}(\mathbf{r}_1,\dots,\mathbf{r}_n)=1-\sum_{i<j}\frac{1}{\nu}\ell^2(r_{ij})+\sum_{i<j<k}\frac{2}{\nu^2}\ell(r_{ij})\ell(r_{jk})\ell(r_{ki})-\dots\, .
\label{eq:exp_gmf}
\end{align}
The factors $1/\nu$ come from the normalization of the exchange operator of Eq. (\ref{eq:exch_central}). 
In particular, producing a two-particle loop, $\ell^2(r_{ij})$, requires one exchange operator $\hat{P}_{ij}$ with the associated 
factor $1/\nu$. For a loop involving $n>2$ particles and n-1 exchange operators, the corresponding factor is $(1/\nu)^{n-1}$. Moreover, there are two possible orderings of the exchange operators producing loops having more than two particles exchanged, bringing and additional factor $2$.

We are now ready to give the general structure of the cluster decomposition for the numerator of Eq. (\ref{eq:num_last}). It is very useful to do this pictorially, introducing the so called ``cluster diagrams''. The diagrammatic rules are the
following:
\begin{itemize}
	\item The diagrams consist of dots (vertices) connected by different kinds of correlation lines. Open dots represent the active (or interacting) particles ($1$ and $2$), while black dots are associated with passive particles, i.e. those in the medium. Integration over the coordinates of a passive particle leads to the appearance of a factor $\rho$.
	\item The dashed lines, representing the  correlations $h(\mathbf{r}_{ij})$ and denoted as ``correlation lines'', cannot be superimposed.
	\item	The statistical factor $-\ell(r_{ij})/\nu$, coming from the expansion of $g_{n}^{MF}(\mathbf{r}_1,\dots,\mathbf{r}_n)$ is represented by an oriented solid  ``exchange line''. The exchange lines must form closed loops and, as can be readily seen from the expansion of $\mathcal{A}$ in terms of the exchange operators of Eq. (\ref{eq:anti_exch}), different loops cannot have common points. Hence, the total exchange pattern consists in one or more non touching exchange loops. 
	\item Each solid point must be reached by at least one correlation line; in fact in Eq. (\ref{eq:num_last}) each	 integration over $\mathbf{r}_i$ is associated with a term $X^{(N)}(\mathbf{r}_1,\mathbf{r}_2;\mathbf{r}_3,\dots,\mathbf{r}_i,\dots,\mathbf{r}_N)$.

\end{itemize}
\begin{figure}[!h]
\begin{center}
\fcolorbox{white}{white}{
  \begin{picture}(300,70)(-40,0)
	\SetWidth{0.5}
	\SetColor{Black}
	\SetScale{0.8}	
        \unitlength=0.8 pt
	\CCirc(0,0){4}{Black}{White}
	\CCirc(30,60){4}{Black}{Black}
	\CCirc(60,0){4}{Black}{White}
	\DashLine(0,0)(30,60){10}
	\DashLine(60,0)(30,60){10}
	\Text(-12,0)[]{1}
	\Text(18,60)[]{3}
	\Text(72,0)[]{2}
	\Text(30,-32)[]{(a)}

	\CCirc(200,0){4}{Black}{White}
	\CCirc(260,0){4}{Black}{White}
	\CCirc(200,60){4}{Black}{Black}
	%\DashLine(200,0)(260,0){10}
	\DashLine(200,0)(200,60){10}
	\ArrowArc(222.5,30)(37.5,130,233)
	\ArrowArc(230,22.5)(37.5,217,323)
	\ArrowLine(260,0)(200,60)
	\CCirc(260,25){4}{Black}{Black}
	\CCirc(260,60){4}{Black}{Black}	
	\Text(188,0)[]{1}
	\Text(188,60)[]{3}
	\Text(272,0)[]{2}
	\DashLine(260,27)(260,57){10}
	\ArrowArc(262.71,42.5)(17.71,112,250)
	\ArrowArc(257.29,42.5)(17.71,-68,68)
	\Text(273,20)[]{4}
	\Text(273,65)[]{5}
        \Text(230,-32)[]{(b)}
    \end{picture}
}
\vspace{1.5cm}
\caption{Examples of cluster diagrams.}
\label{fig:first_ex}
\end{center}
\end{figure}
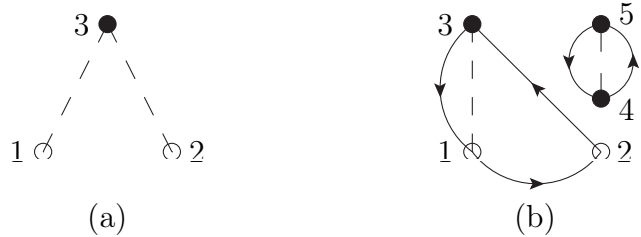\

Fig. \ref{fig:first_ex} shows two examples of cluster diagrams whose analytic expressions read 
\begin{align}
(\ref{fig:first_ex}.a)&=\rho\int d\mathbf{r}_3 h(r_{13})h(r_{23})\, ,\\
(\ref{fig:first_ex}.b)&=-2\nu\Big(\frac{-\ell(r_{12})}{\nu}\Big)h(r_{12})\rho^3\nonumber \\
&\times \int d\mathbf{r}_3d\mathbf{r}_4d\mathbf{r}_5\Big(-\frac{\ell(r_{23})}{\nu}\Big)\Big(-\frac{\ell(r_{13})}{\nu}\Big)
\Big(-\frac{\ell^2(r_{45})}{\nu}\Big)h(r_{13})h(r_{45})
\end{align}

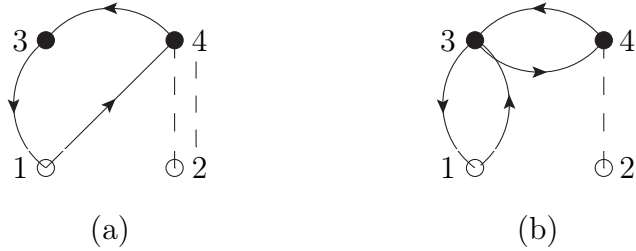
\begin{figure}[!ht]
\begin{center}
\fcolorbox{white}{white}{
  \begin{picture}(300,60)(-50,10)
	\SetWidth{0.5}
	\SetColor{Black}
	\SetScale{0.8}	
        \unitlength=0.8 pt
	
	\CCirc(0,0){4}{Black}{White}
	\CCirc(0,60){4}{Black}{Black}
	\CCirc(60,60){4}{Black}{Black}
	\CCirc(60,0){4}{Black}{White}
	\DashLine(60,0)(60,60){10}
	\DashLine(70,10)(70,50){10}
	\ArrowLine(0,0)(60,60)
	\ArrowArc(22.5,30)(37.5,130,233)
	\ArrowArc(30,37.5)(37.5,40,140)
	\Text(-12,0)[]{1}
	\Text(-12,60)[]{3}
	\Text(72,0)[]{2}
	\Text(72,60)[]{4}
	\Text(30,-30)[]{(a)}
	
	\CCirc(200,0){4}{Black}{White}
	\CCirc(200,60){4}{Black}{Black}
	\CCirc(260,60){4}{Black}{Black}
	\CCirc(260,0){4}{Black}{White}
	\DashLine(260,0)(260,60){10}
	\ArrowArc(222.5,30)(37.5,130,233)
	\ArrowArc(230,37.5)(37.5,40,140)	
	\ArrowArc(178.5,30)(37.5,-50,53)
	\ArrowArc(230,82.5)(37.5,220,320)
	\Text(188,0)[]{1}
	\Text(188,60)[]{3}
	\Text(272,0)[]{2}
	\Text(272,60)[]{4}
	\Text(230,-30)[]{(b)}
    \end{picture}
}
\vspace{1.5cm}
\caption{Examples of not-allowed cluster diagrams.\label{fig:na_diag}}
\label{fig:sec_ex}
\end{center}
\end{figure}

In Fig. \ref{fig:na_diag} two examples of forbidden diagrams have been drawn. Diagram (\ref{fig:na_diag}.a) is not allowed because the solid point $3$ is not reached by any correlation line (moreover, between points $2$ and $4$ there are two superimposed dashed lines); on the other hand in diagram(\ref{fig:na_diag}.b)  there are two touching exchange loops.

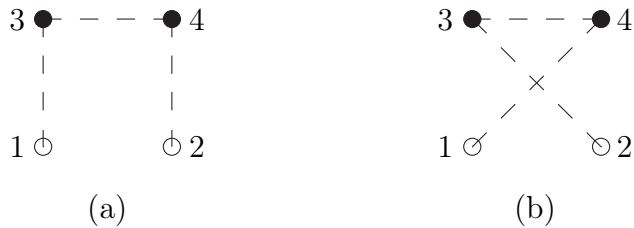
\begin{figure}[!ht]
\begin{center}
\fcolorbox{white}{white}{
  \begin{picture}(300,60)(-40,10)
	\SetWidth{0.5}
	\SetColor{Black}
	\SetScale{0.8}	
        \unitlength=0.8 pt
	\CCirc(0,0){4}{Black}{White}
	\CCirc(0,60){4}{Black}{Black}
	\CCirc(60,60){4}{Black}{Black}
	\CCirc(60,0){4}{Black}{White}
	\DashLine(0,0)(0,60){10}
	\DashLine(0,60)(60,60){10}
	\DashLine(60,60)(60,0){10}
	\Text(-12,0)[]{1}
	\Text(-12,60)[]{3}
	\Text(72,0)[]{2}
	\Text(72,60)[]{4}
	\Text(30,-30)[]{(a)}
	
	\CCirc(200,0){4}{Black}{White}
	\CCirc(200,60){4}{Black}{Black}
	\CCirc(260,60){4}{Black}{Black}
	\CCirc(260,0){4}{Black}{White}
	\DashLine(200,60)(260,60){10}
	\DashLine(200,0)(260,60){10}
	\DashLine(260,0)(200,60){10}
	\Text(188,0)[]{1}
	\Text(188,60)[]{3}
	\Text(272,0)[]{2}
	\Text(272,60)[]{4}
	\Text(230,-30)[]{(b)}
    \end{picture}
}
\vspace{2.0cm}
\caption{Topologically identical clusters diagrams.}
\label{fig:perm_nos}
\end{center}
\end{figure}

The cluster terms have no specific prefactor, except those coming from the exchange rules and a $\rho$ factor for each integration. One
might wonder where the $1/(n-2)!$ of Eq. (\ref{eq:num_last}) ended up. The factor is due to the counting of the permutations of the $n-2$
internal points and it is automatically taken into account by considering only topologically different graphs, or, in other words, by the
fact that the labels of the solid points in the cluster diagrams are dummy indices. The only remnant of that factor is the inverse of what
is usually called ``symmetry factor'', $s$. This counts the permutations of the solid points' labels that, without renaming the integration variables, leave the cluster term unchanged. For instance diagrams (a) and (b) of Fig. \ref{fig:perm_nos}, being topologically identical, give the same contributions
\begin{align}
(\ref{fig:perm_nos}.a)&=\rho^2\int d\mathbf{r}_3 d\mathbf{r}_4 h(r_{13})h(r_{34})h(r_{24})\, ,\\
(\ref{fig:perm_nos}.b)&=\rho^2\int d\mathbf{r}_3 d\mathbf{r}_4 h(r_{14})h(r_{34})h(r_{23})\, .
\end{align}

\begin{figure}[!ht]
\begin{center}
\fcolorbox{white}{white}{
\begin{picture}(300,60)(-120,10)
	\SetWidth{0.5}
	\SetScale{0.8}	
        \unitlength=0.8 pt
	\SetColor{Black}
	\CCirc(0,0){4}{Black}{White}
	\CCirc(0,60){4}{Black}{Black}
	\CCirc(60,60){4}{Black}{Black}
	\CCirc(60,0){4}{Black}{White}
	\DashLine(60,0)(60,60){10}
	\DashLine(0,0)(0,60){10}
	\DashLine(0,60)(60,60){10}
	\DashLine(0,0)(60,60){10}
	\DashLine(0,60)(60,0){10}
	\Text(-12,0)[]{1}
	\Text(-12,60)[]{3}
	\Text(72,0)[]{2}
	\Text(72,60)[]{4}
	\Text(30,-30)[]{(a)}
    \end{picture}
}
\vspace{1.5cm}
\caption{Graph with a prefactor $s=1/2$ associated with.}
\label{fig:perm_s}
\end{center}
\end{figure}
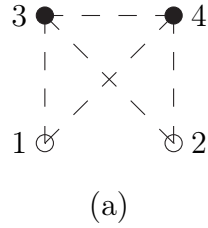

As a consequence we take into account only one of them and no prefactors appear. On the other hand a prefactor $s=1/2$ is associated
with the diagram (a) of Fig. (\ref{fig:perm_s}), because the exchange of points $3$ and $4$ leads to an identical expression, even
without relabelling the dummy variables
\begin{equation}
(\ref{fig:perm_s}.a)=\frac{\rho^2}{2}\int d\mathbf{r}_3 d\mathbf{r}_4 h(r_{13})h(r_{14})h(r_{34})h(r_{23})h(r_{24})\, .
\end{equation}
The reason of this fact lies in the constraint $i<j$ in the expansion of $\mathcal{F}^\dagger \mathcal{F}$ of Eq. (\ref{eq:FF_exp}), implying that a diagram analogous to diagram (\ref{fig:perm_s}.a) with the points $3$ and $4$ exchanged does not appear in the cluster expansion.

\begin{figure}[!ht]
\begin{center}
\fcolorbox{white}{white}{
\begin{picture}(300,60)(-120,15)
  	\SetScale{0.8}	
        \unitlength=0.8 pt
	\SetWidth{0.5}
	\SetColor{Black}
	\CCirc(0,0){4}{Black}{White}
	\CCirc(30,60){4}{Black}{Black}
	\CCirc(60,0){4}{Black}{White}
	\CCirc(90,60){4}{Black}{Black}
	\DashLine(0,0)(30,60){10}
	\DashLine(60,0)(30,60){10}
	\DashLine(30,60)(90,60){10}
	\Text(-12,0)[]{1}
	\Text(18,60)[]{3}
	\Text(72,0)[]{2}
	\Text(102,60)[]{4}
	\Text(30,-30)[]{(a)}
    \end{picture}
}
\vspace{1.5cm}
\caption{Linked and reducible diagram.}
\label{fig:ex_red}
\end{center}
\end{figure}
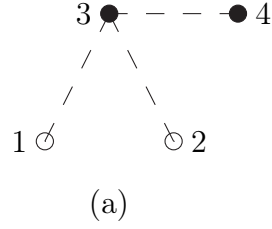

Cluster diagrams may be {\it reducible} or {\it irreducible}. The integrals corresponding to irreducible diagrams cannot be factorized.
Obviously an irreducible diagram must be {\it linked}, i.e. each couple of points must be connected by a sequence of lines, which can be
both correlation and exchange lines. For instance, diagrams (\ref{fig:first_ex}.a), (\ref{fig:perm_nos}.a), (\ref{fig:perm_nos}.b),
(\ref{fig:perm_s}.a) are linked and irreducible, while diagram (\ref{fig:first_ex}.b) is unlinked (and reducible). If we add to the
irreducible diagram (\ref{fig:first_ex}.a) the correlation line $h(\mathbf{r}_{34})$, we obtain the reducible diagram of Fig.
(\ref{fig:ex_red}), the expression of which reads
\begin{align}
(\ref{fig:ex_red}.a)&=\rho^2\int d\mathbf{r}_3 d\mathbf{r}_{4}h(r_{13})h(r_{23})h(r_{34})\, ,\\
&=\rho\int d\mathbf{r}_3 h(r_{13})h(r_{23})\times \rho \int d\mathbf{r}_{34}h(r_{34})\, ,
\end{align}
which is just the cluster term of (\ref{fig:first_ex}.a) multiplied by the cluster term of the added part.

Translational invariance gives rise to a factor $V$ for each unlinked part of the diagram but for the one containing the external points $1$ and $2$. Therefore, the order of magnitude of a cluster diagram is $V^{N_u-1}$, where $N_u$ is the number of unlinked parts. For example, the order of magnitude of the linked diagram (\ref{fig:first_ex}.a) is 1 while the one of diagram (\ref{fig:first_ex}.b), having two unconnected parts, is $V$.

\subsubsection{Expansion of the denominator in cluster diagrams.}
The same procedure followed for the expansion of the numerator can be used for the denominator. However, in this case there are no interacting particles; hence $\hat{\mathcal{F}}^\dagger \mathcal{F}$ can be conveniently expanded as 
\begin{equation}
\hat{\mathcal{F}}^\dag \hat{\mathcal{F}}=\Big(1+\sum_{i<j}X^{(2)}(\mathbf{r}_i,\mathbf{r}_j)+\sum_{i<j<k}X^{(3)}(\mathbf{r}_i,\mathbf{r}_j,\mathbf{r}_k)+\dots\Big)
\label{eq:FF_den}
\end{equation}
where the cluster term $X^{(N)}$ correlates $N$ particles. The explicit expressions for $N=2$ and $N=3$ are
\begin{align}
X^{(2)}(\mathbf{r}_i,\mathbf{r}_j)&=h(r_{ij})\, ,\nonumber \\
X^{(3)}(\mathbf{r}_i,\mathbf{r}_j,\mathbf{r}_k)&=h(r_{ij})h(r_{ik})+h(r_{ik})h(r_{jk})+h(r_{ij})h(r_{jk})+h(r_{ij})h(r_{jk})h(r_{ik})\, .
\end{align}
Substituting the expansion of $\hat{\mathcal{F}}^\dagger \mathcal{F}$ of Eq. (\ref{eq:FF_den}) in the denominator of Eq. (\ref{eq:g2bc_def}) and exploiting the invariance of $|\Psi^{0}_{MF}|^2$ under any two-particle exchange, one finds
\begin{align}
\text{den}=1+\sum_{N=2}^A\frac{\rho^N}{N!}\int d\mathbf{r}_1\dots d\mathbf{r}_N  g_{N}^{c\,MF}(\mathbf{r}_1,\dots,\mathbf{r}_N)X^{(N)}(\mathbf{r}_1,\dots,\mathbf{r}_N)\, ,
\label{eq:den_last}
\end{align}
with $g_{c\,N}^{MF}$ defined in Eq. (\ref{eq:def_corrMF}). 

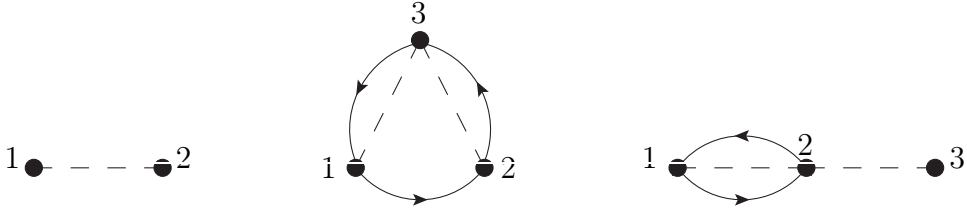
\begin{figure}[!h]
\begin{center}
\fcolorbox{white}{white}{
  \begin{picture}(300,60)(20,5)
  	\SetScale{0.8}	
        \unitlength=0.8 pt
	\SetWidth{0.5}
	\SetColor{Black}
	
	\CCirc(0,0){4}{Black}{Black}
	\CCirc(60,0){4}{Black}{Black}
	\DashLine(0,0)(60,0){10}
	\Text(-10,5)[]{1}
	\Text(70,5)[]{2}

	\CCirc(150,0){4}{Black}{Black}
	\CCirc(180,60){4}{Black}{Black}
	\CCirc(210,0){4}{Black}{Black}
	\DashLine(150,0)(180,60){10}
	\DashLine(210,0)(180,60){10}
	\ArrowArc(189,18)(41.83,100,205)
	\ArrowArc(171,18)(41.83,335,80)
	\ArrowArc(180,22.5)(37.5,220,320)
	\Text(138,0)[]{1}
	\Text(180,73)[]{3}
	\Text(222,0)[]{2}
	
	\CCirc(300,0){4}{Black}{Black}
	\CCirc(360,0){4}{Black}{Black}
	\CCirc(420,0){4}{Black}{Black}
	\DashLine(300,0)(360,0){10}
	\DashLine(360,0)(420,0){10}

	\ArrowArc(330,22.5)(37.5,217,323)
	\ArrowArc(330,-22.5)(37.5,30,150)
	\Text(288,5)[]{1}
	\Text(361,11)[]{2}
	\Text(432,5)[]{3}
	\end{picture}
}
\vspace{1.5cm}
\caption{Examples of cluster diagrams of the denominator.}
\label{fig:first_ex_den}
\end{center}
\end{figure}

Comparing Eqs. (\ref{eq:num_last}) and (\ref{eq:den_last}), we see that the diagrammatic rules for the
denominator are very similar to those of the numerator. The only difference is that the cluster diagrams of the denominator have only solid points, hence their order of magnitude is $V^{N_u}$. In Fig. \ref{fig:first_ex_den} two examples of cluster diagrams coming
from the expansion of the denominator are depicted.

\subsubsection{Two-body distribution function as a sum of cluster diagrams.}
The ratio of Eq. (\ref{eq:g2bc_def}) involves two infinite series of cluster terms, corresponding to the expansion of the numerator and of the denominator.

Let us consider a generic $n$ body linked (reducible or irreducible) cluster diagram, $\mathcal{L}_n$, of the numerator \cite{arias_07}, where each internal point is connected to the points 1 and 2 by at least one continuous path of correlation and/or exchange lines. Each cluster diagram of the numerator can be built as a product of $\mathcal{L}_n$ times a factor $\mathcal{U}_q$, with $q=A-n$, representing the sum of all the $q$ body unlinked diagrams
\begin{equation}
\text{num}=\sum_{n=2}^A\mathcal{L}_n\times\mathcal{U}_{A-n}\,.
\end{equation}

\begin{figure}[!h]
\begin{center}
\fcolorbox{white}{white}{
  \begin{picture}(300,85)(30,-5)
	\SetWidth{0.5}
	\SetColor{Black}
	\SetScale{0.8}	
        \unitlength=0.8 pt
	
	\CCirc(0,0){4}{Black}{White}
	\CCirc(30,60){4}{Black}{Black}
	\CCirc(60,0){4}{Black}{White}
	\DashLine(0,0)(30,60){10}
	\DashLine(60,0)(30,60){10}
	\Text(-12,0)[]{1}
	\Text(18,65)[]{3}
	\Text(72,0)[]{2}
	\Text(90,35)[]{$\quad\times\,\,\,\Bigg[\,\, 1\,\,+$}
	
	\CCirc(130,35){4}{Black}{Black}
	\CCirc(190,35){4}{Black}{Black}
	\DashLine(130,35)(190,35){10}
	\ArrowArc(160,57.5)(37.5,210,330)
	\ArrowArc(160,12.5)(37.5,40,140)

	\Text(210,35)[]{$+$}
	
	\CCirc(230,45){4}{Black}{Black}
	\CCirc(230,25){4}{Black}{Black}
	\CCirc(290,45){4}{Black}{Black}
	\CCirc(290,25){4}{Black}{Black}
	\DashLine(230,25)(290,25){10}
	\DashLine(230,45)(290,45){10}

	\Text(310,35)[]{$+$}

	\CCirc(325,5){4}{Black}{Black}
	\CCirc(355,65){4}{Black}{Black}
	\CCirc(385,5){4}{Black}{Black}
	\DashLine(325,5)(355,65){10}
	\DashLine(385,5)(355,65){10}
	\ArrowArc(364,23)(42.83,100,205)
	\ArrowArc(346,23)(42.83,335,80)
	\ArrowArc(355,27.5)(37.5,220,320)
	
	\Text(400,35)[]{$+$}
	\Text(420,35)[]{$\dots$}	
	\Text(440,35)[]{$\Bigg]$}
	\end{picture}
}
\vspace{0.8cm}
\caption{Series of unlinked diagrams associated with $L_n=(\ref{fig:first_ex}.a)$.}
\label{fig:linked_ev}
\end{center}
\end{figure}
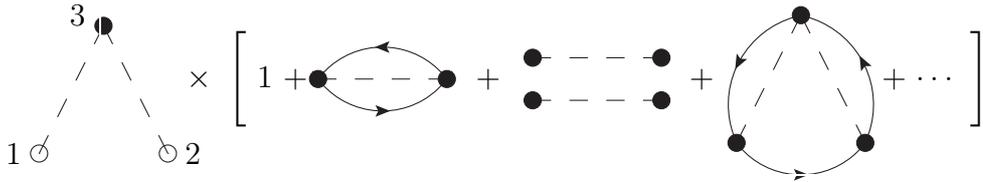

An example of the above equation is depicted in Fig.  \ref{fig:linked_ev}, where the diagram (\ref{fig:first_ex}.a) belongs to $\mathcal{L}_3$ and the sum of the diagrams enclosed in round parenthesis is $\mathcal{U}_{A-3}$.
Considering the expression of the expansion of the denominator, it is readily seen that
\begin{equation}
\text{den}=\sum_{n=1}^A \mathcal{U}_n\, .
\end{equation}

Hence, one might naively think that the denominator cancels the disconnected diagrams of the numerator in the thermodynamic limit, $A\to\infty$, only. However, by using the property of Eq. (\ref{eq:gram_gN}) it is possible to extend up to infinity the upper limits of all the sums appearing in Eqs. (\ref{eq:num_last}) and (\ref{eq:den_last}), yielding
\begin{align}
\text{num}&=\sum_{n=2}^\infty\mathcal{L}_n\times\sum_{q=1}^\infty\mathcal{U}_{q}\nonumber \\
\text{den}&=\sum_{n=1}^\infty\mathcal{U}_n\, ,
\end{align}
that proves the fundamental {\it linked cluster property}
\begin{align}
g_{2}^c(r_{12})&=\sum_{n=2}^\infty \mathcal{L}_n\, .
\label{eq:cluster_property}
\end{align}
Because of this property, which is a common feature of diagrammatic expansion techniques, the divergent terms of the numerator and of the denominator cancel out, giving rise to finite physical quantities. It has to be remarked that for Fermi systems the exact cancellation of unlinked diagrams still holds when spin-isospin dependent correlation are considered.

\subsubsection{Perturbative schemes}
In this subsection we briefly present two possible choices for the smallness parameter of the cluster expansion, namely, the expansion in the number of points and the power series (PS) expansion. 

{\bf Expansion in the number of points} Within this scheme, the order of magnitude of a linked diagram is given by the number of its internal points: expanding the two body distribution function at $n=m$. As a factor $\rho^{n-2}$ is associated with n internal points, the smallness parameter is nothing but the density. 

This was the first scheme adopted for classifying diagrams of Fermi liquids; however the series is rapidly convergent for low-density regimes only. This is not the case for nuclear matter, as found by the authors of Ref. \cite{morales_02}. They have shown that at the equilibrium density, the 1-5 body cluster contributions to the energy of SNM are $22.1$ MeV, $43.7$ MeV, $10.8$ MeV, $3.4$ MeV, and $2.6$ MeV, while those with $n>5$ give $0.8$ MeV. The computational cost of this approach exponentially increases with the number of internal points: actual calculations do not go beyond the three-body cluster contribution in the case of spin-isospin dependent operators.

It has to be noted that the expectation value of the hamiltonian at any finite order of the expansion in the number of internal points is unbound. The main reason for this lies in the normalization of the wave function that is not properly taken into account. Nevertheless, as it will be fully explained at a later stage, a procedure usually employed to obtain the correlation function $f^p(r)$ consists in minimizing the two-body cluster contribution to the energy per particle imposing constraints on the variational parameters of the correlation functions. 
\newline
\newline
{\bf Power series (PS) expansion}
Suppose to multiply each correlation line $h(r_{ij})$ by a parameter $\alpha$, it follows that the term of order $m$ in the PS expansion scheme \cite{fantoni_74} of $g_2(r_{12})$, denoted as $g_{2}^{\alpha_n}$,  corresponds to the terms of order $\alpha^m$ resulting from the sum of Eq. (\ref{eq:cluster_property}). Of course the zeroth order corresponds to the mean field results, $g_{2}^{\alpha_0}=g_{2}^{MF}$.

A given order of the power series expansion mixes different orders of the expansion in the number of points. For instance the first order in PS includes all of the $2$-body diagrams, ten $3$-body diagrams and nine $4$-body diagrams. 

A remarkable feature of the PS scheme is that the sum rule of Eq. (\ref{eq:gc_sumrule}) for the two-body distribution function is satisfied at any order. A very simple argument to proof this statement, that can be found in \cite{fantoni_98}, is the following. For any choice of $\alpha$ the sum rule reads 
\begin{equation}
\rho\int d\mathbf{r}_{12}(\sum_n \alpha^n g_{2}^{c\,\alpha_n}(r_{12})-1)=-1\, .
\end{equation}
Since the mean-field two-body distribution function satisfies the former sum rule, it turns out that
\begin{equation}
\rho\int d\mathbf{r}_{12}\sum_{n>1}\alpha^n g_{2}^{c\,\alpha_n}(r_{12})=0\, .
\end{equation}
The integral of the series is zero for any choice of $\alpha$ within the convergence radius, thus the integral of each of the coefficients $g_{2}^{c\,\alpha_n}$ has to be zero.

The drawback of the PS expansion is that it does not appropriately describe the short-range behavior of the pair function, which is 
crucial for the calculation of the binding energy. The factor $f(r_{12})$, while being set aside in Eq. (\ref{eq:FF_ce}), does not in fact multiply all terms of the expansion.

\subsection{Fermi Hyper-Netted Chain (FHNC) and RFHNC\\ schemes}
\label{sec:RFHNC}

To allow for a good description of both the short and the long range behavior of the pair function, the development of a scheme in which infinite sets of cluster diagrams are summed up is required. This goal is achieved by the FHNC summation procedure, two versions of which had been originally proposed: one by Fantoni and Rosati \cite{fantoni_74,fantoni_74b,fantoni_75} and the other by Krotscheck and Ristig \cite{krotscheck_74,krotscheck_75}.

The two schemes are essentially complementary: the one by Krotscheck and Ristig is better suited to the treatment of long-range correlations, while the FR is more convenient for the evaluation of the energy in the presence of strong short-range correlations. Since in this work have been dealing with the calculation of the expectation value of the potential, we have extensively used the FR approach, that we present for the case of a Fermi liquid and a wave function with Jastrow-type correlations (for which FHNC has been originally developed).

In what follows, we will not assume that the reducible clusters diagrams cancel out, which is strictly true only for a uniform Fermi liquid described by scalar correlations,  as rigorously proved in Ref. \cite{fantoni_74}. In Section \ref{subsec:twothree} we will show how the cancellation mechanism works in the case of the $3$-body cluster contributions.  

The first detailed classification of the cluster diagrams, limited to the bosonic case, can be found in the fundamental work of J.M.J. van Leeuwen, J. Groeneveld and J. de Boer \cite{leeuwen_59}, while the extension to Fermi systems is extensively analyzed in the more recent Refs. \cite{fantoni_98,arias_07}.

\subsubsection{Vertex Corrected Irreducible (VIC) diagrams}
In this subsection we will show that, when scalar correlations only are considered, reducible diagrams can be included in the calculation of the two-body scalar distribution function through vertex corrections to the irreducible diagrams.

\begin{figure}[!ht]
\begin{center}
\fcolorbox{white}{white}{
\begin{picture}(455,50)(-130,0)
	\SetWidth{0.5}
	\SetColor{Black}
	\SetScale{0.8}	
        \unitlength=0.8 pt
	
	\Text(-75,10)[]{$g^c(\mathbf{r}_1)=$}
        \Text(-45,10)[]{$\Big[$}
        \Text(-20,10)[]{$1$}
        \Text(0,10)[]{$+$}
	
	\CCirc(20,10){4}{Black}{White}
	\CCirc(80,10){4}{Black}{Black}
	\DashLine(20,10)(80,10){10}
	\Text(20,22)[]{$1$}
%	\Text(80,25)[]{$4$}
	
	\Text(100,10)[]{$+$}

	\CCirc(0,-45){4}{Black}{White}
	\CCirc(60,-45){4}{Black}{Black}
	\DashLine(0,-45)(60,-45){10}
	\ArrowArc(30,-67.5)(37.5,40,143)
	\ArrowArc(30,-23.5)(37.5,217,323)
	\Text(0,-33)[]{$1$}
%	\Text(60,-35)[]{$4$}

	\CCirc(120,10){4}{Black}{White}
	\CCirc(180,10){4}{Black}{Black}
	\CCirc(240,10){4}{Black}{Black}
	\DashLine(120,10)(180,10){10}
	\DashLine(180,10)(240,10){10}
	\Text(120,22)[]{$1$}

	\Text(260,10)[]{$+$}
	\Text(80,-45)[]{$+$}

	\CCirc(100,-50){4}{Black}{White}
	\CCirc(160,-20){4}{Black}{Black}
	\CCirc(160,-80){4}{Black}{Black}
	\DashLine(100,-50)(160,-20){10}
	\DashLine(100,-50)(160,-80){10}
	\Text(100,-35)[]{$1$}

	\Text(180,-45)[]{$+$}

	\Text(200,-45)[]{$\dots$}
	\Text(220,-45)[]{$\Big]$}

    \end{picture}
}
\vspace{2.5cm}
\caption{Diagrams contributing to $g^c(\mathbf{r}_1)$, i.e. to the vertex correction $\xi_d$. The same diagrams but the first of the second line contribute to $\xi_e$ as well.}
\label{fig_gr}
\end{center}
\end{figure}
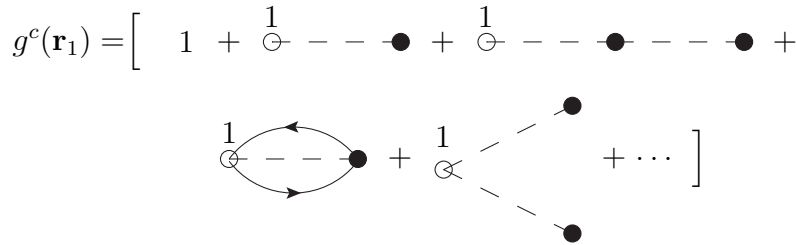

For the sake of introducing the reducible diagrams, the correlation line $h(\mathbf{r}_{34})$ had been attached to the point $3$ of diagram  (\ref{fig:first_ex}.a). It is readily seen that to the point $3$ we can add all the possible linked one-body diagrams,
or, in other words, all the linked diagrams contributing to the one-body correlation function $g(\mathbf{r}_3)$. The net result is that the
sum of all the reducible diagrams having diagram (\ref{fig:first_ex}.a) as the irreducible part, can be
represented by diagram (\ref{fig:first_ex}.a) with the vertex $3$ {\it renormalized}, in the sense that it is vertex corrected by
$\xi_d$, which in the case of translation invariant system is a constant
\begin{equation}
\xi_d=1+\sum(\text{linked one-body cluster terms})=g^c(\mathbf{r})\,.
\label{eq:vertex_correction}
\end{equation}
A pictorial representation of $g^c(\mathbf{r})$ is given in Figure (\ref{fig_gr}).

Note that there are no restrictions on the kind of cluster terms in the sum, because in diagram (\ref{fig:first_ex}.a) no exchange lines
reach the point $3$, which is dubbed ``d'' point (the label $d$ comes from ``dynamical correlations'' which is the name originally given to the correlation lines). Conversely, if the reducibility point is reached by exchange lines, like, for example, the point $3$ of diagram (\ref{fig:first_ex}.b), which is an ``e'' point, then, to avoid the forbidden superposition of exchange lines, the cluster diagrams of the vertex correction needs to be attached to point $3$ through a correlation line. It follows that there are two kinds of vertex corrections: $\xi_d$ and $\xi_e$, for reducibility points of type $d$ and $e$, respectively.

\begin{figure}[!hb]
\begin{center}
\fcolorbox{white}{white}{
\begin{picture}(455,60)(-125,0)
	\SetWidth{0.5}
	\SetColor{Black}
	\SetScale{0.8}	
        \unitlength=0.8 pt
	
	\Text(-75,10)[]{$U_d(\mathbf{r}_1)=$}
        \Text(-45,10)[]{$\Big[$}
       	
	\CCirc(-30,10){4}{Black}{White}
	\CCirc(30,10){4}{Black}{Black}
	\DashLine(-30,10)(30,10){10}
	\Text(-30,22)[]{$1$}
%	\Text(30,25)[]{$4$}
	
	\Text(50,10)[]{$+$}

	\CCirc(70,10){4}{Black}{White}
	\CCirc(130,40){4}{Black}{Black}
	\CCirc(130,-20){4}{Black}{Black}
	\DashLine(70,10)(130,40){10}
	\DashLine(70,10)(130,-20){10}
	\DashLine(130,40)(130,-20){10}
	\Text(70,22)[]{$1$}
	
	\Text(150,10)[]{$+$}

	\CCirc(170,10){4}{Black}{White}
	\CCirc(230,40){4}{Black}{Black}
	\CCirc(230,-20){4}{Black}{Black}
	\DashLine(170,10)(230,40){10}
	\DashLine(170,10)(230,-20){10}
	\DashLine(230,40)(230,-20){10}
	\Text(170,22)[]{$1$}
	\ArrowArc(252.5,10)(37.5,127,233)
	\ArrowArc(207.5,10)(37.5,-53,53)
	\Text(265,10)[]{$+$}
	\Text(285,10)[]{$\dots$}
	\Text(305,10)[]{$\Big]$}

	\Text(-75,-90)[]{$U_e(\mathbf{r}_1)=$}
        \Text(-45,-90)[]{$\Big[$}
       	\CCirc(-30,-90){4}{Black}{White}
	\CCirc(30,-90){4}{Black}{Black}
	\DashLine(-30,-90)(30,-90){10}
	\ArrowArc(0,-112.5)(37.5,37,143)
	\ArrowArc(0,-67.5)(37.5,-143,-37)
	\Text(-30,-78)[]{$1$}
%	\Text(30,-75)[]{$4$}
	
	\Text(50,-90)[]{$+$}

	\CCirc(70,-90){4}{Black}{White}
	\CCirc(130,-60){4}{Black}{Black}
	\CCirc(130,-120){4}{Black}{Black}
	\ArrowArc(113.42,-101.83)(45,64,165)
	\ArrowArc(86.58,-48.17)(45,249,345)
	\DashLine(70,-90)(130,-60){10}
	\DashLine(70,-90)(130,-120){10}
	\DashLine(130,-60)(130,-120){10}
	\Text(70,-78)[]{$1$}
	
	\Text(150,-90)[]{$+$}

	\CCirc(170,-90){4}{Black}{White}
	\CCirc(230,-60){4}{Black}{Black}
	\CCirc(230,-120){4}{Black}{Black}
	\DashLine(170,-90)(230,-60){10}
	\DashLine(170,-90)(230,-120){10}
	\DashLine(230,-60)(230,-120){10}
	\Text(170,-78)[]{$1$}
	\ArrowArc(213.42,-101.83)(45,64,165)
	\ArrowArc(213.42,-78.17)(45,195,296)
	\ArrowArc(207.5,-90)(37.5,-53,53)
	\Text(265,-90)[]{$+$}
	\Text(285,-90)[]{$\dots$}
	\Text(305,-90)[]{$\Big]$}
    \end{picture}
}
\vspace{4cm}
\caption{Diagrams belonging to $U_d(\mathbf{r}_1)$ and to $U_e(\mathbf{r}_1)$.}
\label{fig:Ude1}
\end{center}
\end{figure}

Actually there is a third type of vertex correction, occurring when the reducibility point is an internal point connected to the rest of the diagram containing the external points only through exchange lines. This correction cannot be the full $\xi_e$, since any solid point must be reached by at least one correlation line, hence  in this case  the vertex correction is $\xi_c=\xi_e-1$. 

The above procedure can be applied to the external point of diagram (\ref{fig:first_ex}.a) and to all irreducible diagrams. The
conclusion is that the sum of reducible and irreducible diagrams may be seen as a sum of vertex corrected irreducible diagrams, which are called ``VIC diagrams'', namely irreducible diagrams whose points carry the vertex corrections $\xi_d$, $\xi_e$ and $\xi_c$. Note that, taking into account vertex corrections, diagram (\ref{fig:sec_ex}.a), without the two superimposed $h(\mathbf{r}_{34})$ lines, is allowed, because point $3$ is vertex corrected by $\xi_c$. Consequently, the second diagrammatic rule does not hold for VIC diagrams that may have internal points reached by exchange lines only.

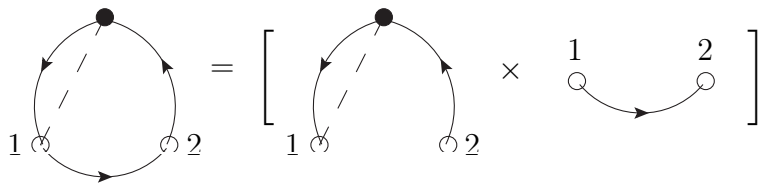
\begin{figure}[!hb]
\begin{center}
\fcolorbox{white}{white}{
\begin{picture}(455,70)(-80,0)
	\SetScale{0.8}	
        \unitlength=0.8 pt
	\SetWidth{0.5}
	\SetColor{Black}	
	\CCirc(0,0){4}{Black}{White}
	\CCirc(60,0){4}{Black}{White}
	\CCirc(30,60){4}{Black}{Black}
	\DashLine(0,0)(30,60){10}
	\ArrowArc(39,18)(41.83,100,205)
	\ArrowArc(21,18)(41.83,335,80)
	\ArrowArc(30,22.5)(37.5,220,320)
	\Text(-12,0)[]{1}
	\Text(72,0)[]{2}
	
	\Text(85,35)[]{$=$}
	\Text(105,32)[]{$\Bigg[$}

	\CCirc(130,0){4}{Black}{White}
	\CCirc(190,0){4}{Black}{White}
	\CCirc(160,60){4}{Black}{Black}
	\DashLine(130,0)(160,60){10}
	\ArrowArc(169,18)(41.83,100,205)
	\ArrowArc(151,18)(41.83,335,80)
	\Text(118,0)[]{1}
	\Text(202,0)[]{2}

	\Text(220,35)[]{$\times$}

	\CCirc(250,30){4}{Black}{White}
	\CCirc(310,30){4}{Black}{White}
	\ArrowArc(280,52.5)(37.5,220,320)
	\Text(250,45)[]{1}
	\Text(311,45)[]{2}
	\Text(335,32)[]{$\Bigg]$}

\end{picture}
}
\vspace{0.5cm}
\caption{A composite diagram decomposed in simple subdiagrams.}
\label{fig:composite}
\end{center}
\end{figure}

For a uniform system like nuclear matter, the one body distribution function is equal to unity, $g_1(\mathbf{r})=1$. This implies the sum rule $\xi_d=1$ that is an useful check for the accuracy of the vertex corrections calculation.

We define a new set of one-body VIC diagrams $U_d(\mathbf{r}_1)$ and $U_e(\mathbf{r}_1)$ in terms of the following equations
\begin{align}
\xi_d&=[1+U_e(\mathbf{r}_1)]\exp[U_d(\mathbf{r}_1)]\\
\xi_e&=\exp[U_d(\mathbf{r}_1)]\,.
\label{eq:vertex}
\end{align}
The external point $1$ of the diagrams belonging to $U_d(\mathbf{r}_1)$  is not reached by any exchange lines; conversely in those forming
 $U_e(\mathbf{r}_1)$ the point $1$ {\it must} be reached by a loop of exchange lines. Moreover diagrams in both $U_d(\mathbf{r}_1)$
and $U_e(\mathbf{r}_1)$ cannot be built by pieces connected only by means of the point $1$. The exponential in the above equations is due to the fact that any number of d-structures forming $U_d$ can be attached to the point $1$. Since the symmetry factor for $n$ topologically identical structures is $1/n!$, the global contribution of $U_d$ is given by $\sum_n(1/n!)U_d(\mathbf{r}_1)^n$.

For the sake of illustration, some of the diagrams belonging to $U_d(\mathbf{r}_1)$ and to $U_e(\mathbf{r}_1)$ are depicted in 
 Fig. \ref{fig:Ude1}. 

\subsubsection{Simple and composite diagrams}
When in an irreducible diagram it is possible to distinguish two or more pieces that are connected with the rest of the diagram by means of the point $i$ and $j$ only, we denote these parts as {\it subdiagrams} \cite{{leeuwen_59}}. Two or more subdiagrams are said to form {\it parallel connections} between the external points $1$ and $2$ when the whole diagram consists of two or more parts which are only connected by means of points $1$ and $2$. As no integration is carried out over the two external points, a factorization of the integral associated with the diagram takes place.

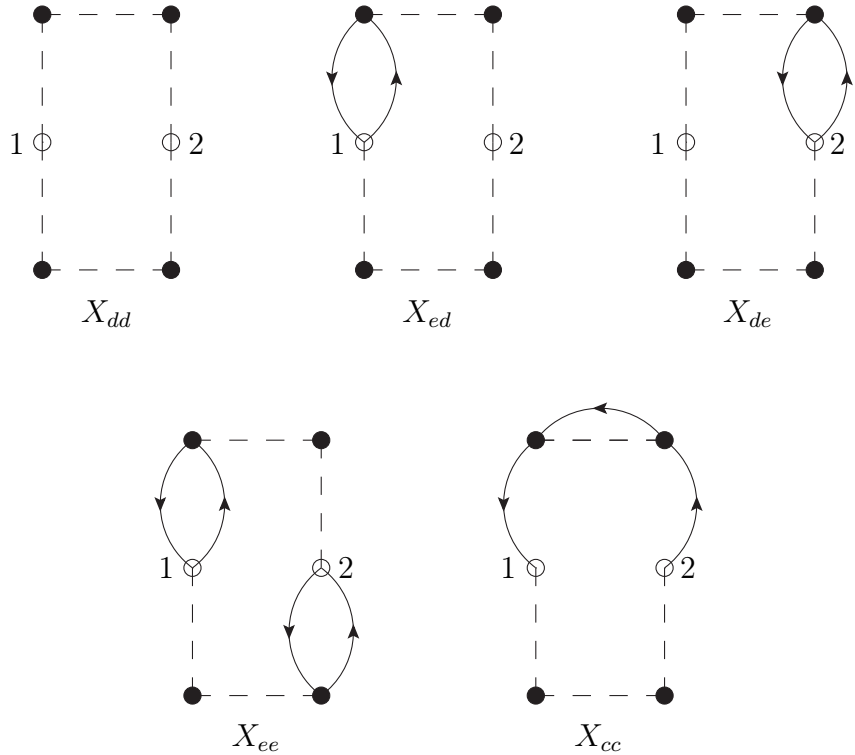
\begin{figure}[!ht]
\begin{center}
\fcolorbox{white}{white}{
\begin{picture}(455,300)(-65,-30)
	\SetWidth{0.5}
	\SetColor{Black}
	\SetScale{0.8}	
        \unitlength=0.8 pt

        \CCirc(0,230){4}{Black}{White}
	\CCirc(60,230){4}{Black}{White}
	\CCirc(0,290){4}{Black}{Black}
	\CCirc(60,290){4}{Black}{Black}
        \DashLine(0,230)(0,290){10}
	\DashLine(0,290)(60,290){10}
	\DashLine(60,290)(60,230){10}
	\Text(-12,230)[]{1}
	\Text(72,230)[]{2}
	\CCirc(0,170){4}{Black}{Black}
	\CCirc(60,170){4}{Black}{Black}
        \DashLine(0,230)(0,170){10}
	\DashLine(0,170)(60,170){10}
	\DashLine(60,170)(60,230){10}
	\Text(30,150)[]{$X_{dd}$}

	\CCirc(150,230){4}{Black}{White}
	\CCirc(210,230){4}{Black}{White}
	\CCirc(150,290){4}{Black}{Black}
	\CCirc(210,290){4}{Black}{Black}
	\DashLine(150,290)(210,290){10}
	\DashLine(210,290)(210,230){10}
	\ArrowArc(172.5,260)(37.5,127,233)
	\ArrowArc(127.5,260)(37.5,307,53)
	\Text(138,230)[]{1}
	\Text(222,230)[]{2}
	\CCirc(150,170){4}{Black}{Black}
	\CCirc(210,170){4}{Black}{Black}
	\DashLine(150,230)(150,170){10}
	\DashLine(150,170)(210,170){10}
	\DashLine(210,170)(210,230){10}
	\Text(180,150)[]{$X_{ed}$}

        \CCirc(300,230){4}{Black}{White}
	\CCirc(360,230){4}{Black}{White}
	\CCirc(300,290){4}{Black}{Black}
	\CCirc(360,290){4}{Black}{Black}
	\DashLine(300,290)(360,290){10}
	\DashLine(300,290)(300,230){10}
	\ArrowArc(382.5,260)(37.5,127,233)
	\ArrowArc(337.5,260)(37.5,307,53)
	\Text(288,230)[]{1}
	\Text(372,230)[]{2}
	\CCirc(300,170){4}{Black}{Black}
	\CCirc(360,170){4}{Black}{Black}
	\DashLine(300,230)(300,170){10}
	\DashLine(300,170)(360,170){10}
	\DashLine(360,170)(360,230){10}
	\Text(330,150)[]{$X_{de}$}

        \CCirc(70,30){4}{Black}{White}
	\CCirc(130,30){4}{Black}{White}
	\CCirc(70,90){4}{Black}{Black}
	\CCirc(130,90){4}{Black}{Black}
	\DashLine(70,90)(130,90){10}
	\DashLine(130,90)(130,30){10}
	\ArrowArc(92.5,60)(37.5,127,233)
	\ArrowArc(47.5,60)(37.5,307,53)
 	\Text(58,30)[]{1}
	\Text(142,30)[]{2}
	\CCirc(70,-30){4}{Black}{Black}
	\CCirc(130,-30){4}{Black}{Black}
	\DashLine(70,30)(70,-30){10}
	\DashLine(70,-30)(130,-30){10}
	\ArrowArc(152.5,0)(37.5,127,233)
	\ArrowArc(107.5,0)(37.5,307,53)
	\Text(100,-50)[]{$X_{ee}$}

	\CCirc(230,30){4}{Black}{White}
	\CCirc(290,30){4}{Black}{White}
	\CCirc(230,90){4}{Black}{Black}
	\CCirc(290,90){4}{Black}{Black}
	\DashLine(230,90)(290,90){10}
	\ArrowArc(252.5,60)(37.5,127,233)
	\ArrowArc(260,67.5)(37.5,37,143)
	\ArrowArc(267.5,60)(37.5,-53,53)
	\Text(218,30)[]{1}
	\Text(302,30)[]{2}
	\CCirc(230,-30){4}{Black}{Black}
	\CCirc(290,-30){4}{Black}{Black}
	\DashLine(230,30)(230,-30){10}
	\DashLine(230,-30)(290,-30){10}
	\DashLine(230,90)(290,90){10}
	\DashLine(290,30)(290,-30){10}
	\Text(260,-50)[]{$X_{cc}$}

\end{picture}
}
\vspace{1cm}
\caption{Classification of composite diagrams based on the kind of the external points.}
\label{fig:comp_class} 
\end{center}
\end{figure}

An irreducible diagram consisting of two or more parallel subdiagrams is called {\it composite} or $X-diagram$. When such division in
parallel subdiagrams is not possible the diagram is called {\it simple}. For example diagram (\ref{fig:first_ex}.a) is simple, as well
as those of Figs. \ref{fig:perm_nos} and \ref{fig:perm_s}. Ignoring the unlinked part, diagram
(\ref{fig:first_ex}.b) is composite, as shown in Fig. \ref{fig:composite}.

As for the vertex corrections, composite diagrams can be classified on the base of the kind of their external points $1$ and $2$. The set of composite diagrams with only  correlation line reaching points $1$ and $2$, like diagram (a) of Fig. \ref{fig:comp_class}, are denoted with $X_{dd}$. When two exchange lines are attached to point $1$ ($2$) the corresponding composite diagrams are labelled with $X_{ed}$ ($X_{de}$), see diagrams (b) and (c) of Fig. \ref{fig:comp_class}. Analogously, diagram  (\ref{fig:comp_class}.d) belongs to the set $X_{ee}$ since two statistical lines arrive at both external points. To build the $ee$ diagrams with the external points in the same statistical loop, it is convenient to define composite diagrams where a statistical open loop starts from the external point 1 and ends to the external point 2, like for instance diagram (e) of Fig. \ref{fig:comp_class} and the one in Fig.  \ref{fig:composite}. Note that the set of these diagrams, denoted by $X_{cc}$, does not directly contribute scalar two-body distribution function.

\subsubsection{Nodal diagrams}
An important concept concerning the diagrams classification is the possible occurrence of a ``node''. Such a node is an internal point through which all possible paths joining the external points $1$ and $2$ pass through. A diagram with one ore more nodes is denoted as ``nodal'' diagram. Diagrams (\ref{fig:first_ex}.a), (\ref{fig:perm_nos}.a) and (\ref{fig:perm_nos}.b) are nodal. 

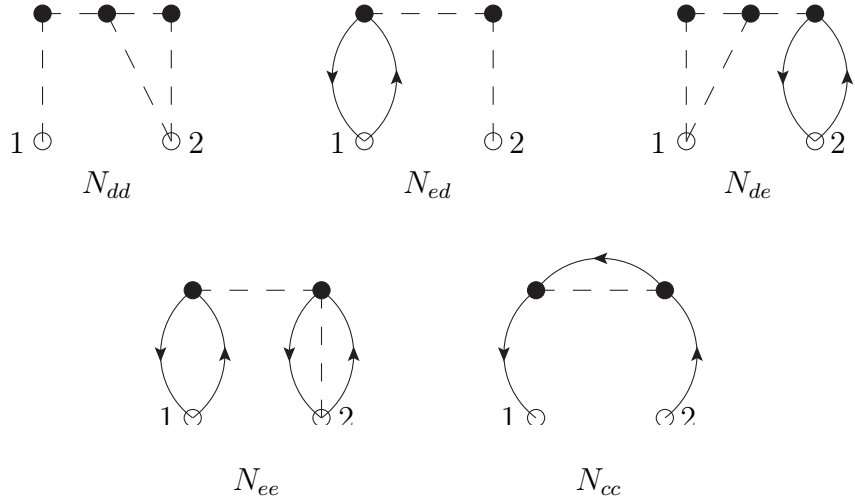
\begin{figure}[!hb]
\begin{center}
\fcolorbox{white}{white}{
\begin{picture}(455,180)(-55,0)
	\SetWidth{0.5}
	\SetColor{Black}
	\SetScale{0.8}	
        \unitlength=0.8 pt

        \CCirc(0,130){4}{Black}{White}
	\CCirc(60,130){4}{Black}{White}
	\CCirc(0,190){4}{Black}{Black}
	 \CCirc(30,190){4}{Black}{Black}
	\CCirc(60,190){4}{Black}{Black}
        \DashLine(0,130)(0,190){10}
	\DashLine(0,190)(60,190){10}
	\DashLine(30,190)(60,130){10}
	\DashLine(60,190)(60,130){10}

	\Text(-12,130)[]{1}
	\Text(72,130)[]{2}
	\Text(30,110)[]{$N_{dd}$}

	\CCirc(150,130){4}{Black}{White}
	\CCirc(210,130){4}{Black}{White}
	\CCirc(150,190){4}{Black}{Black}
	\CCirc(210,190){4}{Black}{Black}
	\DashLine(150,190)(210,190){10}
	\DashLine(210,190)(210,130){10}
	\ArrowArc(172.5,160)(37.5,127,233)
	\ArrowArc(127.5,160)(37.5,307,53)
	\Text(180,110)[]{$N_{ed}$}
	\Text(138,130)[]{1}
	\Text(222,130)[]{2}
	
        \CCirc(300,130){4}{Black}{White}
	\CCirc(360,130){4}{Black}{White}
	\CCirc(300,190){4}{Black}{Black}
	\CCirc(330,190){4}{Black}{Black}
        \CCirc(360,190){4}{Black}{Black}
	\DashLine(300,190)(360,190){10}
	\DashLine(300,190)(300,130){10}
        \DashLine(300,130)(330,190){10}
	\ArrowArc(382.5,160)(37.5,127,233)
	\ArrowArc(337.5,160)(37.5,307,53)
	\Text(330,110)[]{$N_{de}$}
	\Text(288,130)[]{1}
	\Text(372,130)[]{2}

        \CCirc(70,0){4}{Black}{White}
	\CCirc(130,0){4}{Black}{White}
	\CCirc(70,60){4}{Black}{Black}
	\CCirc(130,60){4}{Black}{Black}
	\DashLine(70,60)(130,60){10}
	\DashLine(130,60)(130,0){10}
	\ArrowArc(92.5,30)(37.5,127,233)
	\ArrowArc(47.5,30)(37.5,307,53)
        \ArrowArc(152.5,30)(37.5,127,233)
	\ArrowArc(107.5,30)(37.5,307,53)
	\Text(58,0)[]{1}
	\Text(142,0)[]{2}
	\Text(100,-30)[]{$N_{ee}$}

	\CCirc(230,0){4}{Black}{White}
	\CCirc(290,0){4}{Black}{White}
	\CCirc(230,60){4}{Black}{Black}
	\CCirc(290,60){4}{Black}{Black}
	\DashLine(230,60)(290,60){10}
	\ArrowArc(252.5,30)(37.5,127,233)
	\ArrowArc(260,37.5)(37.5,37,143)
	\ArrowArc(267.5,30)(37.5,-53,53)
	\Text(260,-30)[]{$N_{cc}$}
	\Text(218,0)[]{1}
	\Text(302,0)[]{2}

\end{picture}
}
\vspace{1cm}
\caption{Classification of nodal diagrams according to the kind of the external points. }
\label{fig:nod_class} 
\end{center}
\end{figure}

A nodal diagram is also necessarily a simple diagram, but not all simple diagrams are nodal, as for example diagram (\ref{fig:perm_s}). Non nodal simple diagrams are called ``elementary'' diagrams, or $E-diagrams$. 
Nodal diagrams can be classified according to the same scheme, based on the kind of external points, adopted for composite diagrams. Examples of diagrams belonging to sets $N_{dd}$, $N_{ed}$, $N_{de}$, $N_{ee}$ and $N_{ee}$ are depicted in Fig.  \ref{fig:nod_class}.

\subsection{FHNC equations}
\label{subsec:it_FHNC}
Consider a nodal diagram contributing to $N_{xy}(r_{12})$ and label $3$ the node closest to point $1$. 
All the nodal diagrams can then be found by convoluting the sum of all non-nodal $1-3$ subdiagrams, $X_{xx^\prime}(r_{13})$, with the set of $3-2$ subdiagrams $X_{y^\prime y}(r_{23})+N_{y^\prime y}(r_{23})$ (with or without nodes). This leads to the following integral equation
\begin{align}
N_{xy}(r_{12})=\sum_{x^\prime y^\prime}\rho\int d\mathbf{r}_3 X_{xx^\prime}(r_{13}) \zeta_{x^\prime y^\prime} [X_{y^\prime y}(r_{23})+N_{y^\prime y}(r_{23})]\, ,
\label{eq:fhnc_nodal}
\end{align}
the indexes $x$ and $y$ running over the types $d$ and $e$ of the external point. The coefficients $\zeta_{x^\prime y^\prime}$, accounting for the  vertex corrections and for the proper treatment of the exchange loops, read
\begin{equation}
\zeta_{dd}=\xi_d\quad,\quad \zeta_{de}=\zeta_{ed}=\xi_e\quad,\quad \zeta_{ee}=0\, .
\end{equation}

In order to better understand the effect of the long-range part of the correlation function, it is worth rewriting the convolution Eq. (\ref{eq:fhnc_nodal}) in momentum space
\begin{equation}
\tilde{N}(k)=\rho \xi \tilde{X}(k)[\tilde{X}(k)+\tilde{N}(k)]\, .
\label{eq:iter_Nconv}
\end{equation}
where $\tilde{N}(k)$ and $\tilde{X}(k)$ are the Fourier transforms of the nodal and composite functions, respectively. Note that we have omitted the subscripts referring to the kind of verices, whose presence is irrelevant to the purpose of this discussion. Solving Eq. (\ref{eq:iter_Nconv}) for $\tilde{N}(k)$ yields
\begin{equation}
\tilde{N}(k)=\frac{\rho \xi\tilde{X}(k)}{1-\rho\xi\tilde{X}(k)}\, .
\label{eq:Nk}
\end{equation}
As a further simplification, consider the nodal diagram $N^{n}(r)$ formed by $n$ correlation lines $h(r)$. The analytic expression of $\tilde{N}^{n}(k)$ can be easily derived iterating Eq. (\ref{eq:iter_Nconv}), where $\tilde{X}(k)$ has to be replaced by its first order contribution $\tilde{h}(k)$
\begin{equation}
\tilde{N}^n(k)=(\rho \xi)^{n-1} \,\tilde{h}(k)^n\, .
\end{equation}
Therefore, the sum of all the $N^{n}(r)$ is given by 
\begin{equation}
\sum_n\tilde{N}^n(k)=\frac{\rho\xi \tilde{h}(k)}{1-\rho\xi\tilde{h}(k)}\, ,
\end{equation}
which is in turn a particular case of Eq. (\ref{eq:Nk}). If $h(r)$ is long ranged, say $h(r)\to \alpha/r^2$ for $r\to\infty$, then $\tilde{h}(k) \to -\alpha/(2\pi k)$ for $k\to 0$. Consequently, each $\tilde{N}^n(k)$ is more divergent than $\tilde{h}(k)$ in the long wavelength limit, while their sum is well behaved as it diverges like $\tilde{h}(k)$. From this fact, we can gather that the expansion in the number of points diverges at any finite order, while the chain summations leads to a well behaved long range limit.

The iterative equation for $N_{cc}$, cannot be written in a form analogous to that of Eq. (\ref{eq:fhnc_nodal}). It differs more from those given in ref., where because the cyclic nodal diagrams do not show not all the cancellations occurring when the cancellation of reducible diagrams had been assumed from the very beginning \cite{fantoni_01}. We need to distinguish two different kinds of external points: the point $x$ which is reached by an exchange line and at least one correlation line and the point $p$ which is reached by an exchange line only. Hence four type of nodal cyclic functions, namely $N_{cc}^{xx}$, $N_{cc}^{xp}$, $N_{cc}^{px}$ and $N_{cc}^{pp}$ and correspondingly three type of composite functions, $X_{cc}^{\alpha\beta}$, have to be properly taken into account. For instance, the convolution of $N_{cc}^{xx}$ with any other of the cyclic functions brings a vertex correction $\xi_e$, whereas the convolution of two $pp$ cyclic  functions requires a $\xi_c$ vertex correction. The four nodal equations are given by
\begin{align}
N_{cc}^{xx} (r_{12})&=\rho\int d\mathbf{r}_{3}\xi_e X_{cc}(r_{13})[X_{cc}(r_{32})+ N_{cc}^{xx}(r_{32}) + N_{cc}^{px}(r_{32})] \nonumber \\
N_{cc}^{xp}(r_{12})&=\rho\int d\mathbf{r}_{3}\xi_e X_{cc}(r_{13})\Big[-\frac{1}{\nu}\ell(r_{32})+ N_{cc}^{xp}(r_{32}) + N_{cc}^{pp}(r_{32})\Big] \nonumber \\
N_{cc}^{px}(r_{12})&=-\frac{\rho}{\nu}\int d\mathbf{r}_{3}\xi_e \ell(r_{13})[X_{cc}(r_{32}) +N_{cc}^{xx}(r_{32})] +\xi_c \ell(r_{13})N_{cc}^{px}(r_{32}) \nonumber \\
N_{cc}^{pp}(r_{12}) &= -\frac{\rho}{\nu}\int d\mathbf{r}_{3}\xi_e \ell(r_{13})N_{cc}^{xp}(r_{32})+\xi_c \ell(r_{13})\Big[N_{cc}^{pp}(r_{32})-\frac{1}{\nu}\ell(r_{32})\Big]\, ,
\label{eq:nodal_cyc}
\end{align}
with
\begin{equation}
N_{cc}^{px}(r_{12})=N_{cc}^{xp}(r_{12})\ .
\label{eq:ncc_sym}
\end{equation}

The total cyclic nodal functions is given by
\begin{eqnarray}
N_{cc}(r_{12}) = N_{cc}^{xx}(r_{12}) + N_{cc}^{xp}(r_{12}) + N_{cc}^{px}(r_{12}) + N_{cc}^{pp}(r_{12})\ .
\label{eq:nodal_cyc_tot}
\end{eqnarray}
Summing the Eqs. (\ref{eq:nodal_cyc}), one can solve the following integral equation for the total $N_{cc}$ 
\begin{align}
N_{cc}(\vec{r}_{12}) &= \rho \int_V d\mathbf{r}_{3}  X_{cc}(r_{13})\xi_e\Big[X_{cc}(r_{32}) + N_{cc}(r_{32}) - \frac{\ell(\vec{r}_{32})}{\nu}\Big]\nonumber \\
&- \frac{\ell(r_{13})}{\nu} \xi_e \Big[X_{cc}(r_{32})+\mathcal{P}(r_{32})\Big] \nonumber \\
&- \frac{\ell(r_{13})}{\nu} \xi_c \Big[-\frac{\ell(r_{32})}{\nu} + N_{cc}(r_{32}) - \mathcal{P}(r_{32}) \Big] \ ,
\label{eq:ncc_tot}
\end{align}
where $\mathcal{P}$ is given by
\begin{eqnarray}
\mathcal{P}(r_{12}) = \rho \int_V d\mathbf{r}_{3} \xi_e X_{cc}(r_{13}) \Big[X_{cc}(r_{32}) + N_{cc}(r_{32}) - \frac{\ell(\vec{r}_{32})}{\nu} \Big] \ .
\label{eq:propcc}
\end{eqnarray}

At this stage, it is worth introducing the expressions for the {\it partial} two-body scalar distribution functions 
\begin{align}
g_{dd}^c(r_{12}) &= f^2(r_{12})\exp[N_{dd}(r_{12}) + E_{dd}(r_{12})]\ , \nonumber \\
g_{de}^c(r_{12}) &= g_{ed}^c(r_{12}) =N_{de}(r_{12}) + X_{de}(r_{12})\ , \nonumber \\
g_{ee}^{c\,dir}(r_{12}) &= g_{dd}^c(r_{12})\{N_{ee}(r_{12}) + E_{ee}^{dir}(r_{12}) + [N_{de}(r_{12}) + E_{de}(r_{12})]^2\}\ , \nonumber \\
g_{ee}^{c\,exch}(r_{12}) &= -\nu\,g_{dd}^c(r_{12}) \Big[N_{cc}(r_{12}) - \frac{1}{d}\ell(r_{12}) + E_{cc}(r_{12})\Big]^2  + E_{ee}^{exch}(r_{12}) g_{dd}^c(r_{12}) \ ,\nonumber \\
g_{cc}(r_{12}) &= N_{cc}(r_{12}) + X_{cc}(r_{12}) - \frac{1}{\nu}\ell(r_{12})\ .
\label{eq:distribution}
\end{align}
where $E_{xy}(\vec{r}_{12})$ represent the sum of the $xy$ elementary diagrams. The explanation for the presence of the exponential in the above equations is analogous to the one given for the vertex corrections of Eq. (\ref{eq:vertex}). Note that, as for the vertex corrections, there are no exponentials associated with the $e-$ structures, as two exchange loops cannot be superimposed.

The composite functions, which in turn can be seen as generalized links, are defined as
\begin{align}
X_{dd}(r_{12}) &= g_{dd}^c(r_{12})-N_{dd}(r_{12})-1 \nonumber \\
X_{de}(r_{12}) &= X_{ed}(r_{12}) = g_{dd}^c(r_{12})[N_{de}(r_{12}) + E_{de}(r_{12})] - N_{de}(r_{12})  \ , \nonumber \\
X_{ee}(r_{12}) &= g_{dd}^c(r_{12})\Big\{N_{ee}(r_{12}) + E_{ee}(r_{12}) + [N_{de}(r_{12}) + E_{de}(r_{12})]^2 \nonumber \\
&- \nu\Big[N_{cc}(r_{12}) - \frac{1}{\nu}\ell(r_{12}) + E_{cc}(r_{12})\Big]^2 \Big\} - N_{ee}(r_{12})  \  \nonumber\\
X_{cc}(r_{12}) &= g_{dd}^c(r_{12}) [N_{cc}(r_{12}) - \frac{1}{\nu}\ell(r_{12}) + E_{cc}(r_{12})]-N_{cc}(r_{12}) + \frac{1}{\nu}\ell(r_{12})\,
\label{eq:composite}
\end{align}

The functions $U_{d,e}$ appearing in eq. (\ref{eq:vertex}) and entering the vertex corrections $\xi_{d,e}$ are solutions of the following integral equations
\begin{align}
U_d &= \rho\int_V d\mathbf{r}_{12}
\xi_d [X_{dd}(r_{12}) - E_{dd}(r_{12}) - S_{dd}(r_{12})(g_{dd}(r_{12})-1)] \nonumber \\
&+ \xi_e [X_{de}(r_{12}) - E_{de}(r_{12}) - S_{dd}(r_{12})g_{de}(r_{12}) \nonumber \\
&- S_{de}(r_{12})(g_{dd}(r_{12})-1)] + E_d\ , \nonumber \\
\nonumber\\
U_e &= \rho\int_V d\vec{r}_{12}
\xi_d [X_{ed}(r_{12}) - E_{ed}(r_{12})] + \xi_e [X_{ee}(r_{12}) - E_{ee}(r_{12})] \nonumber \\
-& \xi_d \left[S_{dd}(r_{12})g_{ed}(r_{12})+S_{ed}(r_{12})(g_{dd}(r_{12})-1)  \right] - \xi_e [ S_{ee}(r_{12})(g_{dd}(r_{12})-1) \nonumber \\
+& S_{ed}(r_{12})g_{de}(r_{12}) + S_{dd}(r_{12})g_{ee}(r_{12})+ S_{de}(r_{12})g_{ed}(r_{12})  - 2d S_{cc}(r_{12})g_{cc}(r_{12}) ] \nonumber \\
-& \ell(r_{12})[{\mathcal N}_{cc}^p(r_{12}) - \frac{1}{\nu}\ell(r_{12})] + E_e\ ,
\label{eq:u_vertex}
\end{align}
with
\begin{align}
S_{xy}(\vec{r}_{12}) &= \frac{1}{2} N_{xy}(r_{12}) + E_{xy}(r_{12}) \ , \nonumber \\
{\cal N}_{cc}^x(r_{12}) &= N_{cc}^{xx}(r_{12}) + N_{cc}^{xp}(r_{12}) = \xi_e \rho\int_V d\mathbf{r}_{32} X_{cc}(r_{13})[X_{cc}(r_{32}) + N_{cc}(r_{32}) -\frac{1}{\nu}\ell(r_{32})] \ , \nonumber \\
{\mathcal N}_{cc}^p(r_{12}) &= N_{cc}^{pp}(r_{12}) + N_{cc}^{px}(r_{12}) = N_{cc}(r_{12}) - {\mathcal N}_{cc}^x(r_{12}) \ ,
\label{eq:sdef}
\end{align}
and $E_d$ and $E_e$ are the one--body elementary diagrams  with external point $d$  and $e$ respectively.

The FHNC equations are solved numerically by means of iterative procedures. At the $nth$ step, the nodal diagrams and the 1-body structures resulting from the step $n-1$, namely $N_{xy}(n-1)$ and $U_x(n-1)$, are employed to compute the partial scalar two-body distribution functions as well as the composite functions $X_{xy}(n)$. Hence, the new nodal functions and the 1-body structures are calculated making use of those quantities. 

The iterative procedure is stopped when the difference between the values of a test quantity, e.g. the nodal diagrams or the energy, computed in two successive iterations is smaller than a given convergence parameter. In order for the procedure to start, the nodal diagrams are initially set to zero, while for vertex corrections one sets $\xi_d=\xi_e=1$. 

For dense systems, like liquid helium or nuclear matter, the convergence may be difficult to reach, and one may want to smooth out the iterative process. One of the most used technique consists in multiplying by a ``mixing parameter'', $0<\alpha_{mix}<1$, the nodal diagrams resulting from the step $n-1$ of the iteration procedure and by $1-\alpha_{mix}$ those obtained in the current step $n$. Then all the other quantities at the iteration $n$, like, for instance, the composite diagrams, are obtained using the mixture
\begin{equation}
\alpha_{mix}N_{xy}({n-1})+(1-\alpha_{mix})N_{xy}({n})\, .
\end{equation}

In order to close the FHNC scheme, an iterative equation for the elementary diagrams would be required. However, because of their topological structure, a consistent treatment of the elementary diagrams based on two-body kernel equations, like those for the nodal and composite diagrams, is not feasible. 

The simplest approximation consists in neglecting all the elementary diagrams, setting $E_{xy}=0$ (FHNC/0 approximation). Although elementary diagrams are neglected in this approach, FHNC/0 provides a very good description of the long-range part of the two-body distribution function, and it is also accurate enough at short interparticle distances (although $E_{xy}(r)$ are short-range functions). For instance, in nuclear matter calculation the elementary diagrams' contribution is likely to be small, provided that accurate minimization procedures, like the one described in Section \ref{Variational_SA}, are performed. 

\begin{figure}[!ht]
\begin{center}
\fcolorbox{white}{white}{
\begin{picture}(455,180)(-65,0)
	\SetWidth{0.5}
	\SetColor{Black}
	\SetScale{0.8}	
        \unitlength=0.8 pt

        \CCirc(0,130){4}{Black}{White}
	\CCirc(60,130){4}{Black}{White}
	\CCirc(0,190){4}{Black}{Black}
	\CCirc(60,190){4}{Black}{Black}
        \DashLine(0,130)(0,190){10}
	\DashLine(0,190)(60,190){10}
	\DashLine(0,190)(60,130){10}
        \DashLine(0,130)(60,190){10}
	\DashLine(60,190)(60,130){10}

	\Text(-12,130)[]{1}
	\Text(72,130)[]{2}
	\Text(30,110)[]{$E_{dd}$}

	\CCirc(150,130){4}{Black}{White}
	\CCirc(210,130){4}{Black}{White}
	\CCirc(150,190){4}{Black}{Black}
	\CCirc(210,190){4}{Black}{Black}
	\DashLine(150,130)(210,190){10}
	\DashLine(210,130)(150,190){10}
	\DashLine(150,190)(210,190){10}
	\DashLine(210,190)(210,130){10}
	\ArrowArc(172.5,160)(37.5,127,233)
	\ArrowArc(127.5,160)(37.5,307,53)
	\Text(180,110)[]{$E_{ed}$}
	\Text(138,130)[]{1}
	\Text(222,130)[]{2}
	
        \CCirc(300,130){4}{Black}{White}
	\CCirc(360,130){4}{Black}{White}
	\CCirc(300,190){4}{Black}{Black}
        \CCirc(360,190){4}{Black}{Black}
	\DashLine(300,190)(360,190){10}
	\DashLine(300,190)(300,130){10}
        \DashLine(300,130)(360,190){10}
        \DashLine(300,190)(360,130){10}
        \DashLine(360,190)(360,130){10}
	\ArrowArc(382.5,160)(37.5,127,233)
	\ArrowArc(337.5,160)(37.5,307,53)
	\Text(330,110)[]{$E_{de}$}
	\Text(288,130)[]{1}
	\Text(372,130)[]{2}

        \CCirc(70,0){4}{Black}{White}
	\CCirc(130,0){4}{Black}{White}
	\CCirc(70,60){4}{Black}{Black}
	\CCirc(130,60){4}{Black}{Black}
	\DashLine(70,60)(130,60){10}
	\DashLine(130,60)(130,0){10}
        \DashLine(70,0)(130,60){10}
	\DashLine(130,0)(70,60){10}
	\ArrowArc(92.5,30)(37.5,127,233)
	\ArrowArc(47.5,30)(37.5,307,53)
        \ArrowArc(152.5,30)(37.5,127,233)
	\ArrowArc(107.5,30)(37.5,307,53)
	\Text(58,0)[]{1}
	\Text(142,0)[]{2}
	\Text(100,-30)[]{$E_{ee}$}

	\CCirc(230,0){4}{Black}{White}
	\CCirc(290,0){4}{Black}{White}
	\CCirc(230,60){4}{Black}{Black}
	\CCirc(290,60){4}{Black}{Black}
	\DashLine(230,60)(290,60){10}
	\DashLine(230,0)(290,60){10}
	\DashLine(230,60)(290,00){10}

	\ArrowArc(252.5,30)(37.5,127,233)
	\ArrowArc(260,37.5)(37.5,37,143)
	\ArrowArc(267.5,30)(37.5,-53,53)
	\Text(260,-30)[]{$E_{cc}$}
	\Text(218,0)[]{1}
	\Text(302,0)[]{2}

\end{picture}
}
\vspace{1cm}
\caption{Some 4-body elementary diagrams.}
\label{fig:elementary_4} 
\end{center}
\end{figure}
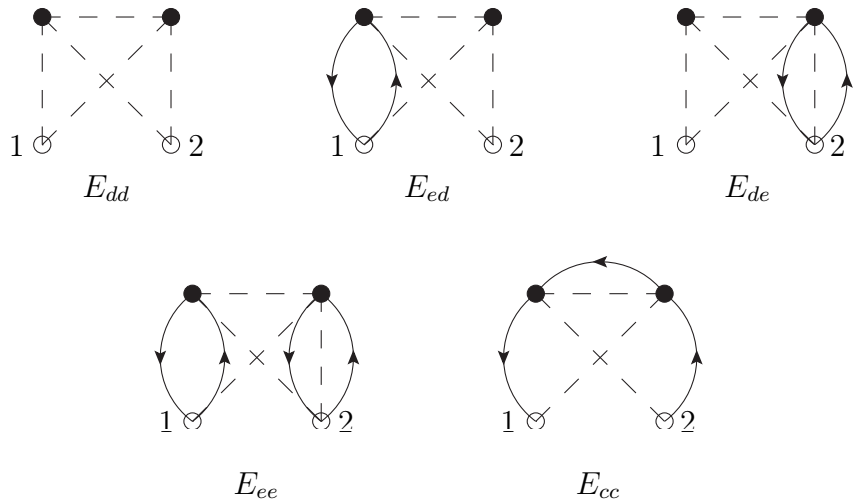

A brute force approach based on the direct calculation of the $n-$body elementary diagrams $E_n$, like the $4-$body diagrams of Fig. \ref{fig:elementary_4}, and the inclusion of their contributions in the FHNC equations, (FHNC/n approximation) does not seem to be a promising strategy. In fact, in the cases where the elementary diagrams give appreciable effects, like in liquid helium, the FHNC/n series shows poor convergence.

A more reliable procedure for bosonic systems, denote as {\it scaling approximation} (HNC/s) \cite{usmani_82} amounts to approximating the full set of elementary diagram by $\alpha E_4$. The scaling parameter, $\alpha$, is determined by matching the expectation values of the kinetic energy obtained following the Pandharipande-Bethe and the Jackson-Feenberg prescriptions, prescriptions, to be discussed in the next Section, which in an exact calculation would give the same results. The existence of a scaling property for fermionic systems is not at all clear, as there are different classes of elementary diagrams, depending on the kind of the external points. Nevertheless, the fermionic generalization of the scaling approximation has been used in liquid $^3$He with some success \cite{manousakis_83}.

Preliminary results obtained using iterative equations with four-body kernels are quite encouraging, as the difference with respect to Variational Monte Carlo results turns out to be much smaller with respect to the FHNC/0 case \cite{wiringa_04}.

The total {\sl scalar} two--body distribution function is given by the sum of the four different kinds of the partial two-body distribution functions, multiplied by the appropriate vertex corrections
\begin{align}
g^c(r_{12}) &= \xi_d^2 g_{dd}^c(r_{12})+\xi_d\xi_e [g_{de}^c(r_{12}) + g_{ed}^c(r_{12})]+ \xi_e^2 [g_{ee}^{c\,dir}(r_{12}) + g_{ee}^{c\,exch}(r_{12})]\,. 
\label{eq:gscalar}
\end{align}

The numerical solution of the coupled FHNC integral equations is not trivial at all. Moreover,the numerical convergence of the solution does not ensure that this solution is acceptable from the physical point of view, as neglecting elementary diagrams could in principle lead to a violation of the variational principle. A useful tool to check both the numerical and the theoretical accuracy of the calculations is the fulfillment of the sum rule of the scalar distribution function of Eq. (\ref{eq:gc_sumrule}). Comparing the values of the kinetic energy expectation value obtained using the Pandharipande-Bethe and the Jackson-Feenberg prescriptions, is also an indicator for the variational principle to be respected. 

Using Eqs. (\ref{eq:potpparcent}) and (\ref{eq:gscalar}) one can evaluate the potential energy contribution to the energy per particle. In addition, the knowledge of the two-body distribution function allows for the calculation of the kinetic energy per particle, to be discussed in the next Section. 

\subsection{Kinetic energy prescriptions}
\label{sec:kin_presc}
In this Section we shall perform a number of manipulations on the kinetic energy expectation value. To keep the discussion general, we do not assume that $\mathcal{F}$ takes the simple scalar Jastrow form of Eq. (\ref{eq:jastrow_corr}) adopted in the FR approach, but only that it is a real function of the relative particle coordinates, so that the correlator of Eq. (\ref{eq:Foperator}) satisfies $\mathcal{F}^\dagger=\mathcal{F}$.

The kinetic energy expectation value is given by
\begin{align}
\langle \hat{T}\rangle &=\frac{\hbar^2}{2m}\sum_i\langle\nabla_{i}^2\rangle=A\frac{\hbar^2}{2m}\langle\nabla_{1}^2\rangle\nonumber \\
\langle\nabla_{1}^2\rangle&=\text{den}^{-1}\int dx_{1,\dots,A} \Psi_0^*\mathcal{F}^\dagger\,\nabla_{1}^2\,\mathcal{F}\Psi_0\, ,
\label{eq:mb_kin}
\end{align}
where den$=\int dx_{1,\dots,A} \Psi_0^*\mathcal{F}\mathcal{F}\Psi_0$ accounts for the normalization of the CBF wave function In the first line we have exploited the symmetry properties of the correlated wave function, as in Eq. (\ref{eq:pot_exp}). It is worth noting that a crucial role in this context is played by the symmetryzation operator $\mathcal{S}$ appearing in Eq. (\ref{eq:Foperator}).

The most straightforward form for the kinetic energy is obtained by applying the laplacian to the right
\begin{align}
\langle\nabla_{i}^2\rangle_{PB}&=\text{num}^{-1}\int dx_{1,\dots,A} \Psi_0^*\mathcal{F}^\dagger(\nabla_{i}^2\mathcal{F}\Psi_0)\nonumber \\
&=\text{num}^{-1}\int dx_{1,\dots,A} \Psi_0^*\mathcal{F}^\dagger[\mathcal{F}(\nabla_{i}^2\Psi_0)+2(\vec{\nabla}_{i}\mathcal{F})\cdot(\vec{\nabla}_i\Psi_0)+(\nabla_{i}^2\mathcal{F})\Psi_0]\, .
\label{eq:kin_PB}
\end{align}
This expression of the kinetic energy is called ``Pandharipande-Bethe'' (PB) \cite{pandha_73}, although it was first implemented by Iwamoto and Yamada \cite{iwamoto_57}. 

The first term in square brackets generates the Fermi gas energy 
\begin{equation}
T_F=\nu\sum_i\frac{\hbar^2k_{i}^2}{2m}=A\frac{3}{5}\frac{\hbar^2k_{F}^2}{2m}\, .
\end{equation}
The full PB kinetic energy is given by
\begin{equation}
\langle T \rangle_{PB} = T_F+W^{kin}+W_F+U+U_F\, .
\end{equation}
Integrals involving $\nabla_{i}^2 F_{ij}$ are included in $W^{kin}$ and are completely analogous to those arising from the two-body potential. Three body terms with derivatives acting on the correlation only, $\vec{\nabla}_{i} F_{ij}\vec{\nabla}_{i} F_{ik}$, are contained in $U$. On the other hand, $W_F$ and $U_F$ accounts for the contributions coming from $\vec{\nabla}_{i} F_{ij}\vec{\nabla}_{i} \ell_{ij}$ and $\vec{\nabla}_{i} F_{ij}\cdot \vec{\nabla}_{i} \ell_{ik}$, respectively.
The different contributions can be readily illustrated in the case of pure Jastrow correlations \cite{manousakis_83}
\begin{align}
W^{kin}&=\frac{\rho}{2}\int d\mathbf{r}_{12} \,g(r_{12}) \Big[-\frac{\hbar^2}{m}\frac{\nabla_{1}^2 f^c(r_{12})}{f^c(r_{12})}\Big]\nonumber\\
U&=\rho^2\int d\mathbf{r}_{12}d\mathbf{r}_{13}g_3(r_{12},r_{13},r_{23})\Big[-\frac{\hbar^2}{2m}\frac{\vec{\nabla}_{1} f^c(r_{12})\cdot \vec{\nabla}_{1}f^c(r_{13})}{ f^c(r_{12}) f^c(r_{13})}\Big]\nonumber \\
W_F&=\rho\int d\mathbf{r}_{12}g_{cc}(r_{12})\Big[-\frac{\hbar^2}{2m}\frac{{f^c\,}'(r_{12})}{f^c(r_{12})}\ell(r_{12})\Big]\nonumber\\
U_F&=\rho^2\int d\mathbf{r}_{12}d\mathbf{r}_{13}\Big[-\frac{\hbar^2}{2m}\frac{{f^c\,}'(r_{12})}{f^c(r_{12})}\ell '(r_{13})\Big]\hat{r}_{12}
\cdot\hat{r}_{13}\nonumber \\
&\times\sum_{exch}\{g_{cc}(r_{13})g_{dy}(r_{12})[g_{dy'}(r_{23})-1]+g_{dd}(r_{13})g_{cc}(r_{23})g_{cc}(r_{13})\}\, .
\end{align}
In order to simplify the notation, cancellation among irreducible diagrams has been assumed; moreover, in the expression for $U_F$ the Abe diagrams appearing in the three-body distribution function have been neglected
\begin{equation}
g_3(r_{12},r_{13},r_{23})=\sum_{exch}g_{xx'}(r_{12})g_{yy'}(r_{23})g_{zz'}(r_{13})\, .
\label{eq:g3_scalar_def}
\end{equation}
Note that terms containing $\vec{\nabla}_i\Psi_0$ gives zero contribution in direct diagrams after summation over $\mathbf{k}_i$. The only terms contributing to $U_F$ are exchange diagrams in which $\sum_i i\mathbf{k}_i \exp(i\mathbf{k}_{ij}\cdot\mathbf{r}_{ij})$ gives $\hat{r}_{ij}\ell(r_{ij})$.

Integrating by parts the last term in square brackets of Eq. (\ref{eq:mb_kin}) and using the identity
\begin{equation}
\sum_i [(\vec{\nabla}_i\Psi_{0}^*)\mathcal{F}^\dagger(\vec{\nabla}_i\mathcal{F})\Psi_{0}-
\Psi_{0}^*(\vec{\nabla}_i\mathcal{F}^\dagger)\mathcal{F}(\vec{\nabla}_i\Psi_{0})]=0
\end{equation}
yields the ``Clark-Westhaus'' form of the kinetic energy
\begin{align}
\langle\nabla_{i}^2\rangle_{CW}&=\text{num}^{-1}\int dx_{1,\dots,A}\Psi_0^*[\mathcal{F}^\dagger\mathcal{F}(\nabla_{i}^2\Psi_0)-(\vec{\nabla}_i\mathcal{F}^\dagger)\cdot(\vec{\nabla}_i\mathcal{F})\Psi_0]\nonumber \\
\langle T \rangle_{CW} &= T_F+W^{kin}_{CW}-U\, .
\label{eq:kin_CW}
\end{align}

In the case of pure scalar Jastrow correlation it turns out that
\begin{equation}
W^{kin}_{CW}=\frac{\rho}{2}\int d\mathbf{r}_{12} \,g(r_{12}) \Big[-\frac{\hbar^2}{m}\Big(\frac{f^c(r_{12})'}{f^c(r_{12})}\Big)^2\Big]\, .
\end{equation}

Integrating once by parts the first line of Eq. (\ref{eq:mb_kin}) leads to $-(\vec{\nabla}_{i}\Psi_{0}^*\mathcal{F}^\dagger)\cdot (\vec{\nabla}_i\mathcal{F}\Psi_0)$, while yet another integration produces an expression in which the laplacian acts on the left, $(\nabla_{i}^2\Psi_{0}^*\mathcal{F}^\dagger)\mathcal{F}\Psi_0$. The Jackson-Feenberg form of the kinetic energy is obtained by averaging these contributions
\begin{align}
\langle\nabla_{i}^2\rangle_{JF}&=\text{num}^{-1}\int dx_{1,\dots,A}\,\frac{1}{4}\Big[ \Psi_0^*\mathcal{F}^\dagger(\nabla_{1}^2\mathcal{F}\Psi_0)-2(\vec{\nabla}_i\Psi_0^*\mathcal{F}^\dagger)\cdot(\vec{\nabla}_1\mathcal{F}\Psi_0)\nonumber \\
&+(\nabla_{1}^2\Psi_{0}^*\mathcal{F}^\dagger)\mathcal{F}\Psi_0\Big]\, \nonumber \\
\langle T \rangle_{JF}&=T_F+W_{B}^{kin}+W_{\phi}+U_{\phi}\, .
\label{eq:kin_JF1}
\end{align}
The $W_B$ two-body integral contains the kinetic contributions involving derivatives on correlations only $(\nabla_{i}^2F_{ij}-\vec{\nabla}_iF_{ij}\cdot \vec{\nabla}_iF_{ij})$. In the case of central correlations it has the form of a two-body integral for a Bose liquid
\begin{equation}
W_B=\frac{\rho}{2}\int d\mathbf{r}_{12} \,g(r_{12}) \Big[-\frac{\hbar^2}{2m}\Big(\frac{f^c(r_{12})\nabla_{1}^2f^c(r_{12})-{f^c}'(r_{12})^2}{f^c(r_{12})^2}\Big)\Big]\, .
\end{equation}
The $W_{\phi}$ and $U_{\phi}$ have a fermionic origin as they result from $\nabla_{i}^2\Psi_{0}^*\Psi_0$ and for a scalar Jastrow correlation read
\begin{align}
W_{\phi}&=\frac{\rho}{2\nu}\int d\mathbf{r}_{12}\Big(-\frac{\hbar^2}{2m}\Big)\Big[(g_{dd}-1)\{[\ell(r_{12})-\nu(N_{cc}+E_{cc})]\nabla_{1}^2\ell(r_{12})+\ell'(r_{12})^2\}\nonumber\\
&-\nu g_{dd}(r_{12})E_{cc}(r_{12})\nabla_{1}^2\ell(r_{12})\Big]\nonumber\\
U_{\phi}&=\frac{\rho}{2}\int d\mathbf{r}_{12}d\mathbf{r}_{13}\Big(-\frac{\hbar^2}{4m}\Big)[(g_{dd}(r_{12})-1)\ell'(r_{12})][(g_{dd}(r_{13})-1)\ell'(r_{13})]\nonumber\\
&\times g_{cc}(r_{23})\hat{r}_{12}\cdot\hat{r}_{13}\, .
\end{align}

Although the three forms of the kinetic energy that we have derived are formally equivalent, each has its own distinctive advantages and disadvantages in actual cluster expansion calculations. The CW form has the remarkable feature of not involving second derivatives and there are no additional terms arising when one goes from a Bose system to a Fermi system. 

Because of the large cancellation among the two-body potential contribution and $W^{kin}+W_{F}$, the PB kinetic energy is rather unsensitive to the short-range uncertainties of $g_c(r_{12})$. The three body term $U_{\phi}$ of the JF procedure is smaller than the three body terms $U$ and $U_F$ of the PB and CW prescriptions, making the JF kinetic energy essentially unaffected by the approximations involved in the three-body distribution function. The drawback of the JF prescription mainly resides in the deficient cancellation occurring between $W_{\phi}+W_{\phi}$ and the two-body potential contribution. Hence, the JF kinetic energy is more affected by the poor knowledge of the two-body distribution function at short distances.

\section{Extension to operators: FHNC/SOC}
\label{sec:fhnc_soc_op}
In this Section we will summarize the extension of the FR cluster expansion scheme, extensively used to deal with spin-isospin dependent correlation operators, introduced in Eq.(\ref{eq:Foperator})
\begin{equation}
\hat{\mathcal{F}}(X)=\Big(\mathcal{S} \prod_{j>i=1}^A \hat{F}_{ij} \Big)=\Big(\mathcal{S} \prod_{j>i=1}^A \sum_{p=1}^6 f^{p}(r_{ij})\hat{O}^{p}_{ij} \Big)\,.
\end{equation}

The operatorial structure of the correlations and their non commutativity make the development of a full FHNC summation scheme for diagrams containing spin-dependent correlation prohibitive. Here, we will briefly discuss the so called Single Operator Chain (FHNC/SOC) summation scheme, a detailed description of which can be found in \cite{pandha_79,arias_07}.

In addition to the function $h(r_{ij})$ of Eq. (\ref{eq:hscalar}), one has to also consider the products
\begin{equation}
2f^{c}(r_{ij})f^{p>1}(r_{ij}) \quad,\quad f^{p>1}(r_{ij})f^{q>1}(r_{ij})\, .
\label{eq:hoperatorial}
\end{equation} 
where the factor $2$ of the first quantity accounts for the term in which the central correlation is on the right of $\hat{O}^{p}_{12}$ while the operatorial one is on the left and for the reversed arrangement. 

For the calculation of the expectation value of the NN potential depending on spin-isospin operators, like the AV18 of Eq. (\ref{eq:av18}), it is worth introducing the two-body state dependent distribution functions, defined analogously to $g^c(\mathbf{r}_1,\mathbf{r}_2)$ of Eq. (\ref{eq:g2bc_def})
\begin{equation}
g^p(\mathbf{r}_1,\mathbf{r}_2)=\frac{A(A-1)}{\rho^2}\frac{\text{Tr}_{12}\int dx_{3,\dots,A} \Psi_0^*(X)\hat{\mathcal{F}}^\dagger \hat{O}^{p}_{12}\hat{\mathcal{F}}\Psi_0(X)}{\int dx_{1,\dots,A}  \ \Psi_0^*(X)\hat{\mathcal{F}}^\dagger \hat{\mathcal{F}} \Psi_0(X)}\, .
\label{eq:g2bop_def}
\end{equation}

The expectation value of the two-body potential can be conveniently rewritten in terms of the two-body state dependent distribution functions
\begin{equation}
\langle \hat{v} \rangle\equiv\sum_{i<j}(0|\hat{v}_{ij}|0)=\frac{A(A-1)}{2}(0|\hat{v}_{12}|0)=\frac{\rho^2}{2}\sum_p\int d\mathbf{r}_{1,2}\,g^p(\mathbf{r}_1,\mathbf{r}_2)v^p(r_{12})\, .
\label{eq:potop_exp}
\end{equation}

Because of translation invariance, the state dependent distribution functions, like the scalar one, depends on the magnitude of the relative distance, $g^p(\mathbf{r}_1,\mathbf{r}_2)\equiv g^p(r_{12})$. Thus, like for the scalar case, the expectation value of the two-body potential diverges with number of particles, while
\begin{equation}
\frac{\langle \hat{v} \rangle}{A}=\frac{\rho}{2}\sum_p\int d\mathbf{r}_{12}\,g^p(r_{12})v^p(r_{12})\, 
\label{eq:tbp_ev}
\end{equation}
is a finite quantity. 

Since the total spin and the total isospin of SNM are both vanishing, the following sum rules are satisfied by $g^p(r_{12})$
\begin{align}
&\rho\int d\mathbf{r}_{12}g^\sigma(r_{12})=-3\nonumber\\
&\rho\int d\mathbf{r}_{12}g^\tau(r_{12})=-3\nonumber\\
&\rho\int d\mathbf{r}_{12}g^{\sigma\tau}(r_{12})=9\, .
\end{align}

As far as the expansion of the numerator is concerned, Eq. (\ref{eq:FF_ce}) can be easily generalized 
\begin{align}
\hat{\mathcal{F}}^\dag \hat{O}^{p}_{12}\hat{\mathcal{F}}&=\hat{X}^{(2)}(x_1,x_2)+\sum_{i\neq1,2}\hat{X}^{(3)}(x_1,x_2;x_i)+\sum_{i<j\neq1,2}\hat{X}^{(4)}(x_1,x_2;x_i,x_j)+\dots\, .
\label{eq:FF_op}
\end{align}

Note that the cluster terms $\hat{X}^{n}$ are operators; for example
\begin{align}
\hat{X}^{(2)}(x_1,x_2)&=\hat{F}_{12}^\dagger \hat{O}^{p}_{12} \hat{F}_{12}\nonumber \\
\hat{X}^{(2)}(x_1,x_2;x_i)&=(\mathcal{S}\hat{F}_{12}^\dagger \hat{F}_{1i}^\dagger\hat{F}_{2i}^\dagger)  \hat{O}^{p}_{12} (\mathcal{S}\hat{F}_{12}\hat{F}_{1i}\hat{F}_{2i})-\hat{F}_{12}^\dagger \hat{O}^p \hat{F}_{12}\, .
\end{align}

The numerator of $g^p(r_{12})$ can be expanded analogously to Eq. (\ref{eq:num_last}) 
\begin{align}
\text{num}=\sum_{N=2}^{A}\frac{\rho^{N-2}}{(N-2)!}\text{Tr}_{12}\int dx_{3,\dots,N}  \hat{X}^{(N)}(x_1,x_2;x_3,\dots,x_N) \hat{g}_{N}^{MF}(x_1,\dots,x_N)\, ,
\label{eq:g_MF_op_exp}
\end{align}
where the operatorial $N-$body mean field distribution function, defined as 
\begin{equation}
\hat{g}_{N}^{MF}(x_1,\dots,x_N)=
\frac{A!}{(A-N)!}\frac{1}{\rho^N}\int dx_{N,\dots,A}\Psi_{0}(X)^*\Psi_{0}(X)\,.
\label{eq:def_corrMF_op}
\end{equation}
can be readily shown to be (see Eq. (\ref{eq:def_corrMF}) and (\ref{eq:gmf_tr})) 
\begin{align}
\hat{g}_{N}^{MF}(x_1,\dots,x_N)=
 \frac{1}{\rho^{\,N}}\sum_{n_1,\dots,n_N}\psi_{n_1}^*(x_1)\dots\psi_{n_N}^*(x_N) \mathcal{A}[\psi_{n_1}(x_1)\dots\psi_{n_N}(x_N)]\, .
\label{eq:gmf_tr_op}
\end{align}

The property of Eq. (\ref{eq:gram_gN}) holds for $\hat{g}_{N}^{MF}$, allowing for the cancellation between the unlinked diagrams of the numerator and the denominator to take place even in the case of operatorial correlations. 

Writing the antisymmetrization operator as in Eq. (\ref{eq:anti_exch}) and summing over the plane wave momenta, for the first terms of $\hat{g}_{N}^{MF}$ we get  
\begin{align}
\hat{g}_{N}^{MF}(x_1,\dots,x_N)&= \frac{1}{\nu^{\,N}}\sum_{\alpha_i}\eta^{*}_{\alpha_1}\dots\eta^{*}_{\alpha_N}\Big[1-\sum_{i<j}\hat{P}^{\sigma\tau}_{ij}\ell^2(r_{ij})\nonumber\\
&+\sum_{i<j<k}(\hat{P}^{\sigma\tau}_{ij}\hat{P}^{\sigma\tau}_{jk}+\hat{P}^{\sigma\tau}_{ij}\hat{P}^{\sigma\tau}_{ik})\ell(r_{ij})\ell(r_{jk})\ell(r_{ki})-\dots\Big]\eta_{\alpha_1}\dots\eta_{\alpha_N}\, .
\label{eq:exp_gmf_op}
\end{align}
This expression is very similar to the one of Eq. (\ref{eq:exp_gmf}), however the sum over the spin-isospin states of the latter equation cannot be directly performed, as $\text{Tr}_{1,\dots,N}$, is not embodied in the definition of $\hat{g}_{N}^{MF}$.

Substituting back the expression of $\hat{g}_{N}^{MF}$ in Eq. (\ref{eq:g_MF_op_exp}), one finds

\begin{align}
\text{num}&=\sum_{N=2}^{A}\frac{\rho^{N-2}}{(N-2)!}\int d\mathbf{r}_{3,\dots,N} \text{CTr}_{1,\dots,N}\Big\{ \hat{X}^{(N)}(x_1,x_2;x_3,\dots,x_N)\Big[1-\sum_{i<j}\hat{P}^{\sigma\tau}_{ij}\ell^2(r_{ij})\nonumber\\
&+\sum_{i<j<k}(\hat{P}^{\sigma\tau}_{ij}\hat{P}^{\sigma\tau}_{jk}+\hat{P}^{\sigma\tau}_{ij}\hat{P}^{\sigma\tau}_{ik})\ell(r_{ij})\ell(r_{jk})\ell(r_{ki})-\dots\Big]\Big\}\, .
\end{align}
The symbol ``CTr'' denotes the normalized trace of the spin-isospin operators, originating from the sum over the spin-isospin states of Eq. (\ref{eq:exp_gmf_op}) and the sum over the spin-isospin degrees of freedom, $\text{Tr}_{1,\dots,N}$, of Eq. (\ref{eq:g_MF_op_exp}). The factor $1/\nu^{\,N}$ accounts for the normalization of the trace, such that CTr$(1)=1$.

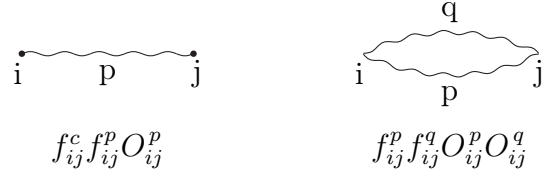
\begin{figure}[!t]
\begin{center}
\fcolorbox{white}{white}{
  \begin{picture}(200,80)(0,0)
	\SetWidth{0.5}
	\SetColor{Black}
        \SetScale{0.8}	
        \unitlength=0.8 pt

	\Photon(0,60)(80,60){1}{5}
	\Vertex(0,60){1.5}
	\Vertex(80,60){1.5}
	\Text(40,15)[]{$f^{c}_{ij}f^{p}_{ij}O^{\,p}_{ij}$}
	\Text(-2,51)[]{i}
	\Text(82,51)[]{j}
	\Text(40,50)[]{p}

	\PhotonArc(200,-20)(90,63,117){1}{6}
	\PhotonArc(200,140)(90,243,297){1}{6}

	\Text(158,51)[]{i}
	\Text(242,51)[]{j}
	\Text(200,80)[]{q}
	\Text(200,40)[]{p}
	\Text(200,15)[]{$f^{p}_{ij}f^{q}_{ij}O^{\,p}_{ij}O^{\,q}_{ij}$}
  \end{picture}
}
\vspace{0.5cm}
\caption{Operatorial correlation bonds.}
\label{fig:op_bonds}
\end{center}
\end{figure}

\subsubsection{Diagrammatic rules}
The diagrammatic rules given in Section \ref{sec:FR_exp} need to be extended to account for $2f^{c}(r_{ij})f^{p>1}(r_{ij})$ and $f^{p>1}(r_{ij})f^{q>1}(r_{ij})$. The former is represented by a single wavy line, the latter by a double wavy line; in both cases a letter indicating the kind of the operator involved in the correlation is placed close to the bond itself, see Fig. \ref{fig:op_bonds}.

A thick solid line, displayed in Fig. \ref{fig:int_op}, has been introduced to represent the interaction term, $\hat{F}_{12}\hat{O}^p_{12}\hat{F}_{12}$, of the operatorial two-body distribution function. 
\begin{figure}[!ht]
\begin{center}
\fcolorbox{white}{white}{
  \begin{picture}(100,40)(-20,-15)
	\SetWidth{0.5}
	\SetColor{Black}
	\SetScale{0.8}	
        \unitlength=0.8 pt
	\COval(35,10)(-3,40)(0){Black}{Black}
	\BCirc(-7,10){3}
	\BCirc(77,10){3}	
	\Text(35,-35)[]{$f^{l}_{12}v^{p}_{12}f^{q}_{12}\hat{O}^{\,l}_{12}\hat{O}^{\,p}_{12}\hat{O}^{\,q}_{12}$}
	\Text(-7,2)[]{1}
	\Text(77,2)[]{2}
	\Text(35,-7)[]{l, p, q}
  \end{picture}
}
\vspace{0.6cm}
\caption{Graphical representation of an operator interaction line.}
\label{fig:int_op}
\end{center}
\end{figure}

Note that the value of the diagrams in general depends on the ordering of the operators, operators; hence all permutations need to be considered.

The diagrammatic classification is analogous to the one used for the FR cluster expansion technique. As already said, disconnected diagrams of the numerator exactly simplify with the denominator and only connected diagrams have to be calculated.

Unlike the Jastrow case, reducible diagrams do not completely cancel out. At a later stage in this Thesis, we will describe the calculation of the two- and three- body cluster contributions to the energy per particle. In those simple examples, we will show how to deal with reducible diagrams.

\subsection{Traces}
As can be realized from Eq. (\ref{eq:g_MF_op_exp}), the calculation of $\langle \hat{v}^p \rangle/A$ requires the evaluation of the traces of spin-isospin dependent operators present in both the potential and the correlations. Since these operators are scalar in the Fock space formed by the product of configuration, spin and isospin spaces the Pauli identity can be written as
\begin{equation}
(\vec{a}\cdot \vec{\sigma_i})(\vec{b}\cdot \vec{\sigma_i})=\vec{a}\cdot \vec{b}+i \vec{\sigma_i}\cdot(\vec{a}\wedge \vec{b})\, ,
\end{equation}
where $\vec{a}$ and $\vec{b}$ are generic vector operators not containing $\vec{\sigma_i}$. The latter equation, which applies for $\vec{\sigma_i}\to\vec{\tau_i}$ also, can be used to express a generic operator product as
\begin{equation}
\prod \hat{O}_{ij}=C+\text{rest}\, 
\end{equation}
where $C$ does not contain any spin-isospin dependent operators while the rest contains terms in which each $\vec{\sigma}_k$ and $\vec{\tau}_k$ occurs at most once. Owing to the fact that Pauli matrices are traceless
\begin{equation}
\text{CTr}(\vec{\sigma}_k)=\text{CTr}(\vec{\tau}_k)=0\, ,
\end{equation}
the only contribution of $\prod \hat{O}_{ij}$ is $C$. In general $C$ depends on the ordering of the operators appearing in $\prod \hat{O}_{ij}$, hence all the possible orderings arising from $(\mathcal{S} \prod  \hat{F} )$ needs to be properly taken into account.

The authors of Ref. \cite{pandha_79} distinguished three different operatorial structures of the cluster term, to the analysis of which we devote the following Sections.

\subsubsection{Product of operators acting on the same pair}
As a first example, consider a cluster term in which the points $i$ and $j$ are joined by two operators, hence
\begin{equation}
\text{CTr}(\hat{O}^{p}_{ij}\hat{O}^{q}_{ij})=A^p\delta_{pq}\, ,
\label{eq:A_def}
\end{equation}
with $A^p=1,3,3,9,6,18$ for $p=1,6$. The CTr of diagrams in which more than two operators insist on the same pair $ij$ can be easily evaluated with the aid of the $K^{pqr}$ matrices, defined as
\begin{equation}
\hat{O}^{p}_{ij}\hat{O}^{q}_{ij}=\sum_rK^{pqr}\hat{O}^{r}_{ij}\, .
\label{eq:op_prod}
\end{equation}
The values of $K^{pqr}$ are given in Table 1 of Ref. \cite{pandha_79}.
Comparing the last two equations it is readily seen that $K^{pq1}=\delta_{pq}A_p$. Using Eq. (\ref{eq:op_prod}) it turns out that
\begin{equation}
\text{CTr}(\hat{O}^{p}_{ij}\hat{O}^{q}_{ij}\hat{O}^{r}_{ij})=\sum_sK^{pqs}\text{CTr}(\hat{O}^{s}_{ij}\hat{O}^{r}_{ij})=K^{pqr}A^r\, .
\label{eq:op_prod2}
\end{equation}
Note that, since operators acting on the same pair of points commute, the order of operator in the previous equation is immaterial, thus
\begin{equation}
K^{pqr}A^r=K^{qpr}A^r=K^{qrp}A^p\,\dots
\end{equation}

\subsubsection{Single operator rings (SOR)}
\begin{figure}[!ht]
\begin{center}
\fcolorbox{white}{white}{
\begin{picture}(150,80)(-55,0)
	\SetWidth{0.5}
	\SetColor{Black}
	\SetScale{0.8}	
        \unitlength=0.8 pt

        \CCirc(0,0){4}{Black}{Black}
	\CCirc(60,0){4}{Black}{Black}
	\CCirc(0,60){4}{Black}{Black}
	\CCirc(60,60){4}{Black}{Black}
        \Photon(0,0)(0,60){1}{5}
	\Photon(0,60)(60,60){1}{5}
	\Photon(60,60)(60,0){1}{5}
	\Photon(0,0)(60,0){1}{5}

	\Text(-12,0)[]{i}
	\Text(72,0)[]{j}
\end{picture}
}
\vspace{1cm}
\caption{Four-body SOR diagram.}
\label{fig:SOR}
\end{center}
\end{figure}
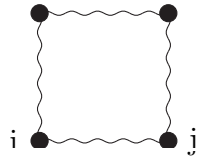

Single operator rings, like the one showed in Fig \ref{fig:SOR}, are characterized by having at most two operators on a given point. The normalized trace of a SOR does not depend on the ordering of the operators having the point $i$ in common. Because of the Pauli identity, the non commuting terms are indeed linear in either $\vec{\sigma_i}$ or $\vec{\tau_i}$, thus their trace vanishes. 

Let $\hat{O}^{p}_{ij}$ and $\hat{O}^{q}_{jk}$ the only two operators arriving at the point $j$. Making use of the Pauli identity it is possible to completely eliminate the operatorial dependence on point $j$. Integrating over the azimuthal angle $\phi_j$ and tracing over the spin-isospin degrees of freedom of particle $j$ yields
\begin{equation}
\text{CTr}_j\int d\phi_j \hat{O}^{p}_{ij} \hat{O}^{q}_{jk}=\sum_r \int d\phi_j \xi^{pqr}_{ijk}\hat{O}^r_{ij}\,.
\label{eq:contr}
\end{equation}
The coefficients $\xi^{pqr}_{ink}$ depends on the internal angles of the triangle $\mathbf{r}_{ij}$, $\mathbf{r}_{jk}$, $\mathbf{r}_{ik}$
\begin{align}
&\xi_{ijk}^{\sigma\sigma r}=\delta_{\sigma r}\nonumber \\
&\xi_{ijk}^{\sigma t r}=\delta_{t r}\frac{1}{2}(3\cos^2\theta_k-1)\nonumber \\
&\xi_{ijk}^{ t\sigma r}=\delta_{t r}\frac{1}{2}(3\cos^2\theta_i-1)\nonumber \\
&\xi_{ijk}^{ t t r}=\delta_{\sigma r}(3\cos^2\theta_j-1)+\delta_{t r}\frac{1}{2}[-9\cos\theta_i\cos\theta_j\cos\theta_k\nonumber \\
&\qquad-3(\cos^2\theta_i+\cos^2\theta_j+\cos^2\theta_k)+2]\nonumber\\
&\xi_{ijk}^{ \tau \tau r}=\delta_{\tau r}\nonumber \\
&\xi_{ijk}^{(p\tau)(q\tau)(r\tau)}=\xi^{pqr}_{ijk}\, , \qquad p,q,r=\sigma,t.
\label{eq:xi_coeff}
\end{align}

The evaluation of SOR diagrams is rather simple: once the operators with one point in common are placed next to each other, e. g., $\hat{O}^p_{ij} \hat{O}^q_{jk} \hat{O}^r_{kl}\dots$, successive contractions over the common points can be made by means of Eq. (\ref{eq:contr}). Every contraction gives a $\xi$ factor until at the end one is left with two operators acting on the same pair, resulting in a factor $A^p$.

\subsubsection{Multiple operator diagrams}
\begin{figure}[!ht]
\begin{center}
\fcolorbox{white}{white}{
\begin{picture}(150,75)(30,0)
	\SetWidth{0.5}
	\SetColor{Black}
	\SetScale{0.8}	
        \unitlength=0.8 pt

        \CCirc(0,0){4}{Black}{Black}
	\CCirc(70,0){4}{Black}{Black}
	\CCirc(35,70){4}{Black}{Black}
	\Photon(0,0)(35,70){1}{5}
	\Photon(35,70)(70,0){1}{5}
	\PhotonArc(35,-80)(87,70,110){1}{5}
	\PhotonArc(35,80)(87,250,290){1}{5}
	\Text(-12,-5)[]{i}
	\Text(80,-5)[]{j}
	\Text(45,75)[]{k}
	\Text(34,15)[]{$p$}
	\Text(34,-18)[]{$q$}
	\Text(0,35)[]{$r'$}
	\Text(70,35)[]{$r''$}
	\Text(34,-40)[]{(a)}

	\CCirc(200,0){4}{Black}{Black}
	\CCirc(270,0){4}{Black}{Black}
	\CCirc(235,70){4}{Black}{Black}
	\PhotonArc(235,-80)(87,70,110){1}{5}
	\PhotonArc(235,80)(87,250,290){1}{5}
	\PhotonArc(286,1)(85,127,178){1}{6}
	\PhotonArc(148,72)(85,307,358){1}{6}
	\Text(188,-5)[]{i}
	\Text(280,-5)[]{j}
	\Text(245,75)[]{k}
	\Text(234,15)[]{$r$}
	\Text(234,-18)[]{$r'$}
	\Text(200,37)[]{$s$}
	\Text(236,36)[]{$s'$}
	\Text(234,-40)[]{(b)}

\end{picture}
}
\vspace{1.5cm}
\caption{Multiple operator diagram}
\label{fig:mult_op}
\end{center}
\end{figure}
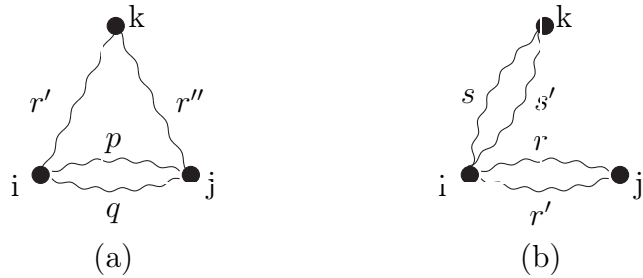

Consider the normalized trace of the diagram (a) of Fig. \ref{fig:mult_op}, where more than two operators arrive at both points $i$ and $j$. In principle, all possible orderings of the operators have to be considered in the evaluation of the normalized trace. However, invariance under cyclic permutations is a general property of the traces. As a consequence, there are only two different orderings of the operators: a ``successive'' order, in which $\hat{O}^p_{ij}$ and $\hat{O}^q_{ij}$ can be placed next to each other, and an ``alternate'' order, in wich either $\hat{O}^r_{ik}$ or $\hat{O}^s_{jk}$ is placed between them. For the successive order, using Eq. (\ref{eq:op_prod}), (\ref{eq:op_prod2}) and (\ref{eq:contr}) it turns out that

\begin{align}
\int d\phi_i \text{CTr}(\hat{O}^p_{ij}\hat{O}^q_{ij}\hat{O}^{r'}_{ik}\hat{O}^{r''}_{jk})&=\sum_r K^{pqr}\int d\phi_k\text{CTr}(\hat{O}^r_{ij}\hat{O}^{r'}_{ik}\hat{O}^{r''}_{jk})\nonumber \\
&=\sum_{r,r'''} K^{pqr}\int d\phi_k \xi_{ikj}^{r'r''r'''}\text{CTr}(\hat{O}^{r}_{ij}\hat{O}^{r'''}_{ij})\nonumber \\
&=\sum_{r} K^{pqr}A^{r}\int d\phi_k \xi_{ikj}^{r'r''r}\, .
\label{eq:mod_norm}
\end{align}

On the other hand, for the alternate order one has
\begin{align}
\int d\phi_i \text{CTr}(\hat{O}^p_{ij}\hat{O}^{r'}_{ik}\hat{O}^q_{ij}\hat{O}^{r''}_{jk})&=\sum_r L^{pqr}\int d\phi_k \xi_{ikj}^{r'r''r}\, .
\label{eq:mod_alt}
\end{align}
To determine the matrix $L^{pqr}$ one has to note that either
\begin{equation}
\text{CTr}(\hat{O}^p_{ij}[\hat{O}^{r'}_{ik},\hat{O}^q_{ij}]\hat{O}^{r''}_{jk})=0\, 
\end{equation}
or
\begin{equation}
\text{CTr}(\hat{O}^p_{ij}\{\hat{O}^{r'}_{ik},\hat{O}^q_{ij}\}\hat{O}^{r''}_{jk})=0\, .
\end{equation}
It can be easily seen that $L^{pqr}=K^{pqr}A^r$ and $L^{pqr}=-K^{pqr}A^r$ in the former and in the latter case, respectively.

Another possibility that needs to be discussed contemplates two SOR meeting at the point $i$, like in the diagram (b) of Fig. \ref{fig:mult_op}. Because of the invariance of the trace upon cyclic exchanges, again there are only two distinct cases. When the two operators acting on the pairs $ij$ and $ik$ are contiguous, it turns out that 
\begin{equation}
\text{CTr}(\hat{O}^{r}_{ij}\hat{O}^{r'}_{ij}\hat{O}^{s}_{ik}\hat{O}^{s'}_{ik} )=\delta_{rr'}A^r\delta_{ss'}A^s\, ,
\label{eq:sor2_norm}
\end{equation}
where we have used $K^{pq1}=\delta_{pq}A_p$. 

In order to deal with the alternate order, we introduce the matrix $D_{rs}$
\begin{equation}
\sum_{\vec{\sigma}_i\vec{\tau}_i}\hat{O}^{r}_{ij}\hat{O}^{s}_{ik}\hat{O}^{r'}_{ij}=\delta_{rr'}A^r(1+D_{rs})\hat{O}^{s}_{ik}\, ,
\end{equation}
where in the case of tensor operators, the above equation implies an integration over the azimuthal angle $\phi$. The entries of $D_{rs}$ depend on the kind of the operators $\hat{O}^{r}$ and $\hat{O}^{s}$
\begin{align}
&D_{\sigma\tau}=0\quad D_{(\sigma\tau,t\tau)(\sigma\tau,t\tau)}=-\frac{8}{9}\nonumber \\
&D_{\sigma\sigma}=D_{\tau\tau}=D_{(\sigma,\tau)(\sigma\tau,t\tau)}-\frac{4}{3}\, .
\end{align}
Thus, for the alternate order trace finds
\begin{align}
\text{CTr}(\hat{O}^{r}_{ij}\hat{O}^{s}_{ik}\hat{O}^{r'}_{ij}\hat{O}^{s'}_{ik})=\delta_{rr'}A^r(1+D_{rs})\delta_{ss'}A^s\, .
\label{eq:sor2_alt}
\end{align}

\subsection{FHNC/SOC approximation}
The technique for summing linked cluster diagrams containing operatorial correlations is made technically difficult because of the their non commutativity, which makes a full FHNC summation prohibitive. 
Diagrams having one or more passive operatorial bonds are calculated at leading order only. Such an approximation is justified by the observation that operatorial correlations are much weaker than the scalar ones. Based on this feature, one would be tempted to conclude that the leading order amounts to dressing the interaction line with all possible FHNC two-body distribution functions. This is not true as, besides the short range behavior, the intermediate range behavior of NN correlations also plays an important role that needs to be taken into account. In particular, tensor correlations, and to some extent also exchange correlations, have a much longer range than the central ones.

In order to handle this problem, summing the class of chain diagrams turns out to be to be of great importance, as remarked in Sec. \ref{subsec:it_FHNC} for the pure central case (see Eq. (\ref{eq:Nk}) and the subsequent discussion). 

The above issue is taken care of by summing up the Single Operator Chains (SOC) in the corresponding FHNC/SOC approximation \cite{pandha_79,wiringa_79}. SOC are chain diagrams in which any single passive bond of the chain has a single operator of the type  $f^c(r_{ij})f^p(r_{ij})\hat{O}^p_{ij}$ or $-h(r_{ij})\ell(k_Fr_{ij})\times P_{ij}$, with $p\leq 6$, or FHNC-dressed versions of them.  Note that if a single bond of the chain is of the scalar type then the spin trace of the corresponding cluster term vanishes, as the Pauli matrices are traceless. Then the SOC is the leading order, and at the same time it includes the main features of the long range behavior of tensor and exchange correlations.

The calculation of SOC, as that of FHNC chains, is based upon the convolution integral of the functions corresponding to two consecutive bonds.  Unlike FHNC chains, however, the SOC have operatorial bonds. Therefore, the basic algorithm is the convolution of two operatorial correlations having one common point of Eq. (\ref{eq:contr}).

The ordering of the operators within an SOC is immaterial, because the commutator $[\hat{O}_{ik},\hat{O}_{kj}]$  is linear in $\vec{\sigma}_k$ and $\vec{\tau}_k$, and Pauli matrices are traceless. The only orderings that matter are those of passive bonds connected to the interacting points $1$ or $2$, discussed in Eqs. (\ref{eq:mod_norm}), (\ref{eq:mod_alt}), (\ref{eq:sor2_norm}) and (\ref{eq:sor2_alt}). 

A second important contribution which is included in FHNC/SOC approximation is the leading order of the vertex corrections. They sum up the contributions of sets of subdiagrams which are joined to the basic diagrammatic structure in a single point, like diagram (b) of Fig. \ref{fig:mult_op}. Therefore, a vertex correction dresses the vertex of all the possible reducible subdiagrams joined to it. In the FHNC/SOC approximation they are taken into account only at the leading order, i.e. including SOR. Vertex corrections play an important role for the fulfillment of the sum rules. 

The full FHNC/SOC equations including the SOR vertex corrections can be found in the reference paper \cite{pandha_79,wiringa_79}. For pedagogical purposes, we limit ourselves to the equations for SOC diagrams, as this eliminates the problem of the reducible diagrams, as all the SOC diagrams are irreducible \cite{arias_07}. 

We will make the further approximation of neglecting elementary diagrams.

Regarding the notation, the symbols for nodal and composite diagrams carry an additional index, specifying the  operatorial dependence: $N_{xy}^p$, $X_{xy}^p$.

The generalization of Eq. (\ref{eq:fhnc_nodal}) accounting for operatorial nodal diagrams reads
\begin{align}
N_{xy}^r(r_{12})=\sum_{p,q=1}^6 \sum_{x^\prime y^\prime}\,\text{CTr}_3\, \rho\int_{V} d\mathbf{r}_3 X_{xx^\prime}^p(r_{13})  \xi^{pqr}_{132}  \zeta_{x^\prime y^\prime} [X_{y^\prime y}^q(r_{23})+N_{y^\prime y}^q(r_{23})]\, .
\label{eq:fhnc_nodal_op}
\end{align}
Since irreducible diagrams only are present, the factor $\zeta_{dd}$ only selects contributions respecting the Pauli principle 
\begin{equation}
\zeta_{dd}= \zeta_{de}=\zeta_{ed}=1\quad,\quad \zeta_{ee}=0\, .
\end{equation}

The partial two-body distribution functions, $g_{xy}^p=N_{xy}^p+X_{xy}^p$, are given by (compare to Eq. (\ref{eq:distribution}))
\begin{align}
g_{dd}^p(r_{12}) &=
h^p(r_{12})h^c(r_{12}) \nonumber \\
g_{de}^p(r_{12}) &= g_{ed}^p(r_{12})=
h^c(r_{12})[h^p(r_{12})N_{de}^c(r_{12})+ f^{c}(r_{12})^2N_{de}^p(r_{12})]\nonumber \\
g_{ee}^{p}(r_{12}) &= h^c(r_{12})\{h^p(r_{12})[N_{ee}^c(r_{12}) + N_{de}^c(r_{12})^2]-
\nu f^{c}(r_{12})^2\mathcal{L}(r_{12})^2\Delta^p\nonumber \\
&+N_{ee}^p(r_{12})+2N_{de}^c(r_{12})N_{de}^p(r_{12})\}\ , \nonumber \\
\label{eq:distribution_op}
\end{align}
where
\begin{align}
h^p(r_{12})&=2f^p(r_{12})f^c(r_{12})+f^{c}(r_{12})^2N_{dd}^p(r_{12})\nonumber \\
h^c(r_{12})&=\exp[N_{dd}(r_{12})]\nonumber \\
\mathcal{L}(r_{12})&=N_{cc}(r_{12})-\ell(r_{12})/\nu
\end{align}
The composite functions can be effortlessly obtained by subtracting the contribution of the nodal diagrams from the partial two-body distribution functions
\begin{equation}
X_{xy}^p(r_{12})=g_{xy}^p(r_{12})-N_{xy}^p(r_{12})\, .
\end{equation}

The total operator distribution function is given by
\begin{equation}
g^p(r_{12})=g_{dd}^p(r_{12})+2g_{de}^p(r_{12})+g_{ee}^{p}(r_{12}) \, .
\end{equation}

\begin{figure}[!ht]
\begin{center}
\fcolorbox{white}{white}{
  \begin{picture}(300,80)(-35,0)
	\SetWidth{0.5}
	\SetScale{0.8}	
        \unitlength=0.8 pt
	
	\CArc(51,30)(60,148,210)
        \CArc(19,30)(60,-30,32)
        \CArc(35,12)(60,56,124)
	\Photon(0,0)(0,60){2}{5}
	\DashLine(0,60)(70,60){5}
	\DashLine(70,60)(70,0){5}
	\CCirc(0,0){3}{Black}{Black}
	\CCirc(0,60){3}{Black}{Black}
	\CCirc(70,0){3}{Black}{Black}
        \CCirc(70,60){3}{Black}{Black}
	\Text(-2,-12)[]{i}
	\Text(74,-12)[]{j}
	\Text(0,72)[]{k}
	\Text(70,72)[]{l}
	\Text(34,-25)[]{(a)}

	\CArc(251,30)(60,148,210)
        \CArc(219,30)(60,-30,32)
        \CArc(235,12)(60,56,124)
	\DashLine(200,60)(270,60){5}
	\Photon(270,0)(270,60){2}{5}
	\DashLine(200,60)(200,0){5}
	\CCirc(200,0){3}{Black}{Black}
	\CCirc(200,60){3}{Black}{Black}
	\CCirc(270,0){3}{Black}{Black}
        \CCirc(270,60){3}{Black}{Black}
	\Text(198,-12)[]{i}
	\Text(274,-12)[]{j}
	\Text(200,72)[]{k}
	\Text(270,72)[]{l}
	\Text(235,-25)[]{(b)}

\end{picture}
}
\vspace{1.5cm}
\caption{Nodal diagrams with a cyclic exchange loop. \label{fig:Ncc_op}}
\end{center}
\end{figure}
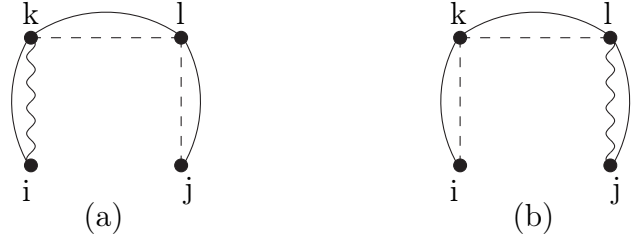

In a generic exchange loop all links but one carry an operator dependence. The only gap is filled by a dynamical operator to complete the operator chain. Within the SOC approximation there are two distinct possibilities: the dynamical operator may be inserted to the left ($L$) or to right ($R$) of the chain, as in the nodal diagrams (a) and (b) of Fig. \ref{fig:Ncc_op}, respectively.  To deal with cyclic exchange, we need a new bond
\begin{align}
X^{p}_{c\,L,R}(r_{12})&=h^c(r_{12})[h^p(r_{12})\mathcal{L}(r_{12})+f^c(r_{12})^2 N^{p}_{c\,L,R}(r_{12})]-N^{p}_{c\,L,R}(r_{12})\, ,
\end{align}
 while the corresponding nodal functions read
\begin{align}
N_{cL}^r(r_{12})&=\sum_{p,q=1}^6 \,\text{CTr}_3\, \rho\int d\mathbf{r}_3 X^{p}_{c\,L}(r_{13})  \xi^{pqr}_{132} \Delta^q [X_{cc}(r_{23})+\mathcal{L}(r_{23})]\nonumber \\
N_{cR}^r(r_{12})&=\sum_{p,q=1}^6 \,\text{CTr}_3\, \rho\int \mathbf{r}_3 X_{cc}(r_{13})\Delta^p  \xi^{pqr}_{132}  [X^{q}_{c\,R}(r_{13})+N^{q}_{c\,R}(r_{13})]\nonumber \\
N_{cc}^r(r_{12})&=N_{cL}^r(r_{12})+N_{cR}^r(r_{12})\, .
\end{align}
Finally, the partial two-body distribution functions with circular exchanges is given by
\begin{equation}
g_{c}^p(r_{12})=g_{cc}^c(r_{12})\Delta^p\, .
\label{eq:distribution_op_cc}
\end{equation}

The $cc$ nodal functions enter as closed SOR in the generalized equations for $X^{c}_{ee}$ and $N^{c}_{cc}$. Moreover, they contribute to the energy expectation value, the full calculation of which will not be reported in this Thesis. The interested reader is again referred to the Refs. \cite{pandha_79,wiringa_79}. Nevertheless, in the next Section, we do present the calculation of the two- and three- body cluster contributions to both the potential and the kinetic energy per particle.

\subsection{Two- and three- body cluster contribution}
\label{subsec:twothree}
In this Section we analyze in detail the two- and the three- body cluster contributions to the energy per particle for the case of a ''static`` potential, without momentum dependent terms. This analysis, although being useful for pedagogical reasons, in particular for the treatment of the reducible diagrams within the FR summation scheme, will be the cornerstone of the effective interaction, that will be developed for the calculation of the response.

With the notation used in Section \ref{sec:fhnc_soc_op}, the two-body contribution of $\mathcal{F}^\dagger v_{12} \mathcal{F}$ can be expressed as
\begin{equation}
\mathcal{F}^\dagger v_{12} \mathcal{F}\,\Big{|}_{2b}\equiv\hat{X}^{2}(x_1,x_2)=\hat{F}_{12}\hat{v}_{12}\hat{F}_{12}\, .
\end{equation}
In what follows, the cluster expansion of the denominator will be disregarded as it is understood that, for fermionic systems, the denominator exactly cancels the disconnected diagrams of the numerator.

The two-body contribution of the potential is then 
\begin{align}
\langle \hat{v} \rangle\Big{|}_{2b}&=\frac{1}{2}\sum_{n_1,n_2}\int dx_1dx_2 
\phi^{*}_{n_1}(x_1)\phi^{*}_{n_2}(x_2)\hat{F}_{12}\hat{v}_{12}\hat{F}_{12} (1-\hat{P}_{12})\phi_{n_1}(x_1)\phi_{n_2}(x_2)\nonumber \\
&=\frac{\rho^2}{2}\int d\mathbf{r}_1d\mathbf{r}_2 \text{CTr}_{12}[\hat{F}_{12}\hat{v}_{12}\hat{F}_{12}(1-\hat{P}^{\,\sigma\tau}_{12}\ell_{12}^2)]\, .
\label{eq:v2bcont}
\end{align}
Because of the translation invariance of the system, it is possible to integrate out the coordinate of the center of mass $\mathbf{R}_{12}=\frac{1}{2}(\mathbf{r}_1+\mathbf{r}_2)$, so that the two-body cluster contribution to the potential energy per particle reads
\begin{equation}
\frac{\langle \hat{v} \rangle}{A}\Big{|}_{2b}=\frac{\rho}{2}\int d\mathbf{r}_{12}\,\text{CTr}_{12}[\hat{F}_{12}\hat{v}_{12}\hat{F}_{12}(1-\hat{P}^{\,\sigma\tau}_{12}\ell^2_{12})]\, .
\end{equation}

The two-body term of the cluster expansion of the kinetic energy, $\hat{T}$,  is
\begin{equation}
-\frac{\hbar^2}{2m}\nabla_{1}^2\mathcal{F}^\dagger\mathcal{F}\,\Big|_{2b}\equiv\sum_{1<i}\hat{X}^{2}(x_1;x_i)=\sum_{1<i}^A\Big(-\frac{\hbar^2}{2m}\hat{F}_{1i}[\nabla_{1}^2,\hat{F}_{1i}]\Big)\, ,
\end{equation}
where the commutator removes the Fermi gas energy, which is a one-body contribution. Using the symmetry of the wave functions one gets
\begin{align}
\langle\hat{T} \rangle \Big{|}_{2b}&=-\frac{\hbar^2}{2m}\sum_{n_1,n_2}\int dx_1dx_2 
\phi^{*}_{n_1}(x_1)\phi^{*}_{n_2}(x_2)\hat{F}_{12}[\nabla_{1}^2,\hat{F}_{12}] (1-\hat{P}_{12})\phi_{n_1}(x_1)\phi_{n_2}(x_2)\, .
\end{align}
In order to remove the term with the product of the gradients acting on both the correlation function $\hat{F}_{12}$ and on the plane wave, it is convenient to integrate by part the latter expression (see Section \ref{sec:kin_presc}), with the result
\begin{align}
\langle \hat{T} \rangle\Big{|}_{2b}&=\frac{\hbar^2}{2m}\sum_{n_1,n_2}\int dx_1dx_2 
\phi^{*}_{n_1}(x_1)\phi^{*}_{n_2}(x_2)(\vec{\nabla}_1\hat{F}_{12})(\vec{\nabla}_{1}\hat{F}_{12})(1-\hat{P}_{12})\phi_{n_1}(x_1)\phi_{n_2}(x_2)\, .
\label{eq:kin_2b_part}
\end{align}
Since $\vec{\nabla}_{1}\hat{F}_{12}=\vec{\nabla}_{12}\hat{F}_{12}$, we can integrate out the coordinate of the center of mass, getting the following expression for the kinetic energy per particle
\begin{equation}
\frac{\langle \hat{T} \rangle}{A}\Big{|}_{2b}=\frac{\hbar^2}{2m}\rho\int d\mathbf{r}_{12}\,\text{CTr}_{12}[(\vec{\nabla}_{12}\hat{F}_{12})(\vec{\nabla}_{12}\hat{F}_{12})(1-\hat{P}^{\,\sigma\tau}_{12}\ell^2_{12})]\, .
\end{equation}

For calculating $(\vec{\nabla}_1\hat{F}_{12})(\vec{\nabla}_{1}\hat{F}_{12})$ one has to account for the fact that the tensor operator depends on $\hat{r}_{12}$, hence
\begin{align}
(\vec{\nabla}_1\hat{F}_{12})(\vec{\nabla}_{1}\hat{F}_{12})&=\sum_{p,q}[(\vec{\nabla}_1f_{12}^p)\hat{O}^{p}_{12}+f_{12}^p(\vec{\nabla}_1\hat{O}^{p}_{12})]
[(\vec{\nabla}_1f_{12}^q)\hat{O}^{q}_{12}+f_{12}^q(\vec{\nabla}_1\hat{O}^{q}_{12})]\nonumber\\
&=\sum_{p,q}[(\vec{\nabla}_1f_{12}^p)\hat{O}^{p}_{12}(\vec{\nabla}_1f_{12}^q)\hat{O}^{q}_{12}+f_{12}^p(\vec{\nabla}_1\hat{O}^{p}_{12})f_{12}^q(\vec{\nabla}_1\hat{O}^{q}_{12})]\, ,
\label{eq:lap_adj}
\end{align}
where the following property of the gradient of the tensor operator has been used
\begin{align}
&\vec{\nabla}_1f(r_{12})\vec{\nabla}_1S_{12}=0\, .
\end{align}
The first term of Eq. (\ref{eq:lap_adj}) can be conveniently written in terms of the derivative with respect the magnitude of $\mathbf{r}_{12}$
\begin{equation}
\sum_{p,q}(\vec{\nabla}_1f_{12}^p)\hat{O}^{p}_{12}(\vec{\nabla}_1f_{12}^q)\hat{O}^{q}_{12}=\sum_{p,q}f_{12}^{\prime\,p}\,\hat{O}^{p}_{12}\,f_{12}^{\prime\,q}\,\hat{O}_{12}^q\, .
\end{equation}
Thanks to the relation
\begin{equation}
(\vec{\nabla}_1 S_{12})(\vec{\nabla}_1 S_{12})=\frac{6}{r_{12}^2}(6+2\sigma_{12}+S_{12})\, ,
\end{equation}
the second term of Eq. (\ref{eq:lap_adj}) turns out to be
\begin{equation}
\sum_{p,q}[f_{12}^p(\vec{\nabla}_1\hat{O}^{p}_{12})f_{12}^q(\vec{\nabla}_1\hat{O}^{q}_{12})]=\frac{6}{r_{12}^2}
(f^{t}_{12}+f^{t\tau}_{12}\tau_{12})^2(6+2\sigma_{12}+S_{12})
\end{equation}

The three-body cluster contribution appearing in the expansion of $\mathcal{F}^\dagger v_{12} \mathcal{F}$ is given by
\begin{align}
\mathcal{F}^\dagger v_{12} \mathcal{F}\,\Big{|}_{3b}&=\sum_{i>2}\hat{X}^{3}(x_1,x_2;x_i)\nonumber\\
&=\sum_{i>2}\Big[\Big(\mathcal{S}\hat{F}_{12}\hat{F}_{1i}\hat{F}_{2i}\Big)\hat{v}_{12}\Big(\mathcal{S}\hat{F}_{12}\hat{F}_{1i}\hat{F}_{2i}\Big)-\hat{F}_{12}\hat{v}_{12}\hat{F}_{12}\Big]\, .
\end{align}
Within the FR diagrammatic scheme, the three-body cluster contribution to $\langle \hat{v}_{12} \rangle$ it is not merely the expectation value of the latter result, unlike the two-body case. As a matter of fact, the reducible diagrams arising form four-body cluster term of  $\mathcal{F}^\dagger v_{12} \mathcal{F}$, the detailed calculations of which can be found in appendix \ref{app:red4b}, needs to be taken into account. 

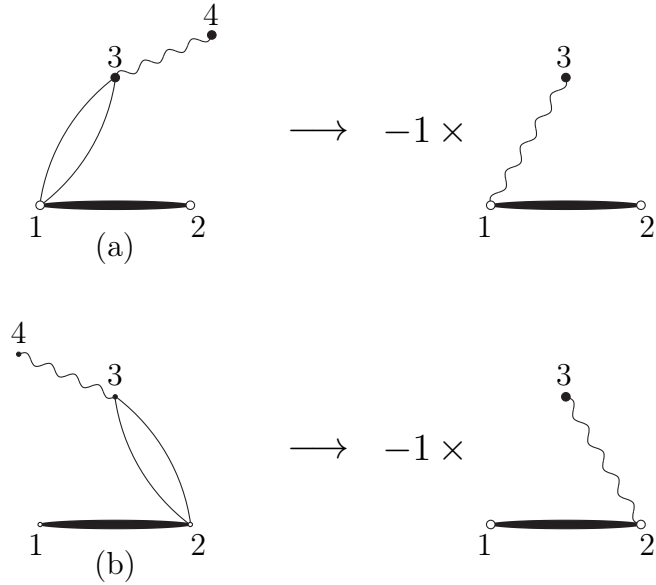
\begin{figure}[!hb]
\begin{center}
\fcolorbox{white}{white}{
  \begin{picture}(300,230)(-40,-140)
	\SetWidth{0.5}
	\SetScale{0.8}	
        \unitlength=0.8 pt

	\CArc(90,-12.2917)(91,127,172)
        \CArc(-55,72.2917)(91,307,353)
	\COval(35,0)(2,32)(0){Black}{Black}
	\Photon(35,60)(80,80){2}{4}
        \BCirc(0,0){2}
	\CCirc(35,60){2}{Black}{Black}
	\BCirc(70,0){2}
        \CCirc(80,80){2}{Black}{Black}
	\Text(-2,-10)[]{1}
	\Text(74,-10)[]{2}
	\Text(35,70)[]{3}
	\Text(80,90)[]{4}
	\Text(35,-20)[]{(a)}

        \Text(130,34)[]{\large{$\longrightarrow$}}
        
        \Text(180,35)[]{\large{$-1\,\times$}}
	\Photon(210,0)(245,60){2}{5}
	\COval(245,0)(2,32)(0){Black}{Black}
        \BCirc(210,0){2}
	\CCirc(245,60){2}{Black}{Black}
	\BCirc(280,0){2}
	\Text(208,-10)[]{1}
	\Text(284,-10)[]{2}
	\Text(245,70)[]{3}
	
%seconda riga di diagrammi	
	\CArc(-20,-162.2917)(91,8,52)
        \CArc(125,-77.7083)(91,188,233)
	\COval(35,-150)(2,32)(0){Black}{Black}
	\Photon(35,-90)(-10,-70){2}{4}
        \BCirc(0,-150){1}
	\CCirc(35,-90){1}{Black}{Black}
	\BCirc(70,-150){1}
        \CCirc(-10,-70){1}{Black}{Black}

	\Text(-2,-160)[]{1}
	\Text(74,-160)[]{2}
        \Text(35,-80)[]{3}
	\Text(-10,-60)[]{4}
	\Text(35,-170)[]{(b)}

        \Text(130,-116)[]{\large{$\longrightarrow$}}
        \Text(180,-115)[]{\large{$-1\,\times$}}

	\Photon(280,-150)(245,-90){2}{5}
	\COval(245,-150)(2,32)(0){Black}{Black}
        \BCirc(210,-150){2}
	\CCirc(245,-90){2}{Black}{Black}
	\BCirc(280,-150){2}
	\Text(208,-160)[]{1}
	\Text(284,-160)[]{2}
	\Text(245,-80)[]{3}
	
\end{picture}
}
\vspace{1cm}
\caption{Four-body reducible diagrams, $v_{4b\to3b}^{\text{dir}}$ and their three-body reduction.\label{fig:red4bdir}}
\end{center}
\end{figure}

The direct term of the three-body cluster contribution in the FR expansion scheme is given by \cite{morales_02}
\begin{align}
\langle \hat{v}_{12} \rangle\Big{|}_{3b}^{\text{dir}}&=\frac{\rho^2}{2}\int d\mathbf{r}_{12}d\mathbf{r}_{13} \text{CTr}_{123}\Big[(\mathcal{S}\hat{F}_{12}\hat{F}_{13}\hat{F}_{23})\hat{v}_{12}(\mathcal{S}\hat{F}_{12}\hat{F}_{13}\hat{F}_{23})\nonumber \\
&\qquad\qquad-\hat{F}_{12}\hat{v}_{12}\hat{F}_{12}(\hat{F}_{13}^2+\hat{F}_{23}^2-1)\Big]\,.
\label{eq:v3b_dir}
\end{align}
It includes the term
\begin{equation}
(\mathcal{S}\hat{F}_{12}\hat{F}_{13}\hat{F}_{23})\hat{v}_{12}(\mathcal{S}\hat{F}_{12}\hat{F}_{13}\hat{F}_{23})-\hat{F}_{12}\hat{v}_{12}\hat{F}_{12}\, 
\end{equation}
from the three-body cluster contribution of $\mathcal{F}^\dagger v_{12} \mathcal{F}$, whereas the reducible four-body diagrams of Fig. \ref{fig:red4bdir} contribute with the factor 
\begin{equation}
-\hat{F}_{12}\hat{v}_{12}\hat{F}_{12}(\hat{F}_{13}^2+\hat{F}_{23}^2-2)\, .
\end{equation}

It is worth remarking that in the pure central Jastrow case, $\hat{F}_{ij}=f^{c}_{ij}$, the four- and three- body reducible diagrams completely cancel, as discussed in Section \ref{sec:RFHNC}, and we obtain the well-known irreducible contribution
\begin{align}
\langle \hat{v}_{12} \rangle\Big{|}_{3b}^{\text{dir-Jastrow}}&=\frac{\rho^2}{2}\int d\mathbf{r}_{12}d\mathbf{r}_{13} {f^c}^2(r_{12})v(r_{12})({f^c}^2(r_{13})-1)({f^c}^2(r_{23})-1)\,.
\label{eq:v3b_dir_jas}
\end{align}

The sum of the diagrams where particles $1$ and $2$ are exchanged gives
\begin{align}
\langle \hat{v}_{12} \rangle\Big{|}_{3b}^{P_{12}}&=-\frac{\rho^2}{2}\int d\mathbf{r}_{12}d\mathbf{r}_{13}\ell^2(r_{12}) \text{CTr}_{123}\Big[(\mathcal{S}\hat{F}_{12}\hat{F}_{13}\hat{F}_{23})\hat{v}_{12}(\mathcal{S}\hat{F}_{12}\hat{F}_{13}\hat{F}_{23})\hat{P}^{\sigma\tau}_{12}\nonumber \\
&\qquad\qquad-\hat{F}_{12}\hat{v}_{12}\hat{F}_{12}(\hat{F}_{13}^2+\hat{F}_{23}^2-1)\hat{P}^{\sigma\tau}_{12}\Big]\,.
\label{eq:v3b_p12}
\end{align}
The corresponding four-body diagram producing the term $\frac{1}{2}\hat{F}_{12}\hat{v}_{12}\hat{F}_{12}(\hat{F}_{13}^2+\hat{F}_{23}^2-2)\hat{P}^{\sigma\tau}_{12}$ is drawn in  Fig. \ref{fig:red4bp12}

\begin{figure}[!ht]
\begin{center}
\fcolorbox{white}{white}{
  \begin{picture}(300,80)(-35,0)
	\SetWidth{0.5}
	\SetScale{0.8}	
        \unitlength=0.8 pt
        
	\CArc(90,-12.2917)(91,127,172)
	\CArc(-20,-12.2917)(91,7,52)
        \CArc(35,71)(80,245,295)
	\COval(35,0)(2,32)(0){Black}{Black}
	\Photon(35,60)(80,80){2}{4}
        \BCirc(0,0){2}
	\CCirc(35,60){2}{Black}{Black}
	\BCirc(70,0){2}
        \CCirc(80,80){2}{Black}{Black}
	\Text(-2,-10)[]{1}
	\Text(74,-10)[]{2}
	\Text(35,70)[]{3}
	\Text(80,90)[]{4}

        \Text(130,34)[]{\large{$\longrightarrow$}}
        
        \Text(180,35)[]{\large{$-1\,\times$}}
	\Photon(210,0)(245,60){2}{5}
	\CArc(245,71)(80,245,295)
        \CArc(245,-71)(80,65,115)

	\COval(245,0)(2,32)(0){Black}{Black}
        \BCirc(210,0){2}
	\CCirc(245,60){2}{Black}{Black}
	\BCirc(280,0){2}
	\Text(208,-10)[]{1}
	\Text(284,-10)[]{2}
	\Text(245,70)[]{3}
\end{picture}
}
\vspace{1.5cm}
\caption{Four-body reducible diagram, $v_{4b\to3b}^{\text{P12}}$, and its three-body reduction.\label{fig:red4bp12}}
\end{center}
\end{figure}

The diagrams in which particles $1$ and $3$ are exchanged contributes with
\begin{align}
\langle \hat{v}_{12} \rangle\Big{|}_{3b}^{P_{13}}&=-\frac{\rho^2}{2}\int d\mathbf{r}_{12}d\mathbf{r}_{13}\ell^2(r_{13}) \text{CTr}_{123}\Big[(\mathcal{S}\hat{F}_{12}\hat{F}_{13}\hat{F}_{23})\hat{v}_{12}(\mathcal{S}\hat{F}_{12}\hat{F}_{13}\hat{F}_{23})\hat{P}^{\sigma\tau}_{13}\nonumber \\
&\qquad\qquad-\hat{F}_{12}\hat{v}_{12}\hat{F}_{12}\hat{F}_{13}^2\hat{P}^{\sigma\tau}_{13}\Big]\,.
\label{eq:v3b_p13}
\end{align}
where the term $\frac{1}{2}\hat{F}_{12}\hat{v}_{12}\hat{F}_{12}(\hat{F}_{13}^2-1)\hat{P}^{\sigma\tau}_{13}$ comes from the four-body reducible diagram of Fig. \ref{fig:red4bp13}.

\begin{figure}[!hb]
\begin{center}
\fcolorbox{white}{white}{
  \begin{picture}(300,80)(-35,0)
	\SetWidth{0.5}
	\SetScale{0.8}	
        \unitlength=0.8 pt
	
	 \CArc(-55,72.2917)(91,307,353)
        \CArc(76,50.125)(91,160,214)
        \CArc(0,41.875)(40,30,105)
	\COval(35,0)(2,32)(0){Black}{Black}
	\Photon(35,60)(-10,80){2}{4}
        \BCirc(0,0){2}
	\CCirc(35,60){2}{Black}{Black}
	\BCirc(70,0){2}
        \CCirc(-10,80){2}{Black}{Black}
	\Text(-2,-10)[]{1}
	\Text(74,-10)[]{2}
	\Text(36,70)[]{3}
	\Text(-10,90)[]{4}

        \Text(130,34)[]{\large{$\longrightarrow$}}
        \Text(180,35)[]{\large{$-1\,\times$}}

	\Photon(210,0)(245,60){2}{5}
	\CArc(300,-12.2917)(91,127,172)
        \CArc(155,72.2917)(91,307,353)
	\COval(245,0)(2,32)(0){Black}{Black}
        \BCirc(210,0){2}
	\CCirc(245,60){2}{Black}{Black}
	\BCirc(280,0){2}
	\Text(208,-10)[]{1}
	\Text(284,-10)[]{2}
	\Text(245,70)[]{3}
\end{picture}
}
\vspace{1.5cm}
\caption{Four-body reducible diagram, $v_{4b\to3b}^{\text{P13}}$, and its three-body reduction.\label{fig:red4bp13}}
\end{center}
\end{figure}

Since the potential is invariant under $x_1\leftrightarrow x_2$. the diagrams with the exchange between particles $2$ and $3$ give the same contribution reported in Eq. (\ref{eq:v3b_p13}). The associated four-body reducible diagram is very similar to the one of Fig. \ref{fig:red4bp13} but with the loop attached to particle $2$ instead of particle $1$.

Consider the diagrams with the circular exchange involving particles $1$, $2$ and $3$. In this case there are no reducible four-body diagrams that partly cancel the reducible part of the three body diagram. In addition, there are no three-body reducible diagrams with circular exchange at all. However, the four-body diagram of Fig. \ref{fig:red4bp13}, with no correlation lines linking particles $1$ and $2$ to the others, can be {\it reduced} to a three-body term, so that the three-body diagram with a circular exchange reads
\begin{align}
\langle \hat{v}_{12} \rangle\Big{|}_{3b}^{cir}&=\rho^2\int d\mathbf{r}_{12}d\mathbf{r}_{13}\ell(r_{12})\ell(r_{13})\ell(r_{23}) \text{CTr}_{123}\Big[(\mathcal{S}\hat{F}_{12}\hat{F}_{13}\hat{F}_{23})\hat{v}_{12}(\mathcal{S}\hat{F}_{12}\hat{F}_{13}\hat{F}_{23})\hat{P}^{\sigma\tau}_{12}\hat{P}^{\sigma\tau}_{13}\nonumber \\
&\qquad\qquad-\hat{F}_{12}\hat{v}_{12}\hat{F}_{12}\hat{F}_{13}^2\hat{P}^{\sigma\tau}_{13}\hat{P}^{\sigma\tau}_{12}\Big]\,.
\label{eq:v3b_cir}
\end{align}

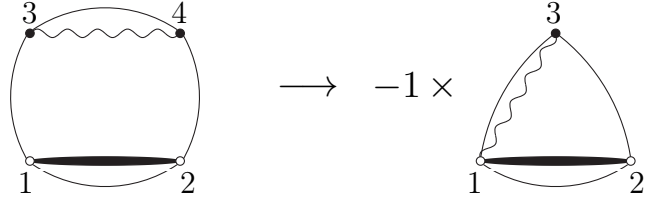
\begin{figure}[!hb]
\begin{center}
\fcolorbox{white}{white}{
  \begin{picture}(300,80)(-35,0)
	\SetWidth{0.5}
	\SetScale{0.8}	
        \unitlength=0.8 pt
	
	\CArc(51,30)(60,148,210)
        \CArc(19,30)(60,-30,32)
        \CArc(35,48)(60,236,304)
        \CArc(35,12)(60,56,124)
	\COval(35,0)(2,32)(0){Black}{Black}
	\Photon(0,60)(70,60){2}{5}
        \BCirc(0,0){2}
	\CCirc(0,60){2}{Black}{Black}
	\BCirc(70,0){2}
        \CCirc(70,60){2}{Black}{Black}
	\Text(-2,-10)[]{1}
	\Text(74,-10)[]{2}
	\Text(0,70)[]{3}
	\Text(70,70)[]{4}

        \Text(130,34)[]{\large{$\longrightarrow$}}
        \Text(180,35)[]{\large{$-1\,\times$}}

	\Photon(210,0)(245,60){2}{5}
	\CArc(300,-12.2917)(91,127,172)
        \CArc(190,-12.2917)(91,7,53)
        \CArc(245,48)(60,236,304)

	\COval(245,0)(2,32)(0){Black}{Black}
        \BCirc(210,0){2}
	\CCirc(245,60){2}{Black}{Black}
	\BCirc(280,0){2}
	\Text(208,-10)[]{1}
	\Text(284,-10)[]{2}
	\Text(245,70)[]{3}
\end{picture}
}
\vspace{1.5cm}
\caption{Four-body diagram, $v_{4b\to3b}^{\text{cir}}$, that contributes to the three-body diagrams having a circular exchange between particles $1$, $2$ and $3$. \label{fig:red4bcir}}
\end{center}
\end{figure}

As explained in Section \ref{sec:kin_presc}, three-body cluster contribution to the PB kinetic energy contains terms of the kind $\nabla_{1}^2(\mathcal{S} \hat{F}_{12}\hat{F}_{13}\hat{F}_{23})$. Their explicit expressions can be obtained from the corresponding equations for the two-body potential by substituting the first term of the normalized traces with
\begin{equation}
\hat{v}_{12}(\mathcal{S} \hat{F}_{12}\hat{F}_{13}\hat{F}_{23})\to-2[\mathcal{S} (\nabla_{1}^2\hat{F}_{12})\hat{F}_{13}\hat{F}_{23}]-2[\mathcal{S} (\vec{\nabla}_{1}\hat{F}_{12})\cdot(\vec{\nabla}_1\hat{F}_{13})\hat{F}_{23}]\, ,
\end{equation}
while
\begin{equation}
\hat{v}_{12} \hat{F}_{12}\to-2 (\nabla_{1}^2\hat{F}_{12})
\end{equation}
for the second term. Following the notation of Section \ref{sec:kin_presc}, terms with $(\nabla_{1}^2\hat{F}_{12})$ are denoted by $W^{kin}$, those having $(\vec{\nabla}_{1}\hat{F}_{12})\cdot(\vec{\nabla}_1\hat{F}_{13})$ are included in $U$. On the other hand, the three-body cluster terms belonging to $W_F$ arise from the diagrams where particles $1$ and $2$ are exchanged 
\begin{align}
\langle \hat{T} \rangle\Big{|}_{3b\,W_F}^{P_{12}}&=\frac{\hbar^2}{m}\rho^2\int d\mathbf{r}_{12}d\mathbf{r}_{13}\ell(r_{12})\ell '(r_{12})\hat{r}_{12}\cdot \text{CTr}_{123}\Big[(\mathcal{S}\hat{F}_{12}\hat{F}_{13}\hat{F}_{23})\times\nonumber \\
&\qquad\qquad[\mathcal{S}(\vec{\nabla}_1\hat{F}_{12})\hat{F}_{13}\hat{F}_{23})\hat{P}^{\sigma\tau}_{12}-\hat{F}_{12}(\vec{\nabla}_{1}\hat{F}_{12})(\hat{F}_{13}^2+\hat{F}_{23}^2-1)\hat{P}^{\sigma\tau}_{12}\Big]
\label{eq:wf_p12}
\end{align}
and from the ones with circular exchange
\begin{align}
\langle \hat{T} \rangle\Big{|}_{3b\,W_F}^{cir}&=-\frac{\hbar^2}{m}\rho^2\int d\mathbf{r}_{12}d\mathbf{r}_{13}\ell(r_{13})\ell(r_{23})\ell '(r_{12})\hat{r}_{12}\cdot \text{CTr}_{123}\Big[(\mathcal{S}\hat{F}_{12}\hat{F}_{13}\hat{F}_{23})\nonumber \\
&\qquad\qquad[\mathcal{S}(\vec{\nabla}_1\hat{F}_{12})\hat{F}_{13}\hat{F}_{23}]\hat{P}^{\sigma\tau}_{12}\hat{P}^{\sigma\tau}_{13}-\hat{F}_{12}(\vec{\nabla}_{1}\hat{F}_{12})\hat{F}_{13}^2\hat{P}^{\sigma\tau}_{13}\hat{P}^{\sigma\tau}_{12}\Big]\,.
\label{eq:wf_cir}
\end{align}

The contributions to $U$ stem from the diagrams with the exchange $P_{13}$
\begin{align}
\langle \hat{T} \rangle\Big{|}_{3b\,U_F}^{P_{13}}&=\frac{\hbar^2}{m}\rho^2\int d\mathbf{r}_{12}d\mathbf{r}_{13}\ell '(r_{13})\hat{r}_{13}\cdot \text{CTr}_{123}\Big[(\mathcal{S}\hat{F}_{12}\hat{F}_{13}\hat{F}_{23})\times\nonumber \\
&\qquad\qquad[\mathcal{S}(\vec{\nabla}_1\hat{F}_{12})\hat{F}_{13}\hat{F}_{23})\hat{P}^{\sigma\tau}_{13}\Big]
\label{eq:uf_p12}
\end{align}
and from those having circular exchange
\begin{align}
\langle \hat{T} \rangle\Big{|}_{3b\,U_F}^{cir}&=-\frac{\hbar^2}{m}\rho^2\int d\mathbf{r}_{12}d\mathbf{r}_{13}\ell(r_{12})\ell(r_{13})\ell '(r_{13})\hat{r}_{23}\cdot \text{CTr}_{123}\Big[(\mathcal{S}\hat{F}_{12}\hat{F}_{13}\hat{F}_{23})\nonumber \\
&\qquad\qquad[\mathcal{S}(\vec{\nabla}_1\hat{F}_{12})\hat{F}_{13}\hat{F}_{23}]\hat{P}^{\sigma\tau}_{12}\hat{P}^{\sigma\tau}_{13}\Big]\,.
\label{eq:uf_cir}
\end{align}
Note that in this case there are no subtraction terms arising from reducible diagrams.

\subsection{Determination of the correlation functions}
\label{Variational_SA}
%%%%%%%%%%%%%%%%%
An upperbound  to the binding energy per particle, $E_V/A$, can be obtained by using the variational method, which amounts to minimizing the energy expectation value $\langle H\rangle/A$ with respect to the variational parameters included in the model.  Its cluster expansion is given by
\begin{equation}
\frac{\langle H\rangle}{A} = T_F +  (\Delta E)_2 + \makebox{higher order terms} \ ,
\label{eq:energy_cluster_cont}
\end{equation}
 where $T_F$ is the energy of the non interacting Fermi gas and $(\Delta E)_2$ denotes the contribution of two-nucleon clusters
\begin{equation}
 (\Delta E)_2=\mathcal{F}^\dagger v_{12} \mathcal{F}\,\Big{|}_{2b}+\langle\hat{T} \rangle \Big{|}_{2b}\, .
\end{equation}
 
Neglecting higher order cluster contributions, the functional minimization of $\langle H\rangle/A$ leads to a set of  six Euler-Lagrange equations, to be solved with proper constraints that force $f^c$ and $f^{(p>1)}$ to ``heal'' at one and zero, respectively. That is most efficiently achieved through the boundary conditions \cite{lagaris_81,pandha_79}
\begin{eqnarray}
f^{p}(r\geq d^p) &=& \delta_{p1}\, , \nonumber \\
\frac{df^p(r)}{dr}\mid_{d^p} &=& 0\, . 
\end{eqnarray}
Numerical calculations are generally carried out using only  two independent ``healing distances'': $d_c~=~d^{p=1 \dots 4}$ and $d_t=d^{5,6}$.
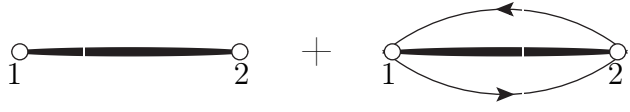
\begin{figure}[!h]
\begin{center}
\fcolorbox{white}{white}{
  \begin{picture}(150,80)(20,-20)
	\SetWidth{0.5}
	\SetColor{Black}
	
	\COval(30,10)(1.5,38)(0){Black}{Black}
%	\ArrowArc(40,10)(-93,0,180)
	\BCirc(-11,10){3}
	\BCirc(71,10){3}
	\Text(-13,2)[]{1}
	\Text(72,2)[]{2}
	
	\Text(100,10)[]{\Large{+}}
	
	\COval(170,10)(1.5,38)(0){Black}{Black}
	\ArrowArc(170,-45)(71,50,130)
	\ArrowArc(170,65)(71,230,310)
	\BCirc(128,10){3}
	\BCirc(212,10){3}
	\Text(127,2)[]{1}
	\Text(212,2)[]{2}
		
  \end{picture}
}
\vspace{0.1cm}
\caption{Diagrammatic representation of the two-body cluster contribution $(\Delta E)_2$ of Eq. (\ref{eq:energy_cluster_cont}). The thick lines represents both the potential and a kinetic contribution, involving derivatives  acting only on the correlation functions. The effect of the other derivatives is included in $T_F$.}
\label{fig:2_body_cc}
\end{center}
\end{figure}

Additional and important variational parameters are the quenching factors $\alpha_p$ whose introduction simulates modifications of the two--body potentials entering in the Euler--Lagrange differential equations arising from the screening induced by the presence of the nuclear medium 
\begin{equation}
\hat{v}_{ij}=\sum_{p=1}^6 \alpha_p v^{p}(r_{ij})\hat{O}^{p}_{ij}\, .
\end{equation}
The full potential is, of course, used in the energy expectation value.
In addition,  the resulting correlation functions $f^p$ are often rescaled according to  
\begin{equation}
\hat{F}_{ij}=\sum_{p=1}^6 \beta_p f^{p}(r_{ij})\hat{O}^{p}_{ij}\; ,
\end{equation}
 
The energy expectation value $\langle H\rangle/A$, calculated in full FHNC/SOC approximation is minimized with respect to variations of $d_c$, $d_t$, $\beta_{p}$, and $\alpha_{p}$.

To determine the best values of the variational parameters we have used a version of the ``Simulated annealing'' algorithm \cite{kirkpatrick_83}. 
In metallurgy the annealing procedure consists in heating and then slowly cooling a metal, to decrease the defects of its structure. During the heating the atoms gain kinetic energy and move away from their initial equilibrium positions, passing through states of higher energy. Afterwards, when the metal slowly cools, it is possible that the atoms freeze in a different configuration with respect to the initial one, corresponding to a lower value of the energy.

In minimization problems the analog of the position of the atoms are the values of the parameters to be optimized, in our case $d_c$, $d_t$, $\beta_{p}$ and $\alpha_p$, while the energy of the system corresponds to the function that has to be minimized, that in our case is the variational energy
\begin{equation}
E_V=E_V(d_c,d_t,\beta_{p},\alpha_p)\, .
\end{equation}

In the simulated annealing procedure, the parameters $d_c$, $d_t$, $\beta_{p}, \alpha_p$ are drawn from the Boltzmann distribution, $\exp(-E_{V}/T)$, where $T$ is just a parameter of the simulated annealing algorithm, having no physical meaning.

We have used a Metropolis algorithm, with acceptance probability of passing from the state $s=\{d_c,d_t,\beta_{p},\alpha_p\}$ to the proposed state $s'=\{d_c',d_t',\beta_{p}',\alpha_p'\}$ given by
\begin{equation}
P_{s,s'}=\exp\Big[-\frac{E(s')-E(s)}{T}\Big]\, ,
\end{equation}
By looking at the distribution of the parameters resulting from the Metropolis random walk, it is possible to find the values $\tilde{d}_c$, $\tilde{d}_t$, $\tilde{\beta}_{p}$ and $\tilde{\alpha}_p$ corresponding to the minimum of $E_V$, e.g. to the maximum of the Boltzmann distribution. As the fictitious temperature $T$ is lowered, the system approaches the equilibrium and the values of the parameters get closer and closer to $\tilde{d}_c$, $\tilde{d}_t$, $\tilde{\beta}_{p},\tilde{\alpha}_p$ .

\section{Auxiliary field diffusion Monte Carlo}
A central issue in many-body physics is the evaluation of multidimensional integrals, like the one of Eq. (\ref{eq:g2bop_def}). An alternative to the cluster expansion technique is represented by stochastic algorithms using the central limit theorem to compute multidimensional integrals, known as ``Monte Carlo methods'' \cite{kalos_84, kalos_09}.  

Using standard numerical integration methods, like the Simpson rule, the computation of a $D$-dimensional integral requires an exponentially growing number of operations. To be definite, in order to estimate the value of a $D$-dimensional integral with an accuracy $\epsilon$, the quantity of operations scales with $\epsilon^{-D}$. 
The central limit theorem guarantees that Monte Carlo methods scale as $\epsilon^{-2}$, regardless from the dimensionality.

\subsection{Variational Monte Carlo}
Variational Monte Carlo (VMC) uses the stochastic integration method for evaluating the expectation values for a chosen trial wave function. In the CBF approach, the trial wave-function is given by $\Psi_T(X) \equiv \langle X |\hat{\mathcal{F}}|\Phi_0\rangle$. In order to comply with the standard notation of Monte Carlo formalism, we do not use the variety of the bra and ket symbols introduced for CBF theory and $|\Phi_0\rangle\equiv |\Psi_0\rangle$. 

The expectation value of any operator $\hat{O}$ on the state $\Psi_T$
\begin{equation}
\langle \Psi_T | \hat{O} |\Psi_T\rangle =\frac{\int dx_{1,\dots,A} \Psi_T^*(X)\hat{O}\Psi_T(X)}{\int dx_{1,\dots,A}  \Psi_T^*(X)\Psi_T(X)}
\end{equation}
can be conveniently written making the spin-isospin sum explicit 
\begin{align}
\sum_{\alpha\beta}\int d\mathbf{r}_{1,\dots,A}\Big[\frac{\Psi_{T\,\alpha}^*(R) \hat{O}_{\alpha\beta}\Psi_{T\,\beta}(R)}{\mathcal{P}(R)_{\alpha\beta}}\Big]\mathcal{P}(R)_{\alpha\beta}=\sum_{\alpha\beta}\int d\mathbf{r}_{1,\dots,A}\hat{\mathcal{O}}_{\alpha\beta}\mathcal{P}(R)_{\alpha\beta}\, ,
\label{eq:O_VMC}
\end{align}
with
\begin{equation}
\mathcal{O}_{\alpha\beta}=\Big[\frac{\Psi_{T\,\alpha}^*(R) \hat{O}_{\alpha\beta}\Psi_{T\,\beta}(R)}{\mathcal{P}(R)_{\alpha\beta}}\Big]\, .
\end{equation}

The VMC algorithm prescribes to sample the configuration $R$ from the probability density 
\begin{equation}
\mathcal{P}(R)_{\alpha\beta}= \Psi_{T\,\alpha}^*(R)\Psi_{T\,\beta}(R)\, 
\end{equation}
and to estimate the integral with the sum 
\begin{equation}
\langle \hat{O} \rangle=\frac{1}{N_c}\sum_{\alpha\beta}\sum_{\{R\}}\mathcal{O}_{\alpha\beta}\, ,
\end{equation}
where $N_c$ is the number of sampled configurations.

VMC can be seen as an alternative to the cluster expansion technique, allowing for controlling the approximation arising from elementary diagrams and SOC approximation. The main drawback of VMC, shared with the CBF, is that the goodness of the result entirely depends on the accuracy of the trial wave function. 

In actual facts, neutron matter calculations with for the energy per particle, have been limited to $14$ nucleons in a box \cite{carlson_03}. This is due to the operator structure of the correlations, implying a sum over the spin-isospin degrees of freedom of $A$ particles. The possible spin states of $A$ nucleons are $2^A$ and since $Z$ of the $A$ nucleons are protons there are $A!/Z!(A-Z)!$ isospin states. Hence the total number of spin-isospin states is 
\begin{equation}
2^A\frac{A!}{Z!(A-Z)!}\, .
\end{equation}

\subsection{Diffusion Monte Carlo}
The diffusion Monte Carlo (DMC) method \cite{grimm_71,kalos_84,kalos_09,foulkes_01}, overcomes the limitation of the variational wave-function by using a projection technique to enhance the true ground-state component of a starting trial wave function. The trial wave-function can be expanded on the complete set of eigenstates of the the full hamiltonian, introduced in Eq. (\ref{eq:exact_h})
\begin{equation}
|\Psi_T\rangle=\sum_n c_n |\Psi_n\rangle\, .
\end{equation}
The evolution in imaginary time $\tau=i t/\hbar$, that project out the true ground state from a trial wave function, provided that it is not orthogonal to the true ground state, i. e. $c_0\neq 0$
\begin{equation}
\lim_{\tau\to\infty}e^{-H\tau}|\Psi_T\rangle = \lim_{\tau\to\infty}\sum_n c_n e^{-E_n \tau}|\Psi_n\rangle=  \lim_{\tau\to\infty}c_0 e^{-E_0 \tau}|\Psi_0\rangle\, .
\label{eq:DMC_exp}
\end{equation}

Diffusion Monte Carlo (DMC) is a stochastic projector method for solving the imaginary-time many-body Schr\"odinger equation
\begin{equation}
-\frac{\partial}{\partial \tau} |\Psi(\tau)\rangle=(\hat{H}-E_T)|\Psi(\tau)\rangle\, \quad \to \quad 
|\Psi(\tau+\Delta\tau)\rangle=e^{-(\hat{H}-E_T)\Delta\tau}|\Psi(\tau)\rangle\, .
\label{eq:sh_im}
\end{equation}
Imposing the initial condition 
\begin{equation}
|\Psi(\tau=0)\rangle=|\Psi_T\rangle \, ,
\end{equation}
it is readily seen that in terms of the imaginary time wave-functions equation (\ref{eq:DMC_exp}) implies
\begin{align}
\lim_{\tau\to\infty}|\Psi(\tau)\rangle&=\lim_{\tau\to\infty}c_0 e^{-(E_0-E_T)\tau}|\Psi_0\rangle\, .
\end{align}
The energy offset $E_T$ is adjusted to be as close as possible to $E_0$ with the aim of making the damping exponential factor constant.

For the sake of simplicity, at this point we limit ourselves in considering the 3N spatial coordinates only. An entire Section will be dedicated to the inclusion of spin-isospin degrees of freedom through the auxiliary fields. Inserting a completeness on the orthonormal basis $\{|R'\rangle\}$ in Eq. (\ref{eq:sh_im}) yields
\begin{align}
|\Psi(\tau+\Delta\tau)\rangle=\int dR' e^{-(\hat{H}-E_T)\Delta\tau}|R'\rangle \Psi(R',\tau)\, .
\end{align}

Projecting on the coordinates $\langle R'|$ leads to
\begin{equation}
\Psi(R,\tau+\Delta\tau)=\int dR' G(R,R',\Delta\tau) \Psi(R',\tau)
\label{eq:gmat_ev}
\end{equation}
where the kernel  
\begin{equation}
G(R,R',\Delta\tau)=\langle R| e^{-(\hat{H}-E_T)\frac{\Delta\tau}{\hbar}}|R'\rangle\, 
\end{equation}
is Green's function of the operator $\hat{H}+\frac{\partial}{\partial\tau}$. Notice that no assumptions on the smallness of $\Delta\tau$ have been made so far. In terms of the wave function the Schr\"odinger Eq. (\ref{eq:sh_im}) reads
\begin{equation}
-\frac{\partial}{\partial \tau}\Psi(R,\tau)=(\hat{H}-E_T)\Psi(R,\tau)\, .
\label{eq:sh_im_wf}
\end{equation}

If we neglect the interaction terms in the Hamiltonian 
\begin{equation}
\hat{H}=\hat{T}=-\frac{\hbar^2}{2m}\sum_i\nabla_{i}^2\, ,
\end{equation}
the imaginary time Schr\"odinger equation (\ref{eq:sh_im_wf}) reduces to the {\it master equation of a diffusion stochastic process} \cite{karlin_81}
\begin{equation}
\frac{\partial}{\partial \tau}\Psi(R,\tau)=\frac{\hbar^2}{2m}\sum\nabla_{i}^2\Psi(R,\tau)\, .
\end{equation}

The associated Green's function is a 3N-dimensional Gaussian for the spatial coordinate having variance $\tau$ in each dimension
\begin{equation}
G_d(R,R',\Delta\tau)=\Big(\frac{m}{2\pi\hbar^2\Delta\tau}\Big)^\frac{3A}{2}e^{-\frac{m}{2\hbar^2\Delta\tau}(R-R')^2}\, 
\end{equation}
describing the Brownian {\it diffusion} of N particles with a dynamic governed by random collisions.

No DMC algorithm is necessary to solve this problem, as the limit $\tau\to \infty$ leading to an uniform wave function can be analytically taken. However, to explain how the DMC algorithm works, let us represent  represent the distribution $\Psi(R,\tau)$ by a set of discrete Brownian sampling points or {\it random walkers}
\begin{equation}
\Psi(R,\tau)=\sum_k\delta(R-R_k)\, .
\end{equation}
Evolving this discrete distribution for an imaginary time $\Delta\tau$ by means of Eq. (\ref{eq:gmat_ev}) we obtain a set of Gaussians, centered in the positions $R_k$
\begin{equation}
\Psi(R,\tau+\Delta\tau)=\sum_k G_d(R,R_k,\Delta\tau)
\end{equation}
In order to get a discrete representation for the positions of the walker at $\tau+\Delta\tau$, each Gaussian is sampled by a new delta function. The procedure of propagation/resampling is then iterated until convergence is reached.

When the full Hamiltonian, including both the potential and the kinetic term, is considered
\begin{equation}
\hat{H}=\hat{T}+\hat{V}\, ,
\end{equation}
the analytic solution of the associated Green's function $\langle R| e^{-(\hat{T}+\hat{V}-E_T)\Delta\tau}|R'\rangle $ is in general not known.  An approximation to the Green's function can be obtained using the Trotter-Suzuki formula
 \begin{equation}
e^{(\hat{A}+\hat{B})\Delta\tau}=e^{-\hat{A}\Delta\tau/2}e^{-\hat{B}\Delta\tau}e^{-\hat{A}\Delta\tau/2}+\mathcal{O}(\Delta\tau^3)\, .
\label{eq:ts_formula}
\end{equation}
In the limit of small $\Delta\tau$ the Green's function can be factorized 
\begin{equation}
\langle R| e^{-(\hat{T}+\hat{V}-E_T)\Delta\tau}|R'\rangle=G_b(R,R',\Delta\tau)G_d(R,R',\Delta\tau)\, .
\end{equation}
with the {\it branching} factor given by
\begin{equation}
G_b(R,R',\Delta\tau)=e^{-\left(\frac{V(R)+V(R')}{2}-E_T\right)\Delta\tau}\, .
\end{equation}

The wave function at $\tau+\Delta\tau$ can then reads
\begin{equation}
\Psi(R,\tau+\Delta\tau)=\Big(\frac{m}{2\pi\hbar^2\Delta\tau}\Big)^\frac{3A}{2} \int dR' e^{-\frac{m(R-R')^2}{2\hbar^2\Delta\tau}} e^{-\frac{V(R)+V(R')}{2}\Delta\tau}e^{E_T\Delta\tau}\Psi(R',\tau)\, .
\label{eq:diff_MC}
\end{equation}
The long imaginary time evolution is obtained by iterating the last equation, which is valid for $\Delta\tau\to 0$, until convergence is reached. Note that since the error in Eq. (\ref{eq:ts_formula}) is $\mathcal{O}(\Delta\tau^3)$, iterating for a finite amount of imaginary time leads to an overall error of the order $\Delta\tau^2$. On the other hand, using the less-refined version of the Trotter-Suzuki formula
\begin{equation}
e^{(\hat{A}+\hat{B})\Delta\tau}=e^{-\hat{A}\Delta\tau}e^{-\hat{B}\Delta\tau}+\mathcal{O}(\Delta\tau^2)\, .
\label{eq:ts_formula2}
\end{equation}
provides a still acceptable bias of the order $\Delta\tau$. 

The integral of Eq. (\ref{eq:diff_MC}) is solved using the Monte Carlo {\it diffusion} algorithm, the main step of which can be summarized as follows \cite{ormoni_11}:
\begin{itemize}
\item The initial set of walkers is sampled from the distribution $\Psi(R',\tau=0)=\Psi_T(R')$ and the starting trial energy $E_T$ is chosen, for instance from a variational calculation.

\item The  coordinates of the walkers are diffused by means of a Brownian motion 
\begin{equation}
R=R'+\xi\qquad
\end{equation}
where $\xi$ is a stochastic variable distributed according to a Gaussian probability density with $\sigma=\hbar\Delta\tau/m$ and zero average so that the walkers are distributed according to $G_d(R,R',\Delta\tau)$.

\item The {\it branching or birth/death algorithm} is applied: the weight 
\begin{equation}
w=G_b(R,R',\Delta\tau)
\label{eq:weight}
\end{equation}
is assigned to each walker and a number of copies of the walker proportional to $w$ is generated. Then the convolution theorem governing the composition of random variables guarantees that the distribution of the walker is $\Psi(R,\tau+\Delta\tau)$. In actual facts, the integer number of copies is given by the branching factor 
\begin{equation}
m=\text{INT}(w+\eta)\, ,
\end{equation}
where INT denotes the integer part of a real number and $\eta$ is a random number drawn from the uniform distribution on the interval $[0,1]$. The energy offset $E_T$ is adjusted to keep the the total population of walkers fluctuating around a desired value.

\item Once convergence is reached, i.e. for large enough $\tau$, the configurations are distributed with a probability density $\Psi(R,\tau)$. Therefore, the ground-state expectation values of observables that commute with the hamiltonian 
\begin{align}
\langle \hat{O}\rangle=\frac{\langle \Psi_0| \hat{O} | \Psi_0 \rangle }{\langle\Psi_0|\Psi_0\rangle}&=
\lim_{\tau\to\infty}\frac{\langle \Psi_T| \hat{O} | \Psi(\tau) \rangle }{\langle\Psi_T|\Psi(\tau)\rangle}
=\lim_{\tau\to\infty}\frac{\int dR \langle \Psi_T| \hat{O} |R\rangle \Psi(R,\tau)}{\int dR\Psi_T(R)\Psi(R,\tau)}
\label{eq:expt_nis}
\end{align}
can be computed by
\begin{equation}
\langle \hat{O}\rangle=\frac{\sum_{\{R\}} \langle R| \hat{O} | \Psi_T\rangle}{\sum_{\{R\}}\langle R | \Psi_T\rangle}=\frac{\sum_{\{R\}} [O\Psi_T](R)}{\sum_{\{R\}}\Psi_T(R)}\,.
\end{equation}
\end{itemize}

\subsection{Importance sampling}

The basic version of the DMC algorithm described in the previous Section is poorly efficient, as the brownian diffusive process ignores the shape of the potential. Hence, the weight of Eq. (\ref{eq:weight}) suffers of large fluctuations from step to step, as, for instance, there is nothing that prevents two-particles from moving very close to each other even in presence of an hard-core repulsive potential.

The idea of the importance sampling technique consists in using the knowledge of the trial wave function $\Psi_T(R)$ \cite{grimm_71,ceperley_79} to guide the diffusive process.

Let us multiply the imaginary time Schr\"odinger equation (\ref{eq:sh_im_wf}) by the trial wave function $\Psi_T(R)$ and introduce a new distribution $f(R,\tau)=\Psi_T(R)\Psi(R,\tau)$. We obtain a non homogenous Fokker-Plank equation
\begin{equation}
-\frac{\partial}{\partial \tau}f(R,\tau)=-\frac{\hbar^2}{2m}\nabla^2f(R,\tau)+\frac{\hbar^2}{m}\vec{\nabla}\cdot[\vec{v}_D(R)f(R,\tau)]+[E_L(R)-E_0]f(R,\tau)\, ,
\label{eq:sc_f_importance}
\end{equation}
where 
\begin{equation}
\vec{v}_D(R)=\frac{\vec{\nabla}\Psi_T(R)}{\Psi_T(R)}
\end{equation}
is the 3A dimensional ``drift velocity'' and 
\begin{equation}
E_L(R)=\frac{H\Psi_T(R)}{\Psi_T(R)}
\end{equation}
is the ``local energy''.

The imaginary time evolution for $f(R,\tau)$ can be conveniently written in terms of a modified Green's function 
\begin{equation}
f(R,\tau+\Delta\tau)=\int \tilde{G}(R,R',\Delta\tau)f(R',\tau)dR'\, .
\label{eq:is_gf}
\end{equation}
Comparing with Eq. (\ref{eq:gmat_ev}) it is immediately found that
\begin{equation}
\tilde{G}(R,R',\Delta\tau)=G(R,R',\Delta\tau)\frac{\psi_T(R)}{\psi_T(R')}\, .
\label{eq:isgf_def}
\end{equation}
It is shown in Appendix \ref{app:isgf} that the importance sampling makes the diffusion driven by the drift velocity, that carries the walkers along in the direction of increasing $\Psi_T$
\begin{equation}
\tilde{G}_d(R,R',\Delta\tau)=\Big(\frac{m}{2\pi\hbar^2\Delta\tau}\Big)^\frac{3A}{2}e^{-\frac{m}{2\hbar\Delta\tau}\left[R-R'-\frac{\hbar^2\Delta\tau}{m}v_D(R')\right]^2}\, .
\label{eq:diff_isgf}
\end{equation}
The branching factor now contains the local energy instead of the potential energy
\begin{equation}
\tilde{G}_b(R,R',\Delta\tau)=e^{-\left(\frac{E_L(R)+E_L(R')}{2}-E_T\right)\frac{\Delta\tau}{\hbar}}\, .
\label{eq:branc_isgf}
\end{equation}
If the trial wave function is sufficiently accurate, the local energy remains close to the ground-state energy throughout the imaginary time evolution.  

%The diffusive process of the DMC algorithm is modified by sampling the walkers position from a drifted %gaussian. Moreover, the weight contains the local energy instead of the potential energy, which is a much less %fluctuating quantity.  

As far as the expectation value of the operator $\hat{O}$ is concerned, from the last term of Eq. (\ref{eq:expt_nis}) it follows
\begin{align}
\langle \hat{O}\rangle=
\lim_{\tau\to\infty}\frac{\int dR  \frac{\langle\Psi_T| \hat{O} |R\rangle}{\Psi_T(R)} f(R,\tau)}{\int dR f(R,\tau)}
\end{align}
Using the central limit theorem, $\langle \hat{O}\rangle$ can be computed by sampling the configurations from $f(R,\tau)$
\begin{equation}
\langle \hat{O}\rangle=\frac{\sum_{\{R\}} \langle R| \hat{O} | \Psi_T\rangle}{\sum_{\{R\}}\langle R | \Psi_T\rangle}=\frac{\sum_{\{R\}} [O\Psi_T](R)}{\sum_{\{R\}}\Psi_T(R)}\,.
\end{equation}

\subsection{Sign problem}
In order to project out the ground state of a given Hamiltonian, the DMC algorithm implies a diffusive process, whose starting distribution of walkers is given by the trial wave function. Hence, for the diffusion interpretation to be applicable, $\Psi_T$ must be positive definite in the whole configuration space. This is the case, for example, of many-bosons system, whose ground-state wave function is positive definite. 

The ground state of a fermionic system on the other hand, is described by an antisymmetric wave function, to which a probability distribution interpretation cannot be given. Let us describe this issue in more details. 

It can be proven that the ground state $\Psi_0(R)$ of a regular hamiltonian $\hat{H}$ is node less. Hence, from a strictly mathematical point of view, the search of an antisymmetric ground state, $\Psi_{0}^A(R)$, corresponds to the search of an excited state of the many-body hamiltonian. In terms of the energy eigenvalues, this corresponds to
\begin{equation}
E_0<E_{0}^A\, ,
\end{equation}
where $E_0$ and $E_{0}^A$ are the ground-state energies for the bosonic and the fermionic system described by $\hat{H}$ \cite{lipparini_08}.

Expanding the trial antisymmetric wave function in terms of eigenstate of the hamiltonian and choosing the energy offset to be $E_{0}^A$, in the limit of large imaginary time we get
\begin{equation}
\lim_{\tau\to\infty}e^{-H\tau}\Psi_T(R) = \lim_{\tau\to\infty}\Big[\sum_n c_n e^{-(E_n-E_{0}^A) \tau}\Psi_n(R)+c_{0}^A\Psi_{0}^A(R)+\dots\Big]
\end{equation}
The sum over $n$ runs over the bosonic eigenfunctions of the hamiltonian having a smaller energy than $E_{0}^A$; when $\tau\to\infty$ these terms diverge. The dots indicate the converging term, i.e. the eigenfunctions (both bosonic and fermionic) with energies larger than $E_{0}^A$ that are exponentially suppressed with respect to $\Psi_{0}^A(R)$.

The exponentially growing component along the symmetric ground state does not affect the expectation of the Hamiltonian. Because of the orthogonality between antisymmetric and symmetric wave functions, it turns out that
\begin{align}
\langle \hat{H} \rangle&=\frac{\langle \Psi_0| \hat{O} | \Psi_0 \rangle }{\langle\Psi_0|\Psi_0\rangle}=\lim_{\tau\to\infty}\frac{ \langle \Psi(\tau)| \hat{H} |\Psi_{T}^A\rangle}{\langle \Psi(\tau)|\Psi_{T}^A\rangle}=\lim_{\tau\to\infty}\frac{\int dR \Psi(R,\tau) [\hat{H}\Psi_{T}^A](R)}{\int dR \Psi(R,\tau)\Psi_{T}^A(R)}\nonumber \\
&=\lim_{\tau\to\infty}\frac{\int dR[\sum_n c_n e^{-(E_n-E_{0}^A) \tau}\Psi_n(R)+c_{0}^A\Psi_{0}^A(R)] [\hat{H}\Psi_{T}^A](R)}{\int dR [\sum_n c_n e^{-(E_n-E_{0}^A) \tau}\Psi_n(R)+c_{0}^A\Psi_{0}^A(R)]\Psi_{T}^A(R)}\nonumber\\
&=\frac{\int dRc_{0}^A\Psi_{0}^A(R)[\hat{H}\Psi_{T}^A](R)}{\int dR c_{0}^A\Psi_{0}^A(R)\Psi_{T}^A(R)}=E_{0}^A\, .
\end{align}
However, the variance of the DMC estimate for the energy expectation value $\sigma^{2}_{E_{0}^A}=|\langle [\hat{H}\Psi_{T}^A]^2\rangle-\langle[\hat{H}\Psi_{T}^A]\rangle^2|$ is exponentially diverging. In particular, the bosonic components dominates the second term of the variance, as the orthogonality which eliminates the symmetric contributions does not apply in the following integral
\begin{align}
\langle \hat{H} \rangle&=\lim_{\tau\to\infty}\frac{\int dR[\sum_n c_n e^{-(E_n-E_{0}^A) \tau}\Psi_n(R)] [\hat{H}\Psi_{T}^A]^2(R)}{\int dR c_{0}^A\Psi_{0}^A(R)\Psi_{T}^A(R)}\, .
\end{align}

We are left with the contradictory statement that the energy converges to exact eigenvalue with an exponentially growing statistical error: the signal to noise ratio exponentially decays.

In order for the DMC to be used for fermionic systems, it is possible to artificially split the configuration space in regions within which the sign of the trial wave function does not change. The $3A-1$ dimensional subset of the configuration space where the trial wave function vanishes is denoted as ``nodal surface''. In the {\it fixed node} approximation \cite{anderson_76}, during the diffusion process the walkers crossing the nodal surface are dropped. In other words, the nodal surface of the ground-state is imposed on the system, as it defined by the constraint $\Psi_T(R)=0$.
It can be proven \cite{reynolds_82} that the energy obtained from a fixed node DMC simulation obeys a variational principle. It is important to note that this variational principle only applies to ground-state calculations while a much weaker variational principle holds for excited-state calculations \cite{foulkes_99}.

In the case of nuclear hamiltonian, the overlap between walkers and the trial-wave function is complex and the  sign problem turns into a phase problem. A generalization of the fixed-node approximation, the so-called ``constrained-path'' \cite{zhang_95} was introduced to deal with complex wave functions.  It amounts in constraining the walkers to diffuse in regions where the overlap with the trial wave function is positive. To this aim, a suitable choice of the drift terms, used in the earlier AFDMC calculations \cite{schmidt_99} is
\begin{equation}
\vec{v}_D(R)=\frac{\vec{\nabla}Re[\Psi_T(R)]}{Re[\Psi_T(R)]}\, .
\end{equation}
To avoid the signal to noise ratio exponentially decay, the constrained-path approximation is realized by imposing 
\begin{equation}
\frac{Re[\Psi_T(R')]}{Re[\Psi_T(R)]}>0\, .
\end{equation}
Thus, the walkers having an overlap with the trial wave function that after a diffusive step changes sign are dropped.

Another approach followed to put under control the sign problem is the ``fixed-phase'' approximation, introduced do deal with hamiltonian containing a magnetic field \cite{ortiz_93}. The walkers are forced to have the same phase as the importance function $\Psi_T$. The drift term is given by 
\begin{equation}
\vec{v}_D(R)=\frac{\vec{\nabla}|\Psi_T(R)|}{|\Psi_T(R)|}\, .
\end{equation}
It can be shown that an additional term in the branching has to be considered; in particular only the real part of the kinetic energy contribution to the local energy has to be kept
\begin{equation}
\frac{\nabla^2\Psi_T(R)}{\Psi_T(R)}\to Re\Big[\frac{\nabla^2\Psi_T(R)}{\Psi_T(R)}\Big]\, .
\end{equation}

Both for the constrained-path and the fixed-phase approximations an accurate trial wave functions would be needed. While in GFMC calculations \cite{carlson_87} the full operator structure of the CBF is taken into account, AFDMC trial wave function only encloses pure central Jastrow correlations, which is positive definite. Hence for AFDMC,  the nodal structure of the nuclear matter wave function is entirely given by the Slater determinant of plane waves.

It is relevant to the purpose of the sign problem discussion to remark that using constrained-path approximation, the DMC algorithm does not necessarily provide an upper bound in the calculation of energy \cite{carlson_99}. Moreover, it has not been proved that the fixed-phase approximation gives an upper bound to the real energy.

For further details concerning constrained path and fixed phase approximations the reader is referred to the original papers and to the exhaustive discussion reported in the PhD Thesis of Paolo Armani \cite{ormoni_11}. 

\subsection{Spin-isospin degrees of freedom and auxiliary fields}
The method we have described needs to be generalized to account for spin-isospin degreed of freedom, that are of major importance in nuclear few- and many- body systems. 

Let us start from an example: within the GFMC approach the eight spin configuration of the $^3H$ nucleus (we neglect for the moment the isospin) are represented by \cite{pieper_98}
\begin{equation}
    |^3 H\rangle = \left(
      \begin{array}{c}
       a_{\,\uparrow\uparrow\uparrow}(R)\\
       a_{\,\uparrow\uparrow\downarrow}(R) \\
       a_{\,\uparrow\downarrow\uparrow}(R)\\
       a_{\,\uparrow\downarrow\downarrow}(R)\\
       a_{\,\downarrow\uparrow\uparrow}(R)\\
       a_{\,\downarrow\uparrow\downarrow}(R) \\
       a_{\,\downarrow\downarrow\uparrow}(R)\\
       a_{\,\downarrow\downarrow\downarrow}(R)\\
      \end{array} \right)
  \end{equation}
  
Each coefficient $a_\alpha$ represent the amplitude of a given many-particles spin configuration; for instance 
\begin{equation}
a_{\,\uparrow\uparrow\downarrow}(R)=\langle \uparrow\uparrow\downarrow|^3 H\rangle\, .
\end{equation}

The many-particles spin configuration space is closed under the action of the operators contained in the hamiltonian. For example, applying $\sigma_{12}=2\hat{P}_{12}^{\sigma}-1$ yields
\begin{equation}
   \hat{\sigma}_{12} |^3 H\rangle = \left(
      \begin{array}{c}
       a_{\,\uparrow\uparrow\uparrow}(R)\\
       a_{\,\uparrow\uparrow\downarrow}(R) \\
       2a_{\,\downarrow\uparrow\uparrow}(R)-a_{\,\uparrow\downarrow\uparrow}(R)\\
       2a_{\,\downarrow\uparrow\downarrow}(R)-a_{\,\uparrow\downarrow\downarrow}(R)\\
       2a_{\,\uparrow\downarrow\uparrow}(R)-a_{\,\downarrow\uparrow\uparrow}(R)\\
       2a_{\,\uparrow\downarrow\downarrow}(R)-a_{\,\downarrow\uparrow\downarrow}(R) \\
       a_{\,\downarrow\downarrow\uparrow}(R)\\
       a_{\,\downarrow\downarrow\downarrow}(R)\\
      \end{array} \right)
      \vspace{0.1cm}
      \label{eq:gfmc_s12}
  \end{equation}

Since the total charge is conserved, for the isospin of the $^3H$ we have $pnn$, $npn$, or $nnp$; thus, the vector describing the whole spin-isospin structure has 24 entries. In the GFMC algorithm the the imaginary time evolution of Eq. (\ref{eq:is_gf}) is need to applied to each of the $2^A\frac{A!}{Z!(A-Z)!}$ spin-isospin configurations and the the imaginary time evolution of Eq. (\ref{eq:gmat_ev}) generalizes to
\begin{align}
a_\alpha(R,\tau+\Delta\tau)=\sum_\beta \int dR' G_{\alpha\beta}(R,R',\Delta\tau)a_\beta(R',\tau)\, .
\label{eq:gfmc_ev}
\end{align}

The Green's function depends on the spin-isospin configuration
\begin{equation}
G_{\alpha\beta}(R,R',\Delta\tau)=\langle R,\alpha |e^{-(\hat{H}-E_T)\Delta\tau}|R',\beta\rangle\, .
\end{equation}

In order to deal with system having a large number of protons and neutrons, like for example medium-heavy nuclei or nuclear matter, GFMC does not seem to be a feasible approach. The idea of AFDMC consists in using a {\it single-particle} wave function, instead of the many-particle wave function of GFMC. For comparing the two methods, the spin structure of $^3H$ in AFDMC approach is 
%\begin{equation}
%  |^3 H\rangle = \left(
%      \begin{array}{c}
 %      c_{1}^\uparrow(R)\\
 %      c_{1}^\downarrow(R) \\
 %      c_{2}^\uparrow(R)\\
 %      c_{2}^\downarrow(R) \\
 %      c_{3}^\uparrow(R)\\
 %      c_{3}^\downarrow(R) \\
%      \end{array} \right)
 %     \vspace{0.1cm}
 %     \label{eq:wf_AFDMC}
 % \end{equation}
 \begin{align}
 |^3 H\rangle = &\Big[c_{1}^\uparrow |\uparrow\rangle_1+c_{1}^\downarrow|\downarrow\rangle_1\Big]\otimes 
\Big[\,c_{2}^\uparrow|\uparrow\rangle_2+c_{2}^\downarrow|\downarrow\rangle_2\Big]\otimes 
\Big[\,c_{3}^\uparrow|\uparrow\rangle_3+c_{3}^\downarrow|\downarrow\rangle_3\Big]
 \end{align}

where complex coefficient $c_{i}^\alpha$ denotes the amplitude for the $i-th$ particle to have spin state $\alpha$. Taking also the isospin degrees of freedom into account, it can be readily shown that the dimension of the vector describing a system with $A$ nucleons is $4A$. 

Already for such a small nucleus like $^3H$ the dimension of the spin-isospin structure of ADMC is a factor $2$ smaller than the one of GFMC. The gain in computational time of AFDMC with respect to GFMC becomes enormous for larger system. However, GFMC is still the best or at least one of the best available method for dealing with hard-core potentials like the Argonne $v_{18}$. The spin-orbit terms and three nucleon forces have not been included in AFDMC algorithm yet, with the notable exception of PNM. 

The main concern of the single-particle wave function is that it is not closed with respect to the application of a quadratic spin (or isospin) operator. Let us again apply the operator $\sigma_{12}$, as we did in Eq. (\ref{eq:gfmc_s12})
 
%\begin{equation}
% \sigma_{12}|^3 H\rangle =2\left(
 %     \begin{array}{c}
  %     c_{2}^\uparrow(R)\\
    %   c_{2}^\downarrow(R) \\
     %  c_{1}^\uparrow(R)\\
 %      c_{1}^\downarrow(R) \\
 %      c_{3}^\uparrow(R)\\
 %      c_{3}^\downarrow(R) \\
  %    \end{array} \right)-
 %\left(
   %   \begin{array}{c}
    %   c_{1}^\uparrow(R)\\
     %  c_{1}^\downarrow(R) \\
    %   c_{2}^\uparrow(R)\\
    %   c_{2}^\downarrow(R) \\
    %   c_{3}^\uparrow(R)\\
    %   c_{3}^\downarrow(R) \\
  %    \end{array} \right)\neq
  %    \left(
 %     \begin{array}{c}
  %     \tilde{c}_{1}^\uparrow(R)\\
   %    \tilde{c}_{1}^\downarrow(R) \\
  %     \tilde{c}_{2}^\uparrow(R)\\
  %     \tilde{c}_{2}^\downarrow(R) \\
%       \tilde{c}_{3}^\uparrow(R)\\
 %      \tilde{c}_{3}^\downarrow(R) \\
%      \end{array} \right)
 %     \vspace{0.1cm}
 %     \label{eq:s12_AFDMC}
%  \end{equation}
 
  \begin{align}
\sigma_{12}|^3 H\rangle =\,2&\Big[c_{2}^\uparrow |\uparrow\rangle_1+c_{2}^\downarrow|\downarrow\rangle_1\Big]\otimes 
\Big[c_{1}^\uparrow|\uparrow\rangle_2+c_{1}^\downarrow|\downarrow\rangle_2\Big]\otimes 
\Big[c_{3}^\uparrow|\uparrow\rangle_3+c_{3}^\downarrow|\downarrow\rangle_3\Big]-\nonumber\\
&\Big[c_{1}^\uparrow |\uparrow\rangle_1+c_{1}^\downarrow|\downarrow\rangle_1\Big]\otimes 
\Big[c_{2}^\uparrow|\uparrow\rangle_2+c_{2}^\downarrow|\downarrow\rangle_2\Big]\otimes 
\Big[c_{3}^\uparrow|\uparrow\rangle_3+c_{3}^\downarrow|\downarrow\rangle_3\Big]\, .
\end{align}
The resulting sum of two single-particle wave functions cannot be expressed as a single particle wave function. Therefore, if using the standard DMC algorithm, the imaginary-time propagator generates a sum of single particles wave functions at each time step. This would be catastrophic, as the number of single particle wave function would soon become enormous.

The idea of AFDMC is to use the Hubbard-Stratonovich transformation to reduce the spin-isospin dependence  from quadratic to linear, making the use of single particle wave functions feasible. In the following we will show how this is done for Argonne potentials incorporating the first six operators. The inclusion of the spin-orbit term is possible in the case of PNM only and it has beed described at length in Ref. \cite{gandolfi_07} and \cite{ormoni_11}. The first six components of the Argonne potential can be rewritten as 
\begin{equation}
\hat{V}=\sum_{i<j}\sum_{p=1}^6v_p(r_{ij})\hat{O}^{p}_{i,j}=\hat{V}_{SI}+\hat{V}_{SD}
\end{equation}
where the {\it spin independent} and {\it spin dependent} contribution read
\begin{align}
\hat{V}_{SI}&=\frac{1}{2}\sum_{ij}v^c(r_{ij})\nonumber \\
\hat{V}_{SD}&=\frac{1}{2}\sum_{i\alpha,j\beta}\sigma_{i\alpha}A^{\sigma}_{i\alpha,j\beta}\sigma_{j\beta}
+\frac{1}{2}\sum_{i\alpha,j\beta}\sigma_{i\alpha}A^{\sigma\tau}_{i\alpha,j\beta}\sigma_{j\beta}\vec{\tau}_i\cdot\vec{\tau}_j
+\frac{1}{2}\sum_{i,j}A^{\tau}_{i,j}\vec{\tau}_i\cdot\vec{\tau}_j\, .
\label{eq:spin_dep_part}
\end{align}
As usual, Latin indices label nucleons, while the Greek indices stay for cartesian components. From Eqs. (\ref{eq:av18}) and (\ref{eq:tens_def}), the $3A\times3A$ matrices $A^{\sigma}$ and $A^{\sigma\tau}$, and the $A\times A$ matrix $A^{\tau}$ in the case of the Argonne $v_{6}^\prime$ potential are readily seen to be \cite{sarsa_03},\cite{gandolfi_07b}
\begin{align}
A^{\sigma}_{i\alpha,j\beta}&=v^\sigma(r_{ij})\delta_{\alpha\beta}+v^t(r_{ij})(3\hat{r}_{ij}^\alpha \hat{r}_{ij}^\beta-\delta_{\alpha\beta})\, , \nonumber \\
A^{\sigma\tau}_{i\alpha,j\beta}&=v^{\sigma\tau}(r_{ij})\delta_{\alpha\beta}+v^{t\tau}(r_{ij})(3\hat{r}_{ij}^\alpha \hat{r}_{ij}^\beta-\delta_{\alpha\beta})\, , \nonumber \\
A^{\tau}_{i,j}\,\,&=v^\tau(r_{ij})\, .
\end{align}
The matrices $A$ are vanishing on the diagonal, as there is no self-interaction in the potential. Moreover, since they are real and symmetric, they have real eigenvalues and orthogonal eigenstates, given by
\begin{align}
&\sum_{j\beta}A^{\sigma}_{i\alpha,j\beta}\psi^{\sigma}_{n,j\beta}=\lambda_n^{\sigma}\psi^{\sigma}_{n,i\alpha}\, , \nonumber \\
&\sum_{j\beta}A^{\sigma\tau}_{i\alpha,j\beta}\psi^{\sigma\tau}_{n,j\beta}=\lambda_n^{\sigma\tau}\psi^{\sigma\tau}_{n,i\alpha}\, , \nonumber \\
&\sum_{j}A^{\tau}_{i,j}\psi^{\tau}_{n,j}=\lambda_n^{\tau}\psi^{\tau}_{n,i}\, .
\label{eq:A_autostati_autovettori}
\end{align}
It is convenient to normalize the eigenstates as follows
\begin{equation}
\sum_{i\alpha}\psi^{p}_{n,i\alpha}\psi^{p}_{m,i\alpha}=\delta_{nm}\, ,
\label{eq:auto_orto}
\end{equation}
for $p=\sigma,\sigma\tau,\tau$.
Using the last equation, $\sigma_{i\alpha}$ can be expanded as
\begin{equation}
\sigma_{i\alpha}=\sum_n \Big(\sum_{j,\beta}\psi^{\sigma}_{n,j\beta}\sigma_{j\beta}\Big) \psi^{\sigma}_{n,i\alpha}\, .
\label{eq:vettore_espanso}
\end{equation}

Substituting the latter result in the first term of Eq. (\ref{eq:spin_dep_part}) yields
\begin{align}
&\frac{1}{2}\sum_{i\alpha,j\beta}\sigma_{i\alpha}A^{\sigma}_{i\alpha,j\beta}\sigma_{j\beta}=\nonumber \\
&\frac{1}{2}\sum_{i\alpha,j\beta}\Big\{\Big[\sum_n \Big(\sum_{l,\delta}\psi^{\sigma}_{n,l\delta}\sigma_{l\delta}\Big) \psi^{\sigma}_{n,i\alpha}\Big]A^{\sigma}_{i\alpha,j\beta}\Big[\sum_m \Big(\sum_{l,\delta}\psi^{\sigma}_{m,l\delta}\sigma_{l\delta}\Big) \psi^{\sigma}_{m,j\beta}\Big]\Big\}=\nonumber \\
&\frac{1}{2}\sum_{i\alpha}\Big\{\Big[\sum_n \Big(\sum_{l,\delta}\psi^{\sigma}_{n,l\delta}\sigma_{l\delta}\Big) \psi^{\sigma}_{n,i\alpha}\Big]\lambda^{\sigma}_{n}\Big[\sum_m \Big(\sum_{l,\delta}\psi^{\sigma}_{m,l\delta}\sigma_{l\delta}\Big) \psi^{\sigma}_{m,i\alpha}\Big]\Big\}=\nonumber \\
&\frac{1}{2}\sum_{n}\Big(\sum_{l,\delta}\psi^{\sigma}_{n,l\delta}\sigma_{l\delta}\Big)^2\lambda^{\sigma}_{n}\, .
\end{align}
The expansion of $\sigma_{i\alpha}\vec{\tau}_i$ reads
\begin{equation}
\sigma_{i\alpha}\vec{\tau}_i=\sum_n \Big(\sum_{j,\beta}\psi^{\sigma\tau}_{n,j\beta}\sigma_{j\beta}\vec{\tau}_j\Big) \psi^{\sigma\tau}_{n,i\alpha}\, ,
\label{eq:vettore_espanso_dif}
\end{equation}
consequently the second term of Eq. (\ref{eq:spin_dep_part}) can be rewritten as
\begin{align}
&\frac{1}{2}\sum_{i\alpha,j\beta}\sigma_{i\alpha}\vec{\tau}_iA^{\sigma\tau}_{i\alpha,j\beta}\sigma_{j\beta}\vec{\tau}_j=\nonumber \\
&\frac{1}{2}\sum_{i\alpha,j\beta}\Big\{\Big[\sum_n \Big(\sum_{l,\delta}\psi^{\sigma\tau}_{n,l\delta}\sigma_{l\delta}\vec{\tau}_l\Big) \psi^{\sigma\tau}_{n,i\alpha}\Big]A^{\sigma\tau}_{i\alpha,j\beta}\Big[\sum_m \Big(\sum_{l,\delta}\psi^{\sigma\tau}_{m,l\delta}\sigma_{l\delta}\vec{\tau}_l\Big) \psi^{\sigma\tau}_{m,j\beta}\Big]\Big\}=\nonumber \\
&\frac{1}{2}\sum_{i\alpha}\Big\{\Big[\sum_n \Big(\sum_{l,\delta}\psi^{\sigma\tau}_{n,l\delta}\sigma_{l\delta}\vec{\tau}_l\Big) \psi^{\sigma\tau}_{n,i\alpha}\Big]\lambda^{\sigma\tau}_{n}\Big[\sum_m \Big(\sum_{l,\delta}\psi^{\sigma\tau}_{m,l\delta}\sigma_{l\delta}\vec{\tau}_l\Big) \psi^{\sigma\tau}_{m,i\alpha}\Big]\Big\}=\nonumber \\
&\frac{1}{2}\sum_{n}\Big(\sum_{l,\delta}\psi^{\sigma\tau}_{n,l\delta}\sigma_{l\delta}\vec{\tau}_l\Big)^2\lambda^{\sigma}_{n}
\end{align}
The same steps can be followed for the term with $A^{(\tau)}$. It is worth introducing a new set of operators written in terms of eigenvectors of matrices $A$
\begin{align}
O_{n}^{\sigma}&=\sum_{j\beta}\sigma_{j\beta}\psi^{\sigma}_{n,j\beta}\nonumber \\
O_{n,\alpha}^{\sigma\tau}&=\sum_{j\beta}\sigma_{j\beta}\tau_{j\alpha}\psi^{\sigma\tau}_{n,j\beta}\nonumber \\
O_{n,\alpha}^{\tau}&=\sum_{j}\tau_{j\alpha}\psi^{\tau}_{n,j}\, .
\label{eq:O_definition}
\end{align}
In terms of these operators, the spin dependent part of the potential reads
\begin{equation}
V_{SD}=\frac{1}{2}\sum_{n=1}^{3A}(O^{\sigma}_n)^2\lambda^{\sigma}_n+\frac{1}{2}\sum_{\alpha=1}^3\sum_{n=1}^{3A}(O^{\sigma\tau}_{n,\alpha})^2\lambda^{\sigma\tau}_n
+\frac{1}{2}\sum_{\alpha=1}^3\sum_{n=1}^{A}(O^{\tau}_{n,\alpha})^2\lambda^{\tau}_n
\end{equation}

Since the operators $\hat{O}_{n}^{p}$ in general do not commute, it turns out that
\begin{equation}
e^{-\frac{1}{2}\sum_n(\hat{O}_n)^2\lambda_n\Delta\tau}=\prod_ne^{-\frac{1}{2}(\hat{O}_n)^2\lambda_n\Delta\tau}+\mathcal{O}(\Delta\tau^2)\,.
\label{eq:prop_diag}
\end{equation}
With the symbol $\hat{O}_n$ we denote the $3A$ $\hat{O}^{\sigma}_n$, the $9A$ $\hat{O}^{\sigma\tau}_n$ and the $3A$ $\hat{O}^{\tau}_n$.

It is now possible to use the Hubbard-Stratonovich transformation, that for a generic operator $\hat{O}$ and a parameter $\lambda$ is defined by
\begin{equation}
e^{-\frac{1}{2}\lambda \hat O^2}=\frac{1}{\sqrt{2\pi}}\int dx e^{-\frac{x^2}{2}+\sqrt{-\lambda}x\hat O}\, .
\end{equation}
The propagator of Eq. (\ref{eq:prop_diag}) can be recasted in the following form
\begin{equation}
e^{-\frac{1}{2}\sum_n(\hat{O}_n)^2\lambda_n\Delta\tau}=\prod_n\int dx_n e^{-\frac{x_{n}^2}{2}+\sqrt{-\lambda}x_n\hat{O}_n}\, +\mathcal{O}(\Delta\tau^2)\,.
\label{eq:prop_hs}
\end{equation}

We define a walker to be the $3A$ spatial coordinates and $\nu A$ spinors, $c_{i}^\alpha$. Hence, within the AFDMC approach, the spin-isospin coordinate $S$ has to be added to the spatial coordinate $R$. The imaginary time evolution (compare with Eq. (\ref{eq:gmat_ev}) and Eq. (\ref{eq:gfmc_ev})) reads
\begin{equation}
\Psi(R,S,\tau+\Delta\tau)=\int dR'dS' G(R,S,R',S',\Delta\tau) \Psi(R',S',\tau)\, ,
\end{equation}
where the AFDMC Green's function, which includes the integration over the {\it auxiliary fields}, is given by
\begin{align}
G(R,S,R',S',\Delta\tau)=&\Big(\frac{m}{2\pi\hbar^2\Delta\tau}\Big)^\frac{3A}{2} e^{-\frac{m(R-R')^2}{2\hbar^2\Delta\tau}} e^{-V_{SI}(R')\Delta\tau}e^{E_T\Delta\tau}\times\nonumber \\
&\langle S |\prod_{n=1}^{15A}\frac{1}{\sqrt{2\pi}}\int dx_n e^{-\frac{x^{2}_n}{2}+\sqrt{-\lambda_n\tau}x_n\hat{O}_n}|S'\rangle \, .\, .
\label{eq:green_standard_afdmc}
\end{align}
An important point to make is that the operator $\hat{O}_n$ contains a sum over particle index $j$, as can be seen from Eq. (\ref{eq:O_definition}). However, these operators commute and we can conveniently represent the exponential of the sum as a product of exponentials, each rotating only one single-particle state. Therefore, the application of the operator $\hat{O}^{p}_n$ to the spin-isospin state $|S'\rangle$, turns into a product of independent rotation. For the spin rotation, generated by $\hat{O}^{\sigma}_n$, one has
\begin{align}
&e^{\sqrt{-\lambda_n\Delta\tau}x_n\hat{O}^{(\sigma)}_n}|S'\rangle=\nonumber \\
&e^{\sqrt{-\lambda_n\Delta\tau}x_n\vec{\sigma}_1\cdot{\vec{\psi}}^{\,(\sigma)}_{n,1}}\Big[c_{1}^\uparrow |\uparrow\rangle_1+c_{1}^\downarrow|\downarrow\rangle_1\Big]\otimes\dots\otimes e^{\sqrt{-\lambda_n\tau}x_n\vec{\sigma}_A\cdot{\vec{\psi}}^{\,(\sigma)}_{n,A}}\Big[c_{A}^\uparrow |\uparrow\rangle_A+c_{A}^\downarrow|\downarrow\rangle_A\Big]\, .
\end{align}

Rotating the $j-th$ single particle state amounts in a change of the coefficients 
\begin{equation}
e^{\sqrt{-\lambda_n\Delta\tau}x_{n}\vec{\sigma}_j\cdot{\vec{\psi}}^{\,\sigma}_{n,j}}\Big[c_{j}^\uparrow |\uparrow\rangle_1+c_{j}^\downarrow|\downarrow\rangle_1\Big]=\Big[c_{j}^{\prime\,\uparrow} |\uparrow\rangle_1+c_{j}^{\prime\,\downarrow}|\downarrow\rangle_1\Big]\, .
\label{eq:rotazione1}
\end{equation}
In order to give an explicit expression to $c_{j}^{\prime\,\uparrow}$ and $c_{j}^{\prime\,\downarrow}$, the following identities have to be used
\begin{align}
e^{i(\vec{a}\cdot\vec{\sigma})}&=\cos(|\vec{a}|)+i\frac{\vec{a}\cdot{\vec{\sigma}}}{|\vec{a}|}\sin(|\vec{a}|)\, 
\label{eq:rot_ll0}
\end{align}
\begin{align}
e^{(\vec{a}\cdot\vec{\sigma})}&=\cosh(|\vec{a}|)+\frac{\vec{a}\cdot{\vec{\sigma}}}{|\vec{a}|}\sinh(|\vec{a}|)\, .
\label{eq:rot_lg0}
\end{align}
where in our case the vector $\vec{a}$ is given by (we omit the index $\sigma$ for brevity)
\begin{equation}
\vec{a}=\sqrt{|\lambda_n|\Delta\tau}x_n\vec{\psi}_{n,j}\, .
\end{equation}
When $\lambda_n<0$, exploiting Eq. (\ref{eq:rot_ll0}), we find
\begin{align}
c_{j}^{\prime\,\uparrow}&=c_{j}^{\uparrow}\Big[\cosh(|\vec{a}|)+\frac{\psi_{n,j}^z}{|\vec{\psi}_{n,j}|}\sinh(|\vec{a}|)\Big]+c_{j}^{\downarrow}\sinh(|\vec{a}|)\Big(\frac{\psi_{n,j}^x-i\psi_{n,j}^y}{|\vec{\psi}_{n,j}|}\Big)\\
c_{j}^{\prime\,\downarrow}&=\beta\Big[\cosh(|\vec{a}|)+\frac{\psi_{n,j}^z}{|\vec{\psi}_{n,j}|}\sinh(|\vec{a}|)\Big]+\alpha\sinh(|\vec{a}|)\Big(\frac{\psi_{n,j}^x+i\psi_{n,j}^y}{|\vec{\psi}_{n,j}|}\Big)
\end{align}

On the other hand, if $\lambda>0$, making use of Eq. (\ref{eq:rot_lg0}) the transformed coefficients read \begin{align}
c_{j}^{\prime\,\uparrow}&=c_{j}^{\uparrow}\Big[\cos(|\vec{a}|)+i\frac{\psi_{n,j}^z}{|\vec{\psi}_{n,j}|}\sin(|\vec{a}|)\Big]+c_{j}^{\downarrow}\sin(|\vec{a}|)\Big[\frac{i\psi_{n,j}^x+\psi_{n,j}^y}{|\vec{\psi}_{n,j}|}\Big]\\
c_{j}^{\prime\,\downarrow}&=c_{j}^{\downarrow}\Big[\cos(|\vec{a}|)+i\frac{\psi_{n,j}^z}{|\vec{\psi}_{n,j}|}\sin(|\vec{a}|)\Big]+c_{j}^{\uparrow}\sin(|\vec{a}|)\Big[\frac{i\psi_{n,j}^x-\psi_{n,j}^y}{|\vec{\psi}_{n,j}|}\Big]
\end{align}

Note that if the integral over the auxiliary fields was computed using the standard methods, like the Simpson rule, we would be left with a sum of rotated spinors, one for each sampled value of the auxiliary fields $x_n$. 

In the first realizations of the AFDMC algorithm a discrete version of the Hubbard-Stratonovich transformation, due to Kooning, was implemented. It essentially consists in replacing the Gaussian by a three point weighted sum. With a probability depending on the weight, only one over these three values has to be used for rotating the spinors.

In more recent works, following the spirit of the Monte Carlo algorithm, the Gaussian has been considered as a probability distribution. One value is sampled directly from the Gaussian and used to rotate the spin-isospin degrees of freedom of the walkers.  

A physical interpretation can be (and has been) given to the auxiliary fields. As can be consistently explained in chiral perturbation theory, nuclear interactions can be explained in terms of pion exchanges. Within the Born-Oppenheimer approximation, the light pions are the fast degrees of freedom, coupled with the slow and more massive nucleons. If the fast meson field is integrated out to give a potential, keeping the nucleonic coordinates fixed,  and solving for its ground-state energy, then meson coordinate corresponds to the Hubbard-Stratonovich auxiliary fields.

Importance sampling can be implemented to the integral over the auxiliary fields. The overlap of the walker with $\Psi_T$ is not generally picked around $x_n=0$. Hence, instead of sampling from the Gaussian, it is more efficient to sample values of $x_n$ where the trial wave-function is thought to be large. One way consists in shifting the Gaussian, introducing a drift term analogous to the one used for the spatial coordinates. For the detailed calculations of the drift term for the Hubbard-Stratonovich variables, the reader is referred to Refs. \cite{sarsa_03,ormoni_11,gandolfi_07}.

The agreement  between Green Function Monte Carlo (GFMC) and AFDMC energies of neutron drops, obtained using the Argonne $v_{8}^\prime$ plus UIX hamiltonian, discussed in Ref.~\cite{gandolfi_11},  supports the validity of PNM calculations carried out within AFDMC  with the Argonne  $v_{8}^\prime$ model. The highly accurate GFMC method has been used to study neutron matter properties in both the normal \cite{carlson_03}  and superfluid \cite{gezerlis_10} phases.  

Moreover, using a fixed-phase like approximation, AFDMC also yields results in very good agreement with those obtained from Green Function Monte Carlo (GFMC) calculations for light nuclei \cite{gandolfi_07b}.

\newpage             
\thispagestyle{empty}       

% Chapter 3 %%%%%%%%%%%%%%%%%%%%%%%%%%%%%%
\chapter{Three-body potential in nuclear matter}
\label{chapt:tbp}
\section{UIX potential within FHNC/SOC approach}
\label{sec:fhncsoc_tbp}
Within CBF approach, the expectation value of a three-body potential, e.g. the UIX model, reads
\begin{equation}
\langle V \rangle =\frac{A!}{(A-3)!3!}\frac{\langle \Psi^\dagger_0 |\hat{\mathcal{F}}^\dagger \hat{V}_{123}\hat{\mathcal{F}}| \Psi_0 \rangle}
{\langle\Psi^\dagger_0|\hat{\mathcal{F}}^\dagger \hat{\mathcal{F}}|\Psi_0\rangle}\, .
\label{eq:exp_three}
\end{equation}

Let us write $\hat{V}_{123}$ as a sum of spin-isospin three--body operators multiplied by scalar functions, depending on the relative distances only
 \begin{equation}
\label{expand:v123}
V_{123} \equiv \sum_P V_{123}^P \hat{O}_{123}^P\, \, .
\end{equation}

As for the case of the two--body distribution functions $g^p_{12}$, it is useful to define three--body distribution functions $g^p_{123}$, reflecting the operatorial structure of $\hat{V}_{123}$
\begin{equation}
g_{123}^P=\frac{A!}{(A-3)!}\frac{\text{CTr}_{123}\int dx_4 \ldots dx_A \Psi^\dagger_0(X)\mathcal{F}^\dagger \hat{O}_{123}^P
\mathcal{F} \Psi_0(X)}{ \rho^3\int dX \  \Psi^\dagger_0(X)\mathcal{F}^\dagger \mathcal{F} \Psi_0(X)}\, .
\label{eq:g3_def}
\end{equation}

Hence, analogously to Eq. (\ref{eq:tbp_ev}) relative to the two-body potential, the expectation value of $\hat{V}_{123}$ can be written as 
 \begin{equation}
\frac{\langle V\rangle}{A}=\frac{\rho^2}{3!} \sum_P \int d\mathbf{r}_{12}d\mathbf{r}_{13} V_{123}^P\,g_{123}^P \, .
\label{eq:3_body_exp_with_g3}
\end{equation}

Neglecting the Abe diagrams, as in Eq. (\ref{eq:g3_scalar_def}), the three-body distribution function can be approximated by a product of the partial two-body distribution function of Eq. (\ref{eq:distribution_op}) and (\ref{eq:distribution_op_cc}), denoted by $Z^{p}_{xy}$ in Ref.~\cite{carlson_83}, 
\begin{align}
g_{123}^P=&\sum_{ex}\sum_{p',p'',p'''}g_{xx'}^{p'}(r_{12})g_{yy'}^{p''}(r_{13})g_{zz'}^{p'''}(r_{23})\times\nonumber\\
& \text{CTr}_{123}\frac{1}{3!}\Big[\hat{O}_{123}^P(O_{12}^{p'}\{O_{13}^{p''},O_{23}^{p'''}\}+ O_{13}^{p''}\{O_{12}^{p'},O_{23}^{p'''}\}+ O_{23}^{p'''}\{O_{12}^{p'},O_{13}^{p''}\})\Big]
\label{eq:tbdf_op}
\end{align}
It can be noted from the previous equation, that the FHNC/SOC approximation has been exploited as at most two-operators arrive at a given point. Moreover, for the sake of brevity, vertex corrections arising from separable diagrams have not been reported. 

\begin{figure}[!ht]
\begin{center}
\fcolorbox{white}{white}{
  \begin{picture}(400,200)(-120,-125)
  	\SetScale{0.9}	
        \unitlength=0.9 pt
	\SetWidth{1.4}
	\SetColor{Black}

	\Line(-10,0)(20,60)
	\Line(50,0)(20,60)
	\SetWidth{0.5}
	\DashCArc(58,3.43)(69,124,182){4}
	\DashCArc(-18,3.43)(69,-3,56){4}
	\PhotonArc(20,55)(63,-120,-60){1}{7}
	\BCirc(-10,0){2}
	\BCirc(20,60){2}
	\BCirc(50,0){2}
	\Text(-12,-10)[]{2}
	\Text(52,-10)[]{1}
	\Text(20,70)[]{3}
	\Text(20,-25)[]{(a)}

	\SetWidth{1.4}
	\SetColor{Black}
	\Line(125,0)(155,60)
	\Line(185,0)(155,60)
	\SetWidth{0.5}
	\PhotonArc(112,1)(74,0,55){1}{8}
	\PhotonArc(198,1)(74,125,180){1}{8}
	\DashCArc(155,55)(63,-120,-60){4}
        \BCirc(125,0){2}
	\BCirc(155,60){2}
	\BCirc(185,0){2}
	\Text(123,-10)[]{2}
	\Text(187,-10)[]{1}
	\Text(155,70)[]{3}
	\Text(155,-25)[]{(b)}

	\SetWidth{1.4}
	\SetColor{Black}
	\Line(-10,-120)(20,-60)
	\Line(50,-120)(20,-60)
	\SetWidth{0.5}
	\PhotonArc(20,-65)(63,-120,-60){1}{7}
	\PhotonArc(63,-119)(74,125,180){1}{8}
	\DashCArc(-18,-117.57)(69,-3,56){4}
	\BCirc(-10,-120){2}
	\BCirc(20,-60){2}
	\BCirc(50,-120){2}
	\Text(-12,-130)[]{2}
	\Text(52,-130)[]{1}
	\Text(20,-50)[]{3}
	\Text(-17,-82)[]{{\large 2$\times$}}
	\Text(20,-145)[]{(c)}

        \SetWidth{1.4}
	\SetColor{Black}
	\Line(125,-120)(155,-60)
	\Line(185,-120)(155,-60)
	\SetWidth{0.5}
	\PhotonArc(155,-65)(63,-120,-65){1}{7}
	\PhotonArc(112,-119)(74,0,55){1}{8}
	\PhotonArc(198,-119)(74,125,180){1}{8}
	\BCirc(125,-120){2}
	\BCirc(155,-60){2}
	\BCirc(185,-120){2}
	\Text(123,-130)[]{2}
	\Text(187,-130)[]{1}
	\Text(155,-50)[]{3}
	\Text(155,-145)[]{(d)}

\end{picture}
}
\vspace{0.3cm}
\caption{Cluster diagrams contributing to the expectation value of  $V^{2\pi}$.\label{fig:tbp_2pi}}

\end{center}
\end{figure}
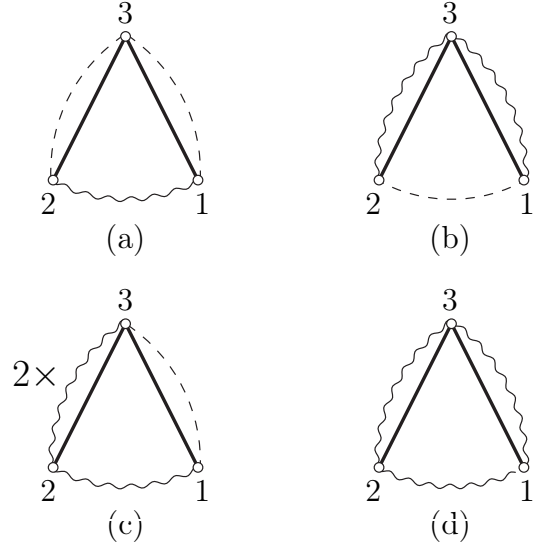

The diagrams involved in the FHNC/SOC calculation of the expectation values of the $V^{2\pi}$ term of the UIX potential are depicted in Fig. \ref{fig:tbp_2pi}. The thick lines represent the potential, while dashed and wavy lines correspond to the partial two-body distribution functions; vertex corrections, although included in the calculations, are not shown. Because of the symmetry properties of the wave function, we can restrict our analysis to the permutation $(3:12)$ . The other three permutations are accounted for in the symmetry factor appearing in front of each of the following expressions
\begin{align}
\label{eq:tbp_2pi.a}
(\ref{fig:tbp_2pi}.a)=&\frac{\rho^2}{2}\sum_{ex}\sum_{p'}\int d^3r_{12}d^3r_{13}g_{x''y'',12}^{p'}\,g_{xy,13}^{c}\,g_{x'y',23}^{c}\text{CTr}\Big[\hat{V}^{2\pi}(3:12)\hat{O}_{12}^{p}\Big]\\ 
\label{eq:tbp_2pi.b}
(\ref{fig:tbp_2pi}.b)=&\frac{\rho^2}{4}\sum_{ex}\sum_{p'',p'''}\int d^3r_{12}d^3r_{13}\,g_{x''y'',12}^{c}\,g_{xy,13}^{p''}\,g_{x'y',23}^{p'''}\text{CTr}\Big[\hat{V}^{2\pi}(3:12)\{\hat{O}_{13}^{p},\hat{O}_{23}^{p'}\}\Big]\, \\
\label{eq:tbp_2pi.c}
(\ref{fig:tbp_2pi}.c)=&\frac{\rho^2}{2}\sum_{ex}\sum_{p',p'''}\int d^3r_{12}d^3r_{13}\,g_{xy,12}^{p'}\,g_{x'y',13}^{c} \,g_{x''y'',23}^{p'''}\text{CTr}\Big[\hat{V}^{2\pi}(3:12)\{\hat{O}_{12}^{p},\hat{O}_{23}^{p''}\}\Big]\, \\
\label{eq:tbp_2pi.d}
(\ref{fig:tbp_2pi}.d)=&\frac{\rho^2}{12}\sum_{ex}\sum_{p',p'',p'''}\int d^3r_{12}d^3r_{13}\,g_{xy,12}^{p'}\,g_{x'y',13}^{p''} \,g_{x''y'',23}^{p'''}\times\nonumber\\
&\text{CTr}\Big[\hat{V}^{2\pi}(3:12)(O_{12}^{p}\{\hat{O}_{13}^{p'},\hat{O}_{23}^{p''}\}+ \hat{O}_{13}^{p'}\{\hat{O}_{12}^{p},\hat{O}_{23}^{p''}\}+ \hat{O}_{23}^{p''}\{\hat{O}_{12}^{p},\hat{O}_{13}^{p'}\})\Big]\,.
\end{align}

\begin{figure}[!ht]
\begin{center}
\fcolorbox{white}{white}{
  \begin{picture}(400,200)(-120,-125)
    	\SetScale{0.9}	
        \unitlength=0.9 pt
	\SetWidth{1.4}
	\SetColor{Black}
	\Line(-10,0)(20,60)
	\Line(50,0)(20,60)
	\SetWidth{0.5}
	\DashCArc(58,3.43)(69,124,182){4}
	\DashCArc(-18,3.43)(69,-3,56){4}
	\DashCArc(20,55)(63,-120,-60){4}
	\BCirc(-10,0){2}
	\BCirc(20,60){2}
	\BCirc(50,0){2}
	\Text(-12,-10)[]{2}
	\Text(52,-10)[]{1}
	\Text(20,70)[]{3}
	\Text(20,-25)[]{(a)}

	\SetWidth{1.4}
	\SetColor{Black}
	\Line(125,0)(155,60)
	\Line(185,0)(155,60)
	\SetWidth{0.5}
	\DashCArc(193,3.43)(69,124,182){4}
	\DashCArc(117,3.43)(69,-3,56){4}
	\PhotonArc(155,55)(63,-120,-60){1}{7}
	\PhotonArc(155,28)(43,-137,-43){1}{7}
	\BCirc(125,0){2}
	\BCirc(155,60){2}
	\BCirc(185,0){2}
	\Text(123,-10)[]{2}
	\Text(187,-10)[]{1}
	\Text(155,70)[]{3}
	\Text(155,-25)[]{(b)}

	\SetWidth{1.4}
	\SetColor{Black}
	\Line(-10,-120)(20,-60)
	\Line(50,-120)(20,-60)
	\SetWidth{0.5}
	\DashCArc(58,-117.43)(69,124,182){4}
	\DashCArc(20,-65)(63,-120,-60){4}
	\PhotonArc(-23,-119)(74,0,55){1}{8}
	\PhotonArc(10,-102.5)(45,-21,76){1}{8}
	\BCirc(-10,-120){2}
	\BCirc(20,-60){2}
	\BCirc(50,-120){2}
	\Text(-12,-130)[]{2}
	\Text(52,-130)[]{1}
	\Text(20,-50)[]{3}
	\Text(-15,-82)[]{{\large 2$\times$}}
	\Text(20,-145)[]{(c)}

        \SetWidth{1.4}
	\SetColor{Black}
	\Line(125,-120)(155,-60)
	\Line(185,-120)(155,-60)
	\SetWidth{0.5}
	\PhotonArc(155,-65)(63,-120,-60){1}{7}
	\PhotonArc(112,-119)(74,0,55){1}{8}
	\PhotonArc(198,-119)(74,125,180){1}{8}
	\BCirc(125,-120){2}
	\BCirc(155,-60){2}
	\BCirc(185,-120){2}
	\Text(123,-130)[]{2}
	\Text(187,-130)[]{1}
	\Text(155,-50)[]{3}
	\Text(155,-145)[]{(d)}

\end{picture}
}
\vspace{0.3cm}
\caption{Same as in Fig. \ref{fig:tbp_2pi}, but for  $V^{R}$.}
\label{fig:tbp_scalar}
\end{center}
\end{figure}
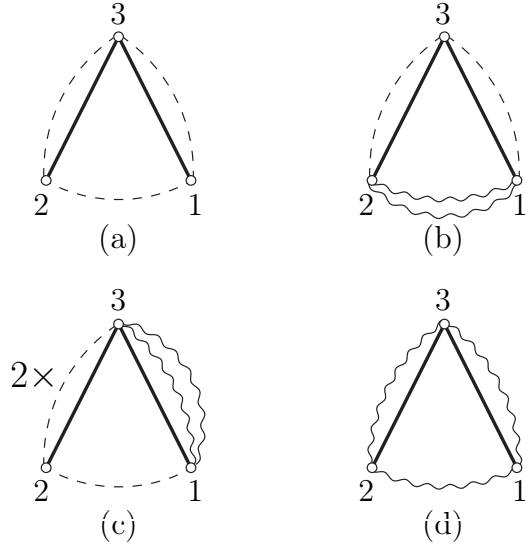

The computation of all of diagrams (\ref{fig:tbp_scalar}.a), (\ref{fig:tbp_scalar}.b) and (\ref{fig:tbp_scalar}.c) and all diagrams of Fig. \ref{fig:tbp_scalar} is outlined in Ref.~\cite{carlson_83},  while the contribution of diagram (3.d), involving three non central correlations was first taken into account by the authors of Ref.~\cite{wiringa_88}.

In addition to the what listed in Eq. (\ref{eq:tbdf_op}), SOR contributions to the three-body distribution function, represented by double wavy lines in Fig. \ref{fig:tbp_scalar} and denoted by $Y_{xy}$, are included in the calculation of the repulsive term of the UIX potential

\begin{align}
\label{eq:tbp_scalar.a}
(\ref{fig:tbp_scalar}.a)=&\frac{\rho^2}{2}\sum_{ex}\int d^3r_{12}d^3r_{13}g_{x''y'',12}^{c}\,g_{xy,13}^{c}\,g_{x'y',23}^{c}V^{R}(3:12)\\ 
\label{eq:tbp_scalar.b}
(\ref{fig:tbp_scalar}.b)=&\frac{\rho^2}{2}\sum_{ex}\int d^3r_{12}d^3r_{13}\,Y_{x''y'',12}\,g_{xy,13}^{c}\,g_{x'y',23}^{c}V^{R}(3:12)\, \\
\label{eq:tbp_scalar.c}
(\ref{fig:tbp_scalar}.c)=&\rho^2\sum_{ex}\int d^3r_{12}d^3r_{13}\,g_{xy,12}^{c}\,Y_{x'y',13}\,g_{x''y'',23}^{c}V^{R}(3:12)\, \\
\label{eq:tbp_scalar.d}
(\ref{fig:tbp_scalar}.d)=&\frac{\rho^2}{2}\sum_{ex}\sum_{p',p'',p'''}\int d^3r_{12}d^3r_{13}\,g_{xy,12}^{p'}\,g_{x'y',13}^{p''} \,g_{x''y'',23}^{p'''}\,\xi_{132}^{p''p'''p'}A^{p'}V^{R}(3:12)\,.
\end{align}

\section{Density dependent effective potential}
\label{sec:ddp}
As shown in Fig. \ref{fig:pieper_nucl_UIX}, the inclusion of the UIX three-body potential in the hamiltonian considerably improves the theoretical estimates  of the energies of the ground and low-lying excited states of nuclei with $A\leq12$ . However, for nuclei heavier than $^3H$, some discrepancies with experimental data persist; moreover the empirical equilibrium properties of nuclear matter are not correctly reproduced. This problem can be largely ascribed to the uncertainties associated with the description of three-nucleon interactions, whose contribution turns out to be significant.

\subsection{Derivation of the effective potential}
\label{ddpot}
\begin{figure}[!ht]
\begin{center}
\fcolorbox{white}{white}{
  \begin{picture}(150,150)(20,-40)
	\SetWidth{0.5}
	\SetColor{Black}
	\SetScale{0.9}	
        \unitlength=0.9 pt
	
	\ArrowArc(40,27.86)(55,100,197)
	\ArrowArc(20,27.86)(55,-20,80)
	\ArrowArc(30,44.55)(55,220,320)
	\COval(30,10)(1.5,38)(0){Black}{Black}
	\COval(10,45)(1.5,38)(60){Black}{Black}
	\COval(50,45)(1.5,38)(-60){Black}{Black}
	\CCirc(-8,55){10}{Black}{Black}
	\CCirc(68,55){10}{Black}{Black}
	\CCirc(30,-10){10}{Black}{Black}
	\Vertex(30,82){3}
	\BCirc(-11,10){3}
	\BCirc(71,10){3}
	\Text(-13,2)[]{1}
	\Text(72,2)[]{2}
	\Text(30,91)[]{3}

	\Text(105,30)[]{{\begin{Large}$\Rightarrow$\end{Large}}}

	\COval(175,10)(1.5,38)(0){Black}{Black}
	\ArrowArc(175,10)(42,0,180)
	\CCirc(175,52){10}{Black}{Black}
	\BCirc(134,10){3}
	\BCirc(216,10){3}
	\Text(132,2)[]{1}
	\Text(217,2)[]{2}

  \end{picture}
}
\vspace{0.1cm}
\caption{Diagrammatic representation of Eq. (\ref{eq:ddp_request}): the two-body density-dependent potential includes the effects of both the bare three-body potential and the correlation and exchange lines. While $g_2$ dresses the line joining particles $1$ and $2$, the dressing being depicted by a line with a big bubble in the middle, $g_3$ dresses the lines $1-2$, $1-3$, and $2-3$. }
\label{fig:g3_g2}
\end{center}
\end{figure}
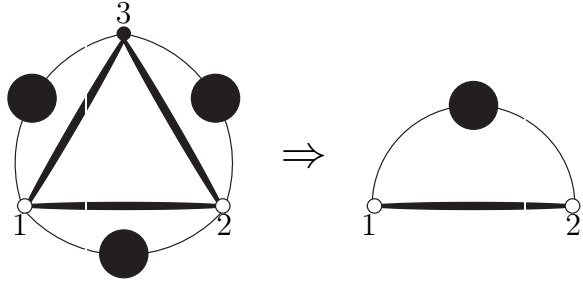

Our work \cite{lovato_11} is aimed at obtaining a two-body density-dependent potential $\hat{v}_{12}(\rho)$ that mimics the three-body 
potential. Hence, our starting point is the requirement that the expectation values of $V_{123}$ and of $\hat{v}_{12}(\rho)$
be the same:
\begin{equation}
\frac{\langle \hat{V} \rangle}{A}=\frac{\langle \hat{v}(\rho)\rangle}{A}\, ,
\end{equation}
implying in turn (compare to  Eqs.(\ref{eq:tbp_ev}) and (\ref{eq:3_body_exp_with_g3})) 
\begin{equation}
\sum_{P}\frac{\rho}{3}\int d\mathbf{r}_3 V_{123}^P\,g_{123}^P=\sum_p v_{12}^{p}(\rho)\,g_{12}^p\, .
\label{eq:ddp_request}
\end{equation}
A diagrammatic representation of the above equation,  which should be regarded as the definition of  the $\hat{v}_{12}(\rho)$, 
is shown in Fig. \ref{fig:g3_g2}. The graph on the left-hand side represents the three-body potential times the three-body correlation function, integrated over the coordinates of particle $3$. Correlation and exchange lines are schematically depicted with a line having a bubble in the middle, while the thick solid lines represent the three-body potential. The diagram in the right-hand side represents the density-dependent two-body potential, dressed with the two-body distribution function. 
Obviously, $v_{12}^\rho$ has to include not only the three-body potential, but also the effects of correlation and 
exchange lines.  

In Section \ref{sec:fhncsoc_tbp} we have examined the left-hand side of Eq.(\ref{eq:ddp_request}), that has been evaluated in \cite{carlson_83} within the FHNC/SOC  scheme. Here we discuss the derivation of the explicit expression of the two-body density-dependent potential appearing in the right-hand side of the equation. The procedure consists of three different step, each corresponding to a different dressing of the diagrams involved in the calculation 

For each of these steps the final result is a density-dependent two-body potential of the form 
\begin{equation}
\hat{v}_{12}(\rho)=\sum_{p} v^p(\rho,r_{12})\hat{O}^{p}_{12}\, ,
\label{eq:ddp_potential}
\end{equation}
where, depending on the step, the $v^p(\rho,r_{12})\equiv v_{12}^p(\rho)$ can be expressed in terms of the functions appearing in the definition of the UIX potential, the correlation functions and of the Slater functions.

\subsubsection{Step I. Bare approximation}%: $\mathbf{g_{123}=g_{12}}$}
As a first step in the derivation of the density-dependent potential one integrates the three-body potential over the coordinate of the third particle
\begin{equation}
\hat{v}_{12}^{\,(I)}(\rho)=\frac{\rho}{3}\int d{x}_3 \hat{V}_{123}\,.
\label{eq:bare_ddp}
\end{equation}
Diagrammatically the above equation implies that neither interaction nor exchange lines linking particle $3$ with particles $1$ and $2$ are included. Only the two-body distribution function is taken into account in the calculation of the expectation value of $V_{123}$
\begin{equation}
\frac{\langle \hat{V} \rangle}{A}=\frac{\rho^2}{3!}\sum_p\int d\mathbf{r}_{12}\Big(\sum_P \int d{x}_3 V_{123}^P\Big)^p g_{12}^p\,.
\end{equation}
Note that only the scalar repulsive term and one permutation of the anticommutator term of the three-body potential provide non vanishing contributions, once the 
trace in the spin--isospin space of the third particle is performed. 

As shown in Fig \ref{fig:compare_potentials}, the contribution of the density-dependent potential to the energy per particle of SNM and PNM $\langle \hat{v}_{12}^{\,(I)}(\rho)\rangle/A$ is more repulsive than the one obtained from the genuine three-body potential UIX. Thus, the scalar repulsive term is dominant when the three-body potential is integrated over particle $3$.

\subsubsection{Step II. Inclusion of statistical correlations}

As a second step we have considered the exchange lines that are present both in $g_{123}$ and $g_{12}$. Their treatment  is somewhat complex, and needs to be analyzed in detail. 

\begin{figure}[!h]
\begin{center}
\fcolorbox{white}{white}{
  \begin{picture}(150,90)(-40,0)
	\SetWidth{0.5}
	\SetColor{Black}
	\SetScale{0.9}	
        \unitlength=0.9 pt
	\ArrowArc(85,7.86)(85,122,178)
	\ArrowArc(-5,7.86)(85,0,60)
	\ArrowArc(40,85)(85,-120,-60)
	\Vertex(40,80){3}
	\BCirc(-1,10){3}
	\BCirc(81,10){3}
	\Text(-2,2)[]{1}
	\Text(82,2)[]{2}
	\Text(40,88)[]{3}
  \end{picture}
}
\vspace{0.6cm}
\caption{Three particle exchange loop.}
\label{fig:3_particle_exchange}
\end{center}
\end{figure}
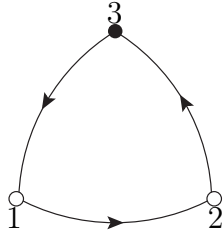

Consider, for example, the diagram associated with the exchange loop involving particles $1$, $2$ and $3$, 
depicted in Fig. \ref{fig:3_particle_exchange}. 
Its inclusion in the calculation of the density-dependent two-body potential would lead to double counting of exchange lines connecting particles 1 and 2, due to the presence of the  exchange operator $\hat{P}_{12}$ in $g_{12}$. 
This problem can be circumvented by noting that the antisymmetrization operator acting on particles $1$, $2$ and $3$ can be written in the form
\begin{align}
&1-\hat{P}_{12}-\hat{P}_{13}-\hat{P}_{23}+\hat{P}_{12}\hat{P}_{13}+\hat{P}_{13}\hat{P}_{23}=\nonumber\\
&\qquad(1-\hat{P}_{13}-\hat{P}_{23})\times(1-\hat{P}_{12})\, ,
\label{eq:exchange_fantoni}
\end{align}
in which the exchange operators contributing to the density-dependent potential only appear in the first term of the 
right-hand side. %They are shown in the the second diagram of Fig. \ref{fig:relevant_diagrams}, the factor $2$ being a 
%symmetry factor due to the fact that $v_{123}$ is symmetric under permutation of particles $1$ and $2$. 

On the other hand, the second term in the right-hand side of Eq. (\ref{eq:exchange_fantoni}) only involves the exchange 
operators $\hat{P}_{12}$, whose contribution is included in $g_{12}$ and must not be taken into account in the calculation 
of $\hat{v}_{12}(\rho)$. 

Two features of the above procedure need to be clarified. First, it has to be pointed out that it is exact only within the SOC approximation that allows one to avoid the calculation of commutators between the exchange operators $\hat{P}_{13}$ and $\hat{P}_{23}$ and the correlation operators acting on particles $1$ and $2$. The second issue is related to the 
treatment of the radial part of the exchange operators. Although it is certainly true that one can isolate the trace over the spin-isospin degrees of freedom of  particle 3, arising from $\hat{P}_{13}$ and $\hat{P}_{23}$, extracting the Slater functions from these 
operators is only possible in the absence of functions depending on the position of particle 3 \cite{pandha_73}.

However, this restriction does not apply to the case under consideration, as both the potential and the correlations depend on $\mathbf{r}_{13}$ and $\mathbf{r}_{23}$. As a consequence, retaining only the $\hat{P}_{13}$ and $\hat{P}_{23}$ exchange operators involves an approximation in the treatment of the the Slater functions, whose validity has been tested by carrying out a numerical calculation.

By singling out  the radial dependence of the exchange operators, and by computing the inverse of the operator $(1-\hat{P}^{\sigma\tau}_{12})$, where $\hat{P}^{\sigma\tau}_{ij}$ denotes the spin-isospin part of $\hat{P}_{ij}$, it is possible to find a ``Slater Exact'' density-dependent potential $v_{12}^{S.E.}(\rho)$ whose calculation does not involve any approximations concerning the Slater functions. 
It can be easily verified that
\begin{equation}
(1-\hat{P}^{\sigma\tau}_{12}\ell_{12}^2)^{-1}=\frac{1+\hat{P}^{\sigma\tau}_{12}\ell_{12}^2}{1+\ell_{12}^4}\, 
\label{eq:inv_exch}
\end{equation}
thus, the ``Slater Exact'' density-dependent potential can be rewritten in the form
\begin{align}
\hat{v}_{12}^{S.E.}(\rho) &=\frac{\rho}{3}\int d{x}_3 \hat{V}_{123}\Big\{1+\frac{1}{1-\ell_{12}^4}\Big[\hat{P}^{\sigma\tau}_{13}(\ell_{12}^3\ell_{13}\ell_{23}-\ell_{13}^2)+\hat{P}^{\sigma\tau}_{23}(\ell_{12}^3\ell_{13}\ell_{23}-\ell_{23}^2)+\nonumber\\
&\qquad \hat{P}^{\sigma\tau}_{12}\hat{P}^{\sigma\tau}_{23}(\ell_{12}\ell_{13}\ell_{23}-\ell_{12}^2\ell_{13}^2)+\hat{P}^{\sigma\tau}_{13}\hat{P}^{\sigma\tau}_{23}(\ell_{12}\ell_{13}\ell_{23}-\ell_{12}^2\ell_{23}^2)\Big]\Big\}\, ,
\label{eq:exact_exchanges}
\end{align}

%%%%%%%%%%%%%%%%%%%%%%%%%%%%%%%%%%%%%%%%%%%%%%%%%%%%%%%%%%%%%%%%%%%%%%%%%%%
\begin{figure}[!t]
\vspace{0.2cm}
\begin{center}
\includegraphics[angle=270,width=9.5cm]{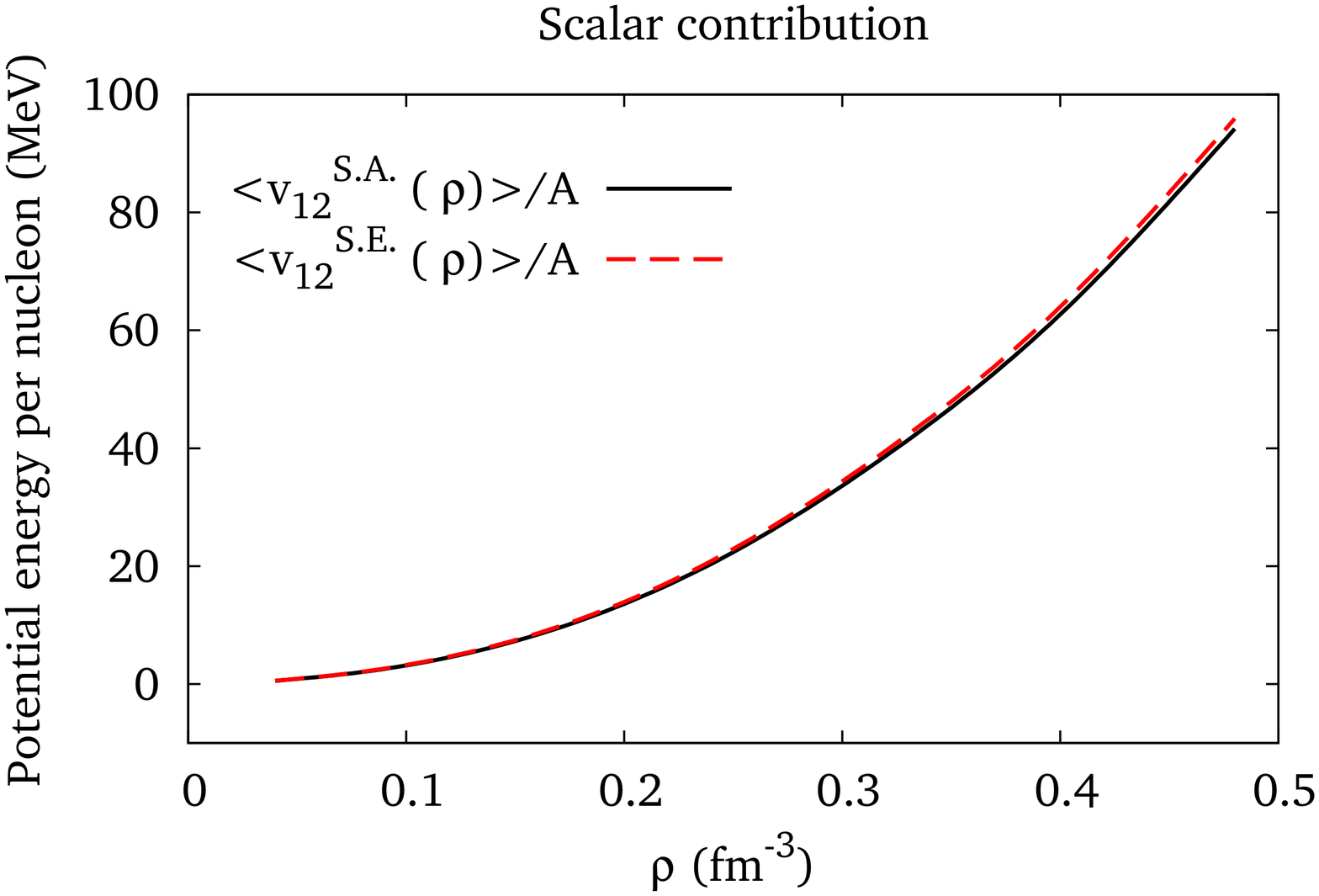}
\vspace{0.1cm}
\includegraphics[angle=270,width=9.5cm]{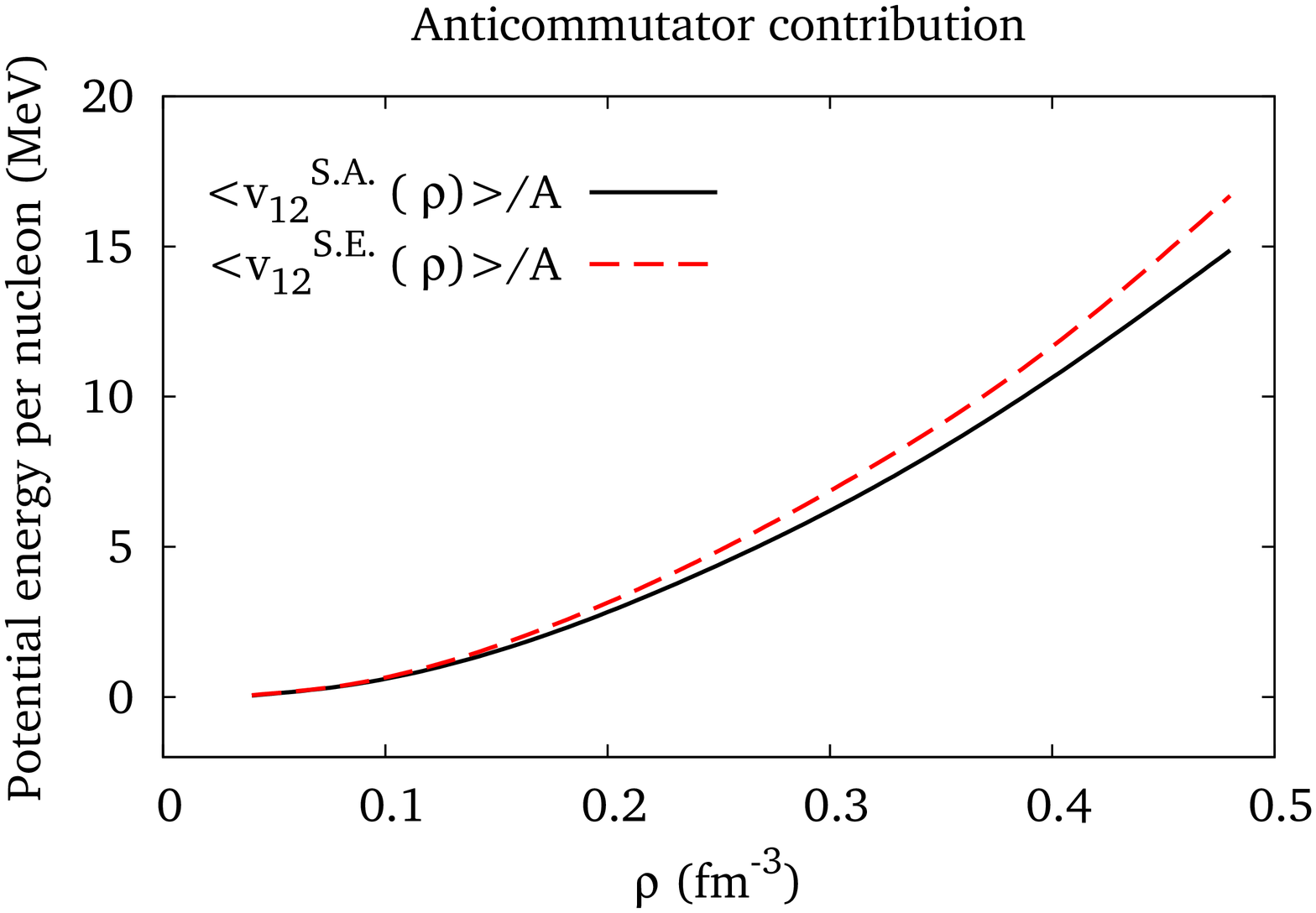}
\caption{Contributions of the density-dependent potential to the energy per particle (see Eqs. (\ref{eq:exact_exchanges}) and (\ref{eq:test_approximate_exchanges})), 
 arising from the scalar term of UIX (upper panel) and from the anticommutator term (lower panel).  \label{fig:exchange_test}}
\end{center}
\end{figure}
%%%%%%%%%%%%%%%%%%%%%%%%%%%%%%%%%%%%%%%%%%%%%%%%%%%%%%%%%%%%%%%%%%%%%%%%%%%
Note that in the above equation we have omitted all correlations functions, whose presence is irrelevant to the purpose of our discussion. The density-dependent potential obtained from Eq.({\ref{eq:exact_exchanges}) must be compared to the one resulting from the approximation discussed 
above, which (again neglecting correlations) leads to the expression
\begin{equation}
\hat{v}_{12}^{S.A.}(\rho)=\frac{\rho}{3}\int d{x}_3 \hat{V}_{123}(1-\hat{P}^{\sigma\tau}_{13}\ell_{13}^3-\hat{P}^{\sigma\tau}_{23}\ell_{23}^2)\, ,
\label{eq:test_approximate_exchanges}
\end{equation}
where ``S. A.'' stands for Slater Approximation.
We have computed $\langle v_{12}^{S.E.}(\rho)\rangle$ and $\langle v_{12}^{S.A.}(\rho)\rangle$ for SNM within the FHNC/SOC scheme, for both the scalar and the anticommutator terms of the UIX potential. 

The results, plotted in Fig. \ref{fig:exchange_test}, clearly show that Eq.(\ref{eq:test_approximate_exchanges}) provides an excellent approximation to the exact result for the exchanges of Eq. (\ref{eq:exact_exchanges}). Hence it has been possible to use Eq. (\ref{eq:test_approximate_exchanges}) also to compute the contribution coming from the commutator of the UIX potential, avoiding the difficulties that would have arisen from an exact calculation of the exchanges. 

The second step in the construction of the density-dependent potential is then 

\begin{equation}
\hat{v}_{12}^{II }(\rho)\equiv \hat{v}_{12}^{S.A.}(\rho)\, 
\end{equation}
which is a generalization of the bare potential of Eq. (\ref{eq:bare_ddp}). 

Figure \ref{fig:compare_potentials} shows that taking exchanges into account slightly improves the approximation of the density-dependent potential. However the differences remain large because correlations have not been taken into account.

\subsubsection{Step III. Inclusion of dynamical correlations}
The third step in the construction of the density-dependent potential amounts to bringing correlations into the game. We have found that the most relevant diagrams are those of Fig. \ref{fig:relevant_diagrams}. 

%%%%%%%%%%%%%%%%%%%%%%%%%%%%%%%%%%%%%%%%%%%%%%%%%%%%%%%%%%%%%%%%%%%%%%%%%%%
\begin{figure}[!ht]
\begin{center}
\fcolorbox{white}{white}{
  \begin{picture}(150,140)(70,-50)
	\SetWidth{0.5}
	\SetColor{Black}
	\SetScale{0.8}	
        \unitlength=0.8 pt
	
	\DashLine(-20,10)(20,80){5}
	\DashLine(60,10)(20,80){5}
	\BCirc(20,80){3}
	\BCirc(-20,10){3}
	\BCirc(60,10){3}
	\GTri(-5,36.25)(-4,47.29)(3,43.29){0}
	\GTri(5,53.75)(-4,47.29)(3,43.29){0}
	\GTri(45,36.25)(44,47.29)(37,43.29){0}
	\GTri(35,53.75)(44,47.29)(37,43.29){0}
	\Text(-22,0)[]{1}
	\Text(62,0)[]{2}
	\Text(20,92)[]{3}

	\DashLine(130,10)(170,80){5}
	\DashLine(210,10)(170,80){5}
	\GTri(145,36.25)(146,47.79)(153,43.29){0}
	\GTri(155,53.75)(146,47.79)(153,43.29){0}
	\GTri(195,36.25)(194,47.79)(187,43.29){0}
	\GTri(185,53.75)(194,47.79)(187,43.29){0}
	\ArrowArc(200,16.43)(71,115,187)
	\ArrowArc(100,73.57)(71,-65,5)
	\BCirc(130,10){3}
	\BCirc(170,80){3}
	\BCirc(210,10){3}
	\Text(128,0)[]{1}
	\Text(212,0)[]{2}
	\Text(170,92)[]{3}
	\Text(115,50)[]{{\large 2$\times$}}

	\DashLine(280,10)(320,80){5}
	\GTri(295,36.25)(296,47.79)(303,43.29){0}
	\GTri(305,53.75)(296,47.79)(303,43.29){0}
	\Photon(360,10)(320,80){1}{8}
	\BCirc(280,10){3}
	\BCirc(320,80){3}
	\BCirc(360,10){3}
	\Text(278,0)[]{1}
	\Text(362,0)[]{2}
	\Text(320,92)[]{3}
\end{picture}
}
\caption{Diagrams contributing to the density-dependent potential. The dashed lines with diamonds represent the first order approximation to $g_{bose}^{NLO}(r_{ij})$, discussed in the text. Only diagrams with at most one operator attached to a given point are taken into account. \label{fig:two_body_bose_corr}}
\label{fig:relevant_diagrams}
\end{center}
\end{figure}
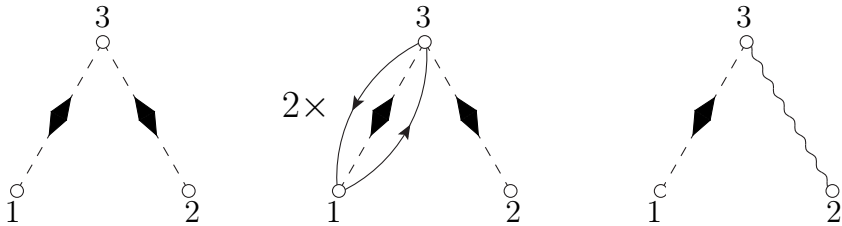
%%%%%%%%%%%%%%%%%%%%%%%%%%%%%%%%%%%%%%%%%%%%%%%%%%%%%%%%%%%%%%%%%%%%%%%%%%%

Note that, in order to simplify the pictures, all interaction lines are omitted. However, it is understood that the three-body potential  is acting on particles $1$, $2$ and $3$. Correlation and exchange lines involving these particles are depicted as if they were passive interaction lines. Moreover, in order to include higher order cluster terms, we have replaced the scalar correlation line ${f^c_{ij}}^2$ with the Next to Leading Order (NLO) %first order
approximation to the bosonic two-body correlation function:
\begin{equation}
{f^{c}_{ij}}^2\rightarrow g_{bose}^{NLO}(r_{ij})={f^{c}_{ij}}^2\Big(1+\rho\int d\mathbf{r}_3 h_{13} h_{23}\Big)\, .
\label{eq:g12_NLO}
\end{equation}
The full bosonic $g_{bose}(r_{ij})$ or $g_{dd}(r_{ij})$ might be used instead of the NLO approximation. However, including higher order terms would have broken our cluster expansion. The correction to ${f^{c}_{ij}}^2$ of Eq. (\ref{eq:g12_NLO}), whose diagrammatic representation is displayed in Fig. \ref{g_2_bose}, can indeed be considered to be of the same order as the operatorial correlations.  

Figure \ref{fig:relevant_diagrams} shows that the vertices corresponding to particles 1 and 2 are
not connected by either correlation or exchange lines. All connections allowed by the diagrammatic rules are 
taken into account multiplying the density-dependent potential by the two-body distribution function, according 
to the definition of Eq.(\ref{eq:ddp_request}).
 
We have already discussed the exchange lines issue, coming to the conclusion that only the exchanges $P_{13}$ and $P_{23}$ have to be taken into account. This is represented by the second diagram, where the factor $2$ is due to the symmetry of the three-body potential, that takes into account both $P_{13}$ and $P_{23}$.  

%%%%%%%%%%%%%%%%%%%%%%%%%%%%%%%%%%%%%%%%%%%%%%%%%%%%%%%%%%%%%%%%%%%%%%%%%%%
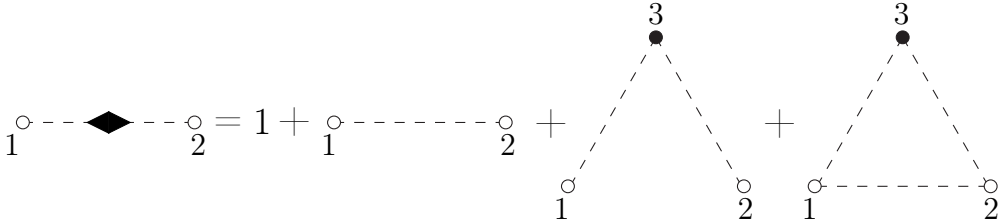
\begin{figure}[!hb]
\begin{center}
\fcolorbox{white}{white}{
  \begin{picture}(150,120)(100,100)
	\SetWidth{0.5}
	\SetColor{Black}
	\SetScale{0.8}	
        \unitlength=0.8 pt
	
	\DashLine(-20,200)(60,200){5}
	\GTri(10,200)(20,205)(20,195){0}
	\GTri(30,200)(20,205)(20,195){0}
	\BCirc(-20,200){3}
	\BCirc(60,200){3}
	\Text(-25,190)[]{1}
	\Text(62,190)[]{2}
	
	\Text(76,200)[]{\large{=}}
	
	\Text(92,201)[]{\large{1}}

	\Text(107,201)[]{\Large{+}}

	\DashLine(125,200)(205,200){5}
	\BCirc(125,200){3}
	\BCirc(205,200){3}
	\Text(123,190)[]{1}
	\Text(207,190)[]{2}
	
	\Text(227,200)[]{\Large{+}}
	
	\DashLine(235,170)(275,240){5}
	\DashLine(315,170)(275,240){5}
	\BCirc(234,170){3}
	\BCirc(316,170){3}
	\CCirc(275,240){3}{Black}{Black}
	\Text(232,160)[]{1}
	\Text(318,160)[]{2}
	\Text(276,252)[]{3}

	\Text(334,200)[]{\Large{+}}
	
	\DashLine(350,170)(430,170){5}
	\DashLine(350,170)(390,240){5}	
	\DashLine(430,170)(390,240){5}	
	\BCirc(349,170){3}
	\BCirc(431,170){3}
	\CCirc(390,240){3}{Black}{Black}
	\Text(348,160)[]{1}
	\Text(433,160)[]{2}
	\Text(391,252)[]{3}

 \end{picture}
}
\caption{NLO approximation to the bosonic two-body correlation function.}
\label{g_2_bose}
\end{center}
\end{figure}
%%%%%%%%%%%%%%%%%%%%%%%%%%%%%%%%%%%%%%%%%%%%%%%%%%%%%%%%%%%%%%%%%%%%%%%%%%%

The explicit expression of $v_{12}^{(III)}(\rho)$ obtained including the diagrams of 
Fig. \ref{fig:relevant_diagrams} can be cast in the form
\begin{align}
\hat{v}^{(III)}_{12}(\rho)&=\frac{\rho}{3}\int d x_3\,\hat{V}_{123}\,\Big[g_{bose}^{NLO}(r_{13}) g_{bose}^{NLO}(r_{23})\nonumber \\
&\times(1-2\hat{P}^{\sigma\tau}_{13}\ell_{13}^2)+4g_{bose}^{NLO}(r_{13})f_{c}(r_{23})\hat{f}(r_{23})\Big]\, ,
\label{eq:ddp_diagrams_expression}
\end{align}
where $\hat{f}(r_{23})$ denotes the sum of non central correlations
\begin{equation}
\hat{f}(r_{23})=\sum_{p\neq 1}^6 f^p(r_{23})\hat{O}^{p}_{ij}\, .
\end{equation}

%%%%%%%%%%%%%%%%%%%%%%%%%%%%%%%%%%%%%%%%%%%%%%%%%%%%%%%%%%%%%%%%%%%%%%%%%%%
\begin{figure}[!hb]
\begin{center}
\includegraphics[angle=270,width=9.5cm]{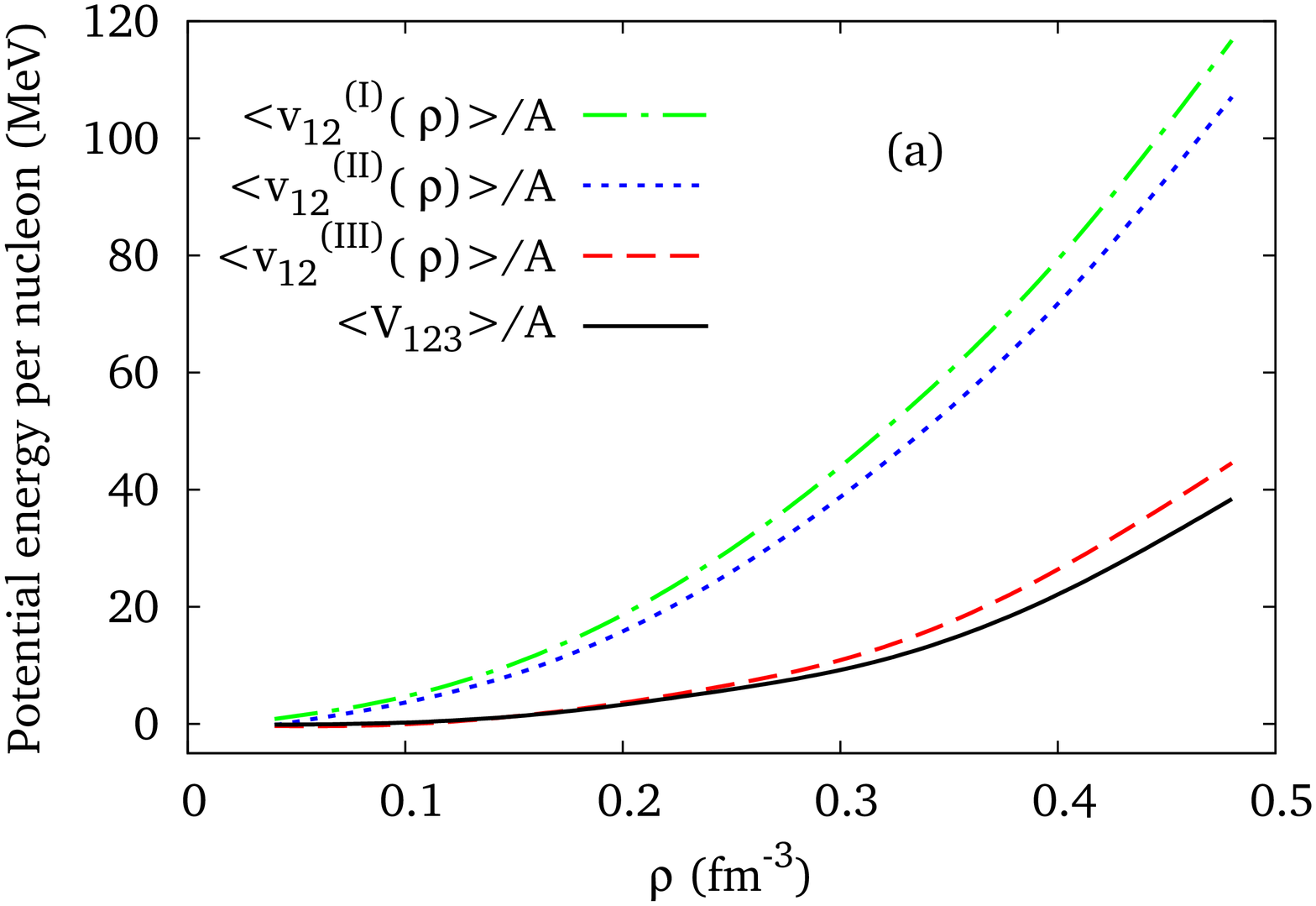}
\vspace{0.1cm}
\includegraphics[angle=270,width=9.5cm]{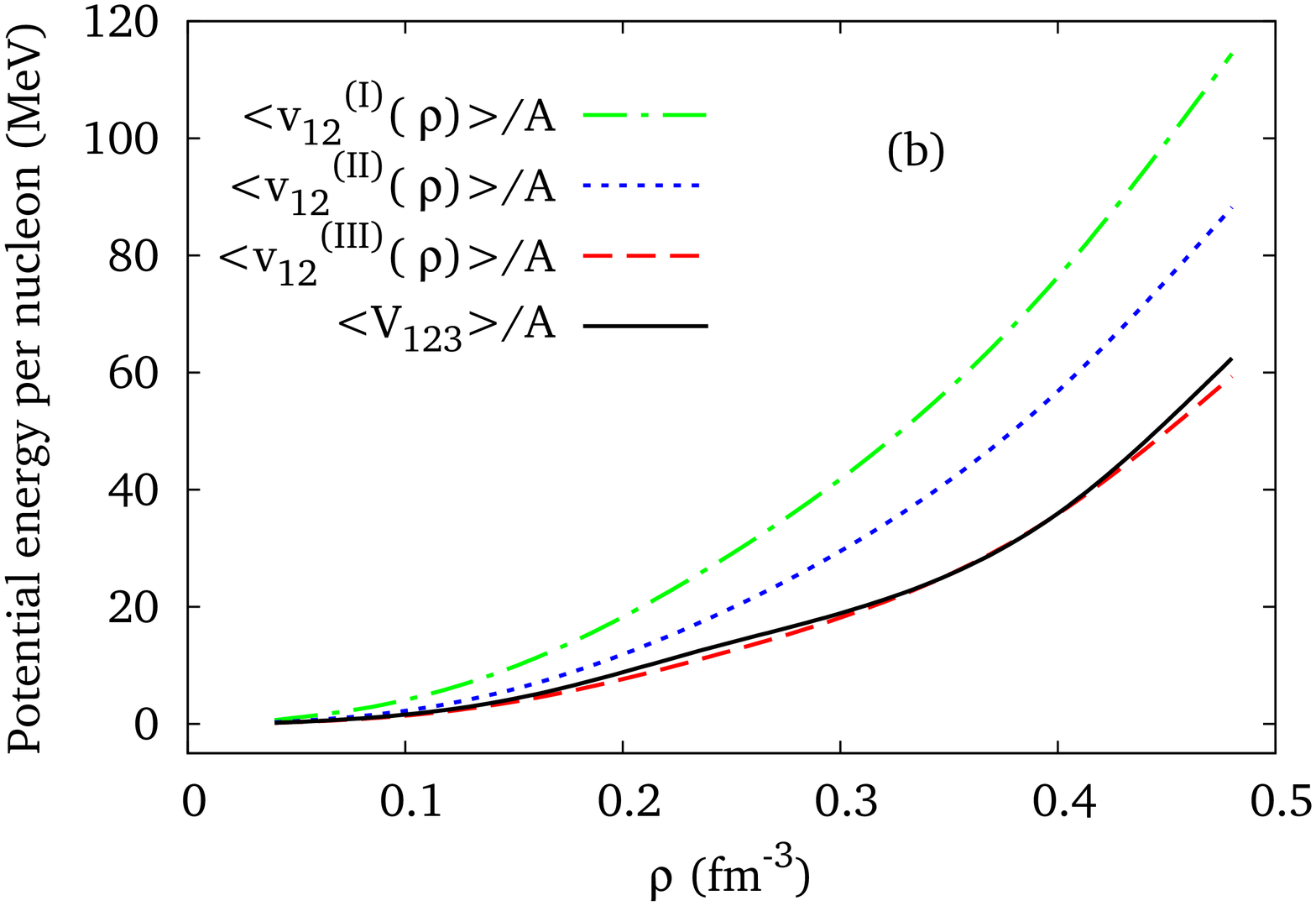}
\vspace{0.1cm}
\caption{Contributions of the density-dependent potential to the energy per particle of SNM (a) and PNM (b), compared to 
the expectation value of the genuine three-body potential UIX: $\langle V_{123}\rangle/A$. \label{fig:compare_potentials}}
\end{center}
\end{figure}
%%%%%%%%%%%%%%%%%%%%%%%%%%%%%%%%%%%%%%%%%%%%%%%%%%%%%%%%%%%%%%%%%%%%%%%%%%%

Note that, in principle,  an additional term involving the anticommutator between the potential and the 
 correlation function should appear in the second line of the above equation. However, due to the structure 
 of the potential it turns out that 
%, which as may be recognized from Eq. (\ref{eq:op_structure_UIX}) is a product of two-body operators, it turns out that
\begin{equation}
\int d{x}_3 \{\hat{V}_{123},\hat{f}(r_{23})\}=2\int d x_{3} \hat{V}_{123} \hat{f}(r_{23})\, .
\end{equation}

The calculation of the right-hand side of  of Eq. (\ref{eq:ddp_diagrams_expression}) requires the evaluation of the traces of commutators and anticommutators of spin-isospin operators, as well as the use of suitable angular functions needed to 
carry out the integration over $\mathbf{r}_3$. 

As for the previous steps, we have computed the contribution of the density-dependent potential $\hat{v}_{12}^{(III)}(\rho)$ to the energy per particle. The results of Fig. \ref{fig:compare_potentials} demonstrate that the density-dependent potential including correlations is able to reproduce the results obtained using genuine three-body UIX to remarkable accuracy. 

To simplify the notation, at this point it is convenient to identify
\begin{equation}
 \hat{v}_{12}(\rho) \equiv \hat{v}_{12}^{(III)}(\rho)\, .
\end{equation}

Note that the above potential exhibits important differences when acting in PNM and in SNM. For example, in SNM $v^p(\rho,r_{12}) \neq 0$ for $p=1,\sigma_{12}\tau_{12},S_{12}\tau_{12}$, while in PNM $v^p(\rho,r_{12}) \neq 0$ for $p=1,\sigma_{12},S_{12}$, as shown in Figs. \ref{fig:effpot}.

%%%%%%%%%%%%%%%%%%%%%%%%%%%%%%%%%%%%%%%%%%%%%%%%%%%%%%%%%%%%%%%%%%%%%%%%%%%
\begin{figure}[!hb]
\begin{center}
\includegraphics[width=9.5cm]{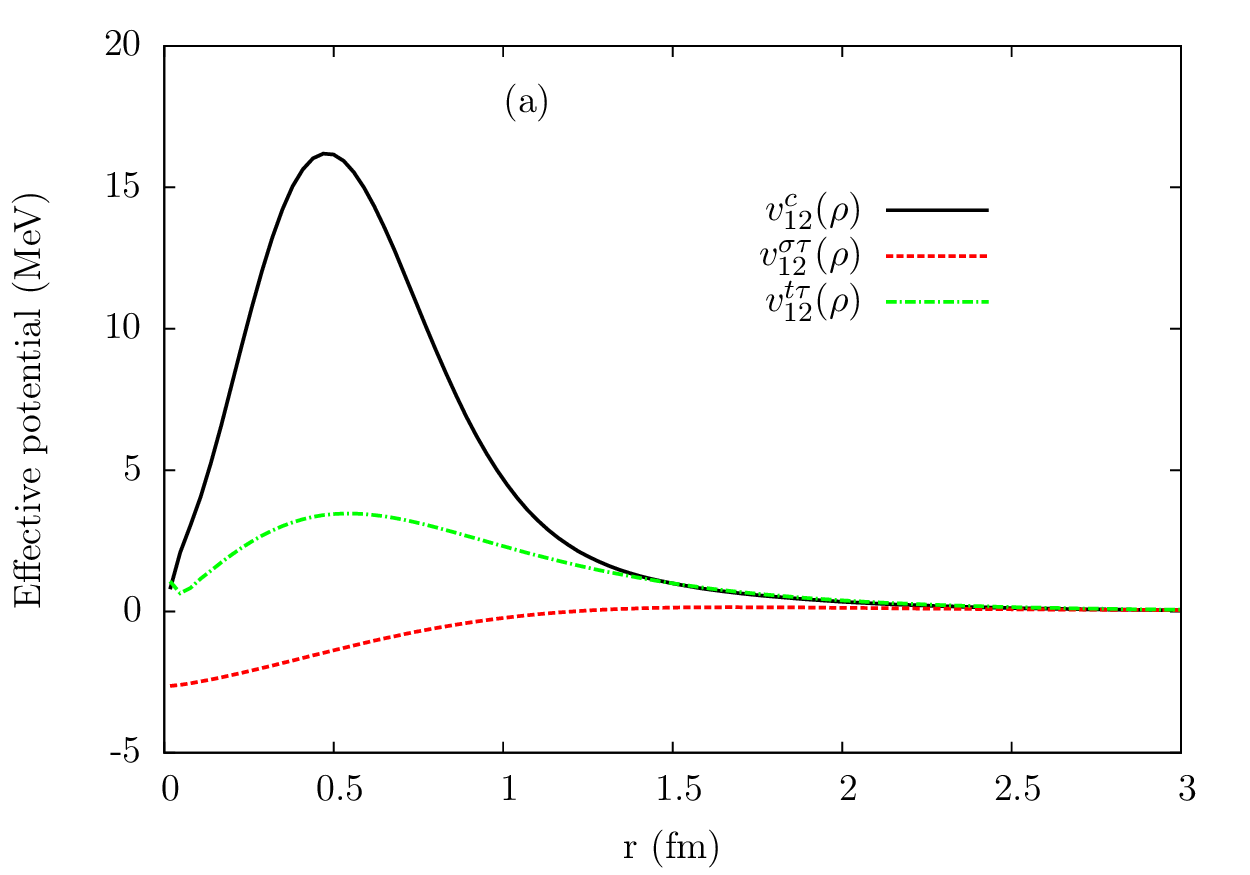}
\vspace{0.1cm}
\includegraphics[width=9.5cm]{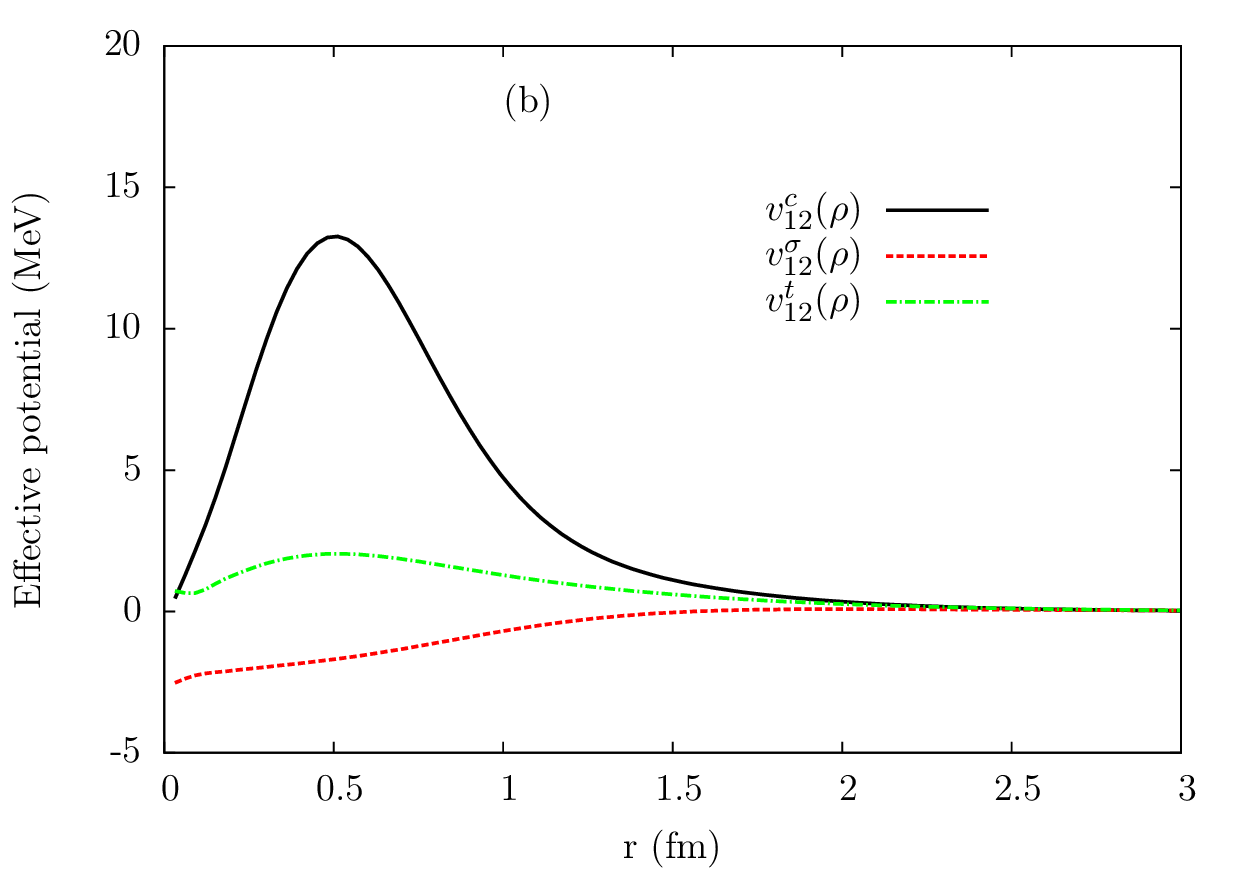}
\vspace{0.1cm}
\caption{Different channels of the effective density dependent potentials calculated for SNM (a) and PNM (b) for $\rho=0.16\,\text{fm}^{-3}$. \label{fig:effpot}}
\end{center}
\end{figure}
%%%%%%%%%%%%%%%%%%%%%%%%%%%%%%%%%%%%%%%%%%%%%%%%%%%%%%%%%%%%%%%%%%%%%%%%%%%

%%%%%%%%%%%%%%%%
\subsection{Numerical Calculations}

A constrained simulated annealing optimization, described in Section \ref{Variational_SA}, has been performed, by imposing the sum rules for the kinetic energy and for the scalar two-body distribution function. In particular the difference between the Pandharipande-Bethe (PB) and the Jackson-Feenberg (JF) kinetic energies has been forced to be less than $10\, \%$ of the Fermi Energy $T_F$ of Eq. (\ref{eq:energy_cluster_cont}), while the sum rule (\ref{eq:gc_sumrule}) for $g^c(r_{12})$ has been satisfied with a precision of $3\, \%$.

In our calculations we have optimized the variational paremeters for four different Hamiltonians, each corresponding to different potential terms: Argonne $v_{8}^\prime$, Argonne $v_{8}^\prime\,+\, $UIX, Argonne $v_{6}^\prime$, and Argonne $v_{6}^\prime\,+\, $UIX.

The energy per particle of SNM and PNM computed adding to the two-body potentials Argonne $v_{8}^\prime$ and Argonne $v_{6}^\prime$ the density-dependent potential of Eq. (\ref{eq:ddp_potential}), have been compared to the results obtained using an hamiltonian with the same two-body potentials and the Urbana IX three-body interaction model. In order to show how much the density dependent potential differs from the original UIX, we compute the expectation values of these potentials with the same correlation functions, i. e. those resulting from the calculation with the genuine three-body potential; no optimization procedure has been performed for the density-dependent potentials.

Both calculations have been consistently carried out within the FHNC/SOC scheme.

It is worth noting that our simulated annealing constrained optimization allows us to: i) reduce the violation of the variational principle due to the FHNC/SOC approximation; ii) perform an accurate scan of the parameter space. As a consequence, our FHNC/SOC calculations provide very close results to those obtained via Monte Carlo calculations, as shown in Figs. \ref{fig:eos_pnm} and \ref{fig:eos_snm}, to be compared with those of Ref. \cite{gandolfi_07b} where the agreement between FHNC and Monte Carlo methods were not nearly as good.

\subsubsection{Auxiliary Field Diffusion Monte Carlo (AFDMC) approach}

In order to check the validity of our variational FHNC/SOC calculations, we carried out AFDMC simulations for both PNM and SNM.

\begin{table}[]
\begin{center}
\caption{Uncorrelated kinetic energy per particle of 14, 38, 66 and 114 neutrons at $\rho~=~0.32\,\text{MeV}$ compared with the limit $A \rightarrow \infty$. \label{tab:fse}}
\vspace{0.3cm}
\begin{tabular}{c c c c c c} 
\hline 
\hline
 	& $A=14$ & $A=38$ & $A=66$ & $A=114$ & $\infty$  \\ 
\hline
$E/A(\text{MeV})$ & 56.51 & 53.50 & 55.43 & 56.58 & 55.71 \\
\hline
\hline
\end{tabular} 
\vspace{0.1cm}
\end{center}
\end{table}

We have computed the equation of state of PNM and SNM using the AFDMC method with the fixed-phase like approximation. We simulated PNM with $A=66$ and SNM with $A=28$ nucleons in a periodic box, as described in \cite{gandolfi_09} and \cite{fantoni_08}.
The finite-size errors in PNM simulations have been investigated in \cite{gandolfi_09} by comparing the Twist Averaged Boundary Conditions (TABC) with the Periodic Box Condition (PBC). It is remarkable that the energies of 66 neutrons computed using either twist averaging or periodic boundary conditions turn out to be almost the same. 
This essentially follows from the fact that the kinetic energy of 66 fermions approaches the thermodynamic limit very well, as can be seen in Table \ref{tab:fse}. The finite-size corrections due to 
the interaction are correctly estimated by including the contributions given by neighboring cells to the simulation box\cite{sarsa_03}.
From the above results for PNM we can estimate that the finite-size errors in the present AFDMC calculations do not exceed 2\% of the asymptotic value of the energy calculated by using TABC. The finite-size effects of SNM calculations can be estimated from the difference of the energies of PNM obtained with 14 neutrons and the TABC asymptotic value, which is of the order of 7\%. Although a calculation with 132 nucleons would be more accurate, performing such a heavy simulation does not appear to be justified in the context of a preliminary model of density dependent potential, and also in view of the fact that, at present, we are not able to simulate SNM with spin-orbit interactions.

The statistical errors, on the other hand, are very small and in the Figures are always hidden by the squares, the triangles and the circles representing the AFDMC energies.

\begin{figure}[!h]
\vspace{0.2cm}
\begin{center}
\includegraphics[angle=270,width=9.5cm]{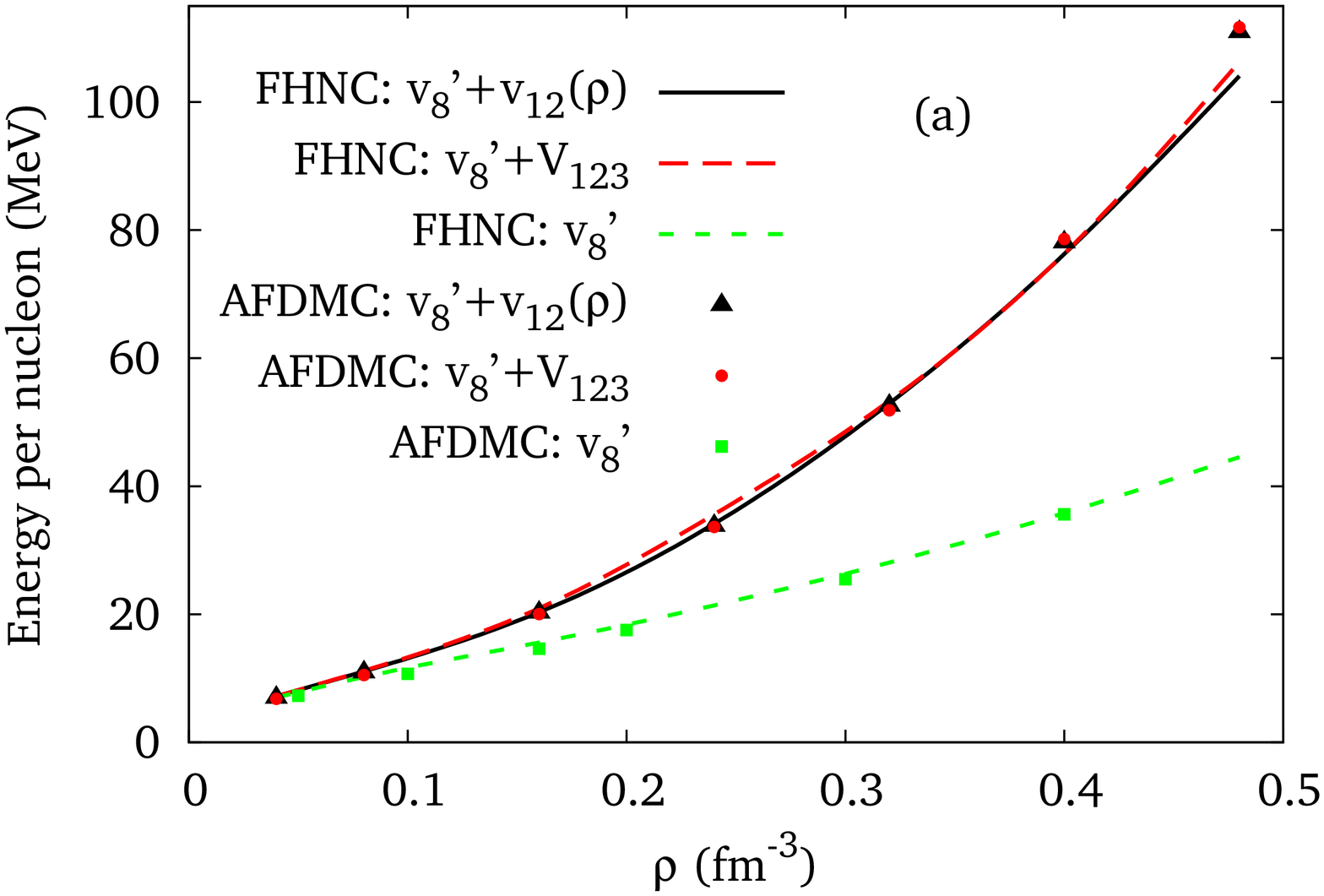}
\vspace{0.1cm}
\includegraphics[angle=270,width=9.5cm]{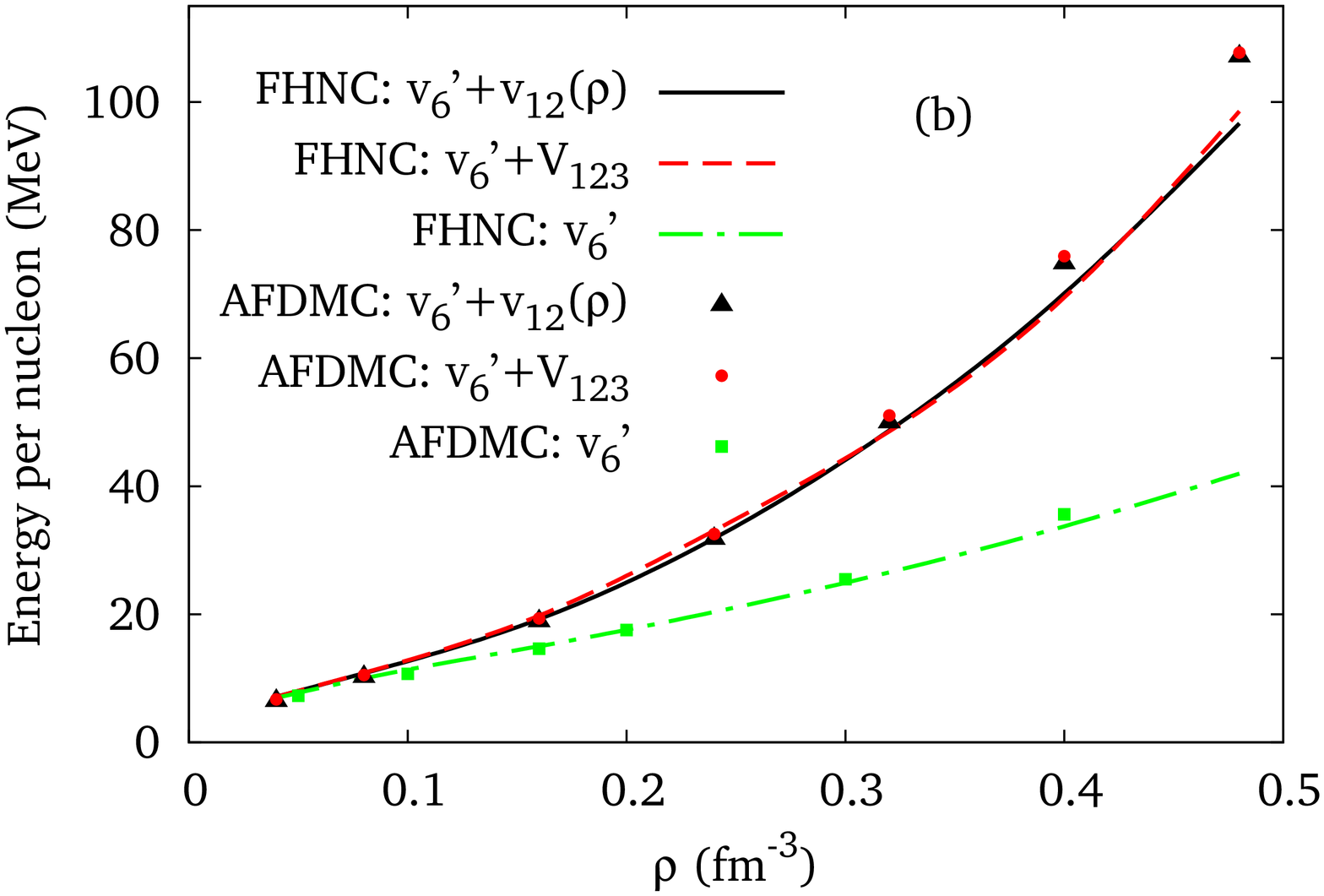}
\vspace{0.1cm}
\caption{Energy per particle for PNM, obtained using the density-dependent potential of Eq. (\ref{eq:bare_ddp}) added to the Argonne $v_{8}^\prime$ (a) and to Argonne $v_{6}^\prime$ (b) potentials. The energies are compared to those obtained from the genuine three-body potential and from 
the two-body potentials alone.  \label{fig:eos_pnm}}
\end{center}
\end{figure}

\subsubsection{PNM equations of state}
In the PNM case (see Fig. \ref{fig:eos_pnm}), the EoS obtained with the three-body potential UIX and using the density-dependent two-body potential are very close to each other. For comparison, in  Fig. \ref{fig:eos_pnm} we also report the results of calculations carried out including the two--body potential only. 
In our approximation, with the exception of the line with diamonds of Fig. \ref{fig:two_body_bose_corr}, we have neglected the cluster contributions proportional to $\rho^2$. One could then have guessed that the curves corresponding to the UIX and density-dependent potential would have slightly moved away from each other at higher densities because, as the density increases, the contributions of higher order diagrams become more important. Probably, in this case a compensation among these second and higher order terms takes place.

The density-dependent potential obtained in the FHNC/SOC framework has been also employed in AFDMC calculations. As can be plainly seen in Fig. \ref{fig:eos_pnm}, the triangles representing the results of  this calculation are very close, when not superimposed, to the circles corresponding to the UIX three-body potential AFDMC results.

\subsubsection{SNM equation of state}
In the EoS of symmetric nuclear matter, the above compensation does not appear to occur, as can be seen in Fig. \ref{fig:eos_snm}. At densities lower than $\rho=0.32 \ \text{fm}^{-3}$, the curves resulting from UIX and the density-dependent potential are very close to one other, while for  $\rho>0.32\ \text{fm}^{-3}$ 
a gap between them begins to develop.

The gap is smaller when the two-body potential Argonne $v_{8}^\prime$ is used, but the reason for this is not completely clear.

%%%%%%%%%%%%%%%%%%%%%%%%%%%%%%%%%%%%%%%%%%%%%%%%%%%%%%%%%%%%%%%%%%%%%%%%%%%
\begin{figure}[!h]
\vspace{0.2cm}
\begin{center}
\includegraphics[angle=270,width=9.5cm]{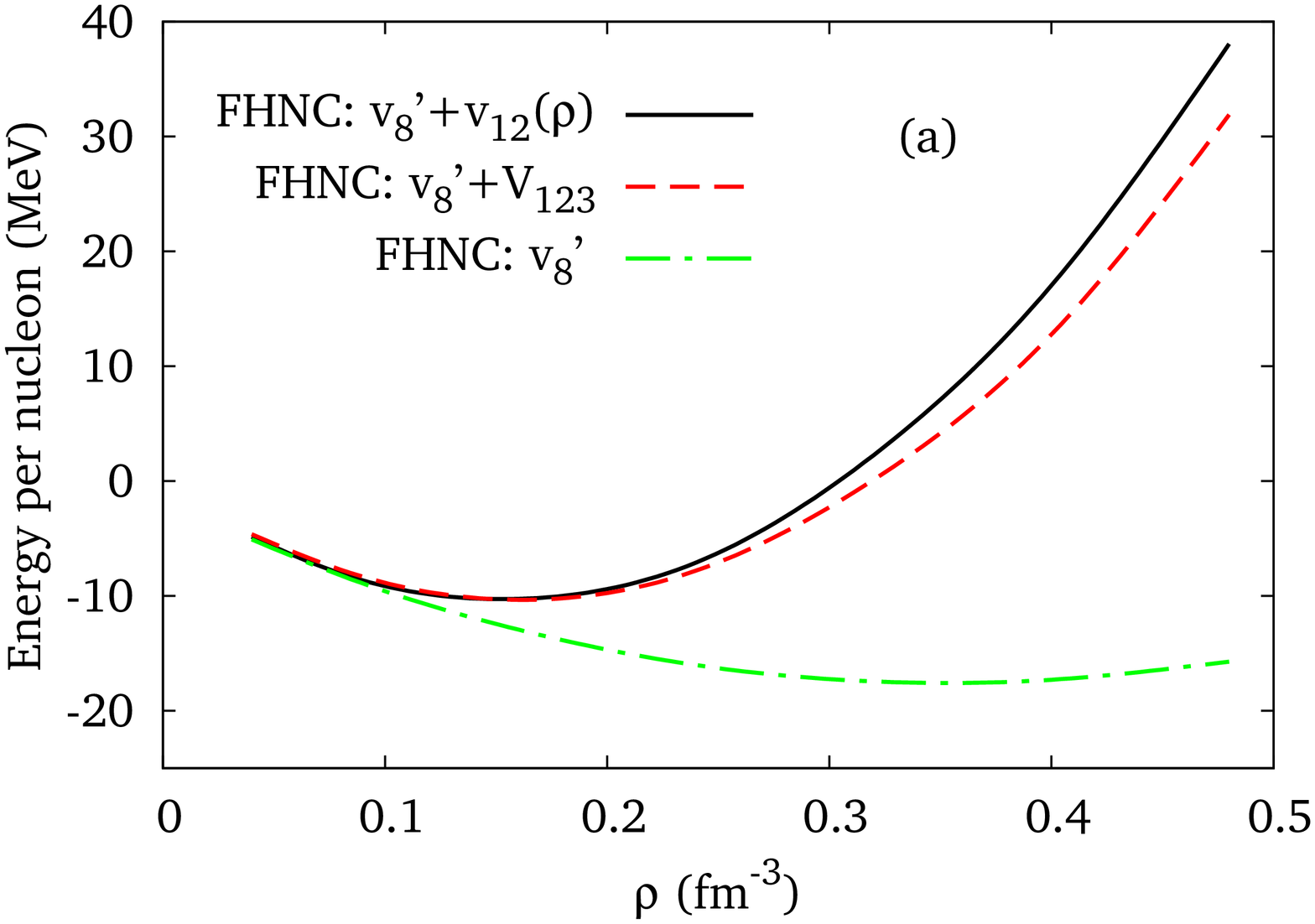}
\vspace{0.1cm}
\includegraphics[angle=270,width=9.5cm]{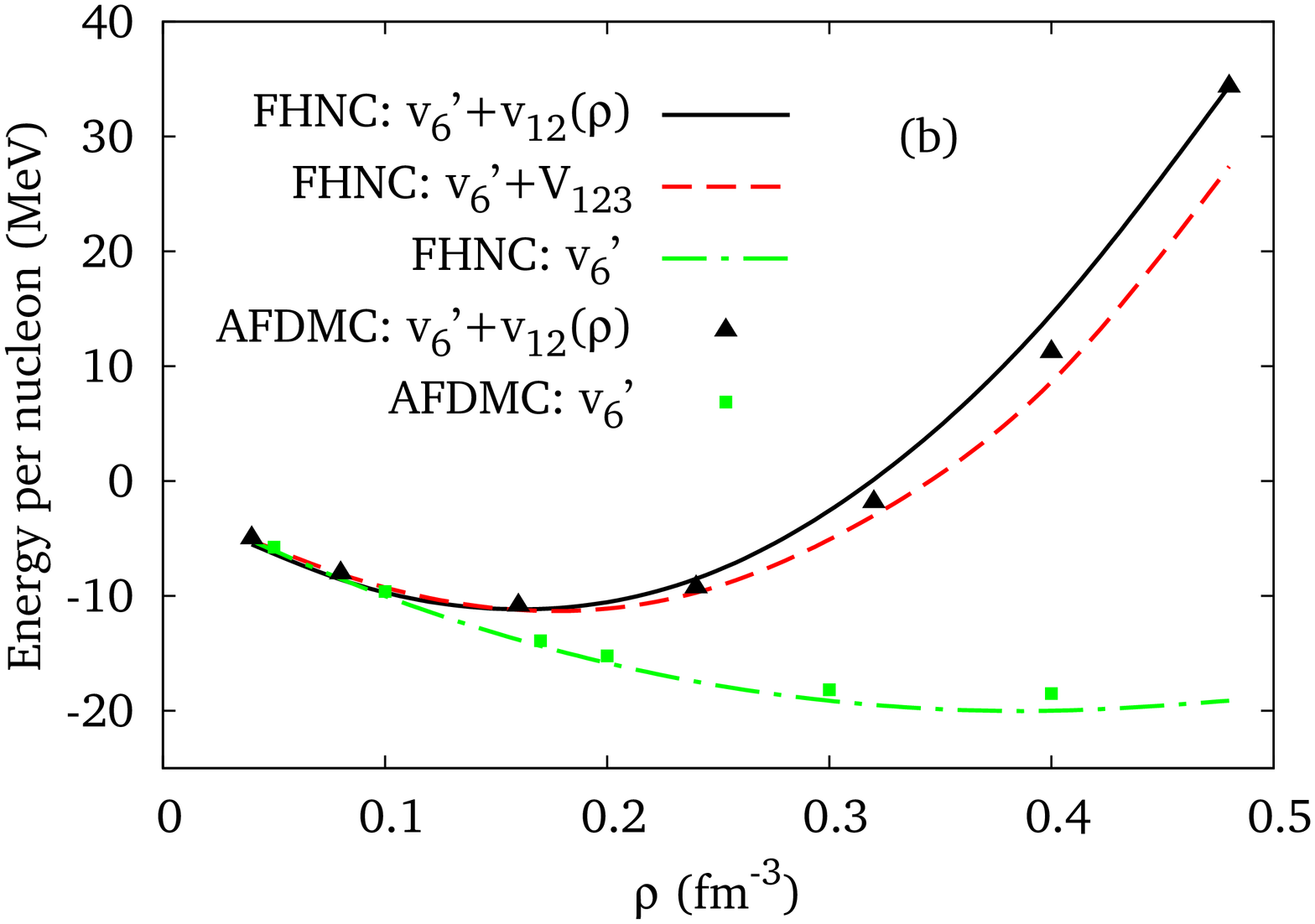}
\vspace{0.1cm}
\caption{Same as in Fig. \ref{fig:eos_pnm}, but for SNM.
%Energy per particle for SNM obtained by using the density-dependent potential of Eq. (\ref{eq:bare_ddp}) added to Argonne $v_{8}^\prime$ (upper graph) 
%and to Argonne $v_{6}^\prime$ (lower graph). The energy are compared to those obtained with the genuine three-body potential and with the two-body potentials 
%alone.  
\label{fig:eos_snm}}
\end{center}
\end{figure}
%%%%%%%%%%%%%%%%%%%%%%%%%%%%%%%%%%%%%%%%%%%%%%%%%%%%%%%%%%%%%%%%%%%%%%%%%%

We have computed the saturation density $\rho_0$, the binding energy per particle $E(\rho_0)$ and the compressibility $K = 9\rho_0 (\partial E(\rho)/\partial\rho)^2$ for all the EoS of Fig. \ref{fig:eos_snm}. The variational FHNC/SOC results are listed in Table \ref{table:parameters_eos}, while those coming from the AFDMC calculation with  $v_{6}^\prime+\hat{v}_{12}(\rho)$ potential are: $\rho_0=0.17\,\text{fm}^{-3}$, $E_0=-10.9\,\text{MeV}$ and K$=201\,\text{MeV}$.
\begin{table}[]
\begin{center}
\caption{Values for the saturation densities, the binding energy per particle, and the compressibility of SNM obtained from the variational FHNC/SOC EoS of Fig. \ref{fig:eos_snm}. \label{table:parameters_eos}}
\vspace{0.3cm}
\begin{tabular}{c c c c c} 
\hline 
 	& $v_{6}^\prime + V_{123}\quad$ & $ v_{6}^\prime + v(\rho) $ & $ v_{8}^\prime + V_{123} $ & $ v_{8}^\prime + v(\rho)$\\ 
\hline
$\rho_0$ (fm$^{-3}$)& 0.17 & 0.16 & 0.16 & 0.15 \\ 

$E_0$ (MeV) &-11.3 & -11.2 & -10.3 & -10.3 \\ 

K (MeV) & 205 & 192 & 189 & 198 \\ 
\hline
 
\end{tabular} 
\end{center}
\vspace{0.1cm}
\end{table}

The saturation densities are quite close to the empirical value $\rho_0=0.16\,\text{fm}^{-3}$. For the genuine three-body potential this is not surprising, since the parameter $U_0$ is chosen to fit the saturation density, as discussed in Section \ref{sec:TBF}. On the other hand,  the fact that the density-dependent potential also reproduces 
this value is remarkable and needs to be emphasized.

The binding energies obtained with $\hat{v}_{12}(\rho)$ are very close to those coming from UIX potential, but they are larger than the empirical value $E_0=-16\,\text{MeV}$. 

As for the compressibility, the experimental value $K\approx 240\,\text{MeV}$ suffers of sizable uncertainties. However, also in this case the result obtained with the density-dependent potential differs from that obtained with the UIX potential by less than $5 \%$.

%%%%%%%%%%%%%%%%%%%%%%%

\section{Chiral inspired three-nucleon potentials in nuclear matter}
\label{many-body}
The work described in this Section, based on Ref. \cite{lovato_12}, is aimed at testing in nuclear matter the different parametrization of the chiral inspired potentials of Ref. \cite{kievsky_10}, introduced in Section \ref{sec:nitbp}. 
%The investigation of uniform nuclear matter may shed light on both the nature and the parametrization of the TNF, although the 
%quantitative description of this system can not be achieved within a mere generalization of the approaches developed for light nuclei. 
%In this Section, we analyze the structure of the contact term of the NNLOL potential of  Ref. \cite{kievsky_10} and discuss the calculation of the TNF contribution to 
%nuclear matter energy. 

\subsection{NNLOL contact term issue}\
\label{sec:cont_issue}
While the  NNLOL chiral interactions provide a fully consistent description of the binding energies of $^3$H and $^4$He, as well as of the scattering length $^2a_{nd}$, some ambiguities emerge when these interactions are used to calculate the nuclear matter EoS. 

For our purposes, it is convenient to rewrite the NNLOL chiral contact term of Eq, (\ref{eq:2pi_3body_conf}) in the form
\begin{equation}
\label{vetau}
\hat{V}_{E}^\tau(3:12)=V_{0}^E\tau_{12}Z_{0}(r_{13})Z_{0}(r_{23})\, .
\end{equation}
where the superscript $\tau$ has a meaning that will be soon clarified. 
The radial function $Z_{0}(r)=m_{\pi}^3/(4\pi) z_0(r)$ approaches the Dirac $\delta$-function in the limit of infinite cutoff. Strictly speaking, the local version of $\hat{V}_E$ is a genuine ``contact term'' in this limit only, while for finite values of the cutoff it  acquires a finite range. 

In addition to $\hat{V}_{E}^\tau$ of Eq.~(\ref{vetau}), the chiral expansion leads to the appearance of six spin-isopin structures in the contact term. For example, the scalar contribution is
\begin{equation}
\hat{V}_{E}^I(3:12)=V_{0}^E Z_{0}(r_{13})Z_{0}(r_{23})\, .
\end{equation}
Within this context,  the superscripts $\tau$ and $I$ identify the $\tau_{12}$ and scalar contact terms. 

In Ref. \cite{epelbaum_02} it has been shown that, once the sum over all cyclic permutation is performed, all contributions to the product between the potential and the antisymmetrization  operator $\mathcal{A}_{123}$ have the same spin-isospin structure. Therefore it is convenient to take into account just one of the contact terms. 
This result  was obtained in momentum space, without the cutoff functions $F_\Lambda$. As a consequence,  in coordinate space it only holds true in the limit of infinite cutoff.
In particular,  for $\hat{V}_{E}^\tau(3:12)$ and $\hat{V}_{E}^I(3:12)$, it turns out that
\begin{equation}
\sum_{cycl}V_{0}^E\delta(\mathbf{r}_{13})\delta(\mathbf{r}_{23})\tau_{12}\mathcal{A}_{123}
=-\sum_{cycl}V_{0}^E\delta(\mathbf{r}_{13})\delta(\mathbf{r}_{23})\mathcal{A}_{123}\, ,
\label{eq:antisimm}
\end{equation}
making this two terms equivalent. The limit of infinite cutoff is crucial, because the radial part of the exchange operator, when multiplied by the Dirac 
$\delta$-functions, is nothing but the identity
\begin{equation}
e^{i\mathbf{k}_{ij}\cdot \mathbf{r}_{ij}}\delta(\mathbf{r}_{ij})=\delta(\mathbf{r}_{ij})\, .
\end{equation}
After the regularization, i.e. with the $\delta$-function replaced by  $Z_0$, the proof is spoiled and the six different structures are no longer equivalent.

In PNM contact terms involving three or more neutrons vanish because of Pauli principle. On the other hand, the expectation value of the contact terms of the NNLOL potential can 
be different from zero.

Let us assume that reproducing the binding energies of light nuclei and $^2a_{nd}$ require a repulsive $V_{E}$ . Then, one has to choose either  $c_{E}^{\tau_{12}}<0$ 
or $c_{E}^{I}>0$. In PNM, as
\begin{equation}
\langle \tau_{12} \rangle_{PNM}=1\,,
\end{equation}
it turns out that $V_{E}^{\tau}$ is attractive and $V_{E}^{I}$ repulsive. This means that fitting the binding energies and the $n-d$ scattering length with either 
$V_{E}^{\tau}$ or $V_E^{I}$ alone leads to an ambiguity in the expectation value of the potential.  

\begin{figure}[!ht]
\begin{center}
\includegraphics[width=8.0cm,angle=270]{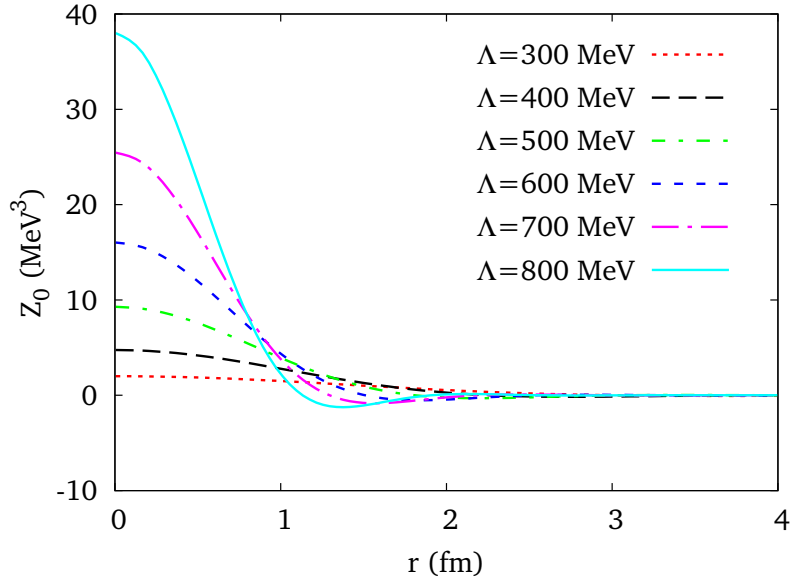}
\caption{(Color online) Radial dependence of the Function $Z_{0}(r)$, appearing in Eq.(\ref{vetau}), plotted for different values of the cutoff $\Lambda$.}
\end{center}
\end{figure}

By expanding the cutoff function
\begin{equation}
F_\Lambda(q^2)=e^{-q^4/\Lambda^4}\sim 1 -\frac{q^4}{\Lambda^4} +
O\Big(\frac{q^8}{\Lambda^8}\Big)\, ,
\end{equation}
one finds
\begin{align}
\hat{V}_{E}^{\tau}(3:12)&=V_{0}^E\tau_{12}\Big[\delta(\mathbf{r}_{13})\delta(\mathbf{r}_{23}
)+O\Big(\frac{q^4}{\Lambda^4}\Big)\Big]\nonumber\\
\hat{V}_{E}^{I}(3:12)&=V_{0}^E\Big[\delta(\mathbf{r}_{13})\delta(\mathbf{r}_{23})+O\Big(\frac{q^4}{
\Lambda^4}\Big)\Big]\, ,
\end{align}
implying that in PNM
\begin{equation}
\langle \hat{V}_{E}^{I,\tau}(3:12)
\rangle_{PNM}=O\Big(\frac{q^4}{\Lambda^4}\Big)\, .
\end{equation}
From the above equation it becomes apparent that the expectation value of the three-nucleon potential, as well as its sign ambiguity, is nothing but a a cutoff effect. Hence, it should be regarded as  a theoretical uncertainty. Note that, since $\Lambda_\chi\simeq \Lambda$, then $\langle V_{E}\rangle_{PNM}$ is of the same order of the next term in chiral expansion. 

To clarify this issue,  let us consider a simple system: a Fermi gas of neutrons, in which correlations among particles are not present. 
The expectation value of the contact interaction reads
\begin{align}
\frac{\langle \hat{V}_{E}^{I,\tau} \rangle_{PNM}^{FG}}{A}=&\frac{\rho^2}{2}V_{0}^E\int
d\mathbf{r}_{12}d\mathbf{r}_{13}
Z_{0}(r_{12})Z_{0}(r_{13})\times\nonumber\\
&\Big(1-\frac{\ell(r_{12})^2}{2}-\frac{\ell^2(r_{13})}{2}
-\frac{\ell^2(r_{23})}{2}+\frac{\ell(r_{12})\ell(r_{13})\ell(r_{23})}{2}\Big)\, ,
\label{eq:VE_PNM_FG}
\end{align}
where $A$ is the number of neutrons. The factor $1/2$ includes the $1/3!$ arising from the unrestricted sum over particle indices $123$, multiplied by a factor $3$ from the cyclic permutations of the potential, all giving the same contribution. The Slater function $\ell(r_{ij})$ is given in Eq. (\ref{eq:slat_def}). It can be  easily seen that, if $\hat{V}_{E}^{I,\tau}(1:23)\propto \delta(\mathbf{r}_{12})\delta(\mathbf{r}_{13})$, then 
\begin{equation}
\frac{\langle \hat{V}_{E}^{I} \rangle_{PNM}^{FG}}{A}=0\, .
\end{equation}
 
Consider now a Fermi gas with equal numbers of protons and neutrons, where
\begin{align}
\frac{\langle \hat{V}_{E}^{\tau} \rangle_{SNM}^{FG}}{A}=&\frac{\rho^2}{2}V_{0}^E\int
d\mathbf{r}_{12}d\mathbf{r}_{13}
Z_{0}(r_{12})Z_{0}(r_{13})\times\nonumber\\
&\Big(-\frac{3}{4}\ell^2(r_{23})+\frac{3}{8}\ell(r_{12})\ell(r_{
13})\ell(r_{23})\Big)
\label{eq:VEtau_SNM_FG}
\end{align}
and
\begin{align}
\frac{\langle \hat{V}_{E}^{I} \rangle_{SNM}^{FG}}{A}=&\frac{\rho^2}{2}V_{0}^E\int
d\mathbf{r}_{12}d\mathbf{r}_{13}
Z_{0}(r_{12})Z_{0}(r_{13})\times\nonumber\\
&\Big(1-\frac{\ell(r_{12})^2}{4}-\frac{\ell(r_{13})^2}{4}-\frac{\ell(r_{23})^2}{4}+\frac{\ell(r_{12})\ell(r_{13})\ell(r_{23})}{8}\Big)\, .
\label{eq:VEI_SNM_FG}
\end{align}
In the limit of infinite cutoff the above equations imply
\begin{align}
\frac{\langle \hat{V}_{E}^{\tau_{12}} \rangle_{SNM}^{FG}}{A}&=-\frac{3}{16}\rho^2
V_{0}^E\nonumber\\
\frac{\langle\hat{V}_{E}^{I} \rangle_{SNM}^{FG}}{A}&=\frac{3}{16}\rho^2 V_{0}^E\, .
\label{eq:aympt_SNM}
\end{align}
As expected from Eq.~(\ref{eq:antisimm}), the two contributions have opposite sign.

We have computed the expectation values of Eqs.~(\ref{eq:VE_PNM_FG}), (\ref{eq:VEtau_SNM_FG}) and (\ref{eq:VEI_SNM_FG}) for different values of the cutoff $\Lambda$  
and density $\rho=0.16\,\text{fm}^{-3}$. The results listed  in Table \ref{tab:contact_cut_scan_PNM} show that for PNM the larger the cutoff the smaller the expectation value of the three nucleon contact term. Note that for $\Lambda = 500$ MeV, the expectation value is still sizably different from the asymptotic limit.

As far as SNM is concerned (see Table \ref{tab:contact_cut_scan_SNM}), as the cutoff increases the possible choices of the three nucleon contact term 
tend to the asymptotic values of Eq.~(\ref{eq:aympt_SNM}). As in the case of PNM, the results corresponding to $\Lambda~=~500$~MeV, are significantly
different from the asymptotic values.

\begin{table}[]
\begin{center}
\caption{Cutoff dependence of the expectation values of the three body contact term of the NNLOL potential in noninteracting PNM. \label{tab:contact_cut_scan_PNM}}
\vspace{0.3cm}
\begin{tabular}{c c} 
\hline 
 $\Lambda\,\text{(MeV)}$ & $
\langle V_{E}^{I,\tau_{12}} \rangle_{PNM}^{FG}/A \,\text{(MeV)}$\\ 
\hline
$300$ &  9.15 \\ 
$400$ &  5.95  \\ 
$500$ &  3.60  \\ 
$600$ &  2.15  \\  
$700$ &  1.30  \\ 
$800$ &  0.81  \\ 
\hline
$\infty$ & 0\\ 
\hline
\end{tabular} 
\end{center}
\end{table}

\begin{table}[]
\begin{center}
\caption{Same as in Table \ref{tab:contact_cut_scan_PNM}, but for SNM.\label{tab:contact_cut_scan_SNM}}
\vspace{0.3cm}
\begin{tabular}{c c c } 
\hline 
 $\Lambda\,\text{(MeV)}$ & $ \langle \hat{V}_{E}^{\tau_{12}} \rangle_{SNM}^{FG}/A
\,\text{(MeV)}$ & $ \langle \hat{V}_{E}^{I} \rangle_{SNM}^{FG}/A \,\text{(MeV)}$\\ 
\hline
$300$ & -2.61 & 10.21  \\ 
$400$ & -3.61 & 8.15   \\ 
$500$ & -4.37 & 6.93   \\ 
$600$ & -4.87 & 6.30   \\  
$700$ & -5.15 & 5.98   \\ 
$800$ & -5.30 & 5.81   \\ 
\hline
$\infty$ & -5.55 & 5.55 \\ 
\hline
\end{tabular} 
\end{center}
\end{table}

We emphasize that the parameter $c_E$ has not been included in this analysis, even though it is  itself cutoff dependent. Unfortunately, the authors of Ref.~\cite{kievsky_10} kept $\Lambda$ fixed to $500\,\text{MeV}$. Had this not been the case, their fit to the experimental data would have resulted in a set of different constants $c_E$, corresponding to different values of $\Lambda$. It would have been interesting to extrapolate the expectation value of $\hat{V}_E$ to the limit of infinite $\Lambda$, where the cutoff effects associated with the regularization procedure are expected to vanish. 

\subsection{FHNC/SOC calculations}
Using the relations for the constants and the radial functions given in Eqs. (\ref{eq:const_rel}) and (\ref{eq:rad_gen_UIX}), the computation of the diagrams of Fig. \ref{fig:tbp_2pi} with the $\hat{V}_3$ and $\hat{V}_4$ terms of both the TM$^\prime$ and NNLOL potentials is the same as that of $\hat{V}^{2\pi}$ reported in Section \ref{sec:fhncsoc_tbp} and in Ref.~\cite{carlson_83}.

Thanks to the identity
 \begin{align}
&(\vec{\sigma}_1\cdot\hat{r}_{13})(\vec{\sigma}_2\cdot\hat{r}_{23})(\hat{r}_{13}\cdot\hat{r}_{23})=\frac{1}{18}\{\sigma_{13}+S_{13},\sigma_{23}+S_{23}\}\, ,
\end{align}
the term $\hat{V}_1$ of Eq.~(\ref{eq:2pi_3body_conf}), appearing in both the TM$^\prime$ and the NNLOL potentials, can be written in the form
\begin{align}
\hat{V}_{1}(3:12)&=\frac{W_0}{36}\frac{r_{13}r_{23}}{(\hat{r}_{13}\cdot \hat{r}_{23})}y(r_{13})y(r_{23})\{\tau_{13},\tau_{23}\}\{\sigma_{13}+S_{13},\sigma_{23}+S_{23}\}\, .
\end{align}

Aside from the radial function, $\hat{V}_1$ is completely equivalent to $\hat{V}_3$, the anticommutator term of the UIX potential. Therefore, we were allowed to use again the results of Section \ref{sec:fhncsoc_tbp} and Ref.~\cite{carlson_83}.

Furthermore, exploiting the identities
\begin{align}
(\vec{\sigma}_1\mathbf{r}_{23})(\vec{\sigma}_2\cdot\mathbf{r}_{23})
&=\frac{r_{23}^2}{6}\{S_{23}+\sigma_{23},\sigma_{13}\}\nonumber\\
(\vec{\sigma}_2\cdot\mathbf{r}_{13})(\vec{\sigma}_1\cdot\mathbf{r}_{13})
&=\frac{r_{13}^2}{6}\{\sigma_{23},S_{13}+\sigma_{13}\}\, ,
\end{align}
we can rewrite the $\hat{V}_D$ term in a form that has again the same spin-isospin structure as the anticommutator contribution of the UIX potential
\begin{align}
\hat{V}_{D}(3:12)&=\frac{W_{0}^D}{4}\{\tau_{13},\tau_{23}\}[\{\sigma_{13},\sigma_{23}\}V_{D}^{YY}(r_{13},r_{23})+\nonumber\\
&\qquad \{S_{13},\sigma_{23}\}V_{D}^{TY}(r_{13},r_{23})+\{\sigma_{13},S_{23}\}V_{D}^{YT}(r_{13},r_{23})]\, ,
\label{eq:V_D_rearrange}
\end{align}
where
\begin{align}
V_{D}^{YY}({r}_{13},{r}_{23})&=Y(r_{13})z_{0}(r_{23})+z_{0}(r_{13})Y(r_{23})\nonumber\\
V_{D}^{YT}({r}_{13},{r}_{23})&=z_0(r_{13})T(r_{23})\nonumber\\
V_{D}^{TY}({r}_{13},{r}_{23})&=T(r_{13})z_0(r_{23})\, .
\end{align}
In conclusion, including $\hat{V}_D$ amounts to properly adding the above radial functions to those already appearing in $\hat{V}_3$. 

The $\hat{V}_E$ term of TM$^\prime$ is completely equivalent to $\hat{V}_R$ (see Eq.~(\ref{eq:E_equiv}) ). This allowed us to use the results of  Section \ref{sec:fhncsoc_tbp} and Ref.~\cite{carlson_83} for the diagrams of Fig. \ref{fig:tbp_scalar}. The same holds true for the chiral contact term $\hat{V}_E$ in PNM, as $\langle\tau_{ij}\rangle_{PNM}=1$, while in SNM the calculation of $\hat{V}_E$ requires the evaluation of the diagrams of Fig. \ref{fig:tbp_2pi}.

Let us start from diagram (\ref{fig:tbp_2pi}.a), whose analytic expression is given in Eq. (\ref{eq:tbp_2pi.a}).  As pointed out  in Ref.~\cite{carlson_83}, integrating and tracing over the radial and spin-isospin variables of particle $3$ leads to the appearance of an effective density dependent interaction 
\begin{align}
\sum_p V_{E,yy',12}^p(\rho)\hat{O}_{12}^{p}&=\rho\sum_{ex}\int d\mathbf{r}_{13}\,g_{xy,13}^{c}\,g_{x'y',23}^{c}\text{CTr}_{3}\Big[\hat{V}_{E}(3:12)\Big]\, ,
\label{eq:dia21_3}
\end{align}
such that
\begin{align}
(\ref{fig:tbp_2pi}.a)=
&\frac{c_E}{2}\rho\sum_{ex}\sum_{p}\int d\mathbf{r}_{12}A^p\,g_{x''y'',12}^{p}V_{E,yy',12}^p(\rho)\, .
\label{eq:dia21_4}
\end{align}
The subscripts $xy$ label exchange patterns at the ends of the generalized correlation lines. In particular, $dd$ correspond stands for 
direct-direct, $de$ for direct-exchange, $ee$ for exchange-exchange, and $cc$ for incomplete circular exchange. 
The matrix $A^p=1,3,3,9,6,18$ has been defined in Eq. (\ref{eq:A_def}).

It turns out that the only nonvanishing term of the density dependent potential is
\begin{align}
V_{E\,yy',12}^{\tau}(\rho)&=W_{0}^E\rho\sum_{ex}\int d\mathbf{r}_{13}\,g_{xy,13}^{c}\,g_{x'y',23}^{c} z_0(r_{13})z_0(r_{23})\, .
\end{align}

As far as diagram  (\ref{fig:tbp_2pi}.b) is concerned, from Eq. (\ref{eq:tbp_2pi.b}), it clearly follows that only $\tau$-type generalized correlation lines contribute. Hence
\begin{align}
(\ref{fig:tbp_2pi}.b)=&\frac{3}{2} c_EW_{0}^E\rho^2\sum_{ex}\int d\mathbf{r}_{12}d\mathbf{r}_{13}\,g_{xy,13}^{\tau}\,g_{x'y',23}^{\tau}\,g_{x''y'',12}^{c}\,z_0(r_{13})z_0(r_{23})\, .
\end{align}

Diagram (\ref{fig:tbp_2pi}.c) does not contribute to $\hat{V}_E$, while the calculation of the spin-isospin trace of diagram (\ref{fig:tbp_2pi}.d), with three  generalized operatorial correlation, yields
\begin{align}
(\ref{fig:tbp_2pi}d)=&\frac{c_E}{2} W_{0}^E\rho^2\int d\mathbf{r}_{12}d\mathbf{r}_{13}[-2g_{12}^{\tau}\,g_{13}^{\tau}\,g_{23}^{\tau}+
9g_{12}^{\sigma\tau}\,g_{13}^{\sigma}\,g_{23}^{\sigma}+9g_{12}^{\sigma}\,g_{13}^{\sigma\tau}\,g_{23}^{\sigma\tau}-\nonumber\\
&6g_{12}^{\sigma\tau}\,g_{13}^{\sigma\tau}\,g_{23}^{\sigma\tau}+18\xi^{\sigma tt}_{231}g_{12}^{t}\,g_{13}^{t\tau}\,g_{23}^{\sigma\tau}-12\xi^{\sigma tt}_{231}\,g_{12}^{t\tau}\,g_{13}^{t\tau}\,g_{23}^{\sigma\tau}+18\xi^{t\sigma t}_{231}\,g_{12}^{t}\,g_{13}^{\sigma\tau}\,g_{23}^{t\tau}+\nonumber\\
&18\xi^{t\sigma t}_{231}g_{12}^{t\tau}\,g_{13}^{\sigma}\,g_{23}^{t}-12\xi^{t\sigma t}_{231}g_{12}^{t\tau}\,g_{13}^{\sigma\tau}\,g_{23}^{t\tau}+9\xi^{tt\sigma}_{231}g_{12}^{\sigma}\,g_{13}^{t\tau }\,g_{23}^{t\tau}+9\xi^{tt\sigma}_{231}\,g_{12}^{\sigma\tau}\,g_{13}^{t}\,g_{23}^{t}-\nonumber\\
&6\xi^{tt\sigma}_{231}g_{12}^{\sigma\tau}\,g_{13}^{t\tau}\,g_{23}^{t\tau}-12\xi^{ttt}_{231}g_{12}^{t\tau}\,g_{13}^{t\tau}\,g_{23}^{t\tau}+18\xi^{ttt}_{231}g_{12}^{t}\,g_{13}^{t\tau}\,g_{23}^{t\tau}+18\xi^{ttt}_{231}g_{12}^{t\tau}\,g_{13}^{t}\,g_{23}^{t}+\nonumber \\
&18\xi^{\sigma tt}_{231}g_{12}^{t\tau}\,g_{13}^{t}\,g_{23}^{\sigma}]z_0(r_{13})z_0(r_{23})\, .
\end{align}

The matrices $\xi_{231}^{pqr}$, depending on the angles formed by the vectors $\mathbf{r}_1$, $\mathbf{r}_2$ and $\mathbf{r}_3$, are defined in Eq. (\ref{eq:xi_coeff}).
Following Ref.~ \cite{wiringa_88} we have considered only the direct term of the generalized operatorial correlations. As a consequence, in the previous equation $g_{ij}^p=g_{dd,ij}^p$.

In order to find the optimal values of the variational parameters, we have employed a procedure similar to simulated annealing, the details of which have been explained in Section \ref{Variational_SA} of this Thesis.

For the calculation of the density dependent potential of Section \ref{sec:ddp}, we constrained the difference between the Pandharipande-Bethe (PB) and the Jackson-Feenberg (JF) kinetic energies to be less than $10\, \%$ of the Fermi Energy $T_F$ and required the sum rule involving the scalar two-body distribution function, $g^c(r_{12})$, to be fulfilled with a precision of $3\, \%$. 

For this comparative study of chiral inspired potential, in variational calculations of SNM we have imposed the further condition, firstly considered in Ref.~\cite{wiringa_88}, that the sum rule of the isospin component of the two-body distribution function
\begin{equation}
\rho\int d\mathbf{r}_{12} g^{\tau}(r_{12}) = -3\, ,
\end{equation} 
be also satisfied to the same accuracy.

Using also the sum rules for the spin and spin-isospin two body distribution functions  leads to a sizable increase of the variational energies, which turn out to be much higher than those obtained releasing the additional constraints, as well as the AFDMC results. The same pattern is observed in the results of variational calculations not including TNF.

For this reason, we have enforced the fulfillment of the sum rules for $g^c(r_{12})$ and $g^{\tau}(r_{12})$ only. 

For potentials other than UIX, it turns out that the variational energies of PNM resulting from our optimization procedure are lower than the AFDMC values at  $\rho > \rho_0$. By carefully analyzing the contributions of the cluster expansion diagrams, we realized that the value of diagram (\ref{fig:tbp_2pi}.a) was unnaturally large. In particular, we have found that a small change in the variational parameters leads to a huge variation of the value of the diagram. Moreover, the minimum of 
the energy in parameter space was reached in a region where the kinetic energy difference was very close to the allowed limit.

To cure this pathology, we have constrained the PB-JF kinetic energy difference to be less than $1$ MeV, regardless of density. The variational energies obtained imposing this new constraint are always larger than the corresponding AFDMC values and the value of diagram (\ref{fig:tbp_2pi}a) is brought under control. For the sake of consistency, the same constraint on the kinetic energies has been also applied to SNM. In addition, the variational energy minimum does not correspond to the maximum allowed violation of the constraints. As a consequence, it would be largely unaffected by a slight modification of the constraints.

\subsection{AFDMC calculations}
We have computed the EoS of PNM using the AFDMC  approach with the TM$^\prime$ and NNLOL chiral potentials combined with the Argonne $v_{8}'$
NN interaction.

An efficient procedure to perform AFDMC calculations with three-body potentials is described in Ref.~\cite{pederiva_04}. Since $V_3$ is equivalent to the anticommutator term of the UIX model (while the commutator, $V_4$, is zero in PNM), and in PNM the $\hat{V}_E$ terms of both the TM$^\prime$ and NNLOL potentials do not show any formal difference with respect to the repulsive term of UIX, the inclusion of these terms reduces to a replacement of constants and radial functions. The authors of Ref.~\cite{pederiva_04} also described how to handle the $V_1$ for the TM model, and no further difficulties arise in the case of the NNLOL potential.

As  the $\hat{V}_D$ term has never been encompassed in AFDMC, it is worthwhile showing how the calculation of this term reduces to a matrix multiplication. The expectation value of $\hat{V}_D$ is given by
\begin{equation}
\langle \hat{V}_D \rangle=\sum_{i<j<k}[\hat{V}_D(i:jk)+\hat{V}_D(j:ik)+\hat{V}_D(k:ij)]
\end{equation}
with $\hat{V}_D(i:jk)=\hat{V}_D(i:kj)$ (otherwise all six permutations need to be summed). Thanks to this property one can write
\begin{equation}
\langle \hat{V}_D \rangle=\sum_{i<k,j}\hat{V}_D(j:ik)\, .
\end{equation}
It is possible to write $\hat{V}_D(j:ik)$ of Eq.~(\ref{eq:V_D_rearrange}) in terms of cartesian components operators
\begin{align}
\hat{V}_D(j:ik)=&(Y_{\alpha i;\beta j}Z_{\gamma j;\delta k}+Z_{\alpha i;\beta j}Y_{\gamma j;\delta k}+T_{\alpha i;\beta j}Z_{\gamma j;\delta k}+Z_{\alpha i;\beta j}T_{\gamma j;\delta k})\{\sigma_{i}^\alpha\sigma_{j}^\beta,\sigma_{j}^\gamma\sigma_{k}^\delta\}\, ,
\end{align}
where
\begin{align}
Y_{\alpha i;\beta j}&=Y(r_{ij})\delta^{\alpha\beta}\nonumber\\
Z_{\alpha i;\beta j}&=z_0(r_{ij})\delta^{\alpha\beta}\nonumber \\
T_{\alpha i;\beta j}&=T(r_{ij})(3\hat{r}_{ij}^\alpha \hat{r}_{ij}^\beta-\delta^{\alpha\beta})\, .
\end{align}
The anticommutation relation $\{\sigma_{i}^\alpha,\sigma_{j}^\beta\}=2\delta^{\alpha\beta}$ makes the expectation value of $\hat{V}_D$ a sum of $3N\times3N$ matrix multiplications
\begin{align}
\langle \hat{V}_D \rangle&=2\sum_{i<k,j}(Y_{\alpha i;\beta j}Z_{\beta j;\delta k}+Z_{\alpha i;\beta j}Y_{\beta j;\delta k}+T_{\alpha i;\beta j}Z_{\beta j;\delta k}+Z_{\alpha i;\beta j}T_{\beta j;\delta k})\sigma_{i}^\alpha\sigma_{k}^\delta\,\nonumber\\
&=2\sum_{i<k}(\{Y,Z\}+\{T,Z\})_{\alpha i;\delta k}\,\sigma_{i}^\alpha\sigma_{k}^\delta\, ,
\end{align}
analogous to those of Ref.~\cite{pederiva_04}. In order to compute the expectation value of $\hat{V}_D$ the former expression has been added to the cartesian matrices 
associated with the two-body potential.

Finally, as the free-gas value obtained with 66 neutrons turns out to be very close to the thermodynamic limit of $73.00$ MeV, the finite size corrections for 66 neutrons tend to be small. }

For the same reasons adduced for the AFDMC calculations of the density dependent potential, we simulated PNM with $A=66$ neutrons in a periodic box, using the fixed-phase approximation. Finite-size effects are expected to be larger when the density is bigger, as the dimension of the box decreases. In order to check the validity of our calculations, at $\rho=0.48\,\text{fm}^{-3}$  we have repeated the calculation with $114$ neutrons.  For all the potentials, the energies per particle obtained with 114 neutrons are higher than those obtained with 66 neutrons. The authors of Ref.~\cite{gandolfi_09} found a similar behavior for PNM at $\rho\leq0.32\,\text{fm}^{-3}$ in the case of the $v_{8}^\prime$ plus UIX hamiltonian, and ascribed part of this difference to the Fermi gas energy, amounting to $72.63$ MeV and at  $74.15$ MeV for 66 and 114 neutrons, respectively.
However, the difference of the energy per particle obtained with $66$ and $114$ neutrons is always within 4 MeV. It is worth noting that once finite-size effects on the Fermi gas energy are accounted for, the residual finite-size effects do not exceed $4 \%$ of the energy per particle.

\section{Nuclear Matter EoS }
\label{results}
\subsection{TM$^\prime$ potential}
\begin{figure}[!ht]
\centering
\includegraphics[width=6.5cm,angle=270]{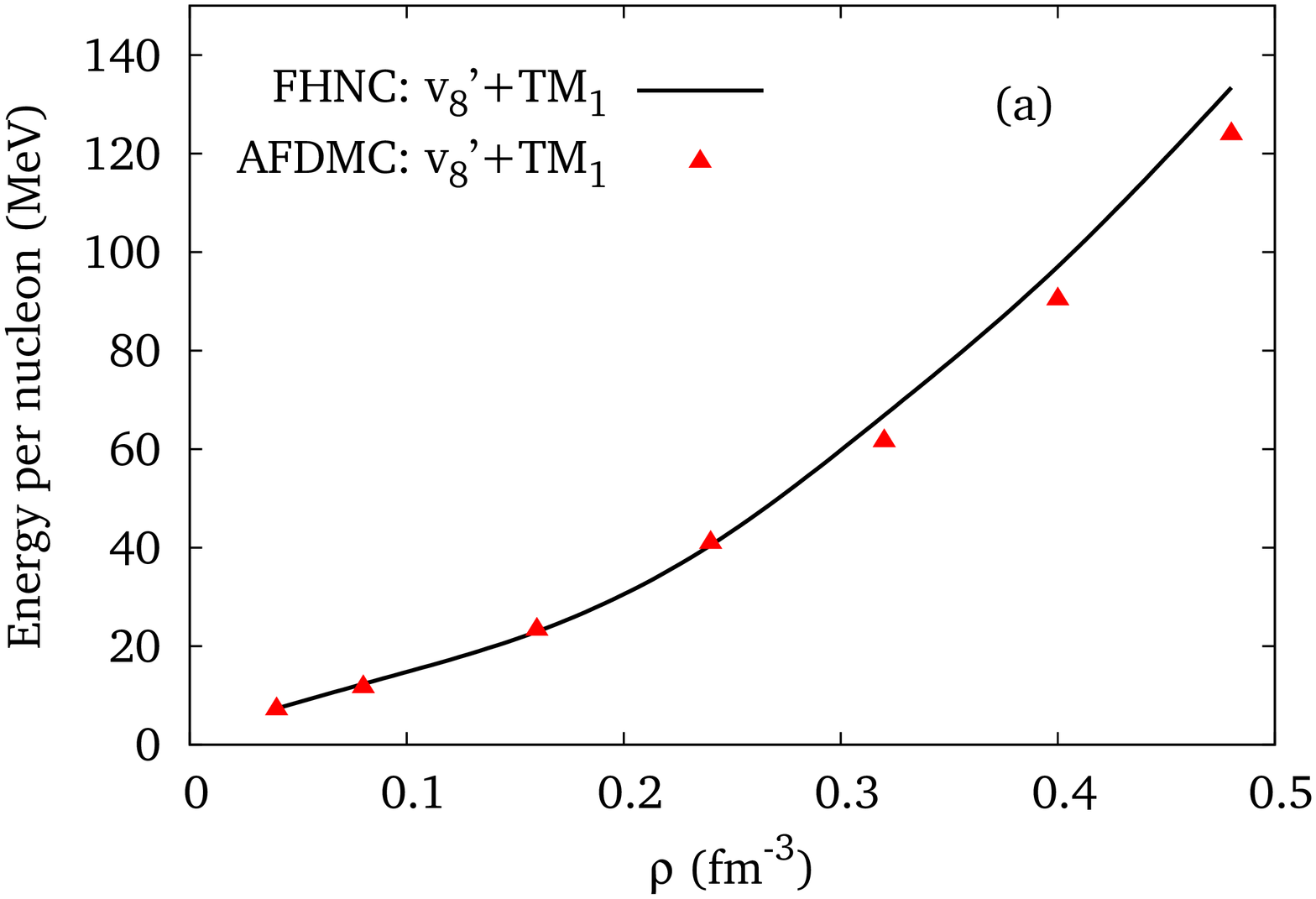}
\hspace{5mm}
\includegraphics[width=6.5cm,angle=270]{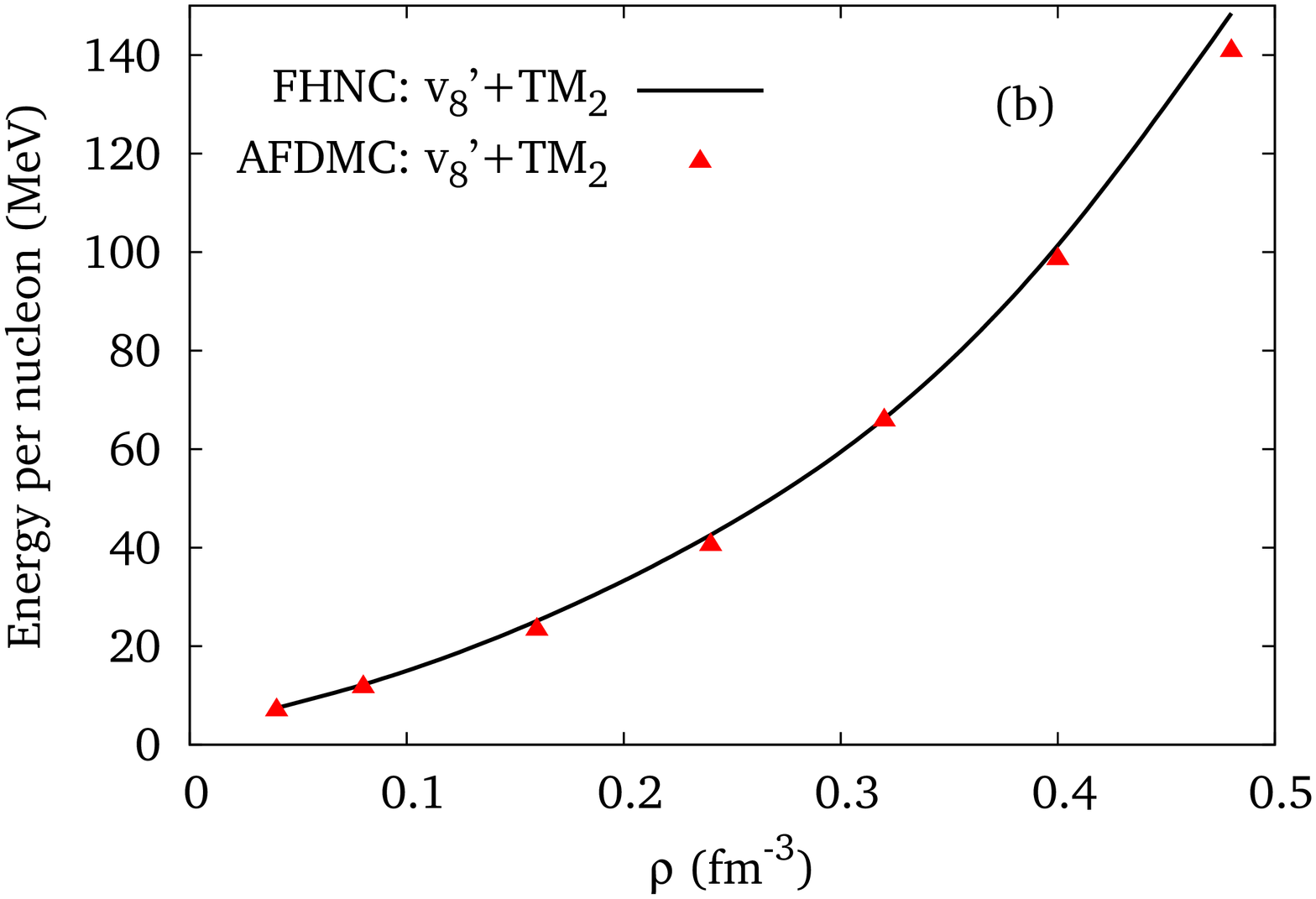}
\hspace{5mm}
\includegraphics[width=6.5cm,angle=270]{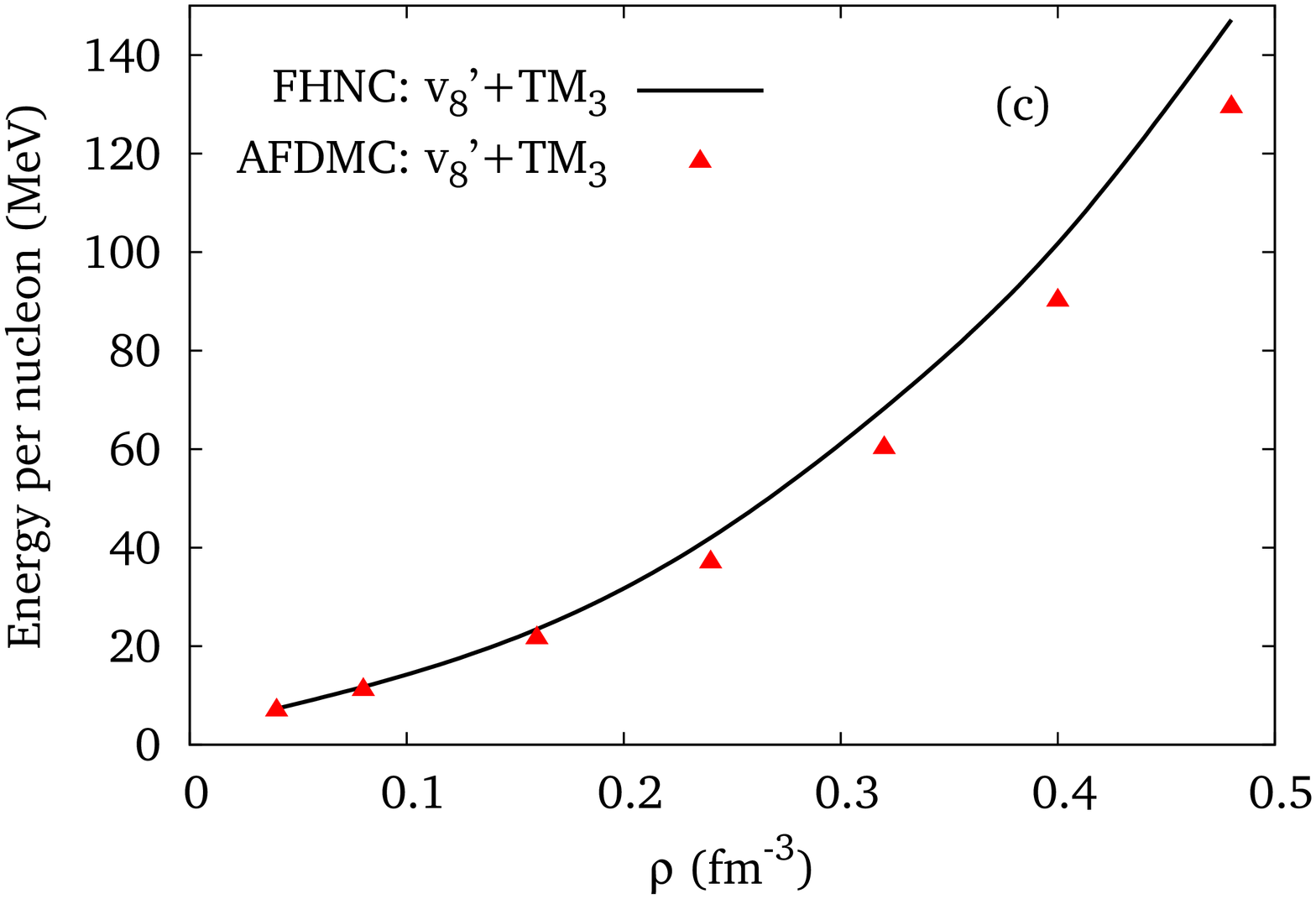}
\caption{(Color online) Equation of state of PNM obtained using the AFDMC (triangles) and FHNC/SOC (solid lines) approaches with the TM$_1^\prime$ (a) and TM$_2^\prime$ (b), TM$_3^\prime$ (c) plus $v_8^\prime$ hamiltonian.\label{fig:tm_compare_PNM}}
\end{figure}

The results of Fig. \ref{fig:tm_compare_PNM}, showing the density dependence of the energy per nucleon in PNM, indicate that,  
once the new constraint on the difference between PB and JF kinetic energies is imposed, the agreement between FHNC/SOC (solid line) and AFDMC (triangles) results 
is very good. 

\begin{figure}[!ht]
\vspace{0.2cm}
\begin{center}
\includegraphics[angle=270,width=9.0cm]{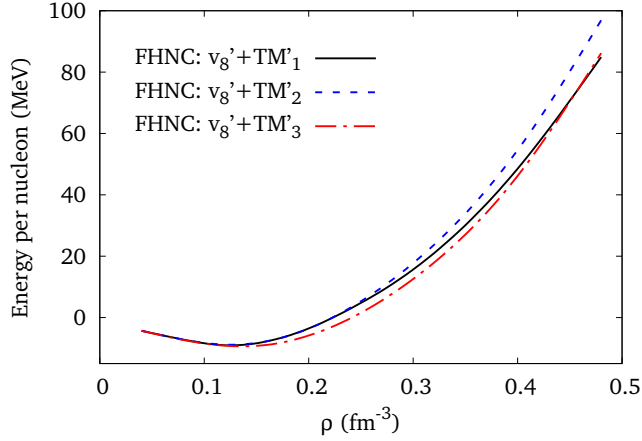}
\caption{(Color online) Equation of state of SNM resulting from FHNC/SOC variational calculations with the TM$^\prime$ plus $v_8^\prime$ hamiltonian.\label{fig:tm_snm}}
\end{center}
\end{figure}

The most striking feature of the results displayed in Fig. \ref{fig:tm_snm} is that, despite the parameters of the three body potentials being different, all SNM EoS obtained from the TM$^\prime$ potential turn out to be very close to each other. This is probably due to the fact that these potentials are designed to reproduce not only the binding energies of $^3$H and $^4$He, but also the n-d doublet scattering length $^2a_{nd}$.

It is remarkable that although the parameters of TM$^\prime$ potentials were not adjusted to reproduce nuclear matter properties, the EoS saturates at densities only slightly lower than 
$\rho_0=0.16\text{fm}^{-3}$, and the compressibilities are in agreement with the experimental value $K\approx 240\,\text{MeV}$. 
On the other hand, the binding energies are larger than the empirical value $E_0=-16\,\text{MeV}$ and rather close to the one obtained from the UIX potential,  $\sim 10\,\text{MeV}$ and shown in Section \ref{sec:ddp}. The numerical values of all these quantities are listed in Table \ref{table:parameters_tm}.

\begin{table}[!ht]
\begin{center}
\caption{Saturation density, binding energy per particle and compressibility of SNM corresponding to the TM$^\prime$ EoS displayed in Fig. \ref{fig:tm_snm}. \label{table:parameters_tm}}
\vspace{0.3cm}
\begin{tabular}{ c c c c} 
\hline 
 	& TM$^\prime_1$ & TM$^\prime_2$ & TM$^\prime_3$\\
\hline
$\rho_0$ (fm$^{-3}$)&  0.12  &  0.13  &  0.14  \\

$E_0$ (MeV) &  -9.0  &  -8.8  &  -9.4 \\

K (MeV) &  266  &  243  &  249   \\ 
\hline
 \end{tabular} 
\end{center}
\end{table}

\vspace{0.1cm}
\clearpage

\subsection{NNLOL chiral potentials}
\label{subsec:nnlolres}
\begin{figure}[!hb]
\includegraphics[width=6.2cm,angle=270]{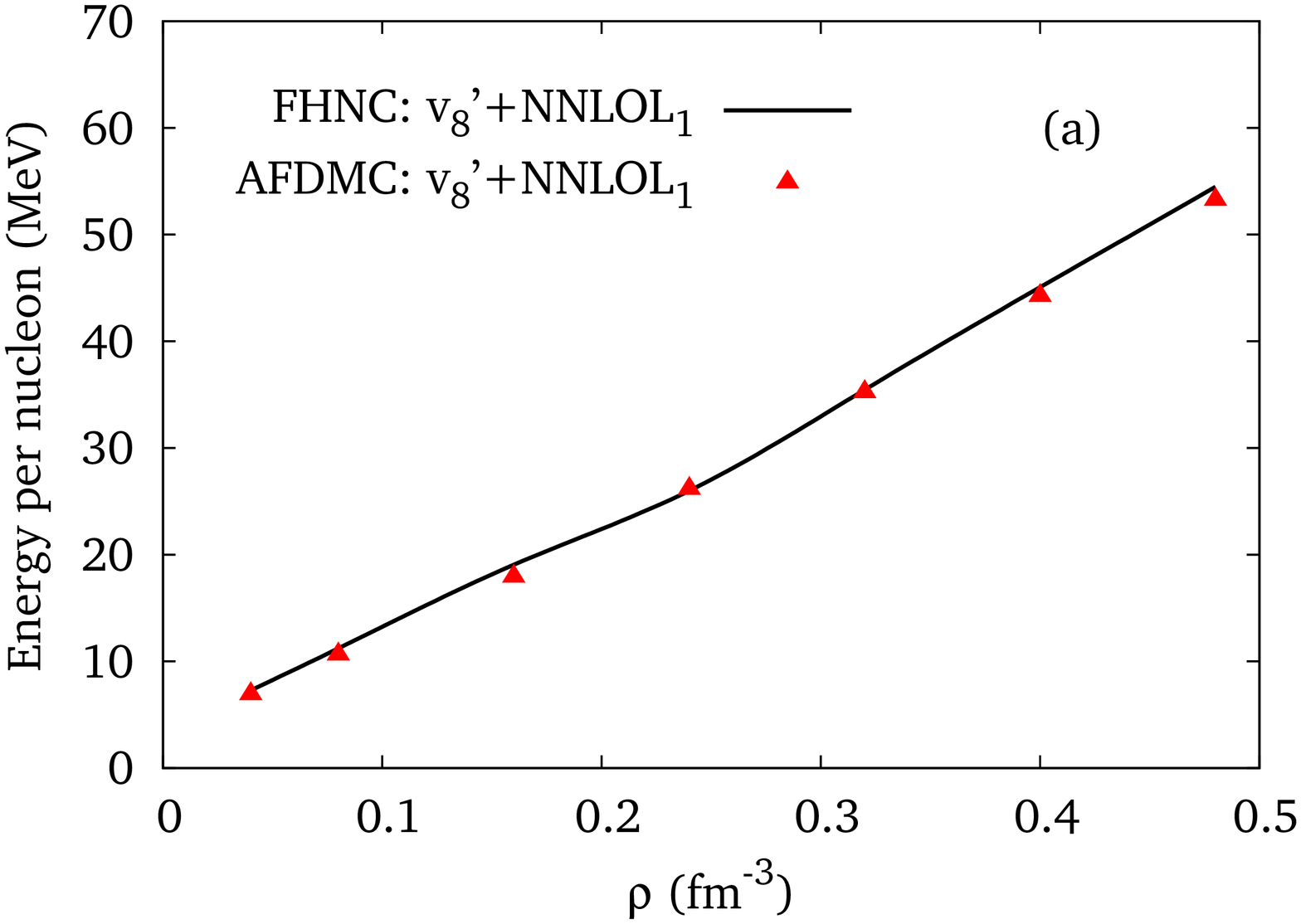}
\includegraphics[width=6.2cm,angle=270]{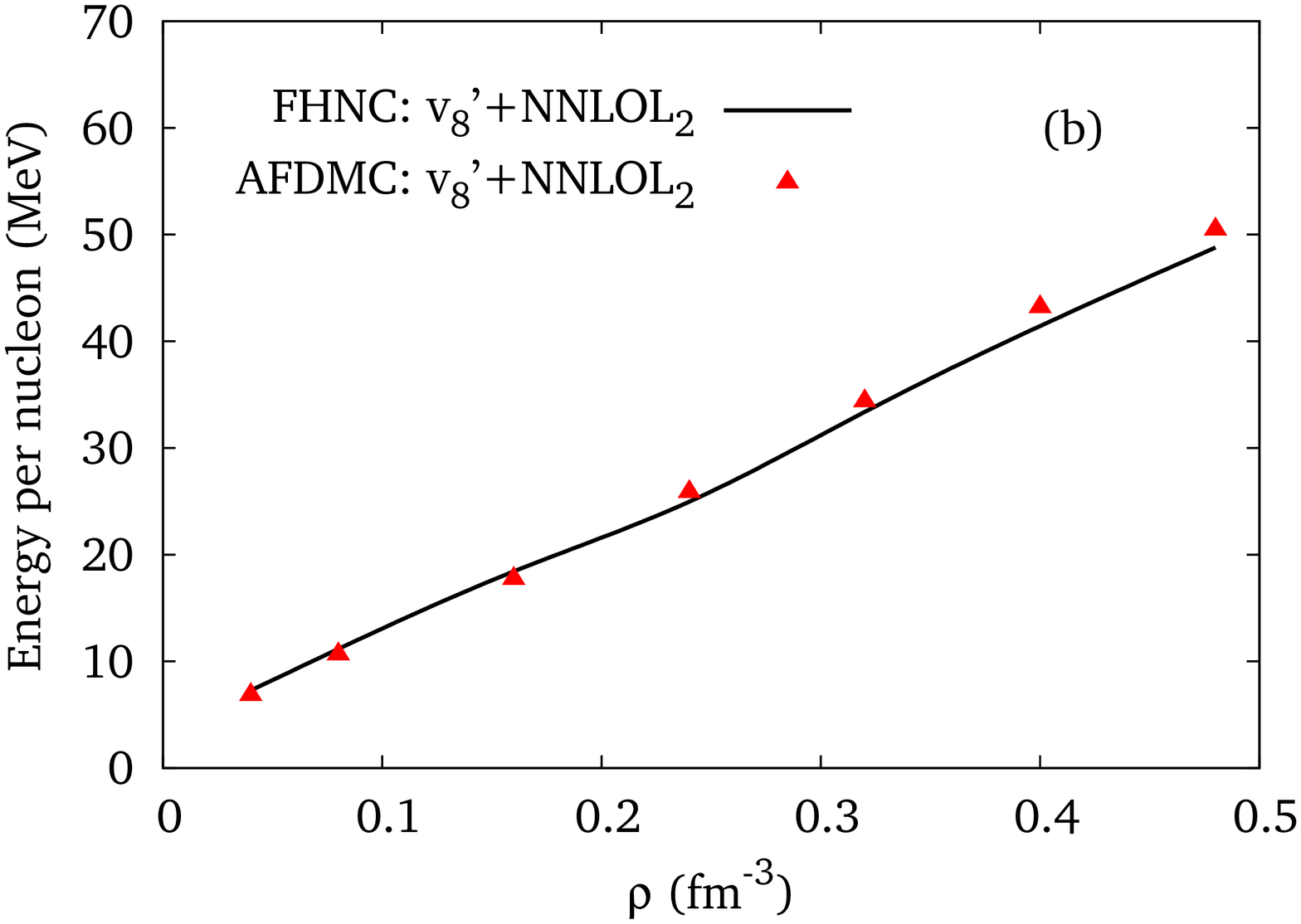}
\includegraphics[width=6.2cm,angle=270]{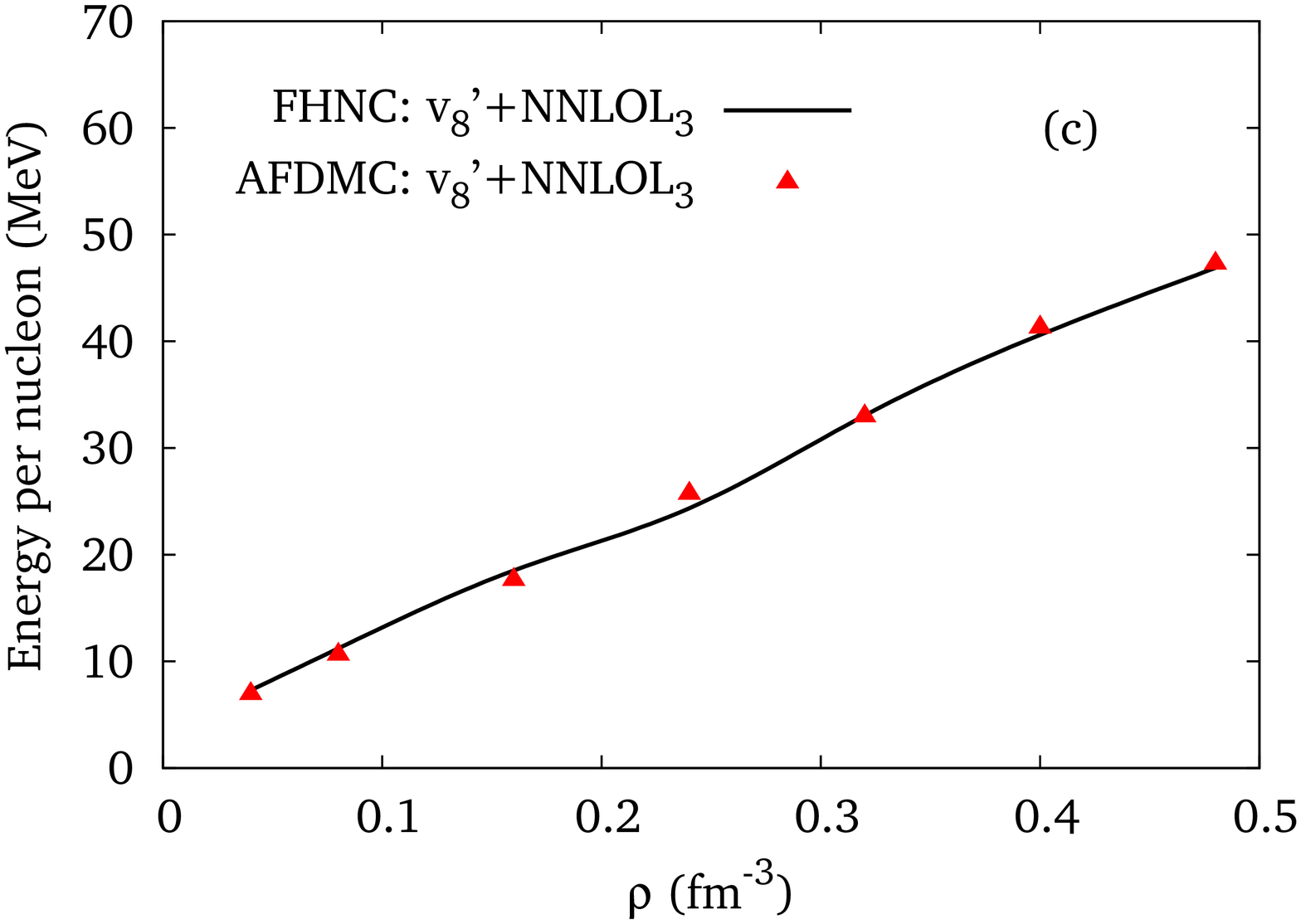}
\includegraphics[width=6.2cm,angle=270]{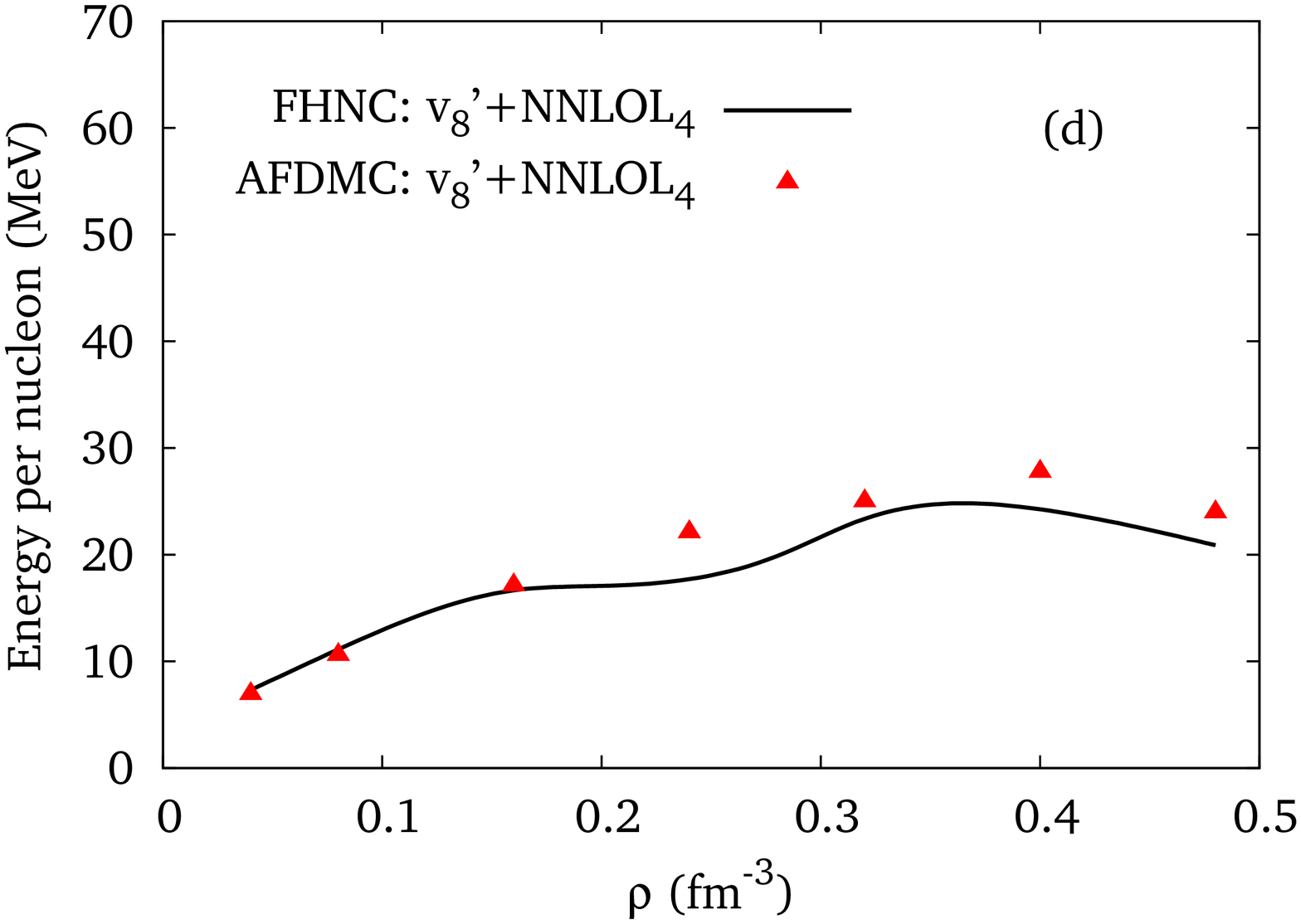}
\caption{(Color online) Same as Fig. \ref{fig:tm_compare_PNM}, but for NNLOL$_1$ (a), for NNLOL$_2$ (b), for NNLOL$_3$ (c) and for NNLOL$_4$ (d) plus $v_8^\prime$ hamiltonian.\label{fig:chiral_compare_PNM}}
\end{figure}

The results displayed in Fig. \ref{fig:chiral_compare_PNM} show that, as in the case of the TM$^\prime$ potentials, the EoS of PNM computed within the AFDMC and FHNC/SOC schemes are very close to each other over the entire density range.

The EoS of Fig. \ref{fig:chiral_compare_PNM} are softer than those obtained from both the TM$^\prime$ (compare to Fig. \ref{fig:tm_compare_PNM}), and  UIX  (see Fig. \ref{fig:eos_pnm}) potentials. This is due to the ambiguity in the term $\hat{V}_E$,  discussed in Section \ref{sec:cont_issue}. 

In the NNLOL$_2$, NNLOL$_3$, and NNLOL$_4$ models the constant $c_E$ is negative. Therefore, the contribution of $\hat{V}_E$ is attractive, making the EoS very soft.  
When $\hat{V}_E$ is repulsive (i.e. $c_E$ is positive), as in the NNLOL$_1$ potential, its contribution is very small and the resulting EoS, while being stiffer than those 
corresponding to the other NNLOL potentials, remains very soft.

The recent astrophysical data of Ref.~ \cite{demorest_10} suggest that the EoS of PNM be at least as stiff as the one obtained with a readjusted version of the effective density-dependent potential of Lagaris and Pandharipande in combination with the Argonne $v_{6}^\prime$ two-body interaction \cite{gandolfi_10}. Therefore, the EoS resulting from chiral NNLOL potentials are not likely to describe a neutron star of mass around $2M_\odot$.

\begin{figure}[!ht]
\vspace{0.2cm}
\begin{center}
\includegraphics[angle=270,width=9.5cm]{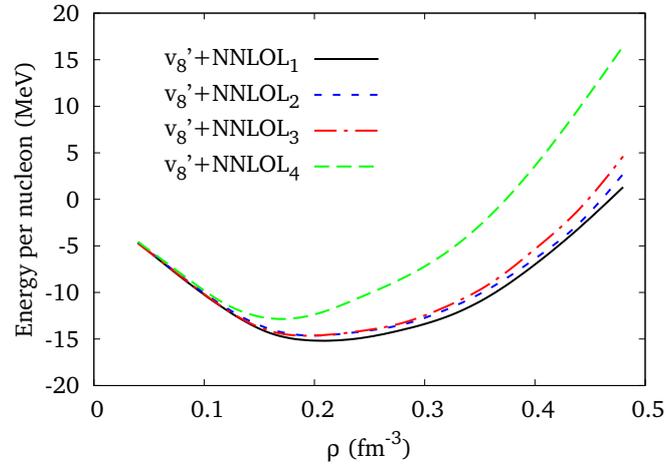}
\caption{(Color online) Same as Fig. \ref{fig:tm_snm}, but for NNLOL plus $v_8^\prime$ hamiltonian.  \label{fig:nnlol_snm}}
\end{center}
\end{figure}

The SNM EoS corresponding to the NNLOL potentials are displayed in Fig. \ref{fig:nnlol_snm}. The fact that the NNLOL$_4$ potential provides the stiffest EoS, while in PNM provided the softest, is not surprising. As discussed in Section \ref{sec:cont_issue}, when the contact term is attractive in PNM, it is repulsive in SNM, and {\em viceversa}. A large cancellation between the repulsive core of the Argonne $v_{8}^\prime$ and the strong attractive contact term contribution of the NNLOL$_4$ potential is observed. This could influence the variational results, which for this particular three-body force could be less accurate than for the other interactions. As the corresponding AFDMC results do not show a similar behavior,  giving a simple physical interpretation to the inflection point at $\rho\simeq0.24$ fm$^{-3}$ resulting from the FHNC/SOC calculations turns out to be difficult.

The results listed in Table \ref{tab:parameters_nnlol} show that none of the chiral NNLOL potentials fulfills the empirical constraints on the SNM EoS. 
All potentials overestimate the saturation density, while the compressibility is compatible with the empirical value only for the NNLOL$_2$ and NNLOL$_3$ models. 
As for the binding energies, they are closer to the experimental value than those obtained using both the UIX and TM$^\prime$ potentials. 

As a final remark, it has to be noticed that using the scalar repulsive term $V_{E}^I$ instead of $V_{E}^\tau$ provides more repulsion, resulting a stiffer EoS. 
As stressed in Section \ref{sec:cont_issue}, this issue needs to be addressed, taking into account all terms that become equivalent in the limit of infinite cutoff only. 
Moreover, since the discrepancies among these terms are of the same order as the NNNLO term of the chiral expansion, other contact terms have to be included  \cite{girlanda_11}.

\begin{table}[!ht]
\begin{center}
\caption{Saturation density, the binding energy per particle, and the compressibility related to the NNLOL Eos displayed in Fig. \ref{fig:nnlol_snm}. \label{tab:parameters_nnlol}}
\vspace{0.3cm}
\begin{tabular}{c c c c c} 
\hline 
 	& NNLOL$_1$ & NNLOL$_2$ & NNLOL$_3$ & NNLOL$_4$ \\
\hline
$\rho_0$ (fm$^{-3}$)&  0.21  &  0.20  &  0.19 & 0.17 \\

$E_0$ (MeV) &  -15.2  &  -14.6  &  -14.6 &  -12.9\\

K (MeV) &  198  &  252  &  220  & 310\\ 
\hline
 
\end{tabular} 
\vspace{0.1cm}
\end{center}
\end{table}

\phantom{AAAAAAA}

\phantom{AAAAAAA}

\phantom{AAAAAAA}

\phantom{AAAAAAA}

\newpage             
\thispagestyle{empty}  

% Chapter 4 %%%%%%%%%%%%%%%%%%%%%%%%%%%%%%
\chapter{Response of nuclear matter}
\label{chapt:resp}
\label{chapt:risposta}

The understanding of neutrino-nucleons interactions is required in both astrophysics, mainly in the context of supernovae explosions and of the cooling of neutron stars, and high energy physics, in particular to reduce the systematic uncertainties of the data analysis. 

In the low-momentum transfer regime ($|\mathbf{q}|$ of the order of tens MeV), the non relativistic limit of the weak current matrix element is expected to be applicable. The nuclear response to weak probes delivering energy $\omega$ and momentum \textbf{q} at leading order in $|\mathbf{q}|/m$ reads
\begin{align}
S(\textbf{q},\omega)=\frac{1}{A}\sum_f |\{ \Psi_f | \hat{O}_{\textbf{q}}| \Psi_0\} |^2 \delta(\omega+E_0-E_n)\,,
\label{eq:resp_def}
\end{align}
In the above equation $A$ is the particle number and $\hat{O}_{\textbf{q}}$ the one--body weak operator that induces a transition from the ground state $|\Psi_0\}$ to the excited state $|\Psi_f\}$, which are eigenvalues of the nuclear hamiltonian with energies $E_0$ and $E_f$, respectively (see Eq. (\ref{eq:exact_h})). 

The non relativistic Fermi (F) and Gamow-Teller (GT) operators describing low energy weak interactions are
\begin{align}
\hat{O}_{\mathbf{q}}^F &= \sum_i \hat{O}_{\mathbf{q}}^F(i)=g_V\sum_i e^{i \mathbf{q} \cdot \mathbf{r}_i}\tau^{+}_i,\\
\hat{O}_{\mathbf{q}}^{GT} &= \sum_i \hat{O}_{\mathbf{q}}^{GT}(i)=g_A\sum_i e^{i \mathbf{q} \cdot \mathbf{r}_i} \vec{\sigma}_i \tau^{+}_i,
\label{eq:f_gt_def}
\end{align}
where $g_V=1.00$ and $g_A=1.26$ are the form factors at zero momentum transfer, while $\tau^{+}_i$ is the isospin-rising operator acting on the $i$-th
nucleon. 

The spin longitudinal and spin transverse components of the Gamow-Teller response functions are defined as the components parallel to $\mathbf{q}$ and orthogonal to $\mathbf{q}$, respectively. They can differ significantly at larger values of $|{\bf q}|$ and, in principle, have to be calculated separately. However, when not otherwise specified, with 
``Gamow-Teller response'' we refer to the total response,  given by the sum over the cartesian components:
\begin{align}
S^{GT}(\textbf{q},\omega)=\sum_{\alpha=x,y,z} \frac{1}{A}\sum_f |\{ \Psi_f | \hat{O}_{\textbf{q}\,\alpha}^{GT}| \Psi_0\} |^2 \delta(\omega+E_0-E_n)\,,
\label{eq:resp_gttot}
\end{align}

The differences between the integrated spin-longitudinal and spin-transverse responses will be discussed in Section \ref{sec:sum_rule}.

Calculations of the charged current weak response have been recently performed in Refs. \cite{cowell_04,benhar_09}. In  the high energy domain the short range correlations of the CBF play a major role, while for low neutrino energy $\leq 10$MeV, the response is mainly affected by long range correlations that lead to excitation of collective modes, taken into account within the correlated Tamm-Dancoff approximation (CTDA).

Even including the spin-isospin dependent correlations of Eq. (\ref{eq:Foperator}), generally the CBF states are not eigenstates of $\hat{H}$. However, it has been argued by Feenberg that the Hamiltonian matrix has smaller non diagonal elements in CBF than in the non interacting FG basis. Hence, perturbative many-body calculations are expected to converge more rapidly in CBF. Neglecting orthogonality corrections, the CBF weak operators matrix elements read
\begin{equation}
\{ \Psi_f | \hat{O}_{\textbf{q}}| \Psi_0 \} \to\frac{\langle \Psi_f | \mathcal{F}^\dagger \hat{O}_{\textbf{q}}  \mathcal{F} | \Psi_0\rangle}{\sqrt{\langle \Psi_0 | \mathcal{F}^\dagger \mathcal{F} | \Psi_0\rangle\langle \Psi_f | \mathcal{F}^\dagger \mathcal{F} | \Psi_f\rangle}}\, .
\end{equation}

Following Refs. \cite{cowell_04,benhar_09}, in this Thesis we only consider transitions between the correlated ground-state and correlated 1particle-1hole ($1p-1h$) excited states. The $n\geq 2$ particle-hole correlated states give a smaller contribution, mainly at large excitation energy, the size of which will be estimated at a later stage, studying the sum rules of
the weak response. 

The CBF matrix element between the ground-state and $1p-1h$ excitation reads
\begin{equation}
(\Psi_{p_m;h_i}|\hat{O}_\mathbf{q}|0)=\frac{\langle \Psi_{p_m; h_i}| \mathcal{F}^\dagger \hat{O}_\mathbf{q}\mathcal{F} | \Psi_0\rangle}{\sqrt{\langle \Psi_0 | \mathcal{F}^\dagger \mathcal{F} | \Psi_0\rangle\langle \Psi_{p_m; h_i}| \mathcal{F}^\dagger \mathcal{F} | \Psi_{p_m; h_i}\rangle}}\, ,
\label{eq:cbf_1p1h}
\end{equation}
where $p_m$ and $h_i$ denote the whole set of quantum numbers of the single nucleon state, namely the momentum, the spin and the isospin projections along the $z-$axis.

This quantity, entering all our calculations of the response function, will allow us to define the effective weak operators, as discussed in the next Section.

\section{Effective weak operators}
\label{sec:ewo}
We define the effective weak operators  $\hat{O}_{\textbf{q}}^{eff}$ through the relation
\begin{equation}
\langle \Psi_{p_m;h_i} | \hat{O}_{\mathbf{q}}^{eff}| \Psi_0 \rangle \equiv \frac{\langle \Psi_{p_m; h_i } | \mathcal{F}^\dagger \hat{O}_{\mathbf{q}}  \mathcal{F} | \Psi_0\rangle}{\sqrt{\langle \Psi_0 | \mathcal{F}^\dagger \mathcal{F} | \Psi_0\rangle\langle \Psi_{p_m; h_i } | \mathcal{F}^\dagger \mathcal{F} | \Psi_{p_m; h_i }\rangle}}
\label{eq:cbf_1p1heff}
\end{equation}

As for the calculation of the hamiltonian expectation value, a cluster expansion of the weak operator correlated matrix element can be performed \cite{cowell_03}. The smallness parameters in this case are $f^{c}_{ij}-1$ and $f^{p}_{ij}$. Following \cite{fantoni_87}, we rewrite the matrix element of Eq. (\ref{eq:cbf_1p1heff}) as 
\begin{equation}
\frac{\langle \Psi_{p_m;h_i}  | \mathcal{F}^\dagger \hat{O}_{\mathbf{q}} \mathcal{F}|\Psi_0\rangle}{\langle \Psi_0 | \mathcal{F}^\dagger \mathcal{F}|\Psi_0\rangle}\cdot\frac{\sqrt{\langle \Psi_0 | \mathcal{F}^\dagger \mathcal{F}|\Psi_0\rangle}}{\sqrt{\langle \Psi_{p_m;h_i}  | \mathcal{F}^\dagger \mathcal{F}| \Psi_{p_m;h_i} \rangle}}\equiv R_a \cdot R_b\, .
\end{equation}
A cluster expansion of both $R_a$ and $R_b$ needs to be performed, so that the order $n$ of the perturbative series in terms of  $f^{c}_{ij}-1$ and $f^{p}_{ij}$ is given by
\begin{equation}
(\Psi_{p_m;h_i}|\hat{O}_{\mathbf{q}}|\Psi_0)^{(n)}=\sum_{l,m} R_{a}^{(l)}\cdot R_{b}^{(m)}\delta_{l+m=n}\, .
\label{eq:wmel}
\end{equation}

\subsection{Cluster expansion of $R_b$}
It is convenient to carry out the cluster expansion of $R_b$ first, since it does not involves the transition operator.
The product of correlation operators can be written as in Eq. \ref{eq:FF_den}, associated with the denominator of the two-body distribution function
\begin{align}
\mathcal{F}^\dag \mathcal{F}&=1+\sum_{i<j}X^{(2)}(x_i,x_j)+\sum_{i<j<k}X^{(3)}(x_i,x_j,x_k)+\dots
\label{eq:FF_exp_r}
\end{align}
Because of the symmetry of the ground state wave functions, one finds 
\begin{align}
\sum_{i<j<\,\dots}\langle \Psi_0 | X^{(N)}(x_i,\dots,x_A)|\Psi_0\rangle&=
\frac{A!}{(A-N)!N!}\langle \Psi_0 | X^{(N)}(x_1,\dots,x_N)|\Psi_0\rangle
\end{align}
and an analogous relation holds for the $1p-1h$ state.

Using the results of Appendix \ref{app:slat}, it is possible to extract $N$ particles from the Slater determinant of the ground state, to obtain
\begin{align}
\langle \Psi_0 | \mathcal{F}^\dag \mathcal{F} |\Psi_0\rangle = &\sum_N \frac{1}{N!}\sum_{n_i}\int dx_{1,\dots,N}
\,\psi^{*}_{n_1}(x_1)\,\dots\,\psi^{*}_{n_N}(x_N)X^{(N)}(x_1,\dots,x_N)\times\nonumber\\
&\mathcal{A}[\psi_{n_1}(x_1)\,\dots\,\psi_{n_N}(x_N)]\, .
\end{align}

In order to make the simplification among disconnected diagrams manifest, it is worth introducing a new index, $\bar{n}_i=1,\dots,h_i-1,h_i+1,\dots,A$, labeling the $A-1$ states of a system lacking both single-particle states $p_m$ and $h_i$. Thus, adding and removing the contribution of the hole state, the $N$-body term of the numerator reads 
\begin{align}
&\langle \Psi_0 | \mathcal{F}^\dagger \mathcal{F}|\Psi_0\rangle=
\sum_{\bar{n}_i}\frac{1}{N!}\int dx_{1,\dots,N}\Big[\nonumber\\
&\quad\psi^{*}_{\bar{n}_1}(x_1)\,\dots\,\psi^{*}_{\bar{n}_N}(x_N)X^{(N)}(x_1,\dots,x_N) \mathcal{A}[\psi_{\bar{n}_1}(x_1)\,\dots\,\psi_{\bar{n}_N}(x_N)]+\nonumber\\
&\quad \psi^{*}_{h_i}(x_1) \dots \psi^{*}_{\bar{n}_N}(x_N)X^{(N)}(x_1,\dots,x_N) \mathcal{A}[\psi_{h_i}(x_1)\dots\psi_{\bar{n}_N}(x_N)]+\,\cdots\,+\nonumber\\
&\quad\psi^{*}_{\bar{n}_1}(x_1) \dots \psi^{*}_{h_i}(x_N)X^{(N)}(x_1,\dots,x_N) \mathcal{A}[\psi_{\bar{n}_1}(x_1)\,\dots\,\psi_h(x_N)]\Big]\, .
\label{eq:Rb_ce}
\end{align}

Although the sum of the cluster terms precisely gives the denominator of the two-body distribution function, the diagrammatic rules stemming from Eq. (\ref{eq:Rb_ce}) slightly differ from those of $g^p(r_{12})$. 
\begin{itemize}
\item[$\bullet$]Wavy lines represent both scalar and operator correlations, $f^{c}_{ij}-1$ and $f^{p}_{ij}$ and two kinds of vertex and exchange lines are present. 
\item[$\bullet$]The bare vertex, arising from $\psi_{\bar{n}_i}(x_i)^*\psi_{\bar{n}_i}(x_i)$, does not involve the hole state $h_j$; it tends to the ``standard'' vertex in the thermodynamic limit only. On the other hand, the direct term of the hole state, $\psi_{h_i}(x_i)^*\psi_{h_i}(x_i)$, is represented by an $h_j$-vertex: a closed loop around the internal point $i$. 
\item[$\bullet$]The hole state is not exchanged in the bare exchange lines, but there is an additional $h_i$-exchange line, coming from $\psi_{h_i}(x_i)^*\psi_{h_i}(x_j)$, which connects points $i$ and $j$. At most one $h_i$ vertex or $h_i$ exchange line can appear in a given diagram.
\end{itemize}

It it possible to factorize the sum of cluster diagrams in two subsets, as shown in Fig. {\ref{fig:Rb_n_fact}. The first subset, denoted with $1+\sum_C (h_i)$, in addition to the unity, contains all the connected diagrams having one $h_i$ vertex or one $h_i$-exchange line. All the remaining diagrams, both connected and disconnected, belong to the second subset. 

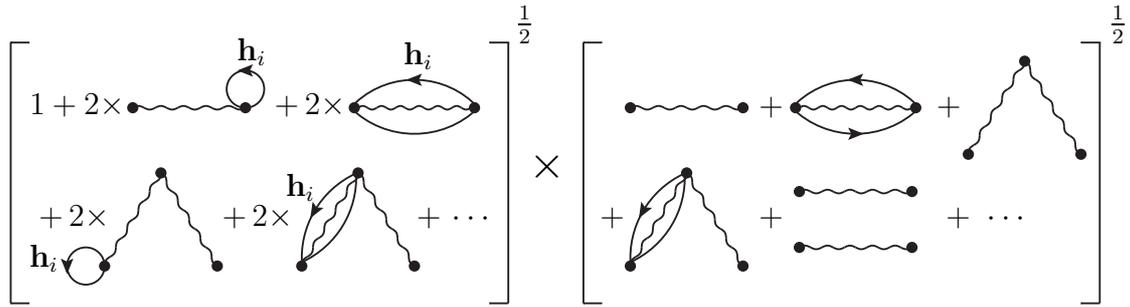
\begin{figure}[!h]
\begin{center}
  \begin{picture}(100,200) (160,-105)
    \SetWidth{1.0}
    \SetColor{Black}
    \SetScale{0.7}	
    \unitlength=0.7 pt
	\Line(-5,70)(-5,-70)
	\Line(-5,70)(5,70)
	\Line(-5,-70)(5,-70)
	\Line(260,70)(260,-70)
	\Line(250,70)(260,70)
	\Line(250,-70)(260,-70)
    \Text(5,30)[lb]{$1+2\times$}
    \Photon(60,35)(120,35){1}{5}
    \Vertex(60,35){3}
    \Vertex(120,35){3}
    \ArrowArc(120,45)(10,270,-90)
    \Text(116,58)[lb]{$\mathbf{h}_i$}
    \Text(136,30)[lb]{$+\,2\times$}
    \Vertex(178,35){3}
    \Vertex(242,35){3}
    \Photon(180,35)(240,35){1}{6}
    \Arc(210,69.286)(48.286,-134.76,-45.24)
    \Text(205,56)[lb]{$\mathbf{h}_i$}
    \ArrowArc(210,3.967)(46.033,42.388,137.612)
    \Text(10,-30)[lb]{$+\,2\times$}
    \Vertex(45,-50){3}
    \Vertex(75,0){3}
    \Vertex(105,-50){3}
    \Photon(45,-50)(75,0){1}{5}
    \Photon(75,0)(105,-50){1}{5}
    \ArrowArc(35,-50)(10,360,0)
    \Text(5,-53)[lb]{$\mathbf{h}_i$}
	\Text(108,-30)[lb]{$+\,2\times$}
    	\Photon(150,-50)(180,0){1}{5}
	\Photon(180,0)(210,-50){1}{5}
	\ArrowArc(200,-45.57)(50,115,185)
	\Arc(129,-2.57)(50,-65,3)
	\Vertex(150,-50){3}
       	\Vertex(180,0){3}
        \Vertex(210,-50){3}
         \Text(142,-15)[lb]{$\mathbf{h}_i$}

	\Text(212,-30)[lb]{$+\,\cdots$}

	\Text(265,70)[lb]{$\frac{1}{2}$}
	\Text(272,-5)[lb]{\Large$\times$}
	\Line(300,70)(300,-70)
	\Line(300,70)(310,70)
	\Line(300,-70)(310,-70)
	\Line(575,70)(575,-70)
	\Line(565,70)(575,70)
	\Line(565,-70)(575,-70)
	\Photon(325,35)(385,35){1}{5}
	\Vertex(325,35){3}
	\Vertex(385,35){3}
	\Text(395,30)[lb]{$+$}
    	\Vertex(413,35){3}
    	\Vertex(477,35){3}
	\Photon(415,35)(475,35){1}{6}
        \ArrowArc(445,69.286)(48.286,-134.76,-45.24)
	\ArrowArc(445,3.967)(46.033,42.388,137.612)
	\Text(490,30)[lb]{$+$}
	\Vertex(505,10){3}
	\Vertex(535,60){3}
	\Vertex(565,10){3}
	\Photon(505,10)(535,60){1}{5}
	\Photon(535,60)(565,10){1}{5}
    	\Text(310,-30)[lb]{$+$}
    	\Photon(325,-50)(355,0){1}{5}
	\Photon(355,0)(385,-50){1}{5}
	\ArrowArc(375,-45.57)(50,115,185)
	\Arc(304,-2.57)(50,-65,3)
	\Vertex(325,-50){3}
       	\Vertex(355,0){3}
        \Vertex(385,-50){3}
    	\Text(395,-30)[lb]{$+$}
    	\Photon(415,-40)(475,-40){1}{5}
	\Photon(415,-10)(475,-10){1}{5}
	\Vertex(415,-40){3}
	\Vertex(415,-10){3}
	\Vertex(475,-10){3}
	\Vertex(475,-40){3}
    	\Text(495,-30)[lb]{$+\,\,\cdots$}
	\Text(582,70)[lb]{$\frac{1}{2}$}

\end{picture}
\end{center}    
\vspace{-1.5cm}
\caption{Factorizations of the sum of cluster diagrams contributing to the numerator of $R_b$. The square brackets on the left enclose $1+\sum_C(h_i)$ from which the denominator, $1+\sum_C(p_m)$, can be obtained replacing $\mathbf{h}_i$ with $\mathbf{p}_m$.\label{fig:Rb_n_fact}}
\label{fig:2b_num_rb}
\end{figure}

The diagrammatic cluster expansion rules for the denominator can be readily obtained from those of the numerator by replacing the hole state  $h_i$ with the particle state $p_m$. The sum of the denominator's cluster diagrams can be factorized in two subsets. The first, $1+\sum_C (p_m)$, is made of the unity and of connected diagrams with one $p_m$ vertex or one $p_m$-exchange line, while the second cancel the corresponding one from the numerator.

Therefore, the following analytic expression for $R_b$ is obtained
\begin{align}
R_b=\sqrt{\frac{1+\sum_C (h_i)}{1+\sum_C (p_m)}}\, ,
\label{eq:Rb_frac_sch}
\end{align}
At first order in $\sum_p f^p-1$, the explicit expression of Eq. (\ref{eq:Rb_frac_sch}), diagrammatically represented in Fig.  \ref{fig:Rb_primo_g}, reads
\begin{align}
R_{b}=\sqrt{\frac{1+\frac{2\rho}{\nu}\int d\mathbf{r}_{12} \sum_{\alpha_1}\langle \alpha_1 \alpha_{h_i}|g_{12}(1-\hat{P}_{12}\ell_{12}e^{i\mathbf{h}_i\cdot\mathbf{r}_{12}})|\alpha_1 \alpha_{h_i}\rangle}{1+\frac{2\rho}{\nu}\int d\mathbf{r}_{12} \sum_{\alpha_1}\langle \alpha_1 \alpha_{p_m}|g_{12}(1-\hat{P}_{12}\ell_{12}e^{i\mathbf{p}_m\cdot\mathbf{r}_{12}})|\alpha_1 \alpha_{p_m}\rangle}}\, .
\end{align}
where $|\alpha_i\rangle$ denotes the spin-isospin state of particle $i$.

\begin{figure}[!h]
\begin{center}
\begin{picture}(100,100) (160,-30)
	\SetWidth{1.0}
	\SetColor{Black}
	\SetScale{0.7}	
	\unitlength=0.7 pt
	\Line(-5,70)(-5,0)
	\Line(-5,70)(5,70)
	\Line(-5,0)(5,0)
	\Line(255,70)(255,0)
	\Line(245,70)(255,70)
	\Line(245,0)(255,0)
	\Text(5,30)[lb]{$1+2\times$}
	\Photon(60,35)(120,35){1}{5}
	\Vertex(60,35){3}
	\Vertex(120,35){3}
	\ArrowArc(120,45)(10,270,-90)
	\Text(116,58)[lb]{$\mathbf{h}_i$}
	\Text(136,30)[lb]{$+\,2\times$}
	\Vertex(178,35){3}	
	\Vertex(242,35){3}
	\Photon(180,35)(240,35){1}{6}
	\Arc(210,69.286)(48.286,-134.76,-45.24)
	\Text(205,58)[lb]{$\mathbf{h}_i$}
	\ArrowArc(210,3.967)(46.033,42.388,137.612)
	\Text(265,58)[lb]{$\frac{1}{2}$}
	\Text(274,28)[lb]{\Large$/$}
	\Line(300,70)(300,0)
	\Line(300,70)(310,70)
	\Line(300,0)(310,0)
	\Line(560,70)(560,0)
	\Line(550,70)(560,70)
	\Line(550,0)(560,0)
	\Text(310,30)[lb]{$1+2\times$}
	\Photon(365,35)(425,35){1}{5}
	\Vertex(365,35){3}
	\Vertex(425,35){3}
	\ArrowArc(425,45)(10,270,-90)
	\Text(415,60)[lb]{$\mathbf{p}_m$}
	\Text(441,30)[lb]{$+\,2\times$}
	\Vertex(483,35){3}
	\Vertex(547,35){3}
	\Photon(485,35)(545,35){1}{6}
	\Arc(515,69.286)(48.286,-134.76,-45.24)
	\Text(510,60)[lb]{$\mathbf{p}_m$}
	\ArrowArc(515,3.967)(46.033,42.388,137.612)
	\Text(570,58)[lb]{$\frac{1}{2}$}
\end{picture}
\end{center}    
\vspace{-1.5cm}
\caption{First order diagrams in terms of $\sum_p f^p-1$ contributing to $R_b$.\label{fig:Rb_primo_g}}
\label{fig:2b_num_fo}
\end{figure}
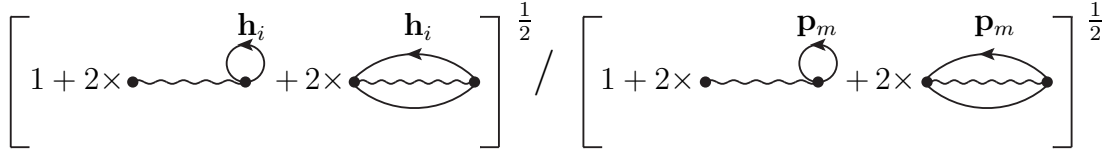

\subsection{Expansion of $R_a$}
Apart form the square root, the denominator of $R_a$ is identical to the numerator of $R_b$, hence the same diagrammatic rules apply. In the numerator, on the other hand, the weak transition operator $\hat{O}_{\mathbf{q}}$ appears. Using the symmetry of the $1p-1h$ and ground-state wave functions, we can write 
\begin{equation}
\langle \Psi_{p_m;h_i}  | \mathcal{F}^\dagger \hat{O}_{\mathbf{q}} \mathcal{F}|\Psi_0\rangle=
A\langle \Psi_{p_m;h_i}  | \mathcal{F}^\dagger \hat{O}_{\mathbf{q}}(1) \mathcal{F}|\Psi_0\rangle\, ,
\end{equation}
with $\hat{O}_{\mathbf{q}}(1)$, defined in Eq. (\ref{eq:f_gt_def}), acts on particle $1$ only. 

The cluster expansion of $\mathcal{F}^\dag \hat{O}_{\mathbf{q}}(1) \mathcal{F}$ reads
\begin{align}
&\mathcal{F}^\dag \hat{O}_{\mathbf{q}}(1) \mathcal{F}=\hat{O}_{\mathbf{q}}(1)+\sum_{1<i}\hat{X}^{(2)}_1(x_1;x_i)+\sum_{1<i<j}\hat{X}^{(3)}_1(x_1;x_i,x_j)+\dots
\label{eq:FFO1_exp}
\end{align}

Using again the results of Appendix \ref{app:slat}, the orthogonality of the Slater minors introduced therein and the properties of the antisymmetrization operator $A$, one obtains 
\begin{align}
&A\langle \Psi_{p_m;h_i}  | \mathcal{F}^\dagger \hat{O}_{\mathbf{q}}(1) \mathcal{F}|\Psi_0\rangle = \sum_N \frac{1}{(N-1)!} \sum_{\bar{n}_i}\int d\mathbf{r}_{1,\dots,N}\Big[\nonumber\\
& \quad\psi^{*}_{p_m}(x_1)\,\dots\,\psi^{*}_{\bar{n}_N}(x_N) X^{(N)}_1(x_1;\dots,x_N)
\mathcal{A}[\psi_{h_i}(x_1)\,\dots\,\psi_{\bar{n}_N}(x_N)]+\,\dots\,+\nonumber\\
&\quad\psi^{*}_{\bar{n}_1}(x_1)\,\dots\,\psi^{*}_{p_m}(x_N)] X^{(N)}_1(x_1;\dots,x_A)
\mathcal{A}[\psi_{\bar{n}_1}(x_1)\,\dots\,\psi_{h_i}(x_N)]\Big]\, .
\label{eq:Ra_exp}
\end{align}
where the index $\bar{n}_i$ has been defined just above Eq. (\ref{eq:Rb_ce}). The diagrammatic rules for $R_b$ need to be modified in order to include the weak transition operator acting on particle $1$ and to deal with the $p-h$ exchange line.

\begin{itemize}
\item  The weak transition operator $\hat{O}_{\mathbf{q}}(1)$, carrying momentum $\mathbf{q}$ and a spin-isospin operator, is attached at point $1$. It is represented by a dashed line and an arrow indicating the flow of the momentum $\mathbf{q}$.
\item[$\bullet$] All internal points but point $1$ must be reached by at least one wavy line.
\item Like in the diagrammatic expansion of $R_b$, bare vertices are due to $\psi^{*}_{\bar{n}_i}(x_i)\psi_{\bar{n}_i}(x_i)$ and are lacking of the hole state; in the bare exchange line the hole state is absent as well. If the diagram is finite in the thermodynamic limit, both vertices and exchange lines tend to the ``standard'' ones of the FR cluster expansion. There are neither $h$- and $p$- vertices, nor $h$- and $p$- exchange lines.
\item There is one single $ph$-exchange line connecting points $i$ and $j$, arising from $\psi^{*}_{p_m}(x_i)\psi_{h_i}(x_j)$, or one single $ph$ vertex, coming from $\psi^{*}_{p_m}(x_i)\psi_{h_i}(x_i)$. In order for a given diagram not to vanish, either the $ph$-exchange line or the $ph$-vertex need to be connected with particle $1$. 
\end{itemize}

As in the case of $R_b$, the sum of cluster diagrams can be factorized in two subsets, shown in Fig. \ref{fig:Ra_n_fact}. The first, indicated with $\sum_C (q,p_m,h_i)$, consists of the connected diagrams with particle $1$ and one $ph$-exchange line or one $ph$- vertex. The second one includes both connected and disconnected diagrams without $ph$-exchange and $ph$-exchange lines. 

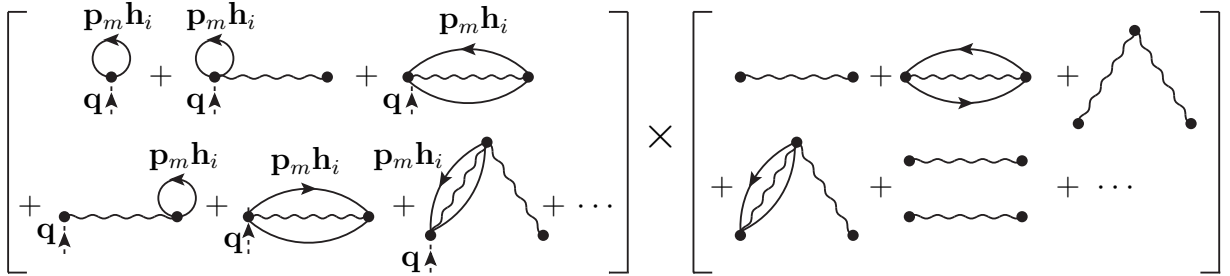
\begin{figure}[!ht]
\begin{center}
  \begin{picture}(100,170) (150,-105)
    \SetWidth{1.0}
    \SetColor{Black}
    \SetScale{0.7}	
    \unitlength=0.7 pt
        \Line(-35,-70)(-35,70)
        \Line(-35,-70)(-25,-70)
        \Line(-35,70)(-25,70)
        \Line(295,-70)(295,70)
        \Line(285,-70)(295,-70)
        \Line(285,70)(295,70)
	\ArrowArc(20,45)(10,270,-90)
	\Vertex(20,35){3}
	\DashArrowLine(20,15)(20,30){3}
	\Text(5,60)[lb]{$\mathbf{p}_m\mathbf{h}_i$}
	\Text(5,15)[lb]{$\mathbf{q}$}
	\Text(40,30)[lb]{$+$}
	\Photon(75,35)(135,35){1}{5}
	\DashArrowLine(75,15)(75,30){3}
	\Text(60,15)[lb]{$\mathbf{q}$}
	\Vertex(75,35){3}
	\Vertex(135,35){3}
	\ArrowArc(75,45)(10,270,-90)	
	\Text(60,60)[lb]{$\mathbf{p}_m\mathbf{h}_i$}
	\Text(150,30)[lb]{$+$}
	\Vertex(178,35){3}
	\Vertex(242,35){3}
	\DashArrowLine(180,15)(180,30){3}
	\Text(165,15)[lb]{$\mathbf{q}$}
	\Photon(180,35)(240,35){1}{6}
	\Arc(210,69.286)(48.286,-134.76,-45.24)
	\Text(195,58)[lb]{$\mathbf{p}_m\mathbf{h}_i$}
	\ArrowArc(210,3.967)(46.033,42.388,137.612)
	\Text(-30,-40)[lb]{$+$}
	\Photon(-5,-40)(55,-40){1}{5}
	\DashArrowLine(-5,-60)(-5,-45){3}
	\Text(-20,-55)[lb]{$\mathbf{q}$}
	\Vertex(-5,-40){3}
	\Vertex(55,-40){3}
	\ArrowArc(55,-30)(10,270,-90)	
	\Text(40,-15)[lb]{$\mathbf{p}_m\mathbf{h}_i$}
	\Text(70,-40)[lb]{$+$}
	\Vertex(93,-40){3}
	\Vertex(157,-40){3}
	\DashArrowLine(93,-60)(93,-35){3}
	\Text(80,-60)[lb]{$\mathbf{q}$}
	\Photon(95,-40)(155,-40){1}{6}
	\Arc(125,-5.7)(48.286,-134.76,-45.24)
	\Text(105,-18)[lb]{$\mathbf{p}_m\mathbf{h}_i$}
	\ArrowArcn(125,-71)(46.033,137.61,42.3882)
	\Text(170,-40)[lb]{$+$}
    	\Photon(190,-50)(220,0){1}{5}
	\Photon(220,0)(250,-50){1}{5}
	\ArrowArc(240,-45.57)(50,115,185)
	\Arc(169,-2.57)(50,-65,3)
	\Vertex(190,-50){3}
       	\Vertex(220,0){3}
        \Vertex(250,-50){3}
       	\DashArrowLine(190,-70)(190,-55){3}
	\Text(175,-70)[lb]{$\mathbf{q}$}
        \Text(160,-15)[lb]{$\mathbf{p}_m\mathbf{h}_i$}
        \Text(250,-40)[lb]{$+\,\cdots$}
	\Text(304,-5)[lb]{\Large$\times$}
        \Line(330,-70)(330,70)
        \Line(330,-70)(340,-70)
        \Line(330,70)(340,70)
        \Line(610,-70)(610,70)
        \Line(610,-70)(600,-70)
        \Line(600,70)(610,70)
	\Photon(355,35)(415,35){1}{5}
	\Vertex(355,35){3}
	\Vertex(415,35){3}
	\Text(425,30)[lb]{$+$}
    	\Vertex(443,35){3}
    	\Vertex(507,35){3}
	\Photon(445,35)(505,35){1}{6}
        \ArrowArc(475,69.286)(48.286,-134.76,-45.24)
	\ArrowArc(475,3.967)(46.033,42.388,137.612)
	\Text(525,30)[lb]{$+$}
	\Vertex(535,10){3}
	\Vertex(565,60){3}
	\Vertex(595,10){3}
	\Photon(535,10)(565,60){1}{5}
	\Photon(565,60)(595,10){1}{5}
    	\Text(340,-30)[lb]{$+$}
    	\Photon(355,-50)(385,0){1}{5}
	\Photon(385,0)(415,-50){1}{5}
	\ArrowArc(405,-45.57)(50,115,185)
	\Arc(334,-2.57)(50,-65,3)
	\Vertex(355,-50){3}
       	\Vertex(385,0){3}
        \Vertex(415,-50){3}
    	\Text(425,-30)[lb]{$+$}
    	\Photon(445,-40)(505,-40){1}{5}
	\Photon(445,-10)(505,-10){1}{5}
	\Vertex(445,-40){3}
	\Vertex(445,-10){3}
	\Vertex(505,-10){3}
	\Vertex(505,-40){3}
    	\Text(525,-30)[lb]{$+\,\,\cdots$}
\end{picture}
\end{center}    
\vspace{-1.5cm}
\caption{Factorization of the diagrams contributing to the numerator of $R_a$; in the square brackets on the left $\sum_C (q,p_mh_i)$ is contained.\label{fig:Ra_n_fact}}
\vspace{1cm}
\end{figure}

As in the calculation of $R_b$, diagrams not containing the weak transition operator, the $ph$-exchange line and the $ph$- vertex cancel with the corresponding ones arising from the denominator. Hence, the following expression for $R_a$ is obtained  
\begin{align}
R_a=\frac{\sum_C (q,p_mh_i)}{1+\sum_C (h_i)}\, .
\label{eq:Ra_frac_sch}
\end{align}

The first order diagrams in $\hat{f}-1$ contributing to $R_a$ are displayed in Fig. \ref{fig:Ra_primo_g}.

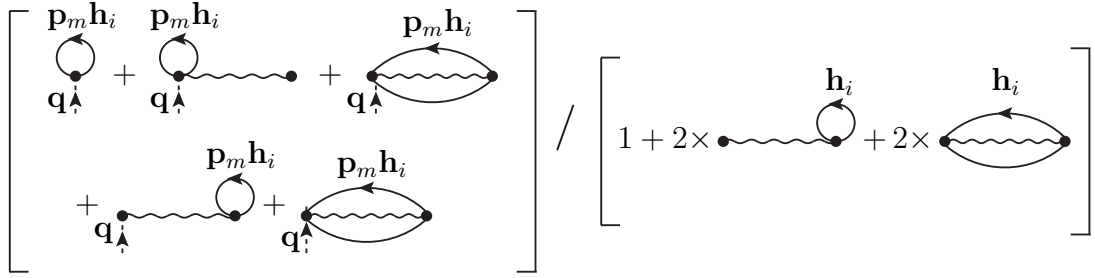
\begin{figure}[!ht]
\begin{center}
  \begin{picture}(100,180) (135,-105)
    \SetWidth{1.0}
    \SetColor{Black}
    \SetScale{0.7}	
    \unitlength=0.7 pt
        \Line(-15,-70)(-15,70)
        \Line(-15,-70)(-5,-70)
        \Line(-15,70)(-5,70)
        \Line(265,-70)(265,70)
        \Line(265,-70)(255,-70)
        \Line(265,70)(255,70)
	\ArrowArc(20,45)(10,270,-90)
	\Vertex(20,35){3}
	\DashArrowLine(20,15)(20,30){3}
	\Text(5,60)[lb]{$\mathbf{p}_m\mathbf{h}_i$}
	\Text(5,15)[lb]{$\mathbf{q}$}
	\Text(40,30)[lb]{$+$}
	\Photon(75,35)(135,35){1}{5}
	\DashArrowLine(75,15)(75,30){3}
	\Text(60,15)[lb]{$\mathbf{q}$}
	\Vertex(75,35){3}
	\Vertex(135,35){3}
	\ArrowArc(75,45)(10,270,-90)	
	\Text(60,60)[lb]{$\mathbf{p}_m\mathbf{h}_i$}
	\Text(150,30)[lb]{$+$}
	\Vertex(178,35){3}
	\Vertex(242,35){3}
	\DashArrowLine(180,15)(180,30){3}
	\Text(165,15)[lb]{$\mathbf{q}$}
	\Photon(180,35)(240,35){1}{6}
	\Arc(210,69.286)(48.286,-134.76,-45.24)
	\Text(195,56)[lb]{$\mathbf{p}_m\mathbf{h}_i$}
	\ArrowArc(210,3.967)(46.033,42.388,137.612)
	\Text(20,-40)[lb]{$+$}
	\Photon(45,-40)(105,-40){1}{5}
	\DashArrowLine(45,-60)(45,-45){3}
	\Text(30,-55)[lb]{$\mathbf{q}$}
	\Vertex(45,-40){3}
	\Vertex(105,-40){3}
	\ArrowArc(105,-30)(10,270,-90)	
	\Text(90,-15)[lb]{$\mathbf{p}_m\mathbf{h}_i$}
	\Text(120,-40)[lb]{$+$}
	\Vertex(143,-40){3}
	\Vertex(207,-40){3}
	\DashArrowLine(143,-60)(143,-35){3}
	\Text(130,-60)[lb]{$\mathbf{q}$}
	\Photon(145,-40)(205,-40){1}{6}
	\Arc(175,-5.7)(48.286,-134.76,-45.24)
	\Text(160,-20)[lb]{$\mathbf{p}_m\mathbf{h}_i$}
	\ArrowArc(175,-71)(46.033,42.388,137.612)
	\Text(276,-5)[lb]{\Large$/$}
        \Line(300,-45)(300,45)
        \Line(300,-45)(310,-45)
        \Line(300,45)(310,45)
        \Line(560,-50)(560,50)
        \Line(560,-50)(550,-50)
        \Line(560,50)(550,50)
    \Text(310,-5)[lb]{$1+2\times$}
    \Photon(365,0)(425,0){1}{5}
    \Vertex(365,0){3}
    \Vertex(425,0){3}
    \ArrowArc(425,10)(10,270,-90)
    \Text(421,25)[lb]{$\mathbf{h}_i$}
    \Text(441,-5)[lb]{$+\,2\times$}
    \Vertex(483,0){3}
    \Vertex(547,0){3}
    \Photon(485,0)(545,0){1}{6}
    \Arc(515,34.286)(48.286,-134.76,-45.24)
    \Text(510,25)[lb]{$\mathbf{h}_i$}
    \ArrowArc(515,-31)(46.033,42.388,137.612)
\end{picture}
\vspace{-1.5cm}
\caption{Diagrammatic representation of the first order in $\hat{f}-1$ of $R_a$.\label{fig:Ra_primo_g}}
\end{center}
\vspace{1cm}
\end{figure}

\subsection{Cluster expansion of the matrix element}
Substituting Eqs. (\ref{eq:Rb_frac_sch}) and (\ref{eq:Ra_frac_sch}) in Eq. (\ref{eq:wmel}), the CBF matrix element of the weak transition operator takes the form 
\begin{align}
(\Psi_{p_m;h_i}| \hat{O}_{\mathbf{q}}|0)=\frac{1+\sum_C (q,p_mh_i)}{\sqrt{1+\sum_C (h_i)}\sqrt{1+\sum_C (p_m)}}\, ,
\label{eq:mat_el_sc}
\end{align}
the diagrammatic representation of which can be easily deduced from Figs. \ref{fig:Rb_n_fact} and \ref{fig:Ra_n_fact}.

For the sake of giving a unified description of matrix elements associated with both the Fermi and Gamow-Teller transitions, it is convenient to distinguish the common radial part from the specific zero momentum form factors and spin-isospin operators, defining
\begin{equation}
\hat{O}_{\mathbf{q}}(1)\equiv e^{i\mathbf{q}\mathbf{r}_1} \hat{O}_{\sigma\tau}(1)\ .
\end{equation}
From Eq. (\ref{eq:f_gt_def}) it follows that $\hat{O}_{\sigma\tau}^{F}(1)=g_V \tau^{+}_1$ and $\hat{O}_{\sigma\tau}^{GT}(1)=g_A\vec{\sigma}_1 \tau^{+}_1$.

\subsubsection{Zero-th order}
The weak operator's matrix element at zeroth order in $\hat{f}-1$, displayed in Fig. \ref{fig:Ra0}, corresponds to the non interacting Fermi gas result
 \begin{equation}
(\Psi_{p_m;h_i}| \hat{O}_{\mathbf{q}}|0)^{(0)}=A\langle \Psi_{p_m;h_i} |  \hat{O}_{\mathbf{q}}(1)|\Psi_0\rangle\equiv R_{a}^{(0)}\equiv O_{N}^{(0)}\, .
\end{equation}

\begin{figure}[!hb]
\begin{center}
  \begin{picture}(100,70) (-40,-50)
    \SetWidth{0.8}
    \SetColor{Black}
    \SetScale{1.1}	
    \unitlength=1.1 pt
	\ArrowArc(0,0)(10,270,-90)
	\Vertex(0,-10){3}
	\DashArrowLine(0,-30)(0,-15){3}
	\Text(-15,13)[lb]{$\mathbf{p}_m\mathbf{h}_j$}
	\Text(-12,-25)[lb]{$\mathbf{q}$}
  \end{picture}
\end{center}    
\vspace{-1cm}
\caption{Zero-th order matrix element $O_{N}^{(0)}$.\label{fig:Ra0}}
\end{figure}
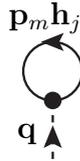

Using Eq. (\ref{eq:Ra_exp}), the right hand side of the above equation can be written in the form
\begin{align}
O_{N}^{(0)} &=\frac{1}{V}\int d \mathbf{r}_1 e^{-i\mathbf{p}_m\cdot\mathbf{r}_1}e^{i\mathbf{q}\mathbf{r}_1}e^{i\mathbf{h}_i\mathbf{r}_1}\langle \alpha_{p_m}|\hat{O}_{\sigma\tau}(1)|\alpha_{h_i}\rangle\nonumber \\
&=\delta_{\mathbf{q},\mathbf{p}_m-\mathbf{h}_i}\langle \alpha_{p_m}|\hat{O}_{\sigma\tau}(1)|\alpha_{h_i}\rangle\, ,
\end{align}
where the discretized momentum conservation is expressed by the Kronecker delta function.
 
The SNM spin-isospin operator matrix elements $\langle \alpha_{p_m}|\hat{O}_{\sigma\tau}(1)|\alpha_{h_i}\rangle$, along with all the other matrix elements appearing in the rest of this Section, are given in Appendix \ref{app:fgtop} for both Fermi and Gamow-Teller transitions. 

\subsubsection{First order}
At first order in $\hat{f}-1$, diagrams arising from both the numerator and the denominator of Eq. (\ref{eq:mat_el_sc}) contribute to the matrix element. In order to include all the first order terms, three-body cluster contributions have to be taken into account. These contributions, neglected in \cite{cowell_03,cowell_04,benhar_09} bring about some inconsistencies, to be discussed later.

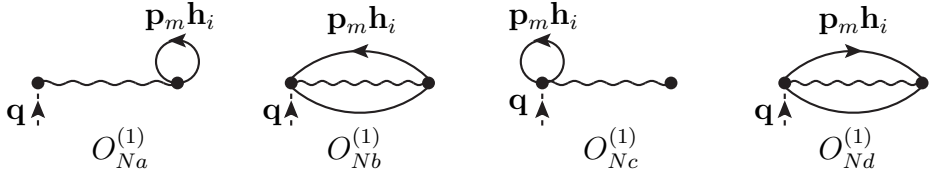
\begin{figure}[!h]
\begin{center}
  \begin{picture}(100,150) (120,-50)
    \SetWidth{1.0}
    \SetColor{Black}
    \SetScale{0.8}	
    \unitlength=0.8 pt
	\Photon(0,35)(65,35){1}{5}
	\DashArrowLine(0,15)(0,30){3}
	\Text(-15,15)[lb]{$\mathbf{q}$}
	\Vertex(0,35){3}
	\Vertex(65,35){3}
	\ArrowArc(65,45)(10,270,-90)	
	\Text(50,60)[lb]{$\mathbf{p}_m\mathbf{h}_i$}
	\Text(25,-5)[lb]{$O_{Na}^{(1)}$}
	\Vertex(118,35){3}
	\Vertex(182,35){3}
	\DashArrowLine(118,15)(118,30){3}
	\Text(105,15)[lb]{$\mathbf{q}$}
	\Photon(120,35)(180,35){1}{6}
	\Arc(150,69.286)(48.286,-134.76,-45.24)
	\Text(135,58)[lb]{$\mathbf{p}_m\mathbf{h}_i$}
	\ArrowArc(150,3.967)(46.033,42.388,137.612)
	\Text(135,-5)[lb]{$O_{Nb}^{(1)}$}
	\Photon(235,35)(295,35){1}{5}
	\DashArrowLine(235,15)(235,30){3}
	\Text(220,20)[lb]{$\mathbf{q}$}
	\Vertex(235,35){3}
	\Vertex(295,35){3}
	\ArrowArc(235,45)(10,270,-90)	
	\Text(220,60)[lb]{$\mathbf{p}_m\mathbf{h}_i$}
	\Text(255,-5)[lb]{$O_{Nc}^{(1)}$}
	\Vertex(348,35){3}
	\Vertex(412,35){3}
	\DashArrowLine(348,15)(348,30){3}
	\Text(335,15)[lb]{$\mathbf{q}$}
	\Photon(350,35)(410,35){1}{6}
	\Arc(380,69.3)(48.286,-134.76,-45.24)
	\Text(365,58)[lb]{$\mathbf{p}_m\mathbf{h}_i$}
	\ArrowArcn(380,4)(46.033,137.612,42.388)
	\Text(365,-5)[lb]{$O_{Nd}^{(1)}$}
  \end{picture}
\end{center}    
\vspace{-1.5cm}
\caption{Two-body diagrams of the first order term in $\hat{f}-1$ coming from the numerator of Eq. (\ref{eq:mat_el_sc}).}
\label{fig:2b_num_ra}
\end{figure}

At first order in $\hat{f}-1$, the two-body cluster term of Eq. (\ref{eq:FFO1_exp}) is given by
\begin{equation}
X^{(2)}(x_1;x_2)=\{\hat{f} -1,\hat{O}_{\mathbf{q}}(1)\}\,.
\label{eq:corr_g_2b}
\end{equation}
The analytic expressions of the two-body diagrams of Fig. \ref{fig:2b_num_ra}, corresponding to the numerator of Eq. (\ref{eq:mat_el_sc}), can be derived by substituting the latter relation in Eq. (\ref{eq:FFO1_exp}). Since all of these diagrams are converging in the thermodynamic limit, the index $\bar{n}_i$ can be safely replaced by the standard $\bar{n}_i$, including the hole state $h_i$. Thus
\begin{align}
O_{Na}^{(1)}
&=\frac{\rho}{\nu}\delta_{\mathbf{q},\mathbf{p}_m-\mathbf{h}_i}\int d\mathbf{r}_{12}e^{-i\mathbf{q}\cdot\mathbf{r}_{12}}\sum_{\alpha_1}\langle \alpha_1 \alpha_{p_m}|\{\hat{f} -1,\hat{O}_{\sigma\tau}(1)\}| \alpha_1 \alpha_{h_i}\rangle\, ,
\end{align}
\begin{align}
O_{Nb}^{(1)}
&=-\frac{\rho}{\nu}\delta_{\mathbf{q},\mathbf{p}_m-\mathbf{h}_i}\int d\mathbf{r}_{12}e^{i\mathbf{p}_m\cdot\mathbf{r}_{12}}\ell_{12}\sum_{\alpha_1}\langle \alpha_1 \alpha_{p_m}| \{\hat{f}_{12}-1,\hat{O}_{\sigma\tau}(1)\}\hat{P}^{\sigma\tau}_{12}| \alpha_1 \alpha_{h_i}\rangle\, ,
\end{align}
\begin{align}
O_{Nc}^{(1)}
&=\frac{\rho}{\nu}\delta_{\mathbf{q},\mathbf{p}_m-\mathbf{h}_i}\int d\mathbf{r}_{12}\sum_{\alpha_2}\langle \alpha_{p_m} \alpha_2| \{\hat{f}_{12}-1,\hat{O}_{\sigma\tau}(1)\}| \alpha_{h_i} \alpha_2\rangle\, ,
\end{align}
\begin{align}
O_{Nd}^{(1)}
&=-\frac{\rho}{\nu}\delta_{\mathbf{q},\mathbf{p}_m-\mathbf{h}_i}\int d\mathbf{r}_{12}e^{-i\mathbf{h}_i\cdot\mathbf{r}_{12}}\ell_{12}\sum_{\alpha_2}\langle \alpha_{p_m} \alpha_2| \{\hat{f}_{12}-1,\hat{O}_{\sigma\tau}(1)\}\hat{P}^{\sigma\tau}_{12}| \alpha_{h_i} \alpha_2\rangle\,.
\, .
\end{align}

\begin{figure}[!ht]
\begin{center}
  \begin{picture}(100,100) (150,-50)
    \SetWidth{1.0}
    \SetColor{Black}
    \SetScale{0.8}	
    \unitlength=0.8 pt
    \Text(10,-5)[lb]{$2\times$}
    \Photon(35,0)(95,0){1}{5}
    \Vertex(35,0){3}
    \Vertex(95,0){3}
    \ArrowArc(95,10)(10,270,-90)
    \Text(90,25)[lb]{$\mathbf{p}_m$}
    \Text(55,-40)[lb]{$O_{Da}^{(1)}$}
    \Text(135,-5)[lb]{$2\times$}
    \Vertex(158,0){3}
    \Vertex(222,0){3}
    \Photon(160,0)(220,0){1}{6}
    \Arc(190,34.286)(48.286,-134.76,-45.24)
    \Text(185,25)[lb]{$\mathbf{p}_m$}
    \ArrowArc(190,-31)(46.033,42.388,137.612)
    \Text(175,-40)[lb]{$O_{Db}^{(1)}$}
    \Text(255,-5)[lb]{$2\times$}
    \Photon(280,0)(340,0){1}{5}
    \Vertex(280,0){3}
    \Vertex(340,0){3}
    \ArrowArc(340,10)(10,270,-100)
    \Text(335,25)[lb]{$\mathbf{h}_i$}
    \Text(300,-40)[lb]{$O_{Dc}^{(1)}$}
    \Text(380,-5)[lb]{$2\times$}
    \Vertex(403,0){3}
    \Vertex(467,0){3}
    \Photon(405,0)(465,0){1}{6}
    \Arc(435,34.3)(48.286,-134.76,-45.24)
    \Text(430,25)[lb]{$\mathbf{h}_i$}
    \ArrowArc(435,-31)(46.033,42.388,137.612)
    \Text(420,-40)[lb]{$O_{Dd}^{(1)}$}
\end{picture}
\caption{First order diagrams stemming from the denominator of Eq. (\ref{eq:mat_el_sc})}
\label{fig:den_primo}
\end{center}
\vspace{-0.5cm}
\end{figure}
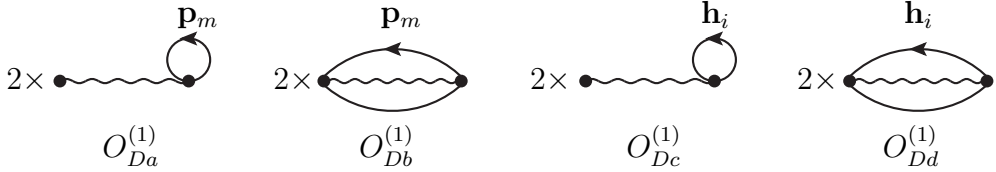

The first order diagrams of the denominator of Eq. (\ref{eq:mat_el_sc}) are displayed if Fig. \ref{fig:den_primo}. By noting that the two-body cluster term of Eq. (\ref{eq:FF_exp_r}) is given by
\begin{equation}
X^{(2)}(x_1,x_2)=2(\hat{f}_{12}-1)\, .
\end{equation}
the following analytic expressions of the diagrams of Fig. \ref{fig:den_primo} can be derived
\begin{align}
O_{Da}^{(1)}&=\frac{2\rho}{\nu}\int d\mathbf{r}_{12} \sum_{\alpha_1}\langle \alpha_1 \alpha_{p_m} |\hat{f}_{12}-1|\alpha_1 \alpha_{p_m}\rangle\nonumber \\
O_{Db}^{(1)}&=-\frac{2\rho}{\nu}\int d\mathbf{r}_{12} e^{i\mathbf{p}_m\cdot\mathbf{r}_{12}} \ell_{12} \sum_{\alpha_1}\langle \alpha_1 \alpha_{p_m} |(\hat{f}_{12}-1)\hat{P}^{\sigma\tau}_{12}|\alpha_1 \alpha_{p_m}\rangle\nonumber\\
O_{Dc}^{(1)}&=\frac{2\rho}{\nu}\int d\mathbf{r}_{12} \sum_{\alpha_1}\langle \alpha_1 \alpha_{h_i} |\hat{f}_{12}-1|\alpha_1 \alpha_{h_i}\rangle\nonumber\\
O_{Dd}^{(1)}&=-\frac{2\rho}{\nu}\int d\mathbf{r}_{12} e^{i\mathbf{h}_i\cdot\mathbf{r}_{12}} \ell_{12} \sum_{\alpha_1}\langle \alpha_1 \alpha_{h_i} |(\hat{f}_{12}-1)\hat{P}^{\sigma\tau}_{12}|\alpha_1 \alpha_{h_i}\rangle\, .
\label{eq:den_1_2b}
\end{align}
Performing a Taylor expansion of the denominator of Eq. (\ref{eq:mat_el_sc}), it can be shown that, up to an error of the order of $(\hat{f}-1)^2$, once multiplied by $O_{N}^{(0)} $, $O_{Da}^{(1)}$ and $O_{Dc}^{(1)}$ cancel $O_{Nc}^{(1)}$, while $O_{Db}^{(1)}$ and $O_{Dd}^{(1)}$ can be interpreted as a correction to $O_{Nb}^{(1)}$ and $O_{Nd}^{(1)}$. 

At first order in $\hat{f}-1$ three body cluster diagrams have to be taken into account. They are associated with the numerator of Eq. (\ref{eq:mat_el_sc}) and arise from the the three body cluster term of Eq. (\ref{eq:FFO1_exp})
\begin{equation}
X^{(3)}(x_1;x_2,x_3)=\{\hat{f}_{23}-1, \hat{O}_{\mathbf{q}}(1)\}\,=2 \hat{O}_{\mathbf{q}}(1)(\hat{f}_{23}-1)\,.
\end{equation}
Because of the property $X^{(3)}_1(x_1;x_2,x_3)=X^{(3)}_1(x_1;x_3,x_2)$, when the latter expression for the three-body cluster term is inserted in Eq. (\ref{eq:Ra_exp}), one finds
\begin{align}
&\sum_{\bar{n}_i}\int dx_1dx_2dx_3\Big[2\psi^{*}_{\bar{n}_1}(x_1)\psi^{*}_{\bar{n}_2}(x_2)\psi^{*}_{p_m}(x_3) \hat{O}_{\mathbf{q}}(1)(\hat{f}_{23}-1)\mathcal{A}[\psi_{\bar{n}_1}(x_1)\psi_{\bar{n}_2}(x_2)\psi_{h_i}(x_3)]+\nonumber \\
&\psi^{*}_{p_m}(x_1)\psi^{*}_{\bar{n}_2}(x_2)\psi^{*}_{\bar{n}_3}(x_3) \hat{O}_{\mathbf{q}}(1)(\hat{f}_{23}-1)\mathcal{A}[\psi_{h_i}(x_1)\psi_{\bar{n}_2}(x_2)\psi_{\bar{n}_3}(x_3)]\Big]\,.
\label{eq:num_3b_g}
\end{align}
The first order three-body cluster diagrams can be grouped in two subgroups; those coming from the first line of Eq.(\ref{eq:num_3b_g}) are depicted in Fig. \ref{fig:num_3b_g_1}, the others in Fig. \ref{fig:num_3b_g_2}.

\begin{figure}[!ht]
\begin{center}
  \begin{picture}(150,230) (70,-150)
    \SetWidth{1.0}
    \SetColor{Black}
    \SetScale{0.8}	
    \unitlength=0.8 pt
	\Photon(30,50)(60,0){1}{5}
	\ArrowArc(30,60)(10,-90,270)
	\Vertex(0,0){3}
       	\Vertex(30,50){3}
        \Vertex(60,0){3}
        	\DashArrowLine(0,-20)(0,-5){3}
	\Text(-15,-15)[lb]{$\mathbf{q}$}
	\Text(15,75)[lb]{$\mathbf{p}_m\mathbf{h}_i$}
	\Text(15,-35)[lb]{$O_{Ne}^{(1)}$}

	\Photon(160,50)(190,0){1}{5}
	\ArrowArc(140,4.43)(50,-5,65)
	\Arc(210,47)(50,175,245)
	\Vertex(130,0){3}
       	\Vertex(160,50){3}
        \Vertex(190,0){3}
        \DashArrowLine(130,-20)(130,-5){3}
	\Text(115,-15)[lb]{$\mathbf{q}$}
	\Text(192,25)[lb]{$\mathbf{p}_m\mathbf{h}_i$}
	\Text(145,-35)[lb]{$O_{Nf}^{(1)}$}

	\Photon(290,50)(320,0){1}{5}
	\Vertex(260,0){3}
       	\Vertex(290,50){3}
        \Vertex(320,0){3}
        	\Arc(290,-40)(50,50,130)
	\Arc(290,40)(50,230,310)
	\ArrowArc(290,60)(10,-90,270)
	\DashArrowLine(260,-20)(260,-5){3}
	\Text(245,-15)[lb]{$\mathbf{q}$}
	\Text(275,75)[lb]{$\mathbf{p}_m\mathbf{h}_i$}
	\Text(275,-35)[lb]{$O_{Ng}^{(1)}$}

	\Photon(30,-70)(60,-120){1}{5}
	\ArrowArc(50,-115.57)(50,115,185)
	\Arc(-21,-72.57)(50,-65,3)
	\Vertex(0,-120){3}
       	\Vertex(30,-70){3}
        \Vertex(60,-120){3}
        	\DashArrowLine(0,-140)(0,-125){3}
	\Text(-15,-135)[lb]{$\mathbf{q}$}
	\Text(-30,-95)[lb]{$\mathbf{p}_m\mathbf{h}_i$}
	\Text(15,-155)[lb]{$O_{Nh}^{(1)}$}

	\Photon(160,-70)(190,-120){1}{5}
	\Arc(140,-115.57)(50,-5,65)
	\ArrowArc(180,-115.57)(50,115,185)
	\Arc(160,-80)(50,230,310)
	\Vertex(130,-120){3}
       	\Vertex(160,-70){3}
        \Vertex(190,-120){3}
        \DashArrowLine(130,-140)(130,-125){3}
	\Text(115,-135)[lb]{$\mathbf{q}$}
	\Text(100,-95)[lb]{$\mathbf{p}_m\mathbf{h}_i$}
	\Text(145,-155)[lb]{$O_{Ni}^{(1)}$}

	\Photon(290,-70)(320,-120){1}{5}
	\ArrowArc(270,-115.57)(50,-5,65)
	\Arc(310,-115.57)(50,115,185)
	\Arc(290,-80)(50,230,310)
	\Vertex(260,-120){3}
       	\Vertex(290,-70){3}
        \Vertex(320,-120){3}
        \DashArrowLine(260,-140)(260,-125){3}
	\Text(245,-135)[lb]{$\mathbf{q}$}
	\Text(320,-95)[lb]{$\mathbf{p}_m\mathbf{h}_i$}
	\Text(275,-155)[lb]{$O_{Nl}^{(1)}$}

\end{picture}
\caption{Three body diagrams emerging from the first line of Eq. (\ref{eq:num_3b_g}).\label{fig:num_3b_g_1}}
\end{center}
\vspace{0.5cm}
\end{figure}
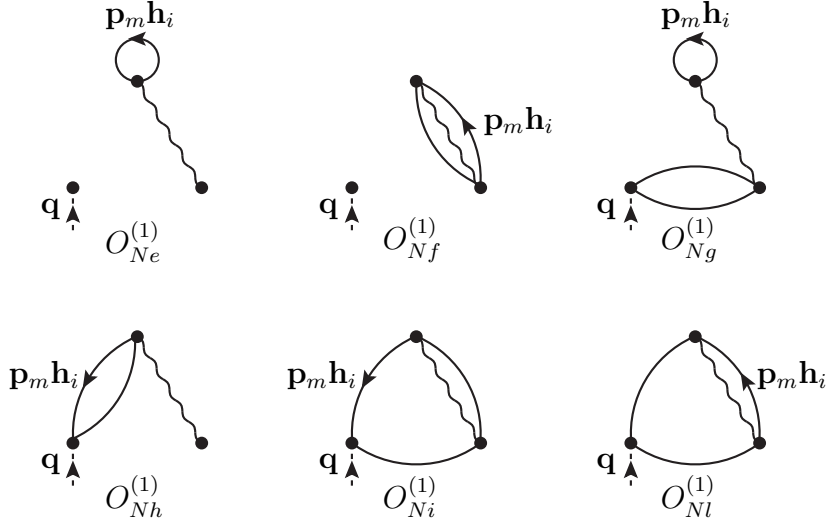

Diagrams $O_{Ne}^{(1)}$ and $O_{Nf}^{(1)}$ vanish, as the part of the diagram containing the weak transition operator is connected neither to the $ph$-vertex nor the the $ph$-exchange line. In the thermodynamic limit diagram $O_{Ng}^{(1)}$ reads
\begin{align}
O_{Ng}^{(1)}=&-\frac{2\rho^2}{\nu^2}\delta_{\mathbf{q},\mathbf{p}_m-\mathbf{h}_i}\int d\mathbf{r}_{12}e^{i\mathbf{q}\cdot\mathbf{r}_{12}}\ell^2_{12} \int d\mathbf{r}_{23}e^{i\mathbf{q}\cdot\mathbf{r}_{23}}\sum_{\alpha_i}\times\nonumber\\
&\,\langle \alpha_1\alpha_2 \alpha_{p_m}|\hat{O}_{\sigma\tau}(1) (\hat{f}_{23}-1)\hat{P}^{\sigma\tau}_{12}|\alpha_1\alpha_2 \alpha_{h_i}\rangle\, .
\end{align}
Note that in the limit of zero momentum transfer, $\mathbf{q}\to 0$, $O_{Ng}^{(1)}$ cancels the contribution of diagram $O_{Na}^{(1)}$. This is an indication that three-body diagrams need to be taken into account.

\begin{figure}[!h]
\begin{center}
  \begin{picture}(150,230) (70,-150)
    \SetWidth{1.0}
    \SetColor{Black}
    \SetScale{0.8}	
    \unitlength=0.8 pt
	\Photon(30,50)(60,0){1}{5}
	\ArrowArc(0,10)(10,-90,270)
	\Vertex(0,0){3}
       	\Vertex(30,50){3}
        \Vertex(60,0){3}
        	\DashArrowLine(0,-20)(0,-5){3}
	\Text(-15,-15)[lb]{$\mathbf{q}$}
	\Text(-15,25)[lb]{$\mathbf{p}_m\mathbf{h}_i$}
	\Text(15,-35)[lb]{$O_{Nm}^{(1)}$}

	\Photon(160,50)(190,0){1}{5}
	\Arc(140,4.43)(50,-5,65)
	\Arc(210,47)(50,175,245)
	\Vertex(130,0){3}
       	\Vertex(160,50){3}
        \Vertex(190,0){3}
        	\ArrowArc(130,10)(10,-90,270)
        \DashArrowLine(130,-20)(130,-5){3}
	\Text(115,-15)[lb]{$\mathbf{q}$}
	\Text(115,25)[lb]{$\mathbf{p}_m\mathbf{h}_i$}
	\Text(145,-35)[lb]{$O_{Nn}^{(1)}$}

	\Photon(290,50)(320,0){1}{5}
	\Vertex(260,0){3}
       	\Vertex(290,50){3}
        \Vertex(320,0){3}
        	\ArrowArc(290,-40)(50,50,130)
	\Arc(290,40)(50,230,310)
	\DashArrowLine(260,-20)(260,-5){3}
	\Text(245,-15)[lb]{$\mathbf{q}$}
	\Text(275,15)[lb]{$\mathbf{p}_m\mathbf{h}_i$}
	\Text(275,-35)[lb]{$O_{No}^{(1)}$}

	\Photon(30,-70)(60,-120){1}{5}
	\ArrowArcn(50,-115.57)(50,185,115)
	\Arc(-21,-72.57)(50,-65,3)
	\Vertex(0,-120){3}
       	\Vertex(30,-70){3}
        \Vertex(60,-120){3}
        	\DashArrowLine(0,-140)(0,-125){3}
	\Text(-15,-135)[lb]{$\mathbf{q}$}
	\Text(-30,-95)[lb]{$\mathbf{p}_m\mathbf{h}_i$}
	\Text(15,-155)[lb]{$O_{Np}^{(1)}$}

	\Photon(160,-70)(190,-120){1}{5}
	\Arc(140,-115.57)(50,-5,65)
	\ArrowArcn(180,-115.57)(50,185,115)
	\Arc(160,-80)(50,230,310)
	\Vertex(130,-120){3}
       	\Vertex(160,-70){3}
        \Vertex(190,-120){3}
        \DashArrowLine(130,-140)(130,-125){3}
	\Text(115,-135)[lb]{$\mathbf{q}$}
	\Text(100,-95)[lb]{$\mathbf{p}_m\mathbf{h}_i$}
	\Text(145,-155)[lb]{$O_{Nq}^{(1)}$}

	\Photon(290,-70)(320,-120){1}{5}
	\Arc(270,-115.57)(50,-5,65)
	\Arc(310,-115.57)(50,115,185)
	\ArrowArc(290,-80)(50,230,310)
	\Vertex(260,-120){3}
       	\Vertex(290,-70){3}
        \Vertex(320,-120){3}
        \DashArrowLine(260,-140)(260,-125){3}
	\Text(245,-135)[lb]{$\mathbf{q}$}
	\Text(275,-125)[lb]{$\mathbf{p}_m\mathbf{h}_i$}
	\Text(275,-155)[lb]{$O_{Nr}^{(1)}$}

\end{picture}
\caption{Three body diagrams emerging from the second line of Eq. (\ref{eq:num_3b_g}).\label{fig:num_3b_g_2}}
\end{center}
\vspace{0.5cm}
\end{figure}
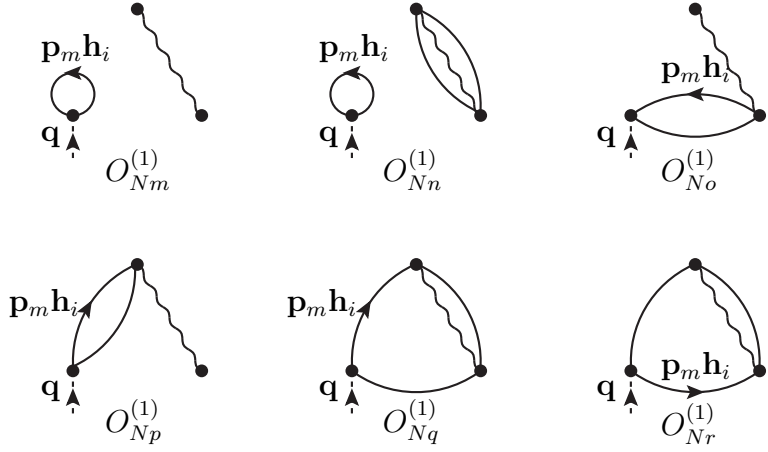

For the diagram $O_{Nh}^{(1)}$ we obtain
\begin{align}
O_{Nh}^{(1)}=&-\frac{2\rho^2}{\nu^2}\delta_{\mathbf{q},\mathbf{p}_m-\mathbf{h}_i}\int d\mathbf{r}_{13}e^{i \mathbf{p}_m\cdot\mathbf{r}_{13}}\ell_{13}\int d\mathbf{r}_{23}\sum_{\alpha_i}\times\nonumber\\
&\,\langle \alpha_1\alpha_2 \alpha_{p_m}|\hat{O}_{\sigma\tau}(1) (\hat{f}_{23}-1)\hat{P}^{\sigma\tau}_{13}|\alpha_1\alpha_2 \alpha_{h_i}\rangle\, .
\end{align}
The latter expression vanishes, because
\begin{equation}
\int d\mathbf{r}_{13}\ell_{13}e^{i \mathbf{p}_m\cdot\mathbf{r}_{13}}=0\, .
\end{equation}

For the same reason we find
\begin{align}
O_{Ni}^{(1)}=&\frac{2\rho^2}{\nu^2}\delta_{\mathbf{q},\mathbf{p}_m-\mathbf{h}_i}\int d\mathbf{r}_{12}e^{i \mathbf{p}_m\cdot\mathbf{r}_{12}}\ell_{12}\int d\mathbf{r}_{23}e^{i \mathbf{p}_m\cdot\mathbf{r}_{23}}\ell_{23}\sum_{\alpha_i}\times\nonumber\\
&\,\langle \alpha_1\alpha_2 \alpha_{p_m}|\hat{O}_{\sigma\tau}(1) (\hat{f}_{23}-1)\hat{P}^{\sigma\tau}_{12}\hat{P}^{\sigma\tau}_{23}|\alpha_1\alpha_2 \alpha_{h_i}\rangle=0\, .
\end{align}

The last diagram coming from the first line of Eq. (\ref{eq:num_3b_g}) gives a non zero contribution
\begin{align}
O_{Nl}^{(1)}=&\frac{2\rho^2}{\nu^2}\delta_{\mathbf{q},\mathbf{p}_m-\mathbf{h}_i}\int d\mathbf{r}_{12}d\mathbf{r}_{23}e^{i \mathbf{p}_m\cdot\mathbf{r}_{13}}e^{-i \mathbf{h}_i\cdot\mathbf{r}_{12}}\ell_{12}\ell_{13}\sum_{\alpha_i}\times\nonumber\\
&\,\langle \alpha_1\alpha_2 \alpha_{p_m}|\hat{O}_{\sigma\tau}(1) (\hat{f}_{23}-1)\hat{P}^{\sigma\tau}_{12}\hat{P}^{\sigma\tau}_{13}|\alpha_1\alpha_2 \alpha_{h_i}\rangle\,.
\end{align}

The first two diagrams of Fig. \ref{fig:num_3b_g_2}, being disconnected, simplify with the denominator. All the remaining diagrams of Fig. \ref{fig:num_3b_g_2} vanish. This can be shown by noting that the bare exchange line is represented by the modified Slater function $\bar{\ell}_{ij}$, that does not contain the hole state
\begin{equation}
\int d\mathbf{r}_{ij}\bar{\ell}_{ij}e^{i \mathbf{h}_i\cdot\mathbf{r}_{ij}}=0\, , 
\end{equation}
and using the matrix elements 
\begin{align}
&\sum_{\alpha_2\neq\alpha_{h_i}\alpha_3}\langle \alpha_{p_m} \alpha_2 \alpha_3 |\hat{O}_{\sigma\tau}(1) (\hat{f}_{23}-1)\hat{P}^{\sigma\tau}_{12}|\alpha_{h_i} \alpha_2\alpha_3 \rangle=0\nonumber\\
&\sum_{\alpha_2\neq \alpha_{h_i}\alpha_3}\langle \alpha_{p_m} \alpha_2\alpha_3 |\hat{O}_{\sigma\tau}(1) (\hat{f}_{23}-1)\hat{P}^{\sigma\tau}_{12}\hat{P}^{\sigma\tau}_{13}|\alpha_{h_i} \alpha_2\alpha_3 \rangle=0\nonumber\\
&\sum_{\alpha_2\alpha_3\neq \alpha_{h_i}}\langle \alpha_{p_m} \alpha_2\alpha_3 |\hat{O}_{\sigma\tau}(1) (\hat{f}_{23}-1)\hat{P}^{\sigma\tau}_{12}\hat{P}^{\sigma\tau}_{23}|\alpha_{h_i} \alpha_2\alpha_3 \rangle=0\, .
\end{align}

\subsubsection{Second order}
As in the first order case, diagrams coming from both numerator and denominator of Eq. (\ref{eq:mat_el_sc}) have to be taken into account. In order to consistently include all the second order terms, in principle up to five-body cluster diagrams  need to be computed. In our calculations \cite{lovato_12b}, we have considered second order two-body cluster terms only, as in Refs.\cite{cowell_03,cowell_04,benhar_09}. Second order three-body cluster diagrams are in fact much smaller than the corresponding two-body ones, as confirmed by a direct computation of some of them.

Considering the second order term, to the two-body cluster of Eq. (\ref{eq:corr_g_2b}) becomes 
\begin{equation}
X^{(2)}(x_1;x_2)=\{\hat{f}_{12}-1, \hat{O}_{\mathbf{q}}(1)\}+(\hat{f}_{12}-1) \hat{O}_{\mathbf{q}}(1)(\hat{f}_{12}-1)\,.
\end{equation}

The additional term $(\hat{f}_{12}-1) \hat{O}_{\mathbf{q}}(1)(\hat{f}_{12}-1)$ brings about new diagrams for the numerator of Eq. (\ref{eq:mat_el_sc}) which are analogous to those of Fig. \ref{fig:2b_num_ra}. As can be seen in Fig.  \ref{fig:2b_num_g2}, the only difference consists in the replacement of the single wavy lines with doubly wave lines, accounting for the second order correlations $(\hat{f}_{12}-1)(\hat{f}_{12}-1)$.

\begin{figure}[!h]
\begin{center}
  \begin{picture}(100,150) (120,-50)
    \SetWidth{1.0}
    \SetColor{Black}
    \SetScale{0.8}	
    \unitlength=0.8 pt
	\Photon(0,36)(65,36){1}{5}
	\Photon(0,34)(65,34){1}{5}
	\DashArrowLine(0,15)(0,30){3}
	\Text(-15,15)[lb]{$\mathbf{q}$}
	\Vertex(0,35){3}
	\Vertex(65,35){3}
	\ArrowArc(65,45)(10,270,-90)	
	\Text(50,60)[lb]{$\mathbf{p}_m\mathbf{h}_i$}
	\Text(25,-5)[lb]{$O_{Na}^{(2)}$}
	\Vertex(118,35){3}
	\Vertex(182,35){3}
	\DashArrowLine(118,15)(118,30){3}
	\Text(105,15)[lb]{$\mathbf{q}$}
	\Photon(120,34)(180,34){1}{6}
	\Photon(120,36)(180,36){1}{6}
	\Arc(150,69.286)(48.286,-134.76,-45.24)
	\Text(135,58)[lb]{$\mathbf{p}_m\mathbf{h}_i$}
	\ArrowArc(150,3.967)(46.033,42.388,137.612)
	\Text(135,-5)[lb]{$O_{Nb}^{(2)}$}
	\Photon(235,34)(295,34){1}{5}
	\Photon(235,36)(295,36){1}{5}
	\DashArrowLine(235,15)(235,30){3}
	\Text(220,20)[lb]{$\mathbf{q}$}
	\Vertex(235,35){3}
	\Vertex(295,35){3}
	\ArrowArc(235,45)(10,270,-90)	
	\Text(221,60)[lb]{$\mathbf{p}_m\mathbf{h}_i$}
	\Text(255,-5)[lb]{$O_{Nc}^{(2)}$}
	\Vertex(348,35){3}
	\Vertex(412,35){3}
	\DashArrowLine(348,15)(348,30){3}
	\Text(335,15)[lb]{$\mathbf{q}$}
	\Photon(350,34)(410,34){1}{6}
	\Photon(350,36)(410,36){1}{6}
	\Arc(380,69.3)(48.286,-134.76,-45.24)
	\Text(365,58)[lb]{$\mathbf{p}_m\mathbf{h}_i$}
	\ArrowArcn(380,4)(46.033,137.612,42.388)
	\Text(365,-5)[lb]{$O_{Nd}^{(2)}$}
  \end{picture}
\end{center}    
\vspace{-1.5cm}
\caption{Two-body diagrams of the second order term in $\hat{f}-1$ coming from the numerator of Eq. (\ref{eq:mat_el_sc}).}
\label{fig:2b_num_g2}
\end{figure}
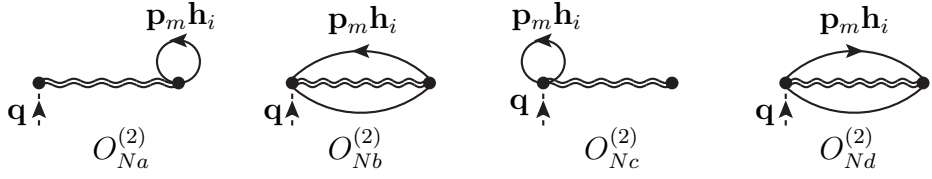

The second order two-body numerator diagrams of Fig. \ref{fig:2b_num_g2} are given by
\begin{align}
O_{Na}^{(2)}=&\frac{\rho}{\nu}\,\delta_{\mathbf{q},\mathbf{p}_m-\mathbf{h}_i}\int d\mathbf{r}_{12}e^{-i\mathbf{q}\mathbf{r}_{12}}
\sum_{\alpha_1}\langle \alpha_1 \alpha_{p_m}|(\hat{f}_{12}-1) \hat{O}_{\sigma\tau}(1)(\hat{f}_{12}-1)| \alpha_1 \alpha_{h_i}\rangle\, ,
\end{align}
\begin{align}
O_{Nb}^{(2)}
=&-\frac{\rho}{\nu} \delta_{\mathbf{q},\mathbf{p}_m-\mathbf{h}_i}\int d\mathbf{r}_{12}e^{i\mathbf{p}_m\cdot\mathbf{r}_{12}}\ell_{12}\times\nonumber\\
&\,\sum_{\alpha_1}\langle \alpha_1 \alpha_{p_m}|(\hat{f}_{12}-1) \hat{O}_{\sigma\tau}(1)(\hat{f}_{12}-1)\hat{P}^{\sigma\tau}_{12}| \alpha_1 \alpha_{h_i}\rangle\, ,
\end{align}
\begin{align}
O_{Nc}^{(2)}&=\frac{\rho}{d}\delta_{\mathbf{q},\mathbf{p}_m-\mathbf{h}_i}\int d\mathbf{r}_{12}\sum_{\alpha_2}\langle \alpha_{p_m} \alpha_2| (\hat{f}_{12}-1) \hat{O}_{\sigma\tau}(1)(\hat{f}_{12}-1)| \alpha_{h_i} \alpha_2\rangle\, ,
\end{align}
\begin{align}
O_{Nd}^{(2)}=&-\frac{\rho}{\nu} \delta_{\mathbf{q},\mathbf{p}_m-\mathbf{h}_i}\int d\mathbf{r}_{12}e^{-i\mathbf{h}_i\cdot\mathbf{r}_{12}}\ell_{12}\times\nonumber\\
&\, \sum_{\alpha_2}\langle \alpha_{p_m} \alpha_2| (\hat{f}_{12}-1) \hat{O}_{\sigma\tau}(1)(\hat{f}_{12}-1)\hat{P}^{\sigma\tau}_{12}| \alpha_{h_i} \alpha_2\rangle\, .
\end{align}

As far as the denominator of Eq. (\ref{eq:mat_el_sc}) is concerned, the two-body second order diagrams, shown in Fig. \ref{eq:mat_el_sc}, can be obtained from those of Fig. (\ref{fig:den_primo})  by again replacing the single wavy lines with doubly wavy lines. As a matter of fact, once the second order term in considered, the two-body cluster term of Eq. (\ref{eq:FF_exp_r}) reads
\begin{equation}
X^{(2)}(x_1,x_2)=2(\hat{f}_{12}-1)+(\hat{f}_{12}-1)^2\, .
\end{equation}

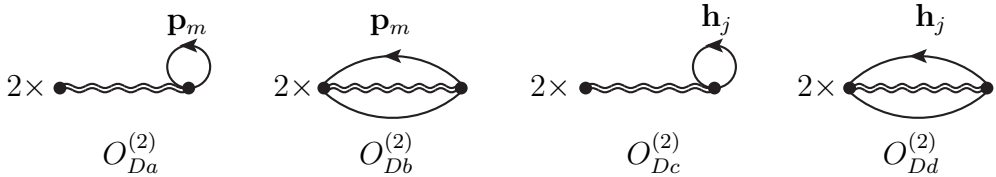
\begin{figure}[!ht]
\begin{center}
  \begin{picture}(100,100) (150,-50)
    \SetWidth{1.0}
    \SetColor{Black}
    \SetScale{0.8}	
    \unitlength=0.8 pt
    \Text(10,-5)[lb]{$2\times$}
    \Photon(35,-1)(95,-1){1}{5}
    \Photon(35,1)(95,1){1}{5}
    \Vertex(35,0){3}
    \Vertex(95,0){3}
    \ArrowArc(95,10)(10,270,-90)
    \Text(85,25)[lb]{$\mathbf{p}_m$}
    \Text(55,-40)[lb]{$O_{Da}^{(2)}$}
    \Text(135,-5)[lb]{$2\times$}
    \Vertex(158,0){3}
    \Vertex(222,0){3}
    \Photon(160,-1)(220,-1){1}{6}
    \Photon(160,1)(220,1){1}{6}
    \Arc(190,34.286)(48.286,-134.76,-45.24)
    \Text(180,25)[lb]{$\mathbf{p}_m$}
    \ArrowArc(190,-31)(46.033,42.388,137.612)
    \Text(175,-40)[lb]{$O_{Db}^{(2)}$}
    \Text(255,-5)[lb]{$2\times$}
    \Photon(280,-1)(340,-1){1}{5}
    \Photon(280,1)(340,1){1}{5}
    \Vertex(280,0){3}
    \Vertex(340,0){3}
    \ArrowArc(340,10)(10,270,-100)
    \Text(335,25)[lb]{$\mathbf{h}_j$}
    \Text(300,-40)[lb]{$O_{Dc}^{(2)}$}
    \Text(380,-5)[lb]{$2\times$}
    \Vertex(403,0){3}
    \Vertex(467,0){3}
    \Photon(405,-1)(465,-1){1}{6}
    \Photon(405,1)(465,1){1}{6}
    \Arc(435,34.3)(48.286,-134.76,-45.24)
    \Text(435,25)[lb]{$\mathbf{h}_j$}
    \ArrowArc(435,-31)(46.033,42.388,137.612)
    \Text(420,-40)[lb]{$O_{Dd}^{(2)}$}
\end{picture}
\caption{Diagrammatic representation of the second order two-body cluster term of the denominator of Eq. (\ref{eq:mat_el_sc}).}
\label{fig:den_2b_g2}
\end{center}
\vspace{0.5cm}
\end{figure}

The analytic expressions of the second order two-body diagrams of Fig. \ref{fig:den_2b_g2} are very similar to those of Eq. (\ref{eq:den_1_2b})
\begin{align}
O_{Da}^{(2)}&=\frac{\rho}{\nu}\int d\mathbf{r}_{12} \sum_{\alpha_1}\langle \alpha_1 \alpha_{h_i} |(\hat{f}_{12}-1)^2|\alpha_1 \alpha_{h_i}\rangle\nonumber \\
O_{Db}^{(2)}&=-\frac{\rho}{\nu}\int d\mathbf{r}_{12} e^{i\mathbf{h}_i\cdot\mathbf{r}_{12}} \ell_{12} \sum_{\alpha_1}\langle \alpha_1 \alpha_{h_i} |(\hat{f}_{12}-1)^2\hat{P}^{\sigma\tau}_{12}|\alpha_1 \alpha_{h_i}\rangle\nonumber \\
O_{Dc}^{(2)}&=\frac{\rho}{\nu}\int d\mathbf{r}_{12} \sum_{\alpha_1}\langle \alpha_1 \alpha_{p_m} |(\hat{f}_{12}-1)^2|\alpha_1 \alpha_{p_m}\rangle\nonumber \\
O_{Dd}^{(2)}&=-\frac{\rho}{\nu}\int d\mathbf{r}_{12} e^{i\mathbf{p}_m\cdot\mathbf{r}_{12}} \ell_{12} \sum_{\alpha_1}\langle \alpha_1 \alpha_{p_m} |(\hat{f}_{12}-1)^2\hat{P}^{\sigma\tau}_{12}|\alpha_1 \alpha_{p_m}\rangle\,.
\end{align}

The results of the second order two-body cluster terms coincide with those obtained by the authors of Refs. \cite{cowell_03,cowell_04}. 

\section{Effective interaction}
Using the formalism of CBF and the cluster expansion technique, the authors of Ref. \cite{cowell_03,cowell_04}, were able to develop an {\it effective interaction}, obtained from the bare Argonne $v_{8}^\prime$ potential, which incorporates the effects of the short-range correlations. In Ref. \cite{benhar_07}, the two-body effective interaction 
of Ref. \cite{cowell_03,cowell_04} was improved with the inclusion of the purely phenomenological density dependent potential of Ref. \cite{lagaris_80}, 
accounting for the effects of interactions involving more than two nucleons. The CBF effective interaction, $v_{12}^{eff}$, suitable for use in Hartree-Fock calculations, is defined through the matrix elements of the hamiltonian in the correlated ground-state
\begin{equation}
\langle\Psi_0|\hat{\mathcal{F}}^\dag \hat{H} \hat{\mathcal{F}}|\Psi_0\rangle\equiv T_F+\langle \Psi_0 | \hat{v}_{12}^{eff} | \Psi_0 \rangle\, .
\label{eq:eff_int}
\end{equation}

As suggested by the above equation, the effective interaction allows one to calculate any nuclear matter observables using perturbation theory in the orthonormal FG basis. However, in general, extracting the effective interaction is a very challenging task, involving difficulties even more severe than those associated with the calculation of the expectation value of the hamiltonian in the correlated ground state.

The procedure developed in Ref. \cite{cowell_04} consists in carrying out a cluster expansion of the {\it lhs} of Eq. (\ref{eq:eff_int})
and keeping only the two-body cluster contribution. The sum of the two-body cluster contribution of the potential and kinetic energies of Eqs. (\ref{eq:v2bcont}) and (\ref{eq:kin_2b_part}) reads
\begin{align}
&\langle\Psi_0|\hat{\mathcal{F}}^\dag \hat{H} \hat{\mathcal{F}}|\Psi_0\rangle\Big{|}_{2b}=\nonumber\\
&\frac{\rho}{2}\int d\mathbf{r}_{12}\text{CTr}\Big[
\Big(\hat{F}_{12}v_{12}\hat{F}_{12}-\frac{\hbar^2}{m}(\vec{\nabla}_1\hat{F}_{12})(\vec{\nabla}_{1}\hat{F}_{12})\Big)(1-\hat{P}_{12}\ell_{12}^2)\Big]\, .
\end{align}
On the other hand, the expectation value of the effective potential is given by
\begin{align}
&\langle\Psi_0|\hat{v}^{eff}_{12}|\Psi_0\rangle\Big{|}_{2b}=\frac{\rho}{2}\int d\mathbf{r}_{12} \text{CTr}\Big[
\hat{v}^{eff}_{12}\Big(1-\hat{P}_{12}\ell_{12}^2)\Big]\, .
\label{eq:eff_pot}
\end{align}
Therefore, the effective potential at two-body cluster level turns out to be
\begin{equation}
\hat{v}_{12}^{eff}\Big|_{2b}=\hat{F}v_{12}\hat{F}-\frac{\hbar^2}{m}(\vec{\nabla}_1\hat{F}_{12})(\vec{\nabla}_{1}\hat{F}_{12})\, .\end{equation}

The effective potential of the above equation slightly differs from the one reported in the literature. The authors of Refs. \cite{cowell_04,benhar_09} have not done the integration by parts leading to Eq. (\ref{eq:kin_2b_part}). As a consequence they have neglected the terms in which the gradient operates on both the correlation function and on the plane waves. In our effective potential, these terms, although small compared to the other contributions, are fully taken into account.

We have improved the effective potential by adding the tree-body cluster contributions of Eqs. (\ref{eq:v3b_dir_jas}-\ref{eq:uf_cir}). This allowed us to consistently include the UIX potential, whose leading order terms emerge at three-body cluster level.

As in the construction of the density dependent potential from UIX, the issue of the exchange pattern has to be carefully analyzed. The distinctive feature of the present calculation is that $v_{12}^{eff}$ contains the correlation between particles $1$ and $2$ making possible to implement the inversion of $\hat{P}^{\sigma\tau}_{ij}$ of Eq. (\ref{eq:inv_exch}) in a straightforward way. To be definite, consider the three-body cluster contribution of the ground-state expectation value of the two-body potential
\begin{align}
&\langle \hat{v}_{12} \rangle\Big{|}_{3b}=\frac{\rho^2}{2}\int dx_{123}\hat{X}(x_1,x_2;x_3)\Big[1-\hat{P}^{\sigma\tau}_{12}\ell^{2}_{12}-2\hat{P}^{\sigma\tau}_{13}\ell^{2}_{13}+2\hat{P}^{\sigma\tau}_{12}\hat{P}^{\sigma\tau}_{13}\ell_{12}\ell_{13}\ell_{23})\Big]\, .
\end{align}

%%%%%%%%%%%%%%%%%%%%%%%%%%%%%%%%%%%%%%%%%%%%%%%%%%%%%%%%%%%%%%%%%%%%%%%%%%%
\begin{figure}[!ht]
\begin{center}
\includegraphics[width=9.0cm]{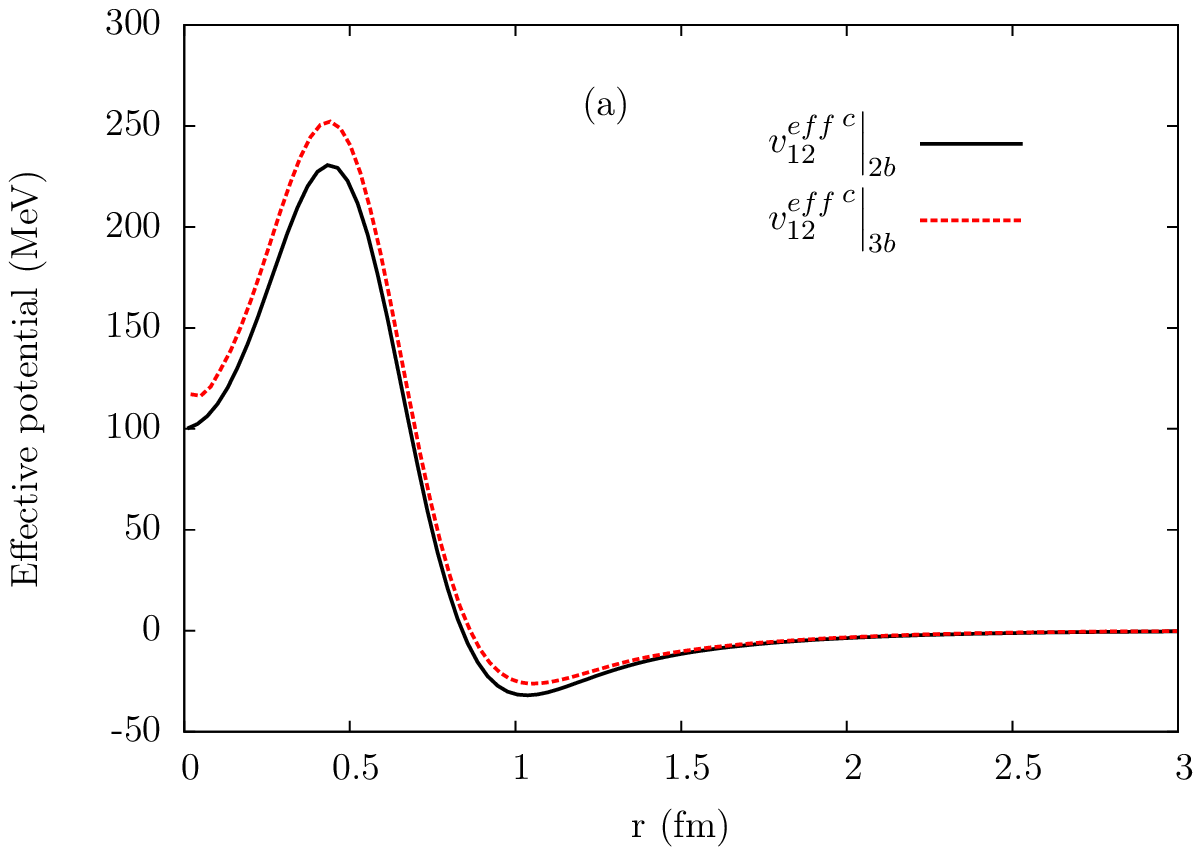}
\vspace{0.1cm}
\\
\includegraphics[width=9.0cm]{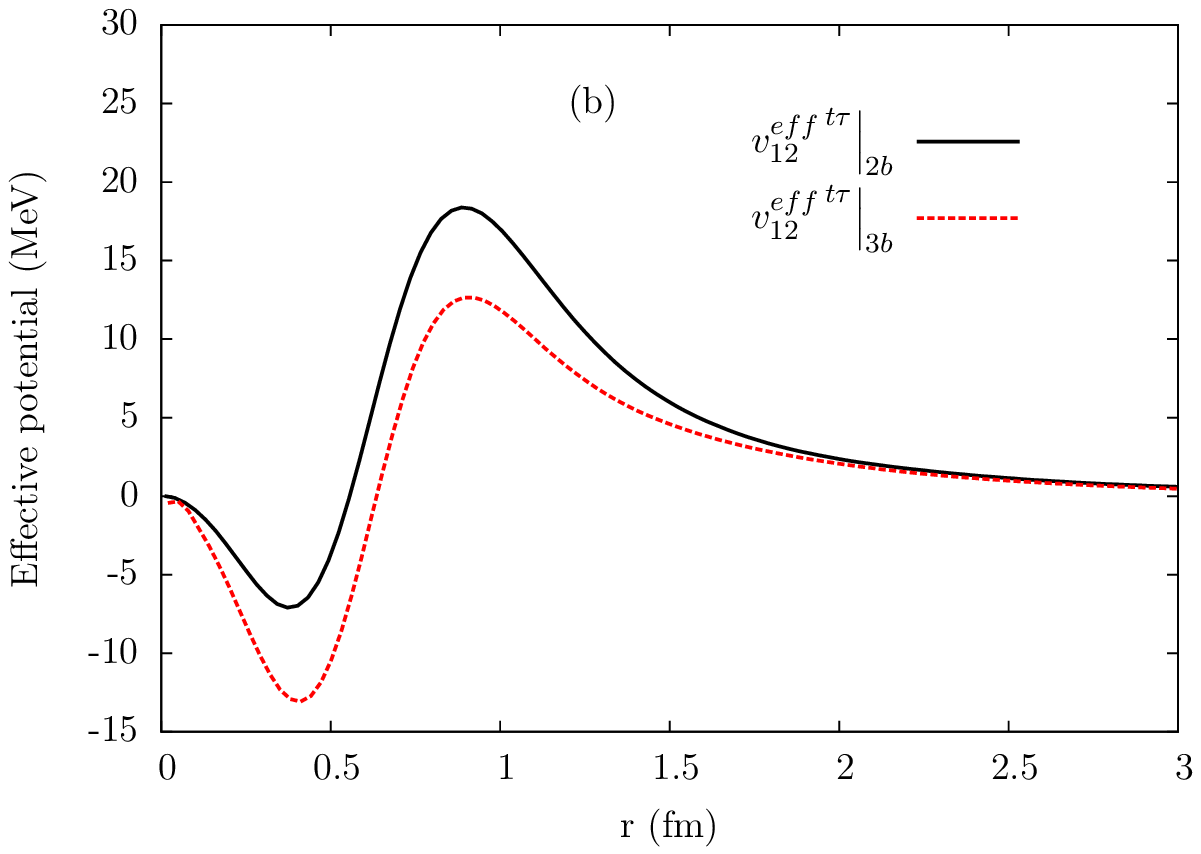}
\vspace{0.1cm}
\caption{Central (a) and the $t\tau$ (b) channels of the effective potentials at two-body and three-body cluster level calculated for SNM at $\rho=0.16\,\text{fm}^{-3}$. \label{fig:eff_pot_comp}}
\end{center}
\end{figure}
%%%%%%%%%%%%%%%%%%%%%%%%%%%%%%%%%%%%%%%%%%%%%%%%%%%%%%%%%%%%%%%%%%%%%%%%%%%

Note that the above equation summarizes the terms corresponding to Eqs. (\ref{eq:v3b_dir}), (\ref{eq:v3b_p12}), (\ref{eq:v3b_p13}) and (\ref{eq:v3b_cir}). A comparison with Eq. (\ref{eq:eff_pot}) immediately leads to 
\begin{align}
\hat{v}_{12}^{eff}\Big|_{3b}=&\rho\int d\mathbf{r}_3 \text{CTr}_3 \Big[ \hat{X}(x_1,x_2;x_3)\Big(1-\hat{P}^{\sigma\tau}_{12}\ell^{2}_{12}-2\hat{P}^{\sigma\tau}_{13}\ell^{2}_{13}+2\hat{P}^{\sigma\tau}_{12}\hat{P}^{\sigma\tau}_{13}\ell_{12}\ell_{13}\ell_{23}\Big)\times\nonumber\\
&(1-\hat{P}_{12}^{\sigma\tau}\ell_{12}^2)^{-1}\Big]\, .
\end{align}
A similar argument holds for the three-body cluster term of the kinetic energy and the three-body potential contributions to the effective potential. 

In Fig. \ref{fig:eff_pot_comp} the central and $t\tau$ components of the effective potentials at two-body and three-body cluster level are compared. The starting bare NN interaction is the Argonne $v_{6}^\prime$; for the three-body cluster results the UIX three-body potential has been included in the calculations. 

Starting from a bare hamiltonian whose only interaction is the Argonne $v_{6}^\prime$ NN potential we have  computed the EoS of SNM for the low-density regime using both the new three-body cluster effective interaction and the older one with only two-body cluster diagrams. The results have been compared with the corresponding FHNC/SOC calculations, displayed as a shaded region Fig. \ref{fig:eos_eff} to account for the PB and JF kinetic energy difference. The curve corresponding to $v_{12}^{eff}\Big{|}_{3b}$ is much closer to the FHNC/SOC results than the one obtained with the older $v_{12}^{eff}\Big{|}_{3b}$. 

In the lower panel of Fig. \ref{fig:eos_eff}, the EoS of SNM are shown for an hamiltonian containing the Argonne $v_{6}^\prime$ NN potential along with the UIX three-body interaction model. Again the curve obtained from the three-body cluster effective potential is close to the full calculation and, it exhibits the saturation, which is a remarkable feature.

%%%%%%%%%%%%%%%%%%%%%%%%%%%%%%%%%%%%%%%%%%%%%%%%%%%%%%%%%%%%%%%%%%%%%%%%%%%
\begin{figure}[!ht]
\begin{center}
\includegraphics[width=9.0cm]{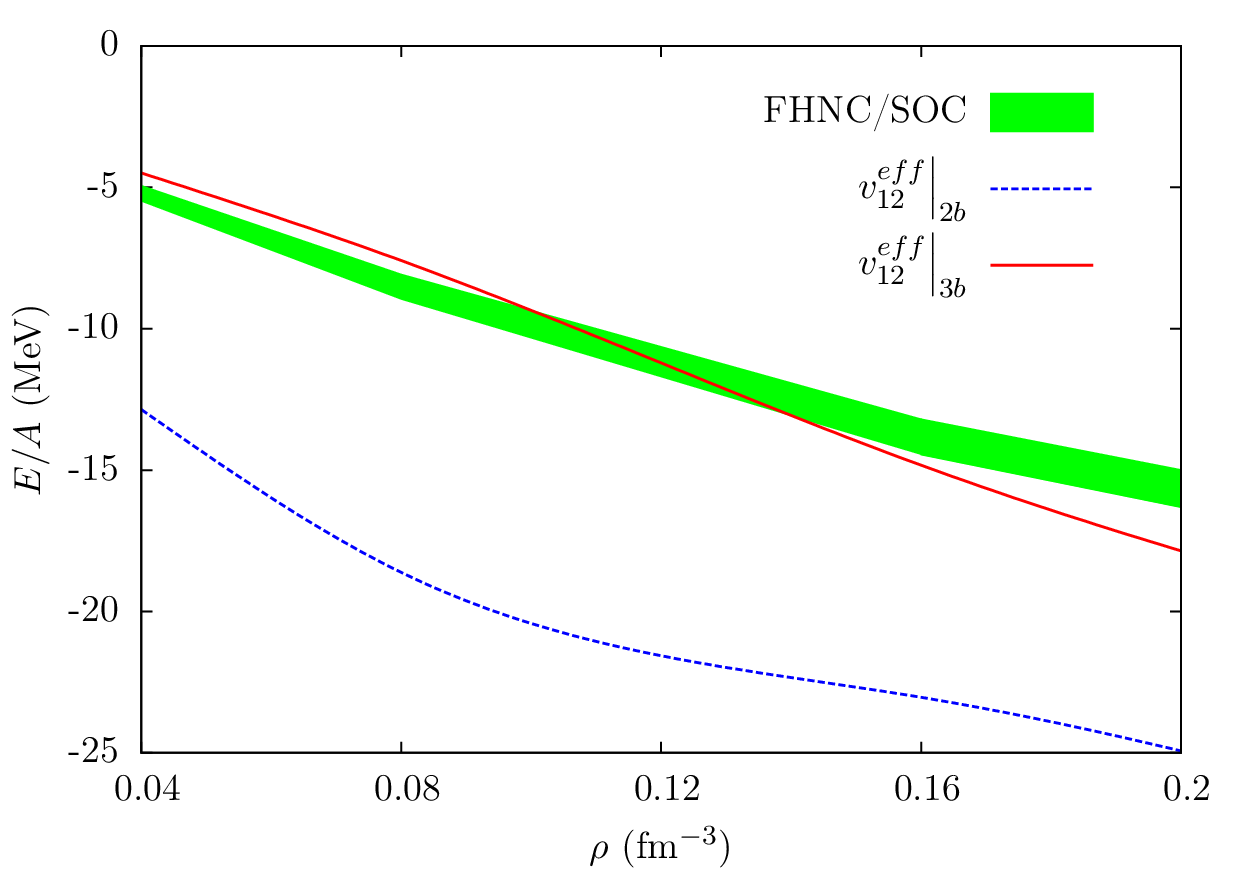}
\vspace{0.1cm}
\\
\includegraphics[width=9.0cm]{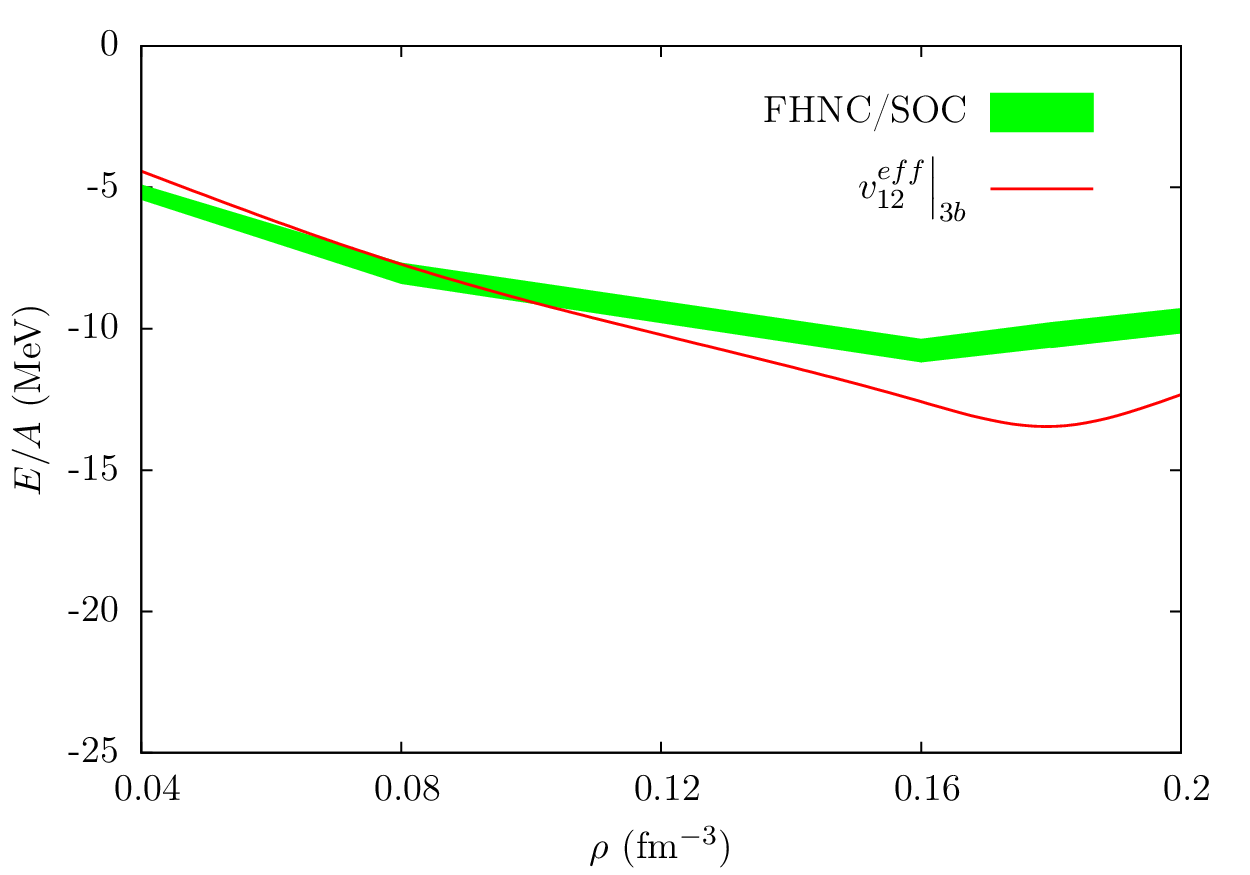}
\vspace{0.1cm}
\caption{EoS of SNM in the low density regime. In the upper panel the bare hamiltonian only contains the Argonne $v_{6}^\prime$ potential, while in the lower panel the UIX three-body interaction model is added to it. The dotted and the solid lines display the three-body and the two-body cluster effective potential results, respectively. The FHNC/SOC calculations are represented by the shaded region, accounting for the PB and JF kinetic energy difference. \label{fig:eos_eff}}
\end{center}
\end{figure}
%%%%%%%%%%%%%%%%%%%%%%%%%%%%%%%%%%%%%%%%%%%%%%%%%%%%%%%%%%%%%%%%%%%%%%%%%%%

\section{Correlated Fermi gas, Hartree-Fock approximations}
In both the correlated Fermi gas (CFG) and correlated Hartree-Fock (CHF) approximations, the weak response of cold SNM, defined in Eq. (\ref{eq:resp_def}), is given by \cite{cowell_04}
\begin{align}
S_{FG}(\mathbf{q},\omega)&=\frac{1}{A}\sum_{p_mh_i} |\langle \Psi_{p_m;h_i} | \hat{O}_{\mathbf{q}}^{eff}| \Psi_0 \rangle|^2 \delta(\omega+\epsilon_{p_m}-\epsilon_{h_i})\, .
\label{eq:cfg}
\end{align}
Within the CGF approximation, the single particle energies are obtained from Eq. (\ref{spe_nm1}) in the case of a non-interacting hamiltonian
\begin{equation}
\epsilon_{n_i}=\frac{\mathbf{k}_{i}^2}{2m}\, .
\label{eq:spe_fg}
\end{equation}

The single particle approximation is retained in the CHF; however the potential enter the calculation of the single particle energies. As mentioned in Section \ref{sec:hfa} the Hartree-Fock approximation is not suitable for nuclear potentials having a repulsive core, like the Argonne models, because it does not encompass the correlations between nucleons. We use 
instead the effective potential described in the previous Section, which is appropriate for mean field calculations. Thus, Eq. (\ref{spe_nm1}) becomes
\begin{equation}
\epsilon_{n_i}=\frac{\mathbf{k}_{i}^2}{2m}+\sum_{n_j=1}^A \int dx_{j}\psi_{n_i}^*(x_i)\psi_{n_j}^*(x_j)v_{ij}^{eff}[\psi_{n_i}(x_i)\psi_{n_j}(x_j)-\psi_{n_j}(x_i)\psi_{n_i}(x_j)]\,.
\end{equation}

\section{Correlated Tamm Dancoff approximation (CTDA)}
The Tamm-Dancoff approximation amounts to expanding the final state of Eq. (\ref{eq:resp_def}) in series of $1p-1h$ excitations
\begin{equation}
|\Psi_f\}^{TDA}=\sum_{mi}\mathcal{C}^{f}_{mi} |\Psi_{p_m;h_i}\rangle\,.
\label{eq:TDA_zero}
\end{equation}

Because the hamiltonian is translationally invariant, the total momentum $\mathbf{q}$ of the state is conserved, and the momenta of the particle, $\mathbf{p}_m$, and the hole, $\mathbf{h}_i$, satisfy the relation $\mathbf{p}_m-\mathbf{h}_i=\mathbf{q}$.

The eigenvalue equation for the effective hamiltonian defines the excitation energy $\omega_f$
\begin{equation}
\hat{H}^{eff}|\Psi_f\}^{TDA}=\Big(\sum_i-\frac{\nabla^{2}_i}{2m}+\sum_{i<j}\hat{v}^{eff}_{ij}\Big)|\Psi_f\}^{TDA}=(E_0+\omega_f)|\Psi_f\}^{TDA}\, .
\label{eq:eigen_TDA}
\end{equation}
Multiplying from the left the previous equation by $\langle  \Psi_{p_n;h_j}|$ and using the orthonormality of the $1p-1h$ states yields
\begin{equation}
\sum_{mi} \langle \Psi_{p_n;h_j}|\hat{H}^{eff}|\Psi_{p_m;h_i}\rangle \mathcal{C}^{f}_{mi} =\mathcal{C}^{f}_{nj}(E_0+\omega_f)\, .
\label{eq:eigen_TDA2}
\end{equation}
Hence, the coefficients $\mathcal{C}^{f}_{mi}$ can be interpreted as the eigenvectors of the hamiltonian $H_{nj;mi}\equiv\langle \Psi_{p_n;h_j}|\hat{H}|\Psi_{p_m;h_i}\rangle$, while the excitation energies $\omega_f$ are the associated eigenvalues. 

In Appendix \ref{app:slat} it is shown that singling out particle $1$ from the $1p-1h$ wave function $\Psi_{p_mh_i}$ yields the result
\begin{align}
\Psi_{p_mh_i}=\frac{1}{\sqrt{A}}\sum_{n_1}(-1)^{n_1+1}\psi_{\tilde{n}_1}(x_1)\Psi^{p_mh_i}_{m \neq \tilde{n}_1}(x_2\,\dots\,x_A)\,.
\label{eq:onep_extr}
\end{align}
The indexes $n_i$ and $\tilde{n}_i$ denote the single particle states of the Fermi gas and of the $1p-1h$ excited state, respectively. Note that, in the above equation, $\tilde{n}_i$ must be regarded 
as a function of $n_i$; in particular it turns out that $\tilde{n}_i(n_i)=n_i$ except for $\tilde{n}_i(h_i)=p_m$.

The minor $\Psi^{p_mh_i}_{m \neq \tilde{n}_1}$ has been obtained from $\Psi_{p_mh_i}$ by removing the column corresponding to the state $\tilde{n}_i$ and the first row, associated with particle $1$.

Extracting particle $2$ from the Slater determinants leads to
\begin{align}
\Psi_{p_mh_i}=\frac{1}{\sqrt{A(A-1)}}\sum_{n_1<n_2}(-1)^{n_1+n_2+1}\mathcal{A}[\psi_{\tilde{n}_1}(x_1)\psi_{\tilde{n}_2}(x_2)]\Psi^{p_mh_i}_{m \neq \tilde{n}_1,\tilde{n}_2}(x_3\,\dots\,x_A)\,,
\label{eq:twop_extr}
\end{align}
where $\Psi^{p_mh_i}_{m \neq \tilde{n}_1,\tilde{n}_2}$, the minor of $\Psi^{p_mh_i}_{m \neq \tilde{n}_1}$, is lacking particle 2 and the state $\tilde{n}_2$.

Being a one-body operator, the kinetic energy, on account of  Eq. (\ref{eq:onep_extr}), contributes only to the diagonal part of $H_{nj;mi}$
\begin{align}
A\,\langle p_m;h_i|-\frac{\nabla_{1}^2}{2m}|p_m;h_i\rangle&=
\sum_{n_1}\int dx_1\psi_{\tilde{n}_1}^*(x_1)\Big(-\frac{\nabla_{1}^2}{2m}\Big)\psi_{\tilde{n}_1}(x_1)\nonumber \\
&=\nu\sum_{|\mathbf{k}|\leq k_F}\frac{k^2}{2m}+\frac{p_{m}^2}{2m}-\frac{h_{i}^2}{2m}\, .
\label{eq:kin_1p1h}
\end{align}

The diagonal matrix elements of the effective potential can be computed employing Eq. (\ref{eq:twop_extr})
\begin{align}
&\frac{A(A-1)}{2} \langle p_m;h_i|v^{eff}_{12}|p_m;h_i\rangle\nonumber\\
\quad &=\frac{1}{2}\sum_{n_i}\int dx_{1,2} \psi_{\tilde{n}_1}^*(x_1)\psi_{\tilde{n}_2}^*(x_2)\hat{v}^{eff}_{12}\mathcal{A}[ \psi_{\tilde{n}_1}(x_1)\psi_{\tilde{n}_2}(x_2)]\nonumber \\
\quad &= \sum_{n_1n_2} \int dx_{1,2}\psi_{n_1}^*(x_1)\psi_{n_2}^*(x_2)\hat{v}^{eff}_{12}\mathcal{A}[ \psi_{n_1}(x_1)\psi_{n_2}(x_2)]\nonumber\\
\quad &+\sum_{n_1}\int dx_{1,2}\psi_{p_m}^*(x_1)\psi_{n_1}^*(x_2)\hat{v}^{eff}_{12}\mathcal{A}[ \psi_{p_m}(x_1)\psi_{n_2}(x_2)]\nonumber \\
\quad &-\sum_{n_1}\int dx_{1,2}\psi_{n_1}^*(x_1)\psi_{h_i}^*(x_2)\hat{v}^{eff}_{12}\mathcal{A}[ \psi_{n_1}(x_1)\psi_{h_i}(x_2)]\nonumber \\
\quad &-\int dx_{1,2} \psi_{p_m}^*(x_1)\psi_{h_i}^*(x_2)\hat{v}^{eff}_{12}\mathcal{A}[ \psi_{p_m}(x_1)\psi_{h_i}(x_2)]\, .
\label{eq:pot_1p1h}
\end{align}

The sum of the first term of Eq. (\ref{eq:kin_1p1h}) and the first of Eq. (\ref{eq:pot_1p1h}) corresponds to the ground state energy of the system
\begin{equation}
E_0=\nu\sum_{|\mathbf{k}|\leq k_F}\frac{k^2}{2m}+
\frac{1}{2}\sum_{n_1n_2} \int dx_{1,2}\psi_{n_1}^*(x_1)\psi_{n_2}^*(x_2)\hat{v}^{eff}_{12}\mathcal{A}[ \psi_{n_1}(x_1)\psi_{n_2}(x_2)]\, .
\end{equation}

By recognizing in the sum of Eq. (\ref{eq:kin_1p1h}) and the Eq. (\ref{eq:pot_1p1h}) the single particle energy of Eq. (\ref{spe_nm1}) one gets for the diagonal part of the $1p-1h$ matrix elements of the hamiltonian
\begin{equation}
\langle \Psi_{p_m;h_i}|\hat{H}^{eff}|\Psi_{p_m;h_i}\rangle=E_0+\epsilon_{p_m}-\epsilon_{h_i}+\langle p_m\,h_i|\hat{v}^{eff}_{12}|h_i\,p_m\rangle\, .
\label{eq:diag_1p1h}
\end{equation}
where we denote with $\langle p_n\,h_i|O_{12}|h_j\,p_m\rangle$ the two-body matrix element of the operator $\hat{O}_{12}$ 
\begin{equation}
\langle p_n\,h_i|\hat{O}_{12}|h_j\,p_m\rangle\equiv\int dx_{1,2} \psi_{p_n}^*(x_1)\psi_{h_i}^*(x_2)\hat{O}_{12}\mathcal{A}[ \psi_{h_j}(x_1)\psi_{p_m}(x_2)]\, .
\end{equation}

The off-diagonal matrix elements of the hamiltonian come from the two-body potential; making again use of Eq. (\ref{eq:twop_extr}) it turns out that
\begin{equation}
\frac{A(A-1)}{2} \langle  \Psi_{p_n;h_j}|\hat{v}^{eff}_{12}| \Psi_{p_m;h_i}\rangle =\langle p_n\,h_i|\hat{v}^{eff}_{12}|h_j\,p_m\rangle\, .
\label{eq:pot_1p1h2}
\end{equation}

Collecting Eqs. (\ref{eq:diag_1p1h}) and (\ref{eq:pot_1p1h2}), one finds the following compact expression \cite{cowell_04} for the matrix elements of the hamiltonian
\begin{equation}
\langle \Psi_{p_n;h_j}|\hat{H}^{eff}|\Psi_{p_m;h_i}\rangle=(E_0+\epsilon_{p_m}-\epsilon_{h_i})\delta_{mn}\delta_{ij}+\langle p_n\,h_i|\hat{v}^{eff}_{12}|h_j\,p_m\rangle\, .
\end{equation}

Finally, substitution in the eigenvalue equation (\ref{eq:eigen_TDA2}) gives
\begin{equation}
\sum_{mi}\Big[(\epsilon_{p_m}-\epsilon_{h_i})\delta_{mn}\delta_{ij}+\langle p_n\,h_i|\hat{v}^{eff}_{12}|h_j\,p_m\rangle\Big]\mathcal{C}^{f}_{mi} =\omega_f\,\mathcal{C}^{f}_{nj}\, .
\label{eq:eigen_TDA3}
\end{equation}

The nuclear hamiltonian commutes with the total isospin, $T$  with the total isospin projection along $z$, $T_z$ and with the total spin, $S$, but not with the total spin projection along $z$, $S_z$ because of the tensor term of the potential.  Solving Eq. (\ref{eq:eigen_TDA2}) would then lead to coefficients $\mathcal{C}^{f}_{mi}$ such that the final states of Eq. (\ref{eq:TDA_zero}) are eigenstates of $S$, $T$ and $T_z$: $|f\rangle\equiv|f\rangle_{TT_zS}$.  

The combinations of particle hole pairs that are eigenstates of the total spin $S$ and its projection along the z-axis $S_z$, that define the particle-hole Clebsch-Gordan coefficients, are shown in Table \ref{tab:s_ph}. The differences between the total spin states of the particle particle pairs, also given in Table \ref{tab:s_ph}, are due to the phase factor appearing in the canonical transformations to particles and holes \cite{fetter_03}. For a more detailed discussion, see Appendix \ref{app:tsphp}. The treatment of the total isospin can be done in complete analogy, replacing the up and the down single particle spin states with the proton and the neutron isospin states, respectively. 
\begin{table}[h!]
\begin{center}
\renewcommand*\arraystretch{1.5}
\caption{Spin configurations for a particle particle pair
and a particle hole pair for spin-1/2 particles. \label{tab:s_ph}}
\vspace{0.3cm}
\begin{tabular}{l c c }
\hline
\hline
Total spin state &    particle particle      &  particle hole \\
\hline
$S=1\,,S_z=1$  & $\uparrow\uparrow$ & $-\uparrow\downarrow$ \\
$S=1\,,S_z=0$  & $\frac{1}{\sqrt{2}}(\uparrow\downarrow+\downarrow\uparrow)$ & $\frac{1}{\sqrt{2}}(\uparrow\uparrow-\downarrow\downarrow)$  \\
$S=1\,,S_z=-1$ & $\downarrow\downarrow$ & $\downarrow\uparrow$ \\
$S=0\,,S_z=0$  & $\frac{1}{\sqrt{2}}(\uparrow\downarrow-\downarrow\uparrow)$ & $\frac{1}{\sqrt{2}}(\uparrow\uparrow+\downarrow\downarrow)$  \\

\hline
\hline
\end{tabular}
\vspace{0.1cm}
\end{center}
\end{table}    

It is possible to reduce the computational complexity of the eigenvalue equation by carrying out the summation over the spin-isospin projections along the $z$-axis of Eq. (\ref{eq:TDA_zero}), grouping the combinations of $|\Psi_{p_m;h_i}\rangle$ with definite $T$, $T_z$, $S$ and $S_z$
\begin{equation}
|\Psi_f\rangle_{TT_zS}=\sum_{\mathbf{p}_m\mathbf{h}_i S_z}C^{f\,TT_zSS_z}_{\mathbf{p}_m\mathbf{h}_i} |\Psi_{\mathbf{p}_m;\mathbf{h}_i}\rangle_{TT_zSS_z}\, ,
\end{equation}
A further simplification arises by noting that the final states of both the Fermi and the Gamow-Teller transitions are characterized by having $T=1$ and $T_z=1$. To simplify the notation, the  isospin indexes may then be omitted
\begin{equation}
|\Psi_f\rangle_{S}=\sum_{\mathbf{p}_m\mathbf{h}_i S_z}C^{f\,SS_z}_{\mathbf{p}_m\mathbf{h}_i} |\Psi_{\mathbf{p}_m;\mathbf{h}_i}\rangle_{SS_z}\, ,
\label{eq:TDA_exp_new}
\end{equation}
and it is understood that $T=1$ and $T_z=1$.

Unlike in Eq. (\ref{eq:TDA_zero}) the sum is now restricted to the momentum of the particle and of the hole and to $S_z$, furthermore the coefficients $C^{f\,TT_zSS_z}_{\mathbf{p}_m\mathbf{h}_i}$ and $\,\mathcal{C}^{f}_{mi}$ are related through the Clebsch-Gordan coefficients of Table \ref{tab:s_ph}. 

The eigenvalue equations for the coefficient $C^{f\,SS_z}_{\mathbf{p}_m\mathbf{h}_i} $ can be obtained from Eq. (\ref{eq:eigen_TDA2}) 
\begin{equation}
\sum_{\mathbf{p}_m\mathbf{h}_i S'_z}\:_{SS_z}\langle\Psi_{\mathbf{p}_n;\mathbf{h}_j}|\hat{H}|\Psi_{\mathbf{p}_m;\mathbf{h}_i}\rangle_{SS'_z}\,C^{f\,SS'_z}_{\mathbf{p}_m\mathbf{h}_i}  =C^{f\,SS_z}_{\mathbf{p}_n\mathbf{h}_j}(E_0+\omega_{f}^{S})\, .
\end{equation}
Thus, finding the coefficient $C^{f\,SS_z}_{\mathbf{p}_m\mathbf{h}_i} $ amounts in diagonalizing the block diagonal hamiltonian for the two subsets of the $1p-1h$ basis having $T=1$, $T_z=1$ corresponding to $S=0$ and to $S=1$ . This is a much less expensive computational task than diagonalizing the hamiltonian in the full $1p-1h$ basis, as it was in Eq. (\ref{eq:eigen_TDA2}). As a consequence, this approach allows for considering a larger number of momentum states.

Since the single particle energies $\epsilon_{p_m}$ and  $\epsilon_{h_i}$ of Eq. (\ref{eq:diag_1p1h}) depend neither on the spin nor on the isospin of the states $p_m$ and ${h_i}$, Eq. (\ref{eq:eigen_TDA3}) can be recasted in the following form
\begin{equation}
\sum_{\mathbf{p}_m\mathbf{h}_i S'_z}\Big[(\epsilon_{\mathbf{p}_m}-\epsilon_{\mathbf{h}_i})\delta_{\mathbf{p}_m\mathbf{p}_n}\delta_{\mathbf{h}_i\mathbf{h}_j}\delta_{S_zS'_z}+\langle \mathbf{p}_n\,\mathbf{h}_i|\hat{v}^{eff}_{12}|\mathbf{h}_j\,\mathbf{p}_m\rangle_{SS'_zS_z}\Big]C^{f\,SS'_z}_{\mathbf{p}_m\mathbf{h}_i}  =\omega_{f}^{S}\,C^{f\,SS_z}_{\mathbf{p}_n\mathbf{h}_j} \, .
\label{eq:eigen_TDA5}
\end{equation}
In the matrix element of the potential $\langle \mathbf{p}_n\,\mathbf{h}_i|\hat{v}^{eff}_{12}|\mathbf{h}_j\,\mathbf{p}_m\rangle_{SS'_zS_z}$ the spin and isospin projections along the $z$ axis of the particle hole pairs $p_m-h_i$ and $p_n-h_j$ are combined as in Table \ref{tab:s_ph} to have definite $S$,  and total spin projections along $z$ equal to $S'_z$ and $S_z$, respectively. We 
recall that the total isospin and its $z$-projection are $T=1$ and $T_z=1$.

The direct and exchange terms for the case $S=0$, relevant to the Fermi transition, for SNM read
\begin{align}
&\langle \mathbf{p}_n\,\mathbf{h}_i|\hat{v}^{eff}_{12}|\mathbf{h}_j\,\mathbf{p}_m\rangle_{000}^{d}\nonumber \\
&\qquad =\frac{1}{2V}\int d\mathbf{r}_{12}e^{-i\mathbf{q}\cdot\mathbf{r}_{12}}[\,
\langle p \uparrow n \uparrow | \hat{v}^{eff}_{12} |n \uparrow p \uparrow\rangle
+\langle p \uparrow n \downarrow | \hat{v}^{eff}_{12} |n \uparrow p \downarrow\rangle\nonumber \\
&\qquad +\langle p \downarrow n \uparrow | \hat{v}^{eff}_{12} |n \downarrow p \uparrow\rangle+
\langle p \downarrow n \downarrow | \hat{v}^{eff}_{12} |n \downarrow p \downarrow\rangle]\nonumber \\
&\qquad = \frac{4}{V}\int d\mathbf{r}_{12}e^{-i\mathbf{q}\cdot\mathbf{r}_{12}}\,v_{12}^\tau \\
\nonumber\\
&\langle \mathbf{p}_n\,\mathbf{h}_i|v_{12}|\mathbf{h}_j\,\mathbf{p}_m\rangle_{000}^{e}\nonumber\\
&\qquad= \frac{1}{2V}\int d\mathbf{r}_{12}e^{i\mathbf{k}_{ij}\cdot\mathbf{r}_{12}}[\,
\langle p \uparrow n \uparrow | \hat{v}^{eff}_{12}\hat{P}_{12} |n \uparrow p \uparrow\rangle
+\langle p \uparrow n \downarrow | \hat{v}^{eff}_{12} \hat{P}_{12}|n \uparrow p \downarrow\rangle\nonumber \\
& \qquad +\langle p \downarrow n \uparrow | \hat{v}^{eff}_{12} \hat{P}_{12}|n \downarrow p \uparrow\rangle+
\langle p \downarrow n \downarrow | \hat{v}^{eff}_{12} \hat{P}_{12}|n \downarrow p \downarrow\rangle]\nonumber \\
&\qquad= \frac{1}{V}\int d\mathbf{r}_{12}e^{i\mathbf{k}_{ij}\cdot\mathbf{r}_{12}}(v^c_{12} - v^\tau_{12} + 3 v^\sigma_{12} - 3v^{\sigma\tau}_{12})\, ,
\end{align}
where $\mathbf{k}_{ij}\equiv\mathbf{h}_i-\mathbf{h}_j$. For the sake of simplicity, the superscript ``${eff}$'' has been omitted where the channels of the effective potential are specified.

For the Gamow Teller transition, the final state has $S=1$; hence it is necessary to compute the nine matrix elements $\langle \mathbf{p}_n\,\mathbf{h}_i|\hat{v}^{eff}_{12}|\mathbf{h}_j\,\mathbf{p}_m\rangle_{SS'_zS_z}$ corresponding to $S_z,S'_z=-1\,,0\,,1$

\begin{align}
&\langle \mathbf{p}_n\,\mathbf{h}_i|\hat{v}^{eff}_{12}|\mathbf{h}_j\,\mathbf{p}_m\rangle_{S=1S'_z=1S_z=1}^{d}\nonumber \\
&\qquad=\frac{1}{V}\int d\mathbf{r}_{12}e^{-i\mathbf{q}\cdot\mathbf{r}_{12}}
\langle p \uparrow n \downarrow | \hat{v}^{eff}_{12} |n \downarrow p \uparrow\rangle\nonumber \\
&\qquad=\frac{2}{V}\int d\mathbf{r}_{12}e^{-i\mathbf{q}\cdot\mathbf{r}_{12}}\Big[
2v^{\sigma\tau}_{12}+v^{t\tau}_{12}\Big(1-\frac{3z_{12}^2}{r_{12}^2}\Big)\Big]\\
\nonumber\\
&\langle \mathbf{p}_n\,\mathbf{h}_i|\hat{v}^{eff}_{12}|\mathbf{h}_j\,\mathbf{p}_m\rangle_{S=1S'_z=1S_z=1}^{e}\nonumber \\
&\qquad=\frac{1}{V}\int d\mathbf{r}_{12}e^{i\mathbf{k}_{ij}\cdot\mathbf{r}_{12}}
\langle p \uparrow n \downarrow | \hat{v}^{eff}_{12}\hat{P}_{12} |n \downarrow p \uparrow\rangle\nonumber \\
&\qquad=\frac{1}{V}\int d\mathbf{r}_{12}e^{i\mathbf{k}_{ij}\cdot\mathbf{r}_{12}}\Big[
v^{c}_{12} - v^{\tau}_{12} - v^{\sigma}_{12} + v^{\sigma\tau}_{12}+(v^{t}_{12}-v^{t\tau}_{12})\Big(1-3\frac{z_{12}^2}{r_{12}^2}\Big)\Big]\\
\nonumber\\
&\langle \mathbf{p}_n\,\mathbf{h}_i|\hat{v}^{eff}_{12}|\mathbf{h}_j\,\mathbf{p}_m\rangle_{S=1S'_z=0S_z=1}^{d}\nonumber \\
&\qquad=-\frac{1}{\sqrt{2}V}\int d\mathbf{r}_{12}e^{-i\mathbf{q}\cdot\mathbf{r}_{12}}[\,
\langle p \uparrow n \uparrow | \hat{v}^{eff}_{12} |n \downarrow p \uparrow\rangle-
\langle p \uparrow n \downarrow |\hat{v}^{eff}_{12} |n \downarrow p \downarrow\rangle
]\nonumber \\
&\qquad=-\frac{6\sqrt{2}}{V}\int d\mathbf{r}_{12}e^{-i\mathbf{q}\cdot\mathbf{r}_{12}}\:v^{t\tau}_{12}
\frac{(x_{12}-iy_{12})z_{12}}{r_{12}^2}\\
\nonumber\\
&\langle \mathbf{p}_n\,\mathbf{h}_i|\hat{v}^{eff}_{12}|\mathbf{h}_j\,\mathbf{p}_m\rangle_{S=1S'_z=0S_z=1}^{e}\nonumber \\
&\qquad=-\frac{1}{\sqrt{2}V}\int d\mathbf{r}_{12}e^{i\mathbf{k}_{ij}\cdot\mathbf{r}_{12}}[\,
\langle p \uparrow n \uparrow | \hat{v}^{eff}_{12}\hat{P}_{12}|n \downarrow p \uparrow\rangle-
\langle p \uparrow n \downarrow | \hat{v}^{eff}_{12} \hat{P}_{12}|n \downarrow p \downarrow\rangle
]\nonumber \\
&\qquad=-\frac{3\sqrt{2}}{V}\int d\mathbf{r}_{12}e^{i\mathbf{k}_{ij}\cdot\mathbf{r}_{12}}\:(v^{t}_{12}-v^{t\tau}_{12})
\frac{(x_{12}-iy_{12})z_{12}}{r_{12}^2}\\
\nonumber\\
&\langle \mathbf{p}_n\,\mathbf{h}_i|\hat{v}^{eff}_{12}|\mathbf{h}_j\,\mathbf{p}_m\rangle_{S=1S'_z=-1S_z=1}^{d}
\nonumber \\
&\qquad=-\frac{1}{V}\int d\mathbf{r}_{12}e^{-i\mathbf{q}\cdot\mathbf{r}_{12}}
\langle p \uparrow n \uparrow | \hat{v}^{eff}_{12} |n \downarrow p \downarrow\rangle\nonumber \\
&\qquad=-\frac{6}{V}\int d\mathbf{r}_{12}e^{-i\mathbf{q}\cdot\mathbf{r}_{12}}\:v^{t\tau}_{12}
\frac{(x_{12}-iy_{12})^2}{r_{12}^2}\,\\
\nonumber\\
&\langle \mathbf{p}_n\,\mathbf{h}_i|\hat{v}^{eff}_{12}|\mathbf{h}_j\,\mathbf{p}_m\rangle_{S=1S'_z=-1S_z=1}^{e}
\nonumber \\
&\qquad=-\frac{1}{V}\int d\mathbf{r}_{12}e^{i\mathbf{k}_{ij}\cdot\mathbf{r}_{12}}
\langle p \uparrow n \uparrow |\hat{v}^{eff}_{12}\hat{P}_{12} |n \downarrow p \downarrow\rangle\nonumber \\
&\qquad=-\frac{3}{V}\int d\mathbf{r}_{12}e^{i\mathbf{k}_{ij}\cdot\mathbf{r}_{12}}\:(v^{t}_{12}-v^{t\tau}_{12})
\frac{(x_{12}-iy_{12})^2}{r_{12}^2}\,.
\end{align}
\clearpage

\begin{align}
&\langle \mathbf{p}_n\,\mathbf{h}_i|\hat{v}^{eff}_{12}|\mathbf{h}_j\,\mathbf{p}_m\rangle_{S=1S'_z=0S_z=0}^{d}
\nonumber \\
&\qquad=\frac{1}{2V}\int d\mathbf{r}_{12}e^{-i\mathbf{q}\cdot\mathbf{r}_{12}}[\,
\langle p \uparrow n \uparrow | \hat{v}^{eff}_{12} |n \uparrow p \uparrow\rangle-
\langle p \uparrow n \downarrow | \hat{v}^{eff}_{12} |n \uparrow p \downarrow\rangle\nonumber\\
&\qquad-\langle p \downarrow n \uparrow | \hat{v}^{eff}_{12} |n \downarrow p \uparrow\rangle+
\langle p \downarrow n \downarrow | \hat{v}^{eff}_{12} |n \downarrow p \downarrow\rangle]
\nonumber \\
&\qquad=\frac{4}{V}\int d\mathbf{r}_{12}e^{-i\mathbf{q}\cdot\mathbf{r}_{12}}\:\Big[v^{\sigma\tau}_{12}-v^{t\tau}_{12}\Big(
1-3\frac{z_{12}^2}{r_{12}^2}\Big)\Big]\,\\
\nonumber\\
&\langle \mathbf{p}_n\,\mathbf{h}_i|\hat{v}^{eff}_{12}|\mathbf{h}_j\,\mathbf{p}_m\rangle_{S=1S'_z=0S_z=0}^{e}\nonumber \\
&\qquad=\frac{1}{2V}\int d\mathbf{r}_{12}e^{i\mathbf{k}_{ij}\cdot\mathbf{r}_{12}}[\,
\langle p \uparrow n \uparrow | \hat{v}^{eff}_{12}\hat{P}_{12} |n \uparrow p \uparrow\rangle-
\langle p \uparrow n \downarrow | \hat{v}^{eff}_{12} \hat{P}_{12}|n \uparrow p \downarrow\rangle\nonumber\\
&\qquad-\langle p \downarrow n \uparrow | \hat{v}^{eff}_{12}\hat{P}_{12} |n \downarrow p \uparrow\rangle+
\langle p \downarrow n \downarrow | \hat{v}^{eff}_{12} \hat{P}_{12}|n \downarrow p \downarrow\rangle]\nonumber \\
&\qquad=\frac{1}{V}\int d\mathbf{r}_{12}e^{i\mathbf{k}_{ij}\cdot\mathbf{r}_{12}}\:\Big[v^{c}_{12}-v^{\tau}_{12}-v^{\sigma}_{12}+v^{\sigma\tau}_{12}-2(v^{t}_{12}-v^{t\tau}_{12})\Big(
1-3\frac{z_{12}^2}{r_{12}^2}\Big)\Big]\,.
\end{align}

Replacing the final state of Eq. (\ref{eq:TDA_exp_new}) in the definition of the response, Eq. (\ref{eq:resp_def}) yields
\begin{align}
S(\mathbf{q},\omega)&=\frac{1}{A}\sum_{f}\sum_{S}|\,_{S}\langle \Psi_f| \hat{O}_\mathbf{q} |\Psi_0\rangle|^2\, \delta(\omega-\omega_f)\nonumber \\
&=\frac{1}{A}\sum_f\sum_{ S}\Big{|}\sum_{\mathbf{p}_m\mathbf{h}_i S_z}C^{f\,SS_z}_{\mathbf{p}_m\mathbf{h}_i}\,_{SS_z}\langle \Psi_{\mathbf{p}_m;\mathbf{h}_i}| \hat{O}_\mathbf{q} |\Psi_0\rangle\Big{|}^2\, \delta(\omega-\omega_f)
\end{align}

\section{Numerical calculation of the response}
We model the infinite system using a cubic box of side $L$ with periodic boundary conditions. Hence, the single particle wave functions are the plane waves of Eq. (\ref{eq:nm_wf}) with the discrete momenta of Eq. (\ref{eq:discrete_mom}). For zero temperature SNM, all single-particle states with $|\mathbf{k}_i|\leq k_F$ are occupied in the ground state. The momenta of the $ 1p-1h$ excitations are such that $|\mathbf{h}_j|\leq k_F$ and $|\mathbf{p}_i=\mathbf{h}_j+\mathbf{q}|> k_F$. For the hole and particle momentum to be on the lattice of allowed momentum states in the box, the momentum transfer must be such that 
\begin{equation}
\mathbf{q}=\frac{2\pi}{L}(n_{q_x},n_{q_y},n_{q_z})  \, .\qquad n_{q_i}=0,\pm 1,\pm 2,\dots  \, .
\end{equation}

Given the magnitude and the direction of the momentum transfer, the side of the box can be determined by inverting the latter equation
\begin{equation}
L=\frac{2\pi}{|\mathbf{q}|}\sqrt{n_{q_x}^2+n_{q_y}^2+n_{q_z}^2}\,.
\end{equation}

The size of the basis, that can be increased by increasing $\sqrt{n_{q_x}^2+n_{q_y}^2+n_{q_z}^2}$, has been determined requiring that the response of a system of noninteracting nucleons computed on the lattice agreed with the analytical result of the FG model \cite{cowell_04,benhar_09}. The FG response is obtained replacing the effective operator with the bare operator in Eq. (\ref{eq:cfg}) and using the single particle energy of Eq. (\ref{eq:spe_fg}). The analytical calculations can be performed using a continuum of momentum states \cite{fetter_03}, while the numerical result consists of collection of discrete delta function peaked at the values of the single particle energies.  For a better representation of the results, as well as for fitting purposes, a gaussian representation of the energy conserving delta function has been adopted 
\begin{equation}
\delta(x-x_0)\to \frac{1}{\sigma\sqrt{\pi}}\exp\Big[-\Big(\frac{x-x_0}{\sigma}\Big)^2\Big]\, .
\end{equation}
For sufficiently small values of the gaussian width $\sigma$, the results become insensitive to it.

All the results that will be shown in this section refer to SNM at equilibrium density $\rho=0.16$ fm$^{-3}$.

\begin{figure}[!hb]
\begin{center}
\includegraphics[width=7.9cm,angle=0]{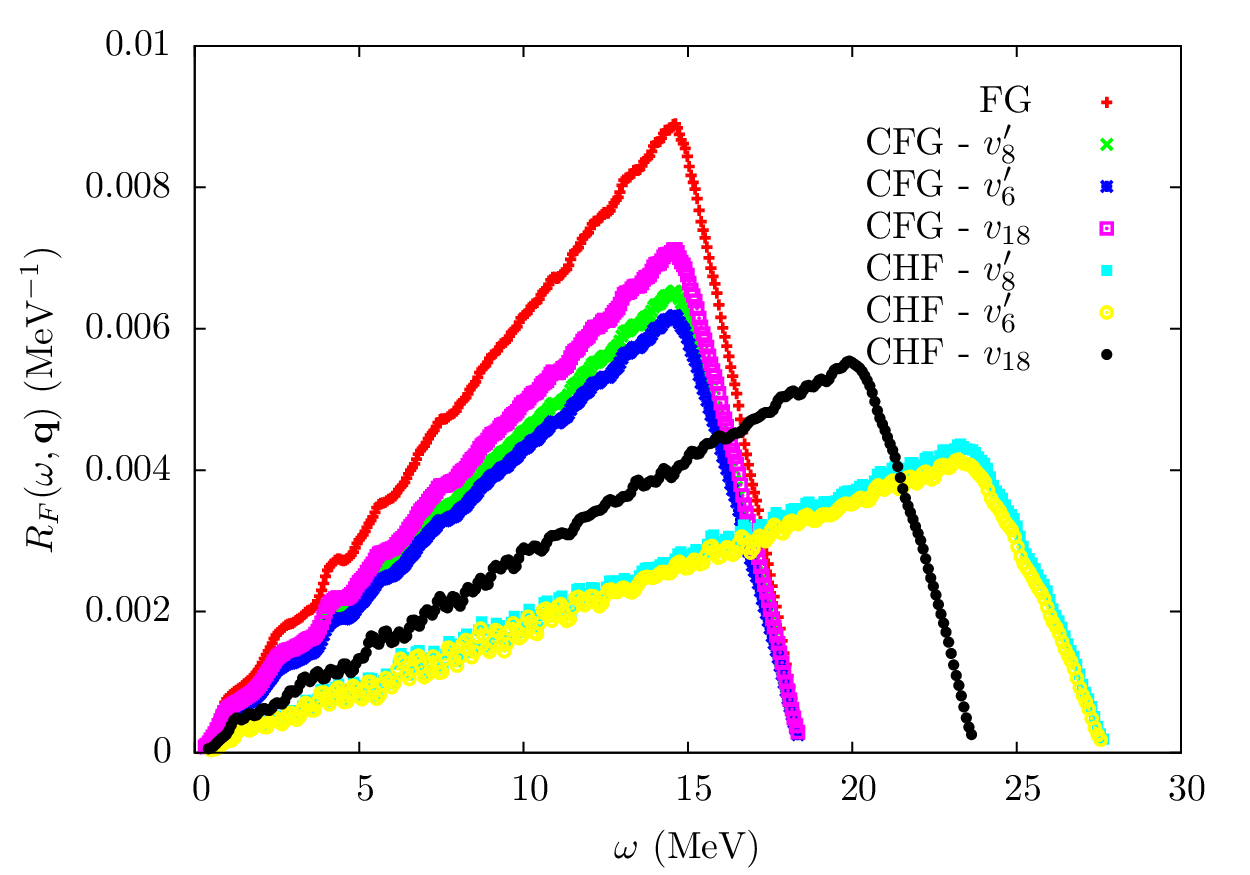}
\includegraphics[width=7.9cm,angle=0]{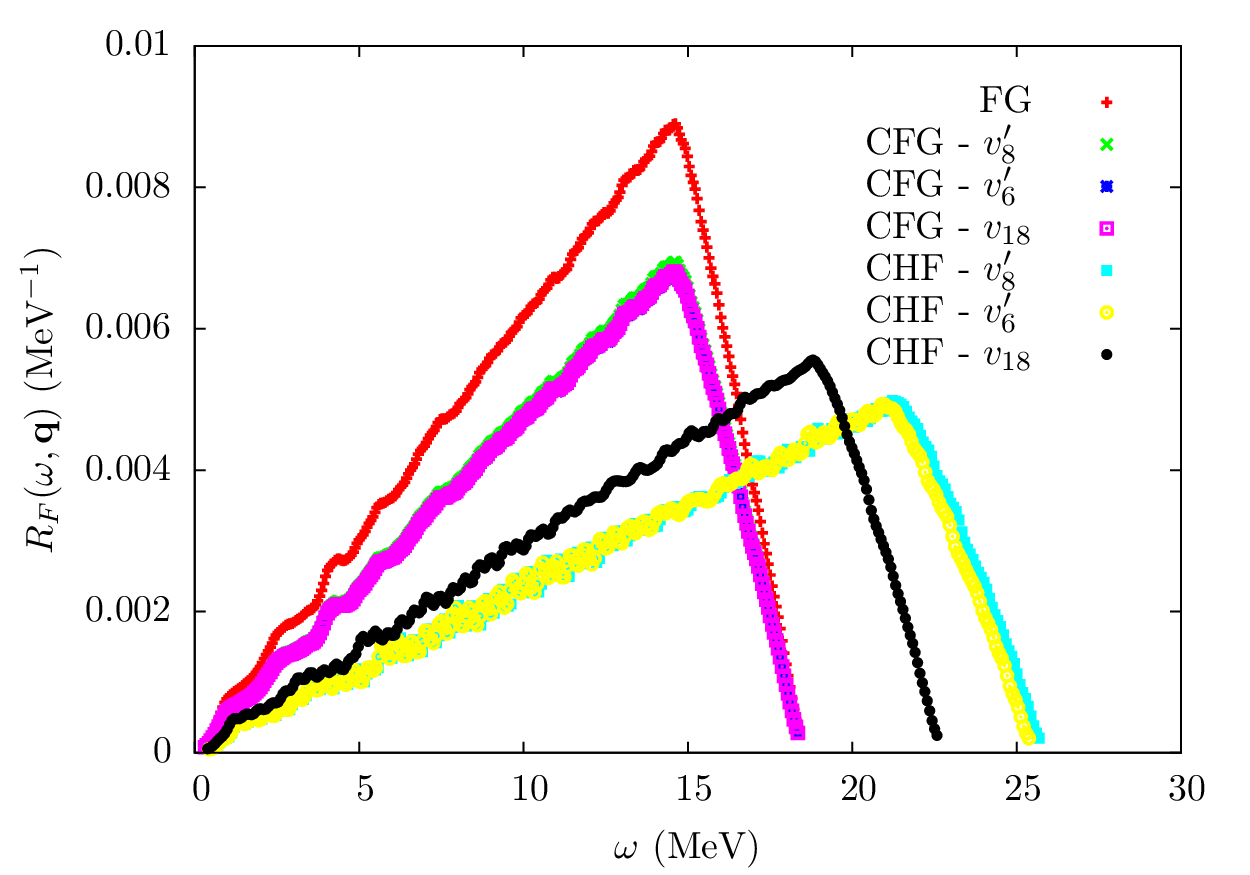}
\end{center}
\caption{Fermi response functions calculated at $q=0.3$ fm$^{-1}$. The left panel shows the response calculated at two-body cluster level for both $O_{\boldsymbol q}^{eff}$ and $v_{12}^{eff}$ for different choices of correlation functions. In the right panel the three body cluster has been included. The plus marks show the response for a non interacting FG. \label{fig:cfg_chf_fermi}}
\end{figure}

\subsection{CFG and CHF}
FHNC/SOC  calculations and their associated minimization procedure, explained in Section \ref{Variational_SA}, provide a set of correlations function, 
corresponding to the minimum of the hamiltonian  expectation value. We have found the best correlation functions for the Argonne $v_{6}^\prime$, $v_{8}^\prime$ two-body potentials, and for comparison we have also considered the correlations of Ref. \cite{akmal_98} corresponding to Argonne $v_{18}$. With these correlations, the Fermi and Gamow-Teller response functions have been evaluated in CFG and CHF approximations.

When only two-body cluster diagrams are considered, as in Refs. \cite{cowell_04,benhar_09}, the CFG response, suppressed by a $20-25\%$ with respect to the FG case, exhibits a sizable dependence on the choice of correlations, as shown in the left panels of Figs. \ref{fig:cfg_chf_fermi} and \ref{fig:cfg_chf_gt}. These figures refer to a transfer momentum 
$\boldsymbol q=|{\bf q}|(4\hat{x}+4\hat{y}+4\hat{z})/\sqrt{48}$ with $|\mathbf{q}|=0.3$ fm$^{-1}$. The folding gaussian function has a width of $0.25$ MeV. 

This unphysical effect is removed once the effective weak transition operator is computed at three-body cluster level. As a matter of fact, the CFG curves in the right panels of the aforementioned figures are very close, when not superimposed, to each other. Therefore our results appear to be more robust than those of Refs. \cite{cowell_04,benhar_09}, as the physical quantities should not  be sensitive to the details of the short range behavior of the correlation functions.

\begin{figure}[!ht]
\begin{center}
\includegraphics[width=7.9cm,angle=0]{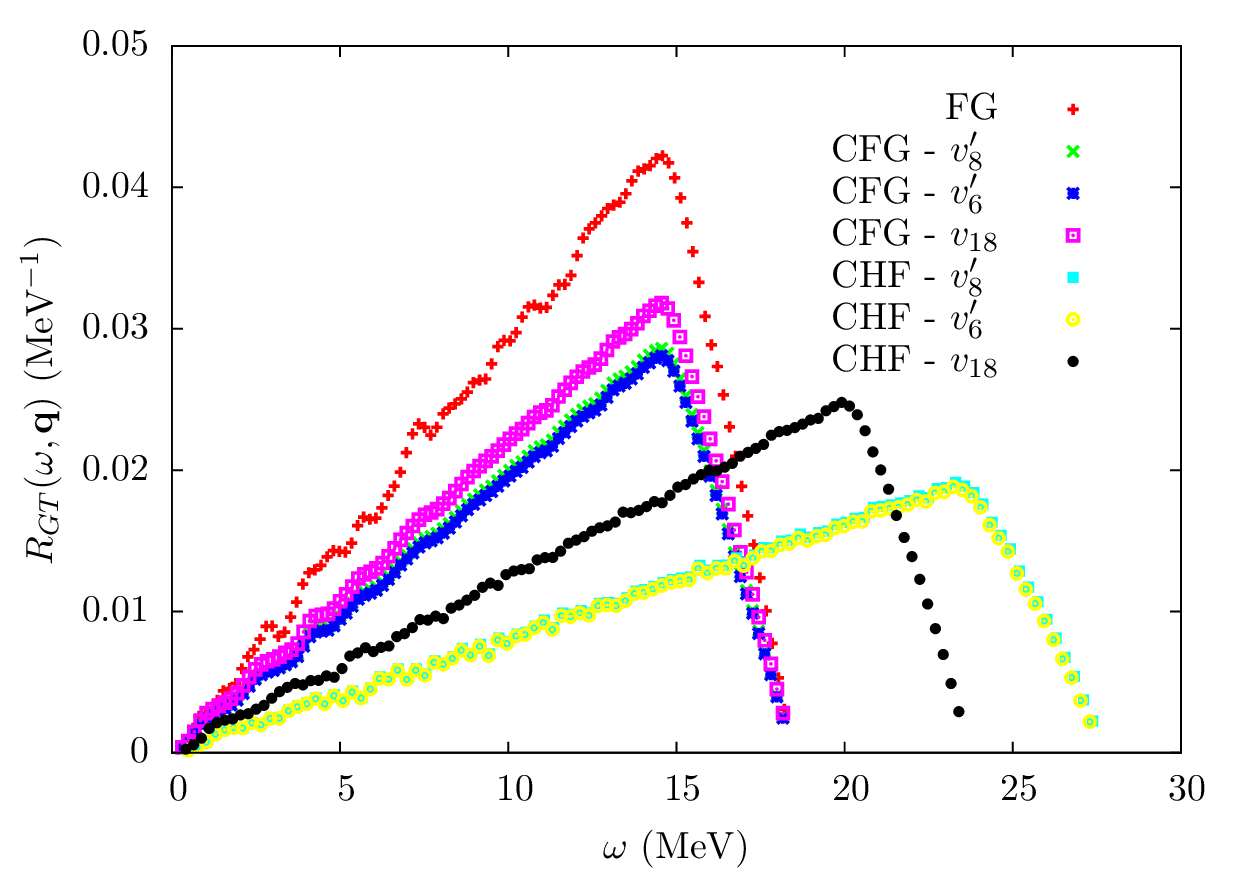}
\includegraphics[width=7.9cm,angle=0]{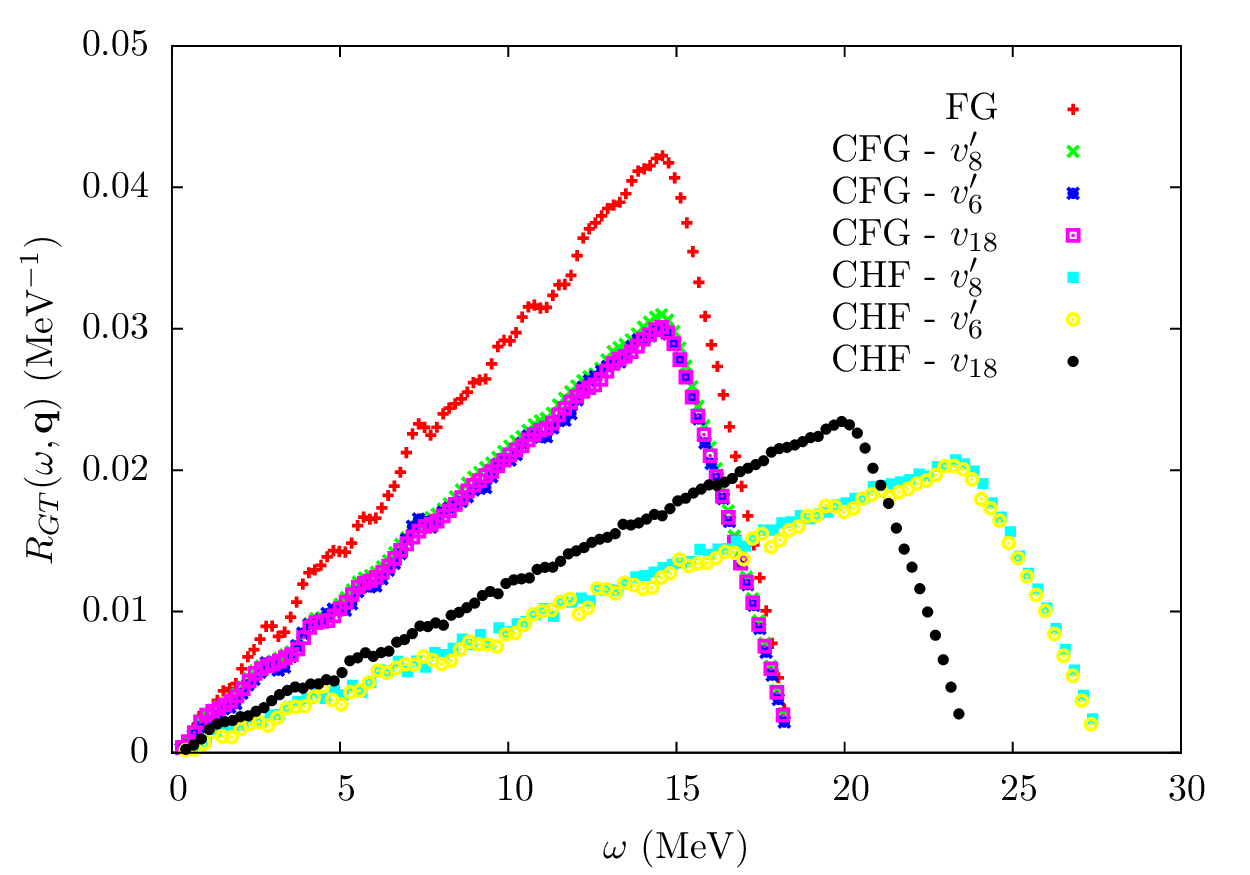}
\end{center}
\caption{Same as Fig. \ref{fig:cfg_chf_fermi} but for Gamow-Teller response functions. \label{fig:cfg_chf_gt}}
\end{figure}

A small dependence on the choice of the correlation is observed in CHF calculations, also shown in Figs. \ref{fig:cfg_chf_fermi} and \ref{fig:cfg_chf_gt}. The reason for this lies in the single particle energies, which depend on the first four components of the NN potential (see Eq. (\ref{eq:spe_nm2})). It turns out that the first four component of the $v_{8}^\prime$ potential are very similar to those of $v_{6}^\prime$, while the same behavior is not observed for the full $v_{18}$.

We observe that the shift of the strength to higher $\omega$ is slightly enhanced by three-body clusters, expecially for Fermi transition.

\subsection{CTDA results}

\begin{figure}[!!h]
\begin{center}
\includegraphics[width=7.9cm,angle=0]{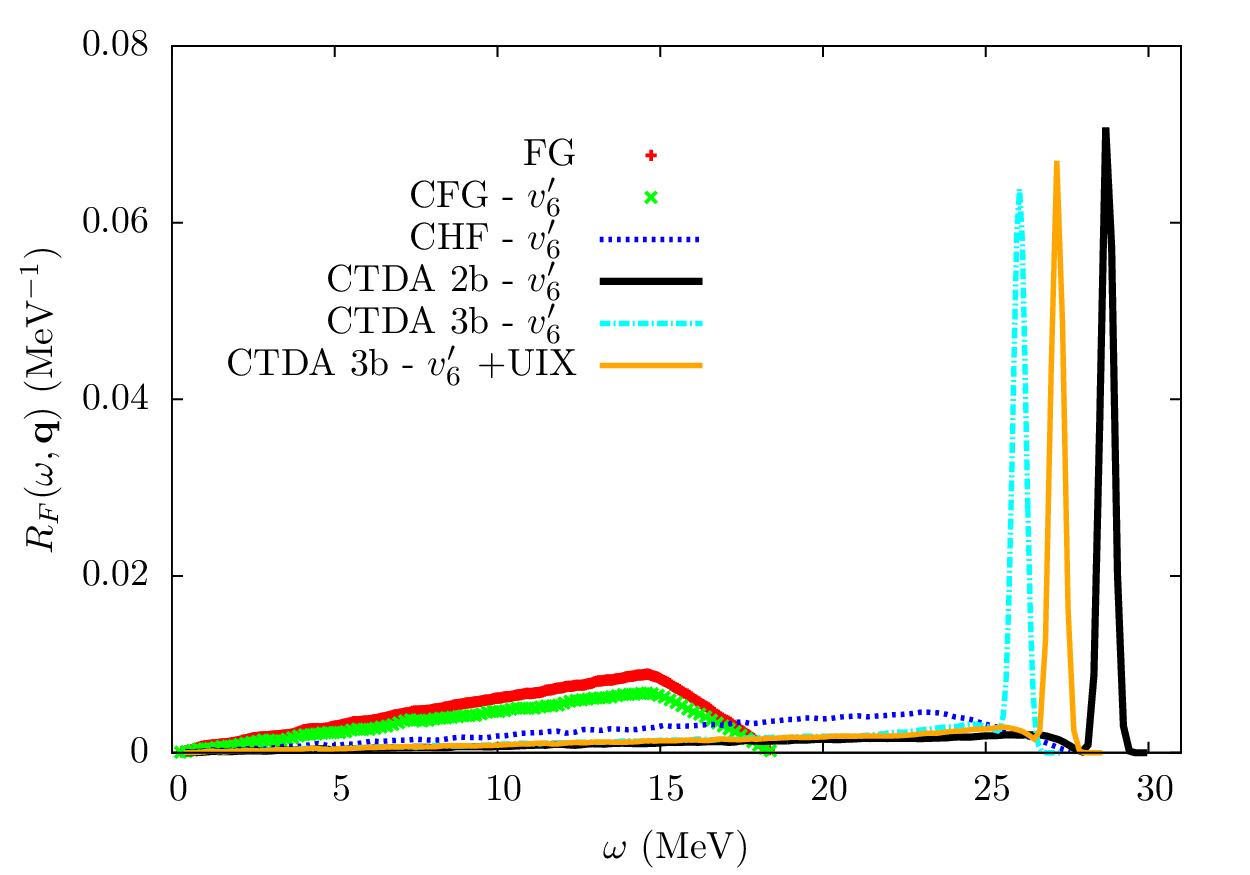}
\includegraphics[width=7.9cm,angle=0]{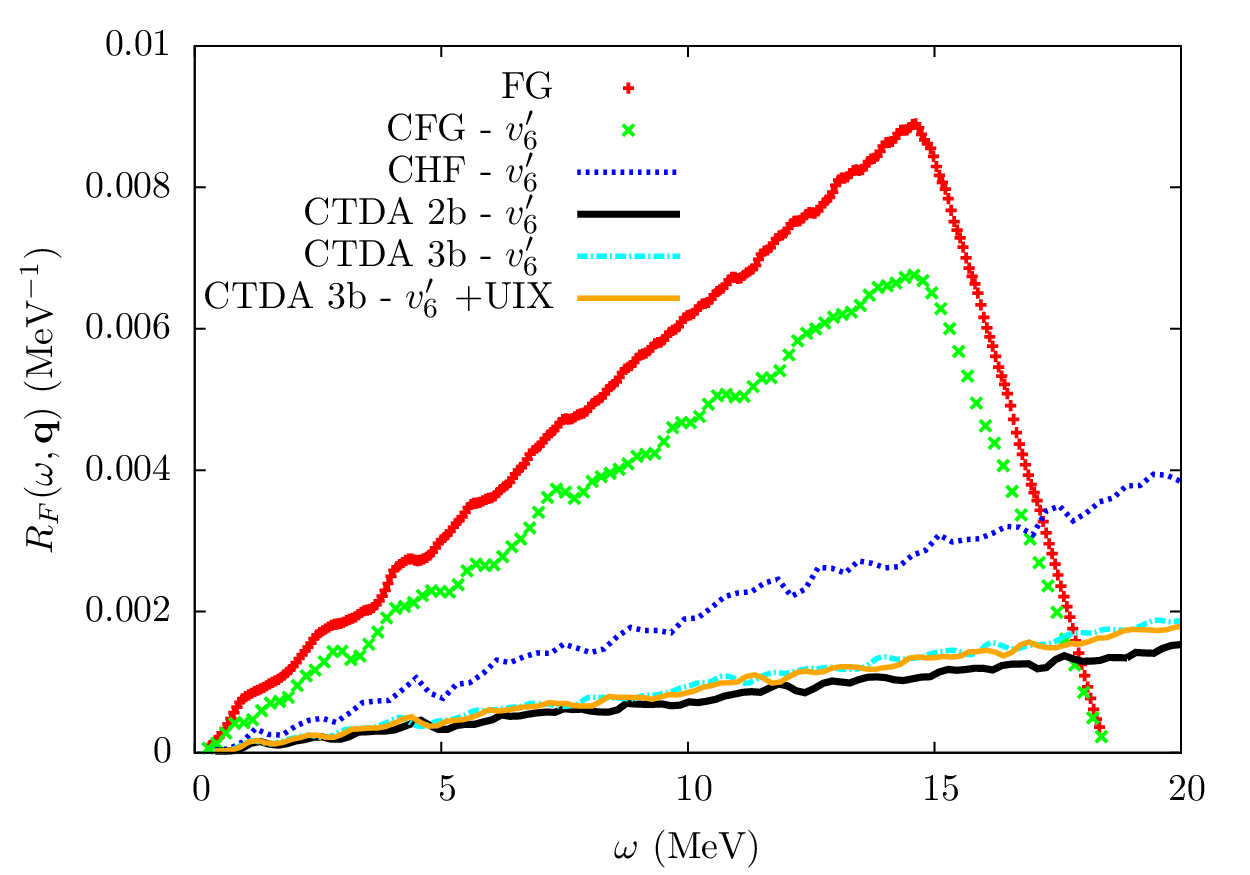}
\end{center}
\caption{Fermi response functions calculated at $q=0.3$ fm$^{-1}$ using $v_{6'}$ and $v_{6'}+UIX$ interaction models. The upper panel shows the full response across all the values of $\omega$, the lower is a magnification of the small $\omega$ region. The plus marks and the cross marks display the response for FG and CFG respectively. The remaining lines refers to CHF approximation and CTDA (see text for details). Responses are folded with a Gaussian of width 0.25 MeV.\label{fig:cfg_ctda_fermi}}
\end{figure}

The nuclear matter response calculated in CTDA for $|\mathbf{q}|=0.30$ MeV is displayed in Figs. \ref{fig:cfg_ctda_fermi} and \ref{fig:cfg_ctda_gt} for Fermi and Gamow-Teller transitions, respectively. The peak corresponding to the collective mode is shifted to lower energies when the tree-body cluster is included. This effect, ascribed to the change of the single particle energies, is mitigated when the UIX potential is included in the hamiltonian. 

When the three-body cluster is included, it produces in a depletion of the Gamow-Teller resonance at  $|\mathbf{q}|=0.30$, particularly apparent when the nuclear hamiltonian is lacking of the three-body potential. This feature can be explained by looking at Fig. \ref{fig:q_scan_fgt}, where the Fermi and Gamow-Teller responses are plotted for different values of  $|\mathbf{q}|$ ranging from $0.10$ fm$^{-1}$ to $0.50$ fm$^{-1}$. The maximum of the peak of the Fermi response is observed at $|\mathbf{q}|\simeq0.40$ fm$^{-1}$, like in the two-body cluster approximation of Ref. \cite{cowell_04}. For the Gamow-Teller case, on the other hand, the position of the 
resonance is shifted  from $|\mathbf{q}|\simeq0.30$ fm$^{-1}$, corresponding to the two-body cluster calculation, to $|\mathbf{q}|\simeq0.25$ fm$^{-1}$. 

\begin{figure}[!!h]
\begin{center}
\includegraphics[width=7.90cm,angle=0]{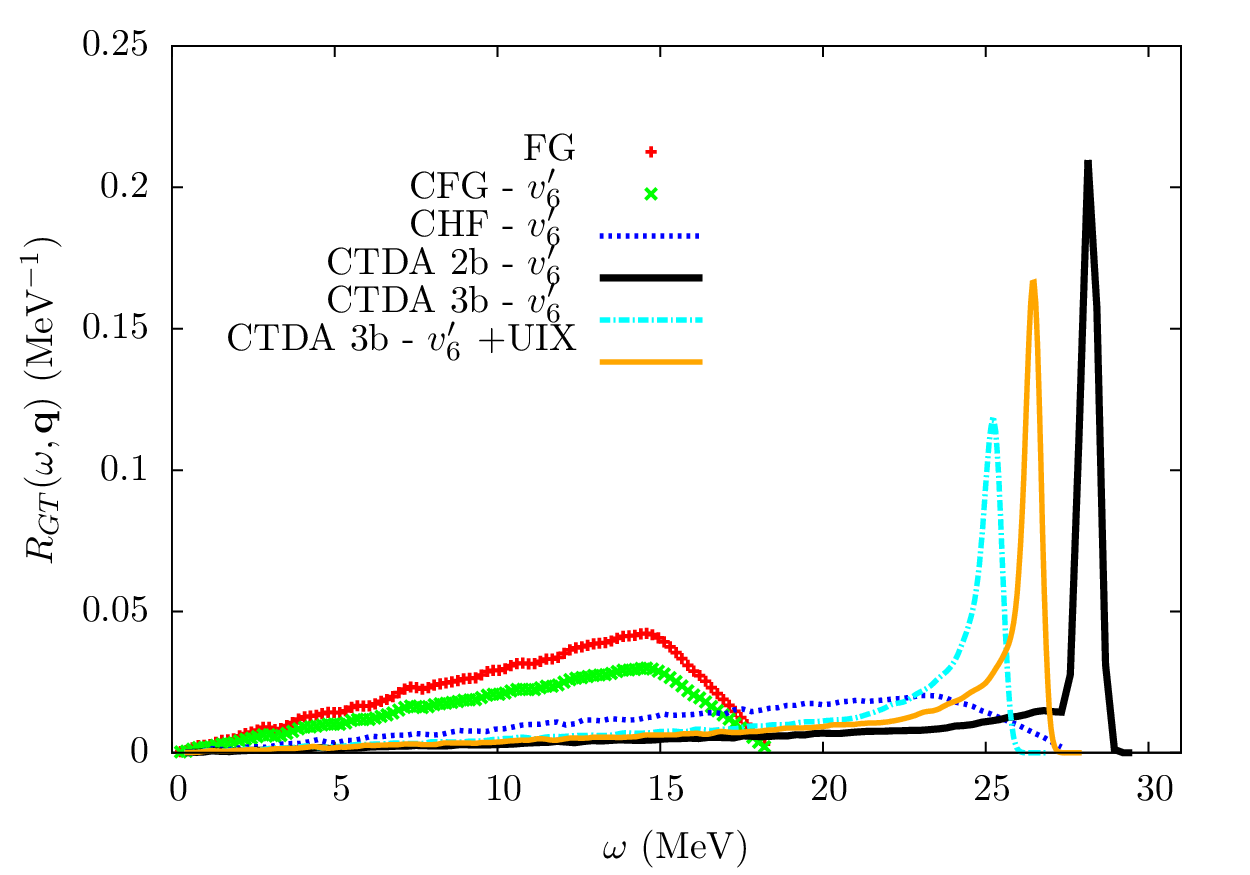}
\includegraphics[width=7.90cm,angle=0]{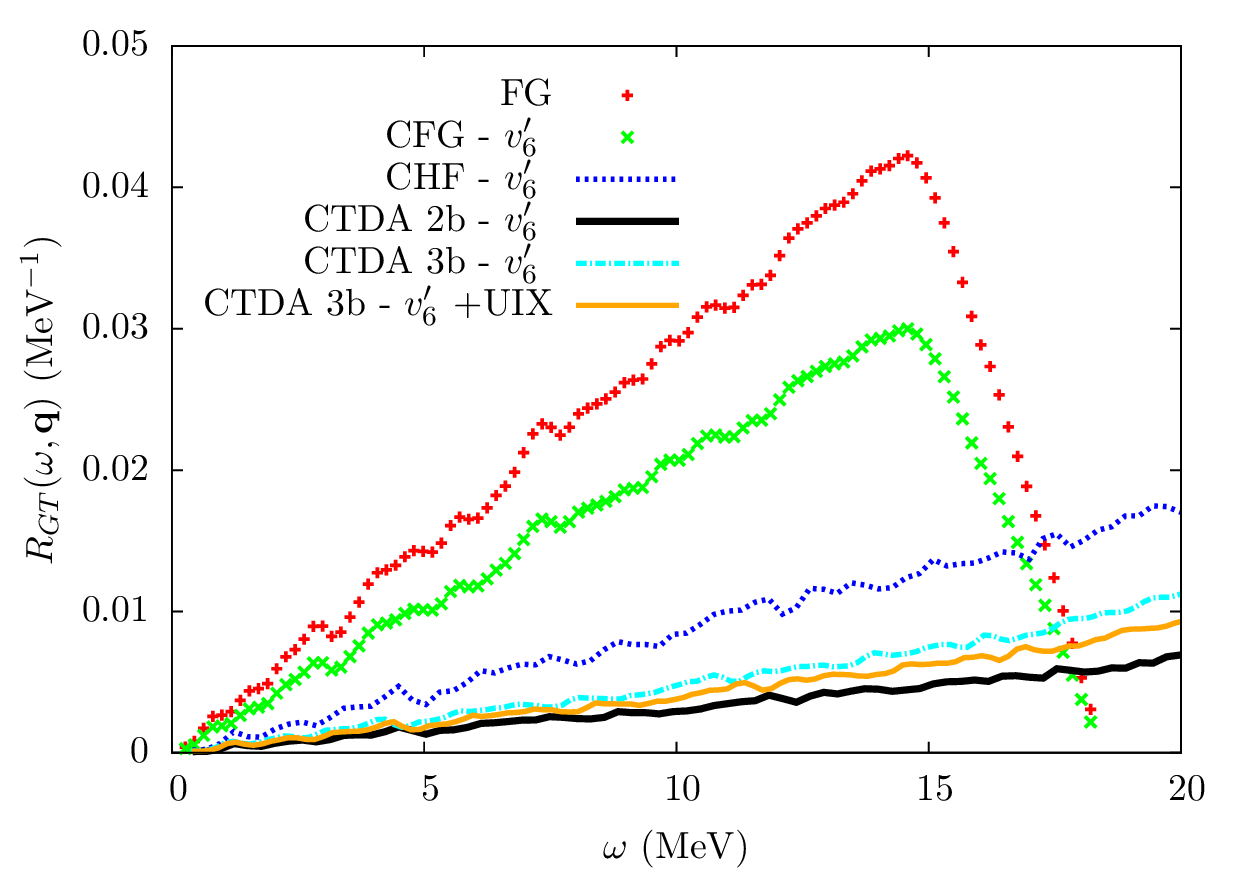}
\end{center}
\caption{The same as Fig. \ref{fig:cfg_ctda_fermi} but for Gamow-Teller response functions.\label{fig:cfg_ctda_gt}}
\end{figure}

\begin{figure}[!!h]
\begin{center}
\includegraphics[width=7.9cm,angle=0]{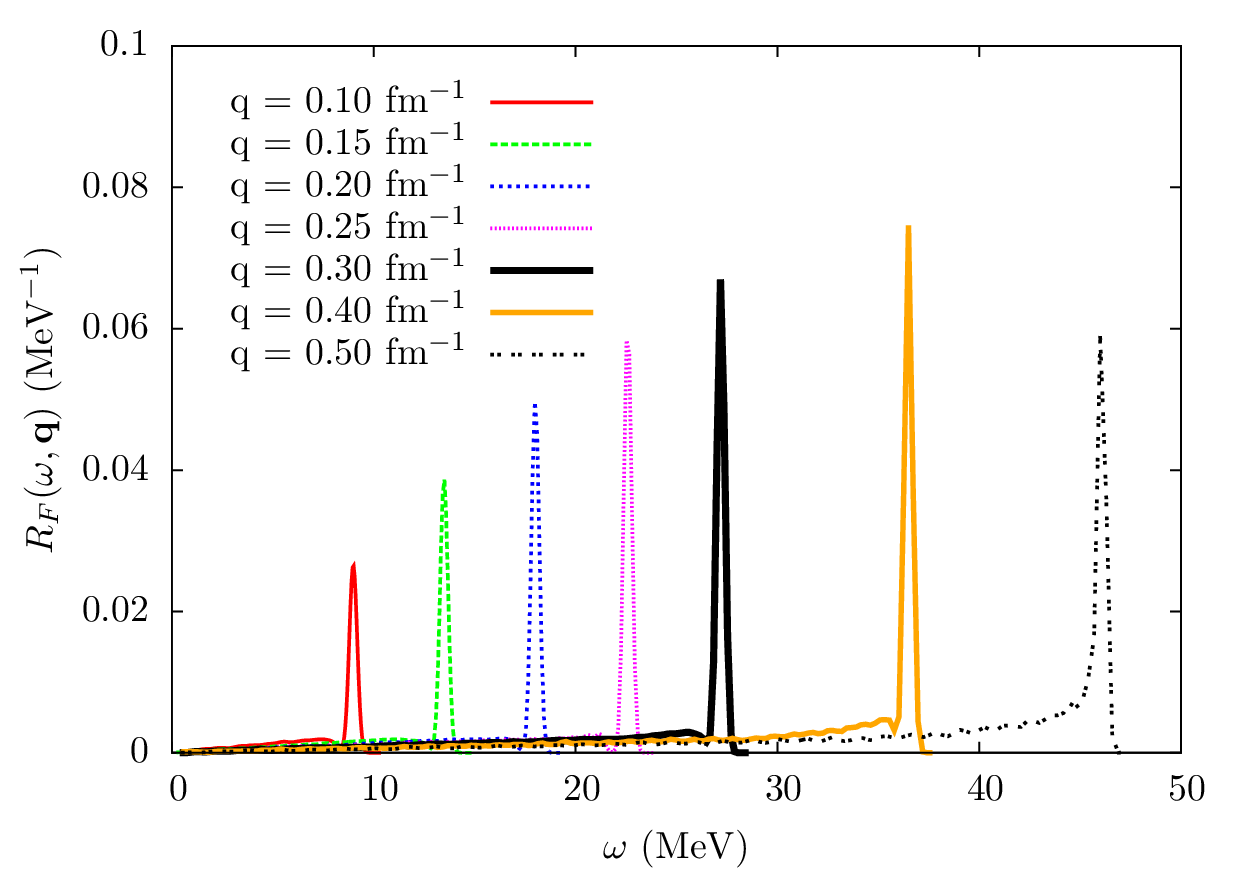}
\includegraphics[width=7.9cm,angle=0]{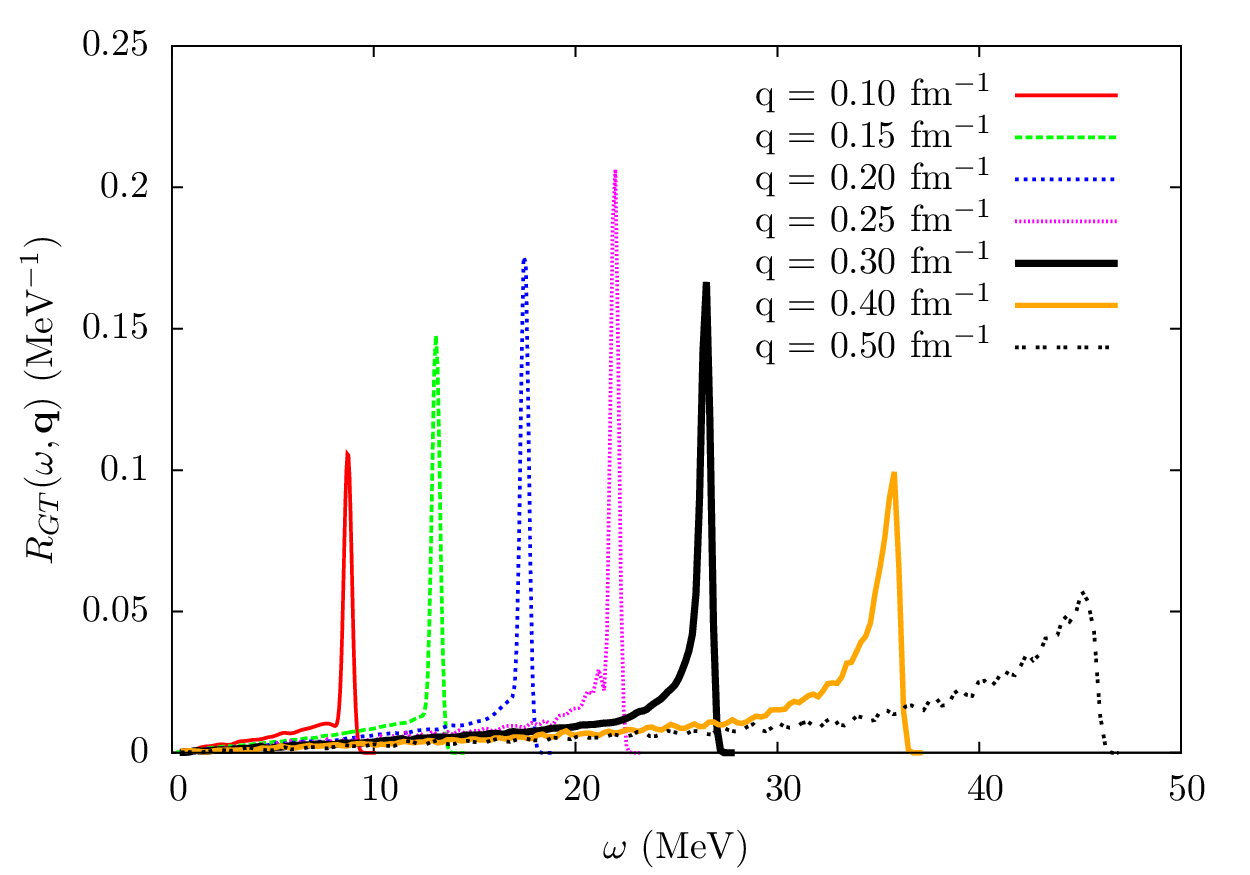}
\end{center}
\caption{Fermi (upper panel) Gamow-Teller (lower panel) response functions calculated at $q=0.10,0.15,0.20,0.25,0.30,0.40,0.50$ fm$^{-1}$ using $v_{6'}\,+\,UIX$  potential and correlations. \label{fig:q_scan_fgt}} 
\end{figure}

\subsection{Sum rules}
\label{sec:sum_rule}

The set of final states in Eq. (\ref{eq:resp_def}) is not exhausted by $1p-1h$ excitations. In principle, transitions to more complex multi $p-h$ states should also be 
considered. So far, the contribution of these states have been neglected; however, an estimate of their importance can be obtained computing the sum rules. 
The static structure function is defined by 
\begin{align}
S(q)&=\int d\omega S(\mathbf{q},\omega)\nonumber\\
&=\frac{1}{A}\int d\omega \sum_f |\{ \Psi_f | \hat{O}_{\textbf{q}}| \Psi_0\} |^2 \delta(\omega+E_0-E_n)\nonumber\\
&=\frac{1}{A} \{ \Psi_0 | \hat{O}_{\textbf{q}}^\dagger  \hat{O}_{\textbf{q}}| \Psi_0\}\, .
\end{align}
While a direct integration of the CTDA response functions allows for the evaluation of $S(q)$, from the last line of the latter equation it is clear that $S(q)$ can be computed by using the variational ground state (VGS) resulting from FHNC-SOC calculations. In particular, the knowledge of the two-body operatorial distribution functions of Eq. (\ref{eq:g2bop_def}) is needed to compute the structure function. 

While the VGS calculations include all the multi $p-h$ excitations in the CBF basis, in the CTDA only the correlated $1p-1h$ states are taken into account. Therefore, multi $p-h$ contributions can be estimated from the difference $S^{VGS}(q)-S^{CTDA}(q)$. Note, however, that an interplay between many-body correlations and multi $p-h$ excitation could in principle take place.  In fact, while VGS includes many-body correlations through the chain summations, in the CTDA of Ref. \cite{cowell_04} only two-body cluster terms have been considered.

\begin{figure}[!!h]
\begin{center}
\includegraphics[width=7.9cm,angle=0]{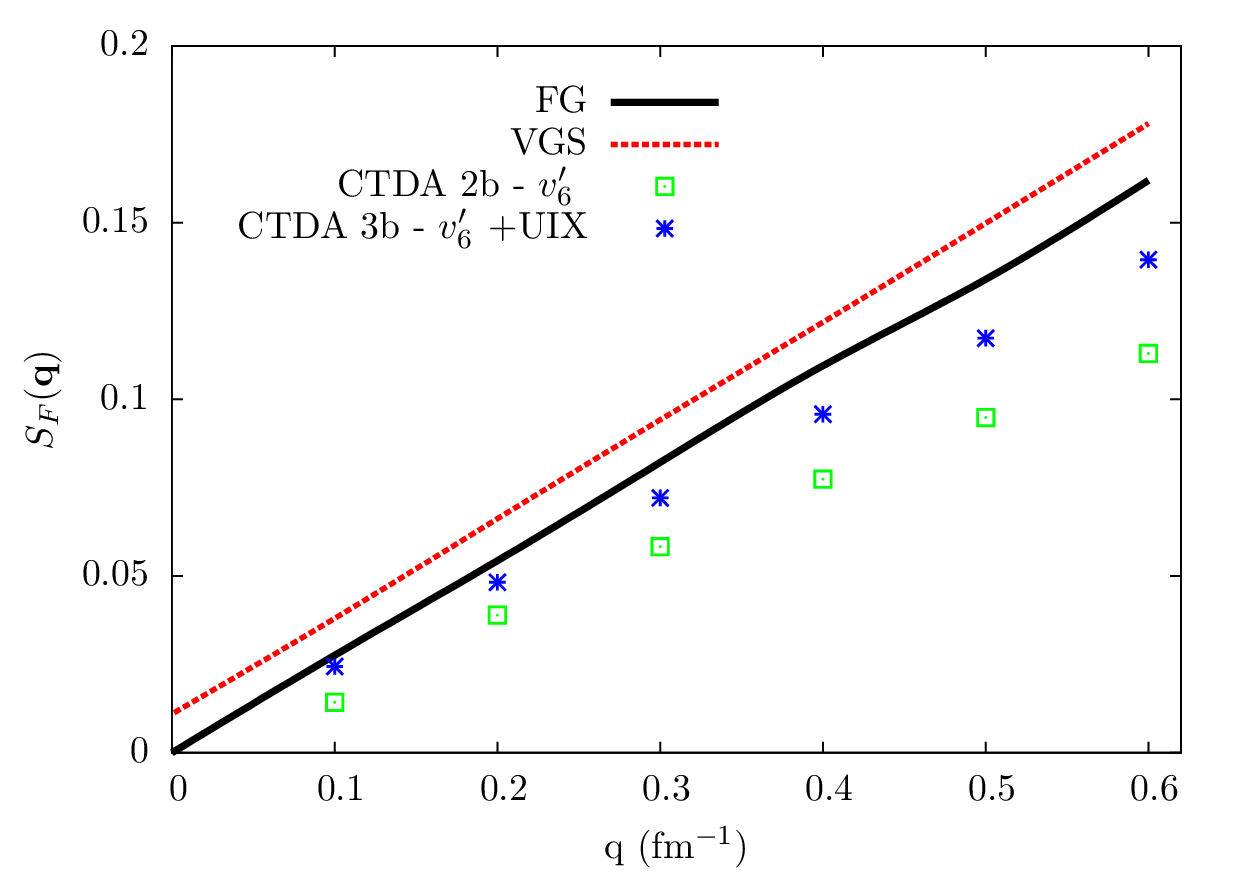}
\includegraphics[width=7.9cm,angle=0]{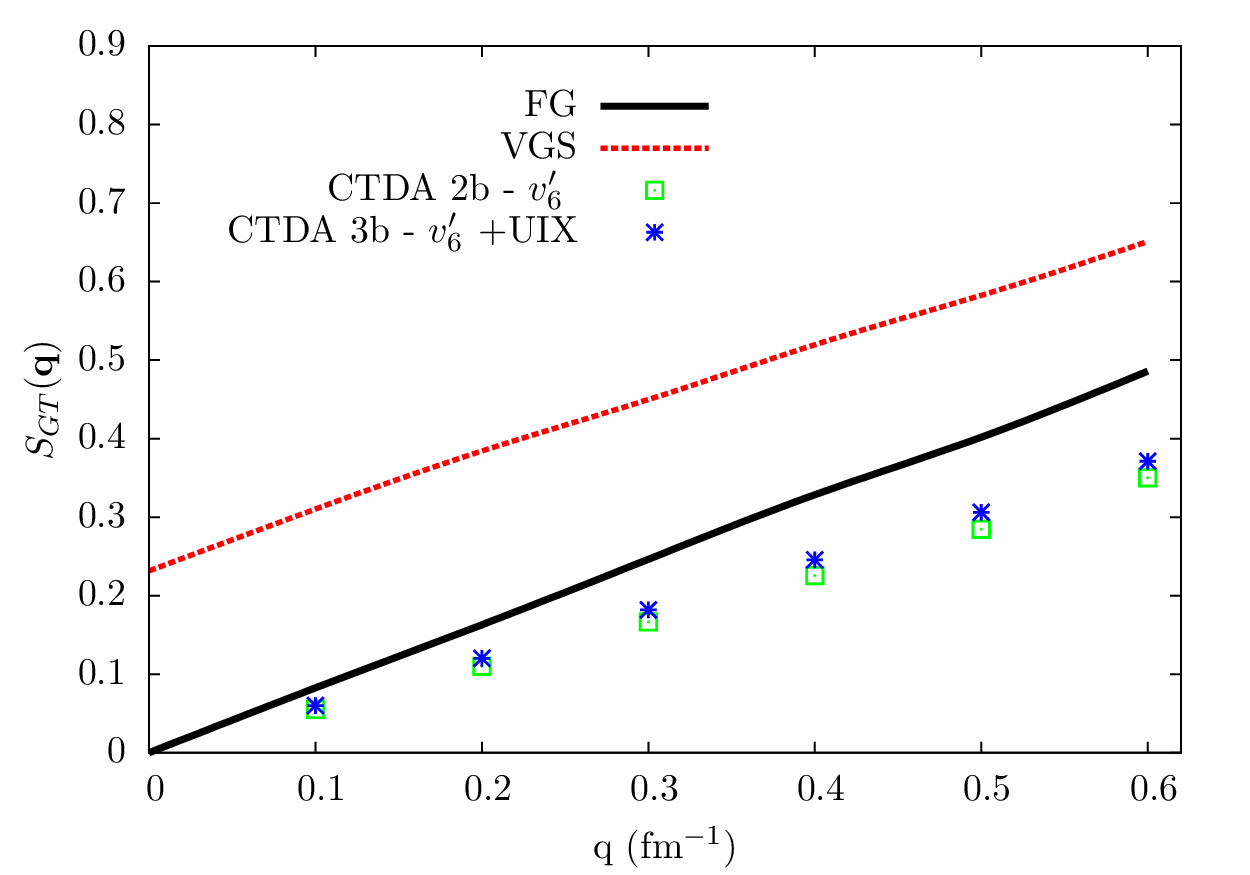}
\end{center}
\caption{Static response function for Fermi (upper panel) and Gamow-Teller (lower panel) transitions. Results are shown for a noninteracting FG (dashed lines), for the integral of the CTDA $S(\mathbf{q},\omega)$ at two-body (squared) and three-body (stars) cluster level and for the VGS (solid lines). \label{fig:sumtot}}
\end{figure}

The static structure functions for the Fermi and Gamow-Teller transitions are displayed in the left and in right panels of  Fig. \ref{fig:sumtot}, respectively. The dashed lines refer to the noninteracting FG, while the squares and the stars represent the two-body and three-body cluster results, respectively. The NN potential of the effective hamiltonian is the Argonne $v_{6}^\prime$, while in the tree-body cluster calculation the UIX interaction has been used. Note that, in the sum rules' calculation, the authors of Ref. \cite{cowell_04} have set to one the factor $g_A$ of the Gamow-Teller transition. For a better comparison with their results, we did the same choice of normalization.

As noted in Ref. \cite{cowell_04}, the variational calculations of Ref. \cite{akmal_97}, because of the approximations involved in FHNC/SOC calculations, do not show the correct behavior, $S(q=0)=0$, required by baryon number conservation. On the other hand, the static structure function obtained within CTDA does exhibit the appropriate low-momentum limit. 

As far as the multi $p-h$ excitations are concerned, the two-body cluster results show that their contribution is smaller than the dominant $1p-1h$ excitation, but it is not negligible. When three-body cluster is accounted for, the VGS and the CTDA curves get closer. The shift turns out to be detectable for the Fermi transition case, while for the total Gamow-Teller response is very small. This is a clear indication that the difference between FHNC/SOC and correlated Tamm-Dancoff results has largely to be ascribed to the multi $p-h$ excitations. 

There are experimental and theoretical indications that the spin longitudinal and spin transverse response functions can differ significantly due to tensor forces. Thus, we studied how the UIX three-body force affects these quantities computing the spin longitudinal and spin transverse static structure functions, defined as
\begin{align}
S^L(q)&=\frac{1}{A}\int d\omega \sum_f |\{ \Psi_f | \hat{q} \cdot \hat{O}_{\textbf{q}}^{GT}| \Psi_0\} |^2 \delta(\omega+E_0-E_n)\\
S^T(q)&=\frac{1}{A}\int d\omega \sum_f |\{ \Psi_f | \hat{q} \wedge \hat{O}_{\textbf{q}}^{GT}| \Psi_0\} |^2 \delta(\omega+E_0-E_n)\\
\end{align}

As shown in Fig. \ref{fig:sum_lt}, the inclusion of the UIX potential brings the CTDA curves for $S^L(q)$ slightly closer to those of VGS across all the values of $|\mathbf{q}|$. As far as the transverse static response function is concerned, the position of the maximum of the CTDA calculations including UIX potential almost coincide with the VGS results. At small momentum transfer however, the three-body cluster contributions move away the $S^T(q)$ obtained from CTDA from the variational results. 

\begin{figure}[!!h]
\begin{center}
\includegraphics[width=7.9cm,angle=0]{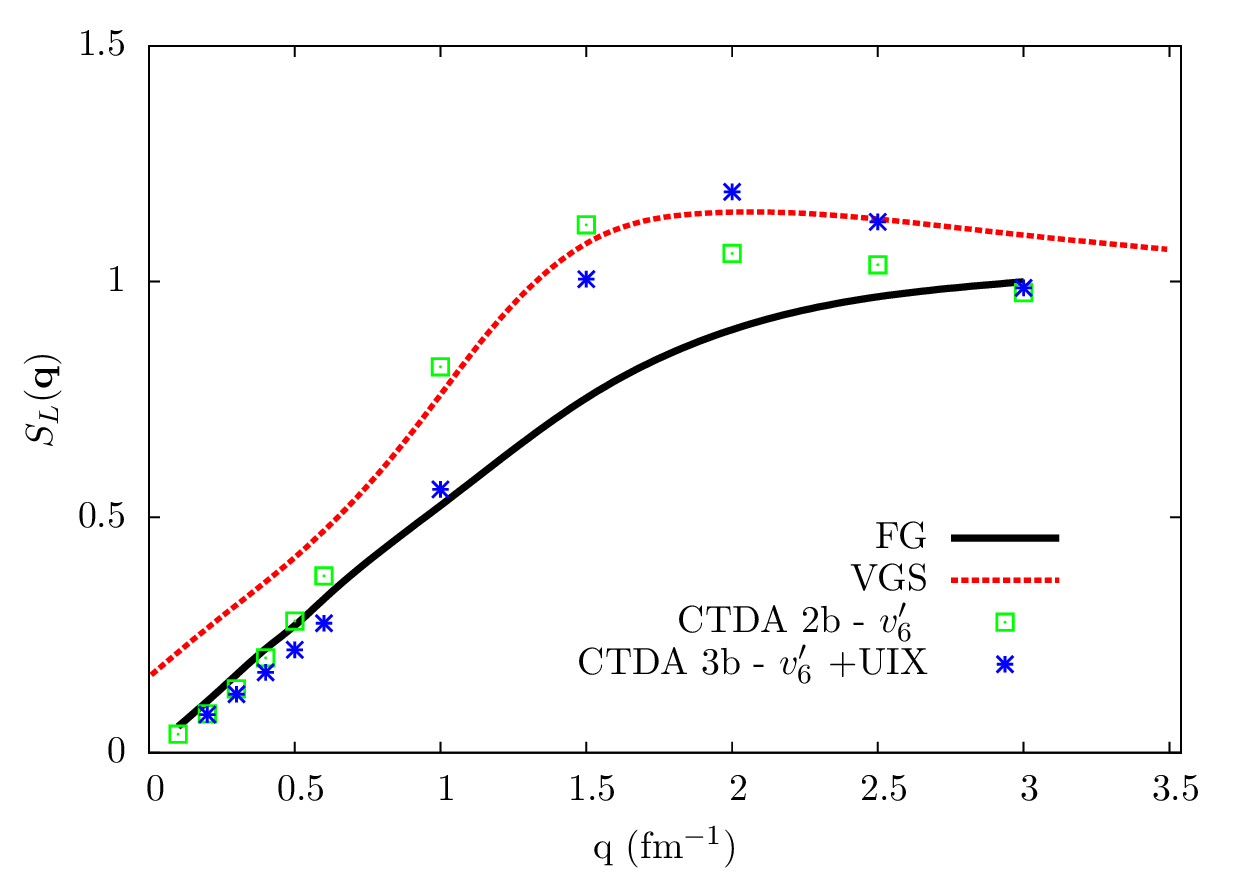}
\includegraphics[width=7.9cm,angle=0]{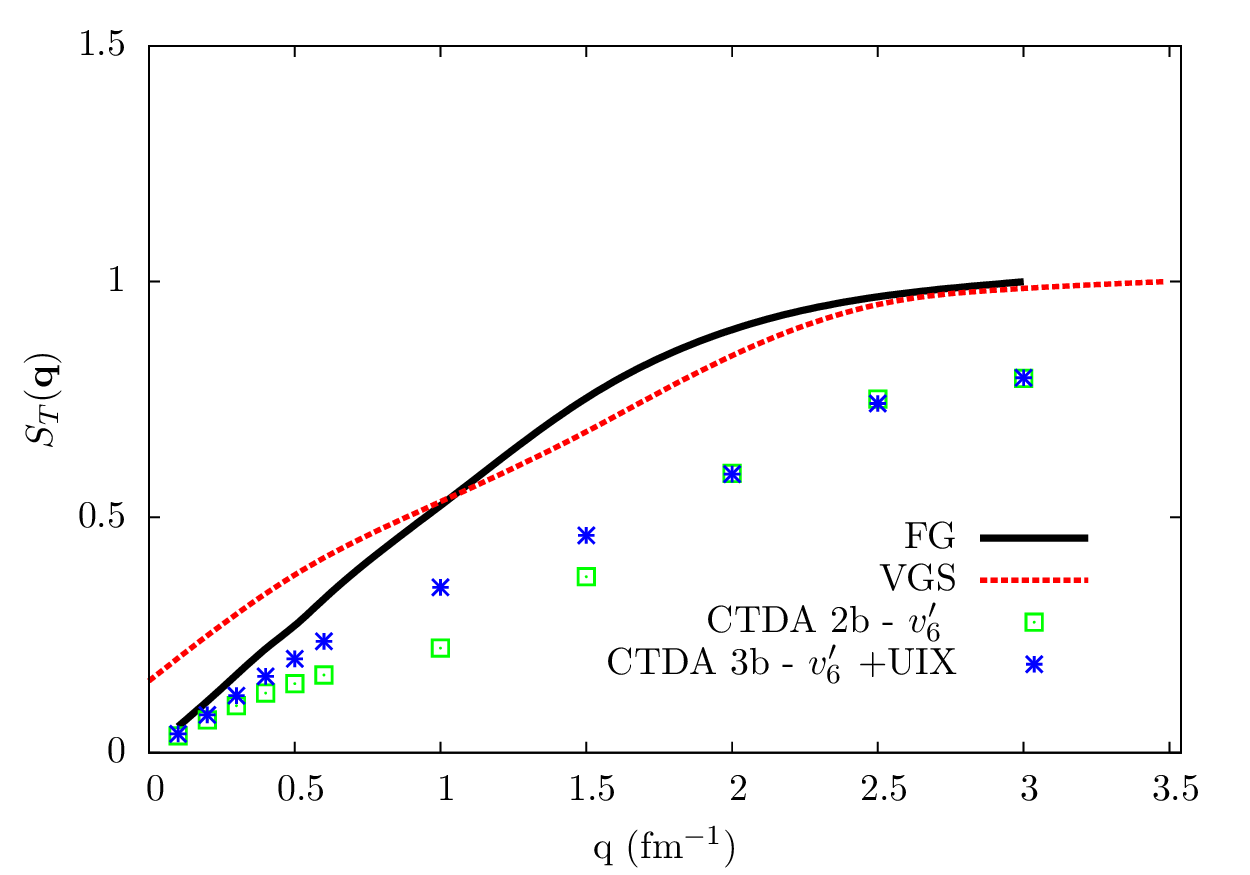}
\end{center}
\caption{Spin longitudinal (left panel) and spin transverse (right panel) static response functions, calculated using VGS (solid lines), two-body CTDA (squares), two-body CTDA (stars), and noninteracting FG (dotted lines). \label{fig:sum_lt}}
\end{figure}

Recently \cite{shen_12}, the sum rules relations (additional sum rules with an increasing power of $\omega$ in the integrand can be defined) has been inverted to obtain the spin-response function of PNM at zero momentum transfer. The authors of Ref. \cite{shen_12} used the formalism of AFDMC to compute the operatorial distribution functions, obtaining very promising results. As a follow up of the present work, we are planning to compute the response function of PNM and compare it with their findings.

\newpage             
\thispagestyle{empty}  

% Conclusioni %%%%%%%%%%%%%%%%%%%%%%%%%
\addcontentsline{toc}{chapter}{Conclusions}
\chapter*{Conclusions}
\markboth{Conclusions}{}
\markright{Conclusions}{}
The main body of this Thesis has been devoted to the discussion of a novel approach, developed in Ref. \cite{lovato_11,lovato_11b}, allowing one to obtain an effective density-dependent 
NN potential taking into account the effects of three-- and many-- nucleon interactions. 

The resulting effective potential can be easily used in calculations of nuclear properties within many-body approaches based on phenomenological hamiltonians, including the effects of strong NN correlations, which can not be treated in {\em standard} perturbation theory in the Fermi gas basis. Moreover, the derivation of the density-dependent NN potential is fully consistent with the treatment of correlations underlying the FHNC and AFDMC approaches. 

While the reduction of $n$-body potentials to a two-body density-dependent potential is reminiscent of the approach of Refs. \cite{lagaris_81,friedman_81}, our scheme significantly improves upon the TNI model, in that i) it is based on a microscopic model of the three nucleon interaction providing a quantitative description  of the properties of few nucleon systems and ii) allows for a consistent inclusion of higher order terms in the density expansion, associated with four- and more-nucleon interactions. 

As shown in Chapter \ref{chapt:tbp}, the results of calculations of the PNM and SNM equation of state carried out using the density-dependent potential turn out to be very close to those obtained with the UIX three-body potential. In this context, a critical role is played by the treatment of both dynamical and statistical correlations, whose inclusion brings the expectation value of the effective potential into agreement with that of the UIX potential (see Fig.~\ref{fig:compare_potentials}). This is a distinctive feature of our approach, as compared to different reduction schemes based on effective interactions,  suitable for use in standard perturbation 
theory \cite{hebeler_10,weise_10}.

Using the density-dependent potential we have been able to carry out, for the first time, a AFDMC calculation of the equation of state of SNM consistently including the effects of three nucleon forces. The results of this calculation show that the $v_6^\prime + $UIX hamiltonian, or equivalently the one including the effective potential, fails to reproduce the empirical data. 

The FHNC results obtained using the $v_8^\prime$ potential indicate that the 5--6~MeV underbinding at equilibrium density can not be accounted for replacing the $v_6^\prime$ with a more refined model, such as $v_{18}$.  Hence, the discrepancy has to be ascribed either to deficiencies of the UIX model or to the effect of interactions involving more than three nucleons. 

In order to improve on the UIX three-nucleon interaction, we have performed nuclear matter calculations using the new generation of chiral inspired three-nucleon potentials in coordinate space \cite{lovato_12}. We have carried out a comparative analysis of the EoS of PNM and SNM obtained using the different parametrizations of the NNLOL potential, as well as the improved versions of the TM model discussed in Ref. \cite{kievsky_10}. 

The calculation of the SNM EoS has been been performed within the variational FHNC/SOC approach. In the case of PNM we have also used the AFDMC computational scheme, the results of which turn out to be in close agreement with the variational FHNC/SOC estimates.

Our analysis shows that the transformation from momentum to coordinate space brings about a cutoff dependence, leading to sizable effects in nuclear matter. As discussed in Section \ref{sec:cont_issue}, the contribution of the contact term, which in PNM would vanish in the $\Lambda \to \infty$ limit, can not be fully determined fitting the low energy observables. Moreover, the NNN contact terms of the NNLOL$_2$ and NNLOL$_3$ models turn out to be attractive in PNM, leading to a strong softening of the EoS. 

An illustrative example of the uncertainty associated with the local form of the NNN contact term is provided by the results of Fig. \ref{fig:nnlol_snm} and Table \ref{tab:parameters_nnlol}. The NNLOL$_4$ model largely overestimates the empirical value of the compressibility modulus of SNM, thus yielding a stiff EoS. On the other hand, as pointed out in Section \ref{subsec:nnlolres}, it predicts a soft EoS of PNM. 

The impact of this is ambiguity is large, since compressibility is a most important property of the EoS. The recent discovery of a $\sim 2$ M$_\odot$ neutron star appears in fact to rule out dynamical models yielding a soft  EoS of $\beta$-stable matter. 

None of the considered three-nucleon potential models simultaneously explains the empirical equilibrium density and binding energy of SNM. However, among the different parametrization that we have analyzed, the NNLOL$_4$ and TM$_3^\prime$ provide reasonable values of $\rho_0$. It has to be emphasized that this is a remarkable result, as, unlike the UIX model, these potential do not involve any parameter adjusted to reproduce $\rho_0$.  

In order to resolve  the inconsistencies involved in the contact term, one should include all contributions to this term arising from the chiral expansion at NNLO. Moreover, as pointed out by the authors of Ref.~\cite{girlanda_11}, due to the choice of the regulator function (see Eq.(\ref{eq:chi_cut})), a fully consistent treatment should also take into account NNNNLO contact contributions. 

In future works we are planning to include the NNNLO contributions to the three-nuclear interactions of Refs. \cite{bernard_08,bernard_11}, as well as the three-body NNNNLO contact terms of Ref. \cite{girlanda_11}. For consistency, at that point also the NNNLO long-range terms calculated in \cite{krebs_12} should be accounted for. 
However, since we are not using the chiral potential in the two-body sector, we need to deal with the issue of  determining the low-energy constants entering the three-body interaction.

The last Chapter of the Thesis reports the results of a calculation of the weak response of nuclear matter, carried out using effective operators and effective interactions derived within 
the framework of the CBF approach. This calculation significantly improves on those of Ref. \cite{cowell_04, benhar_09}, as it explicitly includes the contributions of three-nucleon 
clusters and three-nucleon forces. 

As far as the effective weak operators are concerned, we have shown that the sizable dependence of the response function on the choice of the correlation function is in fact unphysical. 
When three-body clusters diagrams are considered, the transition matrix elements, in which correlations enter only through the effective weak operators, become 
nearly independent of the correlation functions. 

The three-body cluster contributions described in  Section \ref{subsec:twothree} have been consistently included in the construction of the effective interaction. As a first step we have computed the EoS of SNM for an hamiltonian including the two-body potential only. In this case, the three-body cluster effective interaction provides a EoS of SNM much closer to to the one 
resulting from full FHNC/SOC calculations, compared to the one obtained using the two-body cluster effective interaction of Ref. \cite{cowell_04, benhar_07}.
The leading contributions of the UIX potential, emerging at three-body cluster level, have been also included in the effective potential. As a result 
the EoS of SNM exhibits saturation at $\rho\simeq 0.18\,\text{fm}^{-1}$.

Inclusion of the three-nucleon interactions also affects the single particle spectrum, leading in turn to a shift of the CHF response as a function of energy transfer.

The main effect of the three-nucleon force on the TDA response originates from a change of the off-diagonal elements of the effective interaction. As a result, the collective mode associated with the 
Fermi transition at $|\mathbf{q}|=0.3$ MeV turns out to be shifted to lower energy, although its magnitude is nearly unaffected by the three-body cluster contributions. On the other hand, a depletion of the peak is observed for the Gamow-Teller transition. The analysis of the TDA response for different momentum transfer also reveals a sizable effect of three-nucleon cluster
contributions.

The sum rules for the Fermi transition comes closer to the variational results once the three-body cluster cluster is taken into account, thus confirming the importance of many body effects, 
which are included in the variational calculations through the chain summations. The residual discrepancy cannot be accounted for by the $n>4-$body cluster contributions, and is 
likely to be ascribable to the effect of multi $p-h$ excitations, which are taken into account in variational calculations.
This effect appears to be even larger even larger in the structure function obtained from the Gamow-Teller response. In this case the change due to the three-body cluster is in fact very small. 

Some improvements are observed in the sum rules of the longitudinal and the transverse response. As shown in Fig. \ref{fig:sum_lt},  the results of the three-body cluster calculations of $S^L(q)$ carried out within TDA are slightly closer to the variational ones for all the values of $|\mathbf{q}|$,  compared to the two-body cluster case. Moreover, the position of the maximum of the TDA calculations of the static transverse response is almost coincident with that obtained from variational calculations. At small momentum transfer however, the three-body cluster contributions move
the TDA $S^T(q)$ away from the variational result.

% Appendici
\appendix

%Equazioni di Hartree-Fock
\chapter{Hartree-Fock equations}
\label{app:hf}
The starting point to derive the Hartree-Fock equation is the variational principle
\begin{equation}
E_0 \leq E [\Psi] = \langle\Psi |H|\Psi\rangle\, ,
\end{equation}
valid when the Slater determinant has been normalized to unity. At this aim we require that single particle wave functions be orthonormal
\begin{equation}
\int dx \, \psi_{n_i}(x)\psi_{n_j}(x)=\delta_{n_i n_j}\, .
\label{eq:orho}
\end{equation}

It is straightforward to show \cite{bransden_03} that the expectation value of the Hamiltonian is made of three contributions
\begin{equation}
E [\Psi] =\sum_{n_i=1}^AI_{n_i}+\frac{1}{2}\sum_{n_i,n_j =1}^A[J_{n_i n_j}-K_{n_i n_j}]\,.
\end{equation}
The first contribution comes from the kinetic energy, as no genuine one-body potential appears in the nuclear Hamiltonian of Eq. (\ref{eq:hamiltonian})
\begin{equation}
 I_{n_i}= \int dx_i \psi_{n_i}^*(x_i) \frac{\nabla^{2}_i}{2m} \psi_{n_i}(x_i)\, ,
\end{equation}
while the direct term $J_{n_i n_j}$ and the exchange term $K_{n_i n_j}$ arise from the two-body potential
\begin{align}
J_{n_i n_j}&=\int dx_i dx_j \psi_{n_i}^*(x_i)\psi_{n_j}^*(x_j)v_{ij}\psi_{n_i}(x_i)\psi_{n_j}(x_j)\nonumber\\
K_{n_i n_j}&=\int dx_i dx_j \psi_{n_i}^*(x_i)\psi_{n_j}^*(x_j)v_{ij}\psi_{n_j}(x_i)\psi_{n_i}(x_j)\, .
\label{eq:dir_exc_term}
\end{align}

The variational principle is applied by requiring that the functional $E[\Psi]$ being stationary upon variations of the A occupied spin-orbitals $\psi_{n_i}$, with the constraint that the spin-orbitals must remain orthogonal. To achieve this, a set of $A^2$ Lagrange multipliers $\epsilon_{n_i,n_j}$ is introduced in the variational equation
\begin{equation}
\delta E[\Psi]-\sum_{n_i, n_j=1}^A\epsilon_{n_i,n_j} \, \delta\Big[ \int dx_i \psi_{n_i}^*(x_i) \psi_{n_j}(x_i)\Big]=0\,.
\label{eq:var_nd}
\end{equation}
Since the energy is a real quantity, the Lagrange multipliers are elements of an Hermitian matrix, $\epsilon_{n_in_j}=\epsilon_{n_jn_i}^*$. 

The matrix of the Lagrange multipliers can therefore be diagonalized performing a unitary transformation on the spin-orbitals
\begin{equation}
\psi_{n_i}'=\sum_{n_j}U_{n_i,n_j}\psi_{n_j}\, .
\end{equation}
The new Slater determinant differs by a phase factor from the previous one, $\Psi'=det(U)\Psi$, while the functional $E[\psi]$ is not affected by this unitary transformation. 

To simplify the notation, instead of working with the primed indeces, we assume that the diagonalization has been made from the beginning. Therefore Eq. (\ref{eq:var_nd}) can be written as
\begin{equation}
\delta E[\Psi]-\sum_{n_i}\epsilon_{n_i} \, \delta\Big[ \int dx_i \psi_{n_i}^*(x_i) \psi_{n_i}(x_i)\Big]=0\,, 
\label{eq:var_d}
\end{equation}
being $\{\epsilon_{n_i}\}$ the eigenvalues of $\epsilon_{n_in_j}$.

Carrying out the variation $\delta E[\Psi]$ leads to the canonical set of integro-differential Hartree Fock equations
\begin{align}
\Big(-\frac{\nabla^2}{2m}+\sum_{n_j}[\hat{J}_{n_j}-\hat{K}_{n_j}]\Big)\psi_{n_i}(x_i)=\epsilon_{n_i}\psi_{n_i}(x_i)\, .
\label{eq:hf}
\end{align}
The direct and exchange operators, $\hat{J}$ and $\hat{K}$, respectively, result from the differentiation of the direct and of the exchange term defined in Eq. (\ref{eq:dir_exc_term}). While the direct operator is local
\begin{equation}
\hat{J}_{n_j}\psi_{n_i}(x_i)=\int dx_j \psi_{n_j}^*(x_j)v_{ij}\psi_{n_j}(x_j)\psi_{n_i}(x_i)\, ,
\end{equation}
as in order to evaluate their action on a given function at a point in the configurational space, we need to know just the value of the function at the same point. On the other hand, the exchange operator 
\begin{equation}
\hat{K}_{n_j}\psi_{n_i}(x_i)=\int dx_j \psi_{n_j}^*(x_j)v_{ij}\psi_{n_i}(x_j)\psi_{n_j}(x_i)\, ,
\end{equation}
is not local, since we need to know the value of the function $\psi_{n_i}$ in all the configurational space.

Comparing Eq. (\ref{eq:hf}) and (\ref{eq:hf_intro}), it is readily seen that the Hartree-Fock potential is given by
\begin{equation}
\hat{v}^{HF}(x)=\sum_{n_j=1}^A[\hat{J}_{n_j}-\hat{K}_{n_j}]\, .
\end{equation}

The solutions of the eigenvalue equation enter the definition of the Fock hamiltonian, through the direct and exchange operators. For this reason the Hartree-Fock equations of Eq. (\ref{eq:hf}) are often referred to as pseudo-eigenvalue equations and are solved iteratively.
The starting point is building the Fock hamiltonian with a guess of approximate spin-orbitals. Solving the resulting eigenvalue equation provides new spin-orbitals that can be used for the Fock hamiltonian. The procedure must be repeated until the solutions are close enough to the spin-orbitals used for the construction of the operator. This method of solution is often called {\it self-consistent field}.

Using the orthonormality condition (\ref{eq:orho}), $\epsilon_{n_i}$ can be obtained by taking the scalar product of Eq. (\ref{eq:hf}) with $\psi_{n_i}$
\begin{equation}
\epsilon_{n_i}=I_{n_i}+\sum_{n_j}[J_{n_i n_j}-K_{n_i n_j}]\,, 
\label{eq:spe_1}
\end{equation}
or, in Dirac notation,
\begin{equation}
\epsilon_{n_i}=\langle n_i| -\frac{\nabla_{i}^2}{2m}|n_i\rangle +\sum_{n_j}\langle n_i n_j | v_{ij}  |n_i n_j- n_j n_i\rangle\, .
\label{eq:spe_2}
\end{equation}

\thispagestyle{empty}  

%Determinante di Slater
\chapter{Slater determinant}
\label{app:slat}
The groud-state $\Psi_0$ of A non interacting fermions is described by the Slater determinant of single-particle states $\psi_{n_i}(x_i)$
\begin{equation}
\Psi_0=
\begin{vmatrix}
    \psi_{1}(x_1) & \psi_{2}(x_1) & \ldots  & \psi_{N}(x_1) \\
    \psi_{1}(x_2) & \psi_{2}(x_2) & \ldots  & \psi_{N}(x_2) \\
    \vdots & \vdots & \ddots   & \vdots \\
    \psi_{1}(x_N) & \psi_{2}(x_N) & \ldots  & \psi_{N}(x_N)
  \end{vmatrix}\,.
\label{eq:slater_def}
\end{equation}

Isolating the first particle from the determinant amounts in 
\begin{equation}
\Psi_0=\frac{1}{\sqrt{A}}\sum_{n_1} (-1)^{n+1} \psi_{n_1}(x_1) \Psi^{0}_{m \neq n_1}(x_2\,\dots\,x_A)\, ,
\label{eq:prima_isolata}
\end{equation}
where $\Psi^{0}_{m \neq n_1}(x_2\,\dots\,x_A)$ denotes the Slater determinant of system with $A-1$ particles and one hole in the state $n_1$. In other words it is the minor of the matrix defined in Eq. (\ref{eq:slater_def}) from which the first row and the $n_1$th  column have been removed.

Isolating a second particle is a slightly more involves issue, as the sign depends on the ordering of $n_1$ and $n_2$. It turns out that if $n_2<n_1$ the multiplicative factor, $(-1)^{n_2+1}$, is identical to the one of Eq. (\ref{eq:prima_isolata}). When $n_2>n_1$, the fact that the column associated with the sate $n_1$ had been already removed has to be taken into account because the column index of $\Psi^{0}_{m \neq n_1}(x_2\,\dots\,x_A)$ does not correspond to the state index. In this case, the multiplicative factor arising when isolating the $n_2$-th column is then $(-1)^{n_2}$
\begin{align}
\Psi_0=\frac{1}{\sqrt{A(A-1)}}\Big[&\sum_{n_1>n_2} (-1)^{n_1+n_2} \psi_{n_1}(x_1)\psi_{n_2}(x_2) \Psi^{0}_{m \neq n_1,n_2}(x_3\,\dots\,x_A)+\nonumber \\
& \sum_{n_1<n_2} (-1)^{n_1+n_2+1} \psi_{n_1}(x_1)\psi_{n_2}(x_2)\Psi^{0}_{m \neq n_1,n_2}(x_3\,\dots\,x_A)\Big]\, .
\end{align}

With $ \Psi^{0}_{m \neq n_1,n_2}(x_3\,\dots\,x_A)$ we denote the Slater determinant for a system of $A-2$ particles lacking of the states $n_1$ and $n_2$. It corresponds to the minor of the matrix of Eq. (\ref{eq:slater_def}) from which the first two rows and the colums $n_1$ and $n_2$ have been removed. A more compact expression can be given for the latter result
\begin{align}
\Psi_0&=\frac{1}{\sqrt{A(A-1)}}\sum_{n_1<n_2}(-1)^{n_1+n_2+1}\mathcal{A}[\psi_{n_1}(x_1)\psi_{n_2}(x_2)]\Psi^{0}_{m \neq n_1,n_2}(x_3\,\dots\,x_A)\,.
\label{eq:seconda_isolata}
\end{align}

The many-body state $\Psi_{p;h}$ is obtained replacing the $h-$th column of $\Psi_0$ with another column with state index $p$, having energy larger than the Fermi energy. The procedure for isolating particles from $\Psi_{ph}$ follows the same step as the one of $\Psi_0$
\begin{align}
\Psi_{ph}=\frac{1}{\sqrt{N(N-1)}}\sum_{n_1<n_2}(-1)^{n_1+n_2+1}\mathcal{A}[\psi_{\tilde{n}_1}(x_1)\psi_{\tilde{n}_2}(x_2)]\Psi^{ph}_{m \neq \tilde{n}_1,\tilde{n}_2}(x_3\,\dots\,x_A)\,.
\label{eq:estr_gen_ph}
\end{align}
The indexes $n_i$ have the values $1,\dots,h,\dots,A$, while 
\begin{equation}
\left\{
\begin{array}{rl}
\tilde{n}_i=p &  \text{if } i = h,\\
\tilde{n}_i=n_i & \text{otherwise}.
\end{array} \right.
\end{equation}

The generalization to the three-particles case is straightforward 
\begin{align}
\Psi_{0}&=\sqrt{\frac{(A-2)!}{A!}}\sum_{n_1<n_2<n_3}(-1)^{n_1+n_2+n_3+1}\mathcal{A}[\psi_{n_1}(x_1)\psi_{n_2}(x_2)\psi_{n_3}(x_3)]\Psi^{0}_{m \neq \tilde{n}_1,\tilde{n}_2,\tilde{n}_3}(x_4\,\dots\,x_A)\nonumber\\
\Psi_{ph}&=\sqrt{\frac{(A-2)!}{A!}}\sum_{n_1<n_2<n_3}(-1)^{n_1+n_2+n_3+1}\mathcal{A}[\psi_{\tilde{n}_1}(x_1)\psi_{\tilde{n}_2}(x_2)\psi_{\tilde{n}_3}(x_3)]\Psi^{ph}_{m \neq \tilde{n}_1,\tilde{n}_2,\tilde{n}_3}(x_4\,\dots\,x_A)\,.
\label{eq:ph_generic}
\end{align}
\newpage             
\thispagestyle{empty}  

%Diagrammi riducibili a 4 corpi.
\chapter{Reducible diagrams}
\label{app:red4b}
\section{Four-body reducible diagrams}
In this appendix we report the explicit calculations of the four-body reducible diagrams involved in the calculations of the FR three-body cluster contribution of $\langle v_{12} \rangle$.

The analytic expression of diagram (a) drawn in Fig. \ref{fig:red4bdir} is
\begin{align}
v_{4b\to3b}^{\text{dir (a)}}&=-\frac{\rho^4}{2A}\int d\mathbf{r}_{1234}\ell^2(r_{13})\text{CTr}_{1234}\Big[\hat{F}_{12}v_{12}\hat{F}_{12}(\hat{F}_{34}^2-1)\hat{P}_{13}\Big]\, .
\end{align}
Because of the tracing over the spin-isospin variables of particle $4$, only the central part of $\hat{F}_{34}^2$ does not vanish; hence, only the scalar part of the exchange operator survives
\begin{align}
v_{4b\to3b}^{\text{dir (a)}}&=-\frac{\rho^4}{2A}\frac{1}{\nu}\int d\mathbf{r}_{1234}\ell^2(r_{13})\text{CTr}_{1234}\Big[\hat{F}_{12}v_{12}\hat{F}_{12}(\hat{F}_{34}^2-1)\Big]\nonumber \\
&=-\frac{\rho^3}{2}\frac{1}{\nu}\int d\mathbf{r}_{12}d\mathbf{r}_{13}d\mathbf{r}_{34}\ell^2(r_{13})\text{CTr}_{1234}\Big[\hat{F}_{12}v_{12}\hat{F}_{12}(\hat{F}_{34}^2-1)\Big]\nonumber \\
&=-\frac{\rho^2}{2}\int d\mathbf{r}_{12}d\mathbf{r}_{34}\text{CTr}_{1234}\Big[\hat{F}_{12}v_{12}\hat{F}_{12}(\hat{F}_{34}^2-1)\Big]\, ,
\end{align}
where we have used 
\begin{equation}
\int d\mathbf{r}_{12}\ell^2(r_{12})=\frac{\nu}{\rho}
\end{equation}
Renaming the integration variables yields to the expression entering Eq. (\ref{eq:v3b_dir})
\begin{align}
v_{4b\to3b}^{\text{dir (a)}}&=-\frac{\rho^2}{2}\int d\mathbf{r}_{12}d\mathbf{r}_{13}\text{CTr}_{123}\Big[\hat{F}_{12}v_{12}\hat{F}_{12}(\hat{F}_{13}^2-1)\Big]\, .
\end{align}
The calculation of diagram (b) of Fig. \ref{fig:red4bdir} follows analogously.

The contribution of the diagram depicted in Fig. \ref{fig:red4bp12} is  given by
\begin{align}
v_{4b\to3b}^{\text{P12}}&=\frac{\rho^4}{A}\int d\mathbf{r}_{1234}\ell(r_{12})\ell(r_{13})\ell(r_{23})\text{CTr}_{1234}\Big[\hat{F}_{12}v_{12}\hat{F}_{12}(\hat{F}_{34}^2-1)\hat{P}_{12}\hat{P}_{13}\Big]\, .
\end{align}
As before, since only the central part of $\hat{F}_{34}^2$ survives, it turns out that
\begin{align}
v_{4b\to3b}^{\text{P12}}&=\frac{\rho^4}{A}\frac{1}{\nu}\int d\mathbf{r}_{1234}\ell(r_{12})\ell(r_{13})\ell(r_{23})\text{CTr}_{1234}\Big[\hat{F}_{12}v_{12}\hat{F}_{12}(\hat{F}_{34}^2-1)\hat{P}_{12}\Big]\, .
\end{align}
Thanks to the property
\begin{equation}
\int d\mathbf{r}_{13}\ell(r_{13})\ell(|\mathbf{r}_{12}-\mathbf{r}_{13}|)=\frac{\nu}{\rho}\ell(r_{12})\, ,
\label{eq:slat_conv}
\end{equation}
and relabeling the integration variables, one arrives to the following expression
\begin{align}
v_{4b\to3b}^{\text{p12}}&=\rho^2\int d\mathbf{r}_{12}d\mathbf{r}{13}\ell(r_{12})^2 \text{CTr}_{123}\Big[\hat{F}_{12}v_{12}\hat{F}_{12}(\hat{F}_{13}^2-1)\hat{P}_{12}\Big]\, .
\end{align}
Notice that, to recover the contribution to Eq. (\ref{eq:v3b_p12}), one the symmetry $1\leftrightarrow 2$ needs to be exploited. 

Let us now consider the diagram of Fig. \ref{fig:red4bp13}
\begin{align}
v_{4b\to3b}^{\text{p13}}&=\frac{\rho^4}{2A}\int d\mathbf{r}_{1234}\ell(r_{13})\ell(r_{14})\ell(r_{34})\text{CTr}_{1234}\Big[\hat{F}_{12}v_{12}\hat{F}_{12}(\hat{F}_{34}^2-1)\hat{P}_{13}\hat{P}_{34}\Big]\, .
\end{align}
Carrying out the trace over spin-isospin variables of particle $4$ makes only the central part of $(\hat{F}_{34}^2-1)\hat{P}_{34}$ not to vanish. Hence only the scalar part of $\hat{P}_{13}$ contributes
\begin{align}
v_{4b\to3b}^{\text{p13}}&=\frac{\rho^4}{2A}\frac{1}{\nu}\int d\mathbf{r}_{1234}\ell(r_{13})\ell(r_{14})\ell(r_{34})\text{CTr}_{1234}\Big[\hat{F}_{12}v_{12}\hat{F}_{12}(\hat{F}_{34}^2-1)\hat{P}_{34}\Big]\nonumber \\
&=\frac{\rho^3}{2}\frac{1}{\nu}\int d\mathbf{r}_{12}d\mathbf{r}{14}d\mathbf{r}_{34}\ell(r_{13})\ell(r_{14})\ell(r_{34})\text{CTr}_{1234}\Big[\hat{F}_{12}v_{12}\hat{F}_{12}(\hat{F}_{34}^2-1)\hat{P}_{34}\Big]\nonumber \\
&=\frac{\rho^2}{2}\int d\mathbf{r}_{12}d\mathbf{r}_{34}\ell(r_{34})^2\text{CTr}_{1234}\Big[\hat{F}_{12}v_{12}\hat{F}_{12}(\hat{F}_{34}^2-1)\hat{P}_{34}\Big]\,.
\end{align}
where in the last line we have used Eq. (\ref{eq:slat_conv}). A change of integration variables leads to
\begin{align}
v_{4b\to3b}^{\text{p13}}&=\frac{\rho^2}{2}\int d\mathbf{r}_{12}d\mathbf{r}_{13}\ell(r_{13})^2\text{CTr}_{123}\Big[\hat{F}_{12}v_{12}\hat{F}_{12}(\hat{F}_{13}^2-1)\hat{P}_{13}\Big]\,.
\end{align}

The diagram of Fig. \ref{fig:red4bcir} can be written as
\begin{align}
v_{4b\to3b}^{\text{cir}}&=-\frac{\rho^4}{A}\int d\mathbf{r}_{1234}\ell(r_{12})\ell(r_{13})\ell(r_{24})\ell(r_{34})\text{CTr}_{1234}\Big[\hat{F}_{12}v_{12}\hat{F}_{12}(\hat{F}_{34}^2-1)\hat{P}_{12}\hat{P}_{13}\hat{P}_{34}\Big]\nonumber\\
&=-\frac{\rho^4}{A}\frac{1}{\nu}\int d\mathbf{r}_{1234}\ell(r_{12})\ell(r_{13})\ell(r_{24})\ell(r_{34})\text{CTr}_{1234}\Big[\hat{F}_{12}v_{12}\hat{F}_{12}(\hat{F}_{34}^2-1)\hat{P}_{12}\hat{P}_{34}\Big]\nonumber \\
&=-\rho^3\frac{1}{\nu}\int d\mathbf{r}_{12}d\mathbf{r}_{13}d\mathbf{r}_{34}\ell(r_{12})\ell(r_{13})\ell(r_{24})\ell(r_{34})\text{CTr}_{1234}\Big[\hat{F}_{12}v_{12}\hat{F}_{12}(\hat{F}_{34}^2-1)\hat{P}_{12}\hat{P}_{34}\Big]\, ,
\end{align}
having used the fact that only the central term of $(\hat{F}_{34}^2-1)\hat{P}_{34}$ contributes. Since $\ell(r_{24})=\ell(|\mathbf{r}_{34}-\mathbf{r}_{12}+\mathbf{r}_{13}|)$, exploiting the convolution property of Eq. (\ref{eq:slat_conv}), it turns out that
\begin{align}
v_{4b\to3b}^{\text{cir}}&=-\rho^2\int d\mathbf{r}_{12}d\mathbf{r}_{34}\ell(r_{12})\ell(|\mathbf{r}_{34}-\mathbf{r}_{12}|)\ell(r_{34})\text{CTr}_{1234}\Big[\hat{F}_{12}v_{12}\hat{F}_{12}(\hat{F}_{34}^2-1)\hat{P}_{12}\hat{P}_{34}\Big]\nonumber \\
&=-\rho^2\int d\mathbf{r}_{12}d\mathbf{r}_{34}\ell(r_{12})\ell(|\mathbf{r}_{34}-\mathbf{r}_{12}|)\ell(r_{34})\text{CTr}_{1234}\Big[\hat{F}_{12}v_{12}\hat{F}_{12}(\hat{F}_{34}^2-1)\hat{P}_{34}\hat{P}_{12}\Big]\nonumber \\
&=-\rho^2\int d\mathbf{r}_{12}d\mathbf{r}_{13}\ell(r_{12})\ell(r_{13})\ell(r_{23})\text{CTr}_{123}\Big[\hat{F}_{12}v_{12}\hat{F}_{12}(\hat{F}_{13}^2-1)\hat{P}_{13}\hat{P}_{12}\Big]\, .
\end{align}

\newpage             
\thispagestyle{empty}  

%Importance sampling wavefunction.
\chapter{Green's function for importance sampling}
\label{app:isgf}
In this appendix we report the derivation, originally due to Kalos, of the expression for the importance sampling wave function. For the sake of simplicity we will limit ourselves to an accuracy of $\Delta\tau^2$, that can be achieved by using the Trotter Fromula of Eq. (\ref{eq:ts_formula2}). Hence, the ``original'' Green's function is
\begin{equation}
G(R,R',\Delta\tau)=\Big(\frac{m}{2\pi\hbar^2\Delta\tau}\Big)^\frac{3A}{2} e^{-\frac{m(R-R')^2}{2\hbar^2\Delta\tau}} e^{-V(R')\Delta\tau}e^{E_T\Delta\tau}\,. 
\end{equation}

Let us start from Eq. (\ref{eq:isgf_def}), that it is worth rewriting 
\begin{equation}
\tilde{G}(R,R',\Delta\tau)=G(R,R',\Delta\tau)\frac{\psi_T(R)}{\psi_T(R')}\,.
\end{equation}
We want to include the ratio of the trial wave functions in the diffusive part of the Green's function. Assuming that for small $\Delta\tau$ the displacement $|R-R'|$ be small, a first order Taylor series expansion in $R-R'$ yields
\begin{align}
\tilde{G}_d(R,R',\Delta\tau)&\simeq\mathcal{N}\exp\Big[-\frac{m}{2\hbar^2\Delta\tau}(R-R')^2\Big]\Big[1+\frac{\vec{\nabla}\psi_T(R')}{\psi_T(R')}\cdot(R-R')\Big]\, 
\nonumber \\
&\simeq\mathcal{N}\exp\Big[-\frac{m}{2\hbar^2\Delta\tau}(R-R')^2+\vec{v}_D(R')\cdot(R-R')\Big]\nonumber \\
&\simeq\mathcal{N}\exp\Big[-\frac{m}{2\hbar^2\Delta\tau}\Big(R-R'-\frac{\hbar^2\Delta\tau}{m} \vec{v}_D(R')\Big)^2\Big]\, ,
\end{align}
where
\begin{equation}
\mathcal{N}=\Big(\frac{m}{2\pi\hbar^2\Delta\tau}\Big)^\frac{3A}{2}\, .
\end{equation}

As usual, the walkers distribution is represented by a sum of discrete delta functions, centered in the position of each walker
\begin{equation}
\psi(R,\tau_0)=\sum_k\delta(R-R_k)\,. 
\end{equation}

In general, the imaginary time evolution does not preserve the normalization. The branching factor is nothing but the multiplicity of the walkers that at $\tau+\Delta\tau$ generated starting from $R'$. Hence an integration over the possible final positions has to be performed
\begin{equation}
N(R')=\int dR\, \tilde{G}(R,R',\tau)=\int dR\, G(R,R',\tau)\frac{\psi_T(R)}{\psi_T(R')}\, .
\end{equation}
To compute the integral, as before we perform a Taylor series expansion 
\begin{align}
N(R')&\simeq \int dR G(R'\to R,\tau)\Big[1+\frac{\vec\nabla\psi_T(R')}{\psi_T(R)}\cdot(R-R')\nonumber \\
&+\frac{1}{\psi_T(R')}\sum_{i,j=1}^A\sum_{\alpha,\beta=1}^3\Big(1-\frac{\delta_{ij}\delta_{\alpha\beta}}{2}\Big)\frac{\partial^2\psi_T(R')}{\partial r_{i\alpha}\partial r_{j\beta}}(r_{i\alpha}-r'_{i\alpha})(r_{j\beta}-r'_{j\beta})\Big]
\end{align}
where $r_{i\alpha}$ denotes the $\alpha$-th cartesian component of the coordinates of the i-th particle.
The Green's function depends on the integration variable only through the gaussian diffusive term, which is symmetric in the relative coordinate $R-R'$. Therefore, only even terms survive 
\begin{align}
N(R')&\simeq \Big\{1+\frac{1}{2\psi_T(R')}\sum_{i=1}^A\sum_{\alpha=1}^3\frac{\partial^2\psi_T(R')}{\partial^2 r'_{i\alpha}}
\mathcal{N} \int dR\, \exp\Big[-\frac{m(R-R')^2}{2\hbar^2\Delta\tau}\Big](r_{i\alpha}-r'_{i\alpha})^2\Big\}\nonumber\\
&\times\exp[-\Delta\tau(V(R')-E_T)]\, .
\end{align}
Using the following result for the Gaussian integral
\begin{align}
\mathcal{N}\int dR\exp\Big[-\frac{m(R-R')^2}{2\hbar^2\Delta\tau}\Big](r_{i\alpha}-r'_{i\alpha})^2=\Delta\tau\, .
\end{align}
for the multiplicity we obtain 
\begin{align}
N(R')&\simeq\Big[1+\frac{\Delta\tau}{2}\frac{\vec{\nabla}^2\psi_T(R')}{\psi_T(R')}\Big]\times \exp[-\Delta\tau(V(R')-E_T)]\nonumber \\
&\simeq\exp\Big[\Delta\tau\Big(\frac{1}{2}\frac{\vec{\nabla}^2\psi_T(R')}{\psi_T(R')}+E_T-V(R')\Big)\Big]\nonumber \\
&\simeq\exp\Big[-\Delta\tau\Big(\frac{H\psi_T(R')}{\psi_T(R')}-E_T\Big)\Big]\nonumber \\
&\simeq\exp\Big[-\Delta\tau\Big(E_L(R')-E_T\Big)\Big]\, .
\end{align}

Therefore, the branching factor reads
\begin{equation}
\tilde{G}_b(R,R',\Delta\tau)=e^{-\Delta\tau\Big(E_L(R')-E_T\Big)}\, , 
\end{equation}
hence the full Green's function turns out to be
\begin{equation}
\tilde{G}(R,R',\Delta\tau)=\Big(\frac{m}{2\pi\hbar^2\Delta\tau}\Big)^\frac{3A}{2}e^{-\frac{m}{2\hbar^2\Delta\tau}\left(R-R'-\frac{\hbar^2\Delta\tau}{m} \vec{v}_D(R')\right)^2}e^{-\Delta\tau\left(E_L(R')-E_T\right)}\, , 
\end{equation}
which differs from the one of Eq. (\ref{eq:diff_isgf}) by terms of the order $\Delta\tau^2$.

\newpage             
\thispagestyle{empty}  

%Fermi Gamow-Teller matrix elements
\chapter{Matrix elements of Fermi and Gamow-Teller transitions operator}
\label{app:fgtop}
In this appendix we give the results of the Fermi and Gamow-Teller matrix elements of the diagrams computed in section \ref{sec:ewo}. As far as the two-body clusters are concerned, our findings coincide with those of Refs. \cite{cowell_03}, while the three-body contributions have not been computed so far.  

To simplify the notation the subscripts $i$ for the particle and $j$ for the hole states have been omitted. Hence, the spin and isospin states of the particle and of the hole are denoted by $s_h$, $t_h$, $s_p$ and $t_p$.

Note that the Kronecker deltas $\delta_{t_p,p}$ and $\delta_{t_h,n}$ are common to all the matrix elements, showing that in the transition a neutron decays into a proton. The spin-structure, on the other hand, is in general more involved; particularly for the Gamow-Teller transition where we need to distinguish the three cartesian components.

\section{Zeroth order}
\begin{align}
\langle \alpha_p|\hat{O}_{\sigma\tau}(1)|\alpha_h\rangle
\end{align}
\textbf{Fermi}
\begin{align}
g_V\langle \alpha_p|\tau_{1}^{(+)}|\alpha_h\rangle=g_V\delta_{s_p,s_h}\delta_{t_p,p}\delta_{t_h,n}
\end{align}
\textbf{Gamow-Teller}
\begin{align}
g_A\langle \alpha_p|\tau_{1}^{(+)}\sigma_{1}^x |\alpha_h\rangle&=g_A\delta_{t_p,p}\delta_{t_h,n}[\delta_{s_p,\uparrow}\delta_{s_h,\downarrow}+\delta_{s_p,\downarrow}\delta_{s_h,\uparrow}]\nonumber\\
g_A\langle \alpha_p|\tau_{1}^{(+)}\sigma_{1}^y |\alpha_h\rangle&=
g_A\delta_{t_p,p}\delta_{t_h,n}[-i\delta_{s_p,\uparrow}\delta_{s_h,\downarrow}+i\delta_{s_p,\downarrow}\delta_{s_h,\uparrow}]\nonumber\\
g_A\langle \alpha_p|\tau_{1}^{(+)}\sigma_{1}^z |\alpha_h\rangle&=
g_A\delta_{t_p,p}\delta_{t_h,n}[\delta_{s_p,\uparrow}\delta_{s_h,\uparrow}-\delta_{s_p,\downarrow}\delta_{s_h,\downarrow}]
\end{align}
As a general property of the Gamow-Teller transition, the matrix element between the spin states $s_p=\,\downarrow,s_h=\,\downarrow$ is equal to minus the matrix element between $s_p=\uparrow,s_h=\uparrow$. Moreover the matrix element between $s_p=\,\downarrow,s_h=\,\uparrow$ is the complex conjugate of the one between $s_p=\,\uparrow,s_h=\,\downarrow$. For brevity we report the results for $s_p=\uparrow,s_h=\uparrow$ and $s_p=\,\uparrow,s_h=\,\downarrow$, while the ``$\dots$'' symbolize the other contributions.

\section{First order}
{\large $\mathbf{O_{Na}^{(1)}}$}
\begin{align}
\sum_{\alpha_1}\langle \alpha_1 \alpha_p|\{\hat{f}_{12} -1,\hat{O}_{\sigma\tau}(1)\}| \alpha_1 \alpha_h\rangle
\end{align}

\textbf{Fermi}
\begin{align}
g_V\sum_{\alpha_1}\langle \alpha_1 s_p|\{\hat{f}_{12} -1,\tau_{1}^{(+)}\}| \alpha_1 s_h\rangle
=g_V\delta_{t_p,p}\delta_{t_h,n}\delta_{s_p,s_h} 8 f^{\tau}_{12}
\end{align}

\textbf{Gamow-Teller}
\begin{align}
&g_A\sum_{\alpha_1}\langle \alpha_1 s_p|\{\hat{f}_{12} -1,\tau_{1}^{(+)}\sigma_{1}^x \}| \alpha_1 s_h\rangle=g_A\delta_{t_p,p}\delta_{t_h,n}\Big\{\nonumber\\
&\delta_{s_p,\uparrow}\delta_{s_h,\uparrow}\Big[24f_{12}^{t\tau}\frac{x_{12} z_{12}}{r_{12}^2}\Big]+\nonumber\\
&\delta_{s_p,\uparrow}\delta_{s_h,\downarrow}\Big[
8f^{\sigma\tau}_{12} + 8f^{t\tau}_{12}\Big(-1 + \frac{3 x_{12} (x_{12} - i y_{12})}{r_{12}^2}\Big)\Big]+\nonumber\\
&\dots\Big\}
\end{align}

\begin{align}
&g_A\sum_{\alpha_1}\langle \alpha_1 s_p|\{\hat{f}_{12} -1,\tau_{1}^{(+)}\sigma_{1}^y \}| \alpha_1 s_h\rangle=g_A\delta_{t_p,p}\delta_{t_h,n}\Big\{\nonumber\\
&\delta_{s_p,\uparrow}\delta_{s_h,\uparrow}\Big[
24f_{12}^{t\tau}\frac{y_{12} z_{12}}{r_{12}^2}\Big]+\nonumber\\
&\delta_{s_p,\uparrow}\delta_{s_h,\downarrow}\Big[
-8if^{\sigma\tau}_{12} + 8f^{t\tau}_{12}\Big(i + \frac{3 y_{12} (x_{12} - i y_{12})}{r_{12}^2}\Big)\Big]+\nonumber\\
&\dots\Big\}
\end{align}

\begin{align}
&g_A\sum_{\alpha_1}\langle \alpha_1 s_{p_i}|\{\hat{f}_{12} -1,\tau_{1}^{(+)}\sigma_{1}^z\}| \alpha_1 s_{h_j}\rangle=g_A\delta_{t_p,p}\delta_{t_h,n}\Big\{\nonumber\\
&\delta_{s_p,\uparrow}\delta_{s_h,\uparrow}\Big[
8f^{\sigma\tau}_{12} + 8f^{t\tau}_{12}\Big(2- \frac{3 (x_{12}^2+ y_{12}^2)}{r_{12}^2}\Big)\Big]+\nonumber\\
&\delta_{s_p,\uparrow}\delta_{s_h,\downarrow}\Big[
24f_{12}^{t\tau}\frac{(x_{12}-iy_{12}) z_{12}}{r_{12}^2}\Big]+\nonumber\\
&\dots\Big\}\, .
\end{align}

{\large $\mathbf{O_{Nb}^{(1)}}$}
\begin{align}
\sum_{\alpha_1}\langle \alpha_1 \alpha_p| \{\hat{f}_{12}-1,\hat{O}_{\sigma\tau}(1)\}\hat{P}^{\sigma\tau}_{12}| \alpha_1 \alpha_h\rangle
\end{align}

\textbf{Fermi}
\begin{align}
&g_V\sum_{\alpha_1}\langle \alpha_1 s_p|\{\hat{f}_{12} -1,\tau_{1}^{(+)}\}\hat{P}^{\sigma\tau}_{12}| \alpha_1 s_h\rangle
=\nonumber\\
&\qquad g_V\delta_{t_p,p}\delta_{t_h,n}\delta_{s_p,s_h} 2[(f^{c}_{12}-1) + f^{\tau}_{12} + 3 (f^{\sigma}_{12} + f^{\sigma\tau}_{12})]
\end{align}

\textbf{Gamow-Teller}
\begin{align}
&g_A\sum_{\alpha_1}\langle \alpha_1 s_p  |\{\hat{f}_{12} -1,\tau_{1}^{(+)}\sigma_{1}^x \}\hat{P}^{\sigma\tau}_{12}| \alpha_1 s_h\rangle=g_A\delta_{t_p,p}\delta_{t_h,n}\Big\{\nonumber\\
&\delta_{s_p,\uparrow}\delta_{s_h,\uparrow}\Big[6(f_{12}^{t}-f_{12}^{t\tau})\frac{x_{12} z_{12}}{r_{12}^2}\Big]+\nonumber\\
&\delta_{s_p,\uparrow}\delta_{s_h,\downarrow}2\Big[(f_{12}^{c}-1)+f_{12}^{\tau}+f^{\sigma}_{12}+5f^{\sigma\tau}_{12}+(f_{12}^{t}-f^{t\tau}_{12})\Big(-1 + 3\frac{x_{12} (x_{12} - i y_{12})}{r_{12}^2}\Big)\Big]+\nonumber\\
&\dots\Big\}
\end{align}

\begin{align}
&g_A\sum_{\alpha_1}\langle  \alpha_1  s_p|\{\hat{f}_{12} -1,\tau_{1}^{(+)}\sigma_{1}^y \}\hat{P}^{\sigma\tau}_{12}| \alpha_1 s_h  \rangle=g_A\delta_{t_p,p}\delta_{t_h,n}\Big\{\nonumber\\
&\delta_{s_p,\uparrow}\delta_{s_h,\uparrow}\Big[
6(f_{12}^{t}-f_{12}^{t\tau})\frac{y_{12} z_{12}}{r_{12}^2}\Big]+\nonumber\\
&\delta_{s_p,\uparrow}\delta_{s_h,\downarrow}2\Big[-i(f_{12}^{c}-1)-if_{12}^{\tau}-if^{\sigma}_{12}-5if^{\sigma\tau}_{12}-(f_{12}^{t}-f^{t\tau}_{12})\Big(-i - 3\frac{y_{12} (x_{12} - i y_{12})}{r_{12}^2}\Big)\Big]+\nonumber\\
&\dots\Big\}
\end{align}

\begin{align}
&g_A\sum_{\alpha_1}\langle \alpha_1 s_p|\{\hat{f}_{12} -1,\tau_{1}^{(+)}\sigma_{1}^z\}\hat{P}^{\sigma\tau}_{12}| \alpha_1 s_h\rangle=g_A\delta_{t_p,p}\delta_{t_h,n}\Big\{\nonumber\\
&\delta_{s_p,\uparrow}\delta_{s_h,\uparrow}2\Big[(f_{12}^{c}-1)+f_{12}^{\tau}+f^{\sigma}_{12}+5f^{\sigma\tau}_{12}+(f_{12}^{t}-f^{t\tau}_{12})\Big(1 - \frac{2x_{12}^2+2y_{12}^2-z_{12}^2)}{r_{12}^2}\Big)\Big]+\nonumber\\
&\delta_{s_p,\uparrow}\delta_{s_h,\downarrow}\Big[
6(f_{12}^{t}-f_{12}^{t\tau})\frac{z_{12}(x_{12}-iy_{12})}{r_{12}^2}\Big]+\nonumber\\
&\dots\Big\}
\end{align}

{\large $\mathbf{O_{Nc}^{(1)}}$}
\begin{align}
\sum_{\alpha_2}\langle \alpha_p \alpha_2| \{\hat{f}_{12}-1,\hat{O}_{\sigma\tau}(1)\}| \alpha_h \alpha_2\rangle
\end{align}

\textbf{Fermi}
\begin{align}
g_V\sum_{\alpha_2}\langle s_p \alpha_2|\{\hat{f}_{12} -1,\tau_{1}^{(+)}\}|s_h \alpha_2 \rangle
=g_V\delta_{t_p,p}\delta_{t_h,n}\delta_{s_p,s_h} 8(f^{c}_{12}-1)
\end{align}

\textbf{Gamow-Teller}
\begin{align}
&g_A\sum_{\alpha_2}\langle s_p \alpha_2   |\{\hat{f}_{12} -1,\tau_{1}^{(+)}\sigma_{1}^x \}| s_h \alpha_2 \rangle=g_A\delta_{t_p,p}\delta_{t_h,n}\Big\{\nonumber\\
&\delta_{s_p,\uparrow}\delta_{s_h,\downarrow}8\Big[f^{c}_{12}-1\Big]+\nonumber\\
&\dots\Big\}
\end{align}

\begin{align}
&g_A\sum_{\alpha_2}\langle s_p \alpha_2  |\{\hat{f}_{12} -1,\tau_{1}^{(+)}\sigma_{1}^y \}|s_h \alpha_2   \rangle=g_A\delta_{t_p,p}\delta_{t_h,n}\Big\{\nonumber\\
&\delta_{s_p,\uparrow}\delta_{s_h,\downarrow}8i\Big[-f^{c}_{12}+1\Big]+\nonumber\\
&\dots\Big\}
\end{align}

\begin{align}
&g_A\sum_{\alpha_2}\langle s_p \alpha_2 |\{\hat{f}_{12} -1,\tau_{1}^{(+)}\sigma_{1}^z\}|s_h \alpha_2 \rangle=g_A\delta_{t_p,p}\delta_{t_h,n}\Big\{\nonumber\\
&\delta_{s_p,\uparrow}\delta_{s_h,\uparrow}8\Big[f^{c}_{12}-1\Big]+\nonumber\\
&\dots\Big\}
\end{align}

The spin-isospin matrix elements of $O_{Nd}^{(1)}$ are identical to those of $O_{Nb}^{(1)}$ for both Fermi and Gamow-Teller transitions.
\\

{\large $\mathbf{O_{Da}^{(1)}}$}
\begin{align}
\sum_{\alpha_1}\langle \alpha_1 \alpha_p |\hat{f}_{12}-1|\alpha_1 \alpha_p\rangle=4(f^{c}_{12}-1)
\end{align}

{\large $\mathbf{O_{Db}^{(1)}}$}
\begin{align}
\sum_{\alpha_1}\langle \alpha_1 \alpha_p |(\hat{f}_{12}-1)\hat{P}^{\sigma\tau}_{12}|\alpha_1 \alpha_p\rangle=(f^{c}_{12}-1)+3(f^{\tau}_{12}+f^{\sigma}_{12}+3f^{\sigma\tau}_{12})
\end{align}

We do not report the results for the spin-isospin matrix elements of $\mathbf{O_{Dc}^{(1)}}$ and $\mathbf{O_{Dd}^{(1)}}$ as they are identical to those of $\mathbf{O_{Da}^{(1)}}$ and $\mathbf{O_{Db}^{(1)}}$, respectively.
\newline

{\large $\mathbf{O_{Ng}^{(1)}}$}
\begin{align}
\sum_{\alpha_i}\langle \alpha_1 \alpha_2 \alpha_p |\hat{O}_{\sigma\tau}(1)(\hat{f}_{23}-1)\hat{P}^{\sigma\tau}_{12}|\alpha_1 \alpha_2 \alpha_h\rangle
\end{align}

\textbf{Fermi}
\begin{align}
g_V\sum_{\alpha_i}\langle \alpha_1 \alpha_2 s_p |\tau_{1}^{(+)}(\hat{f}_{23}-1)\hat{P}^{\sigma\tau}_{12}| \alpha_1  \alpha_2 s_h \rangle
=g_V\delta_{t_p,p}\delta_{t_h,n}\delta_{s_p,s_h} 4f^{\tau}_{23} 
\end{align}

\textbf{Gamow-Teller}
\begin{align}
&g_A\sum_{\alpha_i}\langle \alpha_1 \alpha_2 s_p |\tau_{1}^{(+)}\sigma_{1}^x (\hat{f}_{23}-1)\hat{P}^{\sigma\tau}_{12}| \alpha_1  \alpha_2 s_h \rangle=g_A\delta_{t_p,p}\delta_{t_h,n}\Big\{\nonumber\\
&\delta_{s_p,\uparrow}\delta_{s_h,\uparrow}\Big[12f^{t\tau}_{23}\frac{x_{23}z_{23}}{r_{23}^2}\Big]+\nonumber\\
&\delta_{s_p,\uparrow}\delta_{s_h,\downarrow}\Big[4f^{\sigma\tau}_{23} + 4f^{t\tau}_{23} \Big(-1 + \frac{3x_{23}(x_{23}-iy_{23})}{r_{23}^2}\Big)\Big]+\nonumber\\
&\dots\Big\}
\end{align}

\begin{align}
&g_A\sum_{\alpha_i}\langle \alpha_1 \alpha_2 s_p |\tau_{1}^{(+)}\sigma_{1}^y (\hat{f}_{23}-1)\hat{P}^{\sigma\tau}_{12}| \alpha_1  \alpha_2 s_h \rangle=g_A\delta_{t_p,p}\delta_{t_h,n}\Big\{\nonumber\\
&\delta_{s_p,\uparrow}\delta_{s_h,\uparrow}\Big[12f^{t\tau}_{23}\frac{y_{23}z_{23}}{r_{23}^2}\Big]+\nonumber\\
&\delta_{s_p,\uparrow}\delta_{s_h,\downarrow}\Big[-4if^{\sigma\tau}_{23} + 4f^{t\tau}_{23} \Big(i + \frac{3y_{23}(x_{23}-iy_{23})}{r_{23}^2}\Big)\Big]+\nonumber\\
&\dots\Big\}
\end{align}

\begin{align}
&g_A\sum_{\alpha_i}\langle \alpha_1 \alpha_2 s_p |\tau_{1}^{(+)}\sigma_{1}^z (\hat{f}_{23}-1)\hat{P}^{\sigma\tau}_{12}| \alpha_1  \alpha_2 s_h \rangle=g_A\delta_{t_p,p}\delta_{t_h,n}\Big\{\nonumber\\
&\delta_{s_p,\uparrow}\delta_{s_h,\uparrow}\Big[4f^{\sigma\tau}_{23}+4f^{t\tau}_{23}\Big(2- 3\frac{(x_{23}^2+y_{23}^2)}{r_{23}^2}\Big)\Big]+\nonumber\\
&\delta_{s_p,\uparrow}\delta_{s_h,\downarrow}\Big[12 f^{t\tau}_{23} \frac{z_{23}(x_{23}-iy_{23})}{r_{23}^2}\Big]+\nonumber\\
&\dots\Big\}
\end{align}
\\
{\large $\mathbf{O_{Nl}^{(1)}}$}
\begin{align}
\sum_{\alpha_i}\langle \alpha_1 \alpha_2 \alpha_p |\hat{O}_{\sigma\tau}(1)(\hat{f}_{23}-1)\hat{P}^{\sigma\tau}_{12}\hat{P}^{\sigma\tau}_{13}|\alpha_1 \alpha_2 \alpha_h\rangle
\end{align}

\textbf{Fermi}
\begin{align}
g_V\sum_{\alpha_i}\langle \alpha_1 \alpha_2 s_p |\tau_{1}^{(+)}(\hat{f}_{23}-1)\hat{P}^{\sigma\tau}_{12}\hat{P}^{\sigma\tau}_{13}| \alpha_1  \alpha_2 s_h \rangle
=g_V\delta_{t_p,p}\delta_{t_h,n}\delta_{s_p,s_h} 4f^{\tau}_{23} 
\end{align}

\textbf{Gamow-Teller}
\begin{align}
&g_A\sum_{\alpha_i}\langle \alpha_1 \alpha_2 s_p |\tau_{1}^{(+)}\sigma_{1}^x (\hat{f}_{23}-1)\hat{P}^{\sigma\tau}_{12}\hat{P}^{\sigma\tau}_{13}| \alpha_1  \alpha_2 s_h \rangle=g_A\delta_{t_p,p}\delta_{t_h,n}\Big\{\nonumber\\
&\delta_{s_p,\uparrow}\delta_{s_h,\uparrow}\Big[6(f^{t}_{23}-f^{t\tau}_{23})\frac{x_{23}z_{23}}{r_{23}^2}\Big]+\nonumber\\
&\delta_{s_p,\uparrow}\delta_{s_h,\downarrow}\Big[(f^{c}_{23}-1)-f^{\tau}_{23}-f_{23}^{\sigma}+f^{\sigma\tau}_{23}
+2(f^{t}_{23}-f^{t\tau}_{23})\Big(-1+3\frac{x_{23}(x_{23}-iy_{23})}{r_{23}^2}\Big)\Big]+\nonumber\\
&\dots\Big\}
\end{align}

\begin{align}
&g_A\sum_{\alpha_i}\langle \alpha_1 \alpha_2 s_p |\tau_{1}^{(+)}\sigma_{1}^y (\hat{f}_{23}-1)\hat{P}^{\sigma\tau}_{12}\hat{P}^{\sigma\tau}_{13}| \alpha_1  \alpha_2 s_h \rangle=g_A\delta_{t_p,p}\delta_{t_h,n}\Big\{\nonumber\\
&\delta_{s_p,\uparrow}\delta_{s_h,\uparrow}\Big[6(f^{t}_{23}-f^{t\tau}_{23})\frac{y_{23}z_{23}}{r_{23}^2}\Big]+\nonumber\\
&\delta_{s_p,\uparrow}\delta_{s_h,\downarrow}\Big[-i((f^{c}_{23}-1)-f^{\tau}_{23}-f^{\sigma}_{23}+f^{\sigma\tau}_{23})+2(f^{t}_{23}-f^{t\tau}_{23})\Big(i+3\frac{y_{23}(x_{23}-iy_{23})}{r_{23}^2}\Big)\Big]+\nonumber\\
&\dots\Big\}
\end{align}

\begin{align}
&g_A\sum_{\alpha_i}\langle \alpha_1 \alpha_2 s_p |\tau_{1}^{(+)}\sigma_{1}^z (\hat{f}_{23}-1)\hat{P}^{\sigma\tau}_{12}\hat{P}^{\sigma\tau}_{13}| \alpha_1  \alpha_2 s_h \rangle=g_A\delta_{t_p,p}\delta_{t_h,n}\Big\{\nonumber\\
&\delta_{s_p,\uparrow}\delta_{s_h,\uparrow}\Big[(f^{c}_{23}-1)-f^{\tau}_{23}-f^{\sigma}_{23}+f^{\sigma\tau}_{23}+2(f^{t}_{23}-f^{t\tau}_{23})\Big(2-3\frac{(x_{23}^2+y_{23}^2)}{r_{23}^2}\Big)\Big]+\nonumber\\
&\delta_{s_p,\uparrow}\delta_{s_h,\downarrow}\Big[6(f^{t}_{23}-f^{t\tau}_{23})\frac{z_{23}(x_{23}-iy_{23})}{r_{23}^2}\Big]+\nonumber\\
&\dots\Big\}
\end{align}

\section{Second order}
In order to write in a compact form the Gamow-Teller matrix elements, it is worth introducing the following quantities
\begin{align}
A_1=&24[f_{12}^{\tau}(f_{12}^{t}+2f_{12}^{t\tau})-f^{\sigma\tau}_{12}(f_{12}^{t}+2f_{12}^{t\tau})
+f_{12}^{t\tau}((f_{12}^{c}-1)-f^{\sigma}_{12}+2(f_{12}^{t}+f_{12}^{t\tau})]\nonumber \\
A_2=&8[f_{12}^{\tau}f^{\sigma}_{12}+(f_{12}^{c}-1)f^{\sigma\tau}_{12}+2f_{12}^{\tau}f^{\sigma\tau}_{12}+2f^{\sigma}_{12} f^{\sigma\tau}_{12}+2{f^{\sigma\tau}_{12}}^2-f_{12}^{\tau}f_{12}^{t} + f^{\sigma\tau}_{12} f_{12}^{t} - \nonumber\\ 
&(f_{12}^{c}-1)f_{12}^{t\tau}-2 f_{12}^{\tau} f_{12}^{t\tau} + f^{\sigma}_{12} f_{12}^{t\tau} + 2 f^{\sigma\tau}_{12} f_{12}^{t\tau} - 4 f_{12}^{t} f_{12}^{t\tau} - 4 {f_{12}^{t\tau}}^2]\nonumber\\
B_1=&6[f_{12}^{t}(f_{12}^{c}-1+3f_{12}^{\tau}+f^{\sigma}_{12}-5 f^{\sigma\tau}_{12}+2f_{12}^{t})+(-(f_{12}^{c}-1)+ f_{12}^{\tau} +7f^{\sigma}_{12}+f^{\sigma\tau}_{12}+\nonumber\\
&4f_{12}^{t})f_{12}^{t\tau}+2{f_{12}^{t\tau}}^2]\nonumber\\
B_2=&(f_{12}^{c}-1)^2+2(f_{12}^{c}-1)f_{12}^{\tau}+{f_{12}^{\tau}}^2+2(f_{12}^{c}-1)f^{\sigma}_{12}-6f_{12}^{\tau}f^{\sigma}_{12} + {f^{\sigma}_{12}}^2+10(f_{12}^{c}-1)f^{\sigma\tau}_{12}+\nonumber\\
&2f_{12}^{\tau}f^{\sigma\tau}_{12}+2f^{\sigma}_{12} f^{\sigma\tau}_{12}+{f^{\sigma\tau}_{12}}^2-2(f_{12}^{c}-1)f_{12}^{t}-
6f_{12}^{\tau}f_{12}^{t}-2f^{\sigma}_{12}f_{12}^{t}+10 f^{\sigma\tau}_{12}f_{12}^{t}-\nonumber \\
&8{f_{12}^{t}}^2+2 (f_{12}^{c}-1)f_{12}^{t\tau}-2f_{12}^{\tau}f_{12}^{t\tau}-14 f^{\sigma}_{12} f_{12}^{t\tau}-2 f^{\sigma\tau}_{12} f_{12}^{t\tau}-16 f_{12}^{t}f_{12}^{t\tau}-8{f_{12}^{t\tau}}^2\nonumber\\
C_1=&24[2f^{\sigma}_{12} f_{12}^{t} + {f_{12}^{t}}^2 - f_{12}^{t\tau} (2 f^{\sigma\tau}_{12} + f_{12}^{t\tau})]\nonumber\\
C_2=&4[(f_{12}^{c}-1)^2-{f_{12}^{\tau}}^2-{f^{\sigma}_{12}}^2+{f^{\sigma\tau}_{12}}^2-4f^{\sigma}_{12}f_{12}^{t}-4{f_{12}^{t}}^2+4f^{\sigma\tau}_{12}f_{12}^{t\tau}+4{f_{12}^{t\tau}}^2]\, .
\end{align}

{\large $\mathbf{O_{Na}^{(2)}}$}
\begin{align}
\sum_{\alpha_1}\langle \alpha_1 \alpha_p|(\hat{f}_{12} -1)\hat{O}_{\sigma\tau}(1)(\hat{f}_{12} -1)| \alpha_1 \alpha_h\rangle
\end{align}

\textbf{Fermi}
\begin{align}
&g_V\sum_{\alpha_1}\langle \alpha_1 s_{p_i}|(\hat{f}_{12} -1)\tau_{1}^{(+)}(\hat{f}_{12} -1)| \alpha_1 s_{h_j}\rangle
=\nonumber\\
&g_V\delta_{t_p,p}\delta_{t_h,n}\delta_{s_p,s_h}8[f^{\tau}_{12} ((f^{c}_{12}-1) + f^{\tau}_{12}) + 3 f^{\sigma\tau}_{12} (f^{\sigma}_{12} + f^{\sigma\tau}_{12}) +6 f_{12}^{t\tau} (f^{t}_{12} + f_{12}^{t\tau})]
\end{align}

\textbf{Gamow-Teller}

\begin{align}
&g_A\sum_{\alpha_1}\langle \alpha_1 s_p|(\hat{f}_{12} -1)\tau_{1}^{(+)}\sigma_{1}^x (\hat{f}_{12} -1)| \alpha_1 s_h\rangle=g_A\delta_{t_p,p}\delta_{t_h,n}\Big\{\nonumber\\
&\delta_{s_p,\uparrow}\delta_{s_h,\uparrow}\Big[A_1\frac{x_{12} z_{12}}{r_{12}^2}\Big]+\nonumber\\
&\delta_{s_p,\uparrow}\delta_{s_h,\downarrow}\Big[
A_2+A_1\frac{x_{12}(x_{12}-iy_{12})}{r_{12}^2}\Big]+\nonumber\\
&\dots\Big\}
\end{align}

\begin{align}
&g_A\sum_{\alpha_1}\langle \alpha_1 s_p|(\hat{f}_{12} -1)\tau_{1}^{(+)}\sigma_{1}^y (\hat{f}_{12} -1)| \alpha_1 s_h\rangle=g_A\delta_{t_p,p}\delta_{t_h,n}\Big\{\nonumber\\
&\delta_{s_p,\uparrow}\delta_{s_h,\uparrow}\Big[
A_1\frac{y_{12} z_{12}}{r_{12}^2}\Big]+\nonumber\\
&\delta_{s_p,\uparrow}\delta_{s_h,\downarrow}\Big[
-iA_2+A_1\frac{y_{12}(x_{12}-iy_{12})}{r_{12}^2}\Big]+\nonumber\\
&\dots\Big\}
\end{align}

\begin{align}
&g_A\sum_{\alpha_1}\langle \alpha_1 s_p|(\hat{f}_{12} -1)\tau_{1}^{(+)}\sigma_{1}^z(\hat{f}_{12} -1)| \alpha_1 s_h\rangle=g_A\delta_{t_p,p}\delta_{t_h,n}\Big\{\nonumber\\
&\delta_{s_p,\uparrow}\delta_{s_h,\uparrow}\Big[
A_2+A_1\frac{z_{12}^2}{r_{12}^2}\Big]+\nonumber\\
&\delta_{s_p,\uparrow}\delta_{s_h,\downarrow}\Big[
A_1\frac{z_{12}(x_{12}-iy_{12})}{r_{12}^2}\Big]+\nonumber\\
&\dots\Big\}\, .
\end{align}

{\large $\mathbf{O_{Nb}^{(2)}}$}
\begin{align}
\sum_{\alpha_1}\langle \alpha_1 \alpha_p|(\hat{f}_{12} -1)\hat{O}_{\sigma\tau}(1)(\hat{f}_{12} -1)\hat{P}^{\sigma\tau}_{12}| \alpha_1 \alpha_h\rangle
\end{align}

\textbf{Fermi}
\begin{align}
&g_V\sum_{\alpha_1}\langle \alpha_1 s_{p_i}(\hat{f}_{12} -1)\tau_{1}^{(+)}(\hat{f}_{12} -1)\hat{P}^{\sigma\tau}_{12}| \alpha_1 s_{h_j}\rangle
=\nonumber\\
&g_V\delta_{t_p,p}\delta_{t_h,n}\delta_{s_p,s_h}[(f^{c\,}_{12}-1)^2 + {f^{\tau\,}_{12}}^2 + 6 f^{\tau}_{12} (f^{\sigma}_{12} + f^{\sigma\tau}_{12})-3(f^{\sigma}_{12}+f^{\sigma\tau}_{12})^2 +\nonumber \\
&\qquad \qquad 2 (f^{c}_{12}-1)(f^{\tau}_{12}+3(f^{\sigma}_{12}+f^{\sigma\tau}_{12})) + 12(f^{t}_{12} + f_{12}^{t\tau})^2]
\end{align}

\textbf{Gamow-Teller}
\begin{align}
&g_A\sum_{\alpha_1}\langle \alpha_1 s_{p_i}|(\hat{f}_{12} -1)\tau_{1}^{(+)}\sigma_{1}^x (\hat{f}_{12} -1)\hat{P}^{\sigma\tau}_{12}| \alpha_1 s_{h_j}\rangle=g_A\delta_{t_p,p}\delta_{t_h,n}\Big\{\nonumber\\
&\delta_{s_p,\uparrow}\delta_{s_h,\uparrow}\Big[B_1\frac{x_{12} z_{12}}{r_{12}^2}\Big]+\nonumber\\
&\delta_{s_p,\uparrow}\delta_{s_h,\downarrow}\Big[
B_2+B_1\frac{x_{12}(x_{12}-iy_{12})}{r_{12}^2}\Big]+\nonumber\\
&\dots\Big\}
\end{align}

\begin{align}
&g_A\sum_{\alpha_1}\langle \alpha_1 s_{p_i}|(\hat{f}_{12} -1)\tau_{1}^{(+)}\sigma_{1}^y (\hat{f}_{12} -1)\hat{P}^{\sigma\tau}_{12}| \alpha_1 s_{h_j}\rangle=g_A\delta_{t_p,p}\delta_{t_h,n}\Big\{\nonumber\\
&\delta_{s_p,\uparrow}\delta_{s_h,\uparrow}\Big[
B_1\frac{y_{12} z_{12}}{r_{12}^2}\Big]+\nonumber\\
&\delta_{s_p,\uparrow}\delta_{s_h,\downarrow}\Big[
-iB_2+B_1\frac{y_{12}(x_{12}-iy_{12})}{r_{12}^2}\Big]+\nonumber\\
&\dots\Big\}
\end{align}

\begin{align}
&g_A\sum_{\alpha_1}\langle \alpha_1 s_{p_i}|(\hat{f}_{12} -1)\tau_{1}^{(+)}\sigma_{1}^z(\hat{f}_{12} -1)\hat{P}^{\sigma\tau}_{12}| \alpha_1 s_{h_j}\rangle=g_A\delta_{t_p,p}\delta_{t_h,n}\Big\{\nonumber\\
&\delta_{s_p,\uparrow}\delta_{s_h,\uparrow}\Big[
B_2+B_1\frac{z_{12}^2}{r_{12}^2}\Big]+\nonumber\\
&\delta_{s_p,\uparrow}\delta_{s_h,\downarrow}\Big[
B_1\frac{z_{12}(x_{12}-iy_{12})}{r_{12}^2}\Big]+\nonumber\\
&\dots\Big\}\, .
\end{align}

{\large $\mathbf{O_{Nc}^{(2)}}$}
\begin{align}
\sum_{\alpha_2}\langle \alpha_p \alpha_2| (\hat{f}_{12} -1)\hat{O}_{\sigma\tau}(1)(\hat{f}_{12} -1)| \alpha_h \alpha_2\rangle
\end{align}

\textbf{Fermi}
\begin{align}
&g_V\sum_{\alpha_2}\langle s_p \alpha_2|(\hat{f}_{12} -1)\tau_{1}^{(+)}(\hat{f}_{12} -1)|s_h \alpha_2 \rangle=
\nonumber\\
&g_V\delta_{t_p,p}\delta_{t_h,n}\delta_{s_p,s_h} 4[(f^{c}_{12}-1)^2 - {f^{\tau}_{12}}^2 + 3{f^{\sigma}_{12}}^2 - 3{f^{\sigma\tau}_{12}}^2 + 6{f^{t}_{12}}^2 - 6 {f^{t\tau}_{12}}^2) 
\end{align}

\textbf{Gamow-Teller}
\begin{align}
&g_A\sum_{\alpha_2}\langle s_p \alpha_2   |(\hat{f}_{12} -1)\tau_{1}^{(+)}\sigma_{1}^x (\hat{f}_{12} -1)| s_h \alpha_2 \rangle=g_A\delta_{t_p,p}\delta_{t_h,n}\Big\{\nonumber\\
&\delta_{s_p,\uparrow}\delta_{s_h,\uparrow}\Big[
C_1\frac{x_{12} z_{12}}{r_{12}^2}\Big]+\nonumber\\
&\delta_{s_p,\uparrow}\delta_{s_h,\downarrow}\Big[
C_2+C_1\frac{x_{12}(x_{12}-iy_{12})}{r_{12}^2}\Big]+\nonumber\\
&\dots\Big\}
\end{align}

\begin{align}
&g_A\sum_{\alpha_2}\langle s_p \alpha_2  |(\hat{f}_{12} -1)\tau_{1}^{(+)}\sigma_{1}^y (\hat{f}_{12} -1)|s_h \alpha_2   \rangle=g_A\delta_{t_p,p}\delta_{t_h,n}\Big\{\nonumber\\
&\delta_{s_p,\uparrow}\delta_{s_h,\uparrow}\Big[
C_1\frac{y_{12} z_{12}}{r_{12}^2}\Big]+\nonumber\\
&\delta_{s_p,\uparrow}\delta_{s_h,\downarrow}\Big[
-iC_2+C_1\frac{y_{12}(x_{12}-iy_{12})}{r_{12}^2}\Big]+\nonumber\\
&\dots\Big\}
\end{align}

\begin{align}
&g_A\sum_{\alpha_2}\langle s_p \alpha_2 |(\hat{f}_{12} -1)\tau_{1}^{(+)}\sigma_{1}^z(\hat{f}_{12} -1)|s_h \alpha_2 \rangle=g_A\delta_{t_p,p}\delta_{t_h,n}\Big\{\nonumber\\
&\delta_{s_p,\uparrow}\delta_{s_h,\uparrow}\Big[
C_2+C_1\frac{z_{12}^2}{r_{12}^2}\Big]+\nonumber\\
&\delta_{s_p,\uparrow}\delta_{s_h,\downarrow}\Big[
C_1\frac{z_{12}(x_{12}-iy_{12})}{r_{12}^2}\Big]+\nonumber\\
&\dots\Big\}
\end{align}

Like in the first order case, the spin-isospin matrix elements of $O_{Nd}^{(2)}$ are identical to those of $O_{Nb}^{(2)}$ for both Fermi and Gamow-Teller transitions.
\\

{\large $\mathbf{O_{Da}^{(2)}}$}
\begin{align}
&\sum_{\alpha_1}\langle \alpha_1 \alpha_p |\hat{f}_{12}-1|\alpha_1 \alpha_p\rangle=4[{(f^{c}_{12}-1)}^2+3({f^{\tau}_{12}}^2 + {f^{\sigma}_{12}}^2 + 3 {f^{\sigma\tau}_{12}}^2 + 2 {f^{t}_{12}}^2 + 6 {f_{12}^{t\tau}}^2)]
\end{align}

{\large $\mathbf{O_{Db}^{(2)}}$}
\begin{align}
&\sum_{\alpha_1}\langle \alpha_1 \alpha_p |(\hat{f}_{12}-1)\hat{P}^{\sigma\tau}_{12}|\alpha_1 \alpha_p\rangle=\nonumber\\
&\{{(f^{c}_{12}-1)\,}^2 + 6 (f^{c}_{12}-1)(f^{\tau}_{12}+f^{\sigma}_{12}+3f^{\sigma\tau}_{12})- 
 3 [{f^{\tau}_{12}}^2 - 6 f^{\sigma}_{12} f^{\tau}_{12} + {f^{\sigma}_{12}}^2 +\nonumber \\
&\qquad \qquad 6f^{\sigma\tau}_{12}(f^{\tau}_{12} + f^{\sigma}_{12})-3{f_{12}^{\sigma\tau}}^2-4 ({f_{12}^{t}}^2 + 6 f^{t}_{12} f_{12}^{t\tau} - 3 {f_{12}^{t\tau}}^2)]\}\end{align}

We do not report the results for the spin-isospin matrix elements of $\mathbf{O_{Dc}^{(1)}}$ and $\mathbf{O_{Dd}^{(1)}}$ as they are identical to those of $\mathbf{O_{Da}^{(1)}}$ and $\mathbf{O_{Db}^{(1)}}$, respectively.
\newline

\newpage             
\thispagestyle{empty}  

%Fermi Gamow-Teller matrix elements
\chapter{Total spin of a particle-hole pair}
\label{app:tsphp}
We have shown that the Hartree-Fock ground state of infinite nuclear matter is a Slater determinant of plane waves. The single particle wave function can be characterized by the quantum number $|\alpha\rangle\equiv |\mathbf{k},s,s_z,t,t_z\rangle$.  with $s=1/2$ and $t=1/2$. Since our Hilbert space is a direct product of spin and isospin spaces, for the sake of simplicity we may consider only the spin, as the treatment of the isospin degree of freedom proceeds analogously.

The general creation and distraction operators are denoted by $c^{\dag}_{\mathbf{k},s_z}$ and $c_{\mathbf{k},s_z}$, where it is understood that $s=1/2$. The particle and hole operators can be defined by the relations 
\begin{IEEEeqnarray}{rCll} 
a^{\dag}_{\mathbf{k},s_z}&\equiv & c^{\dag}_{\mathbf{k},s_z} \qquad  & |\mathbf{k}|> k_F\nonumber \\
b^{\dag}_{\mathbf{k},s_z}&\equiv & S_{-\mathbf{k},-s_z,}c_{-\mathbf{k},-s_z} \qquad  &|\mathbf{k}|\leq k_F\, .
\label{eq:can_anti}
\end{IEEEeqnarray}
The following phase convention has been adopted
\begin{equation}
S_{\mathbf{k},s_z}=(-1)^{1/2-s_z}\, .
\end{equation}

This is a canonical transformation, for it does not affect the anticommutation rules 
\begin{equation}
\{a_{\mathbf{k},s_z},a^{\dag}_{\mathbf{k}',s'_z}\}=\{b_{\mathbf{k},s_z},b^{\dag}_{\mathbf{k}',s'_z}\}= \delta_{\mathbf{k}\mathbf{k}'}\delta_{s_z s'z}\, ,
\end{equation}
while all the the other anticommutators vanish. It can be shown \cite{fetter_03} that the $s_z$-dependent phase operator $b^{\dag}_{\mathbf{k},s_z}$ creates a hole of spin projection along the $z$-axis $s_z$.

Using the canonical transformation we may now compute the total spin of the particle hole pair $|p_m; h_i\rangle$. We start from following the well known relation for the total spin of the particle particle states 
\begin{IEEEeqnarray}{CCl}
a^{\dag}_{\mathbf{p},\uparrow}b^{\dag}_{-\mathbf{h},\uparrow} |0\rangle \quad &\Longrightarrow& \quad S=1,S_z=1 \\
\frac{1}{\sqrt{2}}(a^{\dag}_{\mathbf{p},\uparrow}b^{\dag}_{-\mathbf{h},\downarrow}+a^{\dag}_{\mathbf{p},\downarrow}b^{\dag}_{-\mathbf{h},\uparrow}) |0\rangle \quad &\Longrightarrow& \quad S=1,S_z=0 \\
a^{\dag}_{\mathbf{p},\downarrow}b^{\dag}_{-\mathbf{h},\downarrow} |0\rangle \quad &\Longrightarrow& \quad S=1,S_z=-1 \\
\frac{1}{\sqrt{2}}(a^{\dag}_{\mathbf{p},\uparrow}b^{\dag}_{-\mathbf{h},\downarrow}-a^{\dag}_{\mathbf{p},\downarrow}b^{\dag}_{-\mathbf{h},\uparrow}) |0\rangle \quad &\Longrightarrow& \quad S=0,S_z=0\, .
\label{eq:pp_spin_app}
 \end{IEEEeqnarray}
 
Through the canonical transformation we can relate these states with the corresponding particle-hole states 
\begin{IEEEeqnarray}{rClCl}
a^{\dag}_{\mathbf{p},\uparrow}b^{\dag}_{-\mathbf{h},\uparrow} |0\rangle  &=& 
-c^{\dag}_{\mathbf{p},\uparrow}c_{\mathbf{h},\downarrow} |0\rangle   \\
\frac{1}{\sqrt{2}}(a^{\dag}_{\mathbf{p},\uparrow}b^{\dag}_{-\mathbf{h},\downarrow}+a^{\dag}_{\mathbf{p},\downarrow}b^{\dag}_{-\mathbf{h},\uparrow}) |0\rangle &=&
\frac{1}{\sqrt{2}}(c^{\dag}_{\mathbf{p},\uparrow}c_{\mathbf{h},\uparrow}-c^{\dag}_{\mathbf{p},\downarrow}c_{\mathbf{h},\downarrow}) |0\rangle \\
a^{\dag}_{\mathbf{p},\downarrow}b^{\dag}_{-\mathbf{h},\downarrow} |0\rangle &=&
c^{\dag}_{\mathbf{p},\downarrow}c_{\mathbf{h},\uparrow} |0\rangle   \\
\frac{1}{\sqrt{2}}(a^{\dag}_{\mathbf{p},\uparrow}b^{\dag}_{-\mathbf{h},\downarrow}-a^{\dag}_{\mathbf{p},\downarrow}b^{\dag}_{-\mathbf{h},\uparrow}) |0\rangle &=& 
\frac{1}{\sqrt{2}}(c^{\dag}_{\mathbf{p},\uparrow}c_{\mathbf{h},\uparrow}+c^{\dag}_{\mathbf{p},\downarrow}c_{\mathbf{h},\downarrow}) |0\rangle\, .
\label{eq:ph_spin_app}
 \end{IEEEeqnarray}

By comparing Eq. (\ref{eq:pp_spin_app}) and (\ref{eq:ph_spin_app}), we immediately find the results listed in Table {\ref{tab:s_ph}}.

\newpage             
\thispagestyle{empty}

%Bibliography
\addcontentsline{toc}{chapter}{Bibliography}
\bibliographystyle{unsrt}
\bibliography{biblio}

\begin{thebibliography}{100}

\bibitem{lacombe_80}
M.~Lacombe, B.~Loiseau, J.~M. Richard, R.~Vinh Mau, J.~C\^ot\'e, P.~Pir\`es,
  and R.~de~Tourreil.
\newblock Parametrization of the paris $n-n$ potential.
\newblock {\em Phys. Rev. C}, 21:861--873, Mar 1980.

\bibitem{stoks_94}
V.~G.~J. Stoks, R.~A.~M. Klomp, C.~P.~F. Terheggen, and J.~J. de~Swart.
\newblock Construction of high-quality \textit{NN} potential models.
\newblock {\em Phys. Rev. C}, 49:2950--2962, Jun 1994.

\bibitem{wiringa_95}
R.~B. Wiringa, V.~G.~J. Stoks, and R.~Schiavilla.
\newblock Accurate nucleon-nucleon potential with charge-independence breaking.
\newblock {\em Phys. Rev. C}, 51:38--51, Jan 1995.

\bibitem{machleidt_01}
R.~Machleidt.
\newblock High-precision, charge-dependent {B}onn nucleon-nucleon potential.
\newblock {\em Phys. Rev. C}, 63:024001, Jan 2001.

\bibitem{entem_03}
D.~R. Entem and R.~Machleidt.
\newblock Accurate charge-dependent nucleon-nucleon potential at fourth order
  of chiral perturbation theory.
\newblock {\em Phys. Rev. C}, 68:041001, Oct 2003.

\bibitem{epelbaum_05}
E.~Epelbaum, Gl\"ockle, H., and Mei\ss{}ner.
\newblock The two-nucleon system at next-to-next-to-next-to-leading order.
\newblock {\em Nuclear Physics A}, 747(2-4):362 -- 424, 2005.

\bibitem{stoks_93}
V.~G.~J. Stoks, R.~A.~M. Klomp, M.~C.~M. Rentmeester, and J.~J. de~Swart.
\newblock Partial-wave analysis of all nucleon-nucleon scattering data below
  350 {M}ev.
\newblock {\em Phys. Rev. C}, 48:792--815, Aug 1993.

\bibitem{bergervoet_90}
J.~R. Bergervoet, P.~C. van Campen, R.~A.~M. Klomp, J.-L. de~Kok, T.~A. Rijken,
  V.~G.~J. Stoks, and J.~J. de~Swart.
\newblock Phase shift analysis of all proton-proton scattering data below
  ${\mathit{t}}_{\mathrm{lab}}$=350 {M}ev.
\newblock {\em Phys. Rev. C}, 41:1435--1452, Apr 1990.

\bibitem{vanderleun_82}
C.~Van~Der Leun and C.~Alderliesten.
\newblock The deuteron binding energy.
\newblock {\em Nuclear Physics A}, 380(2):261 -- 269, 1982.

\bibitem{ericson_83}
T.E.O. Ericson and M.~Rosa-Clot.
\newblock The deuteron asymptotic d-state as a probe of the nucleon-nucleon
  force.
\newblock {\em Nuclear Physics A}, 405(3):497 -- 533, 1983.

\bibitem{rodning_90}
N.~L. Rodning and L.~D. Knutson.
\newblock Asymptotic \textit{D} -state to \textit{S} -state ratio of the
  deuteron.
\newblock {\em Phys. Rev. C}, 41:898--909, Mar 1990.

\bibitem{simon_81}
G.G. Simon, Ch. Schmitt, and V.H. Walther.
\newblock Elastic electric and magnetic e-d scattering at low momentum
  transfer.
\newblock {\em Nuclear Physics A}, 364(2-3):285 -- 296, 1981.

\bibitem{bishop_79}
David~M. Bishop and Lap~M. Cheung.
\newblock Quadrupole moment of the deuteron from a precise calculation of the
  electric field gradient in ${\mathrm{d}}_{2}$.
\newblock {\em Phys. Rev. A}, 20:381--384, Aug 1979.

\bibitem{schoen_03}
K.~Schoen, D.~L. Jacobson, M.~Arif, P.~R. Huffman, T.~C. Black, W.~M. Snow,
  S.~K. Lamoreaux, H.~Kaiser, and S.~A. Werner.
\newblock Precision neutron interferometric measurements and updated
  evaluations of the $n-p$ and $n-d$ coherent neutron scattering lengths.
\newblock {\em Phys. Rev. C}, 67:044005, Apr 2003.

\bibitem{shimizu_95}
S.~Shimizu, K.~Sagara, H.~Nakamura, K.~Maeda, T.~Miwa, N.~Nishimori, S.~Ueno,
  T.~Nakashima, and S.~Morinobu.
\newblock Analyzing powers of \textit{p}+\textit{d} scattering below the
  deuteron breakup threshold.
\newblock {\em Phys. Rev. C}, 52:1193--1202, Sep 1995.

\bibitem{demorest_10}
P.~B. Demorest, T.~Pennucci, S.~M. Ransom, M.~S.~E. Roberts, and J.~W.~T.
  Hessels.
\newblock A two-solar-mass neutron star measured using {S}hapiro delay.
\newblock {\em Nature}, 467(7319):1081--1083, 10 2010.

\bibitem{carlson_83}
J.~Carlson, V.R. Pandharipande, and R.B. Wiringa.
\newblock Three-nucleon interaction in 3-, 4- and $\infty$-body systems.
\newblock {\em Nuclear Physics A}, 401(1):59 -- 85, 1983.

\bibitem{coon_81}
Sidney~A. Coon and Walter Gl\"ockle.
\newblock Two-pion-exchange three-nucleon potential: Partial wave analysis in
  momentum space.
\newblock {\em Phys. Rev. C}, 23:1790--1802, Apr 1981.

\bibitem{wiringa_88}
R.~B. Wiringa, V.~Fiks, and A.~Fabrocini.
\newblock Equation of state for dense nucleon matter.
\newblock {\em Phys. Rev. C}, 38:1010--1037, Aug 1988.

\bibitem{akmal_98}
Arya Akmal.
\newblock {\em Variational studies of nucleon matter with realistic
  potentials}.
\newblock PhD thesis, University of Illinois at {U}rbana-Champaign, 1998.

\bibitem{pudliner_95}
B.~S. Pudliner, V.~R. Pandharipande, J.~Carlson, and R.~B. Wiringa.
\newblock Quantum {M}onte {C}arlo calculations of $a\le6$ nuclei.
\newblock {\em Phys. Rev. Lett.}, 74:4396--4399, May 1995.

\bibitem{gandolfi_07b}
Stefano Gandolfi, Francesco Pederiva, Stefano Fantoni, and Kevin~E. Schmidt.
\newblock Quantum {M}onte~{C}arlo calculations of symmetric nuclear matter.
\newblock {\em Phys. Rev. Lett.}, 98:102503, Mar 2007.

\bibitem{bombaci_05}
I.~Bombaci, A.~Fabrocini, A.~Polls, and I.~Vida{\~{n}}a.
\newblock Spin-orbit and tensor interactions in homogeneous matter of nucleons:
  accuracy of modern many-body theories.
\newblock {\em Physics Letters B}, 609(3-4):232 -- 240, 2005.

\bibitem{pieper_01}
Steven~C. Pieper, V.~R. Pandharipande, R.~B. Wiringa, and J.~Carlson.
\newblock Realistic models of pion-exchange three-nucleon interactions.
\newblock {\em Phys. Rev. C}, 64:014001, Jun 2001.

\bibitem{sarsa_03}
A.~Sarsa, S.~Fantoni, K.~E. Schmidt, and F.~Pederiva.
\newblock Neutron matter at zero temperature with an auxiliary field diffusion
  {M}onte {C}arlo method.
\newblock {\em Phys. Rev. C}, 68:024308, Aug 2003.

\bibitem{akmal_98b}
A.~Akmal, V.~R. Pandharipande, and D.~G. Ravenhall.
\newblock Equation of state of nucleon matter and neutron star structure.
\newblock {\em Phys. Rev. C}, 58:1804--1828, Sep 1998.

\bibitem{lovato_11}
Alessandro Lovato, Omar Benhar, Stefano Fantoni, Alexey~Yu. Illarionov, and
  Kevin~E. Schmidt.
\newblock Density-dependent nucleon-nucleon interaction from three-nucleon
  forces.
\newblock {\em Phys. Rev. C}, 83:054003, May 2011.

\bibitem{lovato_11b}
Alessandro Lovato, Omar Benhar, Stefano Fantoni, Alexey~Yu Illarionov, and
  Kevin~E Schmidt.
\newblock Density-dependent nucleon-nucleon interaction from {U}rbana {UIX}
  three-nucleon force.
\newblock {\em Journal of Physics: Conference Series}, 336(1):012016, 2011.

\bibitem{lagaris_81}
I.E. Lagaris and V.R. Pandharipande.
\newblock Variational calculations of realistic models of nuclear matter.
\newblock {\em Nuclear Physics A}, 359(2):349 -- 364, 1981.

\bibitem{friedman_81}
B.~Friedman and V.R. Pandharipande.
\newblock Hot and cold, nuclear and neutron matter.
\newblock {\em Nuclear Physics A}, 361(2):502 -- 520, 1981.

\bibitem{gandolfi_10}
S.~Gandolfi, A.~Yu. Illarionov, S.~Fantoni, J.~C. Miller, F.~Pederiva, and
  K.~E. Schmidt.
\newblock Microscopic calculation of the equation of state of nuclear matter
  and neutron star structure.
\newblock {\em Monthly Notices of the Royal Astronomical Society: Letters},
  404(1):L35--L39, 2010.

\bibitem{fantoni_84b}
S.~Fantoni and V.R. Pandharipande.
\newblock Momentum distribution of nucleons in nuclear matter.
\newblock {\em Nuclear Physics A}, 427(3):473 -- 492, 1984.

\bibitem{fantoni_87}
S.~Fantoni and V.R. Pandharipande.
\newblock Correlated basis theory of nuclear matter response functions.
\newblock {\em Nuclear Physics A}, 473(2):234 -- 266, 1987.

\bibitem{fabrocini_89}
A.~Fabrocini and S.~Fantoni.
\newblock Microscopic calculation of the longitudinal response of nuclear
  matter.
\newblock {\em Nuclear Physics A}, 503(2):375 -- 403, 1989.

\bibitem{benhar_89}
Omar Benhar, Adelchi Fabrocini, and Stefano Fantoni.
\newblock The nucleon spectral function in nuclear matter.
\newblock {\em Nuclear Physics A}, 505(2):267 -- 299, 1989.

\bibitem{benhar_92}
O.~Benhar, A.~Fabrocini, and S.~Fantoni.
\newblock Nuclear-matter green functions in correlated-basis theory.
\newblock {\em Nuclear Physics A}, 550(2):201 -- 222, 1992.

\bibitem{benhar_07}
Omar Benhar and Marco Valli.
\newblock Shear viscosity of neutron matter from realistic nucleon-nucleon
  interactions.
\newblock {\em Phys. Rev. Lett.}, 99:232501, Dec 2007.

\bibitem{benhar_08}
Omar Benhar, Nicola Farina, Salvatore Fiorilla, and Marco Valli.
\newblock Unified description of equation of state and transport properties of
  nuclear matter.
\newblock {\em AIP Conference Proceedings}, 1056(1):248--255, 2008.

\bibitem{epelbaum_02}
E.~Epelbaum, A.~Nogga, W.~Gl\"ockle, H.~Kamada, Ulf-G. Mei\ss{}ner, and
  H.~Wita\l{}a.
\newblock Three-nucleon forces from chiral effective field theory.
\newblock {\em Phys. Rev. C}, 66:064001, Dec 2002.

\bibitem{bernard_08}
V.~Bernard, E.~Epelbaum, H.~Krebs, and Ulf-G. Mei\ss{}ner.
\newblock Subleading contributions to the chiral three-nucleon force:
  Long-range terms.
\newblock {\em Phys. Rev. C}, 77:064004, Jun 2008.

\bibitem{navratil_07}
P.~Navr\'atil.
\newblock Local three-nucleon interaction from chiral effective field theory.
\newblock {\em Few-Body Systems}, 41:117--140, 2007.
\newblock 10.1007/s00601-007-0193-3.

\bibitem{bogner_05}
S.K. Bogner, A.~Schwenk, R.J. Furnstahl, and A.~Nogga.
\newblock Is nuclear matter perturbative with low-momentum interactions?
\newblock {\em Nuclear Physics A}, 763(0):59 -- 79, 2005.

\bibitem{hebeler_11}
K.~Hebeler, S.~K. Bogner, R.~J. Furnstahl, A.~Nogga, and A.~Schwenk.
\newblock Improved nuclear matter calculations from chiral low-momentum
  interactions.
\newblock {\em Phys. Rev. C}, 83:031301, Mar 2011.

\bibitem{hebeler_10}
K.~Hebeler and A.~Schwenk.
\newblock Chiral three-nucleon forces and neutron matter.
\newblock {\em Phys. Rev. C}, 82:014314, Jul 2010.

\bibitem{kievsky_10}
A.~Kievsky, M.~Viviani, L.~Girlanda, and L.~E. Marcucci.
\newblock Comparative study of three-nucleon force models in $a=3,4$ systems.
\newblock {\em Phys. Rev. C}, 81:044003, Apr 2010.

\bibitem{lovato_12}
Alessandro Lovato, Omar Benhar, Stefano Fantoni, and Kevin~E. Schmidt.
\newblock Comparative study of three-nucleon potentials in nuclear matter.
\newblock {\em Phys. Rev. C}, 85:024003, Feb 2012.

\bibitem{cowell_04}
S.~Cowell and V.~R. Pandharipande.
\newblock Neutrino mean free paths in cold symmetric nuclear matter.
\newblock {\em Phys. Rev. C}, 70:035801, Sep 2004.

\bibitem{fabrocini_97}
Adelchi Fabrocini.
\newblock Inclusive transverse response of nuclear matter.
\newblock {\em Phys. Rev. C}, 55:338--348, Jan 1997.

\bibitem{vries_87}
H.~De Vries, C.W.~De Jager, and C.~De Vries.
\newblock Nuclear charge-density-distribution parameters from elastic electron
  scattering.
\newblock {\em Atomic Data and Nuclear Data Tables}, 36(3):495 -- 536, 1987.

\bibitem{farina_09b}
Nicola Farina.
\newblock {\em Weak Response of Nuclear Matter}.
\newblock PhD thesis, Sapienza Universit\`a di Roma, 2009.

\bibitem{krane_87}
K.S. Krane and D.~Halliday.
\newblock {\em Introductory nuclear physics}.
\newblock Wiley, 1987.

\bibitem{fetter_03}
A.L. Fetter and J.D. Walecka.
\newblock {\em Quantum Theory of Many-Particle Systems}.
\newblock Dover Books on Physics. Dover Publications, 2003.

\bibitem{mackie_77}
Frederick~D. Mackie and Gordon Baym.
\newblock Compressible liquid drop nuclear model and mass formula.
\newblock {\em Nuclear Physics A}, 285(2):332 -- 348, 1977.

\bibitem{blaizot_76}
J.P. Blaizot, D.~Gogny, and B.~Grammaticos.
\newblock Nuclear compressibility and monopole resonances.
\newblock {\em Nuclear Physics A}, 265(2):315 -- 336, 1976.

\bibitem{colo_04}
G.~Col\`o, N.~Van~Giai, J.~Meyer, K.~Bennaceur, and P.~Bonche.
\newblock Microscopic determination of the nuclear incompressibility within the
  nonrelativistic framework.
\newblock {\em Phys. Rev. C}, 70:024307, Aug 2004.

\bibitem{cavedon_87}
J.~M. Cavedon, B.~Frois, D.~Goutte, M.~Huet, Ph. Leconte, X.~H. Phan, S.~K.
  Platchkov, C.~N. Papanicolas, S.~E. Williamson, W.~Boeglin, I.~Sick, and
  J.~Heisenberg.
\newblock Measurement of charge-density differences in the interior of pb
  isotopes.
\newblock {\em Phys. Rev. Lett.}, 58:195--198, Jan 1987.

\bibitem{yamazaki_12}
Takeshi Yamazaki, Ken-ichi Ishikawa, Yoshinobu Kuramashi, and Akira Ukawa.
\newblock {Helium nuclei, deuteron and dineutron in 2+1 flavor lattice QCD}.
\newblock 2012.

\bibitem{beane_12}
S.~R. Beane, E.~Chang, W.~Detmold, H.~W. Lin, T.~C. Luu, K.~Orginos,
  A.~Parre\~no, M.~J. Savage, A.~Torok, and A.~Walker-Loud.
\newblock Deuteron and exotic two-body bound states from lattice qcd.
\newblock {\em Phys. Rev. D}, 85:054511, Mar 2012.

\bibitem{aoki_12}
Sinya Aoki et~al.
\newblock {Lattice QCD approach to Nuclear Physics}.
\newblock 2012.

\bibitem{savage_11}
M.J. Savage.
\newblock {Nuclear Physics from Lattice QCD}.
\newblock {\em Prog.Part.Nucl.Phys.}, 67:140--152, 2012.

\bibitem{yukawa_35}
Yukawa H.
\newblock On the interaction of elementary particles.
\newblock {\em Proc. Phys. Math. Soc. Jap.}, 17:48, 1935.

\bibitem{machleidt_11}
R.~Machleidt and D.R. Entem.
\newblock Chiral effective field theory and nuclear forces.
\newblock {\em Physics Reports}, 503(1):1 -- 75, 2011.

\bibitem{weinberg_90}
Steven Weinberg.
\newblock Nuclear forces from chiral lagrangians.
\newblock {\em Physics Letters B}, 251(2):288 -- 292, 1990.

\bibitem{weinberg_91}
Steven Weinberg.
\newblock Effective chiral lagrangians for nucleon-pion interactions and
  nuclear forces.
\newblock {\em Nuclear Physics B}, 363(1):3 -- 18, 1991.

\bibitem{wiringa_02}
R.~B. Wiringa and Steven~C. Pieper.
\newblock Evolution of nuclear spectra with nuclear forces.
\newblock {\em Phys. Rev. Lett.}, 89:182501, Oct 2002.

\bibitem{baldo_12}
M.~Baldo, A.~Polls, A.~Rios, H.-J. Schulze, and I.~Vida{\~{n}}a.
\newblock {Comparative study of neutron and nuclear matter with simplified
  Argonne nucleon-nucleon potentials}.
\newblock 2012.

\bibitem{fujita_57}
Jun ichi Fujita and Hironari Miyazawa.
\newblock Pion theory of three-body forces.
\newblock {\em Progress of Theoretical Physics}, 17(3):360--365, 1957.

\bibitem{bernard_11}
V.~Bernard, E.~Epelbaum, H.~Krebs, and Ulf-G. Mei\ss{}ner.
\newblock Subleading contributions to the chiral three-nucleon force. ii.
  short-range terms and relativistic corrections.
\newblock {\em Phys. Rev. C}, 84:054001, Nov 2011.

\bibitem{krebs_12}
H.~Krebs, A.~Gasparyan, and E.~Epelbaum.
\newblock Chiral three-nucleon force at n${}^{4}$lo: Longest-range
  contributions.
\newblock {\em Phys. Rev. C}, 85:054006, May 2012.

\bibitem{fettes_98}
Nadia Fettes, Ulf-G. Mei\ss{}ner, and Sven Steininger.
\newblock Pion-nucleon scattering in chiral perturbation theory (i):
  Isospin-symmetric case.
\newblock {\em Nuclear Physics A}, 640(2):199 -- 234, 1998.

\bibitem{buttiker_00}
Paul B\"uttiker and Ulf-G. Mei\ss{}nerr.
\newblock Pion-nucleon scattering inside the mandelstam triangle.
\newblock {\em Nuclear Physics A}, 668(1-4):97 -- 112, 2000.

\bibitem{friar_99}
J.~L. Friar, D.~H\"uber, and U.~van Kolck.
\newblock Chiral symmetry and three-nucleon forces.
\newblock {\em Phys. Rev. C}, 59:53--58, Jan 1999.

\bibitem{coon_01}
S.~A. Coon and H.~K. Han.
\newblock Reworking the tucson-melbourne three-nucleon potential.
\newblock {\em Few-Body Systems}, 30:131--141, 2001.
\newblock 10.1007/s006010170022.

\bibitem{hartree_28}
D.~R. Hartree.
\newblock The wave mechanics of an atom with a non-coulomb central field. part
  i. theory and methods.
\newblock {\em Mathematical Proceedings of the Cambridge Philosophical
  Society}, 24(01):89--110, 1928.

\bibitem{fock_30}
V.~Fock.
\newblock N\"aherungsmethode zur l\"osung des quantenmechanischen
  mehrk\"orperproblems.
\newblock {\em Z. Physik}, 61:126--148, Feb 1930.

\bibitem{slater_30}
J.~C. Slater.
\newblock Note on hartree's method.
\newblock {\em Phys. Rev.}, 35:210--211, Jan 1930.

\bibitem{lipparini_08}
E.~Lipparini.
\newblock {\em Modern Many-Particle Physics: Atomic Gases, Nanostructures and
  Quantum Liquids}.
\newblock World Scientific, 2008.

\bibitem{mayer_49}
Maria~Goeppert Mayer.
\newblock On closed shells in nuclei. ii.
\newblock {\em Phys. Rev.}, 75:1969--1970, Jun 1949.

\bibitem{mayer_50}
Maria~Goeppert Mayer.
\newblock Nuclear configurations in the spin-orbit coupling model. i. empirical
  evidence.
\newblock {\em Phys. Rev.}, 78:16--21, Apr 1950.

\bibitem{haxel_49}
Otto Haxel, J.~Hans~D. Jensen, and Hans~E. Suess.
\newblock On the "magic numbers" in nuclear structure.
\newblock {\em Phys. Rev.}, 75:1766--1766, Jun 1949.

\bibitem{jastrow_55}
Robert Jastrow.
\newblock Many-body problem with strong forces.
\newblock {\em Phys.Rev.}, 98:1479--1484, 1955.

\bibitem{brueckner_54}
K.~A. Brueckner, C.~A. Levinson, and H.~M. Mahmoud.
\newblock Two-body forces and nuclear saturation. i. central forces.
\newblock {\em Phys. Rev.}, 95:217--228, Jul 1954.

\bibitem{brueckner_54b}
K.~A. Brueckner.
\newblock Nuclear saturation and two-body forces. ii. tensor forces.
\newblock {\em Phys. Rev.}, 96:508--516, Oct 1954.

\bibitem{brueckner_55}
K.~A. Brueckner and C.~A. Levinson.
\newblock Approximate reduction of the many-body problem for strongly
  interacting particles to a problem of self-consistent fields.
\newblock {\em Phys. Rev.}, 97:1344--1352, Mar 1955.

\bibitem{brueckner_55b}
K.~A. Brueckner.
\newblock Many-body problem for strongly interacting particles. ii. linked
  cluster expansion.
\newblock {\em Phys. Rev.}, 100:36--45, Oct 1955.

\bibitem{bethe_56}
H.~A. Bethe.
\newblock Nuclear many-body problem.
\newblock {\em Phys. Rev.}, 103:1353--1390, Sep 1956.

\bibitem{goldstone_57}
J.~{Goldstone}.
\newblock {Derivation of the Brueckner Many-Body Theory}.
\newblock {\em Royal Society of London Proceedings Series A}, 239:267--279,
  February 1957.

\bibitem{muther_00}
H.~M\"uther and A.~Polls.
\newblock Two-body correlations in nuclear systems.
\newblock {\em Progress in Particle and Nuclear Physics}, 45(1):243 -- 334,
  2000.

\bibitem{baldo_07}
M~Baldo and C~Maieron.
\newblock Equation of state of nuclear matter at high baryon density.
\newblock {\em Journal of Physics G: Nuclear and Particle Physics}, 34(5):R243,
  2007.

\bibitem{clark_79}
John~W. Clark.
\newblock Variational theory of nuclear matter.
\newblock {\em Progress in Particle and Nuclear Physics}, 2(0):89 -- 199, 1979.

\bibitem{feenberg_69}
E.~Feenberg.
\newblock {\em Theory of quantum fluids}.
\newblock Pure and applied physics. Academic Press, 1969.

\bibitem{croxton_78}
C.A. Croxton.
\newblock {\em Progress in liquid physics}.
\newblock Wiley-Interscience Publication. Wiley, 1978.

\bibitem{kalos_81}
M.~H. Kalos, Michael~A. Lee, P.~A. Whitlock, and G.~V. Chester.
\newblock Modern potentials and the properties of condensed $^{4}\mathrm{He}$.
\newblock {\em Phys. Rev. B}, 24:115--130, Jul 1981.

\bibitem{usmani_82}
Q.~N. Usmani, S.~Fantoni, and V.~R. Pandharipande.
\newblock Three-body correlations in liquid $^{4}\mathrm{He}$.
\newblock {\em Phys. Rev. B}, 26:6123--6130, Dec 1982.

\bibitem{moroni_95}
Saverio Moroni, Stefano Fantoni, and Gaetano Senatore.
\newblock Euler {M}onte {C}arlo calculations for liquid $^{4}\mathrm{He}$ and
  $^{3}\mathrm{He}$.
\newblock {\em Phys. Rev. B}, 52:13547--13558, Nov 1995.

\bibitem{feynman_56}
R.~P. Feynman and Michael Cohen.
\newblock Energy spectrum of the excitations in liquid helium.
\newblock {\em Phys. Rev.}, 102:1189--1204, Jun 1956.

\bibitem{benhar_93}
O.~Benhar, V.~R. Pandharipande, and Steven~C. Pieper.
\newblock Electron-scattering studies of correlations in nuclei.
\newblock {\em Rev. Mod. Phys.}, 65:817--828, Jul 1993.

\bibitem{pandha_97}
Vijay~R. Pandharipande, Ingo Sick, and Peter K. A.~deWitt Huberts.
\newblock Independent particle motion and correlations in fermion systems.
\newblock {\em Rev. Mod. Phys.}, 69:981--991, Jul 1997.

\bibitem{clark_66}
John~W. Clark and Paul Westhaus.
\newblock Method of correlated basis functions.
\newblock {\em Phys. Rev.}, 141:833--857, Jan 1966.

\bibitem{fantoni_98}
Stefano Fantoni and Adelchi Fabrocini.
\newblock Correlated basis function theory for fermion systems.
\newblock In Jes\`us Navarro and Artur Polls, editors, {\em Microscopic Quantum
  Many-Body Theories and Their Applications}, volume 510 of {\em Lecture Notes
  in Physics}, pages 119--186. Springer Berlin / Heidelberg, 1998.
\newblock 10.1007/BFb0104526.

\bibitem{kulas_73}
Russell~H. Kullas and William~J. Mullin.
\newblock Theory of dilute solutions of $^3\,${H}e in superfluid $^4\,${H}e. i.
  perturbation theory in a nonorthogonal basis.
\newblock {\em Journal of Low Temperature Physics}, 11:301--319, 1973.
\newblock 10.1007/BF00656555.

\bibitem{morse_53}
P.M.C. Morse and H.~Feshbach.
\newblock {\em Methods of theoretical physics}.
\newblock Number pt. 2 in International series in pure and applied physics.
  McGraw-Hill, 1953.

\bibitem{clark_59}
John~W. Clark and Eugene Feenberg.
\newblock Simplified treatment for strong short-range repulsions in
  $n$-particle systems. i. general theory.
\newblock {\em Phys. Rev.}, 113:388--399, Jan 1959.

\bibitem{fantoni_84}
S.~Fantoni.
\newblock Linked-cluster perturbative expansion in correlated-basis theory.
\newblock {\em Phys. Rev. B}, 29:2544--2550, Mar 1984.

\bibitem{clark_75}
J.W. Clark, P.M. Lam, and W.J.~Ter Louw.
\newblock Perturbation corrections to the jastrow energy for simple models of
  nuclear matter.
\newblock {\em Nuclear Physics A}, 255(1):1 -- 12, 1975.

\bibitem{jackson_82}
A.D. Jackson, E.~Krotscheck, D.E. Meltzer, and R.A. Smith.
\newblock Landau parameters and pairing-on the shores of the nuclear fermi sea.
\newblock {\em Nuclear Physics A}, 386(1):125 -- 165, 1982.

\bibitem{friman_82}
B.~L. Friman and E.~Krotscheck.
\newblock Zero sound, spin fluctuations, and effective mass in liquid
  $^{3}\mathrm{He}$.
\newblock {\em Phys. Rev. Lett.}, 49:1705--1708, Dec 1982.

\bibitem{pandharipande_71}
V.R. Pandharipande.
\newblock Dense neutron matter with realistic interactions.
\newblock {\em Nuclear Physics A}, 174(3):641 -- 656, 1971.

\bibitem{ristig_71}
M.~L. Ristig, W.~J.~Ter Louw, and J.~W. Clark.
\newblock Tensor correlations in nuclear matter.
\newblock {\em Phys. Rev. C}, 3:1504--1513, Apr 1971.

\bibitem{pandharipande_72}
V.R. Pandharipande.
\newblock Variational calculation of nuclear matter.
\newblock {\em Nuclear Physics A}, 181(1):33 -- 48, 1972.

\bibitem{lagaris_80}
I.E. Lagaris and V.R. Pandharipande.
\newblock Variational calculations of ?8 models of nuclear matter.
\newblock {\em Nuclear Physics A}, 334(2):217 -- 228, 1980.

\bibitem{lowdin_50}
Per-Olov L\"{o}wdin.
\newblock On the non-orthogonality problem connected with the use of atomic
  wave functions in the theory of molecules and crystals.
\newblock {\em The Journal of Chemical Physics}, 18(3):365--375, 1950.

\bibitem{fantoni_88}
S.~Fantoni and V.~R. Pandharipande.
\newblock Orthogonalization of correlated states.
\newblock {\em Phys. Rev. C}, 37:1697--1707, Apr 1988.

\bibitem{fantoni_74}
S.~{Fantoni} and S.~{Rosati}.
\newblock {Jastrow correlations and an irreducible cluster expansion for
  infinite boson or fermion systems}.
\newblock {\em Nuovo Cimento A Serie}, 20:179--193, March 1974.

\bibitem{fantoni_75}
S.~{Fantoni} and S.~{Rosati}.
\newblock {The hypernetted-chain approximation for a fermion system}.
\newblock {\em Nuovo Cimento A Serie}, 25:593--615, February 1975.

\bibitem{krotscheck_75}
E.~Krotscheck and M.L. Ristig.
\newblock Long-range jastrow correlations.
\newblock {\em Nuclear Physics A}, 242(3):389 -- 405, 1975.

\bibitem{pandha_76}
V.R. Pandharipande and R.B. Wiringa.
\newblock A variational theory of nuclear matter.
\newblock {\em Nuclear Physics A}, 266(2):269 -- 316, 1976.

\bibitem{pandha_79}
V.~R. Pandharipande and R.~B. Wiringa.
\newblock Variations on a theme of nuclear matter.
\newblock {\em Rev. Mod. Phys.}, 51:821--861, Oct 1979.

\bibitem{mayer_40}
Mayer~Joseph Edward and Mayer~Maria Goeppert.
\newblock {\em Statistical mechanics}.
\newblock J. Wiley \& Sons; Chapman \& Hall, New York; London, 1940.

\bibitem{arias_07}
F.~Arias de~Saavedra, C.~Bisconti, G.~CoÕ, and A.~Fabrocini.
\newblock Renormalized fermi hypernetted chain approach in medium-heavy nuclei.
\newblock {\em Physics Reports}, 450(1-3):1 -- 95, 2007.

\bibitem{morales_02}
J.~Morales, V.~R. Pandharipande, and D.~G. Ravenhall.
\newblock Improved variational calculations of nucleon matter.
\newblock {\em Phys. Rev. C}, 66:054308, Nov 2002.

\bibitem{fantoni_74b}
S.~Fantoni and S.~Rosati.
\newblock Calculation of the two-body correlation function for fermion systems.
\newblock {\em Lettere Al Nuovo Cimento (1971-1985)}, 10:545--551, 1974.
\newblock 10.1007/BF02784779.

\bibitem{krotscheck_74}
E.~Krotscheck and M.L. Ristig.
\newblock Hypernetted-chain approximation for dense fermi fluids.
\newblock {\em Physics Letters A}, 48(1):17 -- 18, 1974.

\bibitem{leeuwen_59}
J.M.J. van Leeuwen, J.~Groeneveld, and J.~de~Boer.
\newblock New method for the calculation of the pair correlation function. i.
\newblock {\em Physica}, 25(7-12):792 -- 808, 1959.

\bibitem{fantoni_01}
S.~Fantoni and K.E. Schmidt.
\newblock Fermi hypernetted chain calculations in a periodic box.
\newblock {\em Nuclear Physics A}, 690(4):456 -- 470, 2001.

\bibitem{manousakis_83}
E.~Manousakis, S.~Fantoni, V.~R. Pandharipande, and Q.~N. Usmani.
\newblock Microscopic calculations for normal and polarized liquid
  $^{3}\mathrm{He}$.
\newblock {\em Phys. Rev. B}, 28:3770--3781, Oct 1983.

\bibitem{wiringa_04}
B.~Wiringa.
\newblock Elementary diagrams in nuclear matter.
\newblock private communication, 2007.

\bibitem{pandha_73}
V.~R. Pandharipande and H.~A. Bethe.
\newblock Variational method for dense systems.
\newblock {\em Phys. Rev. C}, 7:1312--1328, Apr 1973.

\bibitem{iwamoto_57}
Fumiaki Iwamoto and Masami Yamada.
\newblock Cluster development method in the quantum mechanics of many particle
  system, i.
\newblock {\em Progress of Theoretical Physics}, 17(4):543--555, 1957.

\bibitem{wiringa_79}
R.B. Wiringa and V.R. Pandharipande.
\newblock A variational theory of nuclear matter (iii).
\newblock {\em Nuclear Physics A}, 317(1):1 -- 22, 1979.

\bibitem{kirkpatrick_83}
S.~Kirkpatrick, C.D. Gelatt, and M.P. Vecchi.
\newblock {Optimization by Simulated Annealing}.
\newblock {\em Science}, 220:671--680, 1983.

\bibitem{kalos_84}
M.H. Kalos, North Atlantic~Treaty Organization, Centre europ{\'e}en de calcul
  atomique~et mol{\'e}culaire, and North Atlantic Treaty Organization.
  Scientific~Affairs Division.
\newblock {\em {M}onte {C}arlo Methods in Quantum Problems}.
\newblock NATO ASI Series: Mathematical and Physical Sciences. D. Reidel
  Publishing Company, 1984.

\bibitem{kalos_09}
Malvin~H. Kalos and Paula~A. Whitlock.
\newblock {\em Front Matter}, pages I--XII.
\newblock Wiley-VCH Verlag GmbH \& Co. KGaA, 2009.

\bibitem{carlson_03}
J.~Carlson, J.~Morales, V.~R. Pandharipande, and D.~G. Ravenhall.
\newblock Quantum {M}onte {C}arlo calculations of neutron matter.
\newblock {\em Phys. Rev. C}, 68:025802, Aug 2003.

\bibitem{grimm_71}
R.C Grimm and R.G Storer.
\newblock {M}onte-{C}arlo solution of schrodinger's equation.
\newblock {\em Journal of Computational Physics}, 7(1):134 -- 156, 1971.

\bibitem{foulkes_01}
W.~M.~C. Foulkes, L.~Mitas, R.~J. Needs, and G.~Rajagopal.
\newblock Quantum {M}onte {C}arlo simulations of solids.
\newblock {\em Rev. Mod. Phys.}, 73:33--83, Jan 2001.

\bibitem{karlin_81}
S.~Karlin and H.M. Taylor.
\newblock {\em A Second Course in Stochastic Processes}.
\newblock Number v. 2. Academic Press, 1981.

\bibitem{ormoni_11}
Paolo Armani.
\newblock {\em Progress of {M}onte {C}arlo methods in nuclear physics using
  EFT-based NN interaction and in hypernuclear systems}.
\newblock PhD thesis, University of Trento, 2011.

\bibitem{ceperley_79}
D.~M. Ceperley and M.~H. Kalos.
\newblock {\em {M}onte {C}arlo Methods in Statistical Physics}.
\newblock Springer, Berlin, second edition, 1979.

\bibitem{anderson_76}
James~B. Anderson.
\newblock Quantum chemistry by random walk. h [sup 2]p, h[sup + ][sub 3] d[sub
  3h] [sup 1]a[prime][sub 1], h[sub 2] [sup 3] sigma [sup + ][sub u], h[sub 4]
  [sup 1] sigma [sup + ][sub g], be [sup 1]s.
\newblock {\em The Journal of Chemical Physics}, 65(10):4121--4127, 1976.

\bibitem{reynolds_82}
Peter~J. Reynolds, David~M. Ceperley, Berni~J. Alder, and William~A. Lester.
\newblock Fixed-node quantum {M}onte {C}arlo for molecules[sup a) b)].
\newblock {\em The Journal of Chemical Physics}, 77(11):5593--5603, 1982.

\bibitem{foulkes_99}
W.~M.~C. Foulkes, Randolph~Q. Hood, and R.~J. Needs.
\newblock Symmetry constraints and variational principles in diffusion quantum
  {M}onte {C}arlo calculations of excited-state energies.
\newblock {\em Phys. Rev. B}, 60:4558--4570, Aug 1999.

\bibitem{zhang_95}
Shiwei Zhang, J.~Carlson, and J.~E. Gubernatis.
\newblock Constrained path quantum {M}onte {C}arlo method for fermion ground
  states.
\newblock {\em Phys. Rev. Lett.}, 74:3652--3655, May 1995.

\bibitem{schmidt_99}
K.E. Schmidt and S.~Fantoni.
\newblock A quantum {M}onte {C}arlo method for nucleon systems.
\newblock {\em Physics Letters B}, 446(2):99 -- 103, 1999.

\bibitem{ortiz_93}
G.~Ortiz, D.~M. Ceperley, and R.~M. Martin.
\newblock New stochastic method for systems with broken time-reversal symmetry:
  2d fermions in a magnetic field.
\newblock {\em Phys. Rev. Lett.}, 71:2777--2780, Oct 1993.

\bibitem{carlson_87}
J.~Carlson.
\newblock Green's function {M}onte {C}arlo study of light nuclei.
\newblock {\em Phys. Rev. C}, 36:2026--2033, Nov 1987.

\bibitem{carlson_99}
J.~Carlson, J.~E. Gubernatis, G.~Ortiz, and Shiwei Zhang.
\newblock Issues and observations on applications of the constrained-path
  {M}onte {C}arlo method to many-fermion systems.
\newblock {\em Phys. Rev. B}, 59:12788--12798, May 1999.

\bibitem{pieper_98}
Steven Pieper.
\newblock {M}onte carlo calculations of nuclei.
\newblock In Jes\`us Navarro and Artur Polls, editors, {\em Microscopic Quantum
  Many-Body Theories and Their Applications}, volume 510 of {\em Lecture Notes
  in Physics}, pages 337--357. Springer Berlin / Heidelberg, 1998.
\newblock 10.1007/BFb0104530.

\bibitem{gandolfi_07}
Stefano Gandolfi.
\newblock {\em The Auxiliary Field Diffusion {M}onte {C}arlo Method for Nuclear
  Physics and Nuclear Astrophysics}.
\newblock PhD thesis, University of Trento, 2007.

\bibitem{gandolfi_11}
S.~Gandolfi, J.~Carlson, and Steven~C. Pieper.
\newblock Cold neutrons trapped in external fields.
\newblock {\em Phys. Rev. Lett.}, 106:012501, Jan 2011.

\bibitem{gezerlis_10}
Alexandros Gezerlis and J.~Carlson.
\newblock Low-density neutron matter.
\newblock {\em Phys. Rev. C}, 81:025803, Feb 2010.

\bibitem{gandolfi_09}
S.~Gandolfi, A.~Yu. Illarionov, K.~E. Schmidt, F.~Pederiva, and S.~Fantoni.
\newblock Quantum {M}onte {C}arlo calculation of the equation of state of
  neutron matter.
\newblock {\em Phys. Rev. C}, 79:054005, May 2009.

\bibitem{fantoni_08}
Stefano Fantoni, Stefano Gandolfi, Alexey~Yu. Illarionov, Kevin~E. Schmidt, and
  Francesco Pederiva.
\newblock {{M}onte {C}arlo approach to nuclei and nuclear matter}.
\newblock {\em AIP Conf.Proc.}, 1056:233--240, 2008.

\bibitem{pederiva_04}
F.~Pederiva, A.~Sarsa, K.E. Schmidt, and S.~Fantoni.
\newblock Auxiliary field diffusion {M}onte {C}arlo calculation of ground state
  properties of neutron drops.
\newblock {\em Nuclear Physics A}, 742(1-2):255 -- 268, 2004.

\bibitem{girlanda_11}
L.~Girlanda, A.~Kievsky, and M.~Viviani.
\newblock Subleading contributions to the three-nucleon contact interaction.
\newblock {\em Phys. Rev. C}, 84:014001, Jul 2011.

\bibitem{benhar_09}
Omar Benhar and Nicola Farina.
\newblock Correlation effects on the weak response of nuclear matter.
\newblock {\em Physics Letters B}, 680(4):305 -- 309, 2009.

\bibitem{cowell_03}
S.~Cowell and V.~R. Pandharipande.
\newblock Quenching of weak interactions in nucleon matter.
\newblock {\em Phys. Rev. C}, 67:035504, Mar 2003.

\bibitem{lovato_12b}
Benhar~O. Lovato~A., Losa~C.
\newblock Weak response of cold symmetric nuclear matter at three-body cluster
  level.
\newblock In preparation, 2012.

\bibitem{akmal_97}
A.~Akmal and V.~R. Pandharipande.
\newblock Spin-isospin structure and pion condensation in nucleon matter.
\newblock {\em Phys. Rev. C}, 56:2261--2279, Oct 1997.

\bibitem{shen_12}
G.~Shen, S.~Gandolfi, S.~Reddy, and J.~Carlson.
\newblock {Spin Response and Neutrino Emissivity of Dense Neutron Matter}.
\newblock 2012.

\bibitem{weise_10}
J.~W. Holt, N.~Kaiser, and W.~Weise.
\newblock Density-dependent effective nucleon-nucleon interaction from chiral
  three-nucleon forces.
\newblock {\em Phys. Rev. C}, 81:024002, Feb 2010.

\bibitem{bransden_03}
B.H. Bransden and C.J. Joachain.
\newblock {\em Physics of Atoms and Molecules}.
\newblock Pearson Education. Prentice Hall, 2003.

\end{thebibliography}

%Ringraziamenti

\clearpage
\chapter*{Ringraziamenti}
\pagestyle{empty}
\addcontentsline{toc}{chapter}{Ringraziamenti}
Probabilmente le persone che leggeranno i Capitoli centrali della Tesi si conteranno sulle dita di una mano (includendo l'autore, i relatori e il controrelatore); l'Introduzione e le Conclusioni saranno forse fruite da un pubblico appena appena pi\`u vasto. \`E nei Ringraziamenti che l'autore deve dare il meglio di s\'e. Siccome nel mio caso ``meglio'' e ``inglese'' non \`e che vadano molto d'accordo, ho deciso di scriverli in l'italiano, o qualcosa che lo ricordi. 

A proposito di lingua madre, per primi ringrazio i miei genitori che, mentre scrivevo di FHNC/SOC, hanno festeggiato trent' anni di matrimonio. Pur denotando una certa mancanza di fantasia, riporto allora quello che ho scritto sul biglietto d'auguri ``se avessi potuto scegliervi, l'avrei fatto''. Stavolta, per eleganza, estendo la frase a mia sorella. Confesso che, con una prospettiva di precariato dinnanzi, inimicarsi l'unica in famiglia destinata a far quattrini non sarebbe scelta saggia.

Procedendo a ritroso nella genealogia, non posso non essere grato a mia nonna, che nei primi anni e' stata una seconda mamma, pi\`u calorica nella cucina e munifica nelle mancette. Tale abitudine fortunatamente non e' venuta meno con l'avanzare dell'et\`a. A novant'anni suonati, ogni volta che le sento dire <<Ti\`e, con questi te ce compri 'n caff\'e>>, non riesco a non immaginarmi proprietario di un bar del centro.

Sempre della stessa generazione o quasi, vorrei ringraziare Ida e le sue corroboranti polpette che mi hanno sostenuto durante le fasi conclusive della stesura. Ida va ringraziata anche, e soprattutto, per un altro motivo, che tengo per me. 

Voglio qui citare il gruppo storico degli amici di Velletri, in particolare Alessandro, Davide, Roberto, Alessio, Sara, Daniele e Gigi. Ho la certezza che essi siano, e saranno, uno dei punti fermi della mia esistenza, nella loro immutevole mutevolezza, sebbene credano che io avrei dovuto soccombere sotto i colpi di machete dell'ex ministro B.ta (per coerenza ne accorcio il nome). A Diego e ad Annamaria desidero far sapere che hanno fatto decisamente innalzare la mia stima verso l'essere umano: il vostro comportamento \`e per me un insegnamento che non dimenticher\`o. Tuttavia, gi\`a che son qui, ne approfitto per pregare Diego di non investirmi col suo ruggente QUAD.

Spostandomi di 700 Km a Nord-Est, ringrazio gli amici del tennis, tra cui Giacomo, Andrea, Piero e Gabriele: sono stato accolto come uno di loro. Gli allenamenti, le partite e la palestra in loro compagnia sono stati  fondamentali, soprattutto quando ero nero per dei conti che non tornavano, o per delle soluzioni che non si trovavano.

Nella citt\`a adottiva ho avuto il privilegio di conoscere Daniel, un amico sincero, con cui ho condiviso momenti ``epici ed eleganti'' durante questi quattro anni. Se mai sar\`o nonno, egli sar\`a presente nei racconti che far\`o ai miei nipoti per via dei sei giorni in bici da Trieste a Roma, i quattordici da Milano a Santiago di Compostela (con annessi i cinque giorni aggiuntivi nella sua nat\`ia Galizia, terra di Breog\'an), e delle varie gite verso Tarvisio, lo Zoncolan, Venezia e passo Pramollo. Lo ammiro non tanto per essere un ciclista decisamente migliore di me, come dimostrano i minuti di ritardo che accuso appena la strada inizia a salire, ma per le sue qualit\`a di uomo, veramente di rara eleganza (esula da questo discorso la scelta delle scarpe); Epic and Glory!

Per la sua laconica e saggia ironia, che d\`a giusta misura a difficolt\`a a prima vista enormi, ma soprattutto perch\'e ``altrimenti si arrabbia'', sono riconoscente ad Andrea, altres\`i noto come Jens.  Come non nominare anche l'altro compagno di cadute, Castillo alias Edu, non solo perch\'e mi abbia citato nei ringraziamenti della sua Tesi, ma perch\'e d\`a sempre il massimo, anche quando non strettamente necessario (e per questo detiene il record di scalata a piedi dello Zoncolan).

Dalla Galizia alle Asturie, per aver costantemente allenato la mia eloquenza, perch\'e <<La cultura o ce l'hai o non ce l'hai>>, trova giustamente posto in questi ringraziamenti Irene. Non \`e da tutti perseverare nell'errore di frequentarci e di fingere di ridere alle nostre battute, cara Isabelita. A proposito ``{\it Si Evita estuviera viva, Isabel seria soltera... ''} sono certamente in debito con Ignazio per questa canzone e per il suo umorismo sorprendente. 

Bene, mi sto dilungando forse troppo, ma dopo 190 e passa pagine, aggiungerne due o tre non \`e che cambi poi molto la situazione. Certo, avrei potuto cavarmela con un `` mi trovo nella fortunata situazione di avere tante persone da ringraziare e poco spazio per farlo \cite{akmal_98} ''; tuttavia non posso prendere spunto dalla Tesi di Akmal pure per i ringraziamenti! Vado avanti.

Nicola, per l'impegno profuso come rappresentante, per le discussioni alla lavagna e per l'ottimo manzo all'olio. Giorgio e Maurizio, per aver condiviso i primi tempi sotto lo stesso tetto, per la calma e per la stimolante anfetaminicit\`a, rispettivamente. Matteo, Kansas city \`e di per se un motivo bastante per essere qui presente. Poi Ambra, Daniele, Alberto, Aurora (anche per il polpo tramite Michele!), Umberto... 

Ancora Daniele, un altro per\`o, precisamente il mio attuale coinquilino, nonch\'e compagno di giuochi universitari. Grazie per le cene, per i dialoghi notturni, per non aver mai fatto caso alle spigolosit\`a del mio carattere. Se verrai nella mia nuova casa una bella pulizia del piano cottura non te la leva nessuno!  Dovrai portare anche Giorgia (che pu\`o venire anche da sola, eh), solerte compagna di colazioni mattutine, nonch\'e abile autrice della foto sul visto, ottenuta con espedienti fotografici d'avanguardia. Non ti invito solamente per migliorare la danielesca qualit\`a  della pulizia del suddetto piano cottura, non ti preoccupare!

Da Trieste a Trento. Colgo l'occasione per rendere edotti gli sparuti lettori del fatto che queste due citt\`a, che nell'immaginario collettivo sono quasi sovrapposte, forse perch\'e italiane entrambe dal '18, sono assai distanti; metaforicamente, come il mare e la montagna. Paolo ``Ormoni'', senza di te l'AFDMC avrebbe molti pi\`u segreti di quanti ancora non ne abbia, cos\`i come il monte Bondone e il Trentino tutto. Ammiro Diego in particolare per  l'ostinazione e l'intuizione; lo ringrazio per le cose fatte e per quelle che far\`a per risolvere alcuni ``problemucci'' del codice, per non parlare della minuziosa rilettura della presente opera. Elia, per le sciate e per aver ricambiato la visita. Enrico, senza il tuo aiuto (e quello di AURORA che non \`e una persona) dovrei ancora rispondere al referee. Giorgia, per essere la barra di grafite nel reattore della saletta. Tra l'altro vorrei sottolineare l'economicit\`a della pensione ``Roberto'' a Pergine: un buono-caff\`e a notte, ``extra'' inclusi (e che ``extra'').

Giusto per essere il Trip Advisor de noantri, consiglio anche la pensione ``Arianna'' a Barcellona. Squisita ospitalit\`a al costo di un piatto di polpette (vanno bene anche sciape). 

Tempo di seriet\`a. 

Ringrazio Stefano per avermi iniziato all'AFDMC, per l'infinita pazienza dimostratami e per avermi procacciato un posto di lavoro. Non cambiando famiglia, sono in debito con Serena e le birbette, per l'ospitalit\`a in quel di Los Alamos e per il supporto morale seguente. Vorrei calcare i vostri passi.

Senza Francesco Pederiva e Kevin Schmidt il Capitolo sull'AFDMC non avrebbe mai visto la luce, tra l'altro perch\'e  l'AFDMC stesso non sarebbe mai nato. Resto sempre stupito dalla chiarezza, dalla competenza e dalla pazienza con cui cercate di far entrare nel mio cervello quanto i vostri abbiano prodotto. 

E' d'uopo ringraziare Stefano Fantoni, le cui idee mi limito a sviluppare; spero che l'operaio possa imparare dall'architetto. Mi stupisco ancora come abbia potuto, a ragione, credere che le equazioni di FHNC, in apparenza molto intricate e soggette a millemila possibili errori, potessero portare a risultati cos\`i accurati. Credere nelle idee a lungo termine \`e un insegnamento prezioso, anche al di fuori della Fisica.

Omar, dopo quella della mia laurea specialistica, ti ripromisi che mai pi\`u avresti corretto una Tesi ad Agosto. Mi sento un poco responsabile di averti fatto rimangiare tal solenne promessa. Ogni volta che discutiamo di Fisica, o di politica (notare l'uso improprio delle maiuscole), c'\`e sempre qualcosa che dovrei sapere, di ``ovvio'' che non so e che tu invece sai. Ti ringrazio per lasciar trasparire quello che pensi al riguardo, pi\`u dal tuo viso che dalla tua voce. Mi sto anche accorgendo che alcune tue espressioni idiomatiche stanno entrando nel mio parlato. Grazie di cuore. 

E veniamo alla veronese pi\`u terrona della storia; braccio quando c'\`e bisogno di una mente e mente quando  c'\`e bisogno di un braccio: Cristina. L'ultimo Capitolo \`e largamente frutto delle discussioni avute a Natale, a Pasqua e a Ferragosto, pi\`u in quasi tutti gli altri giorni feriali e festivi nel mezzo. Magari avrei da dissentire sui tuoi gusti culinari, forse guidati dal figlio che porti in grembo, a cui sono gi\`a affezionato, ma sei veramente insostituibile, non solo come fisica. Ecco, te l'ho detto, mo' accontentati per i prossimi cinque anni.

Dulcis in fundo, Barbara. Qui mi limito ad un sintetico, ma maiuscolo, ``GRAZIE'', riservandomi il diritto di approfondire in privato.

\end{document}